\documentclass[12pt, a4paper, one column,openright,italian]{report}
 \pdfoutput=1 
\usepackage{latexsym,makeidx}
\usepackage[english]{babel}
\usepackage[applemac]{inputenc}
\usepackage{verbatim}
\usepackage{graphicx}
\usepackage{amsbsy}
\usepackage{amsmath}
\usepackage{amssymb}
\usepackage{color}
\usepackage{soul}
\usepackage{float}
\restylefloat{table}
\usepackage{rotating}
\usepackage{footnote}
\usepackage{indentfirst}
\usepackage{hyperref}
\usepackage{caption}
\usepackage{subcaption}
\usepackage{siunitx}
\usepackage{lscape}
\usepackage{multirow}
\usepackage{mathtools}
\usepackage{amsmath}
\usepackage{slashed}
\usepackage{lmodern}
\usepackage[T1]{fontenc}
\usepackage[nowrite,infront,standard]{frontespizio}

\usepackage{xspace}

\usepackage{hyperref}
\hypersetup{
    colorlinks,
    citecolor=black,
    filecolor=black,
    linkcolor=black,
    urlcolor=blue
}

\usepackage[font=small,labelfont=bf]{caption}
\usepackage[left=2.5 cm, right=2.5 cm, bottom=2.5 cm]{geometry}

\usepackage{fancyhdr}
\renewcommand{\chaptermark}[1]{\markboth{\chaptername\ \thechapter.\ #1}{}}
\usepackage{defs}

\usepackage[backend=biber,style=numeric,sorting=none]{biblatex}
\renewbibmacro{in:}{}

\begin{document}

\begin{frontespizio}
\Istituzione{Universit\`a degli Studi Roma Tre}
\Logo [4cm]{figures/logo.png}

\Dipartimento {Matematica e Fisica}
\Corso [Dottorato di Ricerca]{Fisica\\
XXXII Ciclo}

\Titolo{\Huge{Measurement of Higgs-boson self-coupling with single-Higgs and double-Higgs production channels}}
\Sottotitolo{\textit{\Huge{Eleonora Rossi}}}
\NCandidato{\Large{Supervisore}}{}
  \Candidato{\Large{Prof. Biagio Di Micco}}
  \NRelatore{\Large{Coordinatore}}{}
  \Relatore{\Large{Prof. Giuseppe Degrassi}}
\Annoaccademico{2019-2020}
\end{frontespizio}

\pagestyle{fancy}

\thispagestyle{empty}

\pagenumbering{roman}

\thispagestyle{empty}\null\newpage
\clearpage
\cleardoublepage

\cleardoublepage
\thispagestyle{empty}
\vspace*{\stretch{1}}
\begin{flushright}
\itshape
\begin{tabular}{@{}l@{}}
Alla mente pi\`u ardita, curiosa ed intelligente che abbia mai conosciuto.\\ 
Alle braccia pi\`u aperte e calde tra le quali sono stata stretta ed al cuore\\
pi\`u grande del quale ho avuto la grazia ed il privilegio di essere parte.\\
A Nonno. Nihil obest.
\end{tabular}\\[3pt]
\vspace*{\stretch{0.2}}
\begin{tabular}{@{}l@{}}
To the most bold, curious and clever mind I have ever met. \\ 
To the most open and warm arms I was held in and to the\\
biggest heart I had the grace and the privilege of being part of.\\
To Grandpa. Nihil obest.
\end{tabular}\\[3pt]
\end{flushright}
\vspace{\stretch{2}}
\cleardoublepage

\thispagestyle{empty}\null\newpage

\pagenumbering{arabic}

\renewcommand{\chaptermark}[1]{%
\markboth{\MakeUppercase{%
\chaptername}\ \thechapter.%
\ #1}{}}
\newpage
\clearpage
\chapter*{Abstract}

One of the most important targets of the LHC is to improve the experimental results of the Run 1 and the complete exploration of the properties of the Higgs boson, in particular the Higgs-boson self-coupling. The self-coupling is very loosely constrained by electroweak precision measurements therefore new physics effects could induce large deviations from its Standard Model expectation. 
The trilinear self-coupling can be measured directly using the Higgs-boson-pair production cross section, or indirectly through the measurement of single-Higgs-boson production and decay modes. In fact, at next-to-leading order in electroweak interaction, the Higgs-decay partial widths and the cross sections of the main single-Higgs production processes depend on the Higgs-boson self-coupling via weak loops.\newline
Measurements of $\kappa_\lambda$, \ie the rescaling of the trilinear Higgs self-coupling, are presented in this dissertation. Results are obtained exploiting proton-proton collision data from the Large Hadron Collider at a centre-of-mass energy of 13 TeV recorded by the ATLAS detector in 2015, 2016 and 2017, corresponding to a luminosity of up to 79.8 fb$^{-1}$.\newline
Constraints on the Higgs self-coupling are presented considering the most sensitive double-Higgs channels (HH), $b\bar{b}\tau^+\tau^-$, $b\bar{b}\gamma\gamma$ and $b\bar{b}b\bar{b}$, considering single-Higgs (H) production modes, $ggF$, $VBF$, $ZH$, $WH$ and $t\bar{t}H$, together with $WW^*$, $ZZ^*$, $\tau^+\tau^-$, $\gamma \gamma$ and $b\bar{b}$ decay channels, and combining the aforementioned analyses (H+HH) to improve the sensitivity on $\kappa_\lambda$.\newline
Under the assumption that new physics affects only the Higgs-boson self-coupling, the combined H+HH best-fit value of the coupling modifier is: $\kappa_\lambda = 4.6^{+3.2}_{-3.8}$, excluding values outside the interval $-2.3<\kappa_\lambda<10.3$ at 95\% confidence level. \newline
Results with less stringent assumptions are also provided, decoupling the Higgs-boson self-coupling and the other Standard Model couplings.\newline
The final results of this thesis provide the most stringent constraint on $\kappa_\lambda$ from experimental measurements to date.

\newpage

\clearpage
\tableofcontents

\addcontentsline{toc}{chapter}{Introduction}
\markboth{Introduction}{Introduction} 
\chapter*{Introduction}
\label{sec:introduction}
The Standard Model (SM) of particle physics is the theory that, as of today, best describes matter in terms of elementary particles and interactions, and has been validated with an excellent level of accuracy, thus constituting one of the most successful achievements in modern physics. Among the successes of the SM, it has to be underlined that all the particles the SM predicted have been observed, including the $W$ and $Z$ bosons, the top and bottom quarks, and the Higgs boson, the particle responsible of the Higgs mechanism that allows bosons and fermions to acquire mass in the electroweak gauge theory. 
The search for the Higgs boson has lasted for decades. More than 20 years after the formulation of the Higgs mechanism had to pass until a significant mass range could be probed first with the Large Electron Positron collider (LEP) at CERN and then with the Tevatron proton-antiproton collider. In 2010, the Large Hadron Collider (LHC), a proton-proton and heavy-ion collider, started to take data at unprecedented centre-of-mass energies with the primary goal of searching for this boson.\newline
Thus, in July 2012, the announcement of the discovery of a particle compatible with the SM Higgs boson by the ATLAS and CMS experiments at the LHC, represented a great milestone in the history of particle physics.
After the discovery of the Higgs boson, a new era in understanding the nature of electroweak symmetry breaking, possibly completing the SM and constraining effects from new physics (NP), has opened. One of the main targets of particle physics, and of ATLAS and CMS physics analyses at the LHC, is the precision measurement of the properties of the Higgs boson including spin-parity, couplings and evidence for production mechanisms, which are essential tests of the SM. The complete exploration of the properties of the Higgs boson includes the interactions of the Higgs boson with itself, known as the Higgs-boson self-couplings. The self-couplings determine the shape of the potential which is connected to the phase transition of the early universe from the unbroken to the broken electroweak symmetry and are very loosely constrained by electroweak precision measurements, therefore NP effects could induce large deviations from their SM expectation.\newline
The trilinear Higgs self-coupling can be probed directly in searches for multi-Higgs final states and indirectly via its effects on precision observables or loop corrections to single-Higgs processes, while the quartic self-coupling, being further suppressed with respect to the trilinear self-coupling, is currently not accessible at hadron colliders.\newline
The results presented in this dissertation are obtained using proton-proton collision data from the LHC at a centre-of-mass energy of 13 \TeV\ recorded by the ATLAS detector in 2015, 2016 and 2017.\newline
A description of the Standard Model theoretical framework is reported in Chapter~\ref{sec:SM}, ranging from a summary of the fundamental particles and their properties, to the introduction of the Higgs mechanism, a simple mechanism for the breaking of the electroweak symmetry. Furthermore, this chapter reports a detailed description of the Higgs-boson phenomenology and latest measurements, from production and decay modes to properties like the mass, the couplings and the self-coupling of the Higgs boson itself.
Chapter~\ref{sec:LHC} describes the LHC accelerator complex and the basic concepts of proton-proton collisions, together with the experiments housed in the ring and the periods of operation of the accelerator, while Chapter~\ref{sec:ATLAS} presents the ATLAS experiment, giving details on the sub-detectors composing ATLAS and on the interaction of different particles with the detector materials.\newline
A general overview of the reconstruction of physics objects, consists of combining and interpreting information collected from the sub-detectors described in Chapter~\ref{sec:ATLAS}, is provided in Chapter~\ref{sec:Reco}. Basics concepts of the statical model used to extract the results of this dissertation are reported in Chapter~\ref{sec:stat}.\newline
The work presented in this thesis has the target of probing the sector of the SM that is responsible for electroweak symmetry breaking, focusing on the Higgs potential and on the trilinear Higgs self-coupling. The theoretical models on the basis of which the results of this thesis have been produced are summarised in Chapter~\ref{sec:prob_self} for both double- and single-Higgs productions. \newline
The results coming from the extraction of limits on the rescaling of the trilinear Higgs self-coupling, $\kappa_\lambda$, considering the $gg\rightarrow HH$ production process and the most sensitive double-Higgs channels, $b\bar{b}\tau^+\tau^-$, $b\bar{b}\gamma\gamma$ and $b\bar{b}b\bar{b}$, and exploiting the dependence of the double-Higgs cross section and kinematics on both the coupling of the Higgs boson to the top quark and the Higgs self-coupling, are reported in Chapter~\ref{sec:dihiggs}.\newline
Chapter~\ref{sec:single} exploits the complementary approach to constrain the Higgs self-coupling described in Chapter~\ref{sec:prob_self}, applying next-to-leading order electroweak corrections depending on $\kappa_\lambda$ to single-Higgs processes, combining information from \ggF, \VBF, \ZH, \WH and $t\bar{t}H$ production modes together with $WW^*$, $ZZ^*$, $\tau^+\tau^-$, $\gamma \gamma$ and $b\bar{b}$ decay channels; the limits extracted using this approach are probed to be competitive with double-Higgs limits.\newline
The final results of this dissertation, providing the most stringent constraints on $\kappa_\lambda$ from experimental measurements through the combination of the aforementioned double- and single-Higgs analyses, whose details have been described in Chapters~\ref{sec:dihiggs} and~\ref{sec:single}, are reported in Chapter~\ref{sec:combination}. 

\chapter{The Standard Model of Particle Physics}
\label{sec:SM}
A description of the Standard Model (SM) of particle physics is presented in this chapter. Section~\ref{sec:interactions_SM} introduces the SM as a gauge theory that, currently, is the most accurate theory covering the foundations of particle physics and describes three of the four known fundamental forces.
Section~\ref{Higgs_mechanism} presents the Higgs mechanism, \ie\  a simple mechanism for the breaking of the electroweak symmetry as a consequence of the introduction of an additional scalar field in the SM.
An overview of the Higgs-boson phenomenology, focusing on production and decay channels together with the current status of the couplings of the Higgs boson with other SM particles are reported in Sections~\ref{SM_Higgs_boson} and~\ref{Higgs_boson_property}.
Finally Section~\ref{SM_success_open} is devoted to a brief description of the successes of this theory making more room to remaining open questions of this great even though incomplete model.\newline
The results reported in this thesis represent validations of this theory, looking for deviation of the predicted SM values as possible hints of new physics (NP). \newline
Throughout this chapter, natural units have been used, \ie\ the speed of light in vacuum, $c$, and the reduced Planck constant, $\hslash$, have been set to $c =\hslash = 1$ and the unit of energy is the \GeV.

\section{Fundamental Interactions in the Standard Model}
\label{sec:interactions_SM}
The Standard Model~\cite{Glashow, Weinb, Salam, gla_ilio_maiani, englert, P_Higgs1, P_Higgs2} is currently the quantum field theory, \ie\ a theory having quantum fields as fundamental objects, that better describes the matter in terms of elementary particles and interactions, and constitutes one of the most successful achievements in modern physics; only the gravitational interaction is not included in the theory.\newline
According to the SM, matter is composed of 12 fundamental \textit{fermions}, 4 \textit{vector gauge bosons} (spin = 1), and one \textit{scalar Higgs boson} (spin = 0); fermions are half-integer spin particles obeying Fermi-Dirac statistics and satisfying the Pauli exclusion principle while bosons have integer spin and obey Bose-Einstein statistics. The spin is a quantum number, \ie\ a property describing the values of conserved quantities under transformations of quantum systems, that, in the case of the spin, are rotations.\newline
Fermions are classified in \textit{leptons} and \textit{quarks}, depending on the interaction they are subject to:
\begin{itemize}
\item leptons interact through the electromagnetic and weak forces;
\item neutrinos interact only via the weak force;
\item quarks interact through the electromagnetic, weak and strong forces, thus having an additional quantum number with respect to leptons, related to the strong interaction, the colour charge (\textit{red}, \textit{green} and \textit{blue}).
\end{itemize}
The main experimental difference between leptons and quarks is that quarks cannot be observed as isolated particles as they are confined in colour charge singlets with integer charge, namely hadrons, such as protons and neutrons. Quarks and leptons are further divided into three families, or generations, of increasing mass:
\begin{equation}
\binom{e^-}{\nu_e} \quad \binom{\mu^-}{\nu_\mu} \quad \binom{\tau^-}{\nu_\tau} \nonumber \qquad
\binom{u}{d} \quad \binom{c}{s} \quad \binom{t}{b} \, .
\end{equation}
The electron is the only stable charged lepton while both muon and tau are unstable. Three neutrino flavours match the flavour of the corresponding charged lepton, \ie\ electron, muon, and tau, as indicated by the subscript, for example $\nu_\mu$ matches the muon $\mu$. Within the Standard Model, neutrinos are neutral massless leptons, in contrast with the experimental evidence of their oscillation, which requires a mass different from zero.\newline Each fermion has an anti-particle with identical mass and opposite quantum numbers. This statement is not yet verified for neutrinos, as they might be Majorana particles, namely $\nu=\bar{\nu}$. 
\begin{table}[hbtp]
\begin{center}
 \renewcommand{\arraystretch}{1.3}
\begin{tabular}{|c|c|c|}
\hline
Lepton/quark & $Q/e$ & mass [GeV] \\
\hline
 electron ($e$)& -1 & $0.511 \times 10^{-3}$\\
 electron neutrino ($\nu_e$)& 0 & $< 2\times 10^{-9}$  \\
 muon ($\mu$)& -1 & $0.106$ \\
 muon neutrino ($\nu_\mu$)& 0 & $< 0.19\times 10^{-3}$ \\
 tau ($\tau$) & -1 & $1.777$\\
 tau neutrino ($\nu_\tau$)& 0 & $< 18.2 \times 10^{-3}$ \\
 \hline
 up ($u$) & $\frac{2}{3}$ & $2.2^{+0.5}_{-0.3} \times 10^{-3}$ \\
 down ($d$) & -$\frac{1}{3}$  & $4.7^{+0.5}_{-0.2} \times 10^{-3}$ \\
 charm ($c$) & $\frac{2}{3}$ & $1.27\pm 0.02$ \\
 strange ($s$) & -$\frac{1}{3}$& $93^{+11}_{-5} \times 10^{-3}$ \\
 top ($t$) & $\frac{2}{3}$ & $172.9\pm 0.4$\\
 bottom ($b$) & -$\frac{1}{3}$ & $4.18^{+0.03}_{-0.02}$ \\
\hline
\end{tabular}
\end{center}
\caption{Properties of leptons and quarks: the electric charge $Q$, in units of the electron charge $e$, and the mass (or mass limit), in $\GeV$, are reported~\cite{PDG_particles}. The uncertainties on the mass of charged leptons are omitted, due to the fact that they are several orders of magnitude smaller than the precision adopted in the table.}
\label{tab:fermions}
\end{table}

Thus the SM has 24 fermion fields: 18 of them are quarks, \ie\ 6 types, known as flavours, of quarks (\textit{down}, \textit{up}, \textit{strange}, \textit{charm}, \textit{bottom} and \textit{top}) times 3 colours (\textit{red}, \textit{green} and \textit{blue}), while 6 of them are leptons, 3 charged leptons (electron, muon and tau) and the corresponding neutrinos. Table~\ref{tab:fermions} reports a summary of the aforementioned fermions, along with their charge expressed in units of the electron charge, $e$, and mass (or mass limit): all leptons except for neutrinos have a charge $|Q/e|=1$; hadrons can be composed of three quarks, in which case they are called \textit{baryons} and have half-integer spin, or of a quark-antiquark pair, called \textit{mesons} and being integer-spin; quarks have a fractional charge, $|Q/e|=2/3 \; \text{or} \; 1/3$. \newline
In addition to the direct limits on the masses of neutrinos reported in Table~\ref{tab:fermions}, cosmological observations allowed to set an upper limit on the sum of neutrino masses of 0.12~\eV\ at 95\% confidence level~\cite{Planck_neutrino}. Fermions interact through the exchange of force-carrying particles (mediators), referred to as ``gauge bosons$"$:
\begin{itemize}
\item the photon, $\gamma$, is the spin-1 massless mediator of the electromagnetic interaction between charged particles;
\item the $W$ and $Z$ bosons are the spin-1 massive mediators of the weak interaction, responsible of processes like nuclear decays and processes involving neutrinos; their masses are of order of 100 times the mass of the proton;
\item the gluons, $g$, are the spin-1 massless mediators of the strong interaction, responsible of holding together both quarks in neutrons and protons, and neutrons and protons within nuclei; 
\item the graviton, $G$, is the hypothetical, not existing in the SM neither predicted by a complete quantum field theory, spin-2 massless gauge boson carrying the gravitational interaction, the weakest among the interactions. 
\end{itemize}
The fundamental properties of the bosons, \ie\ their charge, mass, spin and the respective force, are reported in Table~\ref{tab:bosons}.
\begin{table}[hbtp]
\begin{center}
 \renewcommand{\arraystretch}{1.4}
\begin{tabular}{|c|c|c|c|c|}
\hline
Boson & $Q/e$ & mass [GeV] & spin & force\\
\hline
photon ($\gamma$)& 0 &$< 10^{-27}$ & 1& electromagnetic\\
\hline
$W$ boson ($W$)& $\pm 1$& $80.379 \pm 0.012$& 1&  weak\\
$Z$ boson ($Z$)& 0 & $91.1876\pm 0.0021 $ & 1& weak \\
\hline
gluon ($g$) & 0 & $\le 10^{-3}$ & 1&  strong\\
\hline
graviton ($G$)& 0 & $< 6 \times 10^{-41}$& 2& gravitational\\
\hline
\end{tabular}
\end{center}
\caption{Properties of the gauge bosons mediating the four fundamental forces: the electric charge $Q$, in units of the electron charge $e$, the mass (or mass limit), in \GeV, the spin and the type of force are reported~\cite{PDG_particles}; the graviton, $G$, is the hypothetical, not existing in the SM neither predicted by a complete quantum field theory, gauge boson carrying the gravitational force.}
\label{tab:bosons}
\end{table}

\begin{figure}[H]
\begin{center}
\includegraphics[height=10 cm, width =13 cm]{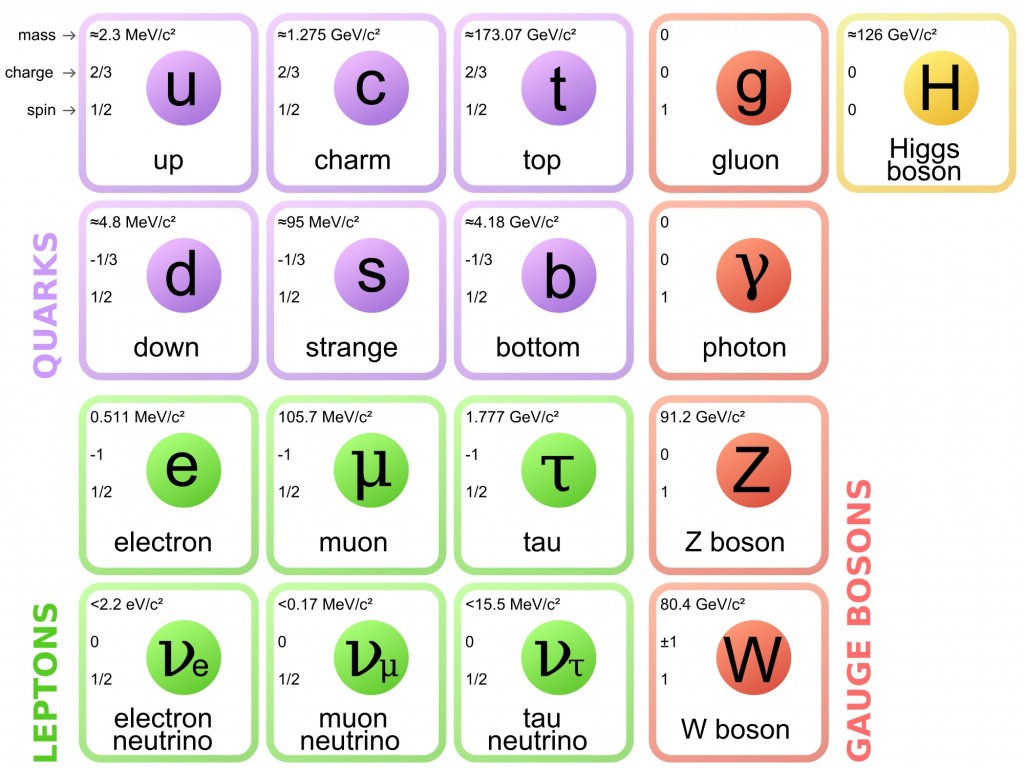}
\end{center}
\caption{Standard Model particles and interaction mediators.}     
\label{SM_particles}
\end{figure}

Finally, the Higgs boson is a neutral fundamental scalar particle introduced in the Standard Model in order to generate the masses of the gauge bosons and of all the other elementary particles considered in the theory, as explained in Section~\ref{Higgs_mechanism}.\newline
Figure~\ref{SM_particles} shows a summary of the SM particles and fundamental interactions.
The construction of the Standard Model has been guided by principles of symmetry: Noether's theorem~\cite{Noether} implies that, if an action is invariant under some group of transformations (symmetry), these symmetries are associated with one or several conserved quantities at the point where the interaction occurs, like the charge and the colour.\newline
Local symmetries, \ie\ when actions are invariant under transformation of parameters depending on the space-time coordinates, are the ones on top of which the SM is defined.\newline
The mathematics of symmetry is provided by group theory; thus the SM is based on three symmetry groups: $\text{SU(3)}_C \otimes \text{SU(2)}_L \otimes \text{U(1)}_Y$, where:
 \begin{itemize}
 \item $\text{SU(3)}_C$ reflects the symmetry of the strong interaction, described by Quantum Chromodynamics (QCD); it represents the non-abelian, \ie\ non-commutative, gauge group, with 8 gauge bosons (gluons); the ``C$"$ letter stands for the colour; 
 \item $\text{SU(2)}_L \otimes \text{U(1)}_Y$ indicates the electroweak symmetry group, which unifies electromagnetic and weak interactions in the so-called ``electroweak theory$"$;  the ``L$"$ letter stands for ``left$"$, involving only left-handed fermion fields, while the ``Y$"$ letter stands for the weak hypercharge.
 \end{itemize}
The foundations of quantum electrodynamics and chromodynamics will be the starting point of the next paragraphs.

\subsection{Quantum Electrodynamics}

\begin{figure}[H]
\begin{center}
\includegraphics[height=4 cm,width =5 cm]{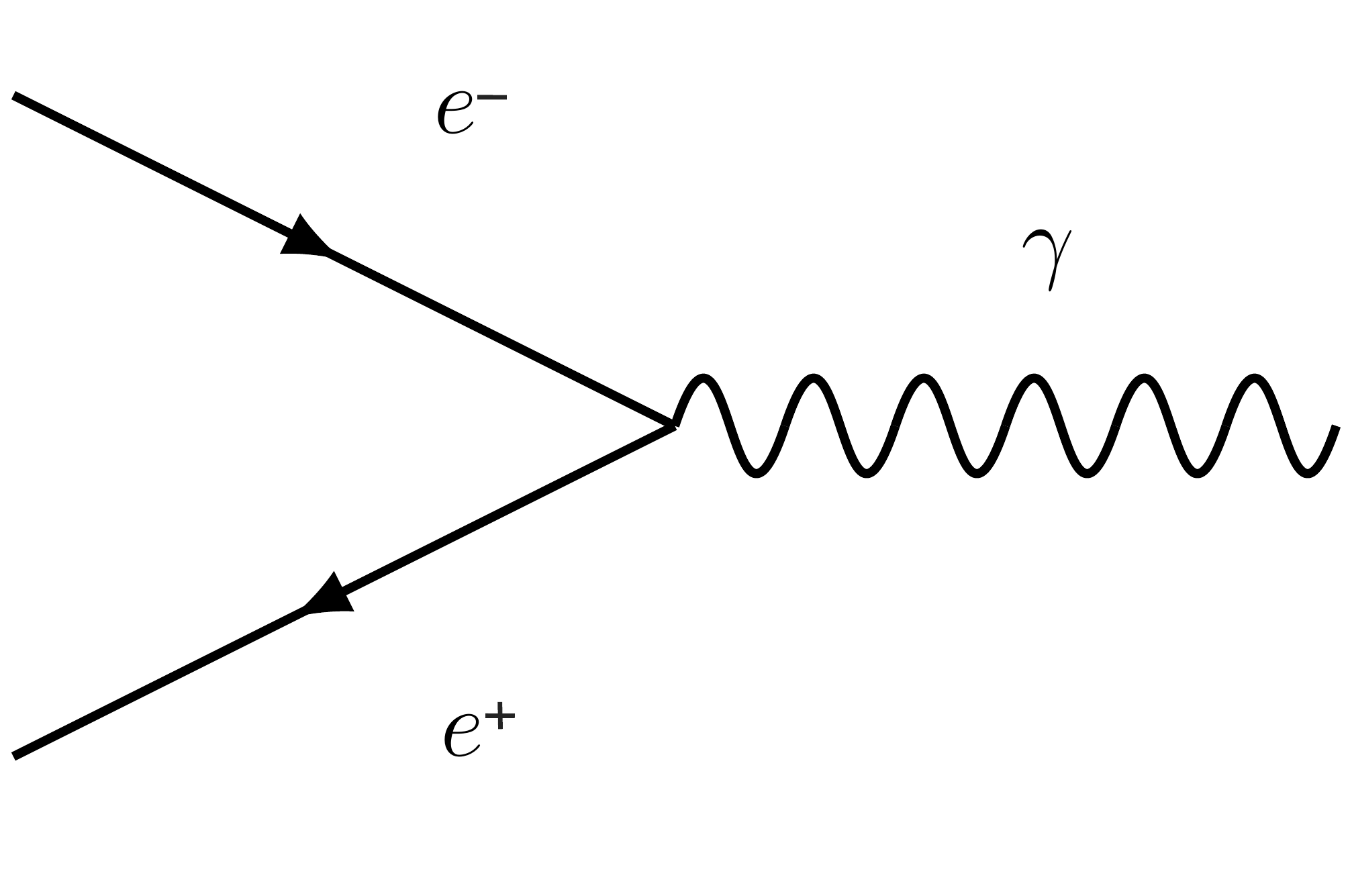}
\end{center}
\caption{Example of an interaction vertex for the QED Lagrangian, showing electron-positron annihilation.}     
\label{QED_vertex}
\end{figure}

Quantum Electrodynamics (QED) is a major success of quantum field theory (QFT) describing the interaction between electrically charged particles and the mediator of the electromagnetic interaction, \ie\ the photon. Mathematically, it is an abelian gauge theory, symmetric with respect to gauge rotations of $\text{U(1)}$ group, while the gauge field is the electromagnetic field.\newline
For a free Dirac fermion of mass $m$, the Lagrangian is:
\begin{equation}\label{dirac_l}
\mathcal{L}=\bar{\psi}(i \slashed{\partial} - m ) \psi
\end{equation}
where $\psi$ represents the fermion field, $\bar{\psi}=\psi^\dagger\gamma^0$, $\slashed{\partial}=\gamma^\mu \partial_\mu$ and $\gamma^\mu$ are the Dirac matrices.
The Lagrangian described in Equation~\ref{dirac_l} is invariant under global $\text{U(1)}$ transformations:
\begin{equation}
\psi \xrightarrow{\text{U(1)}} \psi' =e^{i\vartheta}\psi \nonumber
\end{equation}
where the phase $\vartheta$ is ``global$"$, \ie\ it does not vary for every point in space-time ($\frac{\partial \vartheta}{\partial x}=0$).
If the phase transformation depends on the space time coordinate, \ie\ $\vartheta=\vartheta(x)$, the free Lagrangian is no longer invariant. In order to ensure the local invariance of the Lagrangian, additional terms should be considered, consisting of a gauge field $A_\mu$ transforming as:
\begin{equation}
A_{\mu} \xrightarrow{\text{U(1)}} A_{\mu}' = A_\mu +\frac{1}{e} \partial_{\mu} \vartheta(x)
\end{equation}
and the corresponding covariant derivative through the minimal coupling $e$:
\begin{equation}
D_\mu=\partial_\mu -ie A_\mu \, .
\end{equation}
The Lagrangian for a free gauge field $A_\mu$ is described by:
\begin{equation}
\mathcal{L}=-\frac{1}{4}F_{\mu\nu} F^{\mu\nu}
\end{equation}
where $F_{\mu\nu}=(\partial_\mu A_\nu - \partial_\nu A_\mu)$ is the electromagnetic field strength; a hypothetical mass for the gauge field $A_\mu$ is forbidden because it would violate the local $\text{U(1)}$ gauge invariance. 
After the introduction of the gauge field $A_\mu$, the QED Lagrangian is written as:
\begin{equation}
\mathcal{L}_{QED}= -\bar{\psi}(\gamma^\mu \partial_\mu +m )\psi - \frac{1}{4} F_{\mu\nu}F^{\mu\nu} +ieA_\mu \bar{\psi}\gamma^\mu \psi
\end{equation} 
where the first term describes the free propagation of the $\psi$ fermion field (charged particles), the second term describes the free propagation of the $A_\mu$ field (photons) while the third term describes the interaction of electrons and positrons ($\psi$) with photons ($A_\mu$).\newline
The interaction between the Dirac fermions and the $A_\mu$ gauge field is described, at the lowest order of perturbation theory, by the Feynman diagram shown in Figure~\ref{QED_vertex}.
The electromagnetic coupling constant is the fine structure constant, $\alpha$, expressed at low energies as:
\begin{equation}
\alpha=\frac{e^2}{4\pi\epsilon_0 \hslash c}=\frac{1}{137}
\end{equation}
where $e= 1.602176 \times 10^{-19}$~C is the electron charge, $\epsilon_0 = 8.854187 \times 10^{12}\, \text{F}\cdot \text{m}^{-1}$ is the vacuum dielectric constant, $c = 299792458$~$\text{m} \cdot \text{s}^{-1}$ is the speed of light in vacuum and $\hslash=h/2\pi=1.054571 \times 10^{-34} \,\text{J} \cdot \text{s} $, $h$ being the Planck constant.
The renormalisation of the photon field brings, as a consequence, the fact that the QED coupling constant is not a real constant but a ``running constant$"$ depending on the energy scale and decreasing at large distances given the ``screening effect$"$ of virtual particles in vacuum.

\subsection{Quantum Chromodynamics}

\begin{figure}[hbtp]
\centering
\begin{subfigure}[b]{0.25\textwidth}
\includegraphics[height=3.5 cm,width =4.5 cm]{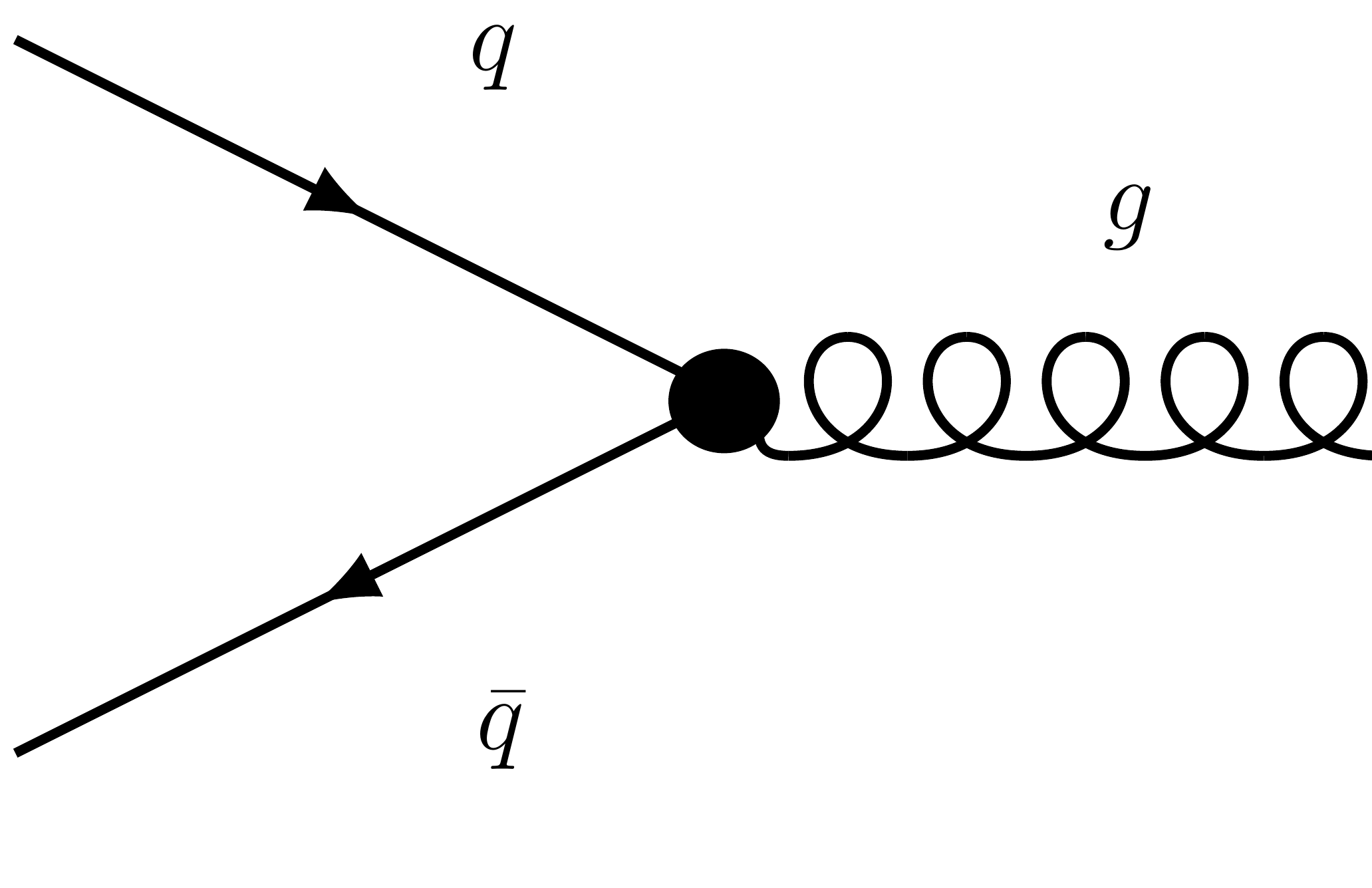}
 \caption{}
\end{subfigure}
\qquad
\begin{subfigure}[b]{0.25\textwidth}
\includegraphics[height=3.5 cm,width =4.5 cm]{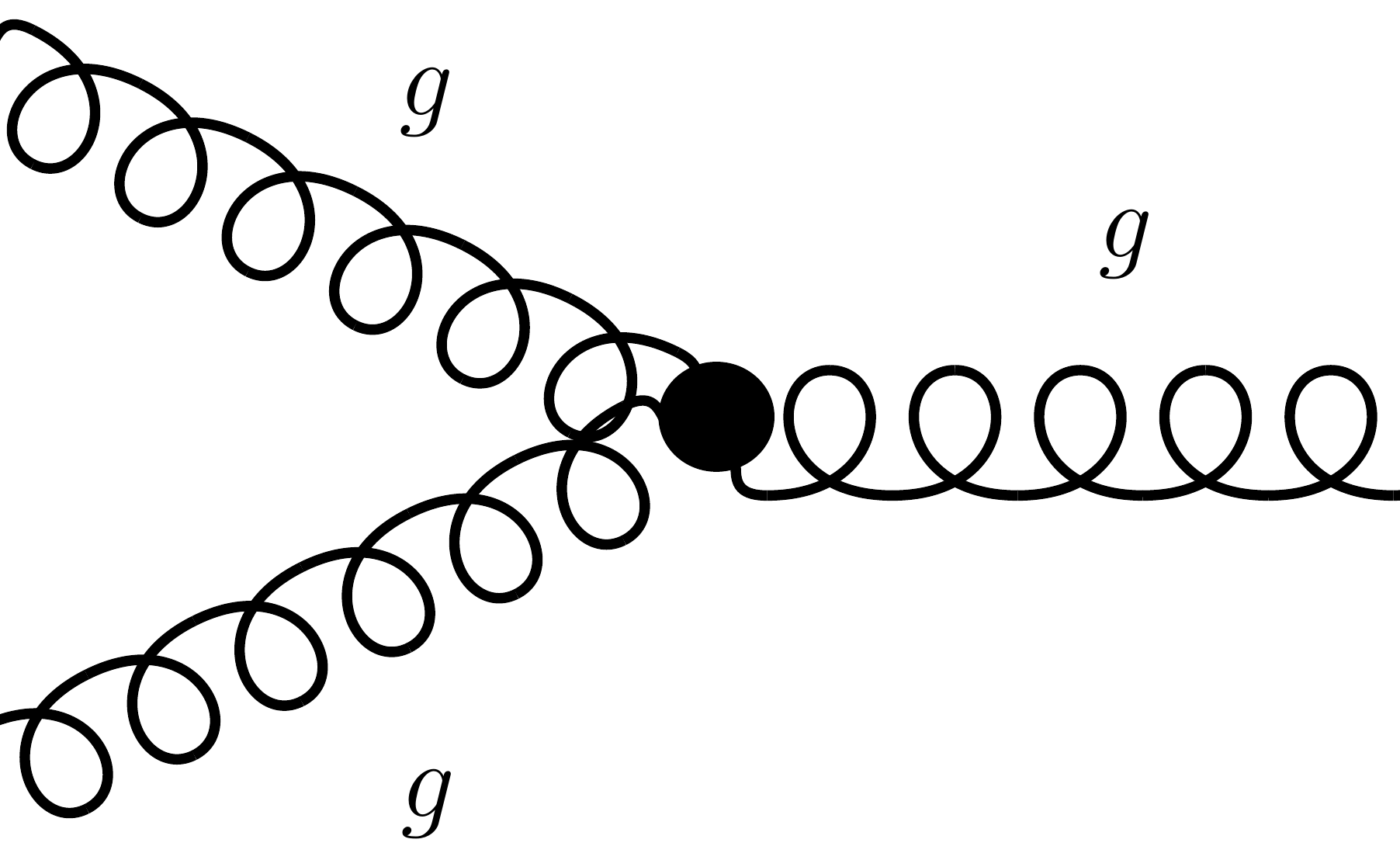}
 \caption{}
\end{subfigure}
\qquad
\begin{subfigure}[b]{0.25\textwidth}
\includegraphics[height=4 cm,width =3.5 cm]{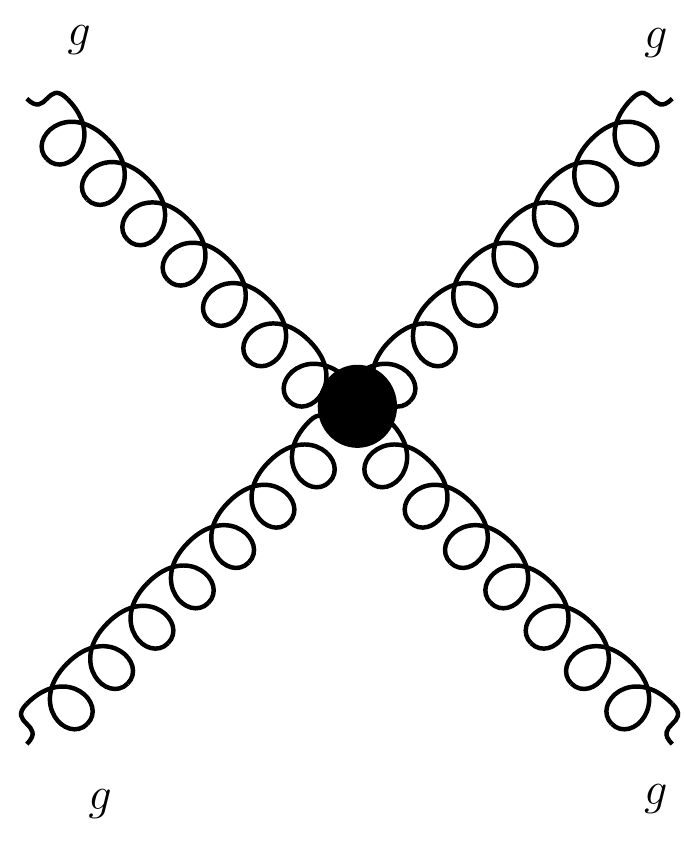}
 \caption{}
\end{subfigure}
\caption{Interaction vertices for the QCD Lagrangian: (a) quark-gluon interaction; (b) three-gluon vertex; (c) four-gluon vertex.}
\label{QCD_vertex}
\end{figure}

Quantum Chromodynamics (QCD) describes the interaction between quarks and the mediators of the strong interaction, \ie\ the gluons. Mathematically, it is a non-abelian gauge theory, symmetric with respect to gauge rotation of the $\text{SU(3)}_C$ group having 8 gauge fields, the gluons. The non-abelian nature of the theory leads to the fact that gluons, having the colour charge unlike photons that are neutral, interact not only with quarks but also among themselves, thus leading to three- or four-gluon vertices.\newline
The free Lagrangian for a quark field of flavour $f$ is~\cite{Pich}:  
\begin{equation} 
\mathcal{L}=\sum_f\bar{q}_f(i\gamma^\mu\partial_\mu -m_f) q_f \, .
\end{equation}
The Lagrangian is invariant under global $\text{SU(3)}_C$ transformations:
\begin{equation}
q_f^\alpha \xrightarrow{\text{SU(3)}} (q_f^\alpha)' =U^\alpha_\beta q_f^\beta \quad \text{being} \quad U=e^{ i\frac{\lambda^a}{2}\vartheta_a }
\end{equation}
where the indices $\alpha$ and $\beta$ run over the colour quantum numbers, the $\frac{1}{2} \lambda^a (a=1,...., 8)$ denotes the generators of the fundamental representation of the $\text{SU(3)}_C$ algebra and $\vartheta_a$ are arbitrary parameters.
The matrices $\lambda^a$ satisfy the commutation relations:
\begin{equation}
 \left [ \frac{\lambda^a}{2}, \frac{\lambda^b}{2}  \right ] = if^{abc} \frac{\lambda^c}{2}
\end{equation}
with $f^{abc}$ being the $\text{SU(3)}_C$ structure constants. The covariant derivative introduced in order to guarantee the local invariance under $\text{SU(3)}_C$ transformations, \ie\ $\vartheta_a=\vartheta_a(x)$, including as additional terms eight different gauge bosons $G^\mu_a(x)$, the gluons, reads:
\begin{equation}
D_\mu=[ \partial_\mu -ig \frac{\lambda^a}{2}G_a^\mu(x) ] =[ \partial_\mu -ig G^\mu(x) ]
\end{equation}
where $g$ is the coupling constant of QCD and $G^\mu(x) \equiv \left ( \frac{\lambda^a}{2} G_a^\mu(x) \right )$.\newline
The field strengths can be generalised for a non-abelian Lie group as:
\begin{equation}
G^{\mu\nu}_a(x)=\partial^\mu G^\nu_a - \partial^\nu G^\mu_a -g f^{abc}G^\mu_bG^\nu_c
\end{equation}
where the last term generates the cubic and quartic gluon self-interactions as a consequence of the non-abelian nature of $\text{SU(3)}_C$.\newline
After the introduction of the gauge fields, the $\text{SU(3)}_C$ invariant QCD Lagrangian can be written as:
\begin{equation}
\mathcal{L}_{QCD}=-\frac{1}{4} G_a^{\mu \nu}G^a_{\mu\nu} +\sum_f \bar{q}_f(i\gamma^\mu D_\mu -m_f)q_f
\end{equation}
where the index $f$ runs over the quark flavour and the index $a$ runs over the colour charge; the first term of the Lagrangian is the gauge boson kinetic term giving rise to three- and four-gluon vertices.
Similarly to the QED Lagrangian, the $\text{SU(3)}_C$ symmetry forbids to add mass terms for the gluon fields, explaining why the gluons are massless bosons in the SM. Interactions between quarks and gluons are shown in the diagrams of Figure~\ref{QCD_vertex}.
\begin{figure}[H]
\begin{center}
\includegraphics[height=8 cm,width =12 cm]{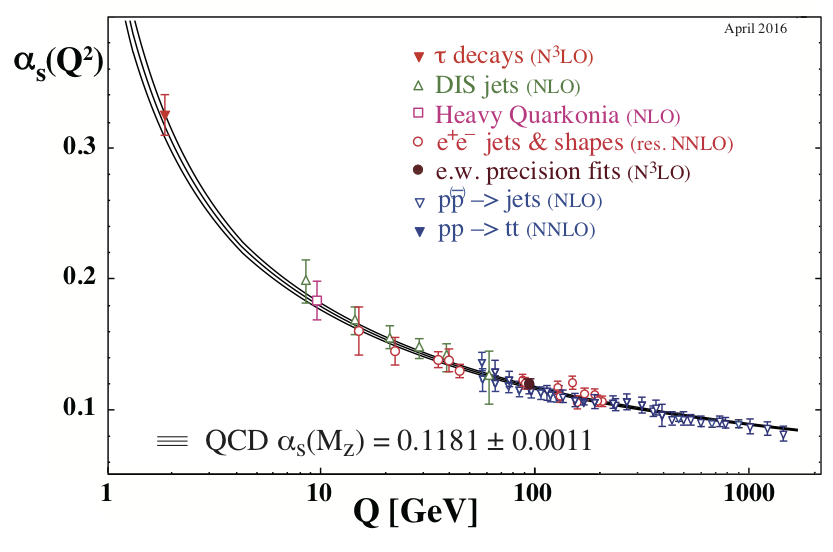}
\end{center}
\caption{Summary of measurements of $\alpha_S$ as a function of the energy scale $Q$. The respective degree of QCD perturbation theory used in the extraction of $\alpha_S$ is indicated within brackets (NLO: next-to-leading order; NNLO: next-to-next-to leading order; res. NNLO: NNLO matched with resummed next-to-leading logs; N3LO: next-to-NNLO)~\cite{world_summary}.}     
\label{alpha_strong}
\end{figure}
The running coupling constant $\alpha_S(Q^2)$ as a function of the energy scale $Q$, exploiting a first order perturbative QCD calculation, strictly valid only if $\alpha_S \ll 1$, is given by:
\begin{equation}
\alpha_s(Q^2)=\frac{12\pi}{(33-2n_f)\text{ln}\frac{Q^2}{\Lambda^2_{QCD}}}
\end{equation}
where $Q$ represents the energy transferred in the interaction, $n_f$ is the number of quark flavours and $\Lambda_{QCD}$ is the energy scale at which the perturbative QCD coupling diverges, $\Lambda_{QCD} \sim $ 0.2 GeV/c.\newline
An opposite effect, compared to the QED case, is present due to vacuum polarisation, thus leading to an ``anti-screening effect$"$ generated by gluon self-interactions.
For a short distance, \ie\ when $Q^2\rightarrow \infty$, the coupling between quarks decreases leading to the famous property of QCD known as asymptotic freedom, \ie\ quarks behaving as free particles; on the other hand, for large distances, the coupling constant increases thus making impossible the detachment of quarks from hadrons, a property known as confinement.\newline
The trend of the running coupling constant as a function of the energy scale $Q$ is shown in Figure~\ref{alpha_strong}.

\subsection{Weak Interactions and Unified Electroweak (EW) Model}
\begin{figure}[H]
\begin{center}
\includegraphics[height=4.5 cm,width =4 cm]{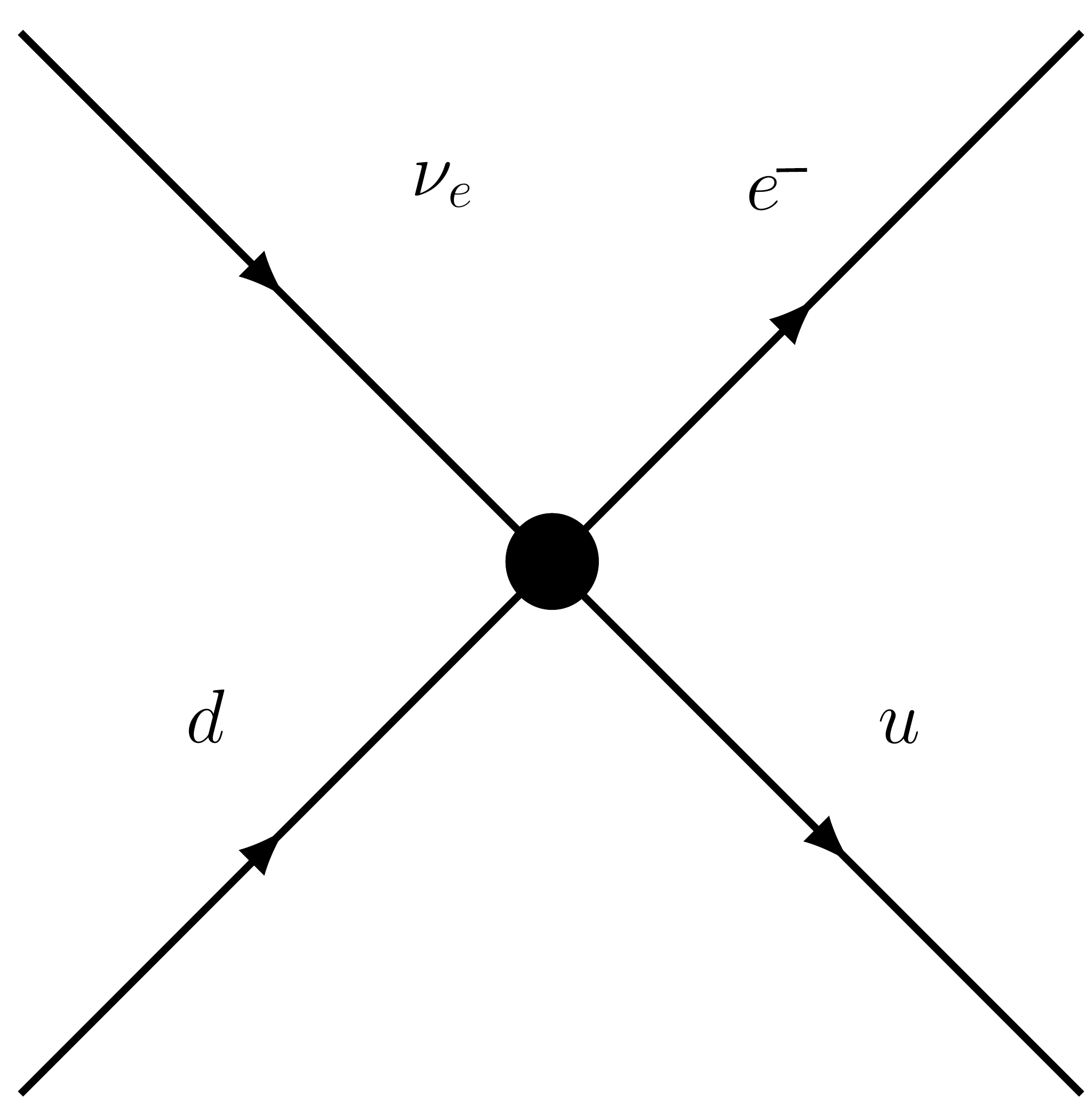}
\end{center}
\caption{Four-fermion interaction describing the $\beta$ decay.}
\label{fermi_vertex}
\end{figure}

In 1932 Enrico Fermi suggested a simple model~\cite{Fermi} to explain the $\beta$ decay, \eg\ $n\rightarrow p$$e^- +\bar{\nu}_e$, an interaction experienced by all SM fermions and characterised by a much smaller intensity with respect to the strong or electromagnetic interactions. The ``weakness$"$ of this interaction can be quantified looking at the lifetimes of the particles weakly decaying that are inversely related to the coupling strengths: the longer muon lifetime $\sim$$10^{-6}$ with respect to $\sim$$10^{-23}$ or $\sim$$ 10^{-16}$ as examples of the strong and electromagnetic interaction typically lifetimes, respectively, reflects a much weaker strength of the interaction.\newline
This interaction was originally explained by Fermi as an effective point-like vectorial current interaction ($V$) between four fermions involving a contact force with no range; Fermi's theory was valid at low energy but did not explain important features of this interaction, like the massive mediators and the parity violation. Driven by the observation that weak interactions violate parity, Fermi's theory was extended introducing to the model an axial ($A$) term which conserves its sign under parity transformations, while the violation of parity arises from the $V-A$ interaction term~\cite{suda,fey_ge}.
Figure~\ref{fermi_vertex} shows the Fermi four-fermion interaction describing the $\beta$ decay.
Particles exists in two helicity states: left-handed or right-handed. Weak interactions are found to involve only left-handed particles or right-handed anti-particles, which are defined as:
\begin{equation}
\psi_L=\frac{1}{2}(1-\gamma^5)\psi \qquad \qquad \psi_R=\frac{1}{2}(1+\gamma^5)\psi \nonumber
\end{equation}
where $\gamma^5=i\gamma^0\gamma^1\gamma^2\gamma^3$. The weak interaction field is invariant under $\text{SU(2)}_L$ transformations, where the subscript ``L$"$ means that only left-handed particles participate to these interaction.
Two types of weak interactions exist, depending on the charge of the interaction mediator: the charged-current interaction mediated by $W^+$ or $W^-$ bosons, carrying an electric charge, and the neutral-current interaction mediated by the $Z^0$ boson. \newline
The weak interaction allows quarks to change their flavour; the transition probability for a quark to change its flavour is proportional to the square of the Cabibbo-Kobayashi-Maskawa (CKM) matrix elements~\cite{CKM_pdg}:
\begin{equation*}
V_{CKM}=
  \begin{pmatrix}
  \renewcommand{\arraystretch}{1.2}
    V_{ud} & V_{us} & V_{ub}  \\
    V_{cd} & V_{cs} & V_{cb}  \\
    V_{td} &  V_{ts} & V_{tb} 
  \end{pmatrix}
  =
    \end{equation*}
\begin{equation*}
=
    \begin{pmatrix}
    0.97446 \pm 0.00010 & 0.22452 \pm 0.00044 & 0.00365 \pm 0.00012  \\
    0.22438 \pm 0.00044 & 0.97359^{+0.00010}_{-0.00011} & 0.04214 \pm 0.00076  \\
    0.00896^{+0.00024}_{-0.00023} & 0.04133 \pm 0.00074 & 0.999105 \pm 0.000032 
  \end{pmatrix}  .
  \end{equation*}

During the 1960's, Weinberg, Salam and Glashow started to work on the unification of the electromagnetic and weak theory~\cite{Glashow, Weinb,Salam}.
In order to develop a unified theory, a symmetry group needs to be identified.\newline
QED is invariant under local gauge transformations of the $\text{U(1)}$ symmetry group; instead of the electric charge for QED, a quantum number called hypercharge, $Y$, is introduced, being related to the electric charge, $Q$, through:
\begin{equation}
Q=T^3+\frac{Y}{2}
\end{equation}
where $T^3$ is the third component of the weak isospin, generating the $\text{SU(2)}$ algebra.\newline
The gauge fields of the $\text{SU(2)}_L\otimes \text{U(1)}_Y$ gauge symmetry group correspond to the four bosons $W^\pm$, $Z^0$ and $\gamma$: they are four massless mediating bosons, organised in a weak isospin triplet $W^1$, $W^2$, $W^3$ ($\text{SU(2)}_L$) and a weak hypercharge singlet $B$ ($\text{U(1)}_Y$).\newline
The free Lagrangian for massless fermions is then written as:
\begin{equation}
\mathcal{L}=i\bar{u}(x) \gamma^\mu \partial_\mu u(x) + i \bar{d}(x) \gamma^\mu \partial_\mu d(x)=\sum_{j=1}^{3} i\bar{\psi}_j (x) \gamma^\mu \partial_\mu \psi_j(x) \, .
\end{equation}
Following the same procedure used for the QED and QCD theories, the EW theory has to be invariant under global and local $\text{SU(2)}_L \otimes \text{U(1)}_Y$ transformations, thus the $\partial_\mu$ derivative has to be replaced with a covariant derivative, that is:
\begin{equation}
D_\mu =\partial_\mu +i g'\frac{Y}{2}B_\mu(x)+ig\frac{\sigma_a}{2} W_\mu^a(x) 
\end{equation}
where $g'$ and $g$ are the coupling constants for $\text{U(1)}_Y$ and $\text{SU(2)}_L$, while $W_\mu^a$ and $B_\mu$ are the gauge bosons of the $\text{SU(2)}_L$ and $\text{U(1)}_Y$ groups, respectively. The Pauli matrices $\sigma_a$ ($a=1,2,3$) and the hypercharge $Y$ represent the generators of such groups.\newline
The electric charge is related to the coupling constants of $\text{SU(2)}_L$ and $\text{U(1)}_Y$ by the equation:
 \begin{equation}
 gsin\theta_W=g'cos\theta_W=e
 \end{equation}
where $\theta_W$ is the weak-mixing angle, also called Weinberg angle, being $sin^2\theta_W=0.23122\pm 0.00017$~\cite{PDG_particles}. The boson field strengths, necessary to build the gauge-invariant kinetic term for the gauge fields are the following:
 \begin{equation}
 B_{\mu \nu} \equiv \partial_\mu B_\nu -\partial_\nu B_\mu
 \end{equation}
 \begin{equation}
 W_{\mu \nu}^i \equiv \partial_\mu W^i_\nu -\partial_\nu W^i_\mu -g \epsilon^{ijk} W_\mu^j W_\nu^k
 \end{equation}
 where $\epsilon^{ijk}$ is the Levi-Civita tensor. Thus the kinetic Lagrangian of the gauge fields becomes:
 \begin{equation}
 \mathcal{L}_{Kin}=-\frac{1}{4}B_{\mu\nu}B^{\mu \nu} -\frac{1}{4} W_{\mu \nu}^i W^{\mu \nu}_i 
 \end{equation}
and the resulting electroweak Lagrangian is:
\begin{equation}
\mathcal{L}_{EW}=\sum_{j=1}^{3} i\bar{\psi}_j (x) \gamma^\mu D_\mu \psi_j(x)-\frac{1}{4}B_{\mu\nu}B^{\mu \nu} -\frac{1}{4} W_{\mu \nu}^a W^{\mu \nu}_a
\end{equation}
where the first term describes fermion propagation and fermion interaction, while the last two terms describe EW free field propagation with the kinetic part for both $W_\mu$ and $B_\mu$ fields and the self-coupling of the $W_\mu$ field.
Since the field strengths, $W_{\mu \nu}^a$, contain a quadratic term, the Lagrangian gives rise to cubic and quartic self-interactions among gauge fields.
The gauge symmetry forbids again to write mass terms for the gauge bosons. The experimental evidences of massive gauge bosons represent one of the elements suggesting the existence of a mechanism which must give mass to these particles (Higgs mechanism, Section~\ref{Higgs_mechanism}).\newline
Fermionic masses are also forbidden, because they would produce an explicit breaking of the gauge symmetry. The electromagnetic interaction and the neutral weak current interaction arise from a mixing of the $W^3$ and $B$ fields, \ie:
\begin{equation}
\binom{A_\mu}{Z_\mu^0}= \binom{cos\theta_W \qquad sin\theta_W}{-sin\theta_W  \qquad  cos\theta_W} \cdot \binom{B_\mu}{W_\mu^3}  \; .
\end{equation}
Furthermore, the $W^\pm$ bosons are linear combinations of $W^1$ and $W^2$:
\begin{equation}
W^\pm=\frac{1}{\sqrt{2}}(W^1\mp iW^2) \, .
\end{equation}

\section{A Spontaneous Symmetry Breaking (SSB): the Higgs Mechanism}
\label{Higgs_mechanism}
\begin{figure}[htbp]
\begin{center}
\includegraphics[height=6 cm,width =12 cm]{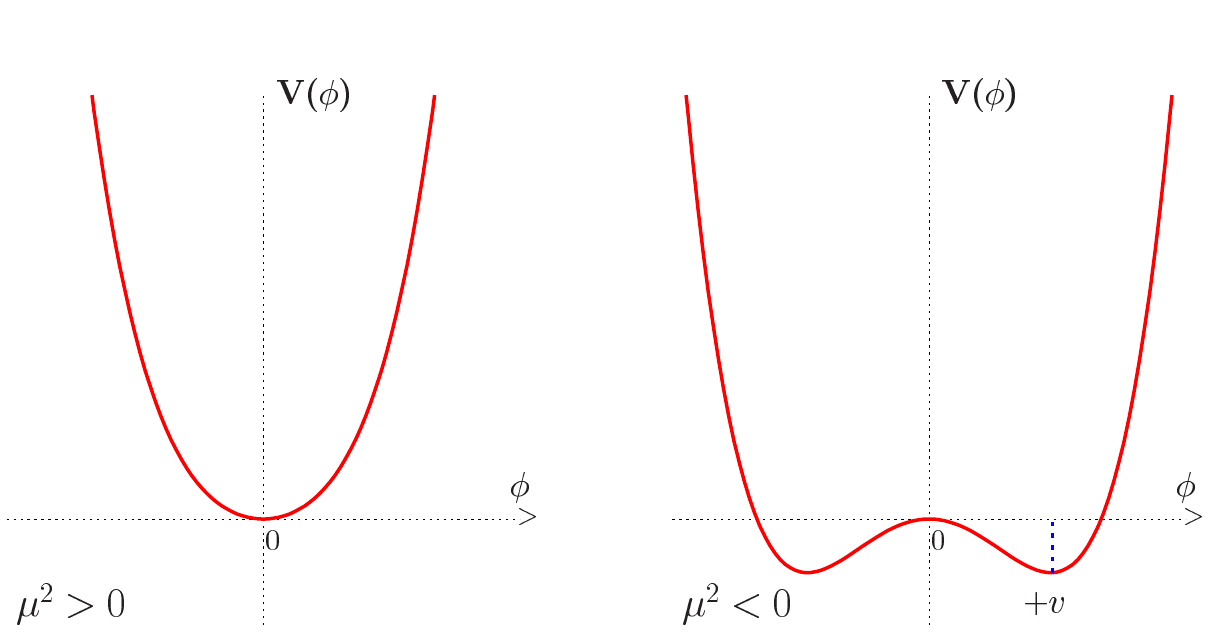}
\end{center}
\caption{The potential $V$ of the scalar field $\phi$ in the case $\mu^2>0$ (left) and $\mu^2<0$ (right)~\cite{potential}.}
\label{potential}
\end{figure}

The Lagrangian for a complex scalar field $\phi(x)$ reads~\cite{Pich}:
\begin{equation}
\mathcal{L}\equiv T-V = \frac{1}{2} (\partial_\mu \phi)^2 -\left (\mu^2\phi^\dagger \phi + \lambda (\phi^\dagger \phi)^2 \right )
\end{equation}
where $\lambda>0$ and the Lagrangian is invariant under the global phase transformations of the scalar field $\text{U(1)}$ already defined. The potential has two possible shapes, depending on the sign of $\mu^2$, as shown in Figure~\ref{potential}:
\begin{enumerate}
\item if $\mu^2>0$, the potential has only the trivial minimum or ground state, identified by $\phi= 0$; this case describes a scalar field with mass $\mu$ and quartic coupling $\lambda$;
\item if $\mu^2<0$, the potential has an infinite number of degenerate states of minimum energy satisfying:
\begin{equation}
| \phi_{0}| =\sqrt{\frac{-\mu^2}{2 \lambda}}\equiv \frac{\nu}{\sqrt{2}}
\end{equation}
where $\phi_{0}$ is the vacuum mean value of the field $\phi$, also called the vacuum expectation value (vev).
\end{enumerate}
For a specific ground state, the original symmetry gets spontaneously broken; in fact, if the perturbation of the ground state is parameterised in terms of $\phi_1$ and $\phi_2$, where $\phi_1$ and $\phi_2$ are real fields, as:
\begin{equation}
\phi(x)=\nu+\frac{1}{\sqrt{2}} (\phi_1(x)+i\phi_2(x)),
\end{equation}
the potential becomes:
\begin{equation}
V(\phi)=V(\phi_0)-\mu^2\phi_1^2 + \lambda \nu \phi_1 (\phi_1^2 + \phi_2^2) +\frac{\lambda}{4} (\phi_1^2+\phi_2^2)^2 \, .
\end{equation}
Thus the $\phi_1$ describes a state with mass $m^2_{\phi_1}= 2\lambda\nu^2=-2\mu^2$ while $\phi_2$ describes a massless state: the Lagrangian does not possess the original symmetry. This is the simplest example of spontaneously broken symmetry.\newline
The fact that massless particles appear when a global symmetry is spontaneous broken is known as the Goldstone theorem~\cite{Goldstone1,Goldstone2}: given a Lagrangian that is invariant under a group of continuous transformations with $N$ generators, if $M$ of $N$ generators are spontaneously broken, in the particle spectrum of the theory, developed around the vacuum expectation value, there will be $M$ massless particles. \newline
This approach has to be extended to the non-abelian case of the SM, where masses for the three gauge bosons $W^\pm$ and $Z$ have to be generated, while the photon should remain massless: the resulting theory must still include QED with its unbroken $\text{U(1)}$ symmetry. The Higgs mechanism is used in order to introduce the mass terms~\cite{englert, P_Higgs1,P_Higgs2}; first of all, a $\text{SU(2)}$ doublet of complex scalar field is introduced:
\begin{equation}
\phi=\binom{\phi^+}{\phi^0}=\frac{1}{\sqrt{2}} \binom{\phi_1+i\phi_2}{\phi_3+i\phi_4}
\end{equation}
with the corresponding Lagrangian being:
\begin{equation}
\mathcal{L}_{scalar}=(D_\mu\phi)^\dagger(D^\mu\phi)-V(\phi)
\end{equation}
where $D_\mu$ is the covariant derivative associated to $\text{SU(2)}_L \otimes \text{U(1)}_Y$ and $V(\phi)$,
\begin{equation}
V(\phi)=\mu^2(\phi^\dagger \phi)+\lambda(\phi^\dagger \phi)^2 \, ,
\end{equation}
is the quartic potential associated to the new scalar field. The parameter $\lambda$ of the potential is assumed to be positive.\newline
When $\mu^2<0$, there is not a single vacuum located at 0, but the two minima in one dimension correspond to a continuum of minimum values in $\text{SU(2)}$. 
The canonical solution for the Higgs potential ground state is:
\begin{equation}
\phi_1=\phi_2=\phi_4=0, \quad \phi_3=-\frac{\mu^2}{\lambda}=\nu , \quad \text{being} \quad \nu=\left ( \frac{-\mu^2}{\lambda} \right )^{\frac{1}{2}} 
\end{equation}
and the corresponding vacuum state is:
\begin{equation}
\phi_0=\frac{1}{\sqrt{2}}\binom{0}{\nu} \, .
\end{equation}
The $\phi$ field can be expanded around the vacuum expectation value by a perturbation:
\begin{equation}
\phi(x)=\frac{1}{\sqrt{2}}\binom{0}{\nu+H(x)}
\end{equation}
where $H(x)$ is a physical scalar Higgs field and the unitarity gauge is chosen in order to set the Goldstone boson components in the scalar field to zero.\newline
The scalar Lagrangian can be expanded including the gauge Lagrangian expressed in terms of the physical gauge fields:
\begin{equation}
\begin{split}
\mathcal{L}_{Higgs}={}&
  \frac{1}{2} \partial_\mu H \partial^\mu H + \frac{g^2}{4}(\nu+H)^2 \left ( W_\mu^+W^{\mu -} + \frac{1}{2cos^2\theta_W} Z_\mu Z^\mu \right ) \\
 & +\mu^2H^2-\lambda_{HHH}\nu H^3- \frac{\lambda_{HHHH}}{4}H^4 
 \end{split}  
\end{equation}
where the first three terms describe the kinetic and the mass terms of the $W$ and $Z$ fields together with the interaction between these fields and the Higgs field. The last two terms describe the self-couplings of the Higgs scalar field. 
Writing explicitly the Lagrangian, the $W$ and $Z$ boson masses and the self-interactions of the Higgs boson can be expressed as:
\begin{itemize}
\item $M_W=g\nu/ 2$
\item $M_Z=g\nu/2cos\theta_W$
\item $\nu=(\sqrt{2}G_F)^{-1/2} \sim 246$ \GeV\ 
\item $\lambda_{HHH}=3M_H^2/\nu$ and $\lambda_{HHHH}=3M_H^2/\nu^2$.
\end{itemize}
The Higgs mass is given by $M_H=\sqrt{-2\mu^2}=\sqrt{2\lambda}\nu$, where $\nu$ is known, while $\lambda$ is a free parameter of the theory. Thus the SM does not predict the Higgs-boson mass value.\newline
A fermionic mass term, $m\bar{\psi}\psi$, is prohibited since it would break the gauge symmetry. Thus new terms, involving the so-called \textit{Yukawa coupling}, need to be added to the Lagrangian in order to generate the masses of charged leptons:
\begin{equation}
\mathcal{L}_{Yukawa}^{Leptons}=-y_t ( \bar{\psi}_L\phi \psi_R +  \bar{\psi}_R\phi^\dagger \psi_L )
\end{equation}
where $y_t$ is the Yukawa coupling. After spontaneous symmetry breaking, the Yukawa Lagrangian becomes:
\begin{equation}
\mathcal{L}^{Leptons}_{Yukawa}=-\frac{y_t}{\sqrt{2}} \nu( \bar{ \ell }_ L \ell_R+ \bar{ \ell }_ R \ell_L ) -\frac{y_t}{\sqrt{2}} \nu( \bar{\ell}_L\ell_R+ \bar{\ell}_R\ell_L )H
\end{equation}
where lepton masses of value $M_\ell=y_t \nu/\sqrt{2}$ are generated.
The Higgs interaction with quarks can be described by:
\begin{equation}
\mathcal{L}^{Quarks}_{Yukawa}=-y_d\bar{Q}_L\phi d_R  -y_u\bar{Q}_L\tilde{\phi}^c u_R + h.c. 
\end{equation}
where $Q_L=\binom{u}{d}_L$, $\tilde{\phi}^c\equiv i\sigma_2\phi^*$, \ie\ the $\mathcal{C}$-conjugate scalar field, $y_{u,d}$, \ie\ the Yukawa couplings, are matrices introducing the mixing between different quark flavours and $M_{u,d} =$$ y_{u,d}\, \nu/\sqrt{2}$. The generic form of the Yukawa Lagrangian reads:
\begin{equation}
\mathcal{L}_{Yukawa}=- \left ( 1+\frac{H}{\nu} \right )  m_f \bar{f}f \, .
\end{equation}
The total SM Lagrangian is represented by the sum of the following terms:
\begin{equation}
\mathcal{L}_{SM}=\mathcal{L}_{QCD} + \mathcal{L}_{EW} + \mathcal{L}_{Higgs} + \mathcal{L}_{Yukawa} \, .
\end{equation}
It is important to note that lepton and quark masses are free parameters of the theory; moreover, neutrinos, that do not have right-handed states, remain massless. Following the experimental evidences of their oscillation which requires a mass different from zero, the three right-handed neutrinos with the corresponding mass terms can be added in a minimal extension of the Standard Model.\newline
The coupling of the Higgs boson to fermions is proportional to the mass, thus leading to very different values of the strengths of these couplings, given the huge mass range considered. 

\section{The Standard Model Higgs Boson}
\label{SM_Higgs_boson}
The search for the particle responsible of the Higgs mechanism, the Higgs boson, has lasted for decades. More than 20 years after the formulation of the Higgs mechanism had to pass until a significant mass range could be probed first with the Large Electron Positron Collider (LEP) at CERN~\cite{LEP} and then with the Tevatron~\cite{Tevatron} proton-antiproton collider.\newline
In 2010 the Large Hadron Collider (LHC)~\cite{Lindon}, whose description is reported in Chapter~\ref{sec:LHC}, started to take data at unprecedented centre-of-mass energies with the primary goal of confirming the existence of this boson.\newline
Finally, in July 2012, the discovery of a particle compatible with the SM Higgs boson by the ATLAS~\cite{Atlas, Higgs_Atlas} and CMS~\cite{Cms, Higgs_CMS} experiments at the LHC represented a great milestone in the history of particle physics.
After the discovery of the Higgs boson, a new era in understanding the nature of electroweak symmetry breaking, possibly completing the SM and constraining effects from NP, has opened. One of the main focus of ATLAS and CMS physics analyses is the precision measurement of the properties of the Higgs boson including spin-parity, couplings and evidence for production mechanisms, which are essential tests of the SM.\newline
In the following sections, these properties, ranging from the main production modes and main decay channels in proton-proton collisions, to the mass and width of this particle, are described.

\subsection{Higgs-Boson Production}
\begin{figure}[hbtp]
\centering
\begin{subfigure}[b]{0.49\textwidth}
\includegraphics[height=7 cm,width =8 cm]{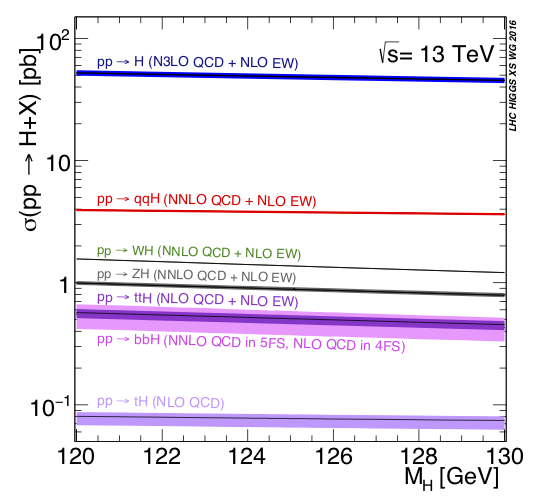}
 \caption{}
\end{subfigure}
\begin{subfigure}[b]{0.49\textwidth}
\includegraphics[height=7.2 cm,width =8 cm]{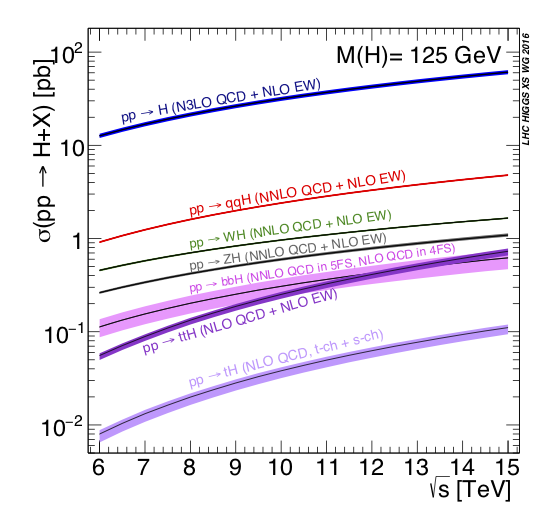}
 \caption{}
\end{subfigure}
\caption{The SM Higgs-boson production cross section at $\sqrt{s} = 13$ \TeV\ in proton-proton collisions as a function of the Higgs-boson mass (a) and as a function of the LHC centre-of-mass energy (b)~\cite{Higgs_CS}.}     
\label{higgs_production}
\end{figure}

The main mechanisms to produce the Higgs boson are the following: through the fusion of gluons (gluon-fusion, or \ggF); through the fusion of weak vector bosons (\VBF); in association with a $W$ or a $Z$ boson (\WH or \ZH, \VH to identify both \WH and \ZH), or in association with one or more top quarks ($t\bar{t}H$+ $tH$).
The size of the respective cross sections depends both on the type of colliding hadrons and on the collision energy; the ranking of different mechanisms at the LHC is shown in Figure~\ref{higgs_production} as a function of the Higgs-boson mass $(a)$ and as a function of LHC centre-of-mass energy $(b)$.

\subsubsection{The gluon-fusion production mode}
 \begin{figure}[hbtp]
\centering
\begin{subfigure}[b]{0.49\textwidth}
\includegraphics[height=3 cm,width =6.2 cm]{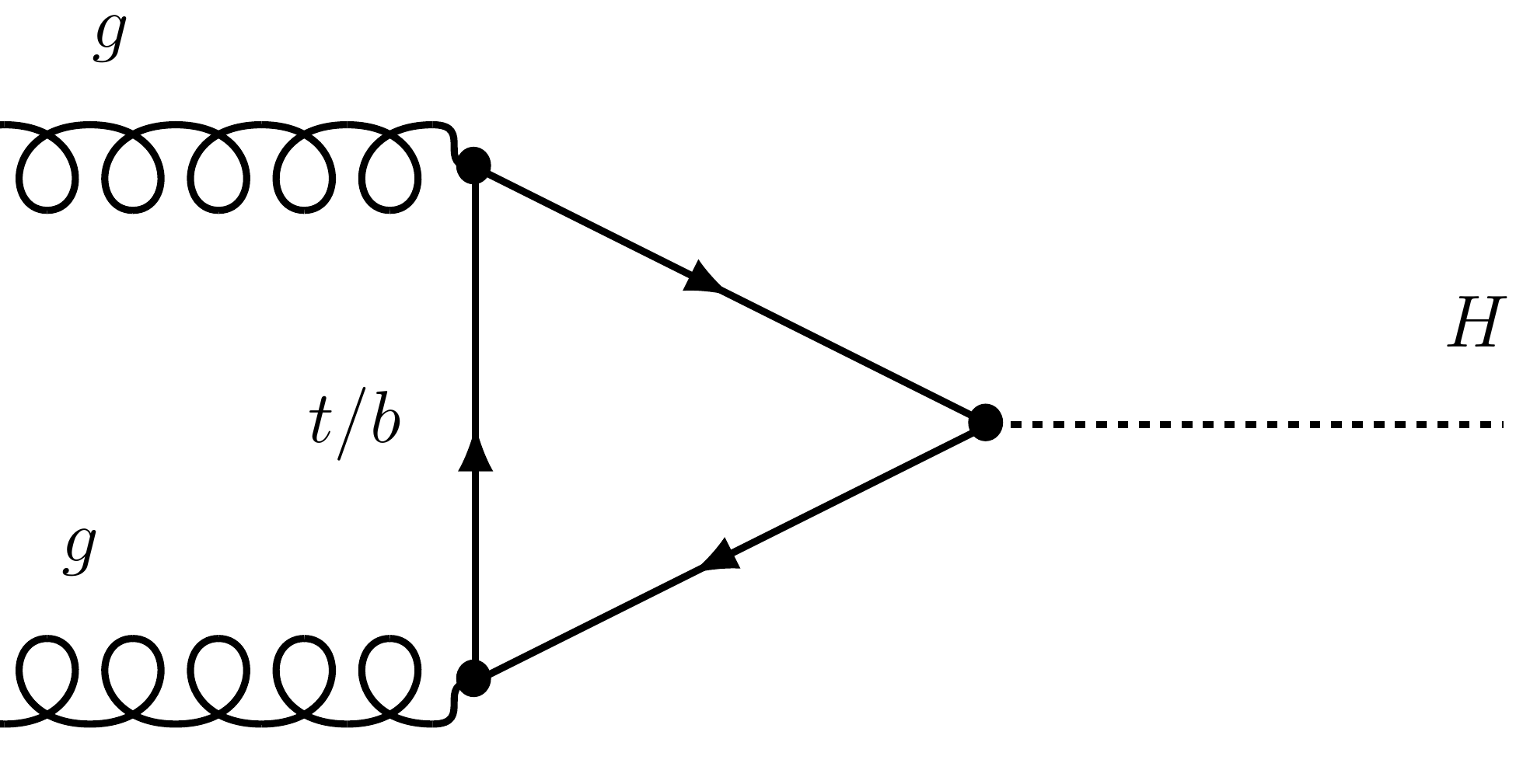}
 \caption{}
\end{subfigure}
\begin{subfigure}[b]{0.49\textwidth}
\includegraphics[height=5.5 cm,width =6 cm]{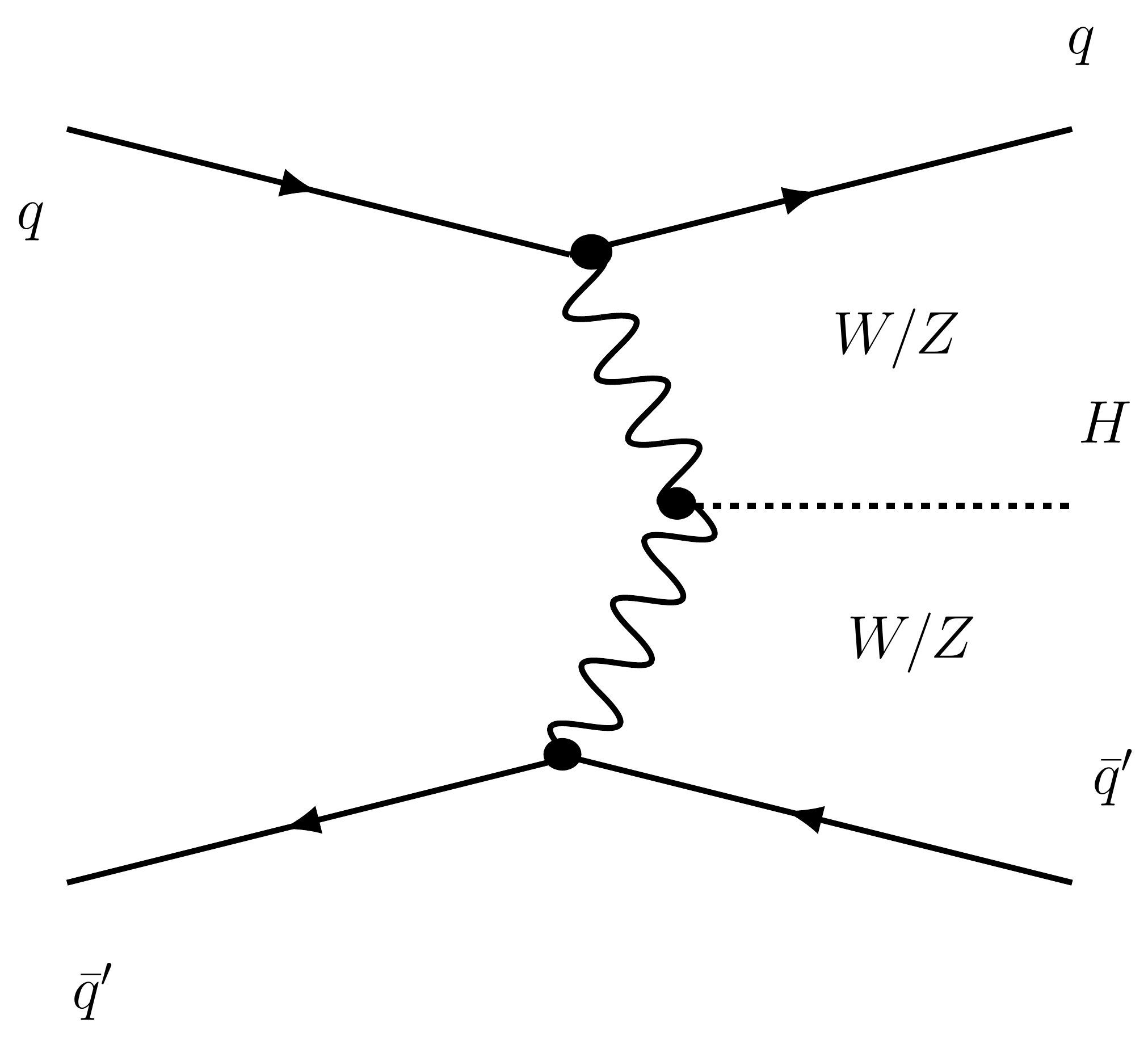}
 \caption{}
\end{subfigure}
\caption{Leading-order diagrams for the gluon fusion (a) and vector-boson fusion (b) initiated production of the SM Higgs boson.}     
\label{ggF_VBF}
\end{figure}
Due to the high flux of gluons in high-energy proton-proton collisions, the gluon-fusion process is the production mode having the largest cross-section at the LHC: two gluons combine mediated by a loop of virtual quarks. Due to the dependence of the Higgs-boson couplings to quarks on the square of the quark mass, this process is more likely for heavier quarks, thus it is sufficient to consider virtual top and bottom loops. When the Higgs boson is produced through this production mode, there are no additional particles in the final state except for the products of the decays of the Higgs boson itself other than any additional QCD radiation.\newline
The current best prediction for the inclusive \ggF\ cross section of a Higgs boson with a mass $M_H= 125$ \GeV\ at the LHC, considering a centre-of-mass energy of $\sqrt{s}=13$ $\TeV$, is~\cite{Higgs_CS}: 
\begin{equation}
\sigma_{ggF}=48.6\; \text{pb}^{+2.2\; \text{pb}\, (+4.6 \%)}_{+3.3\; \text{pb}\, (+6.7 \%)} \; (\text{theory}) \pm 1.6\; \text{pb}\, (3.2\%)\; (\text{PDF} + \alpha_S)
\end{equation}
where the total uncertainty is divided into contributions from theoretical uncertainties, ``theory$"$, and from parametric uncertainties due to parton-distribution-function (PDF) uncertainties and $\alpha_S$ computation uncertainties, ``PDF + $\alpha_S$$"$.\newline
This cross section is at least an order of magnitude larger than the other production cross sections.
Figure~\ref{ggF_VBF} $(a)$ shows the leading-order (LO) diagram for the gluon-fusion production mode.

\subsubsection{The vector-boson-fusion production mode}

The Higgs-boson production mode with the second largest cross section at the LHC is vector-boson fusion. It proceeds through the scattering of two quarks or anti-quarks mediated by the exchange of a virtual $W$ or $Z$ boson, which radiates the Higgs boson. This production mode has a clear signature consisting in two energetic jets, coming from the fragmentation of the quarks; they appear in the forward region of the detector close to the beam pipe, in addition to the products of the Higgs-boson decay.
The \VBF production mode represents $\sim$$10 \%$ of the total production cross section for a Higgs boson with a mass $M_H=125$ $\GeV$.
The leading-order diagram for \VBF is shown in Figure~\ref{ggF_VBF} $(b)$.

\subsubsection{Higgs-strahlung: $WH$ and $ZH$ associated production mechanism}

\begin{figure}[hbtp]
\centering
\begin{subfigure}[b]{0.49\textwidth}
\includegraphics[height=3 cm,width =6 cm]{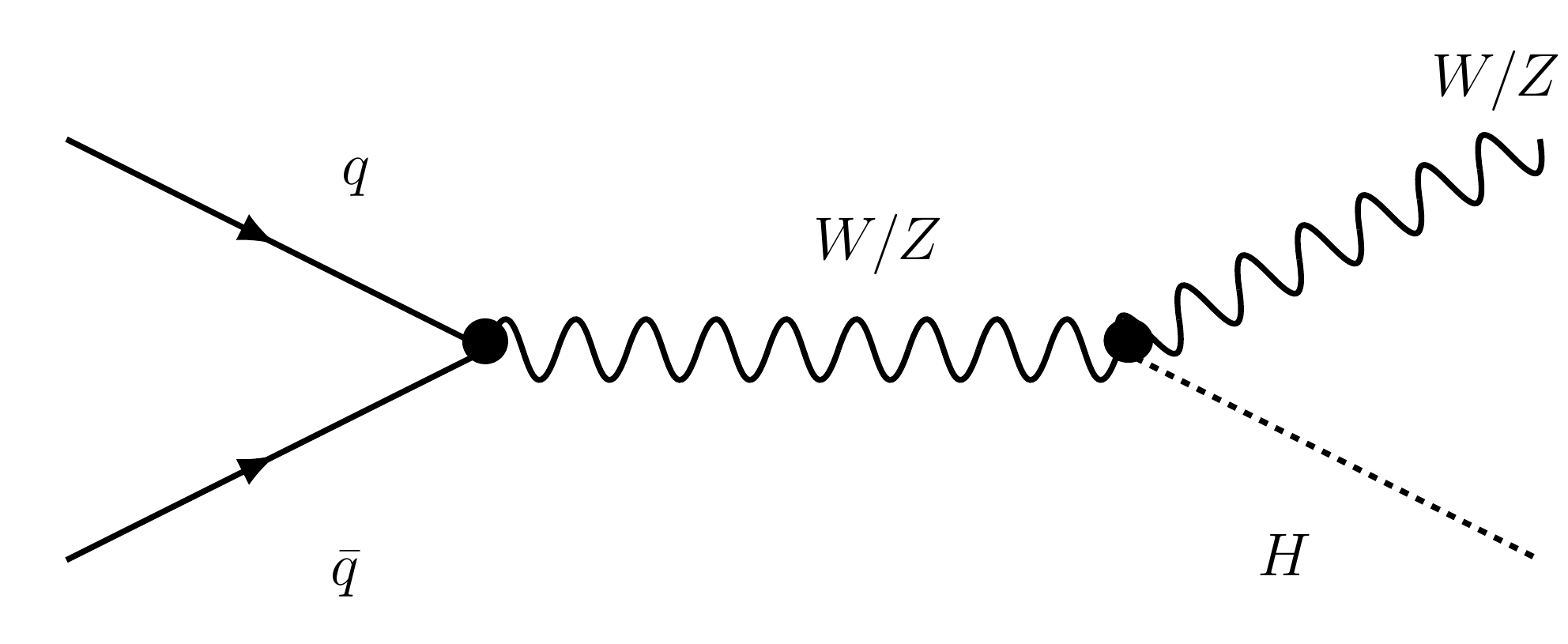}
 \caption{}
\end{subfigure}
\begin{subfigure}[b]{0.49\textwidth}
\includegraphics[height=3 cm,width =6 cm]{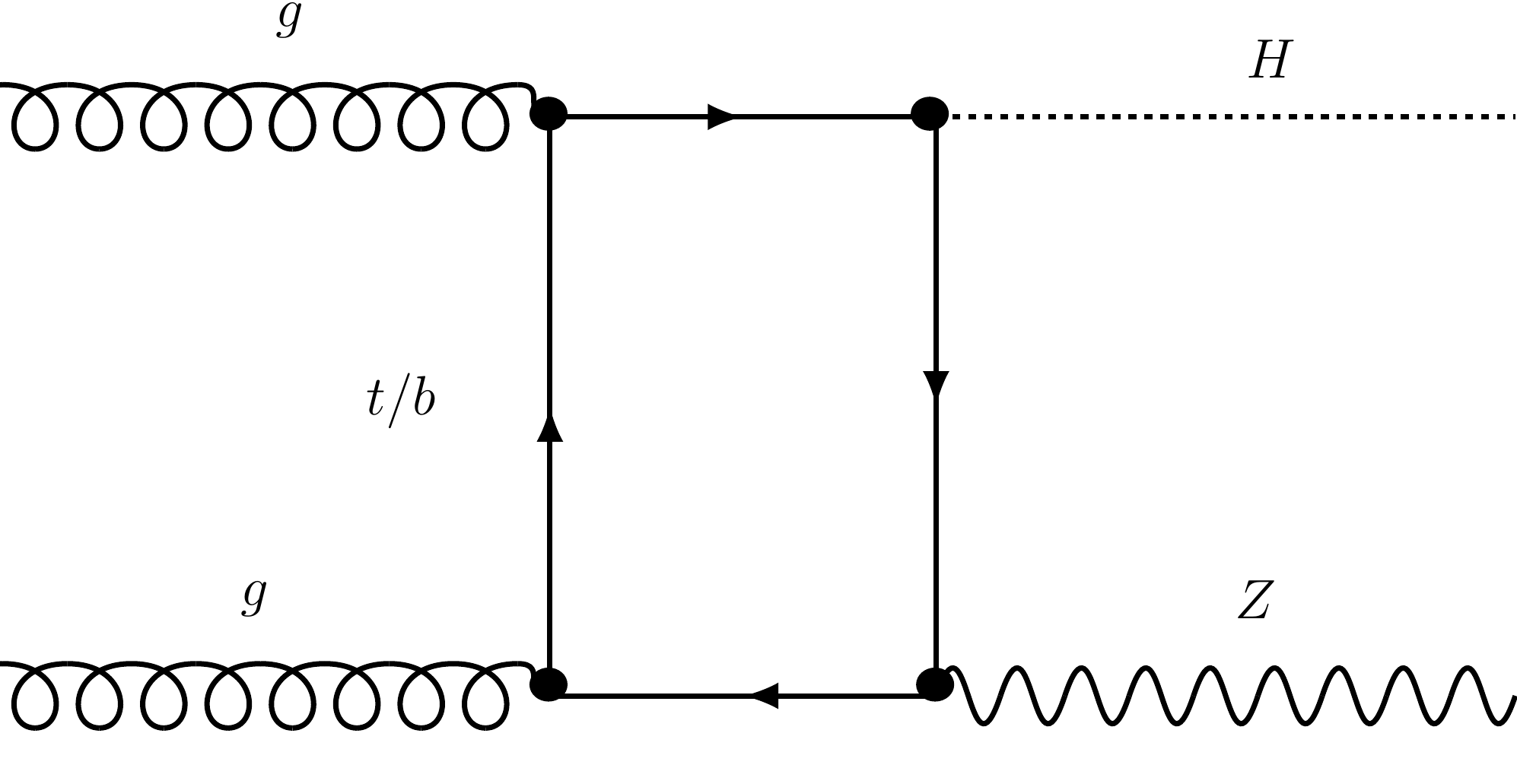}
 \caption{}
\end{subfigure}
\begin{subfigure}[b]{0.49\textwidth}
\includegraphics[height=3 cm,width =6 cm]{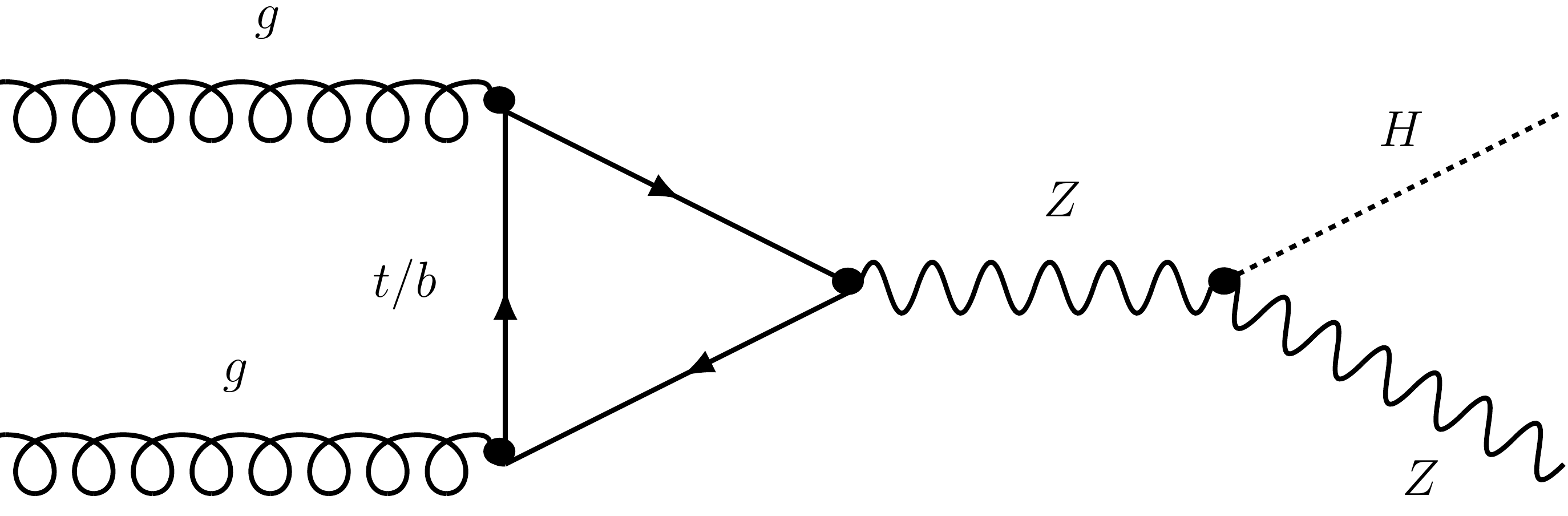}
 \caption{}
\end{subfigure}
\caption{Leading-order diagrams for the production of a Higgs boson in association with a vector boson.}     
\label{VH_ZH}
\end{figure}

The next most relevant Higgs-boson production mechanism is the associated production with an electroweak vector boson $W$ or $Z$, also called Higgs-strahlung. Most of the contribution to this production mode comes from the annihilation of quarks even if, for the \ZH production, there are also gluon-gluon contributions that produce the Higgs and the $Z$ bosons through a top-quark loop.
Figure~\ref{VH_ZH} shows Feynman diagrams for $qq$- and $gg$-initiated \VH processes.

\subsubsection{Higgs production in association with top quarks}
\begin{figure}[hbtp]
\centering
\begin{subfigure}[b]{0.49\textwidth}
\includegraphics[height=5.5 cm,width =6 cm]{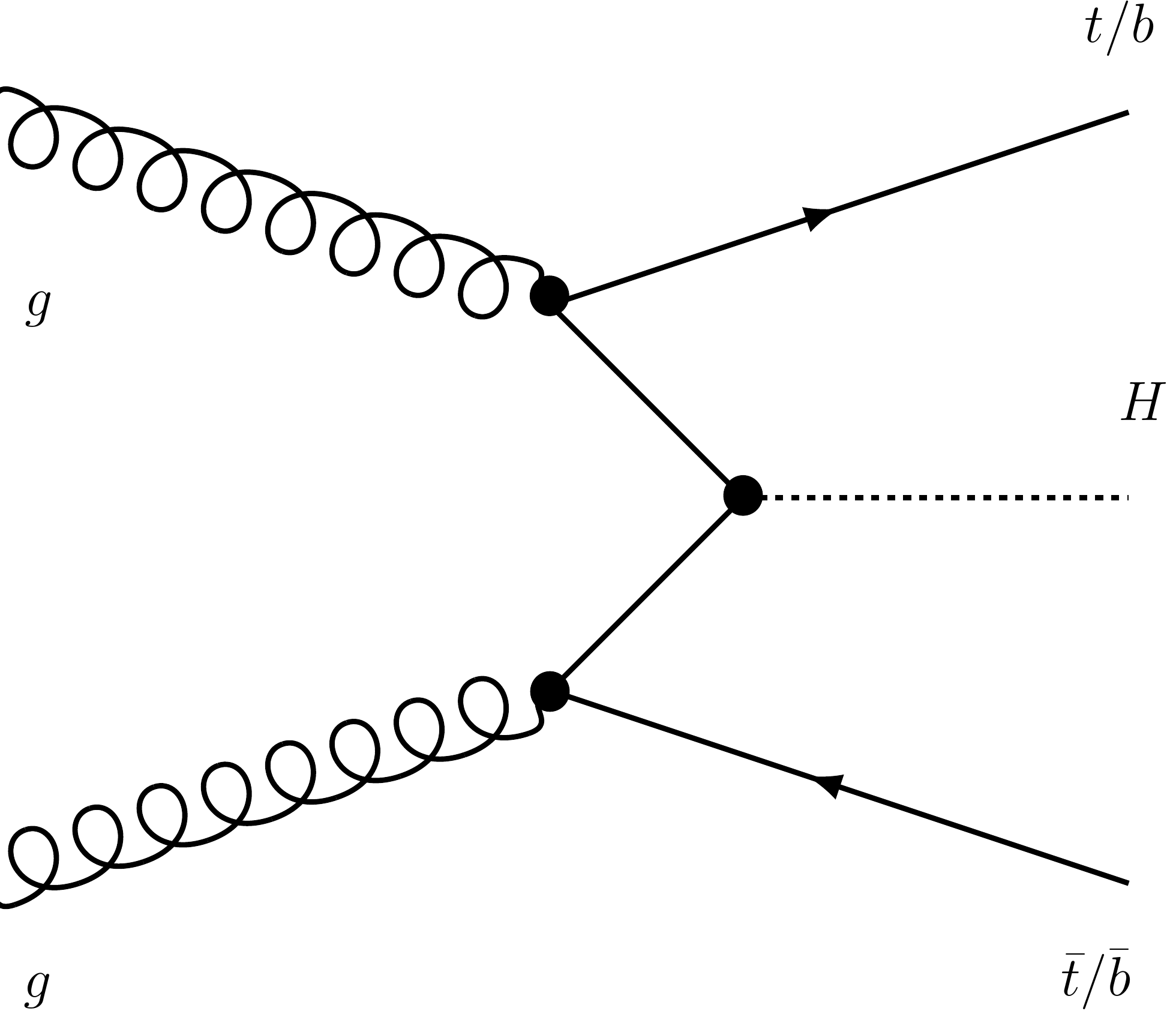}
 \caption{}
\end{subfigure}

\begin{subfigure}[b]{0.49\textwidth}
\includegraphics[height=6.5 cm,width =6 cm]{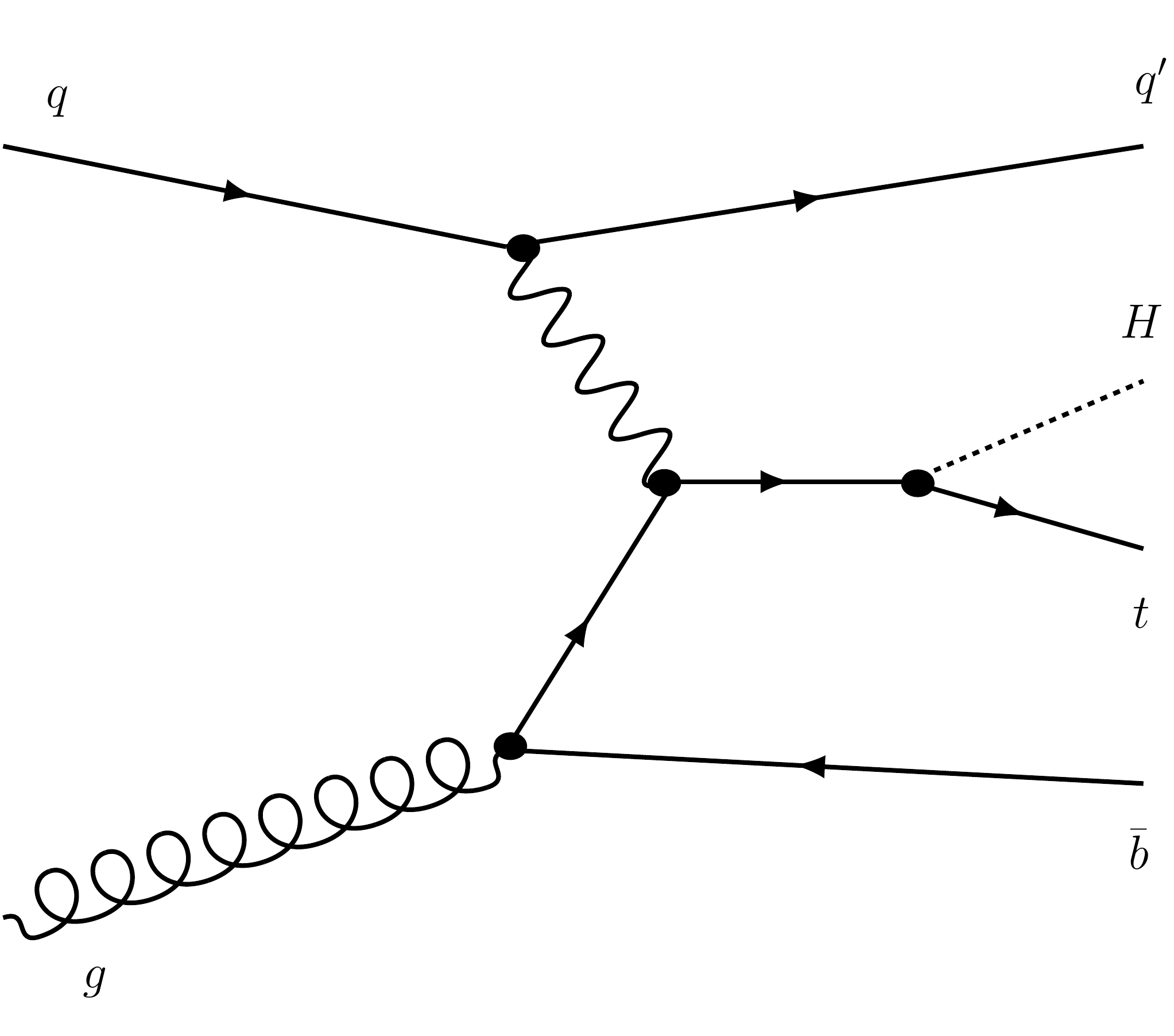}
 \caption{}
\end{subfigure}
\begin{subfigure}[b]{0.49\textwidth}
\includegraphics[height=6.5 cm,width =6 cm]{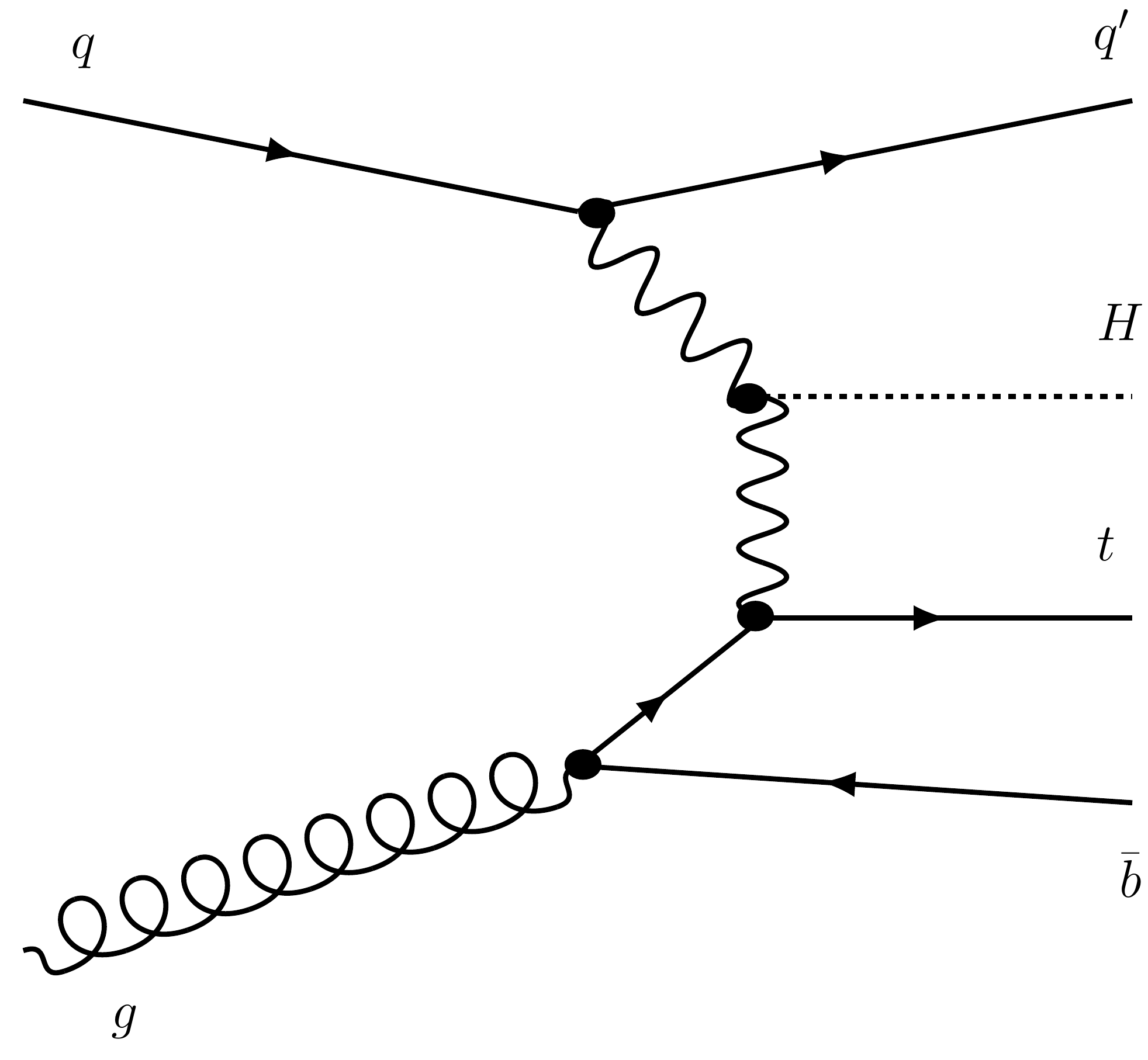}
 \caption{}
\end{subfigure}
\begin{subfigure}[b]{0.49\textwidth}
\includegraphics[height=3.9 cm,width =5.5 cm]{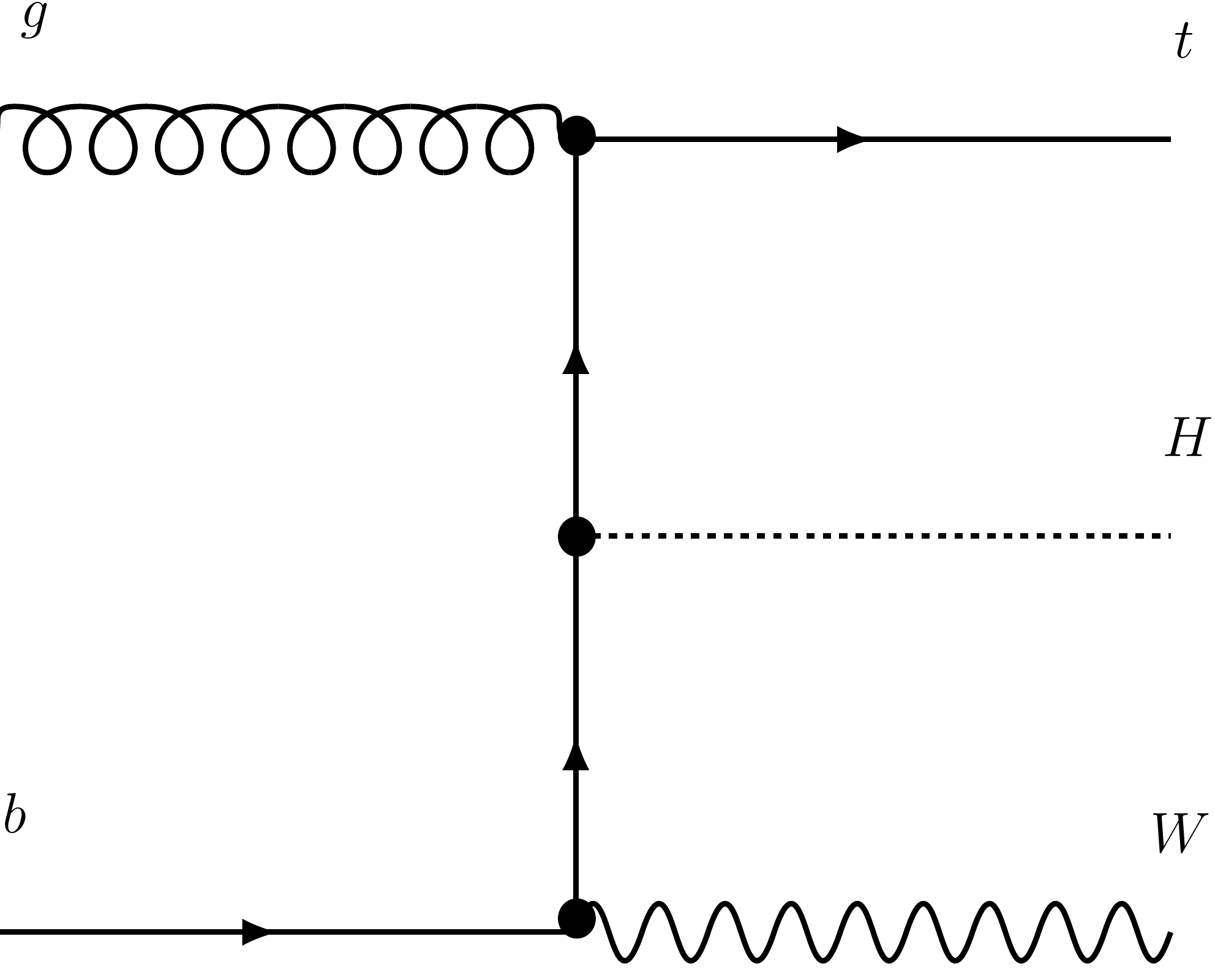}
 \caption{}
\end{subfigure}
\caption{Leading-order diagrams for the production of a Higgs boson in association with top quarks.}     
\label{top_feynman}
\end{figure}

Finally, the Higgs production in association with top quarks represents one of the rarest Higgs-boson production modes. Nevertheless, this production mode can provide important information on the Yukawa coupling and its relative sign ($tH$), since it involves the direct coupling of the Higgs boson to the top quark.
As can be seen from the set of Feynman diagrams shown in Figure~\ref{top_feynman}, the $t\bar{t}H$ and $tH$ processes have very complex final states, thus increasing the experimental challenge of isolating them. The presence of other tagging objects (either $b$-jets, jets or leptons), in addition to the Higgs-boson decay products, allows to reduce the background and to reach a good sensitivity despite the low cross section of this process.

\subsection{Higgs-Boson Decays}
\begin{figure}[hbtp]
\centering
\begin{subfigure}[b]{0.49\textwidth}
\includegraphics[height=6 cm,width =8 cm]{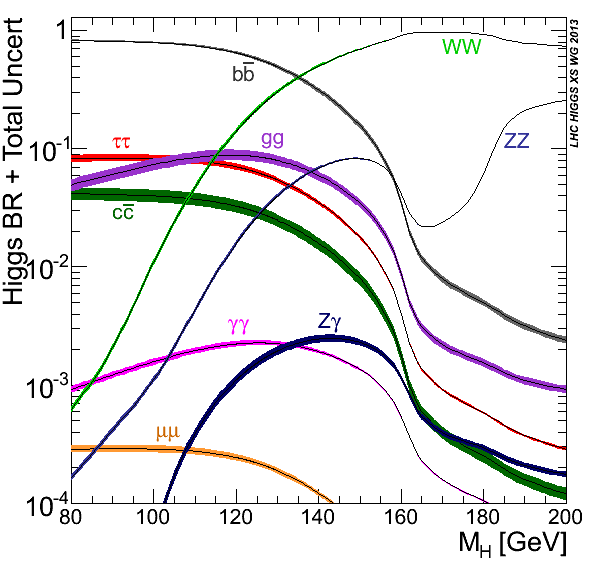}
 \caption{}
\end{subfigure}
\begin{subfigure}[b]{0.49\textwidth}
\includegraphics[height=6 cm,width =8 cm]{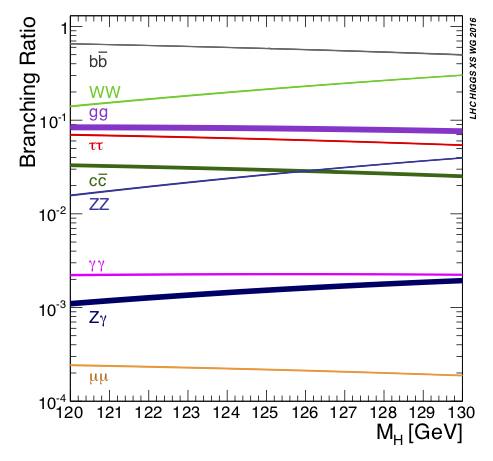}
 \caption{}
\end{subfigure}
\caption{The SM Higgs-boson branching fractions as a function of the Higgs-boson mass as a function of the Higgs-boson mass over an extended mass range from 80 to 200 \GeV\ $(a)$ and in a zoomed range 5 $\GeV$ within the best-fit measured mass $(b)$, $120-130$ $\GeV$~\cite{Higgs_CS}.}     
\label{higgs_BR}
\end{figure}
The branching fraction of a certain final state $f$ is defined as the fraction of the time that a particle decays into that certain final state; it is related to the partial and the total width through:
\begin{equation}
BR(H\rightarrow X_f)=\frac{\Gamma(H\rightarrow X_f)}{\sum_f \Gamma(H\rightarrow X_f)} \, .
\end{equation}
The SM prediction for the branching fractions of the different decay modes of the Higgs boson depends on the value of the Higgs-boson mass as it is shown in Figure~\ref{higgs_BR}. The branching fractions are reported as a function of the Higgs-boson mass over an extended mass range from 80 to 200~\GeV\ $(a)$ and in a zoomed range 5~$\GeV$ within the best-fit measured mass $(b)$, $120-130$~$\GeV$. Table~\ref{BR_table} reports the branching fractions for a SM Higgs boson with $M_H$ = 125.09~$\GeV$, \ie\ Run 1 ATLAS+CMS best-fit combined result~\cite{Run1_combined}. As a general rule, like it was made explicit in previous sections, the Higgs boson is more likely to decay into heavy fermions than light fermions, because of the fact that the strength of fermion interaction with the Higgs boson is proportional to fermion mass.
In case of a Higgs boson heavier than the one that was discovered in 2012 with a mass of $\sim$125~\GeV, the most common decay should be into a pair of $W$ or $Z$ bosons. However, given the measured mass, the SM predicts that the most common decay is into a $b\bar{b}$ pair ($H\rightarrow b\bar{b}$), accounting for $\sim$$58\%$ of the total decays. Due to the large QCD background, the gluon fusion production mode is really difficult to be detected but other production modes, like \VH, can be used to achieve the evidence for this decay channel.
$H\rightarrow WW^*$ represents the second most common decay mode, with a branching fraction of $\sim$$21\%$. The $W$ bosons subsequently decay into a quark and an antiquark or into a charged lepton and a high transverse momentum neutrino; the decays into quarks are difficult to distinguish from the background and the decays into leptons cannot be fully reconstructed due to the presence of neutrinos. A cleaner signal is given by the decay into a pair of $Z$ bosons when each of the bosons subsequently decays into a pair of charged leptons (electrons or muons) that are easy to be detected and result in almost no background contributions; despite the really low production rate, this channel is the so-called ``golden channel$"$, as it has the clearest and cleanest signature among all the possible decay modes and has a good invariant mass resolution (1-2\%).
\begin{table}[htbp]
\begin{center}
{\def\arraystretch{1.4}
\begin{tabular}{|c|c|}
\hline
Decay channel & Branching fraction\\
\hline
$H\rightarrow b\bar{b}$ & $5.81 \times 10^{-1}$\\
$H\rightarrow W^+W^-$ & $2.15 \times 10^{-1}$ \\
$H\rightarrow gg $ & $8.18 \times 10^{-2}$ \\
$H\rightarrow \tau^+\tau^-$ & $6.26 \times 10^{-2}$ \\
$H\rightarrow ZZ$ & $2.64 \times 10^{-2}$ \\
$H\rightarrow \gamma \gamma$ & $2.27 \times 10^{-3}$ \\
$H\rightarrow Z\gamma$ & $1.54 \times 10^{-3}$ \\
$H\rightarrow \mu^+\mu^-$ & $2.17 \times 10^{-4}$ \\
\hline
\end{tabular}}
\end{center}
\caption{Branching fractions for a SM Higgs boson with $M_H$ = 125.09 $\GeV$~\cite{LHC_XS_WG}.}
\label{BR_table}
\end{table}

Decays into massless gauge bosons (\ie\ gluons or photons) are also possible, but require intermediate loop of virtual heavy quarks (top or bottom) for gluons and photons, and massive gauge bosons ($W^\pm$ loops) for photons.
The most common process is the decay into a pair of gluons through a loop of virtual heavy particles occurring $\sim$$9\%$ of the times; it is really difficult to distinguish such a decay from the QCD background, typical of a hadron collider.\newline
The decay into a pair of photons, proceeding via loop diagrams with main contributions from $W$ boson and top quark loops, has a small branching fraction, $\sim$$0.23\%$, but provides the highest signal sensitivity to a SM Higgs boson signal followed by the $ZZ^*$ and $WW^*$ channels, due to two high energetic photons that form a very narrow invariant mass peak; at the same time, it has a good mass resolution (1-2\%).

\subsubsection{Double-Higgs production}
\begin{figure}[hbtp]
\centering
\begin{subfigure}[b]{0.49\textwidth}
\includegraphics[height=2.5 cm,width =6.5 cm]{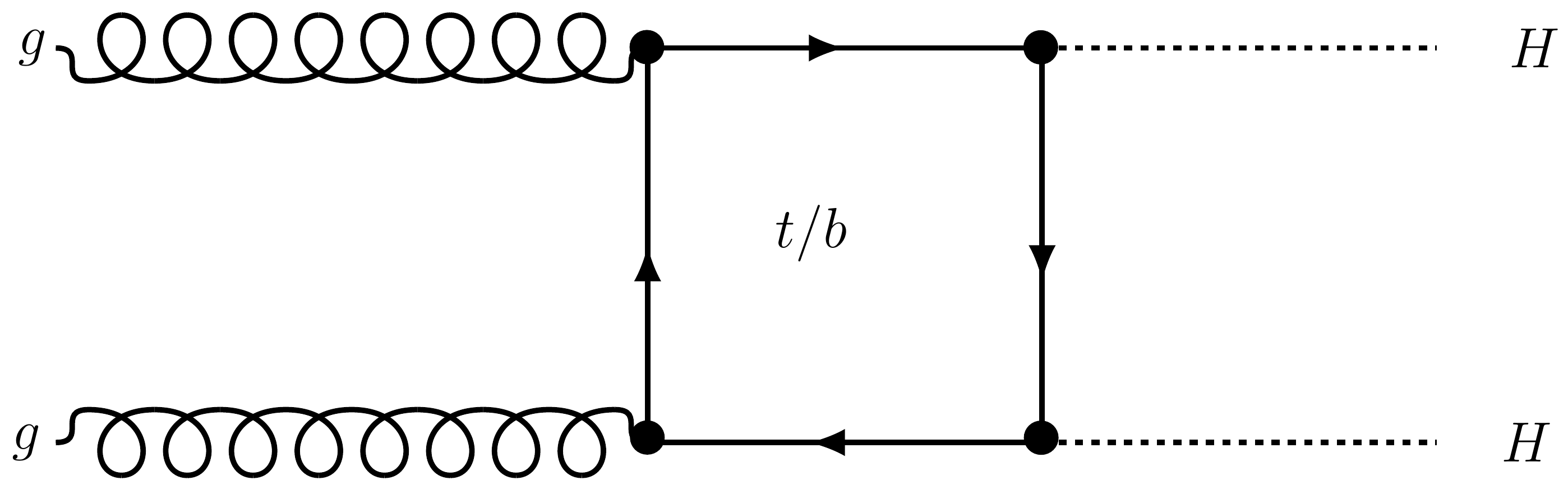}
 \caption{}
\end{subfigure}
\begin{subfigure}[b]{0.49\textwidth}
\includegraphics[height=2.5 cm,width =6.5 cm]{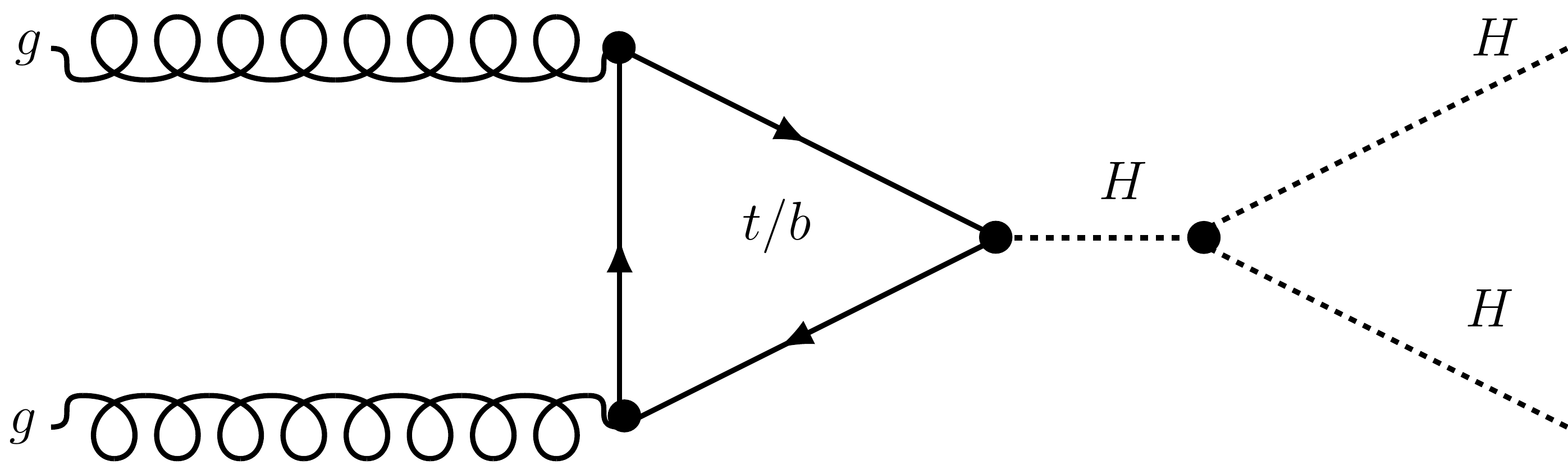}
 \caption{}
\end{subfigure}
\caption{Feynman diagrams for box (a) and triangle (b) topologies contributing to Higgs-boson pair production via gluon fusion at leading order.}
\label{HH_box_triangle}
\end{figure}

The main interest in the double-Higgs production comes from the fact that it provides information on the Higgs potential; in particular, it gives direct access to the Higgs cubic self-interaction and to the quartic couplings among two Higgs bosons and a pair of gauge bosons or of top quarks. At hadron colliders, Higgs pairs are dominantly produced via the following processes: gluon fusion (\ggF), vector-boson fusion (\VBF), associated production of Higgs pairs with a $W$ or a $Z$ boson ($VHH$) and $t\bar{t}HH$ associated production. While searches in the \ggF production mode are more sensitive to deviations in the Higgs self-interactions, the \VBF production mode is particularly sensitive to $c_{2V}$, \ie\ the quartic coupling between the Higgs bosons and vector bosons (di-vector-boson di-Higgs-boson $VVHH$). The $c_{2V}$ coupling is significantly constrained by ATLAS excluding a region that corresponds to $c_{2V}<$-1.02 and $c_{2V}>$2.71 thanks to a search for double-Higgs production via vector-boson fusion (VBF) in the $b\bar{b}b\bar{b}$ final state~\cite{vbf_hh}.\newline
The most relevant production is gluon fusion $gg\rightarrow HH$, accounting for more than 90\% of the total Higgs-boson pair production cross section and proceeding via virtual top and bottom quarks, \ie\ box and triangle diagrams, as shown in Figure~\ref{HH_box_triangle}, like single-Higgs \ggF\ production.

\begin{figure}[hbtp]
\centering
\begin{subfigure}[b]{0.49\textwidth}
\includegraphics[height=7 cm,width =8 cm]{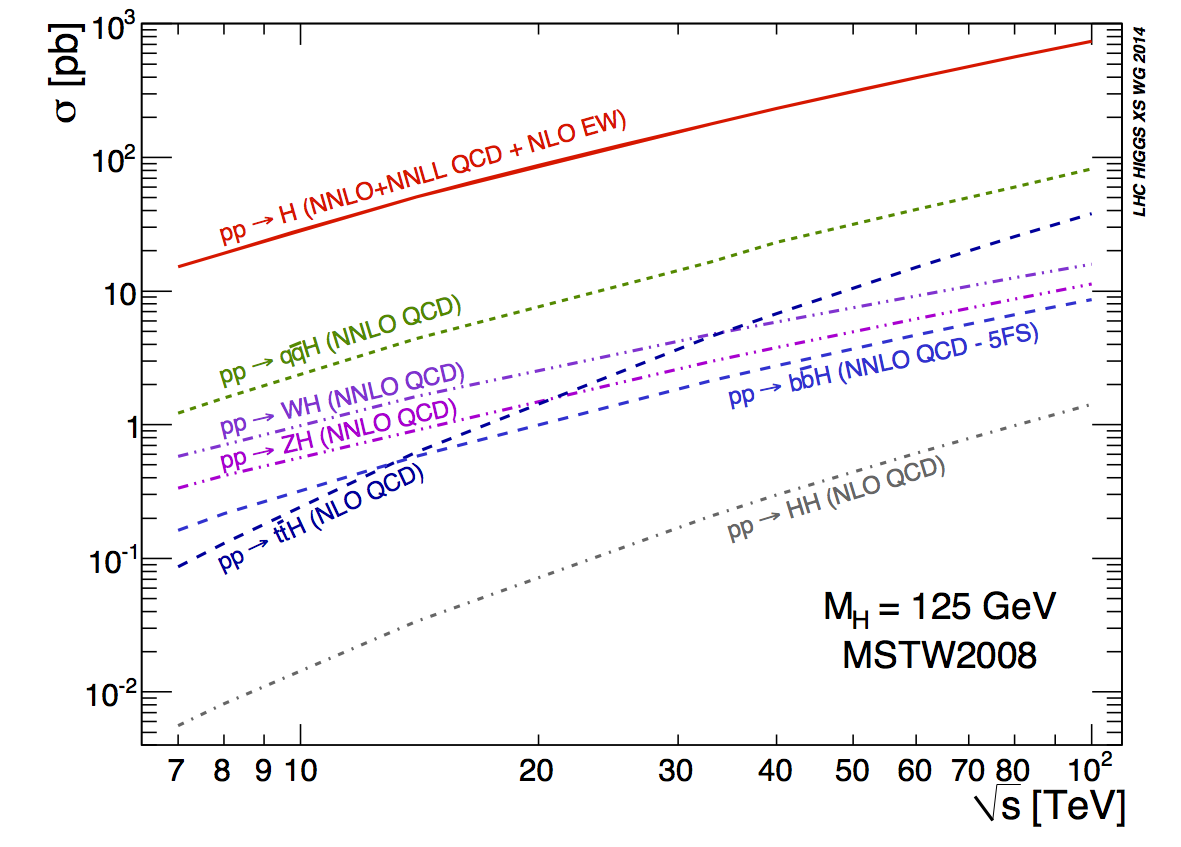}
 \caption{}
\end{subfigure}
\begin{subfigure}[b]{0.49\textwidth}
\includegraphics[height=7.3 cm,width =8 cm]{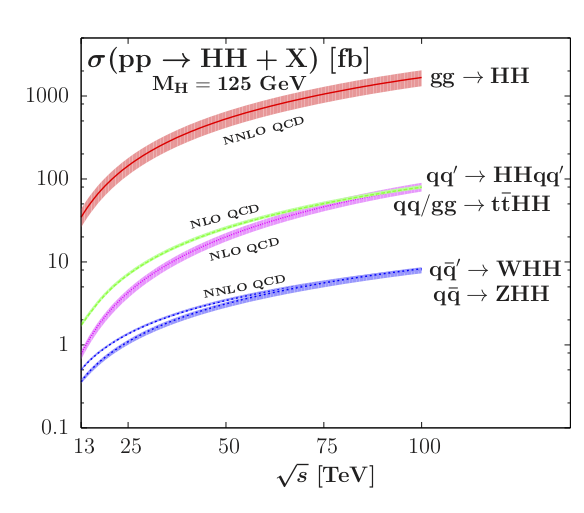}
 \caption{}
\end{subfigure}
\caption{Higgs-boson production cross sections as a function of centre-of-mass energies with $M_H$ = 125 $\GeV$ including double-Higgs production~\cite{LHC_XS_WG} (a); total cross sections for Higgs-pair production at a proton-proton collider in the main production channels as a function of the centre-of-mass energy with $M_H$ = 125 $\GeV$, including higher-order corrections~\cite{HH_XS} (b).}     
\label{HH_xs}
\end{figure}

The interference between the diagrams leads to the small cross-section value which is a thousand times smaller than the single-Higgs cross section as shown in Figure~\ref{HH_xs} $(a)$ reporting the cross sections of the different production modes including double-Higgs production. Figure~\ref{HH_xs} $(b)$ shows the current total cross sections for Higgs pair production at a proton-proton collider, including higher-order corrections.
The current best prediction for the inclusive \ggF cross section for Higgs-boson pair production, considering a Higgs boson with a mass $M_H= 125$ \GeV\ and a centre-of-mass energy of $\sqrt{s}=13$ TeV, is~\cite{lhc_xs_hh}:
\begin{equation}
\sigma_{pp\rightarrow HH}^{ggF}=31.05\; \text{fb} ^{(+2.2\%)}_{(-5.0 \%)} \;\text{(scale)}\; \pm 3.0 \% \; (\text{PDF}+\alpha_S) \; \pm 2.6 \% \; (\text{m}_{top} \; \text{unc})
\end{equation}
where ``scale$"$  stands for the QCD renormalisation and factorisation scale, ``PDF+$\alpha_S$$"$ stands for uncertainties on the PDFs and  on the $\alpha_S$ computation and ``m$_{top}$ unc$"$ represents the uncertainties related to missing finite top-quark mass effects~\cite{top_mass}.\newline
Table~\ref{HH_br} reports the branching fractions for the leading double-Higgs final states. The largest contribution comes from the $b\bar{b}b\bar{b}$ decay channel, accounting for $\sim$$34\%$ of the total decays but affected by a large QCD background.
\begin{table}[htbp]
\begin{center}
{\def\arraystretch{1.4}
\begin{tabular}{|c|c|}
\hline
Decay channel & Branching fraction\\
\hline
$HH\rightarrow b\bar{b}b\bar{b}$ & $3.37 \times 10^{-1}$ \\
$HH\rightarrow b\bar{b}W^+W^-$ & $2.50 \times 10^{-1}$ \\
$HH\rightarrow b\bar{b}\tau^+\tau^-$ & $7.27 \times 10^{-2}$ \\
$HH\rightarrow W^+W^- W^+W^-$ & $4.63 \times 10^{-2}$ \\
$HH\rightarrow b\bar{b}\gamma\gamma$ & $2.64 \times 10^{-3}$ \\
$HH\rightarrow W^+W^- \gamma \gamma$ & $9.77 \times 10^{-4}$\\
\hline
\end{tabular}}
\end{center}
\caption{Double-Higgs branching fractions considering a Higgs boson with $M_H=125.09$~$\GeV$.}
\label{HH_br}
\end{table} 

\begin{figure}[H]
\begin{center}
\includegraphics[height=9 cm,width =12 cm]{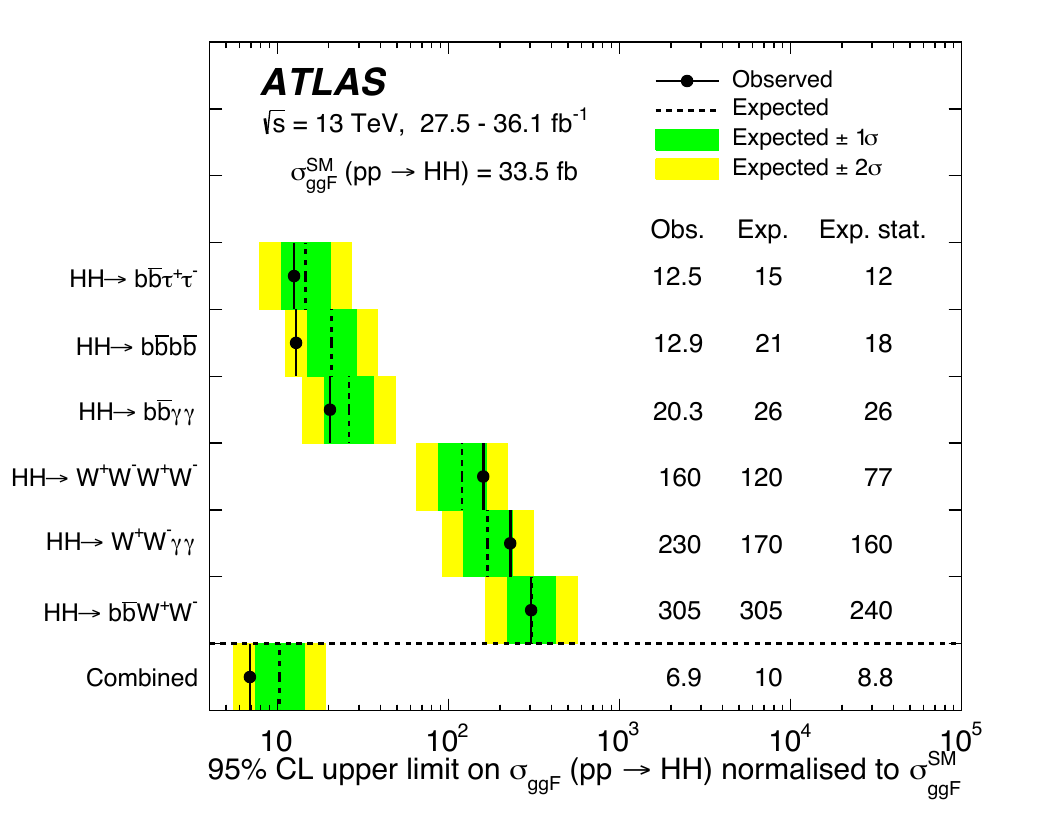}
\end{center}
\caption{Upper limits at 95\% CL on the cross section of the \ggF SM HH production normalised to its SM expectation $\sigma^{\ggF}_{SM}(pp\rightarrow HH)$ from the $b\bar{b}b\bar{b}$, $b\bar{b}W^+W^-$, $b\bar{b}\tau^+\tau^-$, $W^+W^-W^+W^-$, $b\bar{b}\gamma \gamma$ and $W^+W^-\gamma \gamma$ searches, and their statistical combination. The column ``Obs.$"$ lists the observed limits, ``Exp.$"$ the expected limits with all statistical and systematic uncertainties, and ``Exp. stat.$"$ the expected limits obtained including only statistical uncertainties in the fit~\cite{Paper_hh}.}     
\label{ggF_hh}
\end{figure}

The most sensitive final states are chosen according to a compromise between the largeness of the Higgs branching fractions and their cleanliness with respect to the backgrounds~\cite{white_paper}.\newline
Thus they involve one Higgs boson decaying into a pair of $b$-quarks and one decaying into either two tau-leptons ($HH\rightarrow b\bar{b}\tau^+\tau^-$), another pair of $b$-quarks ($HH\rightarrow b\bar{b}b\bar{b}$) or two photons ($HH\rightarrow b\bar{b}\gamma\gamma$).
Despite the low branching fraction, $\sim$$0.26 \%$, the sensitivity of the $b\bar{b}\gamma \gamma$ final state arises from the fact that it has a clean signal and an excellent diphoton mass resolution due to the small background.\newline
Latest results from the ATLAS experiment setting limits on the gluon fusion $gg\rightarrow HH$ production process exploiting up to 36.1 fb$^{-1}$ of proton-proton collision data, have been produced combining six analyses searching for Higgs boson pairs in the $b\bar{b}b\bar{b}$, $b\bar{b}W^+W^-$, $b\bar{b}\tau^+\tau^-$, $W^+W^-W^+W^-$, $b\bar{b}\gamma \gamma$ and $W^+W^-\gamma \gamma$ final states. Upper limits at the 95\% confidence level are shown in Figure~\ref{ggF_hh}: the combined observed (expected) limit at 95\% confidence level on the non-resonant Higgs-boson pair production cross section is 6.9 (10) times the predicted SM cross section~\cite{Paper_hh}.

\clearpage
\section{Higgs-Boson Property Measurements}
\label{Higgs_boson_property}

\subsubsection{Higgs-boson mass measurements}
\begin{figure}[hbtp]
\centering
\begin{subfigure}[b]{0.49\textwidth}
\includegraphics[height=7.4 cm,width =8 cm]{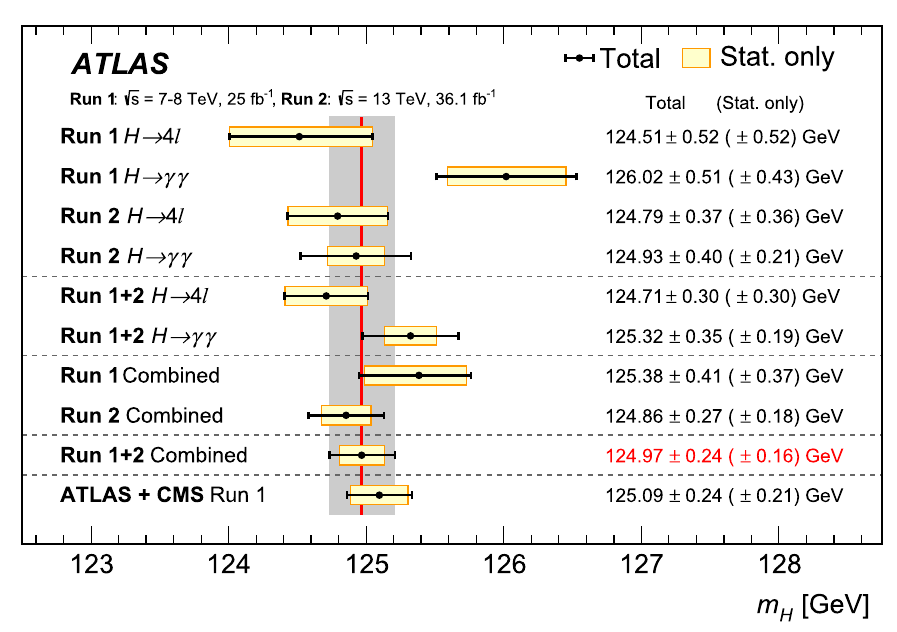}
 \caption{}
\end{subfigure}
\begin{subfigure}[b]{0.49\textwidth}
\includegraphics[height=7.6 cm,width =8 cm]{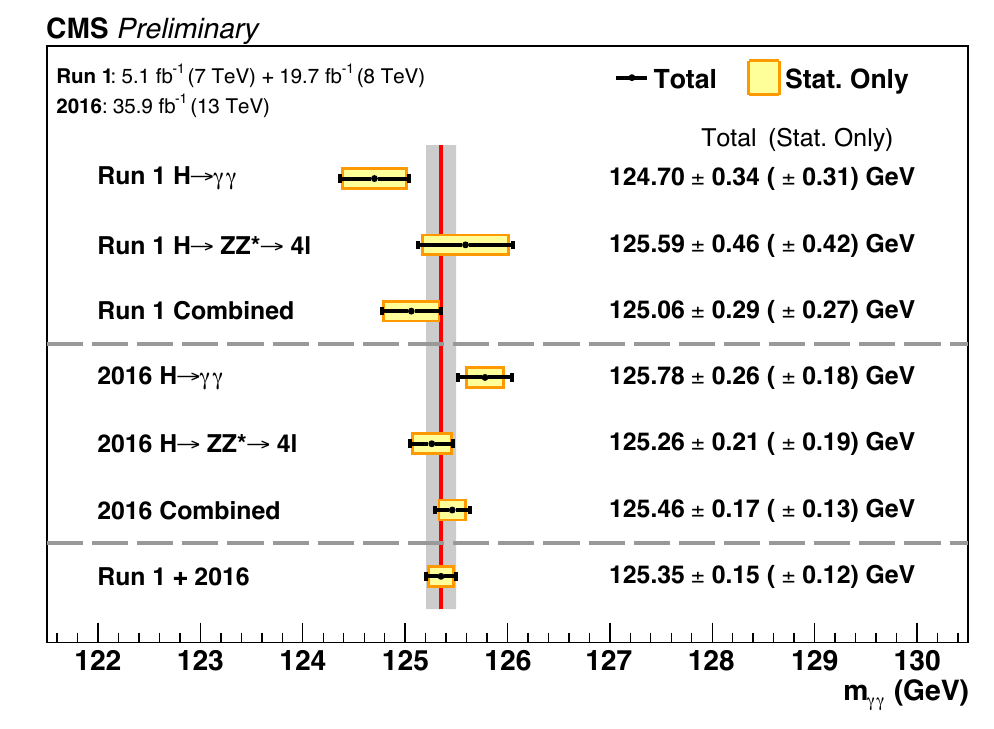}
 \caption{}
\end{subfigure}
\caption{Summary of the Higgs-boson mass measurements from the individual and combined analyses, compared to the combined Run 1 measurement by ATLAS and CMS. The statistical-only (horizontal yellow-shaded bands) and total (black error bars) uncertainties are indicated. The (red) vertical line and corresponding (grey) shaded column indicate the central value and the total uncertainty of the combined ATLAS Run 1+2 measurement $(a)$ and CMS Run 1+2 measurement $(b)$~\cite{combination_mass, combination_mass_cms}.}     
\label{mass_higgs}
\end{figure}

In order to measure the mass of the Higgs boson, the ATLAS and CMS experiments rely on the two high mass resolution and sensitive channels, $\gamma \gamma$ and $ZZ^*$, with a typical resolution of 1-2\%, while the other channels have significantly worse resolutions up to $\sim$20\%. 
The results from each of the four individual measurements, as well as various combinations, along with the LHC Run 1 result, are summarised in Figure~\ref{mass_higgs} for both experiments.

The combination of CMS Run 1 and Run 2 measurements leads to a mass value~\cite{combination_mass_cms}:
\begin{equation}
M_H= 125.35 \pm 0.15 \; \text{GeV}= 125.35 \pm 0.12 \; \text{(stat.)} \pm 0.09\; \text{(syst.)}\; \text{GeV}
\end{equation}
where ``stat.$"$ stands for the statistical uncertainty and ``syst.$"$ for systematic uncertainties. The combination of the ATLAS Run 1 and Run 2 measurements yields a mass~\cite{combination_mass}:
\begin{equation}
M_H= 124.97 \pm 0.24 \; \text{GeV}= 124.97 \pm 0.16 \; \text{(stat.)} \pm 0.18\; \text{(syst.)}\; \text{GeV} \, .
\end{equation}
The CMS mass measurements represent the most precise $M_H$ to date.
\subsubsection{Higgs-boson width measurements}
In the Standard Model, the Higgs-boson width is very precisely predicted once the Higgs-boson mass is known. For a Higgs boson with a mass of $125$ GeV, the width is 4.1~MeV~\cite{Higgs_CS}. It is dominated by the fermionic decay partial width at approximately 75\%, while the vector-boson modes are suppressed and contribute at 25\% only.\newline
Direct on-shell measurements of the Higgs-boson width are limited by detector resolution and have much larger errors than the expected SM width, reaching a sensitivity of $\sim$1~GeV. Indirect measurements exploiting off-shell production of the Higgs boson have a substantial cross section at the LHC, due to the increased phase space as the vector bosons ($V=W, Z$) and top-quark decay products become on-shell with the increasing energy scale~\cite{width_ATLAS}.
Both ATLAS and CMS have exploited the combination of on- and off-shell measurements to set the best limits on the Higgs-boson width.
The ATLAS limits, determined using $ZZ^*\rightarrow 4\ell$ and $ZZ^*\rightarrow 2\ell 2\nu$ final states using data corresponding to an integrated luminosity of $36.1$~fb$^{-1}$, are~\cite{width_ATLAS}:
\begin{equation}
\Gamma_H < 14.4 \; \text{MeV} \, .
\end{equation}
The CMS limits for the Higgs-boson width from on-shell and off-shell Higgs boson production in the four-lepton final state using an integrated luminosity of $80.2$~fb$^{-1}$, under the assumption of SM-like couplings, are~\cite{width_CMS}:
\begin{equation}
0.08 <\Gamma_H < 9.16 \; \text{MeV} \, .
\end{equation}
The CMS lower bound on the Higgs width comes from the different fit procedure that has been used with respect to ATLAS measurement, \ie\ profile-likelihood technique vs CLs method, respectively, explained in Chapter~\ref{sec:stat}.
\subsubsection{Higgs-boson coupling and signal-strength measurements}
\begin{figure}[hbtp]
\centering
\begin{subfigure}[b]{0.49\textwidth}
\includegraphics[height=10 cm,width =8 cm]{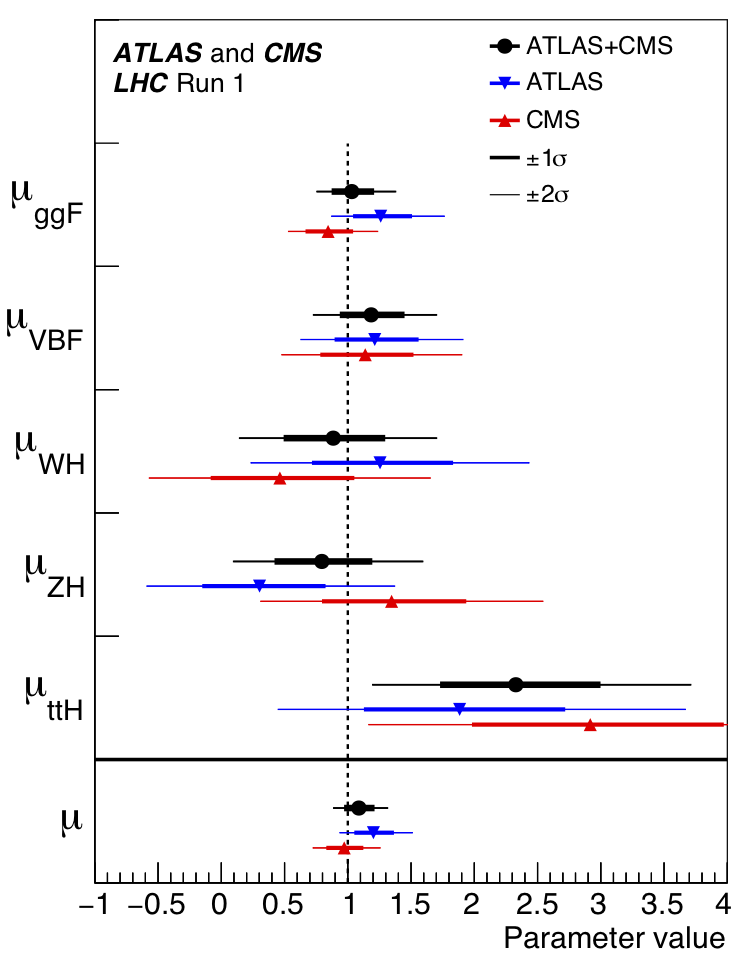}
 \caption{}
\end{subfigure}
\begin{subfigure}[b]{0.49\textwidth}
\includegraphics[height=9.95 cm,width =8 cm]{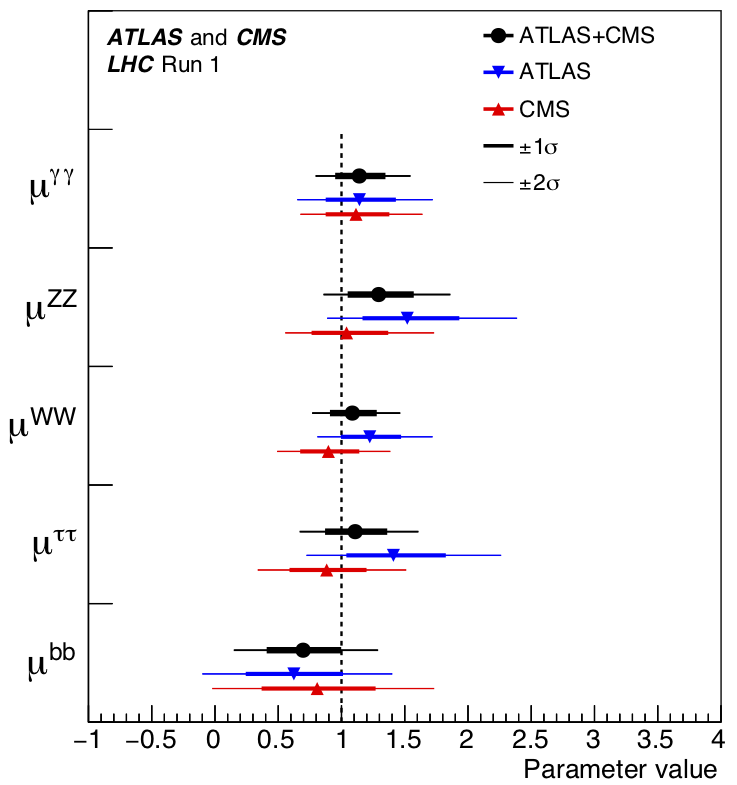}
 \caption{}
\end{subfigure}
\caption{Best-fit results for the production (a) and decay (b) signal strengths for the combination of ATLAS and CMS Run 1 data. Results from each experiment together with the global signal strength $\mu$ are also shown. The error bars indicate the $1\sigma$ (thick lines) and $2\sigma$ (thin lines) intervals~\cite{Coupling_run1}.}     
\label{production_decay_run1}
\end{figure}

The Higgs-boson cross sections and branching fractions are often presented in terms of the modifier $\mu$, called ``signal strength$"$ and defined as the ratio of the measured Higgs-boson yield, \ie the total cross section times the branching fraction, to its SM expectation value:
\begin{equation}
\mu=\frac{\sigma\times BR}{\sigma_{SM} \times BR_{SM}} \, .
\end{equation}
For a specific production mode $i$ and decay final state $f$, the signal strengths $\mu_{if}$ are defined as:
\begin{equation}
\mu_{if}=\frac{\sigma_i}{\sigma_i^{SM}} \times \frac{BR_f}{BR_f^{SM}}
\end{equation}
where $i$=\ggF, \VBF, \WH, \ZH, $t\bar{t}H$ production modes and $f=\gamma \gamma, \, ZZ^*, \, WW^*, \, \tau^+\tau^-, \, b\bar{b}$ decay channels. In the SM hypothesis, $\mu_i=\mu_f=1$.\newline
The best-fit value of the global signal strength obtained by ATLAS and CMS with the full Run 1 dataset is~\cite{Coupling_run1}:
\begin{equation}
\mu = 1.09^{+0.11}_{-0.10}=1.09^{+0.07}_{-0.07}\; \text{(stat.)}^{+0.04}_{-0.04} \; \text{(exp.)}^{+0.07}_{-0.06}\; \text{(sig. th.)}^{+0.03}_{-0.03}\; \text{(bkg. th.)}
\end{equation}
where ``stat.$"$ stands for the statistical uncertainty, ``sig. th.$"$ and ``bkg. th.$"$ account for signal theory and background theory uncertainties, respectively. Finally, ``exp.$"$ contains the contributions of all the experimental systematic uncertainties.\newline
Figure~\ref{production_decay_run1} shows the best-fit results for the production $(a)$ and decay $(b)$ signal strengths for the Run 1 combination of ATLAS and CMS data.

\begin{figure}[H]
\begin{center}
\includegraphics[height=12 cm,width =9 cm]{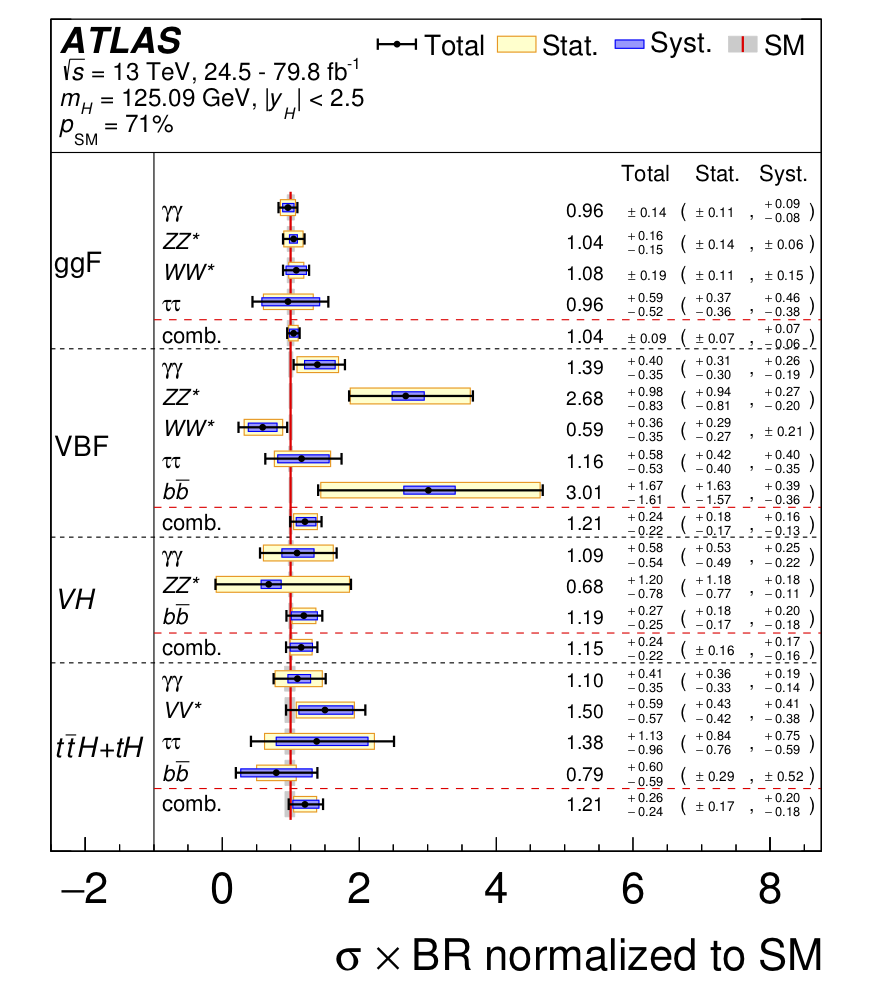}
\end{center}
\caption{Cross section times branching fraction for $ggF$, $VBF$, $VH$ and $t\bar{t}H+tH$ production in each relevant decay mode normalised to their SM predictions, measured by the ATLAS experiment. The values are obtained from a simultaneous fit to all channels. The cross sections of the $ggF$, $H\rightarrow b\bar{b}$, $VH$, $H\rightarrow WW^*$ and $VH$, $H\rightarrow \tau^+\tau^-$ processes are fixed to their SM predictions. Combined results for each production mode are also shown, assuming SM values for the branching fractions into each decay mode~\cite{Coupling_run2}.}     
\label{sigma_BR_run2}
\end{figure}

\begin{figure}[htbp]
\begin{center}
\includegraphics[height=8 cm,width =9 cm]{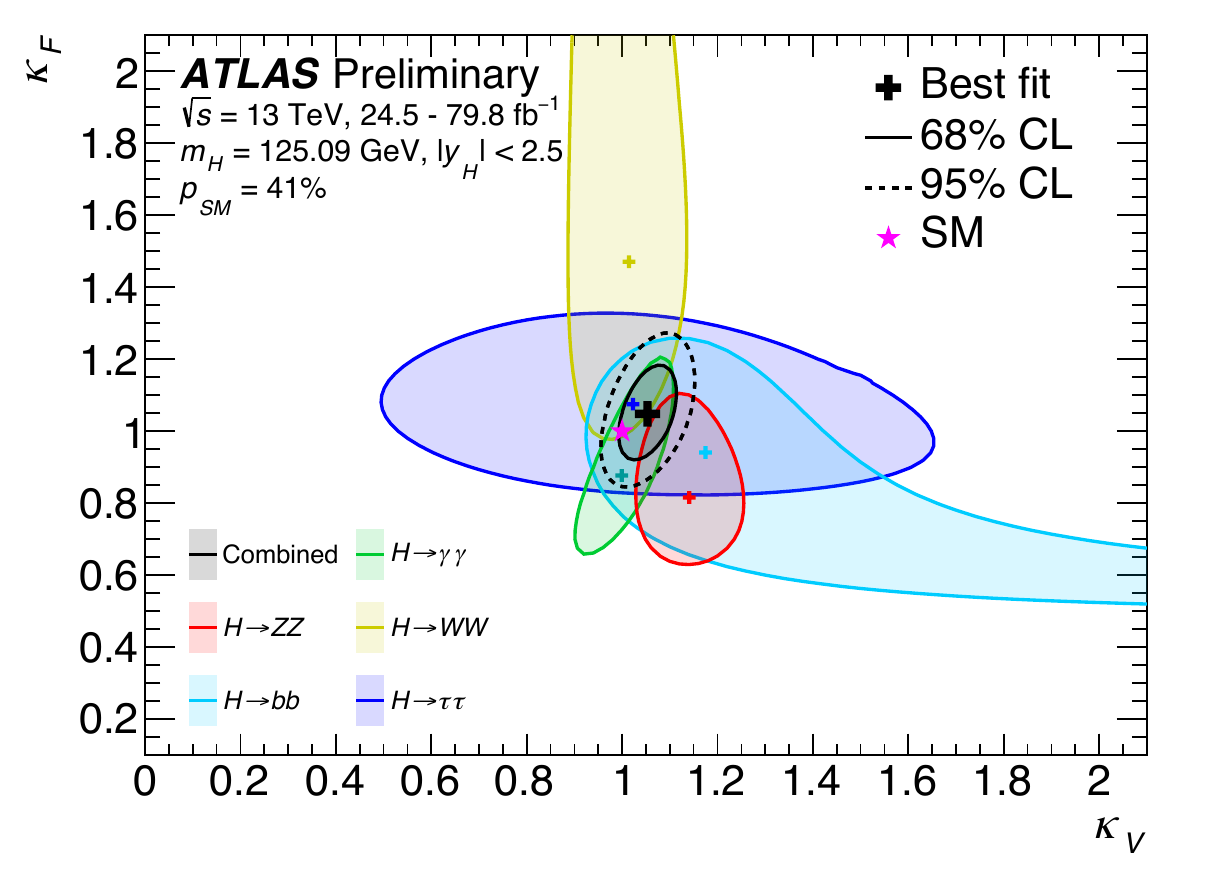}
\end{center}
\caption{Negative log-likelihood contours at 68\% and 95\% confidence level in the ($\kappa_V^f, \kappa_F^f$) plane for the individual decay modes and their combination (black) assuming the coupling strengths to fermions and vector bosons to be positive. No contribution from invisible or undetected Higgs boson decays is assumed. The best-fit value for each measurement is indicated by a cross while the SM hypothesis is indicated by a star~\cite{Coupling_run2}.}     
\label{contour_run2}
\end{figure}

Preliminary Run 2 measurements of Higgs-boson production cross sections and branching fractions have been performed using up to 79.8 fb$^{-1}$ of proton-proton collision data produced by the LHC at a centre-of-mass energy of $\sqrt{s}$ = 13 TeV and recorded by the ATLAS detector. The best-fit value of the global signal strength obtained by ATLAS is:
\begin{equation}
\mu=1.11^{+0.09}_{-0.08}=1.11^{+0.05}_{-0.05}\; \text{(stat.)}^{+0.05}_{-0.04} \; \text{(exp.)}^{+0.05}_{-0.04}\; \text{(sig. th.)}^{+0.03}_{-0.03}\; \text{(bkg. th.)} \, ;
\end{equation}
the standalone ATLAS measurement with the partial Run 2 dataset is already better than the combined ATLAS and CMS Run 1 result, mainly due to the reduction of statistical uncertainties.
Figure~\ref{sigma_BR_run2} shows the signal strengths $\mu_{if}$ with $i$=\ggF, \VBF, \VH and $t\bar{t}H + tH$ production in each relevant decay mode $f=\gamma \gamma,\, ZZ^*, \,WW^*, \, \tau^+ \tau^-, \, b\bar{b}$ using a luminosity of up to $79.8\; \text{fb}^{-1}$ recorded with the ATLAS detector. The values are obtained from a simultaneous fit to all channels. No significant deviation from the Standard Model predictions is observed.
In order to parameterise the Higgs coupling deviations from the SM, a simple parameterisation (the so called $\kappa$-framework) has been introduced in Reference~\cite{k_framework}, based on the leading-order contributions to each production and decay modes; using the zero-width approximation, the signal cross section can be decomposed in the following way for all channels:
\begin{equation}
(\sigma \cdot BR)(i\bar{i}\rightarrow H\rightarrow f\bar{f})=\frac{\sigma_{i\bar{i}}\cdot \Gamma_{f\bar{f}}}{\Gamma_H}
\end{equation}
where $\sigma_{i\bar{i}}$ is the production cross section through the initial state $i\bar{i}$, $\Gamma_{f\bar{f}}$ the partial decay width into the final state $f\bar{f}$ and $\Gamma_H$ the total width of the Higgs boson. Higgs-boson production cross sections and decay rates for each process are thus parameterised via coupling-strength modifiers $\kappa$ in the following way:
\begin{equation}
 \kappa_i^2=\frac{\sigma_{i\bar{i}}}{\sigma_{i\bar{i}}^{SM}}\qquad \text{or} \qquad \kappa_f^2=\frac{\Gamma_{f\bar{f}}}{\Gamma_{f\bar{f}}^{SM}} \, .
\end{equation}
The SM expectation corresponds by definition to $\kappa_i=\kappa_f=1$.\newline
Leading-order-coupling-scale-factor relations for Higgs-boson cross sections and partial-decay widths, relative to the SM and used in the results reported in this thesis, are reported in Table~\ref{k_coupling_production_gamma}.
\begin{table}[htbp]
\begin{center}
 \renewcommand{\arraystretch}{1.4}
\begin{tabular}{|c|c|}
\hline
Production Mode & Resolved modifiers\\
\hline
$\sigma(ggF$) & $1.04\,\kappa_t^2+0.002\,\kappa_b^2-0.04\,\kappa_t\kappa_b$\\
$\sigma(VBF$) & $0.73\,\kappa_W^2 + 0.27\,\kappa_Z^2$\\
$\sigma(qq/qg\rightarrow ZH$) & $\kappa_Z^2$ \\
$\sigma(gg\rightarrow ZH$) & $2.46\,\kappa_Z^2 +0.46\,\kappa_t^2 -1.90\,\kappa_Z\kappa_t$\\
$\sigma(WH$) & $\kappa_W^2$\\
$\sigma(t\bar{t}H$) & $\kappa_t^2$\\
$\sigma(tHW$) & $2.91\,\kappa_t^2 + 2.31\,\kappa_W^2 -4.22\, \kappa_t\kappa_W$\\
$\sigma(tHq$) & $2.63\, \kappa_t^2 +3.58\, \kappa_W^2-5.21\,\kappa_t\kappa_W$ \\
$\sigma(b\bar{b}H$) & $\kappa_b^2$\\
\hline
Partial decay width & Resolved modifiers\\
\hline
$\Gamma^{bb}$ & $\kappa_b^2$ \\
$\Gamma^{WW}$ & $\kappa_W^2$ \\
$\Gamma^{gg}$ & $1.11 \,\kappa_t^2+0.01\,\kappa_b^2 -0.12\,\kappa_t\kappa_b$ \\
$\Gamma^{\tau\tau}$ & $\kappa_b^2$ \\
$\Gamma^{ZZ}$ & $\kappa_Z^2$ \\
$\Gamma^{cc}$ & $\kappa_c^2(=\kappa_t^2)$ \\
$\Gamma^{\gamma\gamma}$ & $1.59\,\kappa_W^2+0.07\,\kappa_t^2 -0.67\,\kappa_W\kappa_t$ \\
$\Gamma^{Z\gamma}$ & $1.12\,\kappa_W^2 -0.12 \kappa_W\kappa_t$ \\
$\Gamma^{ss}$ & $\kappa_s^2(=\kappa_b^2)$ \\
$\Gamma^{\mu\mu}$ & $\kappa_\mu^2$ \\
\hline
\end{tabular}
\end{center}
\caption{Parameterisations of Higgs-boson production cross sections $\sigma_i$ and partial decay widths $\Gamma_f$, normalised to their SM values, as functions of the coupling-strength modifiers $\kappa$~\cite{Coupling_run2}.}
\label{k_coupling_production_gamma}
\end{table}
The ratio of the observed couplings to the SM expectation is conventionally indicated by $\kappa_V$ for vector bosons and $\kappa_F$ for fermions.\newline
Figure~\ref{contour_run2} shows the results of the combined fit in the ($\kappa_V, \kappa_F$) plane as well as the contributions of the individual decay modes. Both coupling modifiers $\kappa_V$ and $\kappa_F$ have been measured to be compatible with the SM expectation.

\subsection{Higgs self-coupling}

\begin{figure}[H]
\begin{center}
\includegraphics[height=8 cm,width =12 cm]{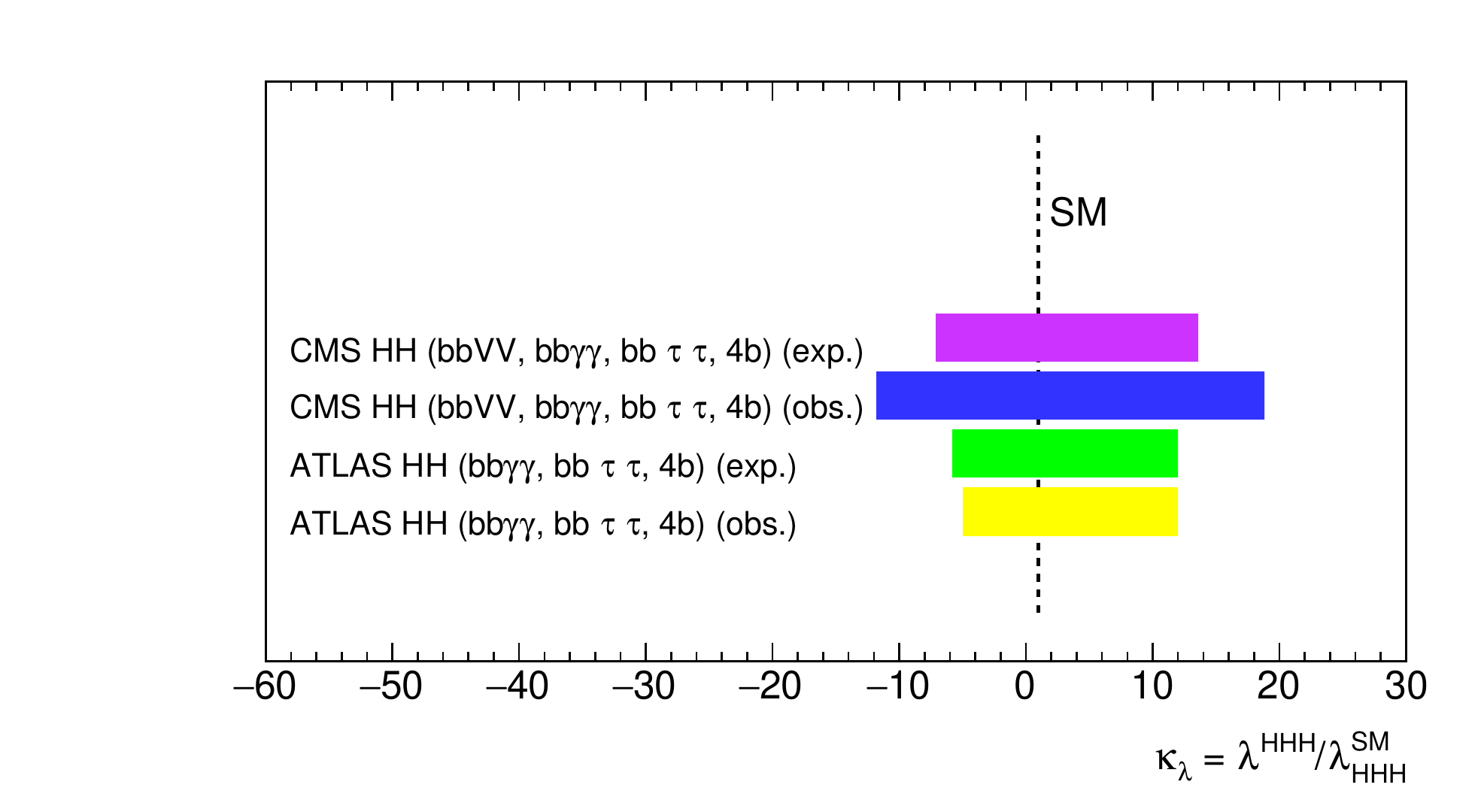}
\end{center}
\caption{Summary of recent constraints on the Higgs self-coupling by the ATLAS and CMS experiments~\cite{Paper_hh, Paper_hh_CMS}.}     
\label{kl_constraints}
\end{figure}

One of the most important targets of the LHC is to improve the experimental results of the Run 1 and the complete exploration of the properties of the Higgs boson, in particular the self-interactions. This is the only way to reconstruct the scalar potential of the Higgs doublet field $\phi$, that is responsible for spontaneous electroweak symmetry breaking,
\begin{equation}
V_H=\mu^2\phi^\dagger\phi +\frac{1}{2}\lambda (\phi^\dagger\phi)^2 \qquad \lambda=\frac{M_H^2}{v^2} \qquad \mu^2=-\frac{1}{2}M_H^2
\end{equation}
with $\nu= 246$ GeV. In the SM, the potential is fully determined by only two parameters, the vacuum expectation value, $v = (\sqrt{2}G_F)^{-1/2}$, and the coefficient of the ($\Phi^\dagger\Phi)^2$ interaction, $\lambda$. Considering the Standard Model an effective theory, $\lambda$ stands for two otherwise free parameters, the trilinear ($\lambda_{HHH}$) and the quartic ($\lambda_{HHHH}$) self-couplings:
\begin{equation}
\lambda_{HHH}\;  (\text{or}\; \lambda_3)=\frac{3M_H^2}{\nu} \; , \qquad \lambda_{HHHH}\;  (\text{or}\; \lambda_4)=\frac{3M_H^2}{\nu^2} \, .
\end{equation}
The self-couplings determine the shape of the potential which is connected to the phase transition of the early universe from the unbroken to the broken electroweak symmetry.\newline
Large deviations of the trilinear and quartic couplings, $\lambda_3$ and $\lambda_4$, are possible in scenarios beyond the SM predictions (BSM).
As an example, in two-Higgs doublet models where the lightest Higgs boson is forced to have SM-like couplings to vector bosons, quantum corrections may increase the trilinear Higgs-boson coupling by up to 100\%~\cite{trilinear}. 
Examples of two-Higgs doublet models modifying the value of the trilinear Higgs coupling are the Gildener-S.Weinberg (GW)~\cite{GW} models of electroweak symmetry breaking: they are based on an extension of Coleman-Weinberg~\cite{Coleman} theory of radiative corrections as the origin of spontaneous symmetry breaking, and involve a broken scale symmetry to generate a light Higgs boson in addition to a number of heavy bosons. The scalar couplings can acquire values larger than in the Standard Model at one-loop level of the Coleman-E.Weinberg expansion. In a two-Higgs doublet model of the GW mechanism, the trilinear Higgs self-coupling $\lambda_{HHH}$ is typically $1.5-3.0$ times its SM value~\cite{HDBS_plenary}.\newline
Anomalous Higgs-boson self-couplings also appear in other BSM scenarios, such as models with a composite Higgs boson~\cite{trilinear_1}, or in Little-Higgs models~\cite{trilinear_3,trilinear_4,trilinear_5}. \newline
The trilinear Higgs self-coupling can be probed directly in searches for multi-Higgs final states and indirectly via its effect on precision observables or loop corrections to single-Higgs production; the quartic self-coupling instead, being further suppressed by a power of $\nu$ compared to the trilinear self-coupling, is currently not accessible at hadron colliders~\cite{quartic}. \newline
Preliminary Run 2 results of the Higgs self-coupling from direct searches for Higgs pairs of the ATLAS and CMS collaborations have been performed using up to 36.1 fb$^{-1}$ and 35.9~fb$^{-1}$ of proton-proton collision data produced by the LHC at a centre-of-mass energy of $\sqrt{s}$ = 13 TeV and recorded by the ATLAS and CMS detectors, respectively.  Results are reported in terms of the ratio of the Higgs-boson self-coupling to its SM expectation, \ie $\kappa_\lambda = \lambda_{HHH}/\lambda_{HHH}^{SM}$.
Latest constraints coming from the combination of the most sensitive final states, \ie\ $b\bar{b}\tau^+\tau^-$,  $b\bar{b}b\bar{b}$ and $b\bar{b}\gamma\gamma$ (and $b\bar{b}VV$ for CMS), are shown in Figure~\ref{kl_constraints} and Table~\ref{tab:klambda_hh} where the limits from single channels are reported.
Details on the channels used in the ATLAS combination and the methodology exploited in order to extract $\kappa_\lambda$ intervals are reported in Chapter \ref{sec:dihiggs}.\newline
The best final states for the $\kappa_\lambda$ limit are the $b\bar{b}\tau^+\tau^-$ and $b\bar{b}\gamma \gamma$ channels for ATLAS and CMS, respectively.
Differences between ATLAS and CMS sensitivities in each channel come from different optimisations of the analysis strategies.
\begin{table}[h]
\begin{center}
{\def\arraystretch{1.4}
\begin{tabular}{|c|c|c|c|}
\hline
Channels & Collaboration & $\kappa_\lambda$  [95\% CL] (obs.) & $\kappa_\lambda$  [95\% CL] (exp.)\\ 
\hline
\multirow{2}{*}{$HH\rightarrow b\bar{b}\tau^+\tau^-$}  & ATLAS~\cite{Paper_hh} &  $[-7.4, 15.7]$ &  $[-8.9, 16.8]$ \\
                                      &  CMS~\cite{white_paper} &$[-18, 26]$ &  $[-14, 22]$ \\
\multirow{2}{*}{$HH\rightarrow b\bar{b}b\bar{b}$}  & ATLAS~\cite{Paper_hh} & $[-10.9, 20.1]$ &  $[-11.6, 18.8]$ \\
                                          &  CMS~\cite{white_paper} &$[-23, 30]$ &  $[-15, 23]$ \\
                               
\multirow{2}{*}{$HH\rightarrow b\bar{b}\gamma \gamma$} & ATLAS~\cite{Paper_hh} &  $[-8.1, 13.1]$ &  $[-8.1, 13.1]$ \\
                                     &  CMS~\cite{white_paper} &$[-11, 17]$ &  $[-8.0, 11.4]$ \\
 \hline
\multirow{2}{*}{Combination}  & ATLAS~\cite{Paper_hh} &  $[-5.0, 12.0]$ &  $[-5.8, 12.0]$ \\
                                        &  CMS~\cite{Paper_hh_CMS} &$[-11.8, 18.8]$ &  $[-7.1, 13.6]$ \\
\hline
\end{tabular}
}
\caption{Allowed $\kappa_\lambda$ intervals at 95\% CL for the $b\bar{b}\tau^+\tau^-$,  $b\bar{b}b\bar{b}$ and $b\bar{b}\gamma\gamma$ final states and their combination for both ATLAS and CMS experiments. The column ``obs.$"$ lists the observed results while the column ``exp.$"$ reports the expected results obtained including all statistical and systematic uncertainties in the fit~\cite{white_paper,Paper_hh,Paper_hh_CMS}.}
\label{tab:klambda_hh}
\end{center}
\end{table}
\begin{figure}[H]
\begin{center}
\includegraphics[height=9 cm,width =13 cm]{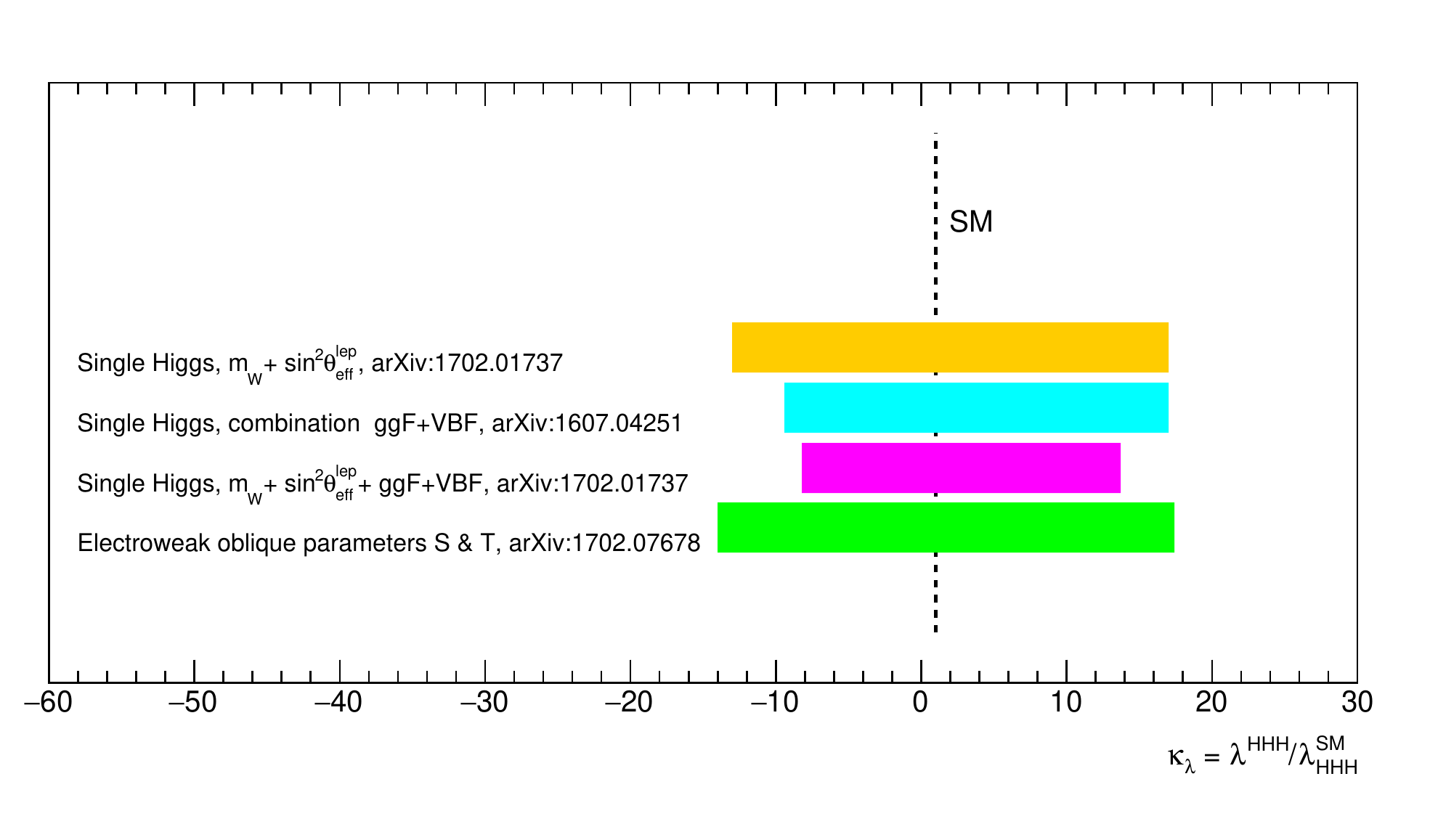}
\end{center}
\caption{Summary of constraints on the Higgs self-coupling from precise observable measurements~\cite{Degrassi_precision, EWK_kl,Degrassi}.}     
\label{kl_precision}
\end{figure}
Figure~\ref{kl_precision} shows constraints on the trilinear Higgs self-coupling from precision observables, like the mass of the $W$ boson, $m_W$, the effective Weinberg angle, $sin^2\theta_{\text{eff}}^{\text{lep}}$~\cite{Degrassi_precision}, the electroweak oblique parameters~\cite{EWK_kl}, and loop corrections to single-Higgs production, the best $\kappa_\lambda$ interval coming from the combination of \ggF and \VBF production mode~\cite{Degrassi}. 
Theoretical models describing the extraction of the Higgs self-coupling either from double-Higgs production measurements or from single-Higgs production measurements are reported in Chapter~\ref{sec:prob_self}.\newline
This thesis is dedicated to the improvement of experimental constraints on the Higgs-boson self-coupling with the ATLAS detector. These results are presented in Chapters~\ref{sec:dihiggs},~\ref{sec:single},~\ref{sec:combination}.

\section{The Standard Model: successes and open issues}
\label{SM_success_open}
The discovery of the Higgs boson by the ATLAS~\cite{Higgs_Atlas} and CMS~\cite{Higgs_CMS} experiments in 2012 is considered as the last milestone in the long history of the Standard Model of particle physics, a highly predictive and rigorously tested model that has been validated with an excellent level of accuracy throughout the years and shows an impressive agreement between theory prediction and experimental measurements. Among the successes of the SM, it has to be underlined that all the particles the SM predicted have been observed, including the $W$ and $Z$ bosons, as well as the top and bottom quarks and the Higgs boson. Furthermore, other successes are related to predictions of particle properties, like the electron ``anomalous$"$ magnetic dipole moment, which is one of the most accurately measured properties of an elementary particle, and one of the properties of a particle that can be most accurately predicted by the SM. \newline
However, at the same time, there are indications of the incompleteness of the SM that cannot be explained in terms of minor or negligible deviations of some measured observables from their theory predictions due to insufficient precision of the measurements or of the theoretical calculations.
Here is a list of the main issues remaining opened in particle physics:
\begin{itemize}
\item According to the SM, neutrinos are massless particles; however, there are experimental evidences, \ie\ neutrino oscillations, predicted by Pontecorvo in 1957 and observed for the first time in 1998, that prove the fact that neutrinos do have mass. Neutrino mass terms can be added introducing at least nine more parameters: three neutrino masses, three real mixing angles, and three CP-violating phases.
\item The SM does not explain why fundamental particles are divided in three generations of leptons and three of quarks with properties that are very similar to the first generation, as well as it does not explain the hierarchy of the Yukawa couplings.
\item The SM has 18 free parameters, \ie\  3 lepton masses, 6 quark masses, 3 CKM angles and 1 CKM CP-violation phase, 3 gauge couplings, the Higgs mass and the Higgs vacuum expectation value, that are not predicted by the theory but are numerically established by the experiments.
\item The hierarchy problem~\cite{nature} in the SM arises from the fact that the electroweak symmetry breaking scale ($\sim$100~$\GeV$) and the Planck scale ($\sim$10$^{19}$~$\GeV$) are separated by many orders of magnitude. The Higgs mass is modified by one-loop radiative corrections coming from its couplings to gauge bosons, from Yukawa couplings to fermions and from its self-couplings, resulting in a quadratic sensitivity to the ultraviolet cutoff, \ie\ the scale below which QFTs are valid. For the Standard Model, this scale can go to the Planck scale, and so the QFT expectation for the Higgs mass is much higher than the experimental result.
\item The SM does not include the gravitational interaction, one of the four fundamental forces; this inclusion would require the gravity to be quantised. Since the gravity strength is much smaller than the other strengths, quantum gravitational effects would become important at length scales near the Planck scale, \ie\ $10^{19}$~$\GeV$, not accessible at any experimental facilities.
\item The SM does not explain the matter-antimatter asymmetry in the universe, \ie\ the imbalance between baryonic and antibaryonic matter; in fact, the measured CP violation and deviation from equilibrium during electroweak symmetry breaking are both too small, thus making unlikely that baryogenesis, \ie\ the physical process that could produce baryonic asymmetry, is possible within the SM theoretical framework~\cite{matter_anti}.
\item The SM describes the ordinary matter surrounding us that accounts just for the 5\% of the mass/energy content of the universe; it does not fully describe the nature of dark matter or dark energy, even if, from cosmological observations, they contribute to approximately 27\% and to 68\% of this content, respectively.
\end{itemize}

\chapter{The Large Hadron Collider}
\label{sec:LHC}
The \textbf{L}arge \textbf{H}adron \textbf{C}ollider (\textbf{LHC})~\cite{Lindon} is a two-ring hadron accelerator and collider with superconducting magnets built by the European Organisation for Nuclear Research (CERN); it was installed in the existing 26.7~km tunnel situated at a mean depth of 100~m underground, that was constructed between 1984 and 1989 for the CERN LEP machine.\newline
Beams of particles travel in opposite directions, kept separated in two ultra-high vacuum pipes and bent in the accelerator ring by a magnetic field of up to 8.33 T produced by superconducting electromagnets which operate at the temperature of 1.9~K.\newline
The LHC was designed to reach the highest energy ever explored in particle physics,  \ie centre-of-mass collision energies of up to 14 \TeV\, with the primary purpose of discovering new particles, like the Higgs boson, as well as revealing physics beyond the Standard Model. To this end several detectors were placed in the accelerator ring. The four largest experiments at the LHC are ALICE~\cite{Alice}, ATLAS~\cite{Atlas}, CMS~\cite{Cms} and LHCb~\cite{Lhcb}.\newline
In this Chapter, Sections~\ref{sec:acc_complex} and~\ref{sec:lhc_experiments} report details on the accelerator complex and the LHC experiments placed along the beam line, respectively. The most important beam and machine parameters are summarised in Section~\ref{sec:lhc_parameters} while Section~\ref{sec:lhc_operation} describes the scheduled periods of operation and shutdown.

\section{Accelerator Complex}
\label{sec:acc_complex}
\begin{figure}[H]
\begin{center}
\includegraphics[height=15 cm, width=15 cm]{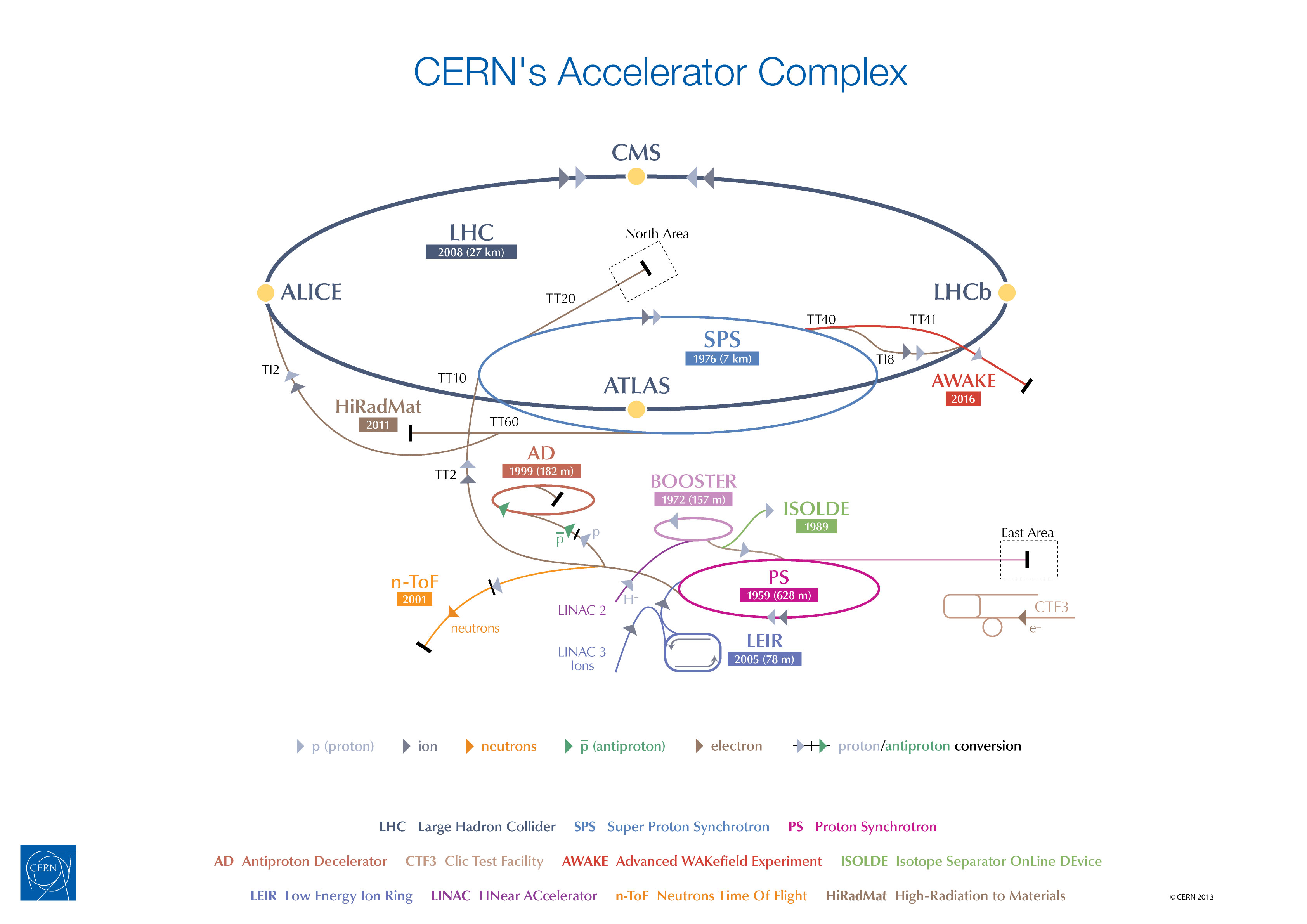}
\end{center}
\caption{Accelerator Complex and Experiments~\cite{complex}.}     
\label{accelerator}
\end{figure}

The LHC is the last accelerator in a complex chain of machines, a scheme of which is shown in Figure~\ref{accelerator}. The primary proton source is a bottle of hydrogen gas connected to a metal cylinder that strips off the electrons leaving just protons.\newline
Before being injected in the LHC, protons are accelerated through a series of accelerators that gradually increases their energy:
\begin{itemize}
\item \textbf{Linac2}: is a linear accelerator that uses radiofrequency cavities to charge cylindrical conductors and small quadrupole magnets to focus protons in a tight beam, accelerating them to an energy of 50 \MeV;
\item \textbf{Proton Synchrotron Booster (PSB)}: is made of four superimposed synchrotron rings that receive beams of protons at 50 \MeV\ and accelerate them to 1.4 \GeV;
\item \textbf{Proton Synchrotron (PS)}: is CERN's first synchrotron and has 277 conventional electromagnets, including 100 dipoles to bend the beams round the ring; it pushes the beam to 25 \GeV;
\item \textbf{Super Proton Synchrotron (SPS)}: is the second largest machine in CERN's accelerator complex measuring nearly 7 km in circumference; it has 1317 conventional electromagnets, including 744 dipoles to bend the beams round the ring, and it operates at up to 450 \GeV.
\end{itemize}
Protons are finally injected into the LHC beam pipes, with beam circulating both clockwise and anticlockwise. 

\section{The LHC Experiments}
\label{sec:lhc_experiments}
\begin{figure}[H]
\begin{center}
\includegraphics[height=8 cm,width = 12 cm]{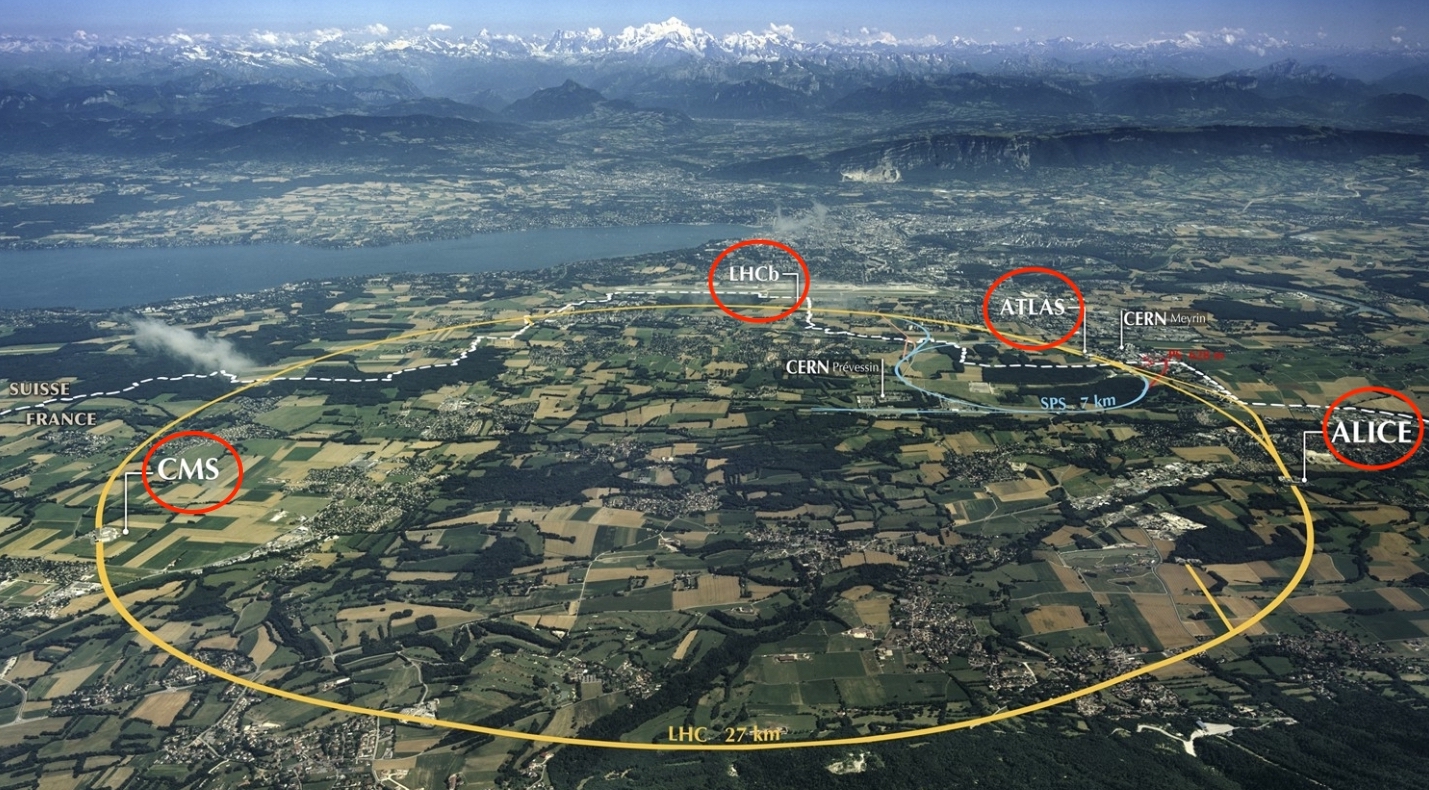}
\end{center}
\caption{The LHC ring and its main experiments around the IPs.}     
\label{ring}
\end{figure}

The experiments located around the IPs are:
\begin{itemize}
\item \textbf{ALICE}~\cite{Alice} (A Large Ion Collider Experiment), a general-purpose, heavy-ion detector which is designed to address the physics of strongly interacting matter and the quark-gluon plasma;
\item \textbf{ATLAS}~\cite{Atlas} (A Toroidal LHC ApparatuS), the largest, multi-purpose particle detector experiment designed to explore a wide range of physics processes;
\item \textbf{CMS}~\cite{Cms} (Compact Muon Solenoid), a general-purpose detector designed to target the same processes of ATLAS while using different and complementary technologies;
\item \textbf{LHCb}~\cite{Lhcb} (Large Hadron Collider beauty experiment), an experiment dedicated to heavy flavour physics; its primary goal is to look for indirect evidences of new physics in CP violation and rare decays of beauty and charm hadrons.
\end{itemize}
Two additional experiments, TOTEM and LHCf, are much smaller in size. They are designed to focus on ``forward particles$"$ (protons or heavy ions):
\begin{itemize}
\item \textbf{TOTEM}~\cite{Totem} is an experiment that studies forward particles and is focused on physics that is not accessible to the general-purpose experiments; it measures the total $pp$ cross section with the luminosity-independent method and studies elastic and diffractive scattering at the LHC;
\item \textbf{LHCf}~\cite{LHCf} is an experiment dedicated to the measurement of neutral particles emitted in the very forward region of LHC collisions. The physics goal is to provide data for calibrating the hadron interaction models that are used in the study of extremely high-energy cosmic rays.
\end{itemize}

\section{Luminosity}
\label{sec:lhc_parameters}
\begin{figure}[H]
\centering
\begin{subfigure}[b]{0.49\textwidth}
\includegraphics[width=\textwidth]{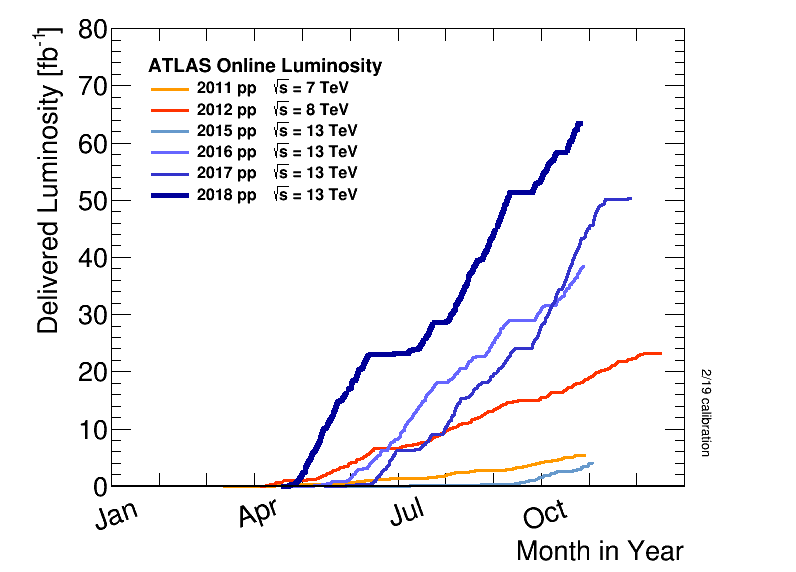}
\caption{}
\end{subfigure}
\begin{subfigure}[b]{0.49\textwidth}
\includegraphics[width=\textwidth]{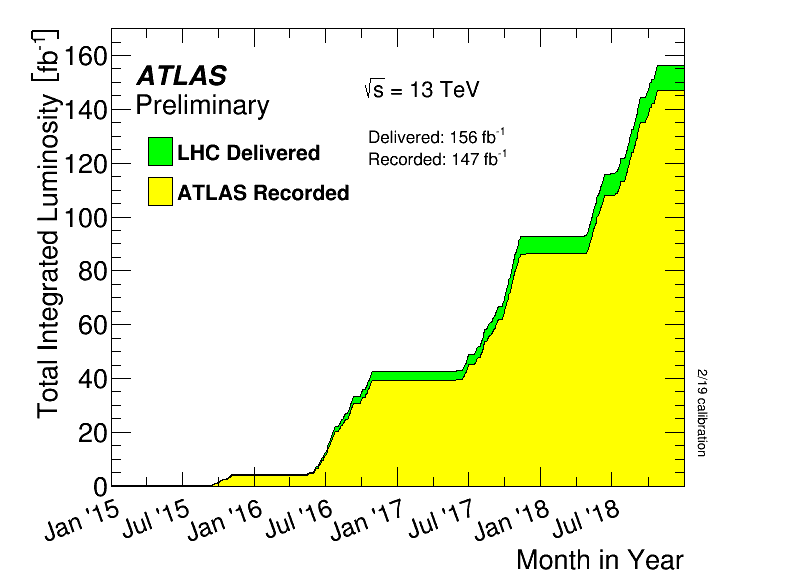}
 \caption{}
\end{subfigure}
\caption{(a) Cumulative luminosity versus time delivered to ATLAS for high energy $pp$ collisions and (b) cumulative luminosity versus time delivered to ATLAS (green) and recorded by ATLAS (yellow) during stable beams for $pp$ collisions at 13 TeV centre-of-mass energy in LHC Run 2~\cite{lumi}.}     
\label{luminosity}
\end{figure}

In the LHC collisions, the rate of produced events ($R_{event}$), \ie\ the number of events produced per second, is given by:
\begin{equation}
R_{event}=\frac{dN_{event}}{dt}=\mathcal{L}\cdot \sigma_{event}
\end{equation}
where $\mathcal{L}$ is the instantaneous luminosity of the accelerator (machine luminosity) and $\sigma_{event}$ is the cross section of the corresponding physics process.
Thus, in order to produce a significant amount of interesting/rare physics events and increase the discovery opportunity, high luminosity is a crucial achievement.\newline
In the case of two Gaussian beams colliding head-on, the machine luminosity [cm$^{-2}$s$^{-1}$] can be expressed in terms of the beam parameters as~\cite{Lindon}:
\begin{equation}
\mathcal{L}=\frac{N_b^2n_bf_{rev}\gamma_r}{4\pi\epsilon_n\beta^*}F
\end{equation}
where:
\begin{itemize}
\item $N_b$ is the number of particles per bunch: protons do not flow as a continuous beam inside the machine but are packed into bunches;
\item $n_b$ is the number of bunches per beam;
\item $f_{rev}$ is the revolution frequency;
\item $\gamma_r$ is the relativistic gamma factor of the protons;
\item $\epsilon_n$ is the normalised transverse beam emittance, that is a measure of the average spread of particles in the beam;
\item $\beta^*$ is the beta function at the collision point relating the beam size to the emittance, $\beta=\pi \sigma^2/\epsilon$, determined by the accelerator magnet configuration (basically, the quadrupole magnet arrangement) and powering;
\item $F$ is the geometric luminosity reduction factor due to the crossing angle at the interaction point (IP).
\end{itemize}
The geometric reduction factor $F$, assuming round beams and equal beam parameters for both beams, is in turn expressed in terms of $\theta_c$, the full crossing angle at the IP, $\sigma_z$, the RMS bunch length, and $\sigma^*$, the transverse RMS beam size at the IP, as:
\begin{equation}
F=\left ( 1+ \left ( \frac{\theta_c\sigma_z}{2\sigma^*} \right )^2  \right )^{-1/2} \, .
\end{equation}
The nominal LHC peak luminosity $\mathcal{L}=10^{34}$ cm$^{-2}$s$^{-1}$ corresponds to a nominal bunch spacing of 25 ns, $\beta^*$= 0.55 m, a full crossing angle $\theta_c$= 300 $\mu$rad, and bunch population, N$_b$= $1.1\times10^{11}$, while the RMS beam size and the geometric reduction factor are $\sigma^*$$=16.7~\mu$m and $F$=0.836, respectively~\cite{param}.\newline
The instantaneous luminosity is not constant over a physics run, indeed the peak luminosity is achieved at the beginning of stable beams, \ie the phase of actual physics data taking in the LHC cycle, but decreases due to the degradation of the intensities of the circulating beams, according to the following law:
\begin{equation}
\mathcal{L}(t)=\frac{\mathcal{L}_0}{(1+t/\tau_{nuclear})^2} \qquad \text{with} \qquad \tau_{nuclear}=\frac{N_{tot,0}}{\mathcal{L}_0\sigma_{tot}k},
\end{equation}
where:
\begin{itemize}
\item $\tau_{nuclear}$ is the initial decay time of the bunch intensity due to the beam loss from collisions;
\item $N_{tot,0}$ is the initial beam intensity;
\item $\mathcal{L}_{0}$ is the initial luminosity;
\item $\sigma_{tot}$ is the total cross section ($\sigma_{tot}= 10^{-25}$cm$^{2}$ at 13 TeV);
\item $k$ is the number of IPs with luminosity $\mathcal{L}_{0}$.
\end{itemize}
Further contributions to beam losses come from a blow-up of the transverse emittance related to the intra-beam scattering, to synchrotron radiation and noise effects and from particle-particle collisions within a bunch.\newline 
Assuming the LHC nominal parameters and combining the different contributions, the length of a luminosity run is estimated as $\tau_L \sim$15 h.\newline
Typical values of the most important beam and machine parameters are reported in Table~\ref{beam_par}~\cite{lumi_param}; the design machine luminosity of $\mathcal{L}=10^{34}$~cm$^{-2}$s$^{-1}$ has already been surpassed in 2016 when the instantaneous luminosity has reached the value of $\mathcal{L}$$=1.3 \times$$10^{34}$~cm$^{-2}$s$^{-1}$. Considering a luminosity of $\mathcal{L}=10^{34}$~cm$^{-2}$s$^{-1}$ and an inelastic cross section of $\sim$80~mb~\cite{Totem_xs}, an estimation of the expected rate of events at the LHC can be made, thus leading to $R_{event}=\mathcal{L}\sigma_{event}\sim 8\times 10^8$ events/s.
\begin{table}
\small
\scalebox{0.97}{
\begin{tabular}{|l|cccc|}
\hline
Parameter & 2015 & 2016 & 2017 & 2018 \\ 
\hline
Maximum number of colliding bunch pairs ($n_b$) & 2232 & 2208 & 2544/1909 & 2544 \\
Bunch spacing (ns) & 25 & 25 & 25/8b4e & 25 \\
Typical bunch population (10$^{11}$ protons) & 1.1 & 1.1 & 1.1/1.2 & 1.1 \\
$\beta^*$ (m) & 0.8 & 0.4 & 0.3 & 0.3-0.25 \\
Peak luminosity $\mathcal{L}_{peak}$ (10$^{33}$ cm$^{-2}$s$^{-1}$) & 5 & 13 & 16 & 19 \\
Peak number of inelastic interactions/crossing ($<\mu >$) & $\sim $ 16 &   $\sim $ 41 &  $\sim $ 45/60 &  $\sim $ 55 \\
Luminosity-weighted mean inelastic interactions/crossing & 13 & 25 & 38 & 36 \\
Total delivered integrated luminosity (fb$^{-1}$) & 4.0 & 38.5 & 50.2 & 63.4 \\
 \hline
\end{tabular}
}
 \caption{Selected LHC parameters for $pp$ collisions at $\sqrt{s}=$13 TeV in 2015--2018. The values shown are representative of the best accelerator performance during normal physics operation~\cite{lumi_param}.}
 \label{beam_par}
\end{table}
The actual figure of merit of the luminosity is the so-called integrated luminosity which directly relates the number of events to the cross section; it is defined integrating the instantaneous luminosity over the time of operation $T$:
\begin{equation}
L=\frac{\text{number of events of interest}}{\sigma_{event}}=\int_{0}^{T}\mathcal{L}dt \, .
\end{equation}
The results presented in this thesis are based on data collected by the ATLAS detector at $\sqrt{s}$ = 13 TeV, corresponding to an integrated luminosity of up to 79.8~fb$^{-1}$.\newline
Figure \ref{luminosity} shows, for the ATLAS detector, the delivered luminosity, defined as the luminosity made available by the LHC machine, and the recorded luminosity, defined as the luminosity recorded by the detector. 
ATLAS and CMS are the high-luminosity LHC experiments, both designed to aim at a peak luminosity of $\mathcal{L} =10^{34}$ cm$^{-2}$s$^{-1}$ for proton operation; moreover, two low-luminosity experiments are present: LHCb aiming at a peak luminosity of $\mathcal{L}$$= 10^{32}$ cm$^{-2}$s$^{-1}$, and TOTEM aiming at a peak luminosity of $\mathcal{L} = 2\times10^{29}$ cm$^{-2}$s$^{-1}$.\newline The LHC has also one dedicated heavy-ion experiment ($p-Pb$ or $Pb-Pb$), ALICE, aiming at a peak luminosity of $\mathcal{L} =10^{27}$ cm$^{-2}$s$^{-1}$~\cite{Lindon}.
\begin{figure}[H]
\begin{center}
\includegraphics[height=6 cm,width = 9 cm]{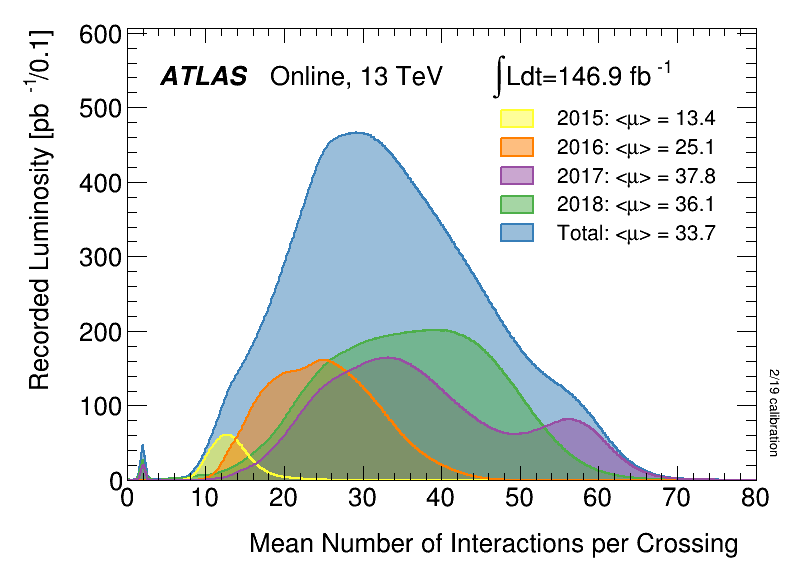}
\end{center}
\caption{Luminosity-weighted distribution of the mean number of interactions per crossing for the 2015--2017 $pp$ collision data at 13~TeV centre-of-mass energy~\cite{lumi}.}     
\label{mu}
\end{figure}
The average number of interactions per bunch crossing, whose distribution is shown in Figure \ref{mu} for the 2015, 2016, 2017 and 2018 data, is given by the pile-up $\langle \mu \rangle$, related to the instantaneous luminosity by the following formula:
\begin{equation}
 \langle \mu \rangle=\frac{\mathcal{L}\sigma_{tot}}{f_{rev}n_b} \, .
\end{equation}
The pile-up is therefore proportional to the luminosity and constitutes a challenge from the detector side for resolving the individual collisions and thus a limit to the increase of luminosity of a collider.
The peak value for the pile-up in 2016 data taking has been $\mu\sim 50$, considering a total cross section $\sigma_{tot}= 10^{-25}$~cm$^{2}$ at 13~TeV, a peak luminosity $\mathcal{L}=1.3 \times 10^{34}$~cm$^{-2}$s$^{-1}$, a number of bunches $n_b\sim$ 2200 and a revolution frequency $f_{rev}=$$11.245$~kHz.
\section{LHC Operation}
\label{sec:lhc_operation}
\begin{figure}[H]
\begin{center}
\includegraphics[width=\textwidth]{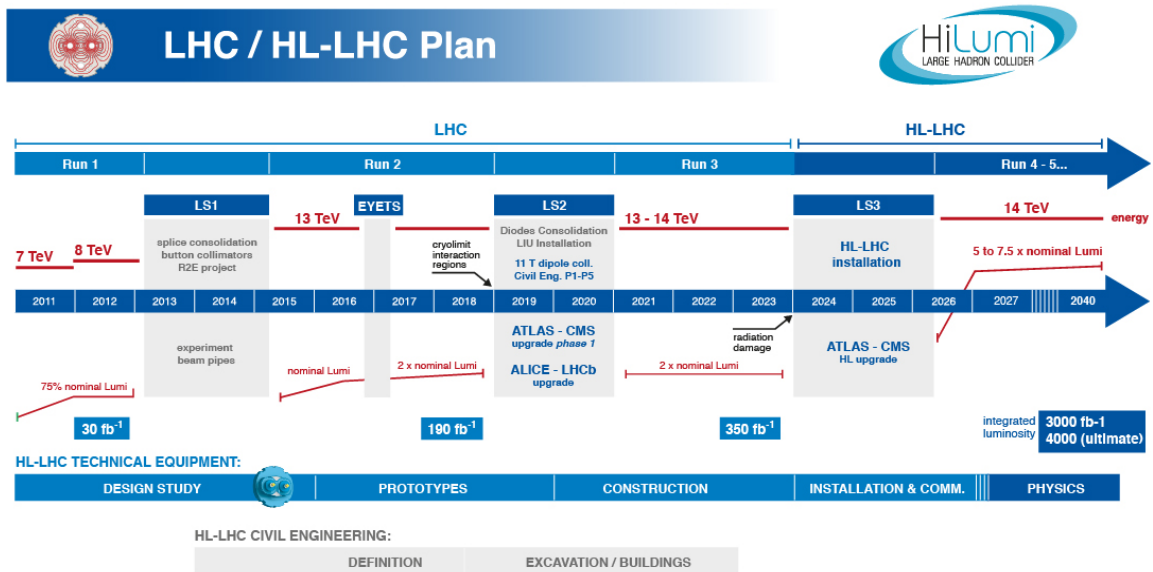}
\end{center}
\caption{LHC schedule involving active periods and technical shutdown, from $\text{Run}$$\; \text{1}$ to the future upgrade to High-Luminosity LHC~\cite{HL1}.}     
\label{LHC_HL}
\end{figure}
The scheduled periods of operation of the LHC and the shutdown periods are shown in Figure~\ref{LHC_HL}, together with future developments and upgrades~\cite{HL1}.\newline
The LHC began operation for data taking in 2009, with the first operational run called ``Run 1$"$: beams were injected in both rings and stable beam collisions were performed at 450 GeV (900 GeV centre-of-mass energy). By the end of the year, beams were accelerated to 1.18 TeV (2.36 TeV centre-of-mass energy) per beam.\newline
In 2010 the centre-of-mass energy was successfully increased to 7 TeV and the LHC continued to run during 2010 and 2011 at $\sqrt{s}$= 7 TeV, delivering a cumulative luminosity of 5.46~fb$^{-1}$, corresponding to a recorded luminosity, for the ATLAS experiment, of $5.08$~fb$^{-1}$.\newline
During 2012, an increase in beam energy from 3.5 to 4 TeV per beam was made, corresponding to a centre-of-mass energy of 8 TeV, thus leading to a total recorded integrated luminosity of 22.8~fb$^{-1}$. The first operational run therefore collected  $\sim$25~fb$^{-1}$ of ``good for physics$"$ data~\cite{LumiRun1}.\newline
After a long shutdown, necessary to upgrade the magnet interconnects and safety systems for a centre-of-mass energy of 13~TeV, the second operational run of the LHC, called ``Run 2$"$, started in 2015 and ended in 2018; LHC accelerated protons up to an energy of 6.5 TeV, corresponding to a centre-of-mass energy of 13 TeV. The peak instantaneous luminosity achieved was 2.1$\times 10^{34}$~cm$^{-2}$ s$^{-1}$. The total integrated luminosity delivered to ATLAS during the second run was 156~fb$^{-1}$, corresponding to $\sim$140~fb$^{-1}$ of data good for physics analyses.\newline
The 2015--2017 ATLAS data-taking period, corresponding to an integrated luminosity of up to 79.8~fb$^{-1}$, has been exploited for the results presented in this thesis.\newline
A long shutdown period, LS2, has just started (2019); this period will be devoted to the consolidation and the upgrades of the detectors and to start testing some new systems and technologies that will be essential to further pushing the LHC machine beyond its limits.\newline
After 2020, the statistical gain in running the accelerator without a significant luminosity increase will become marginal.\newline
A key element for further increasing the luminosity is a new linear accelerator, the Linac4, that is replacing the Linac2 in providing protons to the LHC, accelerating them to an energy of 160~MeV. \newline
Furthermore, the LHC will undergo a major upgrade, Phase-2 Upgrade, to a High-Luminosity LHC (HL-LHC) expected to start operations in 2026, after collecting a total dataset of approximately 400~fb$^{-1}$ by the end of Run 3 (in 2023). \newline
The two main goals of the HL-LHC project will be the following~\cite{HL1}: 
\begin{itemize}
\item a peak luminosity from 5 to 7 $\times 10^{34}$~cm$^{-2}$ s$^{-1}$ with levelling, allowing:
\item an integrated luminosity of 300/350~fb$^{-1}$ per year with an ultimate goal of 4000~fb$^{-1}$ within twelve years. This integrated luminosity is about ten times the expected luminosity reach of the first twelve years of the LHC lifetime.
\end{itemize}

\chapter{The ATLAS Experiment at the Large Hadron Collider}
\label{sec:ATLAS}
\begin{figure}[H]
\begin{center}
\includegraphics[width =\textwidth]{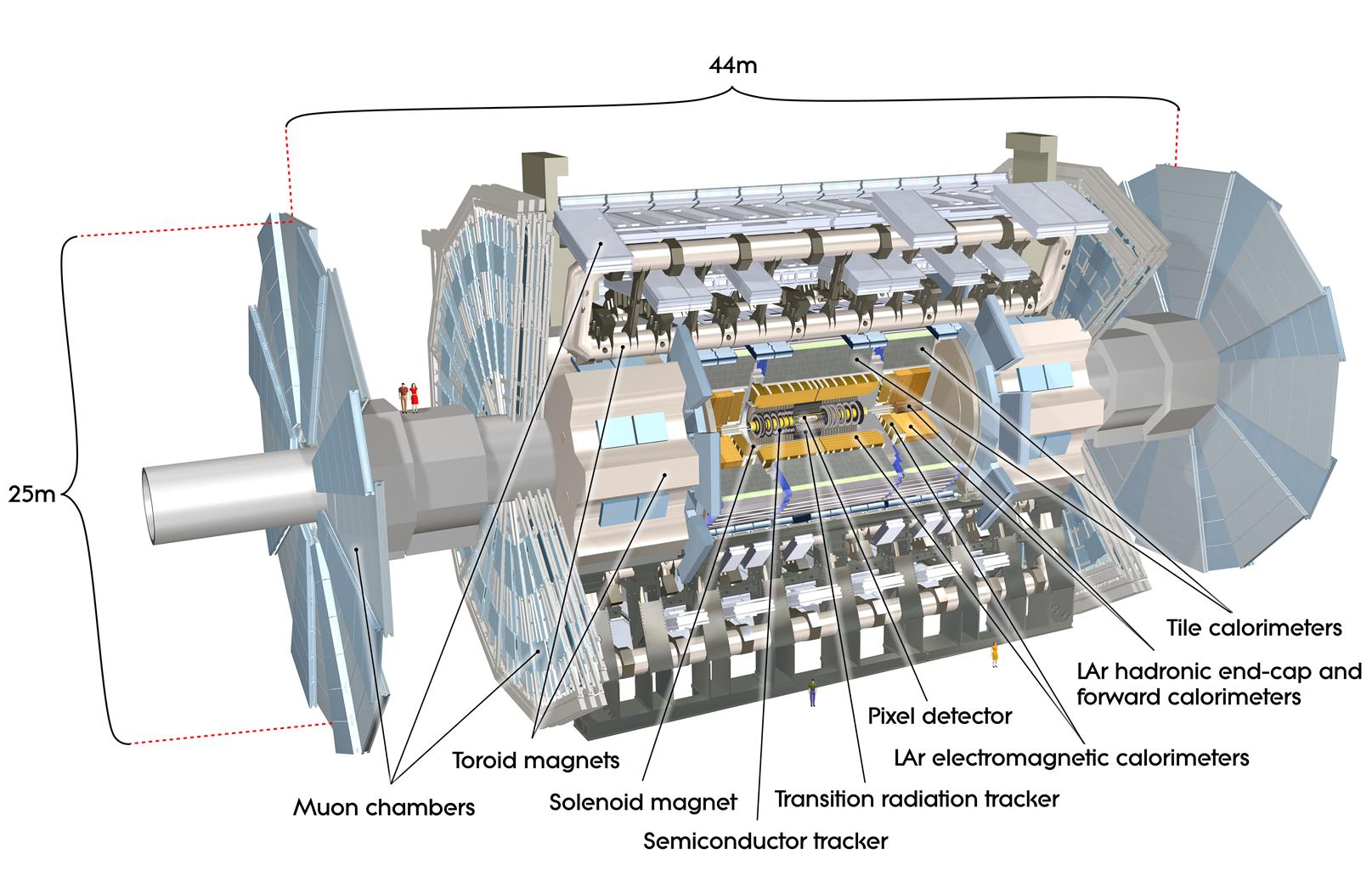}
\end{center}
\caption{Cut-away view of the whole ATLAS detector and its sub-systems~\cite{Atlas}.}     
\label{detector}
\end{figure}

\textbf{ATLAS} (\textbf{A} \textbf{T}oroidal \textbf{L}HC \textbf{A}pparatu\textbf{S})~\cite{TDR1,TDR2} is a general-purpose detector, \ie a detector capable of addressing a huge range of physics processes and observing all possible decay products of the $pp$ interactions; furthermore, it is the largest-volume detector ever built for a particle collider. The ATLAS detector is forward-backward symmetric with respect to the interaction point, covering almost the entire 4$\pi$ solid angle with a cylindrical shape. It is 44~m long and 25~m high; it weighs over 7000~tons and it sits in a cavern $\sim$100~m underground placed at Point 1 (one of the LHC Interaction Points). Figure \ref{detector} shows a schematic representation of the ATLAS detector.
The high luminosity and high centre-of-mass energy of the LHC $pp$ collisions allow to explore physics at the TeV scale and are needed because of the small cross sections expected for many of the following processes, the ATLAS detector has been designed to target:
\begin{itemize}
\item the Higgs-boson search and the measurement of its fundamental properties;
\item high precision tests of QCD, electroweak interactions, and flavour physics;
\item exotic searches, \eg searches for new heavy gauge bosons or extra dimensions;
\item precision measurements of the top-quark properties, like its mass, coupling and spin;
\item the search for supersymmetry-like extensions of the SM.
\end{itemize}
In order to handle a high rate of events, $\sim$$6 \times 10^8$ events/s for inelastic $pp$ interactions, as discussed in Chapter~\ref{sec:LHC}, as well as a high rate of bunch crossing, $\sim$40~MHz, the ATLAS detector was designed to fulfil these general requirements~\cite{Atlas}:
\begin{itemize}
\item fast, radiation-hard electronics and sensor elements together with a high detector granularity needed to handle particle fluxes and to reduce the influence of overlapping events;
\item large acceptance in pseudorapidity with almost full azimuthal angle coverage;
\item accurate tracking of charged-particle, \ie good momentum resolution and reconstruction efficiency in the inner tracker as well as precise reconstruction of secondary vertices in order to identify $\tau$-leptons and $b$-jets;  
\item accurate electromagnetic calorimetry to identify electrons and photons, complemented by full-coverage hadronic calorimetry for precise jet and missing transverse energy measurements;
\item good muon identification and momentum resolution over a wide range of momenta.
\end{itemize}
The detector is constituted by a central part, called ``barrel$"$, and two side parts, called ``end-caps$"$.\newline
ATLAS sub-detectors and coordinate system are introduced in Sections~\ref{sec:ATLAS_sub} and~\ref{sec:ATLAS_coor}, respectively. Section~\ref{sec:ATLAS_magnet} describes the magnet system, necessary to make accurate track reconstruction and momentum measurement. A comprehensive description of each sub-detector is reported in Sections~\ref{sec:ATLAS_inner},~\ref{sec:ATLAS_calo},~\ref{sec:ATLAS_muon} and~\ref{sec:ATLAS_lumi}. The Trigger and Data Acquisition System is described in Section~\ref{sec:ATLAS_trigger}.

\section{Detector Sub-Systems}
\label{sec:ATLAS_sub}

ATLAS is composed by different sub-detectors that are arranged in an onion-like layered structure to provide an angular uniform coverage around the beam pipe; going from the closest to the beam pipe to the external one, it is composed by the \textbf{Inner Detector}~\cite{Inner_Detector}, the \textbf{Electromagnetic Calorimeter}, the \textbf{Hadronic Calorimeter}, the \textbf{Forward Calorimeter}~\cite{Calorimeter_liqui, Calorimeter_tile}, the \textbf{Muon Spectrometer}~\cite{MS} and the \textbf{luminosity detectors}~\cite{fabbri}:
\begin{itemize}
\item the Inner Detector is immersed in a 2~T magnetic field parallel to the beam axis; it measures the direction, momentum, and charge of electrically-charged particles and reconstructs the interaction vertices;
\item the Calorimeters absorb photons, electrons and hadrons and measure their energy; they are able to stop most known particles except muons and neutrinos;
\item the Muon Spectrometer is the outermost part of the ATLAS detector and measures the energy and trajectory of the muons with high accuracy. To this end, the particles are deflected in a strong magnetic field, which is generated by superconducting magnetic coils;
\item the forward detectors LUCID (Luminosity measurement Using Cherenkov Detectors), followed by ZDC (Zero Degree Calorimeter) and ALFA (Absolute Luminosity For ATLAS), measure the online luminosity.
\end{itemize}
\begin{figure}[H]
\begin{center}
\includegraphics[height=9 cm,width =10 cm]{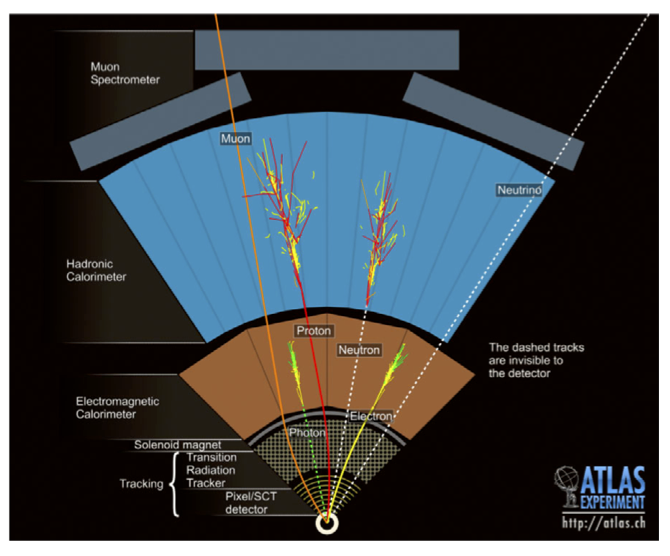}
\end{center}
\caption{Sketch representing how particles interact with different layers and subsystems of the detector~\cite{ipatia}.}     
\label{cipolla}
\end{figure}
Particles are reconstructed according to their interactions with detector materials. \newline
A complete representation of the particles reconstructed and identified in the ATLAS detector is shown in Figure \ref{cipolla}. Charged particles leave ionisation signatures in the innermost part of ATLAS where the Inner Detector is present, whereas neutral particles such as neutrons are invisible to it. A magnetic field bends charged particles and allows to reconstruct their momentum and measure their charge according to the direction they bend towards. All particles (bar neutrinos) deposit a fraction (or all) of their energy in the Electromagnetic and Hadronic Calorimeters, the former targeting photons and electrons and the latter hadrons; photons, electrons, protons and neutrons create showers in the calorimeters and are stopped there. Muons cross all the sub-systems depositing only a small fraction of their energy throughout their path before being stopped in the Muon Spectrometer. Neutrinos, due to their really elusive nature and small interaction cross section, can't be detected by ATLAS; their presence is deduced by looking for missing momentum in the momentum balance of the event.\newline
A comprehensive description of particle reconstruction is given in Chapter 4, while details on the different sub-detectors are reported in the following sections.\newline
The main performance goals of the detector sub-systems are listed in Table~\ref{Atlas_requirements}.
\begin{table}[H]
\begin{center}
\scalebox{1.2}{
{\def\arraystretch{1.2}
\begin{tabular}{|c|c|c|c|}
\hline
 \multicolumn{1}{|c|}{\textbf{{\scriptsize Detector component}}} &
 \multicolumn{1}{c|}{\textbf{{\scriptsize Required resolution}}} &
 \multicolumn{2}{c|}{{\scriptsize $\mathbf{\eta}$ \textbf{coverage}}} \\
 \hline
 & & {\scriptsize Measurements} & {\scriptsize Trigger} \\
\hline
 {\scriptsize Tracking} & {\scriptsize $\sigma_{p_T}/p_T = 0.05\% / p_T \oplus 0.01 \%$} & {\scriptsize $\pm 2.5$} & \\
\hline
 {\scriptsize EM calorimetry} & {\scriptsize$\sigma_{E}/E = 10\% / \sqrt{E} \oplus 0.7 \%$} & {\scriptsize$\pm 3.2$} & {\scriptsize$\pm 2.5$} \\
\hline
 {\scriptsize Hadronic calorimetry (jets)} &  &  &  \\
 {\scriptsize barrel and end-cap} & {\scriptsize $\sigma_{E}/E = 50\% / \sqrt{E} \oplus 3 \%$} & {\scriptsize $\pm 3.2$} & {\scriptsize $\pm 3.2$} \\
 {\scriptsize forward} & {\scriptsize $\sigma_{E}/E = 100\% / \sqrt{E} \oplus 10 \%$} & {\scriptsize $3.1 < |\eta| < 4.9$} & {\scriptsize $3.1 < |\eta| < 4.9$} \\
\hline
 {\scriptsize Muon spectrometer} & {\scriptsize $\sigma_{p_T}/p_T = 10\%$ at $p_T = 1\ TeV$} & {\scriptsize $\pm 2.7$} & {\scriptsize $\pm 2.4$} \\
\hline
\end{tabular}}}
\newline
\caption{General performances of the various components of the ATLAS detector~\cite{Atlas}.}
\label{Atlas_requirements}
\end{center}
\end{table}

\section{Coordinate System}
\label{sec:ATLAS_coor}
\begin{figure}[H]
\begin{center}
\includegraphics[height= 6 cm, width =11 cm]{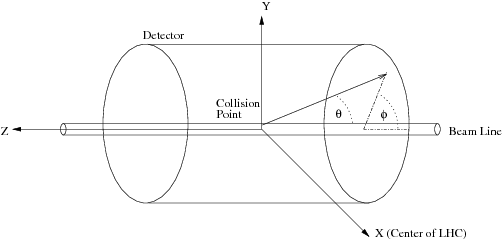}
\end{center}
\caption{ATLAS and CMS coordinate system~\cite{coor}.}     
\label{coordinate}
\end{figure}
\begin{figure}[H]
\begin{center}
\includegraphics[height= 4 cm,width =4 cm]{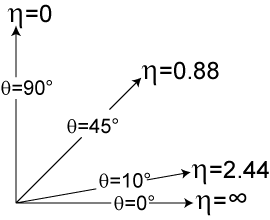}
\end{center}
\caption{Pseudorapidity $\eta$ and corresponding $\theta$ values.}     
\label{pseudo}
\end{figure}

ATLAS uses a right-handed coordinate system with the origin in the nominal interaction point while the beam direction defines the $z$-axis; the $x$-axis points to the centre of the LHC ring and the $y$-axis points vertically upwards, thus the $x-y$ plane is transverse to the beam, as shown in the sketch of Figure \ref{coordinate}. Given the symmetry of the detector, a system of cylindrical coordinates ($R$, $\phi$, $\theta$) can be used, where $R=\sqrt{x^2+y^2}$, the polar angle $\theta$ is the angle from the beam axis and $\phi$ is the azimuthal angle measured around the beam ($z$) axis. The rapidity $y$ is defined as:
\begin{equation}
y=\frac{1}{2}ln \left ( \frac{E+p_z}{E-p_z}\right )
\end{equation}
where $E$ and $p_z$ are the energy and the $z$-axis momentum component of the particle. Differences in rapidity are invariant under Lorentz transformations along the $z$-axis.\newline
In case of particles with a mass negligible with respect to the energy, $y$ corresponds to the pseudorapidity $\eta$, shown graphically in Figure \ref{pseudo} and often used to measure angular distances:
\begin{equation}
\eta=-ln \left ( tan \frac{\theta}{2} \right ) \, .
\end{equation}
Transverse momentum and transverse energy are defined in the $x-y$ plane as $p_T=$$\sqrt{p_x^2+p_y^2}$ and $E_T = E sin\theta$, respectively.\newline
$\Delta R$ is the distance in the ($\eta-\phi$) space between particles defined as:
\begin{equation}
\Delta R=\sqrt{\Delta\eta^2+\Delta \Phi^2}
\end{equation}
where $\Delta\eta$ and $\Delta \phi$ are the differences in pseudorapidity and azimuthal angles between the particles.
\section{Magnet System}
\label{sec:ATLAS_magnet}
A strong magnetic field represents the key element to provide sufficient bending power to make accurate track reconstruction and momentum measurement. 
The radius of curvature $\rho$ of a particle with charge $q$ and momentum $p$ entering perpendicularly a magnetic field $B$, follows from the Lorentz force:
\begin{equation}
\rho=\frac{p}{q\cdot B} \; .
\end{equation}
Thus, in order to determine the momentum of a charged particle, the curvature of its trajectory through the tracking detectors, placed in magnetic fields, is measured.
\begin{figure}[H]
\begin{center}
\includegraphics[height= 7 cm,width =11 cm]{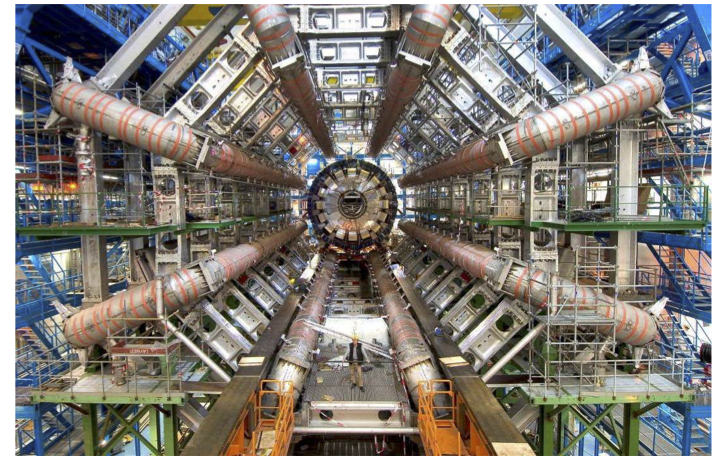}
\end{center}
\caption{Barrel toroid as installed in the underground cavern~\cite{Atlas}.}     
\label{magnet}
\end{figure}
\begin{figure}[H]
\begin{center}
\includegraphics[height= 6 cm,width =10 cm]{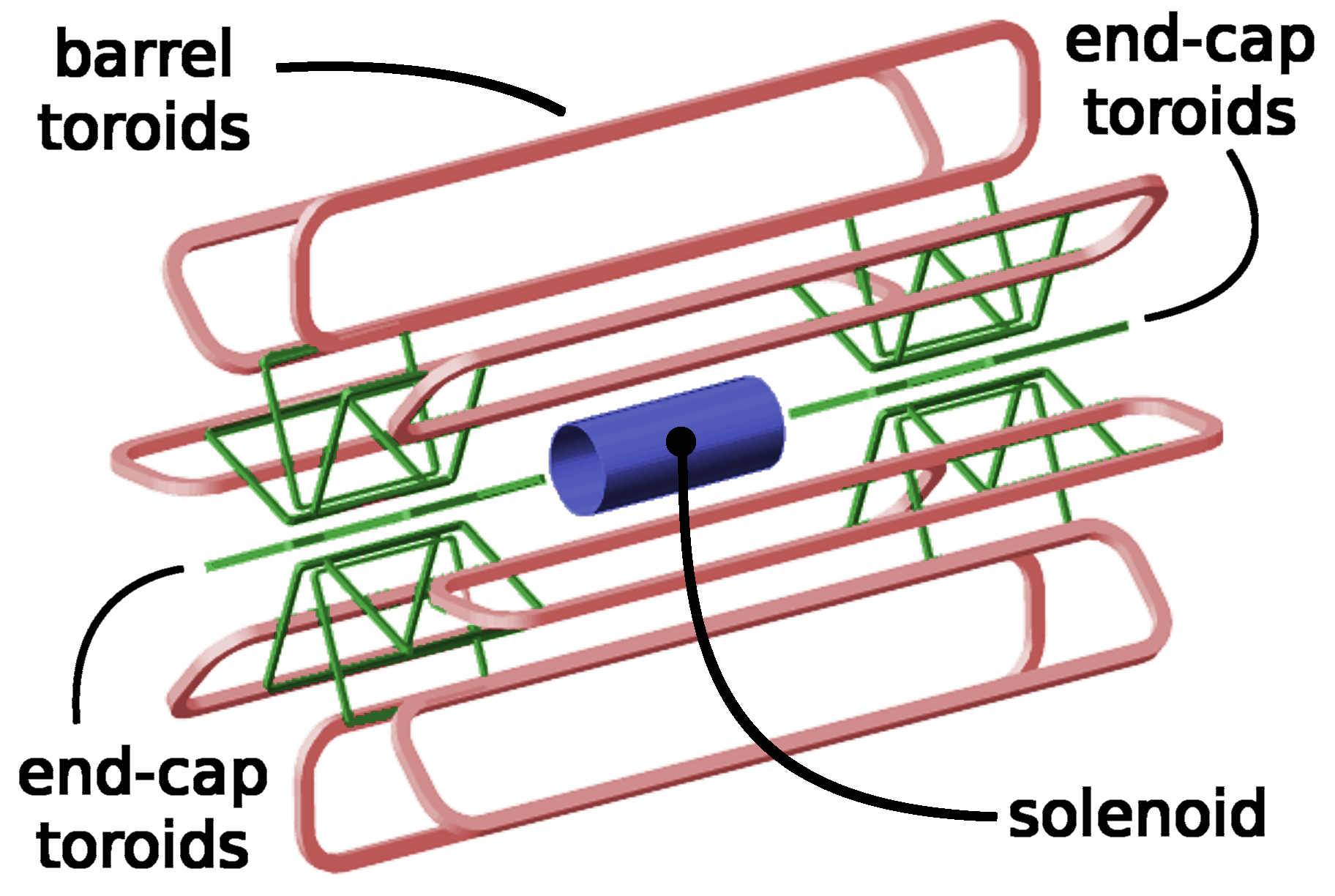}
\end{center}
\caption{Layout of the ATLAS Magnet System.}     
\label{magnet1}
\end{figure}
Differently from CMS, which uses a single solenoid magnet to provide a 4~T magnetic field, the ATLAS design includes two separate magnetic systems~\cite{Magnet} composed by the following four large superconducting magnets:
\begin{itemize}
\item a Central Solenoid (CS), which is aligned on the beam axis and is located between the Inner Detector and the Electromagnetic Calorimeter; it has a diameter of 2.4~m and a length of 5.3~m and provides a 2~T axial magnetic field for the inner detector, while minimising the radiative thickness in front of the barrel electromagnetic calorimeter;
\item a Barrel Toroid (BT), shown in Figure \ref{magnet}, 25~m long (inner core 9.4~m, outer diameter 20.1~m), and two End-Cap Toroids (ECT), 5~m long (inner core 1.64~m, outer diameter 10.7~m), which produce a toroidal magnetic field of 4~T in the Muon Spectrometer volume mostly orthogonal to muon trajectories.
\end{itemize}
The ATLAS magnet system layout is shown in Figure \ref{magnet1}. The whole magnetic system is cooled at liquid helium temperature ($\sim$4.8~K).

\section{Inner Detector}
\label{sec:ATLAS_inner}
 \begin{figure}[H]
\begin{center}
\includegraphics[height=9 cm, width =13 cm]{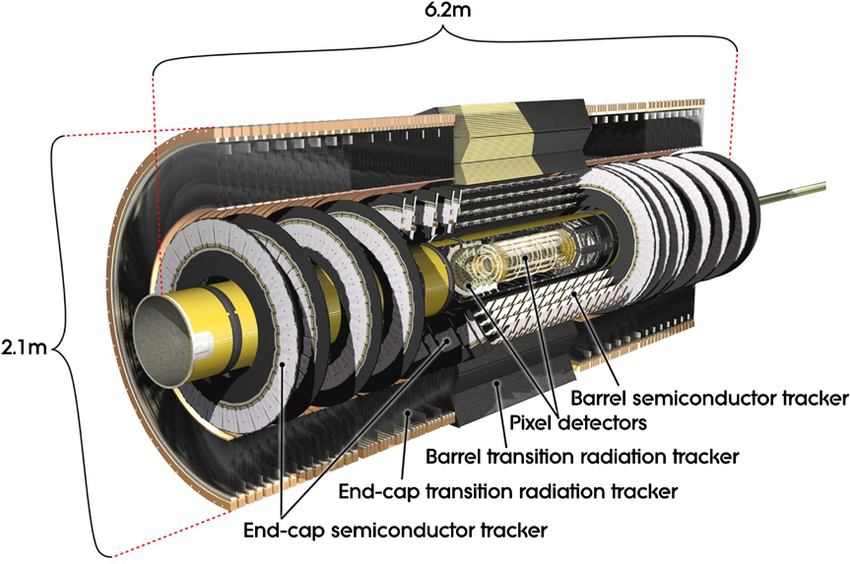}
\end{center}
\caption{Cut-away view of the ATLAS Inner Detector~\cite{Atlas}.}     
\label{inner1}
\end{figure}
\begin{figure}[H]
\begin{center}
\includegraphics[height=10 cm, width =12 cm]{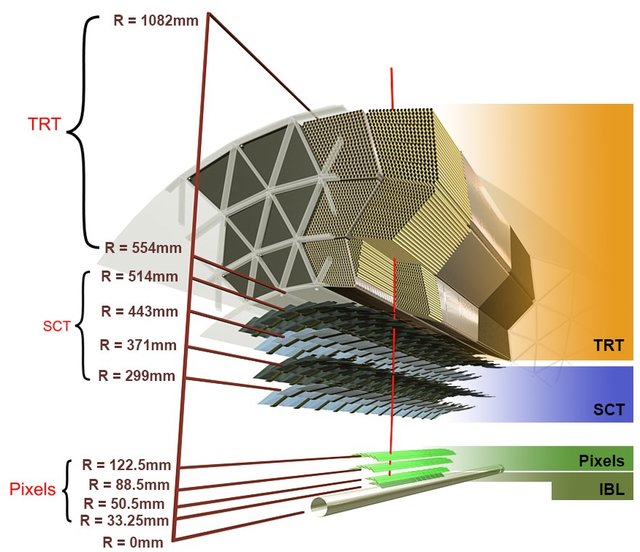}
\end{center}
\caption{Sketch of a segment of the ATLAS ID barrel modules, showing the radial layout of the detection sub-systems. The grey little cylinder is the LHC beam-pipe. The IBL pixel layer, that has been added for Run 2, is visible in the sketch.} 
\label{inner}
\end{figure}

The ATLAS Inner Detector (ID), shown in Figure~\ref{inner1}, has been designed to provide hermetic and robust pattern recognition, excellent momentum resolution and both primary and secondary vertex measurements for charged tracks, thus contributing, together with the calorimeter and muon systems, to the electron, photon and muon identification. The ID is composed of concentric layers of detecting material, divided into a barrel and two end-caps, and its acceptance covers the pseudorapidity range $| \eta | <$ 2.5. \newline
It is immersed in a 2~T solenoidal magnetic field, has a total radius of 1.1~m and length of 6.2~m, and it is constituted of high granularity detectors, needed to perform high-precision track parameter measurements and event vertex reconstruction. The ID consists of several independent but complementary sub-detectors going from layers of high resolution silicon detectors at inner radii to gaseous tracking detectors at higher radii; starting from the inner layer and following the sketch shown in Figure~\ref{inner}, they are: the Insertable B-Layer, the Pixel and the Silicon microstrip of the Semi Conductor Trackers, used in conjunction with the Transition Radiation Tracker.\newline
The magnetic field is essential to measure the charge and the momentum of particles from their bendings.
 \begin{figure}[H]
\begin{center}
\includegraphics[width =\textwidth]{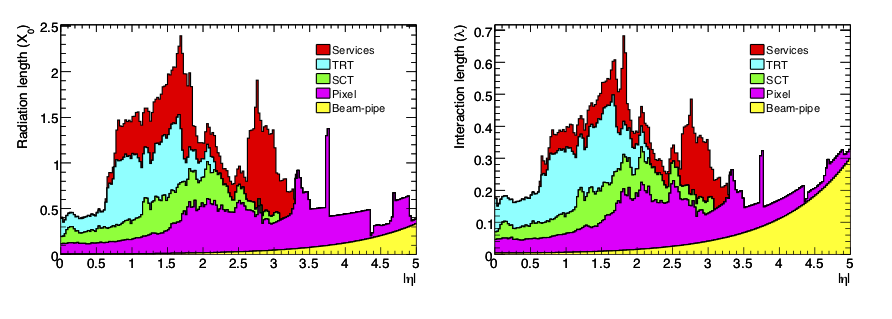}
\end{center}
 \caption{Material distribution ($X_0$, $\lambda$) at the exit of the ID envelope, including the services and thermal enclosures. The distribution is shown as a function of $| \eta |$ and averaged over $\phi$~\cite{Atlas}.}
 \label{material_ID}
\end{figure}
The performance requirements of the ATLAS ID are more stringent than any tracking detector built so far for operation at a hadron collider. In order to achieve high granularity and include readout and cooling system, it is necessary to introduce a significant amount of material in the ID; Figure~\ref{material_ID} shows the material distribution in terms of the radiation length X$_0$, where X$_0$ represents the average path the particle needs to travel to reduce its initial energy by a factor $1/e$, and in terms of the nuclear interaction length $\lambda$, defined as the mean free path between interactions; due to this material budget, photons may convert into electron-positron pairs before reaching the electromagnetic calorimeter and electrons may loose part of their energy through bremsstrahlung emissions affecting the resolution of the energy measured by the calorimeter system.\newline
Table~\ref{id} summarises the main features of each ID sub-detector, \ie element size and resolution of each sub-system as well as the radii of the barrel layers.

\begin{table}
\small
\begin{center}
{\def\arraystretch{1.2}
\begin{tabular}{|c|c|c|c|}
\hline
Sub-detector & Element size [\SI{}{\mu \meter}] & Intrinsic resolution [\SI{}{\mu \meter}] & Radius of the barrel layers [mm]\\
\hline
IBL & $50\times 250$ & $8\times 40$ & 33.2\\
Pixel & $50\times 400$ & $10\times 115$ & 50.5, 88.5, 122.5\\
SCT & 80 & 17 & 299, 371, 443, 514\\
TRT & 4000 & 130 & 554 - 1082\\
\hline
\end{tabular}}
\end{center}
\caption{Summary of the main features of the ID sub-detectors. The intrinsic resolution of the IBL and the Pixel is reported along $R-\phi$ and $z$ and for SCT and TRT along $R-\phi$. For SCT and TRT the element sizes refer to the spacing of the readout strips and the diameter of the straw tubes, respectively~\cite{Inner_Detector}.}
\label{id}
\end{table}

\subsection{The Insertable B-Layer (IBL)}
\begin{figure}[H]
\begin{center}
\includegraphics[height=7 cm,width =10 cm]{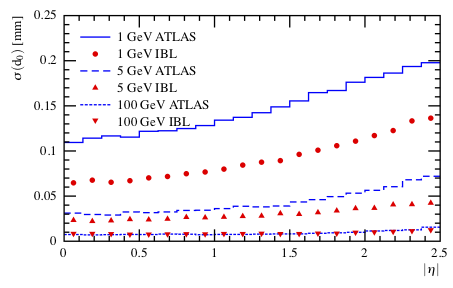}
\end{center}
 \caption{Track parameter resolution of the impact parameter $d_0$ for single muons at 1, 5 and 100~GeV as a function of $\eta$ for the original ID and for the ID with IBL~\cite{IBL}.}
 \label{d0_IBL}
\end{figure}
The Insertable B-Layer (IBL)~\cite{IBL} is the ATLAS sub-detector closest to the beam pipe, being on average 33.2~mm away.
One of the main concerns before Run 2 was that increasing luminosity, significant radiation damage could occur to the ID; this would lead to a loss in tracking efficiency, especially affecting $b$-tagging. To cope with this issue, the IBL has been added to the ATLAS detector. It was installed before the start of Run 2 data taking, between a new beryllium beam pipe with a reduced inner diameter (47~mm instead of 59~mm) to fit inside the IBL and the first layer of the original Pixel detector. It consists of a cylindrical layer 3.5~m long. Due to its position close to the beam pipe, the IBL pixel sensors have a small size (50$\,\times\,$\SI{250}{\mu \meter}) to reduce occupancy at high luminosity, and are radiation hard. The IBL also provides a full $\phi$ coverage.\newline
The biggest contribution of the IBL to the precision of the measurement, is the improvement, shown in Figure~\ref{d0_IBL}, of the resolution of the impact parameter $d_0$, that is defined as the distance in the $x-y$ plane between the track closest point to the $z$ axis and the $z$ axis itself and it also the crucial parameter that affects the $b$-tagging performance. The significant gains are due to the additional layer at smaller radius, and to the smaller $z$ pitch of the IBL compared to the present Pixel detector. The contribution from the IBL to the measurement of the track curvature is small as the overall track length is nearly unchanged. Furthermore, the IBL improves the quality of the vertex reconstruction and of the $b$-tagging performance.

\subsection{The Pixel Detector}
The Pixel Detector provides critical tracking information for pattern recognition near the collision point, measuring the particle impact parameter resolution and largely determining the ability of the Inner Detector to find secondary vertices.
The system provides three or more precision measurement points for tracks with pseudorapidity $| \eta | <$ 2.5 and it has a full $\phi$ coverage. Due to its position close to the barrel, where the particle density is at its highest, this system has to achieve a high granularity. It is composed by three barrel layers placed at the radii of 51~mm, 89~mm, and 123~mm respectively, centred around the beam axis ($z$), and by two end-caps, each end-cap having three disk layers at $| z |$=495, 580 and 650~mm; the silicon pixel sensors have a minimum pixel size in ($R-\phi \times z$) of  $\SI{50}{\mu \meter} \times \SI{400}{\mu \meter}$ in both the barrel and the end-cap positions. The dimensions are chosen in order to maximise the probability that a particle crossing one layer will cross also the other two layers. The intrinsic precisions of the Pixel Detector in the barrel are \SI{10}{\mu \meter} ($R-\phi$) and \SI{115}{\mu \meter} ($z$) and in the disks are \SI{10}{\mu \meter} ($R-\phi$) and \SI{115}{\mu \meter} (R). The Pixel Detector has approximately $80.4$$\; \text{million}$ readout channels.

\subsection{The SemiConductor Tracker (SCT)}
The SemiConductor Tracker (SCT) is placed in the intermediate range of the ID and employs the same semiconductor technology used by the Pixel Detector, providing the pseudorapidity coverage in the range $| \eta | <$ 2.5 replacing pixels with silicon microstrips having a 120 mm $\times$ 60~mm size in $\phi \times z$ and completing the precision tracking of the Pixel detector in the measurement of momentum, impact parameter and vertex position. The SCT is composed by four layers in the barrel (299~mm $ < R < 514$~mm) and nine in each end-cap (850~mm $< z <$ 2730~mm).\newline
It is arranged in twenty-two layers: four cylindrical barrel layers and eighteen disk layers, nine on each of the end-caps. The barrel layers are organised in 4 cylinders made of two layers of sensors, placed at approximate radial distances of 30, 37, 44 and 51~cm from the interaction point, to provide 4 additional space points for each tracks in the $R-\phi$ and $z$ coordinates. Each layer is made of $p-n$ silicon semiconductor modules of nominal size 6.36~cm $\times$ 6.40~cm with 780 readout strips. Each strip is 12~cm long and has a constant pitch of \SI{80}{\mu \meter}. The end-cap modules have a very similar structure, but exploit tapered strips, where one set is aligned radially. The SCT has a total of 6.3 million readout channels.
The intrinsic measurement precisions of the SCT per module are \SI{17}{\mu \meter} for the $R-\phi$ plane and \SI{580}{\mu \meter} for the $z(R)$-coordinate in both the barrel and the end-caps. 

\subsection{The Transition Radiation Tracker (TRT)}

The Transition Radiation Tracker (TRT) is placed in the outermost part of the ID; it is a combination of a tracker (based on straw tubes) working as a drift chamber measuring the charge drift time, and a transition radiation detector for the pattern recognition; the transition radiation detector allows to discriminate between light and heavy particles. It exploits the fact that particles emit transition radiation according to the speed they have passing through several layers of material with different refraction indices; thus high relativistic particles (typically electrons) have a higher probability of emitting transition radiation photons with respect to the other particles.\newline
A single TRT component is made of Polyamide straw tubes of 4~mm diameter and long up to 144~cm in the barrel region; at the centre of each straw tube, there is a \SI{31}{\mu \meter} diameter tungsten wire, the anode, directly connected to the front-end electronics and kept at ground potential. The gap between the straw and the wire is filled by a mixture of gases. The passage of ionising particles induces a low amplitude signal on the anodes. At the same time, some particles crossing polypropylene fibres cause transition radiation emission, absorbed by the Xenon present in the gas mixture; this last process leads to a high amplitude signal in the TRT electronics that can be distinguished from low amplitude ionisation signal.\newline
The TRT only provides $R-\phi$ information, for which it has an intrinsic measurement accuracy of \SI{130}{\mu \meter} per straw.

\section{Calorimeters}
\label{sec:ATLAS_calo}
\begin{figure}[H]
\begin{center}
\includegraphics[height=8 cm, width =16 cm]{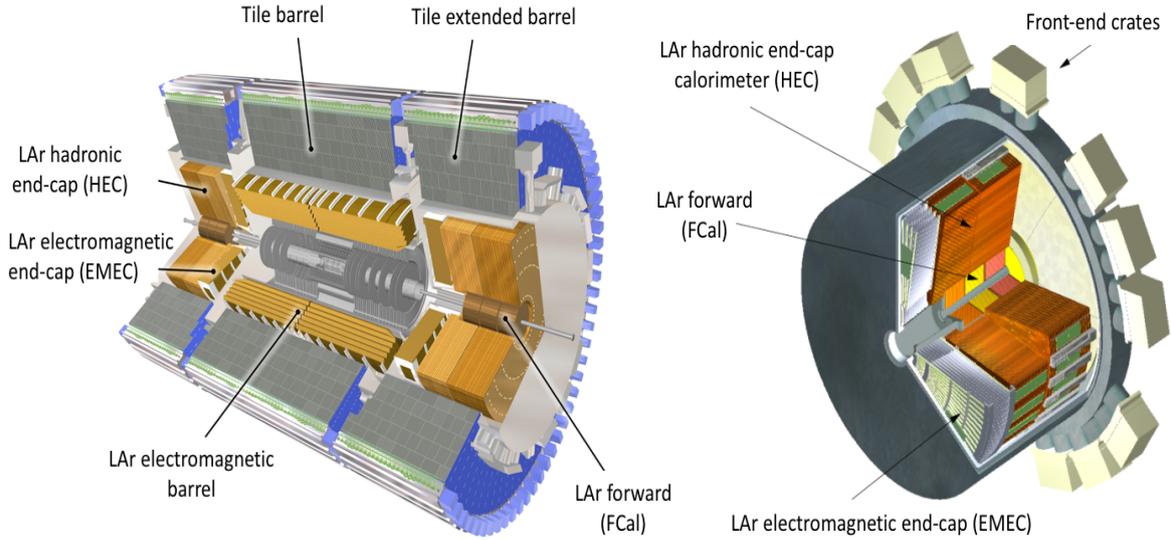}
\end{center}
 \caption{Left: sketch showing the sub-systems of the ATLAS calorimeter; right: cut-away view of an end-cap cryostat showing the positions of the three end-cap calorimeters. The outer radius of the cylindrical cryostat vessel is 2.25~m and the length of the cryostat is 3.17 m~\cite{Atlas}.}
 \label{calorimeter}
\end{figure}

The ATLAS calorimetry system consists of the Electromagnetic Calorimeter (ECal) that covers the pseudorapidity range $|\eta| <$ 3.2, the Hadronic Calorimeter (HCal) which covers the pseudorapidity range $|\eta| <$ 3.9 and the Forward Calorimeter (FCal) covering the pseudorapidity range $3.1 < |\eta| <$ 4.9; it is structured in three cryostats, one barrel and two end-caps, is finely segmented in $\eta$ and $\phi$ and covers the full azimuthal range, as shown in Figure \ref{calorimeter}.\newline
The calorimeters' main purpose is to fully contain and measure destructively the energy and direction of incident electrons and photons, producing electromagnetic showers in the ECal and FCal electromagnetic part, and hadrons, interacting via the strong force in the HCal and FCal hadronic part; particles that can interact either electromagnetically (other than the muon) or strongly, deposit their energy in all calorimeters. The system gives also a fundamental contribution in measuring the missing transverse energy.
\begin{figure}[H]
\begin{center}
\includegraphics[height=7 cm, width =11 cm]{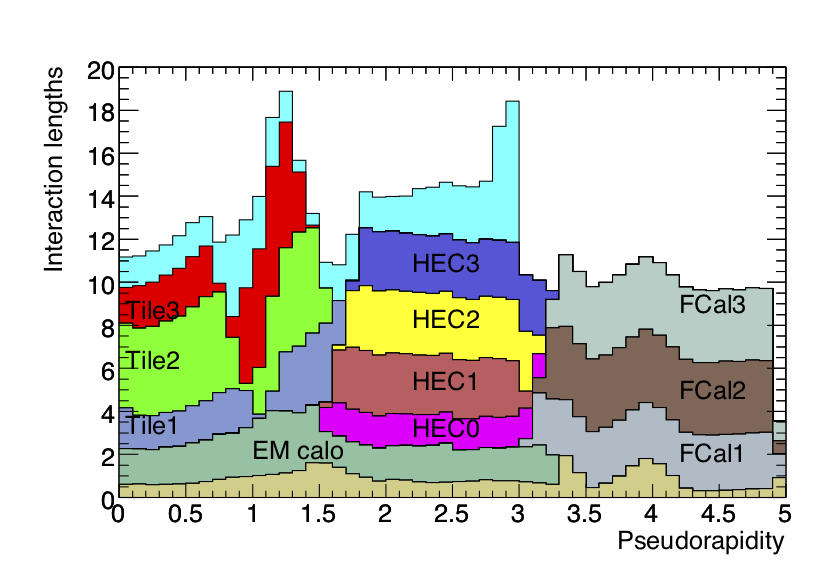}
\end{center}
 \caption{Cumulative amount of material as a function of the pseudorapidity in units of interaction length $\lambda$ considering different positions: in front of the electromagnetic calorimeters, in the electromagnetic calorimeters themselves, in each hadronic layer, and total amount of material at the end of the active calorimetry. The total amount of material in front of the first active layer of the muon spectrometer (up to $| \eta | < $ 3.0) is also shown~\cite{Atlas}.}
 \label{material}
\end{figure}

The ATLAS calorimeters are sampling calorimeters, \ie they are made of alternating layers of ``passive$"$ material that degrades the particle energy and ``active material$"$ that provides a measurable signal and collects the energy of particles via ionisation or scintillation; lead (Pb), copper, or iron are used as passive materials and Liquid Argon (LAr) or polystyrene scintillator as active materials.\newline
The total thickness of the electromagnetic calorimeter is more than 22~X$_0$ in the barrel and more than 24~X$_0$ in the end-caps. The total interaction length $\lambda$ of the entire system is $\sim$10~$\lambda$, adequate to provide good resolution for high-energy jets. The numbers of interaction lengths as a function of the pseudorapidity in front of and in the ECal, HCal and FCal are shown in Figure \ref{material}, while the main features of the calorimeters are described in the following sections.
\subsection{The Electromagnetic Calorimeter}

\begin{figure}[H]
\begin{center}
\includegraphics[height=7 cm, width =11 cm]{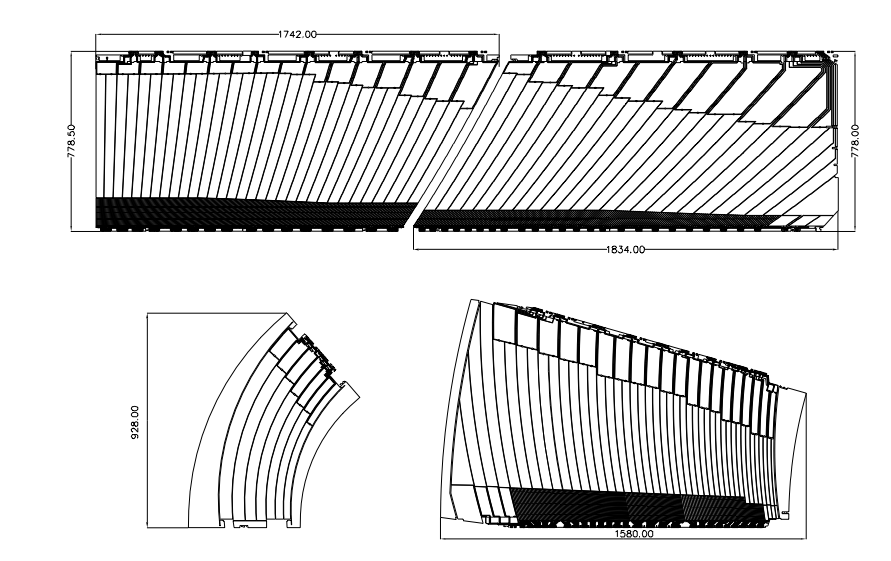}
\end{center}
 \caption{Layout of the signal layer for the four different types of electrodes before folding. The two top electrodes are for the barrel and the two bottom electrodes are for the end-cap inner (left) and outer (right) wheels. Dimensions are in millimetres. The drawings are all at the same scale~\cite{Atlas}.}
 \label{accordion}
\end{figure}

The ATLAS Electromagnetic Calorimeter is 6.65 m long and has an outer radius of 2.25~m. The main part of the ATLAS ECal is a lead-Liquid Argon (LAr) detector with accordion-shaped Kapton electrodes and lead absorber plates over its full coverage, as shown in Figure~\ref{accordion}. The LAr serves as the active material and was chosen due to its intrinsic linear behaviour, its stability of response over time and its intrinsic radiation-hardness; the lead absorber plates act as the passive material.
The accordion geometry provides complete $\phi$ symmetry without azimuthal cracks. The ECal is divided into a barrel part (EMB) ($|\eta | <$ 1.475) and two end-cap components (EMEC) (1.375 $< |\eta | <$ 3.2).
The overall segmentation allows for high precision spatial measurements, providing a geometry that helps in identifying photons coming from a primary vertex. In order to completely contain the electromagnetic shower, the EM calorimeter has a thickness of more than 22 $X_0$ in the barrel and more than 24 $X_0$ in the end-caps.\newline
The nominal Electromagnetic Calorimeter resolution is:
\begin{equation}
\frac{\sigma(E)}{E}=\frac{10 \%}{\sqrt{E}}\oplus 0.7 \% 
\end{equation}
where $E$ is expressed in $\GeV$.
\subsection{The Hadronic Calorimeter}
The Hadronic Calorimeter is placed directly outside the ECal and it is composed by a barrel and two end-caps; it is 6.10 m long and has an external radius of 4.25~m.
The central barrel region, called Tile Calorimeter, is a sampling calorimeter using iron as passive material and scintillating tiles as active material; it covers the pseudorapidity range $| \eta | < $ 1.7 and is divided in cells of size $\Delta\eta \times \Delta \phi = 0.1 \times 0.1$.
The Hadronic End-Cap Calorimeter (HEC), covering the range 1.5 $< | \eta | < $ 3.2, uses LAr as active medium and consists of two independent wheels per end-cap, located directly behind the end-cap ECal and sharing the same LAr cryostats; its cells have a granularity of $\Delta\eta \times \Delta \phi = 0.1 \times 0.1$ or  $0.2 \times 0.2$ depending on $\eta$. The overall thickness of the HCal is 11 $\lambda$ for $\eta=0$.
The nominal energy resolution for hadronic jets (combined with the ECal) is:
\begin{equation}
\frac{\sigma(E)}{E}=\frac{50 \%}{\sqrt{E}}\oplus 3 \% \, .
\end{equation}

\subsection{Forward Calorimeter}
The Forward Calorimeter is an electromagnetic and hadronic calorimeter, with a total thickness of 10 $\lambda$.
The FCal modules are located at high $\eta$, at a distance of approximately 4.7~m from the interaction point, so they are exposed to high particle fluxes; the positioning of these systems results in a quite hermetic design, which minimises energy losses in cracks between the calorimeter systems and limits the backgrounds that reach the muon system. Each FCal is split into three 45 cm deep modules: one electromagnetic module (FCal1) and two hadronic modules (FCal2 and FCal3). Both the electromagnetic part and the hadronic parts use LAr as active material while copper and tungsten are used as passive materials. To optimise the resolution and the heat removal, copper was chosen as the absorber for FCal1, while mainly tungsten was used in FCal2 and FCal3, to provide containment and minimise the lateral spread of hadronic showers. The FCal provides a measurement of both electromagnetic and hadronic showers.
The typical energy resolution of the FCal is:
\begin{equation}
\frac{\sigma(E)}{E}=\frac{100 \%}{\sqrt{E}}\oplus 10 \% \, .
\end{equation}

\section{Muon Spectrometer}
\label{sec:ATLAS_muon}
\label{sec:MS}
\begin{figure}[H]
\begin{center}
\includegraphics[height= 15 cm,width =10 cm]{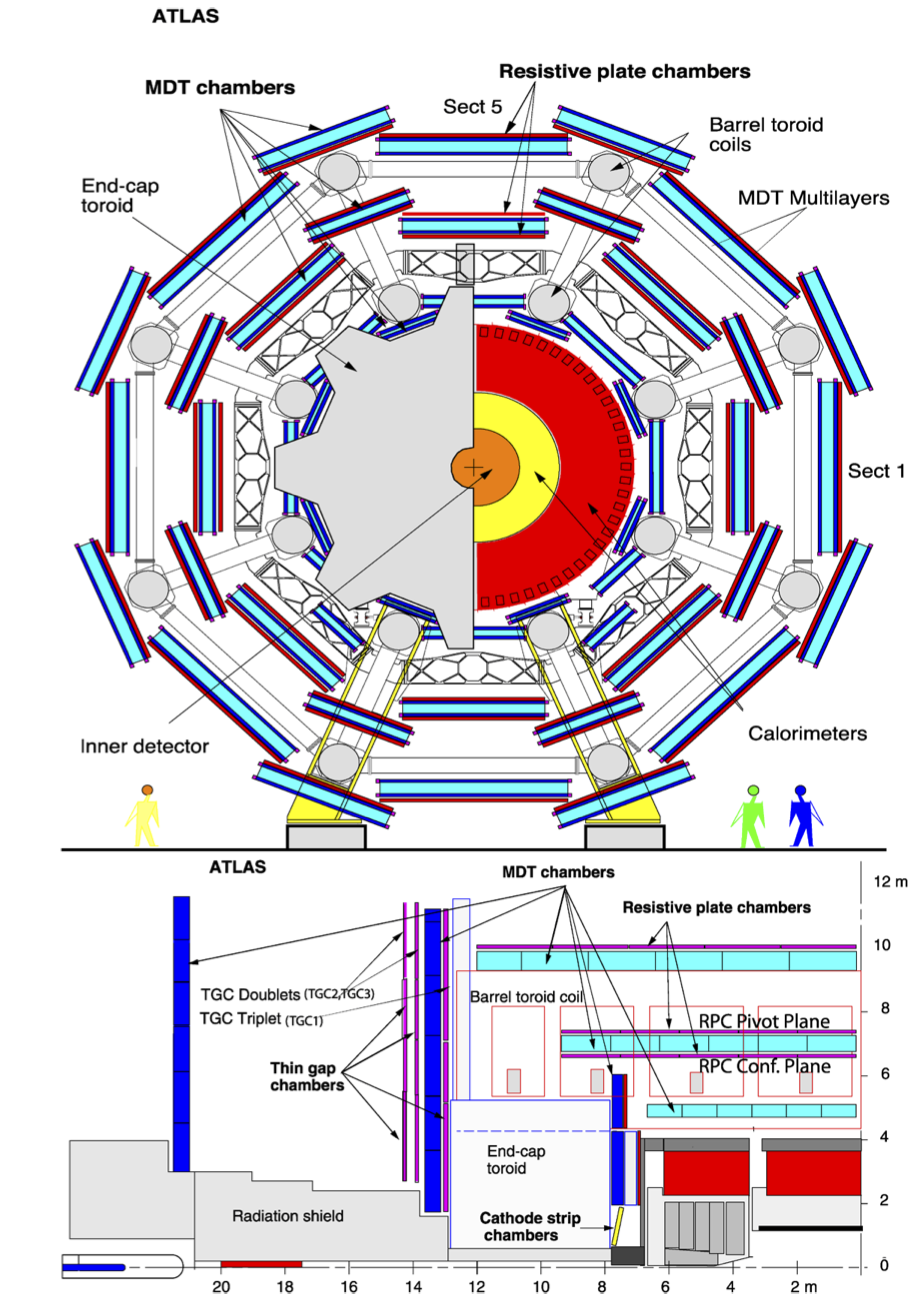}
\end{center}
\caption{Schematic view of the Muon Spectrometer in the $x-y$ (top) and $z-y$ (bottom) projections~\cite{Muonspectro}.}
 \label{spectro}
\end{figure}
The choice for the Muon Spectrometer (MS) to be the outermost part of the ATLAS detector comes from the fact that muons travel in the detector much more than the other charged particles generated in the collisions; furthermore, they loose just few MeV/mm in calorimeter electromagnetic interactions, since they have a low interaction cross section with the materials, they radiate bremsstrahlung far less than the electrons due to the larger mass and are long-lived particles. All other particles, except neutrinos, are expected to not escape the hadronic calorimeters.\newline 
The MS is designed to reconstruct muons and measure their momentum in the pseudorapidity range $| \eta | <$ 2.7 as well as trigger on these particles in the pseudorapidity range $| \eta | <$ 2.4; the toroidal magnets surrounding the calorimeters generate a magnetic field perpendicular to the beam and orthogonal to the solenoid field in the ID, causing muons to bend in the $R-z$ plane. Thus muons are reconstructed by exploiting the combination of information obtained both from the ID and from the MS that makes an independent measurement of the momentum.\newline
Precision chambers, the Monitored Drift Tubes (MDT) and the Cathode Strip Chambers (CSC), are used to reconstruct the trajectory of the muons.\newline
An essential design criterion of the muon system is the capability to trigger on muon tracks. The precision-tracking chambers have therefore been complemented by a system of fast trigger chambers capable of delivering track information within a few tens of nanoseconds after the passage of the particle. Trigger chambers are fast muon momentum measurement detectors consisting of the Resistive Plate Chambers (RPC) and the Thin Gap Chambers (TGC).\newline
A schematic layout of the MS in the $x-y$ and $z-y$ projections is shown in Figure \ref{spectro}.\newline
Most important parameters regarding MS sub-detectors are reported in Table \ref{muon_spec_param}.
\begin{table} [H]
\scalebox{0.8}{
{\def\arraystretch{1.4}
\begin{tabular}{|c|l|c|c|c|c|c|c|c|c|c|}
\hline
\multicolumn{1}{|c|}{\textbf{{Type}}}&
\multicolumn{1}{c|}{\textbf{{Function}}} &
\multicolumn{3}{c|}{\textbf{{Chamber resolution in}}} &
\multicolumn{2}{c|}{\textbf{{Measurements/track}}} &
\multicolumn{2}{c|}{\textbf{{Number of}}} \\  
\hline
& & $z/R$ & $\phi$ & time & barrel & end-cap & chambers & channels \\
\hline
\textbf{MDT} & tracking & $35$~$\mu$m ($z$) & $-$ & $-$  & 20 &  20 &  1150 & 354k \\
\textbf{CSC} & tracking & $40$~$\mu$m ($R$) & 5~mm & 7~ns & $-$  & 4 & 32 & 30.7k \\
\textbf{RPC} & trigger & $10$~mm ($z$) & 10~mm & 1.5~ns & 6 & $-$  & 606 & 373k \\
\textbf{TGC} & trigger& $2-6$~mm ($R$) & 3-7~mm & 4~ns  & $-$  & 9 & 3588 & 318k\\
\hline
\end{tabular}}}
\caption{Parameters of the four sub-systems of the muon detector. The quoted spatial resolution (columns 3, 4) does not include chamber-alignment uncertainties. Column 5 lists the intrinsic time resolution of each chamber type, to which contributions from signal-propagation and electronics contributions need to be added~\cite{Atlas}.}
\label{muon_spec_param}
\end{table}

\subsection{Tracking Chambers}
\begin{figure}[H]
\begin{center}
\includegraphics[height=5 cm, width =8 cm]{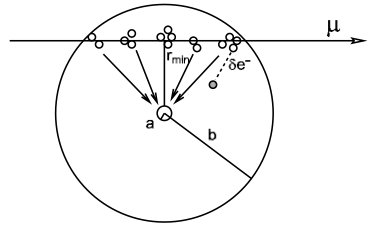}
\end{center}
 \caption{Cross section of a charged particle passing through a drift tube. The charged particle ionises electrons in the gas which drift to the anode wire at the centre of the tube~\cite{Atlas}.}
 \label{drift_tube}
\end{figure}

\begin{figure}[H]
\begin{center}
\includegraphics[height=5 cm, width =8 cm]{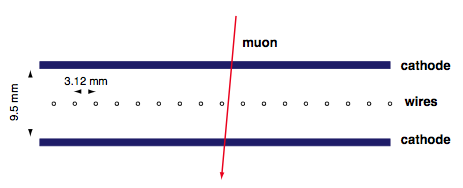}
\end{center}
 \caption{Diagram of the gas gap in a Cathode Strip Chamber.}
 \label{csc}
\end{figure}

A brief description of the muon tracking chambers is reported in the following lines.
\begin{itemize}
\item The \textbf{Monitored Drift Tubes} (MDT) measure only the $z$ coordinate in the barrel region and in the end-cap region up to $| \eta | <$ 2.7 except for the innermost layer where they are replaced by the CSC.
The basic elements of the MDT chambers are pressurised drift tubes, that act as the cathode, with a diameter of 30~mm, operating with Ar/CO$_2$ gas (93/7, \ie 93\% Ar and 7\% CO$_2$) at 3 bar, and central tungsten-rhenium wires, which act as the anode, with a diameter of \SI{50}{\mu \meter}, at a potential of $\sim$3~kV. Muons ionise the gas mixture in the tubes to create electrons (which are attracted to the wire) and positive ions (which drift towards the cathode), as shown in Figure \ref{drift_tube}. Each MDT chamber is made of $3-8$ layers of drift tubes, has  an average resolution of \SI{80}{\mu \meter} per tube and \SI{35}{\mu \meter} per chamber; the limiting factor at high luminosity is a typical drift time of 700~ns.
\item The \textbf{Cathode Strip Chambers} (CSC) are multi-wire proportional chambers consisting of arrays of positively-charged ``anode$"$ wires oriented in the radial direction crossed with negatively-charged copper ``cathode$"$ strips measuring muon momentum in the forward region, \ie in the pseudorapidity range $2 < | \eta | <$ 2.7. The wires are kept at a voltage of 1.9~kV and the space enclosing the wires is filled with a gas mixture of Ar/CO$_2$ (80/20). The CSC combine high spatial, time and double track resolution with high-rate capability and low neutron sensitivity. The whole CSC system consists of two disks each with eight chambers (eight small and eight large), each chamber containing four CSC planes.
This design provides a resolution of \SI{60}{\mu \meter} in the bending plane and 4~mm in the transverse plane while the time resolution is about 7~ns. A sketch of the gas gap in a CSC is shown in Figure \ref{csc}.
 \end{itemize}
The precision of the momentum measurement for a high-$p_T$ muon track depends on the resolution of the sagitta, namely the deviation in the $R-z$ plane with respect to a straight line. For a high-momentum track ($p_T\sim 1$ TeV), the typical sagitta is around \SI{500}{\mu \meter}.\newline
The ATLAS muon system, in particular the MDTs, provides a momentum measurement with a $\sigma_{p_{T}}/p_T$ resolution between $2-3$\% and $\sim$$10$\% in the $p_T$ range between 10~\GeV\ and 1~$\TeV$.

\subsection{Triggering Chambers}
The trigger chambers of the muon system provide fast information on muon tracks traversing the detector, allowing the trigger logic to recognise their multiplicity and approximate energy range. Details on the trigger system are reported in the next section.\\
A brief description of the muon trigger system is reported in the following lines.
\begin{itemize}
\item The \textbf{Resistive Plate Chambers} (RPC) are fast gaseous detectors and consist of two parallel plates, a positively-charged anode and a negatively-charged cathode, both made of a very high resistivity plastic material and separated by a gas volume, as shown in Figure \ref{rpc}; each plate is read by two orthogonal series of strips: the $\eta$-strips are parallel to the MDT wires and provide the bending view of the trigger detector, the $\phi$-strips are orthogonal to the MDT wires and provide the second-coordinate measurement. The resistive plates are kept at a potential difference of 9.8~kV and the chamber is filled with a gas mixture of C$_2$H$_2$F$_4$, C$_4$H$_{10}$ and SF$_6$ (94.7/5/0.3 respectively). They cover the pseudorapidity range $ | \eta | <$ 1.05 and combine a good spatial resolution with a time resolution of just 1~ns.
\begin{figure}[H]
\begin{center}
\includegraphics[height=5 cm, width =8 cm]{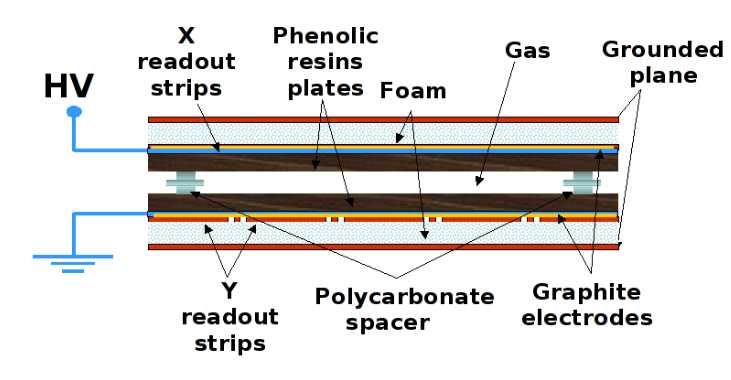}
\end{center}
\caption{Structure of a Resistive Plate Chamber.}
 \label{rpc}
\end{figure}
\begin{figure}[H]
\begin{center}
\includegraphics[height=5 cm, width =8 cm]{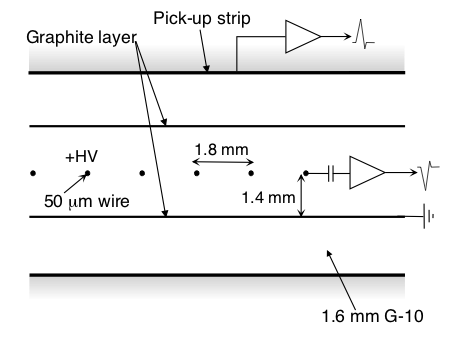}
\end{center}
\caption{TGC structure showing anode wires, graphite cathodes, G-10 layers and a pick- up strip, orthogonal to the wires~\cite{Atlas}.}
 \label{tgc}
\end{figure}
\item The \textbf{Thin Gap Chambers} (TGC) consist of planes of closely spaced wires maintained at positive high voltage, sandwiched between resistive grounded cathode planes covering the pseudorapidity range $1.05 < | \eta | <$ 2.4. The operational gas is a mixture of CO$_2$/n-C$_5$H$_{12}$ (55/45 respectively) (n-pentane). The anode wires, arranged parallel to the MDT wires, provide trigger signal together with readout strips arranged orthogonal to the wires. The TGC can provide spatial resolution better than \SI{100}{\mu \meter}. Their spatial resolution is mainly determined by the readout channel granularity: several wires (the number depending on the desired granularity at that $\eta$ location) are ganged together to provide an anode signal.
Furthermore, the TGC provide good time resolution and high rate capability.
\end{itemize}

\section{Trigger and Data Acquisition System}
\label{sec:ATLAS_trigger}
\begin{figure}[H]
\begin{center}
\includegraphics[height=10 cm,width=12 cm]{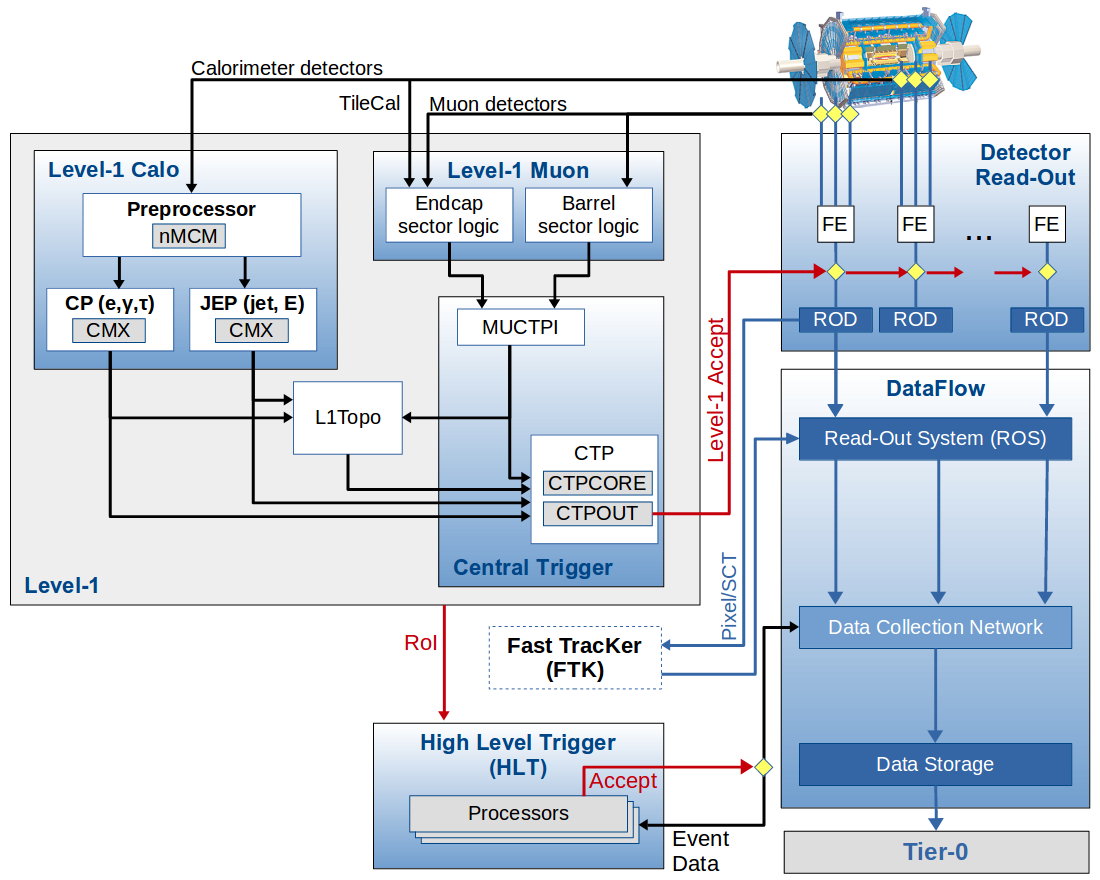}
\end{center}
\caption{Scheme of the ATLAS Trigger and Data Acquisition system in Run 2 with specific focus given to the components of the L1 Trigger system~\cite{trigger}.}
 \label{trigger}
\end{figure}
\begin{figure}[H]
\begin{center}
\includegraphics[height=6 cm,width=11 cm]{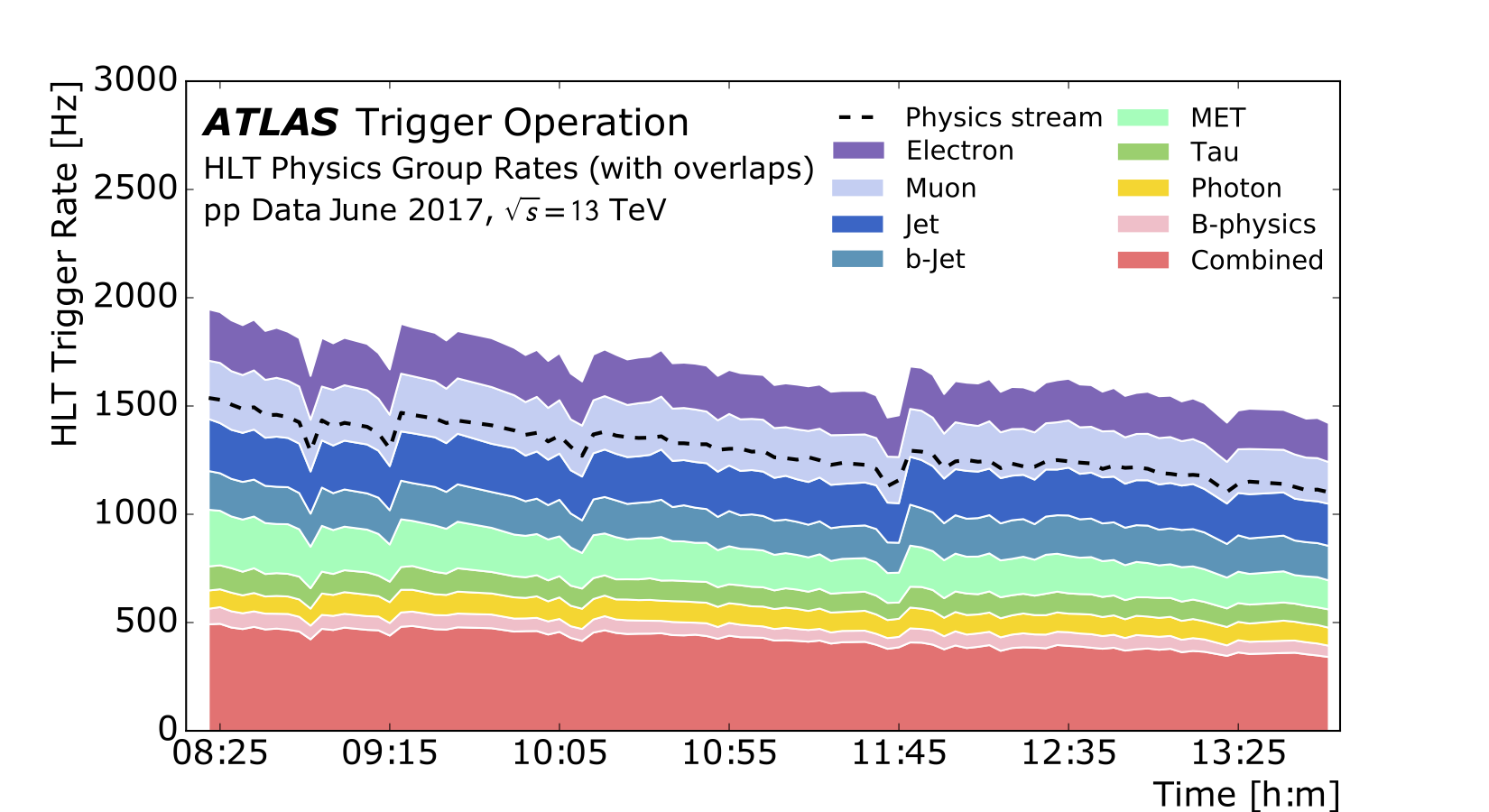}
\end{center}
\caption{Physics trigger group rates at the High Level Trigger (HLT) as a function of time in a fill taken in June 2017 with a peak luminosity of $\mathcal{L} = 1.53\times 10^{34}~$cm$^{-2}s^{-1}$ and a peak pile-up of $\mu$ = 43. Presented are the rates of the individual trigger groups specific to trigger physics objects~\cite{trigger1}.}
 \label{trigger1}
\end{figure}

The Trigger and Data Acquisition system (TDAQ)  is an essential component of the ATLAS experiment because it is responsible for deciding whether or not to save a given collision as an interesting physics process for the offline analysis among the large number of data collected. The main challenges this system has to face are the unprecedented rate of events, $R_{event} \sim$$8\times 10^8$ events/s as discussed in Chapter~\ref{sec:LHC}, the need to select rare predicted physics processes with high efficiency while rejecting much higher-rate background processes, as well as acquire information from large and complex detectors with huge numbers of channels O(107)~\cite{Atlas}. The system has to decrease the event rate from the nominal bunch crossing rate of 40~MHz to a rate of about 1~kHz, that is the maximum reachable rate in order to process data.\newline
Higher luminosity, increased collision energy and higher pile-up have led to an increase of the rates as compared to the Run 1 trigger selections by up to a factor five; therefore, after the LHC long shutdown (2013 -- 2014), the ATLAS trigger system was upgraded to reduce the amount of data~\cite{trigger_run2}. Starting from the three levels of Run 1, Level-2 (L2) and Event Filter (EF) triggers have been merged into a single ``High Level Trigger$"$ (HLT) farm. The Trigger System in Run 2 consists of a hardware-based First Level Trigger (Level-1)~\cite{L1} and a software-based High Level Trigger (HLT)~\cite{HLT}, as schematically shown in Figure \ref{trigger}:
\begin{itemize}
\item \textit{Level-1 Trigger}: the initial selection is made by the Level-1 Trigger, which reduces the event rate from the LHC bunch crossing of $\sim$40~MHz to $\sim$100~kHz, with an overall latency of less than \SI{2.5}{\mu \second}; it uses custom electronics to determine Regions-of-Interest (RoIs) in the detector, the size of such regions depending on the type of object being triggered, taking as input coarse granularity calorimeter and muon detector information. There are two types of Level-1 triggers: calorimeter Level-1 and muon Level-1: as an example, the Level-1 muon trigger receives inputs from the RPCs in the barrel region and from the TGCs in the end-cap region;
\item \textit{High Level Trigger}: the second stage of the Trigger System, the HLT, further reduces the event rate up to 1~kHz with a processing time of 200~ms; the RoIs formed at Level-1 are sent to the HLT in which sophisticated offline selection and reconstruction algorithms are run using full granularity detector information in either the RoI defined by the L1 trigger or the whole event; it is used to refine the Level-1 decision and it is the responsible for the final physics selection for the following offline analyses.
\end{itemize}
Figure \ref{trigger1} shows an example of HLT trigger rates grouped by trigger signature in a 2017 run with a peak luminosity of $\mathcal{L} = 1.53\times 10^{34}~$cm$^{-2}s^{-1}$ and a peak pile-up of $\mu$ = 43~\cite{trigger1}.

\section{Luminosity Detectors}
\label{sec:ATLAS_lumi}
\begin{figure}[H]
\begin{center}
\includegraphics[width=\textwidth]{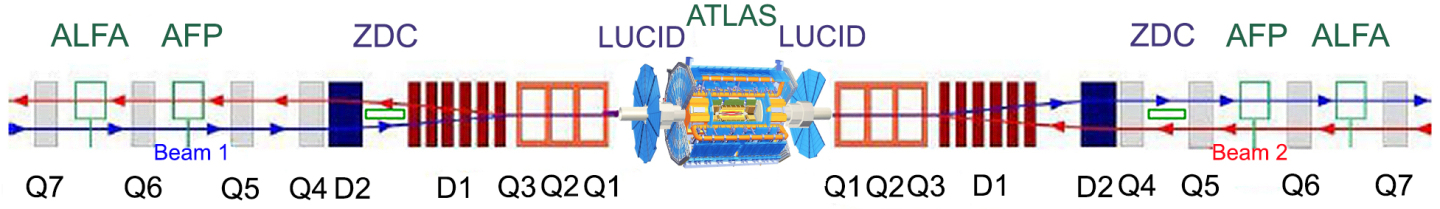}
\end{center}
\caption{ATLAS luminosity detectors~\cite{Forward}.}
 \label{forward_det}
\end{figure}
The main target of the ATLAS forward detectors is to extend ATLAS program by detecting particles in the high rapidity region.\newline
The ATLAS strategy to understand and control the systematic uncertainties affecting the luminosity determination is to compare the measurements of several luminosity detectors~\cite{fabbri} shown in Figure \ref{forward_det}:
\begin{itemize}
\item \textbf{LUCID}, \textbf{L}uminosity \textbf{M}easurements \textbf{U}sing \textbf{C}herenkov \textbf{I}ntegrating \textbf{D}etector, is the official ATLAS luminosity monitor since the beginning of Run 2 and the luminosity detector closest to the IP, located at a distance of $\pm$ 17~m from the IP, covering a pseudorapidity range $5.6 < |\eta| < 6$; its main purpose is to monitor inelastic $pp$ scattering rate in the forward direction with sufficient efficiency and low sensitivity to the background, counting the mean number of inelastic $pp$ collisions through the number of charged particles that are produced in each collision within the LUCID acceptance; it can both measure the integrated luminosity and provide online monitoring of the instantaneous luminosity.
\item \textbf{ZDC}, \textbf{Z}ero \textbf{D}egree \textbf{C}alorimeter is a system of calorimeters designed for relative luminosity measurements during $pp$ and heavy ion runs, placed at a distance of $\pm$$140$~m from the IP and covering a pseudorapidity range $8.3 < |\eta|$, about zero degree to the beam; the primary purpose of this system is to detect forward neutrons in heavy-ion collisions and $pp$ collisions.
\item \textbf{ALFA}, the \textbf{A}bsolute \textbf{L}uminosity \textbf{F}or \textbf{A}TLAS is based on scintillator fibers placed in Roman Pots (RP)~\cite{Roman} close to the LHC proton beam at 240~m from the IP, covering a pseudorapidity range $10.6< |\eta| <13.5$. Its target is to measure the $pp$ scattering at very small angles in order to determine, at the same time, the absolute luminosity and the total $pp$ cross-section $\sigma_{tot}$.
\end{itemize}

\chapter{Reconstruction of physics objects}
\label{sec:Reco}
\begin{figure}[hbtp]
\begin{center}
\includegraphics[height=6 cm,width =10 cm]{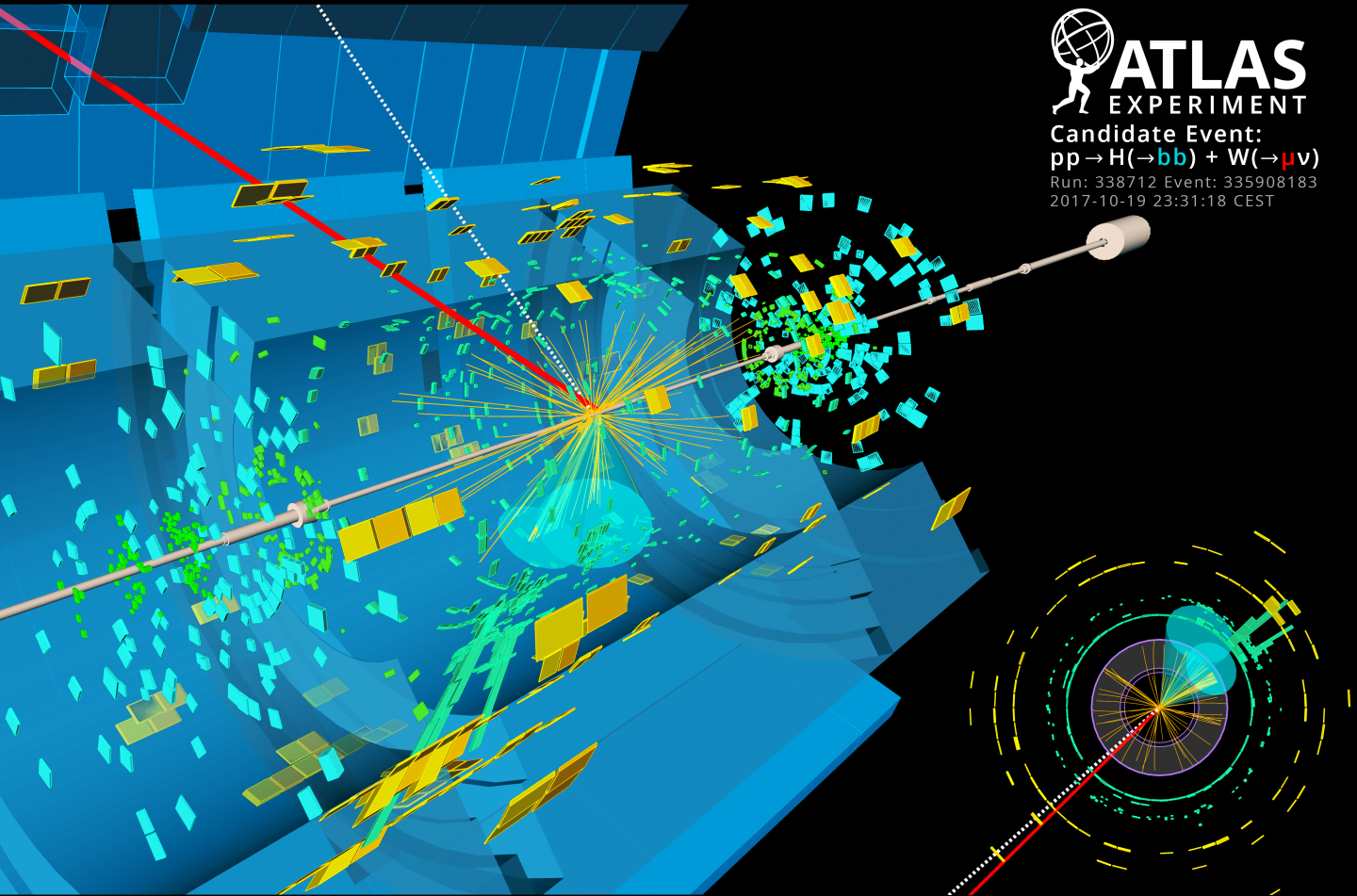}
\end{center}
\caption{A candidate event display for the production of a Higgs boson decaying to two $b$$-$quarks (blue cones), in association with a $W$ boson decaying to a muon (red) and a neutrino~\cite{Higgs_photo}.}     
\label{Higgs_ev} 
\end{figure}

This chapter presents a general overview of the reconstruction of physics objects, a procedure that consists of combining information collected from the sub-detectors described in Chapter~\ref{sec:ATLAS} and using it in order to reconstruct interaction vertices (Section~\ref{sec:tracks}), to identify, from tracks and calorimeter clusters, electrons and photons (Section~\ref{sec:electron}), jets (Section~\ref{sec:jet}), muons (Section~\ref{sec:muon}) and tau leptons (Section~\ref{sec:tau}), and to measure the global properties of the event, like the total transverse energy and, through the energy balance, the so-called ``missing transverse energy$"$ (Section~\ref{sec:missing}) attributed to neutrinos; tau leptons, due to their short lifetime of $2.9\times 10^{-13}\SI{}{\second}$ ($c\tau=$ \SI{87}{\mu\metre}) decay inside the beam pipe, so they are not identified as tau reconstructed objects, but they are reconstructed and identified through their decay products, from leptonic and hadronic decay modes.\newline
Figure~\ref{Higgs_ev} shows a candidate event display for the \VH$\rightarrow b\bar{b}$ channel where the Higgs boson decays to two $b$-quarks and the $W$ boson to a muon and a neutrino; the latter leaves the detector without being seen and is thus reconstructed through the missing transverse energy represented by the dashed line.   

All physics analyses need to define their objects of interest; given the fact that the results presented in this thesis come from the combination of many analyses targeting Higgs-boson decay channels, a brief description of the several types of reconstruction will be reported in the following sections.

\section{Reconstruction of tracks and vertices}
\label{sec:tracks}
Tracks are used both to identify the particles produced in the collisions and to locate the primary vertex, by extrapolating their path to the beam line; they are reconstructed from individual particle interaction with the detector using a sequence of algorithms~\cite{Track_reco}:
\begin{itemize}
\item track reconstruction begins with the formation of space-points: the first step consists in exploiting detector information in order to create clusters in the Pixel and SCT, and drift circles in the TRT. Then, clusters and drift circles are transformed into 3D space-points, \ie three dimensional representations of detector measurements. A space-point corresponds to a hit in the IBL and Pixel detector, while the SCT space-points correspond to hits on both sides of the module. Clusters created by charge deposits from one particle are called single-particle clusters. Clusters created by charge deposits from multiple particles are called merged clusters; the different types of clusters are illustrated in Figure~\ref{cluster}.
\begin{figure}[hbtp]
\begin{center}
\includegraphics[height=6 cm,width =14 cm]{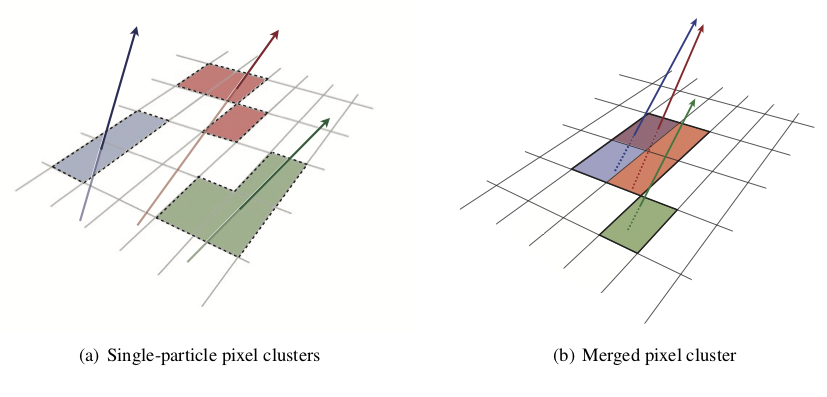}
\end{center}
\caption{Illustration of (a) single-particle pixel clusters on a pixel sensor and (b) a merged pixel cluster due to very collimated charged particles. Different colours represent energy deposits from different charged particles traversing the sensor and the particles trajectories are shown as arrows~\cite{Track_reco}.}     
\label{cluster} 
\end{figure}
\item Track seeds are formed from sets of three space-points, either Pixel-only, SCT-only or Mixed seed. Seeds which pass the initial transverse momentum ($p_T$) and impact parameter resolution cuts are also required to match a fourth space-point that is compatible with the particle's trajectory estimated from the seed. A combinatorial Kalman filter~\cite{Kalman} is then used to build track candidates from the chosen seeds by incorporating additional space-points from the remaining layers of the pixel and SCT detectors which are compatible with the preliminary trajectory.
\item When all combinations of space-points have been made, there are a number of track candidates where space-points overlap, or have been incorrectly assigned. This necessitates an ambiguity-solving stage: track candidates are ranked based on track score, favouring tracks with a higher score. Track candidates are rejected if they fail to meet basic quality criteria. An artificial neural network (NN) clustering algorithm~\cite{Neural} is trained to identify merged clusters and separate multiple particles within a merged cluster in dense environments.
\item Finally, the track candidates selected through this procedure are extended to the TRT if there is a valid set of matching drift circles. Then, a high-resolution fit is performed using all available information. Fitted tracks which pass through the ambiguity solver without modification are added to the final track collection.
\end{itemize}
The reconstruction of the primary vertex, \ie the location where the $pp$ interaction takes place, is essential for physics analyses. 
The correct assignment of charged particle trajectories to their source vertex, together with an accurate reconstruction of the number and positions of interaction vertices, is essential to reconstruct the full kinematic properties of the hard-scatter and separate the effects of additional collisions.\newline
The reconstruction of vertices can be split into two main stages~\cite{Vertex}: vertex finding, \ie\ the association of reconstructed tracks to a given vertex candidate, and vertex fitting, \ie\ the reconstruction of the actual vertex position.
\begin{figure}[hbtp]
\begin{center}
\includegraphics[height=8 cm,width =10 cm]{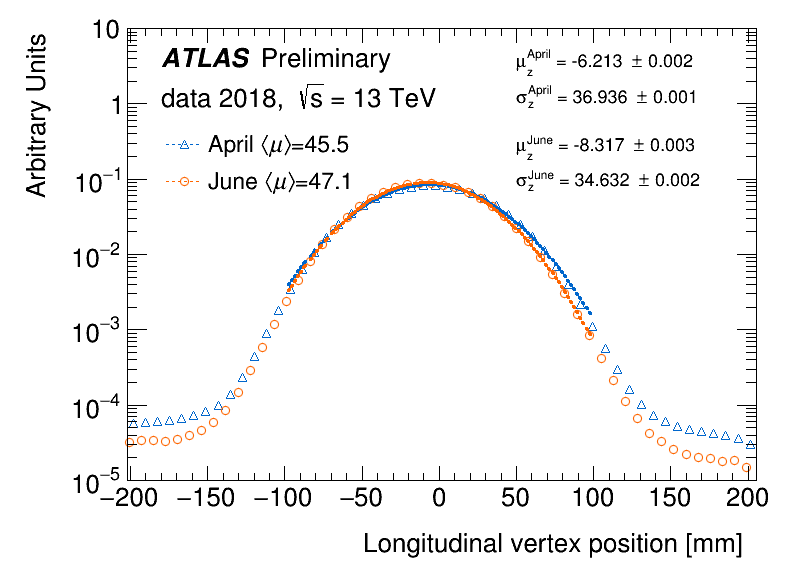}
\end{center}
\caption{Normalised distribution of the longitudinal position of primary vertices in two fills with different average $\mu$ taken at different points in 2018. The non-gaussian tails at large values of $z$ are due to fake vertices. Fakes vertices, that are expected in normal operation, are due to combinatorics and become relevant in high pile-up environments. A gaussian function is fitted in the range (-100, 100)~mm to extract the mean value and the sigma of the distribution~\cite{primary}.}
\label{primary} 
\end{figure}

It consists of the following steps~\cite{Vertex1}:
\begin{itemize}
\item a set of tracks passing the vertex selection criteria is defined;
\item a seed position for the first vertex is selected;
\item the tracks and the seed are used to fit the best-vertex position, exploiting an iterative procedure; in each iteration, less compatible tracks are down-weighted and the vertex position is recomputed;
\item after the determination of the vertex position, tracks that are found incompatible with the vertex are removed and are used in the determination of another vertex; vertices are required to have at least two associated tracks;
\item this procedure is repeated until no track in the event is left unassociated or no additional vertex can be found in the remaining set of tracks.
\end{itemize}
The output of the vertex reconstruction algorithm is a set of three dimensional vertex positions. 
Figure~\ref{primary} shows the normalised distribution of the longitudinal position of primary vertices in two fills with different average $\mu$, taken at different points in 2018.\newline
The vertex reconstruction efficiency is determined from data by taking the ratio between events with a reconstructed vertex and events with at least two reconstructed tracks.

\section{Electron and photon reconstruction and identification}
\label{sec:electron}
The reconstruction of electrons and photons is based on combining information from the tracking and calorimeter systems. The interactions of photons and electrons with the ATLAS ECal produce similar electromagnetic showers, depositing a significant amount of energy in a restricted number of neighbouring calorimeter cells; thus their reconstruction proceeds in parallel through the following steps~\cite{Photon_reco}:
\begin{enumerate}
\item \textit{Topo-cluster reconstruction}: the preparation of the clusters uses a topo-cluster reconstruction algorithm~\cite{topo, topo1} whose first step is the formation of proto-clusters in the electromagnetic and hadronic calorimeters using a set of noise thresholds in which the cell initiating the cluster is required to have significance $|\varsigma_{cell}^{EM}| \ge 4$, being the significance defined as:
\begin{equation}
\varsigma_{cell}^{EM}=\frac{E_{cell}^{EM}}{\sigma_{noise, cell}^{EM}}
\end{equation}
where $E_{cell}^{EM}$ is the cell energy at the EM scale and $\sigma_{noise, cell}^{EM}$ is the expected cell noise that includes the known electronic noise and an estimation of the pile-up noise corresponding to the average instantaneous luminosity expected for Run 2.
In this initial stage, cells from the pre-sampler and the first LAr EM calorimeter layer are excluded from initiating proto-clusters, to suppress the formation of noise clusters. Then the proto-clusters collect neighbouring cells with significance $|\varsigma_{cell}^{EM}| \ge 2$; each neighbour cell that passes this threshold becomes a seed cell in the next iteration, collecting each of its cell neighbours in the proto-cluster. After all nearby cells have been collected, a final set of neighbouring cells with $|\varsigma_{cell}^{EM}| \ge 0$ is added to the cluster.
 \item \textit{Track reconstruction}~\cite{El_reco}: topo-clusters are associated to tracks reconstructed in the ID. The track information from the ID is extracted by using both pattern recognition and track fit. The pattern recognition algorithm uses the pion hypothesis for energy loss from interactions of the particle with the detector material, or the electron hypothesis if the track seeds have transverse momentum above 1 \GeV\ and the track seed cannot be extended to a full track (with at least 7 hits) using the pion hypothesis; the modified pattern recognition, designed to better account for energy loss of charged particles in materials, uses an optimised Gaussian-sum filter (GSF)~\cite{GSF} fit while the standard fit uses the ATLAS Global $\chi^2$ Track Fitter~\cite{chi2fitter}. The photon conversion reconstruction proceeds in a similar way~\cite{Photon_reco}: a candidate particle is reconstructed as a photon if there are no tracks with at least four hits in the silicon detector matched to the calorimeter cluster; tracks loosely matched to fixed-size clusters serve as input to the reconstruction of the conversion vertex. Both tracks with silicon hits (denoted Si tracks) and tracks reconstructed only in the TRT (denoted TRT tracks) are used for the conversion reconstruction. Two-track conversion vertices are reconstructed from two opposite-charged tracks forming a vertex consistent with that of a massless particle, while single-track vertices are essentially tracks without hits in the innermost sensitive layers. To increase the converted-photon purity, requirements on the TRT tracks and on double-track Si conversions are applied.
\item \textit{Super-cluster reconstruction}: 
the reconstruction of electron and photon super-clusters proceeds independently, each in two stages: in the first stage, EM topo-clusters are tested in order to be used as seed cluster candidates, which form the basis of super-clusters; in the second stage, EM topo-clusters near the seed candidates are identified as satellite-cluster candidates, which may emerge from bremsstrahlung radiation or topo-cluster splitting. If satellite clusters satisfy the necessary selection criteria, they are added to the seed candidates to form the final super-clusters. The super-clustering algorithm, shown in the sketch of Figure~\ref{super}, has replaced the sliding-window algorithm~\cite{topo1} to search for cluster ``seeds$"$ previously exploited in ATLAS for the reconstruction of fixed-size clusters of calorimeter cells.
\begin{figure}[hbtp]
\begin{center}
\includegraphics[height=6 cm,width =10 cm]{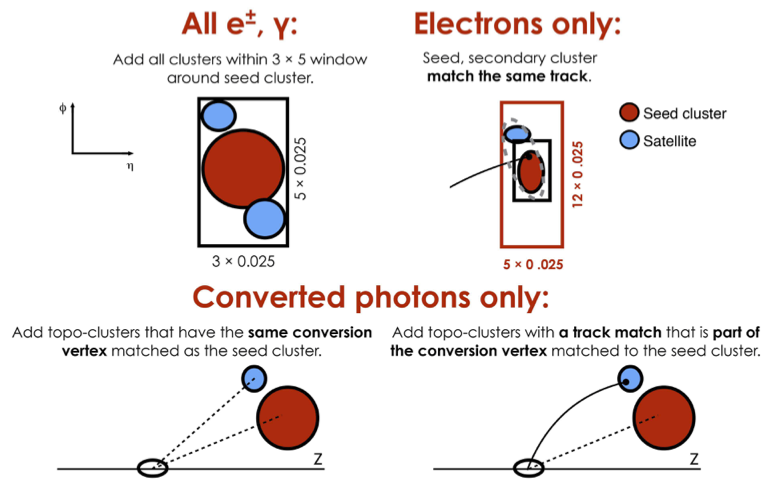}
\end{center}
\caption{Sketch of the super-clustering algorithm for electrons and photons. Seed clusters are shown in red, satellite clusters in blue~\cite{Photon_reco}.}     
\label{super} 
\end{figure}
\item \textit{Creation of electrons and photons for analysis}: after building electron and photon super-clusters, initial energy calibration and position correction are applied, and tracks are matched to electron super-clusters and conversion vertices to photon super-clusters. The matching is performed in the same way the matching to EM topo-clusters was performed, but using super-clusters. After this matching, analysis-level electrons and photons are built and discriminating variables used to separate electrons or photons from background are added.
\end{enumerate}
\subsection{Electron identification and isolation}
\begin{figure}[hbtp]
\begin{center}
\includegraphics[height=8 cm,width =9 cm]{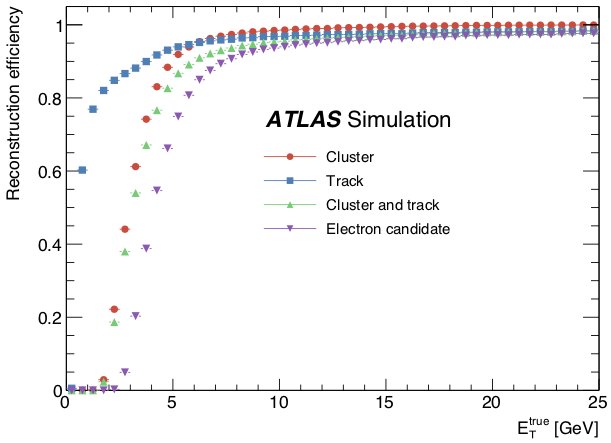}
\end{center}
\caption{Cluster, track, cluster and track, and electron reconstruction efficiencies as a function of the electron $E_T$~\cite{Photon_reco}.}     
\label{el_eff}
\end{figure}

\begin{figure}[hbtp]
\begin{center}
\includegraphics[height=9 cm,width =9 cm]{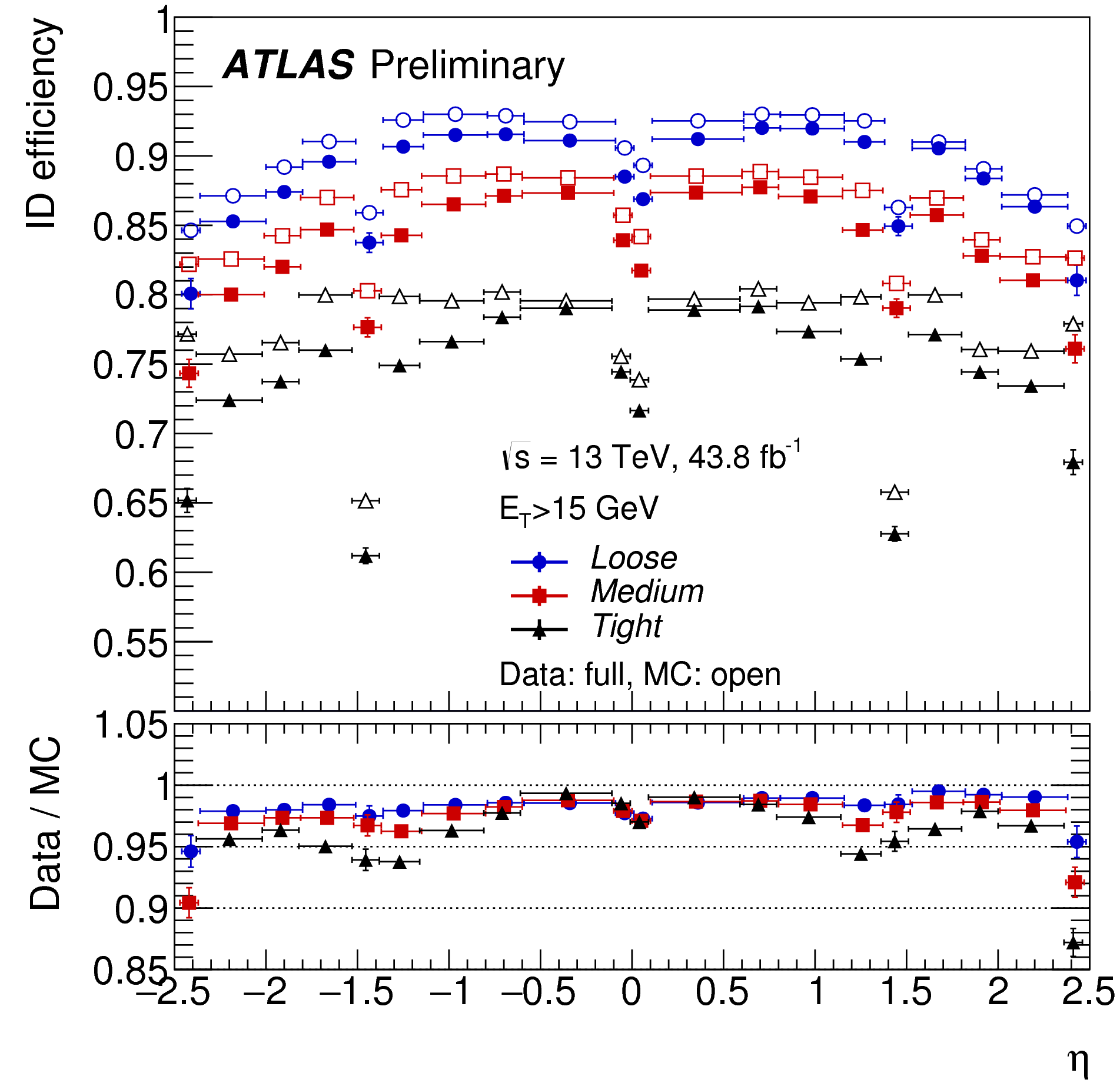}
\end{center}
\caption{Electron identification efficiencies in $Z\rightarrow e^+e^-$ events as a function of the pseudorapidity $\eta$ for electrons with $E_T>15$~$\GeV$. The efficiencies are shown in data and MC for the three operating points, Loose, Medium and Tight. The data efficiencies are obtained by applying data/MC efficiency ratios that were measured in $J/\Psi \rightarrow e^+e^-$ and $Z\rightarrow e^+e^-$ events to MC simulation. A dataset corresponding to an integrated luminosity of $43.8$~fb$^{-1}$ that was recorded by the ATLAS experiment in the year 2017 at a centre-of-mass energy of $\sqrt{s}=13$~\TeV\ was used. The total statistical and systematic uncertainty is shown~\cite{El_reco_site}.}     
\label{eff_eta_LMT}
\end{figure}

The baseline identification algorithm for electrons is the likelihood-based (LH) method, that is a multivariate technique evaluating signal versus background probability density functions. For each electron candidate, these probabilities are combined into a discriminant $d_L$, defined as:
\begin{equation}
d_L=\frac{L_S}{L_S+L_B} \qquad \qquad \text{with} \qquad \qquad L_{S(B)}(\vec{x})= \prod_{i=1}^{n} P_{s(b),i}(x_i) 
\end{equation}
where:
\begin{itemize}
\item $\vec{x}$ is the set of discriminating variables, like track conditions from the ID measurement and track-cluster matching;
\item $P_{s(b),i}(x_i)$ are the values of the signal/background pdfs for quantity $i$ at value $x_i$;
\item $L_{S(B)}(\vec{x})$ are the likelihood functions for signal and background.
\end{itemize}
Three fixed values of the LH discriminant are used to define three operating points corresponding to increasing thresholds for the LH discriminant; they are referred to as Loose, Medium, and Tight.\newline
In order to further reject hadronic jets misidentified as electrons, most analyses require electrons to pass some isolation requirements in addition to the identification requirements described above, where the isolation is a measurement of the detector activity around a candidate. The two main isolation variables are: calorimeter-based isolation and track-based isolation. The total efficiency $\epsilon_{total}$ for a single electron can be factorised as a product of different efficiency terms~\cite{El_reco} that can be estimated directly from data using tag-and-probe methods~\cite{tag1,tag2}:
\begin{equation}
\epsilon_{total}=\epsilon_{EMclus}\times \epsilon_{reco}\times \epsilon_{id}\times \epsilon_{iso}\times \epsilon_{trig}=
\end{equation}
\begin{equation}
= \left ( \frac{N_{cluster}}{N_{all}} \right) \times \left ( \frac{N_{reco}}{N_{cluster}} \right) \times \left ( \frac{N_{id}}{N_{reco}} \right) \times \left ( \frac{N_{iso}}{N_{id}} \right) \times \left ( \frac{N_{trig}}{N_{iso}} \right) \nonumber
\end{equation}
where:
\begin{itemize}
\item the efficiency to reconstruct EM-cluster candidates, $\epsilon_{EMclus}$, is given by the number of reconstructed EM calorimeter clusters, $N_{cluster}$, divided by the number of produced electrons, $N_{all}$;
\item the reconstruction efficiency, $\epsilon_{reco}$, is given by the number of reconstructed electron candidates, $N_{reco}$, divided by the number of EM-cluster candidates, $N_{cluster}$;
\item the identification efficiency, $\epsilon_{id}$, is given by the number of identified and reconstructed electron candidates, $N_{id}$, divided by $N_{reco}$;
\item the isolation efficiency, $\epsilon_{iso}$, is calculated as the number of identified electron candidates satisfying the isolation, identification, and reconstruction requirements, $N_{iso}$, divided by $N_{id}$;
\item the trigger efficiency, $\epsilon_{trigger}$, is calculated as the number of triggered (and isolated, identified, reconstructed) electron candidates, $N_{trig}$, divided by $N_{iso}$.
\end{itemize}
The reconstruction efficiency, $\epsilon_{reco}$, is shown in Figure~\ref{el_eff} as a function of the true (generator) transverse energy $E_T$ for each step of the electron-candidate formation. Figure~\ref{eff_eta_LMT} reports electron identification efficiencies as a function of the pseudorapidity for the three operating points in $Z\rightarrow e^+e^-$ events, used to benchmark the expected electron efficiencies and to define the electron identification criteria, exploiting the 2017 ATLAS dataset that corresponds to an integrated luminosity of 43.8~fb$^{-1}$~\cite{El_reco_site}. 

\clearpage
\subsection{Photon identification and isolation}
\begin{figure}[hbtp]
\centering
\begin{subfigure}[b]{0.49\textwidth}
\includegraphics[height=7 cm,width =8 cm]{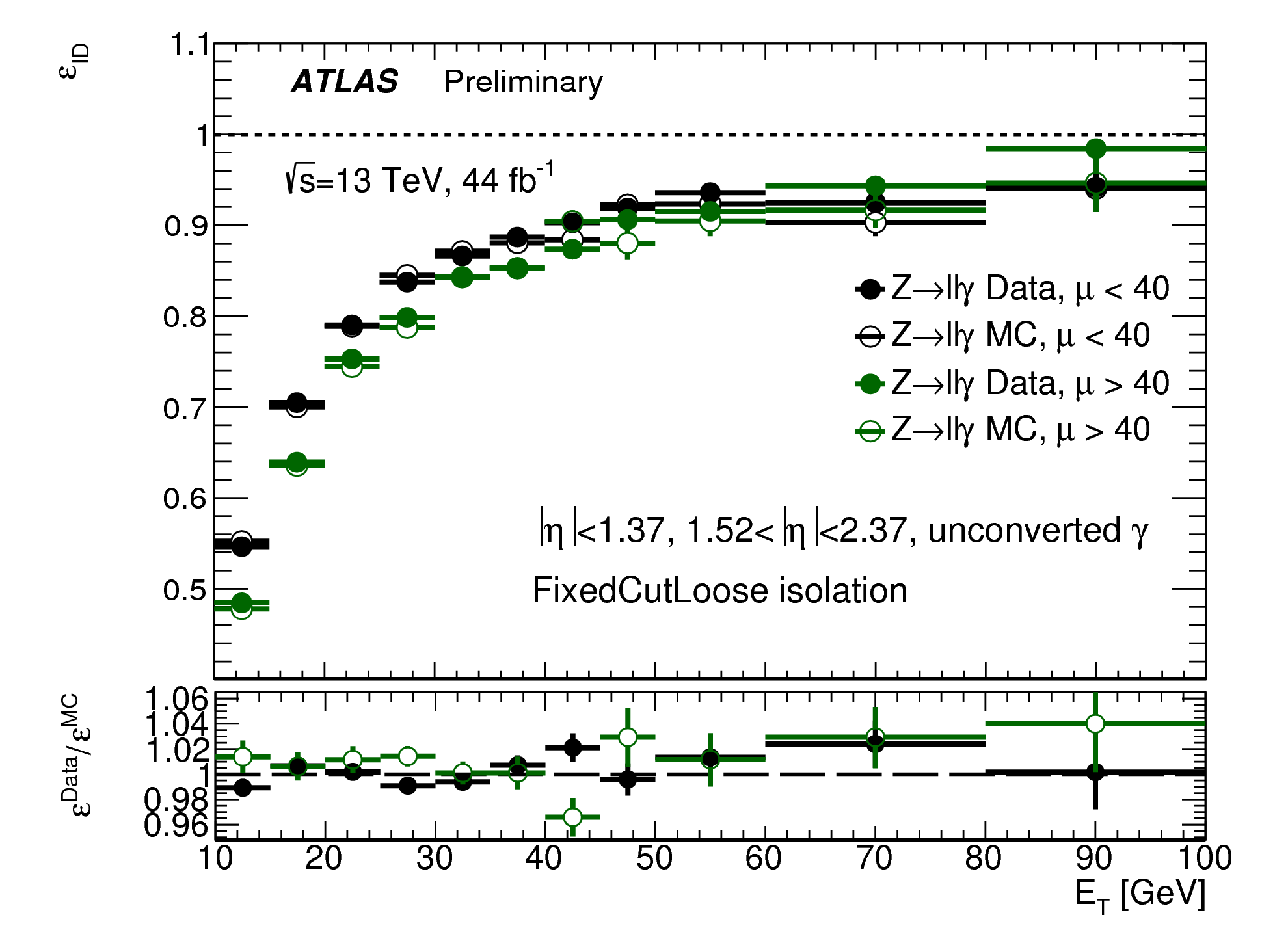}
 \caption{}
\end{subfigure}
\begin{subfigure}[b]{0.49\textwidth}
\includegraphics[height=7 cm,width =8 cm]{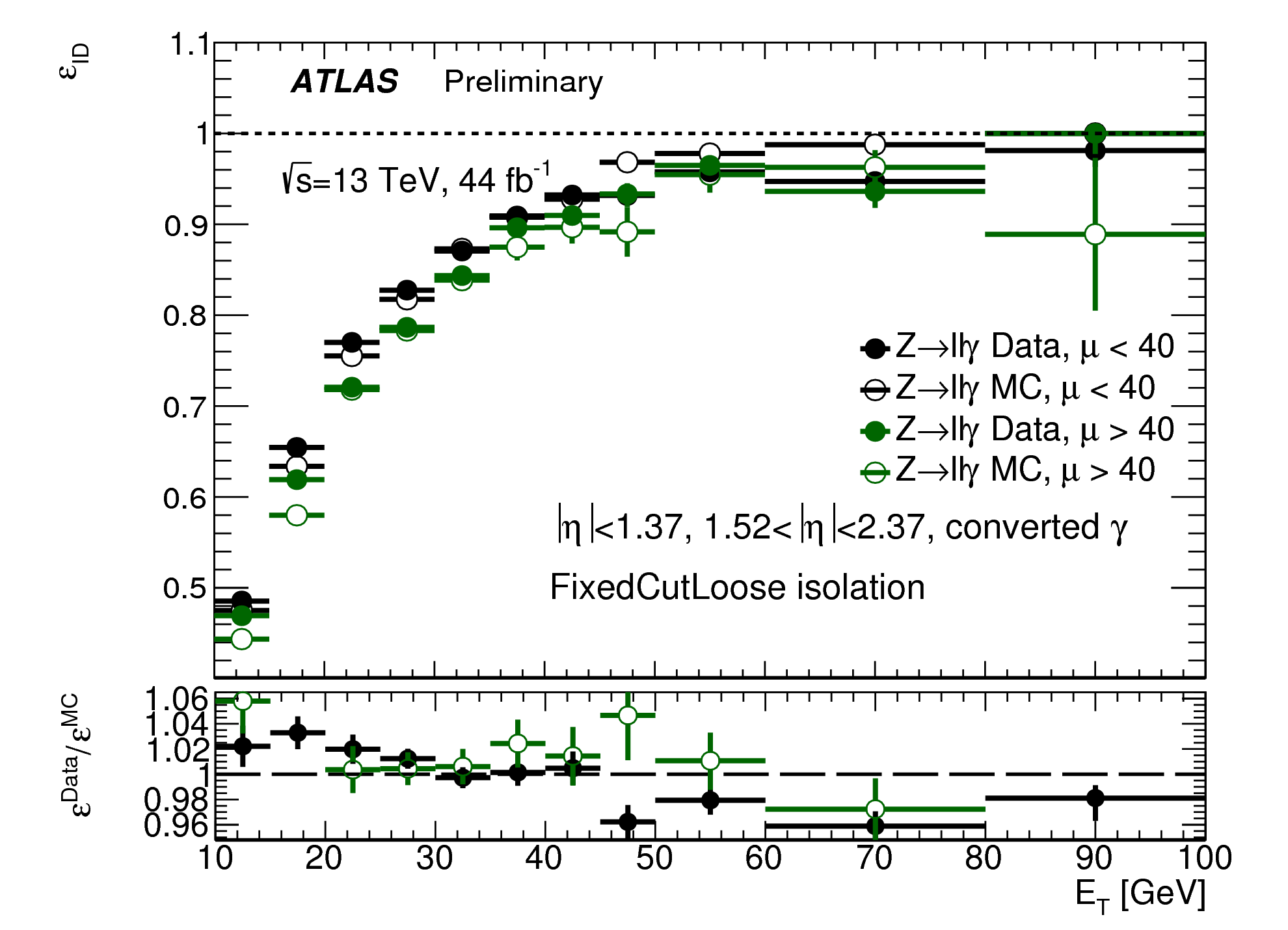}
 \caption{}
\end{subfigure}
\caption{Efficiency of the tight identification requirement for unconverted (a) and converted photons (b) from radiative $Z$ decays as a function of the transverse energy $E_T$, averaged over pseudorapidity. Loose photon isolation requirement is applied. The efficiencies have been measured using 44.3~fb$^{-1}$ of data recorded by the ATLAS experiment in 2017 at $\sqrt{s}$=13~\TeV\ for $\mu <$$ 40$, and are compared to the 2017 data with $\mu >$$40$. Only the statistical uncertainties are shown~\cite{El_reco_site}.}     
\label{eff_photons}
\end{figure}

The photon identification criteria are designed to efficiently select prompt, isolated photons and reject backgrounds from hadronic jets. The photon identification is constructed from one-dimensional selection criteria, or a cut-based selection, using shower shape variables like the lateral and the total shower widths in different EM layers~\cite{Photon_reco}. The variables using the EM first layer play a particularly important role in rejecting $\pi^0$ decays into two highly collimated photons.\newline
Three identification selections are identified: the primary one is labelled as Tight, with less restrictive selections called Medium and Loose, which are used for trigger algorithms. \newline
In order to better discriminate signal vs background, additional information is exploited quantifying the activity near photons from the tracks of nearby charged particles, or from energy deposits in the calorimeters; requirements on the calorimeter and track isolation variables are applied, and three photon isolation operating points are defined, \ie\ FixedCutLoose, FixedCutTight and FixedCutTightCaloOnly operating points.\newline
Figure~\ref{eff_photons} shows the efficiency of the tight identification requirement for unconverted $(a)$ and converted photons $(b)$ from radiative $Z$ decays as a function of the transverse energy $E_T$ using 44.3~fb$^{-1}$ of data recorded by the ATLAS experiment in 2017~\cite{El_reco_site}.

\section{Jet and $b$-jet reconstruction}
\label{sec:jet}

Hadronic collisions in the ATLAS experiment produce a variety of particles, including quarks or gluons, called ``partons$"$. Due to colour confinement, partons cannot exist individually and they re-combine, \ie\ ``hadronise$"$, with quarks and anti-quarks spontaneously created from the vacuum, in order to form hadrons. A  jet can thus be defined as a collimated shower of stable particles arising from fragmentation and hadronisation of a parton (quark or gluon) after a collision.
Jet reconstruction algorithms combine calorimeter objects grouping individual clusters in an ordered way. The jets provide a link between the observed colourless stable particles and the underlying physics at the partonic level~\cite{review_jet}.\newline
Jets can be reconstructed through the following steps~\cite{jet1}:
\begin{enumerate}
\item the inputs of the jet reconstruction are topo-clusters, \ie topologically-grouped noise-suppressed clusters of calorimeter cells. They are formed from seed cells through the procedure already described in Section~\ref{sec:electron}. The process concludes by adding all calorimeter cells adjacent to the topo-cluster, irrespective of their energy;
\item a jet finding algorithm is then typically used, and the standard ATLAS solution is the anti-$k_t$~\cite{anti_kt} algorithm, whose application is shown in Figure~\ref{jet_antikt}, that sequentially recombines clusters exploiting the following procedure:
\begin{figure}[hbtp]
\begin{center}
\includegraphics[height=7 cm,width =7 cm]{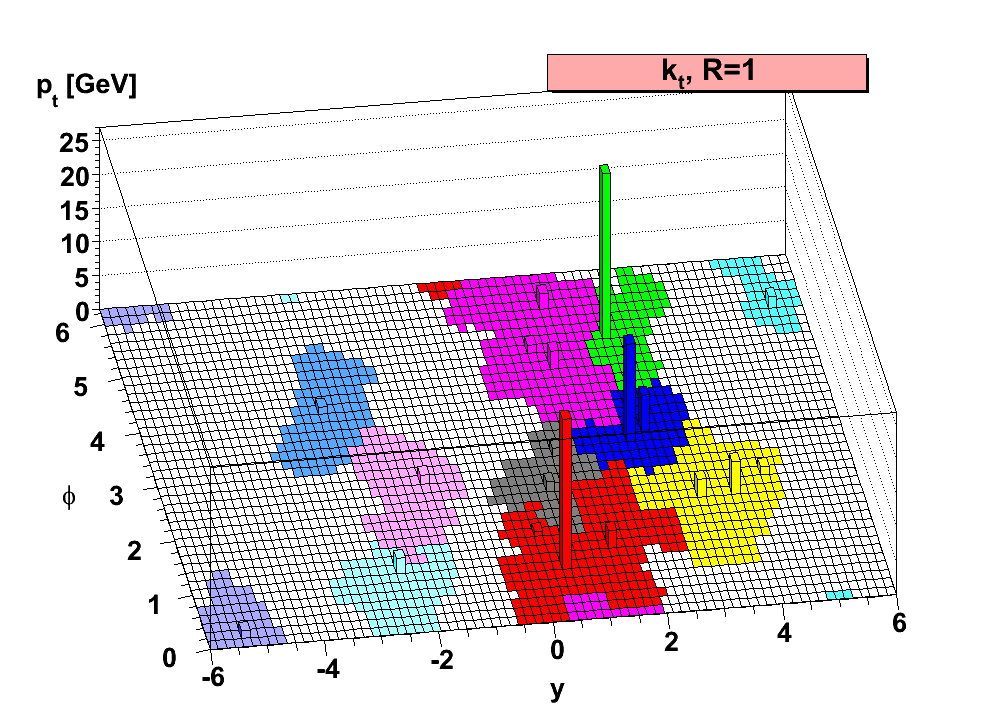}
\end{center}
\caption{A sample parton-level event, together with many random soft ``ghosts$"$, clustered with the ant-$k_t$ algorithm, illustrating the ``active$"$ catchment areas of the resulting hard jets~\cite{anti_kt}.}     
\label{jet_antikt}
\end{figure}
\begin{itemize}
\item the distance $d_{ij}$ between entities (particles, pseudo-jets) $i$ and $j$ is evaluated:
\begin{equation}
d_{ij}=min(k_{ti}^{2p}, k_{tj}^{2p})\frac{\Delta_{ij}^2}{R^2}  
\end{equation}
where $\Delta_{ij}=(y_i-y_j)^2+(\phi_i-\phi_j)^2$ and $y_i$, $\phi_i$ and $k_{ti}$,  are respectively the rapidity, the azimuthal coordinate and the transverse momentum of the $i$ particle, $R$ represents the size of the jet, while $p$ is a parameter of the anti-$k_t$ algorithm fixed equal to -1.
\item For each entity $i$, the distance from the beam is estimated:
\begin{equation}
d_{iB}=k_{ti}^{2p} \, .
\end{equation}
\item The minimum distance between $d_{ij}$ and $d_{iB}$ is identified.
\item If $d_{ij}$ is the minimum distance, $i$ and $j$ are combined into a single pseudo-jet and the procedure is repeated from the first step; if this is not the case, $i$ is considered as a final state and is removed from the list of entities; the distances are recalculated and the procedure is repeated until no entities are left. 
\end{itemize}
Two different distance parameters $R$ are typically used: jets representing quarks and gluons are typically called ``small-$R$$"$ jets, and are reconstructed with $R=0.4$, while jets representing hadronically decaying massive particles are typically called ``large-$R$$"$ jets, and are reconstructed with $R=1.0$.
\item Using large-$R$ jets, results in a substantially increased sensitivity to pile-up effects due to the larger fraction of the calorimeter enclosed within the jet volume; this kind of jets are however necessary to fully contain the hadronic massive particle decays. Additionally, pile-up is randomly distributed so it can obscure the angular structure within the jet, representing the key element in order to identify massive particle decays. To get around these limitations, large-$R$ jets are typically groomed, where grooming is a class of algorithms that take a jet and throw away constituents following a defined strategy and rebuilding the final jet from the remaining constituents. 
\end{enumerate}

\begin{figure}[hbtp]
\begin{center}
\includegraphics[height=6 cm,width =16 cm]{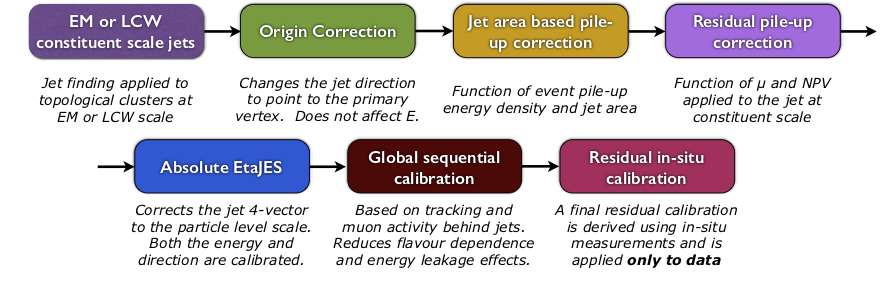}
\end{center}
\caption{Overview of the ATLAS jet calibration scheme~\cite{jet_cali}.}     
\label{jet_calibration}
\end{figure}
After jets have been built, they have to be calibrated to account for several effects as, for example, the fact that the energy scale of reconstructed jets does not correspond to the truth-particle jet energy scale (JES), defined as the energy of jets built from all stable Monte Carlo particles from the hard interaction only, including the underlying event activity. A dedicated jet energy calibration is then needed to calibrate, on average, the reconstructed jet energy to that of the corresponding truth-particle jet. The energy scale calibration needs to also correct for pile-up effects, that add energy deposits to the jets from the hard-scatter event and create additional jets (pile-up jets). Furthermore, the jet energy calibration has to bring the energy scale of jets in data and simulation to the same footing~\cite{cali_jet}. Figure~\ref{jet_calibration} shows an overview of the jet calibration used in the ATLAS experiment.

\subsection{$b$-tagging algorithms}

\begin{figure}[hbtp]
\begin{center}
\includegraphics[height=7 cm,width =7 cm]{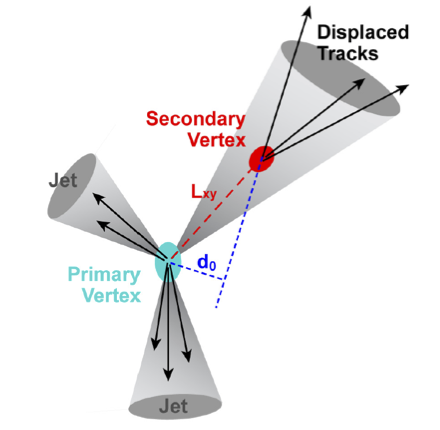}
\end{center}
\caption{Schematic view of a jet containing a secondary vertex and displaced tracks, signature of a $b$-jet. Tracks are represented by arrows and the circles mark the primary (grey/blue) and secondary (dark grey/red) vertices. The transverse decay length $L_{xy}$ and the transverse impact parameter distance $d_0$ which characterise the secondary vertex are indicated by dashed lines~\cite{btag}.}     
\label{btag}
\end{figure}

\begin{figure}[hbtp]
\begin{center}
\includegraphics[height=7 cm,width =16 cm]{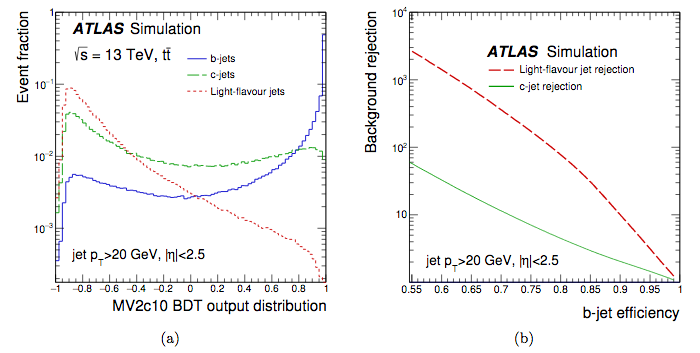}
\end{center}
\caption{The MV2c10 output for $b$-jets (solid-line), $c$-jets (dashed line) and light flavour jets (dotted line) in simulated $t\bar{t}$ events; (b) the light-flavour jet (dashed line) and $c$-jet rejection factors (solid line) as a function of the $b$-jet tagging efficiency of the MV2c10 $b$-tagging algorithm~\cite{btag_eff}.}
\label{btag_eff}
\end{figure}

The long lifetime of hadrons with $b$-quarks ($\sim$$1.5 \times 10^{-12}$ s) results in a typical decay topology with at least one vertex displaced from the primary vertex coming from the hard-scattering collision as shown in Figure~\ref{btag}.
The identification of $b$-quark jets is important for many physics analyses, including the $H\rightarrow b\bar{b}$ channel and all the double-Higgs channels studied in this thesis, \ie\ $b\bar{b}b\bar{b}$, $b\bar{b}\tau^+\tau^-$ and $b\bar{b}\gamma\gamma$; it is based on three fundamental algorithms~\cite{btag1}:
\begin{itemize}
\item an impact-parameter-based algorithm;
\item an inclusive secondary vertex reconstruction algorithm, based on the reconstruction of the distance of the transverse decay length, $L_{xy}$, of the $b$-hadron which is the vector pointing from the primary vertex to the $b$-hadron decay vertex;
\item a decay chain multi-vertex reconstruction algorithm (JetFitter), exploiting the topological features of weak $b$- and $c$-hadron decays inside the jet and trying to reconstruct the full $b$-hadron decay chain.
\end{itemize}
The outputs of these three algorithms are combined into a Boosted Decision Tree (BDT), \ie\ a machine-learning technique combining linear cuts on input discriminant observables in order to maximise the separation between two or more processes, and the multivariate discriminant tagger is called MV2~\cite{btag_eff}, capable to provide the best separation between the different jet flavours. The training of the BDT is performed using $t\bar{t}$ events with $b$$-$jets as signal, and $c$$-$jets and light-flavour jets as background.  Different MV2 taggers can be defined, depending on the fraction of $c$$-$jets used in the training.  Different $b$$-$jets efficiency working points are defined, corresponding to different cuts on the BDT output score of the $b$-tagging algorithm. Since the majority of physics analysis are limited by charm rather than light-flavour jet rejection, the $c$$-$jet fraction is set in such a way to enhance charm rejection keeping a good light-flavour rejection as well. The MV2c10 tagger background composition is made of 93\% light-flavour jets and 7\% $c$$-$jets; its output for $b$$-$jets, $c$$-$jets and light-flavour jets in a $t\bar{t}$ sample is presented in Figure~\ref{btag_eff} (a). The rejection rates for light-flavour jets and $c$-jets are defined as the inverse of the efficiency for tagging a light-flavour jet or a $c$$-$jet as a $b$$-$jet, respectively.  Figure~\ref{btag_eff} (b) shows the corresponding light-flavour jet and $c$$-$jet rejection factors as a function of the $b$$-$jet tagging efficiency.

\section{Muon reconstruction and identification}
\label{sec:muon}
\begin{figure}[H]
\begin{center}
\includegraphics[height=8 cm,width =9 cm]{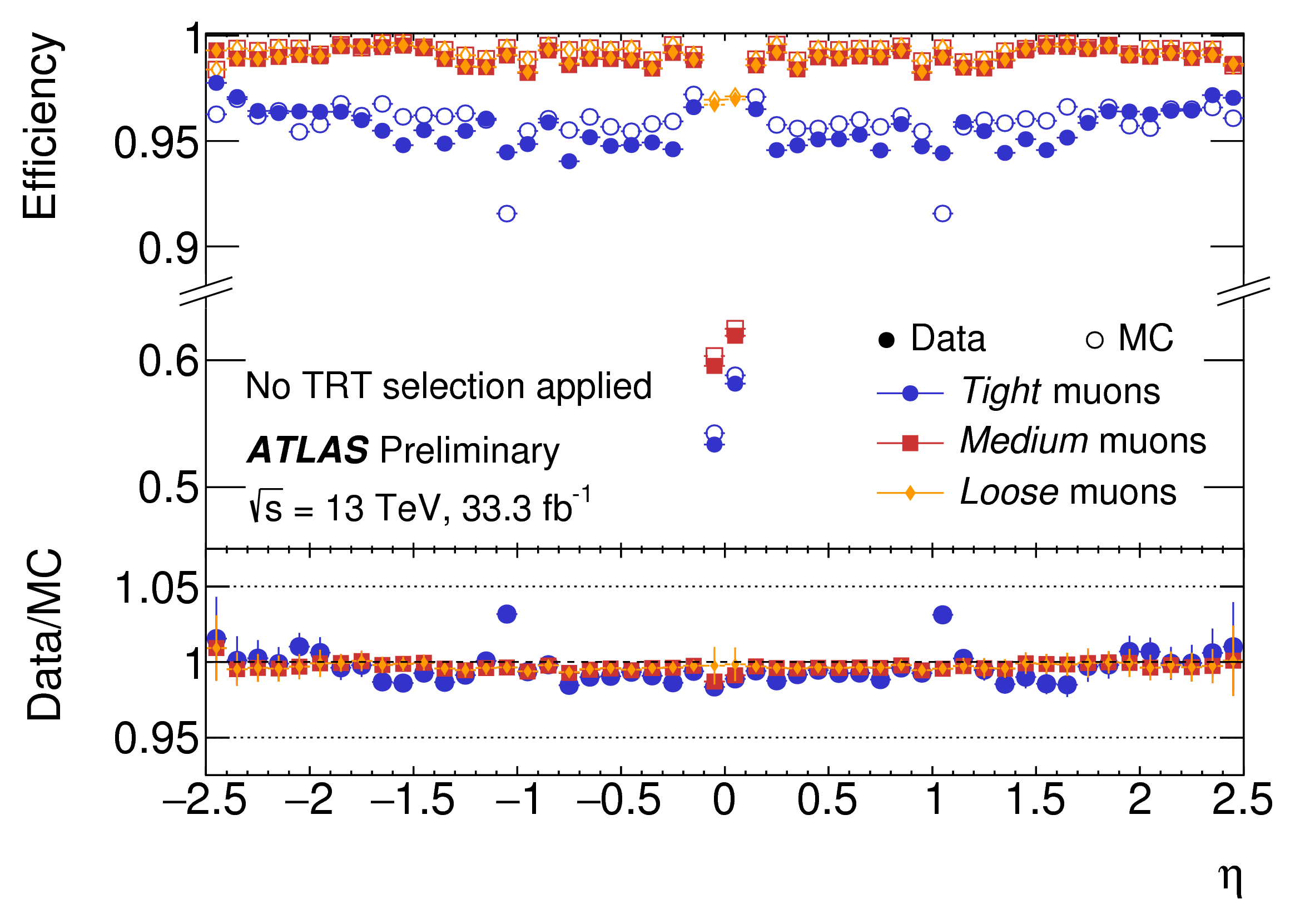}
\end{center}
\caption{Muon reconstruction efficiencies with the full 2016 dataset for the Loose/Medium/Tight identification algorithms measured in $Z\rightarrow \mu^+ \mu^-$ events as a function of the muon pseudorapidity for muons with $p_T$$ > 10$~\GeV. The prediction by the detector simulation is depicted as open circles, while filled dots indicate the observation in collision data with statistical errors. The bottom panel shows the ratio between expected and observed efficiencies, \ie\ the efficiency scale factor. The errors in the bottom panel represent the quadratic sum of statistical and systematic uncertainties~\cite{Muon_reco_site}.}     
\label{muon_eff_2018}
\end{figure}

Muon reconstruction is first performed independently in the ID and in the MS, described in Chapter~\ref{sec:ATLAS}. The information from individual sub-detectors is then combined to form the muon tracks that are used in physics analyses. In the ID, muons are reconstructed following the general track reconstruction described in Section~\ref{sec:tracks}. In the MS, muon reconstruction starts searching for hit patterns from the MDT and trigger chambers through a Hough transform algorithm~\cite{hough} to form segments. This algorithm finds at least two seed-segments in the middle layers of the MDT; then the muon tracks are reconstructed by performing a straight line fit, which takes seed-segments and hits found in each layer as inputs. The RPC or TGC hits measure the coordinate orthogonal to the bending plane. In the CSC chambers, segments are built searching in the $\eta$ and $\phi$ planes. Muon track candidates are then built by fitting together hits from segments in different layers~\cite{Muon_reco1}.\newline 
The combined information of ID, MS and calorimeter system leads to the definition of four types of reconstructed muons, a sketch of whom is shown in Figure~\ref{muon_all_reco}, depending on the sub-detector used in the reconstruction~\cite{Muon_reco}:
\begin{figure}[hbtp]
\begin{center}
\includegraphics[height=6.5 cm,width =7.7 cm]{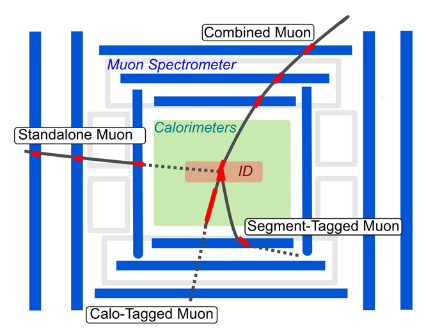}
\end{center}
\caption{Schematic drawing of the detector parts used for muon reconstruction and of the different types of muon reconstruction~\cite{Nick_muon}.}     
\label{muon_all_reco}
\end{figure}
\begin{enumerate}
\item Stand-Alone (SA) muons: the muon trajectory reconstruction is based only on the MS track and on a loose requirement on compatibility with tracks originating from the IP. In general, the muon has to cross at least two layers of MS chambers to provide a track measurement, but three layers are required in the forward region. SA muons are mainly used to extend the acceptance to the pseudorapidity range $2.5<|\eta|< 2.7$ which is not covered by the ID.
\item Combined (CB) muons (main type of reconstructed muons): track reconstruction is performed independently in the ID and in the MS, and a combined track is formed through a global fit procedure that uses hits from both ID and MS sub-detectors. In order to maximise the fit quality, MS hits may be added to or removed from the track. The acceptance of CB muons is limited by the ID coverage, \ie $|\eta| < 2.5$.
\item Segment-tagged (ST) muons: a track in the ID is classified as a muon if, once extrapolated to the MS, it is associated with at least one local track segment in the MDT or CSC chambers. ST muons can be used to increase the acceptance in cases in which the muon has crossed only one layer of MS chambers, either because of its low $p_T$ or because it falls in regions with reduced MS acceptance.
\item Calorimeter-tagged (CT) muons: these muons are reconstructed by matching a track in the ID and an energy deposit in the calorimeter compatible with a minimum ionising particle. This type of reconstructed muons has the lowest purity of all the muon types but it recovers acceptance in the uninstrumented regions of the MS.
\end{enumerate}
Overlaps between different types of reconstructed muons are resolved in the following way: when two muon types share the same ID track, preference is given to CB muons, then to ST, and finally to CT muons. The overlap with SA muons in the muon system is resolved by analysing the track hit content and selecting the track with better fit quality and larger number of hits.\newline
Muon identification is based on quality requirements suppressing background muons, that come mainly from pion and kaon decays, while selecting prompt muons with high efficiency, ensuring at the same time a precise momentum measurement. 
The variables, necessary to discriminate between background and prompt muons, are the following~\cite{Muon_reco1}:
\begin{itemize}
\item $q/p$ significance, defined as the absolute value of the difference between the ratio of the charge and momentum of the muons measured in the ID and in the MS divided by the sum in quadrature of the corresponding uncertainties;
\item $\rho'$, defined as the absolute value of the difference between the transverse momentum measurements in the ID and in the MS divided by the $p_T$ of the combined track;
\item the normalised $\chi^2$ of the combined track fit.
\end{itemize}
In order to provide a robust momentum measurement, additional specific requirements on the number of hits in the ID and in the MS are used.\newline
The four muon identification selections optimised for different physics analyses are: Loose, Medium, Tight, and High-$p_T$. The muon reconstruction efficiency is obtained with the tag-and-probe method for muons in the region $|\eta| <$ 2.5, using $J/\Psi \rightarrow \mu^+ \mu^-$ and $Z\rightarrow \mu^+ \mu^-$ decays for low- ($<10$ \GeV) and high-$p_T$ muons, respectively.\newline
The reconstruction efficiencies for signal and background, considering all the identification selections, are reported in Table~\ref{reco_muon_table} for prompt muons from $W$ decays and hadrons decaying in flight, categorised according to the MC truth information.
\begin{table}[H]
\begin{center}
{\def\arraystretch{1.4}
\begin{tabular}{|c|lc|cl|}
\hline
\multicolumn{1}{|c|}{\textbf{{\scriptsize}}} & 
\multicolumn{2}{c|}{{ $4 < p_T < 20 $ \GeV\ }} &
\multicolumn{2}{c|}{{ $20 < p_T < 100 $ \GeV\ }} \\
\hline
\textbf{Selection} & $\epsilon_\mu^{MC}$ [\%] & $\epsilon_{Hadrons}^{MC}$ [\%]  & $\epsilon_\mu^{MC}$ [\%] & $\epsilon_{Hadrons}^{MC}$ [\%] \\
\hline 
Loose & 96.7 & 0.53 & 98.1 & 0.76 \\
Medium & 95.5 & 0.38 & 96.1& 0.17 \\
Tight & 89.9 & 0.19 & 91.8 & 0.11 \\
High p$_T$ & 78.1 & 0.26 & 80.4 & 0.13\\
\hline
\end{tabular}
}
\end{center}
\caption{Efficiency for prompt muons from $W$ decays and hadrons decaying in flight and misidentified as prompt muons computed using a $t\bar{t}$ MC sample. The results are shown for the four identification selection criteria separating low momentum ($4 < p_T < 20$ \GeV) and high momentum ($20 < p_T < 100$~\GeV) muons for candidates with $|\eta|<$ 2.5~\cite{Muon_reco1}.}
\label{reco_muon_table}
\end{table}
Figure~\ref{muon_eff_2018} shows the reconstruction efficiencies with the full 2016 dataset for the Loose/Medium/Tight identification algorithms measured in $Z\rightarrow \mu^+ \mu^-$ events as a function of the muon pseudorapidity for muons with $p_T>$$10$~$\GeV$.
Another powerful tool that can be exploited in order to reject background, is the measurement of muon isolation. As already mentioned in the electron case, two variables are used to assess the muon isolation: a track-based variable and a calorimeter-based isolation variable.

\section{Tau reconstruction}
\label{sec:tau}
The tau lepton is the heaviest lepton and, due to its short lifetime, it decays inside the beam pipe without reaching any detector. \newline
It is the only lepton which decays both leptonically, \ie $\tau_{lep}\rightarrow l \nu_l\nu_\tau,\, l = e, \mu$  (BR $\sim$35\%) and hadronically, \ie $\tau_{had}\rightarrow hadrons +\nu_\tau$ (BR $\sim$65\%). Since the leptonic tau decay products are nearly indistinguishable from prompt electrons and muons, they will not be treated in this section.\newline
Most of hadronic tau decays are characterised by one ($\sim$72\% of the cases) or three charged pions ($\sim$22\% of the cases) together with neutral pions ($\sim$68\% of all hadronic decays), thus the typical signature of such a decay is a narrow jet or spray of particles in the calorimeter, associated to one or three tracks in the ID. \newline
The main background comes from jets of energetic hadrons produced via the fragmentation of quarks and gluons that are present both at trigger level and during the event reconstruction.
The tau reconstruction proceeds through the following steps~\cite{Tau_reco}:
\begin{itemize}
\item hadronic jets are reconstructed starting from their energy deposits in the calorimeter cells, using the anti-$k_t$ algorithm already described in Section~\ref{sec:jet}, with a distance parameter $R = 0.4$ (small-R); additionally, tau candidates are required to have  $p_T >$$10$~\GeV\ and $|\eta| < $ 2.5; they constitute the jet seeds used in later steps;
\item the primary tau production vertex (TV) is identified among the possible candidates by choosing the candidate track vertex with the largest fraction of momentum from tracks associated ($\Delta R <$ 0.2) with the jet; the tracks must have $p_T >$1 \GeV\ and pass requirements on the number of hits, \ie at least two hits in the pixel detector and seven in the total pixel and SCT layers; 
\item the last step of the reconstruction is achieved using a track association algorithm, imposing additional requirements on the shortest distance from the track to the tau vertex in the transverse plane, $|d_0| <$1~mm, and the shortest distance in the longitudinal plane, $|\Delta z_0 sin(\theta)|< $1.5~mm, where $\theta$ is the polar angle of the track and $z_0$ is the point of closest approach along the longitudinal axis.
\end{itemize}
\begin{figure}[H]
\begin{center}
\includegraphics[height=8 cm,width =16 cm]{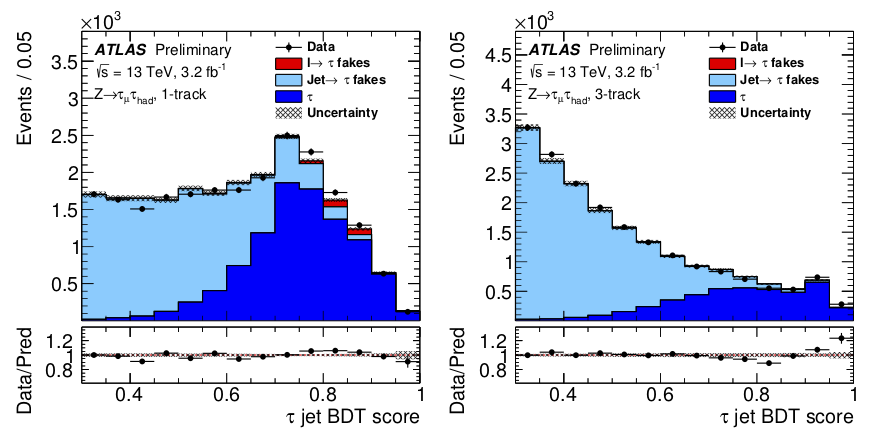}
\end{center}
\caption{The jet discriminant BDT output distribution for one track (left) and three tracks (right) $\tau_{had-vis}$ candidates. The uncertainty band contains only the statistical uncertainty~\cite{Tau_reco}.}     
\label{tau_BDT}
\end{figure}
During the reconstruction process, no attempt is made in order to separate tau leptons from quark- and gluon-initiated jets. The tau identification algorithm is designed to address the rejection of these backgrounds through BDT-based methods~\cite{BDT1,BDT2}. The BDT is separately trained for tau candidates with one or three associated tracks with simulated $Z/\gamma^*\rightarrow \tau^+\tau^-$ for signal and dijet events for background. In order to distinguish tau candidates from jets, a set of discriminating variables based on the shower in the calorimeter, the number of associated tracks and the displaced tau-lepton-decay vertex is used.\newline
Figure~\ref{tau_BDT} shows the jet BDT score distribution, for one and three tracks $\tau_{had-vis}$ candidates, \ie neutral and charged hadrons stemming from the tau-lepton decay that make up the visible part of the tau lepton.\newline
The performance of online and offline tau identification are measured using a tag-and-probe method applied to events enriched in the $Z\rightarrow \tau^+\tau^-$ process, with one tau lepton decaying to muon and neutrinos, $\tau_\mu$ (tag), and the other decaying to hadrons and neutrino, $\tau_{had}$ (probe).
Three working points are provided: Loose, Medium and Tight; they correspond to different tau identification efficiency values, with the efficiency designed to be independent of $p_T$. \newline

\section{Missing transverse momentum}
\label{sec:missing}
\begin{figure}[H]
\begin{center}
\includegraphics[height=9 cm,width =9 cm]{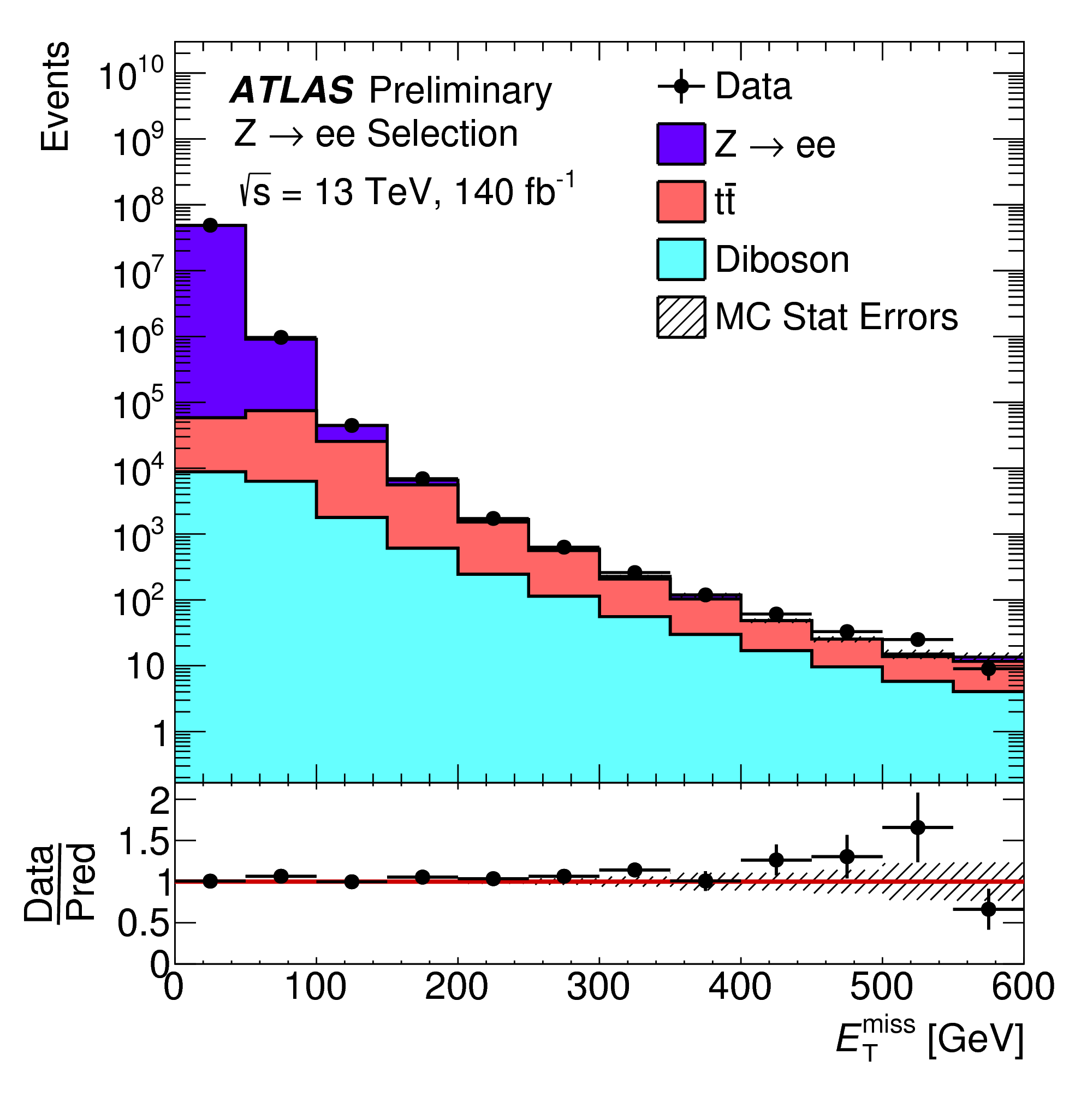}
\end{center}
\caption{The TST $E_T^{miss}$ distribution is shown for the complete dataset, 2015--2018, with an integrated luminosity of 140 fb$^{-1}$, and it is compared to Monte Carlo simulation. A $Z\rightarrow e^+e^-$ selection has been applied to the data. The tight $E_T^{miss}$ working point is used: this requires that jets have transverse momentum ($p_T$) greater than 30 \GeV\ and $|\eta| > $2.5. The $E_T^{miss}$ distribution is shown for data and compared to simulation which is broken up in the contribution from each physics process~\cite{Etmiss_site}.}
\label{etmiss_2018}
\end{figure}

Momentum conservation in the plane transverse to the beam axis ($x-y$) implies that the vector sum of the transverse momenta of all particles in the final state should be zero. An imbalance in the sum of the transverse momenta, known as missing transverse
momentum $E_T^{miss}$, indicates the presence of undetectable particles like SM neutrinos but also new particles that do not interact with the detector materials such as particles included in BSM models.\newline
The reconstruction of the $E_T^{miss}$~\cite{Missing_reco} in ATLAS is a challenge involving all detector sub-systems; the $E_T^{miss}$ is characterised by two contributions:
\begin{enumerate}
\item a \textit{hard term}, \ie a contribution from the hard-event signals comprising fully reconstructed and calibrated particles and jets (hard objects); the reconstructed particles are electrons, photons, $\tau$-leptons, and muons;
\item a \textit{soft term}, \ie a contribution from the soft-event signals consisting of reconstructed charged-particle tracks (soft signals) associated with the hard-scatter vertex but not associated with all reconstructed hard objects; the soft component can be estimated through two main algorithms, the Calorimeter Soft Term (CST) algorithm, accounting for both neutral and charged particles, and the Track Soft Term (TST) algorithm, where the missing transverse momentum is reconstructed entirely from tracks avoiding pile-up contamination.
\end{enumerate}
Considering dedicated variables corresponding to specific objects, the full $E_T^{miss}$ is built as the negative vectorial sum in the transverse plane of missing transverse momentum terms $E_T^{miss,\, p}$, with $p \in \{ e, \gamma, \tau_{had}, \mu, \text{jet} \}$ reconstructed from the $p_T$ = ($p_x$, $p_y$) of particles and jets, and the corresponding soft term, $E_T^{miss,\, soft}$, from the soft-event signals~\cite{Missing_reco}:
\begin{equation}
\boldsymbol{E}_T^{miss}= \underbrace{\underbrace{- \sum_{\mathclap{\substack{electrons}}} \boldsymbol{p}_T^e }_{E_T^{miss,e}}\quad \underbrace{ - \quad \sum_{\mathclap{\substack{photons}}} \boldsymbol{p}_T^\gamma }_{E_T^{miss,\gamma}} \quad \underbrace{ -  \quad  \sum_{\mathclap{\substack{\tau-leptons}}} \boldsymbol{p}_T^{\tau_{had}}}_{E_T^{miss,\tau_{had}}}  \quad \underbrace{- \quad\sum_{\mathclap{\substack{muons}}} \boldsymbol{p}_T^\mu }_{E_T^{miss,\mu}}}_{\text{hard term}} \quad  \underbrace{\underbrace{- \quad \sum_{\mathclap{\substack{unused \\ tracks}}} \boldsymbol{p}_T^{track} }_{E_T^{miss,soft}}}_{\text{soft term}} \, .
\end{equation}
The $E_T^{miss}$ reconstruction performance is assessed by comparing a set of reconstructed $E_T^{miss}$-related observables in data and MC simulations for the same final-state selection, with the same object and event selections applied. Systematic uncertainties in the $E_T^{miss}$ response and resolution are derived from these comparisons and are used to quantify the level of understanding of the data from the physics models~\cite{Missing_reco}. The performances of $E_T^{miss}$ reconstruction may be quantified by the observed width of the $E_T^{miss}$ distribution, an example of which is shown in Figure~\ref{etmiss_2018} for the soft term exploiting a TST algorithm for the complete Run 2 dataset, 2015-2018~\cite{Etmiss_site}.


\chapter{Statistical Treatment}
\label{sec:stat}
This chapter is dedicated to a brief description of the profile-likelihood technique (Section~\ref{sec:profile}) together with all the statistical ingredients, like the definition of the sensitivity and of the $p-\text{value}$ (Section~\ref{sec:pvalue}), of the asymptotic limit and Asimov dataset (Section~\ref{sec:asimov}) and of the confidence intervals (Section~\ref{sec:Neym}), necessary to extract the results reported in Chapters~\ref{sec:dihiggs},~\ref{sec:single} and~\ref{sec:combination}.

\section{Profile-likelihood technique}
\label{sec:profile}
\begin{figure}[H]
\begin{center}
\includegraphics[height=7 cm, width =11 cm]{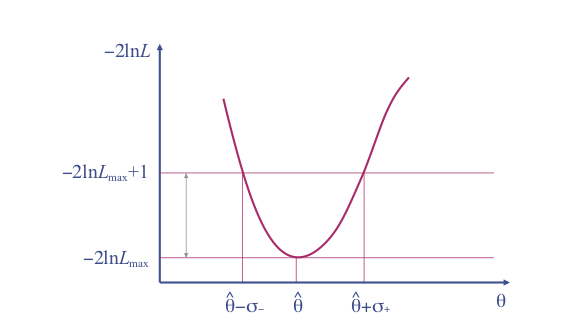}
\end{center}
\caption{Scan of -2$\ln{L}$ in order to determine asymmetric 1$\sigma$ errors~\cite{Lista}.}     
\label{scan_asy}
\end{figure}

The outcome of an experiment can be modelled as a set of random variables, $\boldsymbol{x}=$$x_1, \cdot \cdot \cdot \cdot, x_n$, whose distribution takes into account both intrinsic physics randomness (theory) and detector effects (\eg resolution, efficiency, etc.), also called statistical and systematic uncertainties. Theory and detector effects can be described according to some auxiliary parameters, $\boldsymbol{\theta}=\theta_1, \cdot\cdot \cdot , \theta_m$, whose values are, in most of the cases, unknown and have to be fitted to data.\newline
The overall Probability Density Function (PDF), evaluated for the parameters of interest (POI) and including the uncertainties which enter as nuisance parameters (NP), is called likelihood function; the global likelihood function is obtained as the product of the likelihoods of the input analyses.
If a sample consisting of $N$ independent measurements, typically each corresponding to a collision event, is considered, the likelihood function can be written as~\cite{Lista}:
\begin{equation}
L(\boldsymbol{x},\boldsymbol{\theta}(x))=\prod_{i=1}^N f(x_1^i, \cdot \cdot \cdot x_n^i; \boldsymbol{\theta})
\end{equation}
where the number of events $N$ is treated as fixed.\newline
Usually it is convenient to use -$\ln{L}$ or -2 $\ln{L}$ rather than $L$ in numerical calculations and computations because of the properties of the logarithms.\newline
In event-counting experiments, the actual number of observed events $N$ is a quantity of interest and the probability of observing them depends on the $\boldsymbol{\theta}$ parameters. Thus, if $N$ follows a Poisson distribution with mean $m$ and all the $\boldsymbol{x}$ values follow $f(\boldsymbol{x}$; $\boldsymbol{\theta}$) the likelihood function becomes an ``extended$"$ likelihood:
\begin{equation}
L(\boldsymbol{x},\boldsymbol{\theta})=\frac{m^N}{N!}e^{-m} \prod_{i=1}^N f(x_1^i, \cdot \cdot \cdot x_n^i; \boldsymbol{\theta}) \, .
\end{equation}
For a Poissonian process that is given by the sum of a signal plus a background process, summing over all the NPs ($r$) the extended likelihood function may be written as:
\begin{equation}
L(\boldsymbol{x}; s,b,\boldsymbol{\theta})= \frac{e^{-(s+b)}}{N!}\prod_{i=1}^{N} \left ( sP_s(x_i;\boldsymbol{\theta})+bP_b(x_i;\boldsymbol{\theta}) \right ) \times \prod_{j=1}^{r} \rho_j (\theta_j)
\end{equation}
 where:
\begin{itemize}
\item $s$ and $b$ are the signal and background expected yields;
\item $P_s$ and $P_b$ are the PDFs of the variable $x$ for signal and background, respectively;
\item $\rho_j (\theta_j)$ represents the functional distributions assumed for the nuisance parameters.
\end{itemize}
The NPs describe the systematic uncertainties that can affect the normalisation of the samples, the shape of the final discriminants or both normalisation and shape.
Two different types of nuisance parameters are usually considered: unconstrained normalisation factors determined only from data and parameters associated to systematics that have external constraints and use information from auxiliary measurements, like experimental and modelling uncertainties.
In the case of gaussian distributed systematics, like for experimental systematic uncertainties, $\rho(\theta)=e^{-(\vartheta -\theta)^2/2}/\sqrt{2\pi}$, where $\vartheta$ represents the central value of the measurement and $\theta$ the associated nuisance parameter for a given systematic uncertainty. For normalisation systematic uncertainties, where the $\theta$ parameter can only assume positive values and cannot be well described by a Gaussian distribution, a log-normal distribution is adopted, $\rho(\theta)=(1+\epsilon)^\theta$, being $\epsilon$ the value of the uncertainty in question~\cite{HWW_stat}. Finally, in order to describe systematic uncertainties associated to finite Montecarlo-sample size or to the number of observed events in a data control sample, a gamma distribution is adopted.\newline
The correlation between nuisance parameters is implemented in the fit by associating the systematic uncertainties to the same nuisance parameter in the global likelihood.\newline
Usually, the strengths of the signal process, $\boldsymbol{\mu}$, are introduced as the vector of parameters of interest of the model, while in the case of the analyses presented in this dissertation, being the signal strengths parameterised in terms of the rescaling of the Higgs self-coupling, $\kappa_\lambda$, or in terms of the coupling modifiers to the Higgs boson, identified by a generic $\kappa$, the substitution $\mu \rightarrow \kappa$ is made. When the likelihood function depends on many parameters, the achievable constraints on $\boldsymbol{\kappa}$ might be weak so that the true values of both the POIs and of the NPs can't be estimated; the main aim of the fit procedure is never determining the true values of $\boldsymbol{\theta}$, but rather obtaining tight intervals for $\boldsymbol{\kappa}$. To this end, the profile likelihood ratio is considered as test statistic:
\begin{equation} 
q_{\boldsymbol{\kappa}}=-2\,\ln{\lambda(\boldsymbol{\kappa})}=-2 \ln {\left ( \frac{L(\boldsymbol{\kappa},\hat{\hat{\boldsymbol{\theta}}})}{L(\hat{\boldsymbol{\kappa}},\hat{\boldsymbol{\theta}})} \right )}
 \end{equation}
 where:
 \begin{itemize}
 \item $\hat{\hat{\boldsymbol{\theta}}}$ denotes the value of $\boldsymbol{\theta}$ that maximises $L$ for the specified $\kappa$, \ie it is the conditional Maximum-Likelihood (ML) estimator of $\boldsymbol{\theta}$;
 \item $\hat{\boldsymbol{\kappa}}$ and $\hat{\boldsymbol{\theta}}$ are the vectors of ML estimator of the unconditional likelihood.
 \end{itemize}
The maximum-likelihood method takes as best-fit values of the unknown parameter the values that maximise the likelihood function. The choice of the POIs depends on the tested model under consideration, while the remaining parameters are ``profiled$"$, \ie\ they are set to the values that maximise the likelihood function.\newline
The maximisation of the likelihood function can be performed analytically only in the simplest cases, while a numerical study of the likelihood in the $\boldsymbol{\theta}$ parameter space is needed in most of the realistic case to localise the $\hat{\boldsymbol{\theta}}$ point that minimises -$2\ln{L}$; this procedure is called a Maximum-Likelihood fit. MINUIT~\cite{Minuit} is historically the most widely used minimisation software engine in High Energy Physics. 
It is conceived as a tool to find the minimum value of a multi-parameter function and analyse the shape of the function around the minimum. Minuit mainly works on $\chi^2$ or log-likelihood functions, to compute the best-fit-parameter values and uncertainties, including correlations between the parameters. One of the most important Minuit program that has been widely exploited to provide the results reported in this thesis, is the Minuit  processor MINOS that can calculate parameter errors taking into account both parameter correlations and non-linearities, thus leading, in general, to asymmetric error intervals, an shown in the example of Figure~\ref{scan_asy}.

\section{Sensitivity and $p-\text{value}$}
\label{sec:pvalue}

\begin{figure}[H]
\begin{center}
\includegraphics[height=6 cm,width =8 cm]{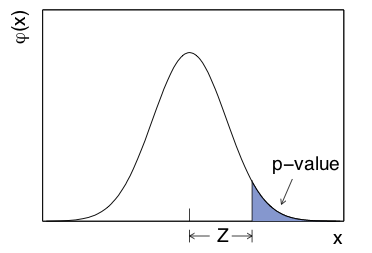}
\end{center}
\caption{The standard normal distribution $\phi(x)=1/\sqrt{2\pi}e^{-x^2/2}$ showing the relation between the significance Z and the $p-\text{value}$~\cite{Cowan}.}     
\label{pvalue}
\end{figure}

In order to discover a new process, the statistical significance of an observation can be quantified by evaluating the compatibility of the observed data with a given hypothesis $H$; testing different hypotheses has the target of answering to the question whether some observed data samples are more compatible with one theory model or with an alternative one. The level of agreement of the observed data with a given hypothesis $H$ is quantified by computing a $p-\text{value}$:
\begin{equation}
p_H=\int_{q_{obs}}^{\infty} f(q|H) dq
\end{equation}
where $q_{obs}$ is the observed value in data of the test statistic $q$ and $f(q|H)$ represents the PDF of $q$ under the assumption of the $H$ hypothesis.\newline
In particle physics two hypotheses are usually considered: 
\begin{enumerate}
\item the null hypothesis, $H_0$, describing only known processes, and designated as background ($\mu=0$); 
\item the alternative hypothesis $H_1$, which includes both background and signal ($\mu=1$). 
\end{enumerate}
Therefore, exploiting the definitions of these two hypotheses, the $p-\text{value}$ is the probability, assuming $H_0$ to be true, of getting a value of the test statistic as result of the test at least as extreme as the observed test statistic; the significance level, denoted $\alpha$, is the probability, assuming $H_0$ to be true, of rejecting $H_0$; an hypothesis can be rejected if its $p-\text{value}$ is observed below a specific threshold, \ie $p < \alpha$. Another quantity, related to the $p-\text{value}$, that is used to exclude/confirm a signal/background hypothesis, is the equivalent significance, $Z$. It is defined as the number of normal Gaussian standard deviations ($\sigma$) above which the mean of the Gaussian has an upper-tail probability equal to $p$~\cite{Cowan}: 
\begin{equation}
Z=\Phi^{-1}(1-p)
\end{equation}
where $\Phi^{-1}$ is the quantile (inverse of the cumulative distribution) of the standard Gaussian. As a convention, in the particle physics community, a threshold $\alpha$ of 0.05 for the $p-\text{value}$, which corresponds to $Z=1.64$, is used to exclude a signal hypothesis, while the discovery threshold corresponds to $Z=5$, \ie $p-\text{value}$=$2.87\times10^{-7}$. Figure~\ref{pvalue} shows how the significance is related to the $p-\text{value}$.
In the case where a little or no sensitivity occurs to some models, the $CL_s$ technique~\cite{CLs} is used as the standard technique to set exclusion limits; the $CL_s$ is defined as:
\begin{equation}
CL_s=\frac{p_{\mu}}{1-p_b}
\end{equation}
where $p_b$ is the $p-\text{value}$ for the background-only hypothesis and $p_\mu$ is the $p-\text{value}$ for the signal plus background hypothesis. In the usual formulation of the method, these two quantities are computed using the same test statistic $q_\mu$, and the definition of $CL_s$ above assumes this statistic is continuous~\cite{pdg}. A point in a model's parameter space is regarded as excluded if one finds $CL_s\le \alpha$: for example, if the $CL_s$ is below 5\%, a hypothesis is excluded at 95\% CL.
Results can be presented providing confidence intervals, that are constructed so as to cover the true value of a parameter with a specified probability; how these intervals are constructed is reported in Section~\ref{sec:Neym}.

\section{Asymptotic limit and Asimov Dataset}
\label{sec:asimov}
In the limit of large statistics, thanks to Wilks' theorem~\cite{Wilks} and Wald's asymptotic approximation~\cite{Wald}, the test statistic can be approximated as:
\begin{equation}
-2\ln{\lambda(\kappa)}=\frac{(\kappa-\hat{\kappa})^2}{\sigma^2} + \mathcal{O}(1/\sqrt{N})
\end{equation}
where $\hat{\kappa}$ is distributed according to a Gaussian with average $\kappa'$ and standard deviation $\sigma$ and $N$ represents the data sample size. The standard deviation is obtained from the covariance matrix of the estimators for all the parameters, $V_{ij}=cov[ \hat{\theta_i}, \hat{\theta_j} ]$, where here the $\theta_i$ represent both the parameter of interest, $\kappa$, as well as the nuisance parameters.
In the asymptotic approximation, the covariance matrix is given by:
\begin{equation}
V^{-1}_{ij}= \left \langle \frac{\partial^2\ln{L}}{\partial \theta_i \partial \theta_j} \right \rangle_{\kappa=\kappa'} 
\label{covariance}
\end{equation}
where $\kappa'$ is assumed as the value of the $\kappa$ parameter.\newline
When several parameters of interest are considered and can be identified with a subset of the parameters $\boldsymbol{\theta}_n=(\theta_1,\cdot \cdot \cdot, \theta_n)$, the test statistics can be generalised as~\cite{Cowan}:
\begin{equation}
-2\ln{\lambda(\boldsymbol{\kappa})}=\sum_{i,j=1}^{n} (\theta_i - \theta_i^{'}) \tilde{V}_{ij}^{-1} (\theta_j - \theta_j^{'}) + \mathcal{O}(1/\sqrt{N})
\end{equation}
where $\tilde{V}_{ij}^{-1}$ is the inverse of the submatrix obtained restricting to the parameters of interest the full covariance matrix defined in Equation~\ref{covariance}.
Thus -2 $\ln{\lambda(\boldsymbol{\kappa},\theta)}$ is approximately distributed as a $\chi^2$ variable with $n$ degrees of freedom, where $n$ equals the number of parameters of interest in the model, \ie\ the dimensionality of the vector $\boldsymbol{\kappa}$.\newline
Asymptotic approximation can be written in terms of an Asimov dataset~\cite{Cowan} that is a dataset obtained by replacing all observable (random) variables with their expected values.

\section{Neyman's confidence intervals}
\label{sec:Neym}
A procedure to determine frequentist confidence intervals, constructed to include the true value of the parameter/parameters of interest with a probability greater than or equal to a specified level, has been elaborated by Neyman~\cite{Neiman} and it is described in the following lines~\cite{Lista}:
\begin{enumerate}
\item first of all a scan of the allowed range of the unknown parameter of interest $\kappa$ has to be made;
\item given a value $\kappa_0$ of $\kappa$, the interval $[x_1(\kappa_0), x_2(\kappa_0)]$ that contains $x$ with a probability 1-$\alpha$ (confidence level, or CL) equal to 68.3\% (or 90\%, 95\%) is computed; if $x$ is discrete, the integral is replaced by the corresponding sum; 
\item finally the confidence interval obtained for $x$ has to be inverted in order to find the corresponding interval $[\kappa_1(x), \kappa_2(x)]$.
\end{enumerate}
By construction, a fraction of the experiments equal to 1-$\alpha$ will measure $x$ such that the corresponding confidence interval $[\kappa_1(x), \kappa_2(x)]$ contains the true value of $\kappa$, \ie :
\begin{equation}
1-\alpha=P(x_1(\kappa)< x < x_2(\kappa))=P(\kappa_2(x)<\kappa<\kappa_1(x)) \, .
\end{equation}
If a Gaussian distribution with known parameter $\sigma$=1 is considered, the inversion gives a central value $\hat{\kappa}= x$ and a confidence interval $[\kappa_1,\kappa_2] = [x-\sigma$, $x+\sigma]$. The result can be quoted as $\kappa = x \pm \sigma$.\newline
An equivalent method of constructing confidence intervals consists in considering a test statistics, like the profile likelihood ratio $q_{\kappa}$. All values of $\kappa$ where the hypothesis would be rejected at a significance level less than $\alpha$, are excluded; thus, the confidence interval is given by the interval $[\kappa_1,\kappa_2]$ for which all $\kappa$ satisfy $q_{\kappa}<\lambda_\alpha=\chi^2_{1,\alpha}$, where $\alpha$ denotes the confidence level and $\lambda_\alpha$ is computed from a $\chi^2$ distribution with one degree of freedom. The values of $\lambda_\alpha$ for different confidence levels are reported in Reference~\cite{pdg}. As an example, for a confidence interval at 95\% CL, considering one degree of freedom, $\lambda_\alpha$ is $\lambda_{0.95}=3.84$. 
Thus the 95\% CL intervals, used to extract the results of this dissertation, are identified requiring $-2\ln L< 3.84$, when one parameter of interest is considered.\newline 
The likelihood function can be easily written in terms of a vector of parameters of interest, thus the concept of confidence interval can be extended to the one of confidence regions, built through a scan of the phase space defined by the $n$ parameters of interest and assuming a $\chi^2$ distribution with $n$ degrees of freedom.\newline
The results presented in Chapters~\ref{sec:dihiggs},~\ref{sec:single} and ~\ref{sec:combination}, are based on the statistical tools described in this chapter, particularly on the profile-likelihood evaluation; 68\% as well as 95\% CL intervals are given in the asymptotic approximation~\cite{Cowan}.

\chapter{Probing the Higgs self-coupling}
\label{sec:prob_self}
This chapter describes the theoretical models on the basis of which the results of the following chapters have been produced. Section~\ref{theory_hh} summarises how the trilinear Higgs self-coupling enters in double-Higgs processes and how the dependence on the Higgs self-coupling and top Yukawa coupling can be implemented in double-Higgs observables. Section~\ref{sec:HL_LHC_hh} reports projections of $\kappa_\lambda$ constraints coming from double-Higgs measurements and considering a luminosity of 3000 fb$^{-1}$. Section~\ref{theory_single} describes a complementary approach in order to extract limits on the Higgs self-coupling exploiting Next-to-Leading-Order (NLO) EW corrections to single-Higgs production and decay processes; furthermore, the parameterisations used to produce the results of this thesis are reported. Finally, Section~\ref{sec:HL_LHC_theory} reports projections of $\kappa_\lambda$ constraints coming from single-Higgs measurements and considering a luminosity of 300 and 3000 fb$^{-1}$.

\section{Higgs self-coupling through direct Higgs-boson pair searches}
\label{theory_hh}
\begin{figure}[htbp]
  \centering
    \begin{subfigure}[b]{0.34\textwidth}
\includegraphics[width =\textwidth]{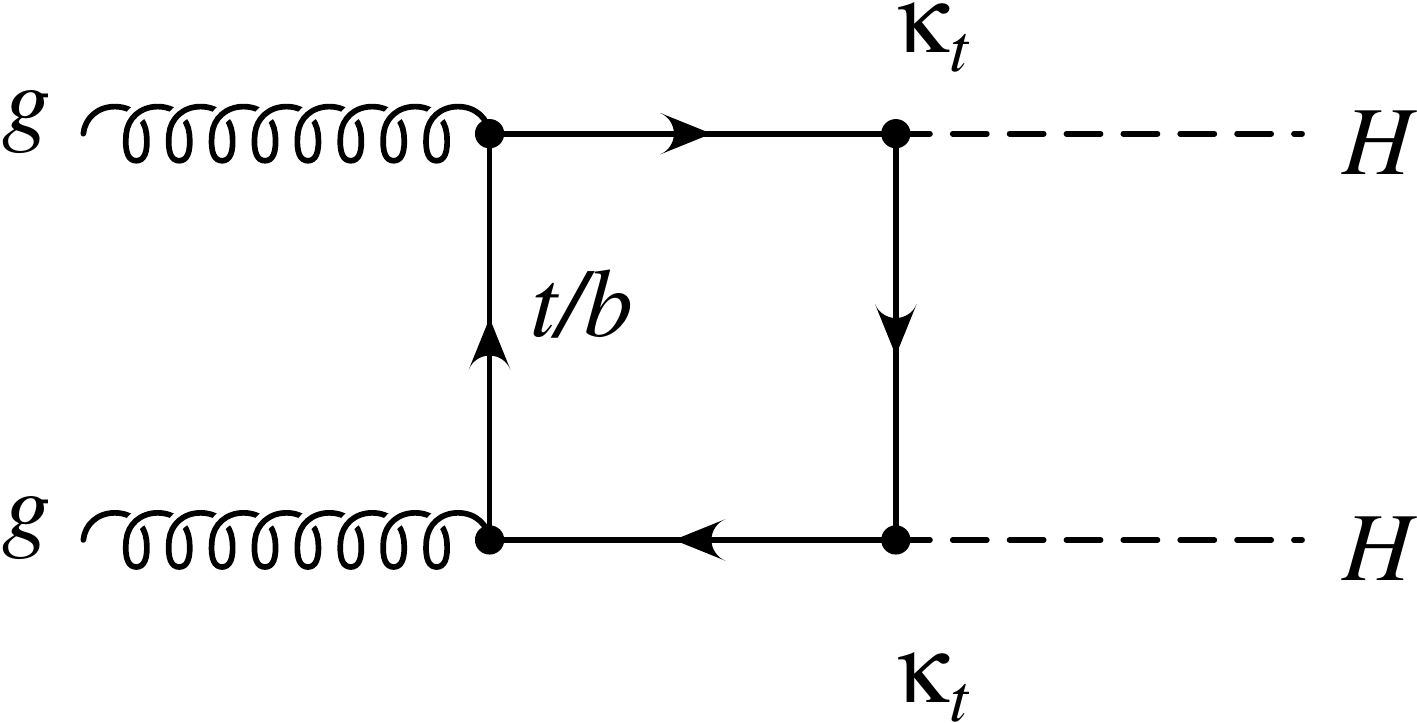}
 \caption{}
\end{subfigure}
\qquad
  \begin{subfigure}[b]{0.34\textwidth}
\includegraphics[width =\textwidth]{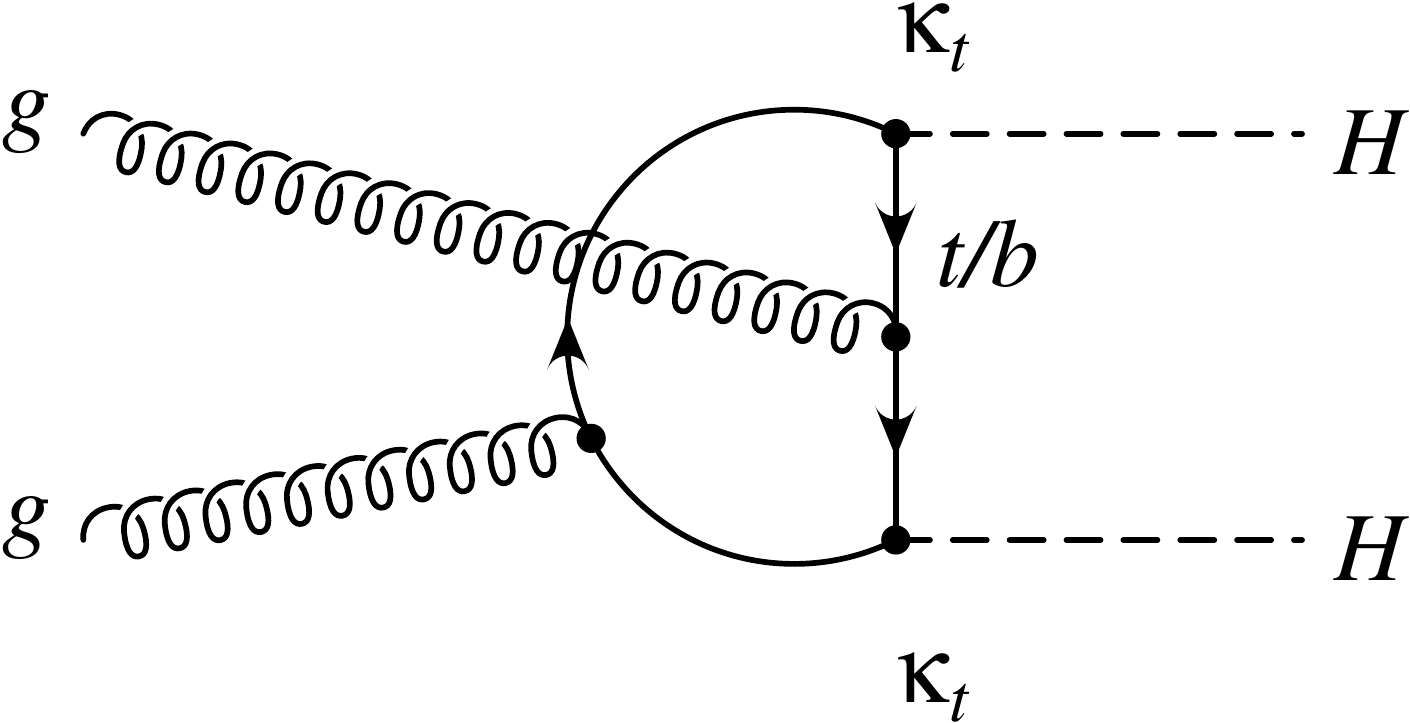}
 \caption{}
\end{subfigure}
  \begin{subfigure}[b]{0.34\textwidth}
\includegraphics[width =\textwidth]{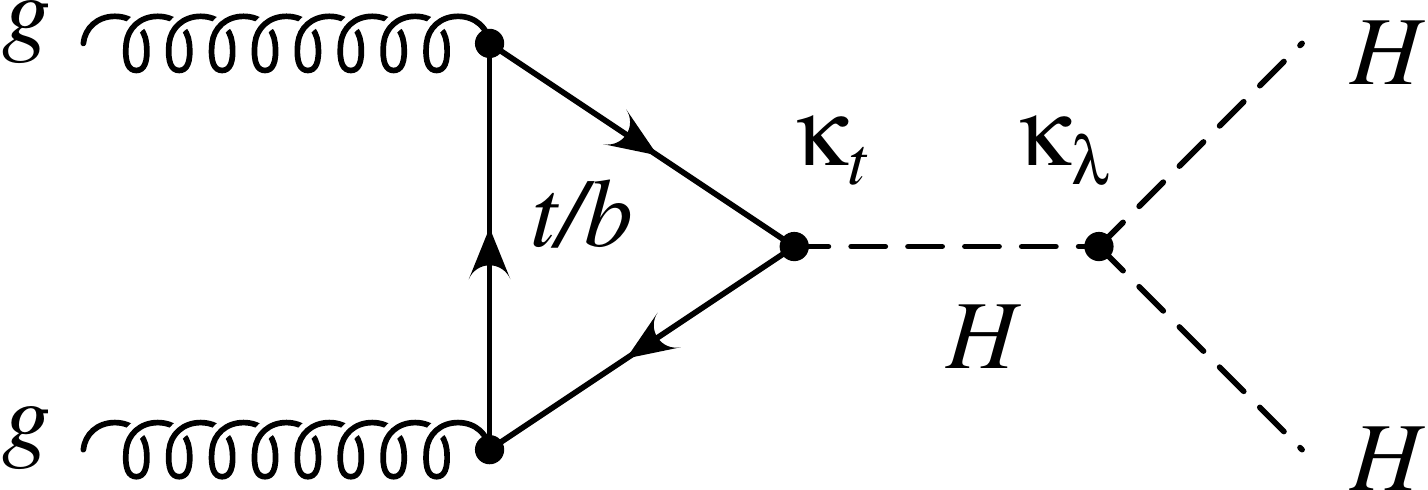}
 \caption{}
\end{subfigure}
    \caption{Examples of leading-order Feynman diagrams for double-Higgs \ggF production.}
  \label{hh_diagrams}
\end{figure}

Double-Higgs processes are directly sensitive to the trilinear Higgs self-coupling at the lowest order in electroweak expansion, like it is shown in the diagram $(c)$ of Figure~\ref{hh_diagrams}; in the SM, the \ggF double-Higgs production process, that is mediated by top quark loops with a negligible contribution from bottom quark loops and is the only double-Higgs production process studied in this thesis, accounts for more than 90\% of the total production cross section, while the second-largest production mechanism is vector-boson fusion ($qq\rightarrow HH qq$). Due to the fact that the \VBF double-Higgs cross section is an order of magnitude smaller than the \ggF one and has a less sensitive topology, the \VBF contribution to the estimation of the self-coupling is negligible~\cite{vbf_hh}.\newline
The gluon-fusion mechanism proceeds via two amplitudes, $\mathcal{A}_1$, proportional to the square of the Higgs coupling to the top quark, $y_t$, and represented by the diagrams $(a)$ and $(b)$ of Figure~\ref{hh_diagrams}, and $\mathcal{A}_2$, proportional to the product of the Higgs coupling to the top quark and the Higgs self-coupling, represented by the diagram $(c)$ of Figure~\ref{hh_diagrams}.\newline
In the SM, the interference between these two amplitudes is destructive and yields an overall cross-section value which is $\sim$$10^3$ times smaller than the corresponding \ggF single-Higgs production, \ie\ 31.05 fb, according to recent calculations~\cite{lhc_xs_hh}. This cross section can be enhanced in the case of BSM physics modifying the relative sign of the amplitudes $\mathcal{A}_1$ and $\mathcal{A}_2$, and increasing their contributions through modifications of the aforementioned couplings.\newline
For BSM scenarios affecting $y_t$ and $\lambda_{HHH}$, defining the coupling modifier to the top quark as $\kappa_t=y_t^{BSM}/y_t^{SM}$ and to the Higgs self-coupling as $\kappa_\lambda=\lambda_{HHH}^{BSM}/\lambda_{HHH}^{SM}$, the total amplitude can be written as:
\begin{equation}
\mathcal{A}(\kappa_t, \kappa_\lambda) = \kappa_t^2\mathcal{A}_1 + \kappa_t \kappa_\lambda \mathcal{A}_2\,. 
\label{eq:amplitude}
\end{equation}

\begin{figure}[htbp]
\begin{center}
\includegraphics[width=0.5\textwidth]{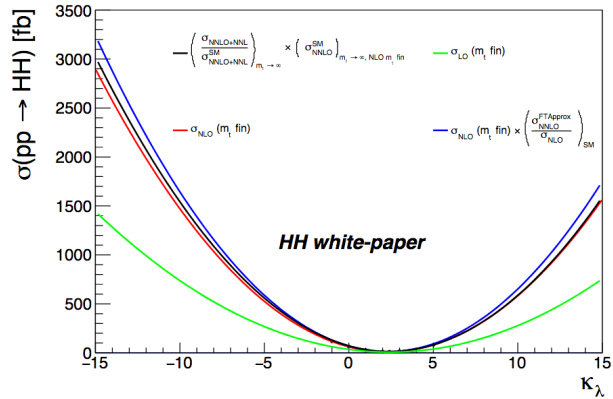}
\end{center}
\caption{Double-Higgs \ggF production cross section as a function of $\kappa_\lambda$ including different cross-section computations~\cite{white_paper}.}
\label{xs_hh_white}
\end{figure}

Higher order QCD corrections do not add further $t\bar{t}H$ or $HHH$ vertices to the diagrams shown in Figure \ref{hh_diagrams}, implying that Equation~\ref{eq:amplitude} is applicable to any order in QCD (\ie also when the amplitudes $\mathcal{A}_1$ and $\mathcal{A}_2$ are modified to include their higher order QCD corrections)~\cite{Confnote_comb}.\newline
From Equation~\ref{eq:amplitude}, after integrating over the final-state phase space and over the PDFs, the \ggF double-Higgs cross section $\sigma_{\mathrm{\ggF}}(pp \to HH)$ can be written in terms of $\kappa_\lambda$ and $\kappa_t$ as:
\begin{equation}
\sigma_{\mathrm{\ggF}}(pp \to HH) 
\propto  \int_{}   \kappa_t^4  \left [|\mathcal{A}_1|^2 +
  2\left(\frac{\kappa_\lambda}{\kappa_t}\right)\Re{(\mathcal{A}_1^*\mathcal{A}_2)} 
+\left(\frac{\kappa_\lambda}{\kappa_t}\right)^2|\mathcal{A}_2|^2\right]\,. 
\label{eq:pph}
\end{equation}
Expression~\ref{eq:pph} makes clear that the kinematic distributions and, consequently, the signal acceptance, depend only on $\kappa_\lambda/\kappa_t$, while the $\kappa_t^4$ factor affects only the total cross section. The effects of $\kappa_b$ are negligible.\newline
Assuming that new physics affects only the Higgs-boson self-coupling, the differential and inclusive \ggF $pp\rightarrow HH$ cross section can be expressed as a second degree polynomial in $\kappa_\lambda$, \ie~\cite{white_paper}:
\begin{equation}
\frac{d\sigma}{d\Phi}=A+B\kappa_\lambda +C\kappa_\lambda^2
\end{equation}
being $d\Phi$ the infinitesimal phase-space volume. In a first approach, this feature can be used to simulate MC samples for any values of $\kappa_\lambda$ combining three samples generated for three different values of $\kappa_\lambda$; the procedure consists in solving the system of three equations depending on the value of $\kappa_\lambda$, computing the dependence of the coefficients $A$, $B$, $C$ from the differential cross section in a given phase space volume, and inverting the matrix above to obtain the coefficients reported in Reference~\cite{white_paper}. As a natural choice $\kappa_\lambda= 0, 1$ are chosen, corresponding to the box-only and the SM cases, while the third value can be chosen close to the expected exclusion region, in order to optimise the signal generation. For the analyses reported in this dissertation, this corresponds to $\kappa_\lambda=20$.\newline 
Furthermore, the three samples need to be properly normalised to the best cross-section computations, as the ones shown in Figure~\ref{xs_hh_white}~\cite{white_paper}: leading order (LO); next-to-next-to-leading order (NNLO)+ next-to-next-to-leading-logarithmic (NNLL) SM cross section obtained in the limit of heavy top quarks rescaled with the NNLO+NNLL SM cross section obtained including finite top quark mass NLO contribution and NNLO corrections in the limit of heavy top quarks; finite top quark mass NLO for all $\kappa_\lambda$ values rescaled with the NNLO SM cross section obtained with the FTApprox method (partial finite top quark mass).

\begin{figure}[htbp]
  \centering
    \begin{subfigure}[b]{0.42\textwidth}
\includegraphics[height=5.46 cm,width =\textwidth]{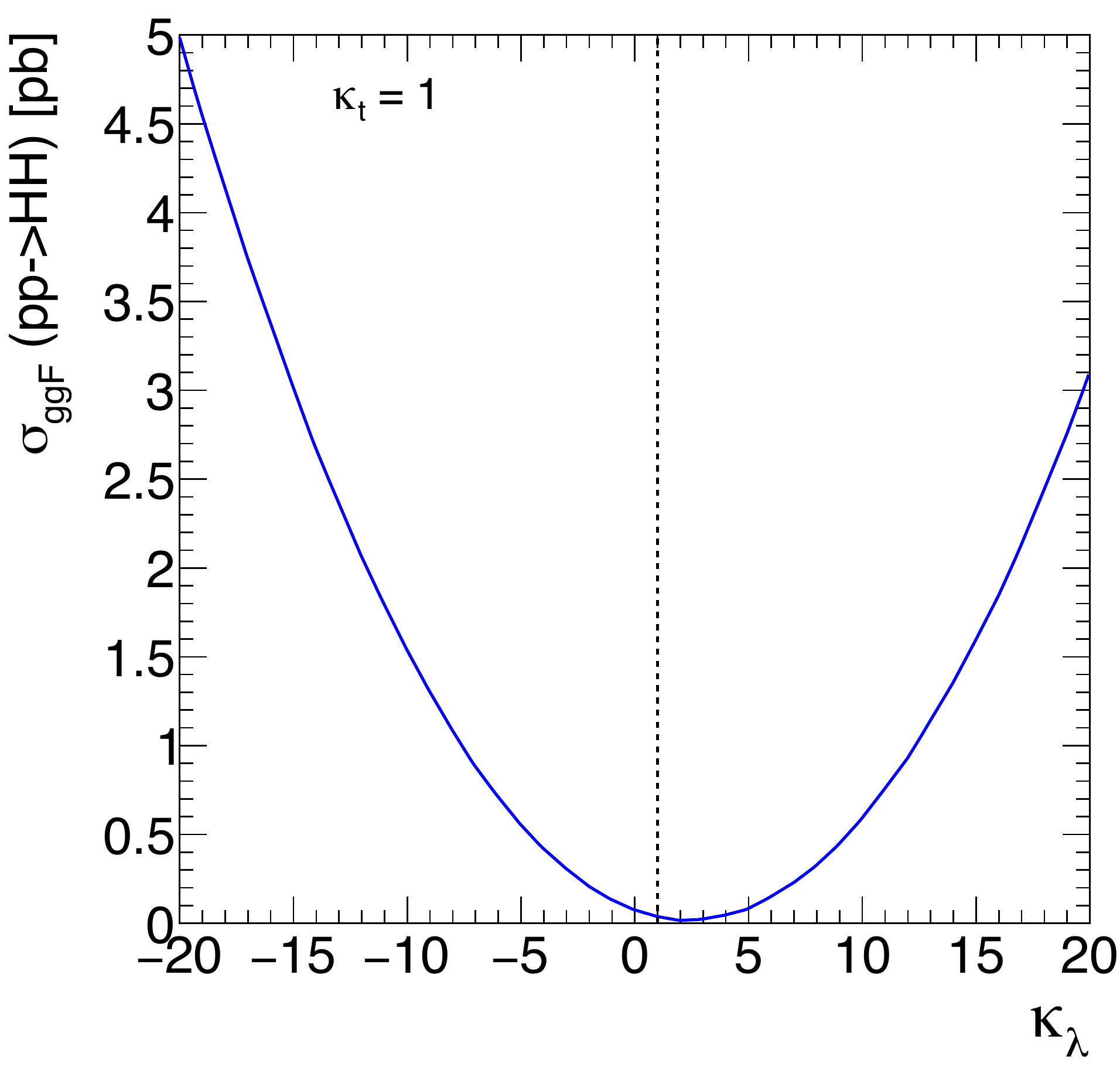}
 \caption{}
\end{subfigure}
    \begin{subfigure}[b]{0.428\textwidth}
\includegraphics[height=5.6 cm, width =\textwidth]{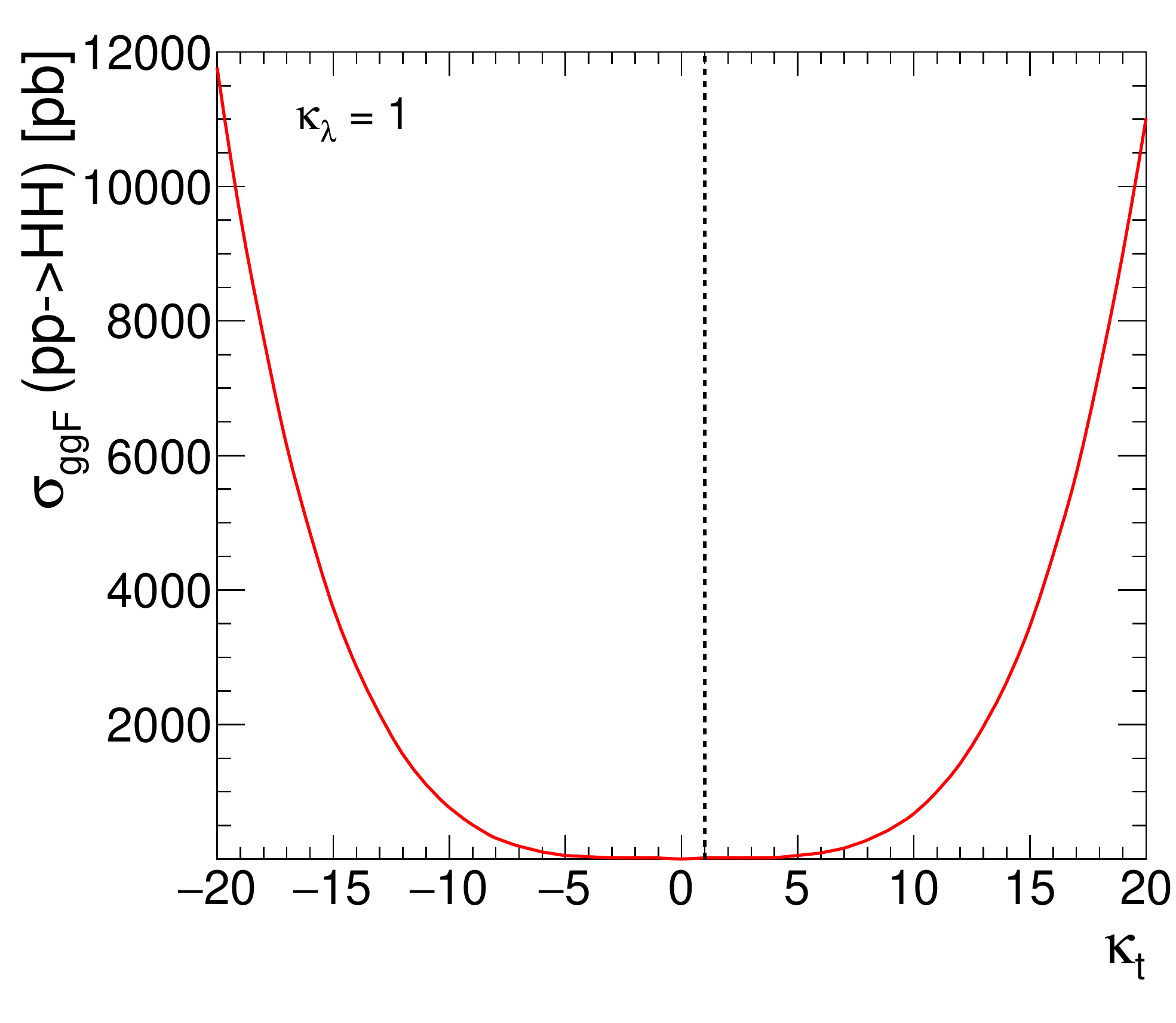}
 \caption{}
\end{subfigure}
 \caption{Double-Higgs \ggF $pp\rightarrow HH$ cross section as a function of $\kappa_\lambda$ setting $\kappa_t=1$ (a) and as a function of $\kappa_t$ setting $\kappa_\lambda=1$; the vertical dashed lines represent the SM case, \ie $\kappa_\lambda=1$ (a) and $\kappa_t=1$ (b).}
  \label{prova_xs_hh}
\end{figure}

Figure~\ref{prova_xs_hh} shows the dependence of the \ggF $pp\rightarrow HH$ cross section on $\kappa_\lambda$ setting $\kappa_t=1$ $(a)$ and on $\kappa_t$ setting $\kappa_\lambda=1$ $(b)$, given by a second order and a fourth order polynomial, respectively; the cross section as a minimum at $\kappa_\lambda/\kappa_t=2.4$, \ie\ where the maximal destructive interference between the two diagrams occurs.

 \section{HL-LHC projections for double-Higgs processes}
\label{sec:HL_LHC_hh}

\begin{figure}[H]
\begin{center}
\includegraphics[height=6.5 cm, width=0.9\textwidth]{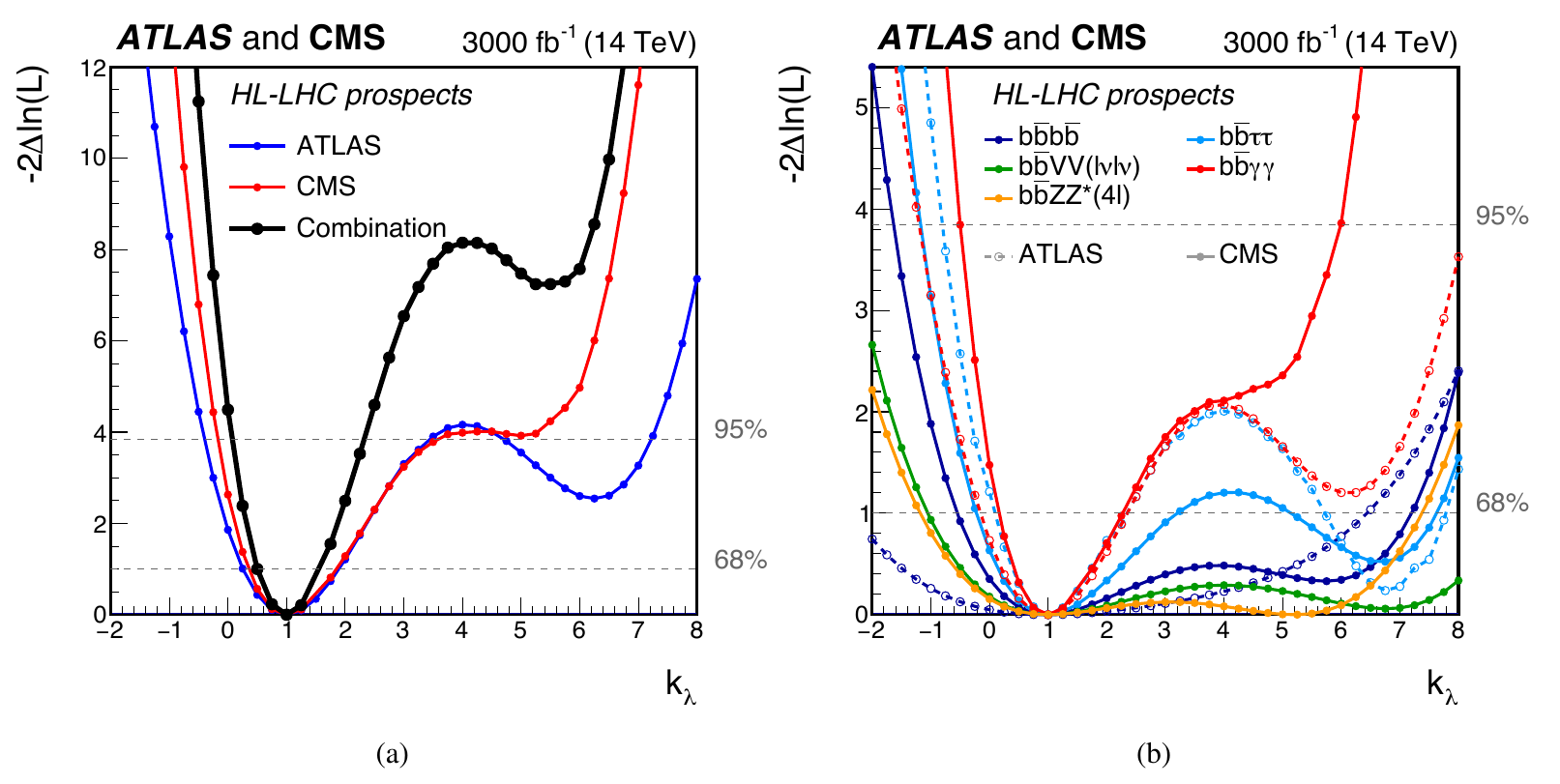}
\end{center}
\caption{Value of $-2 \ln{\Lambda(\kappa_\lambda)}$ as a function of $\kappa_\lambda$ for an Asimov dataset generated under the SM hypothesis for a luminosity of 3000 fb$^{-1}$ at $\sqrt{s}=14$ TeV; (a) black solid line: combined ATLAS and CMS results; blue and red solid lines: ATLAS and CMS standalone results, respectively; (b) the different colours correspond to the different channels, the solid lines correspond to the CMS results while the dashed lines correspond to the ATLAS results~\cite{HL-LHC}.}
\label{HL_hh}
\end{figure}

The High Luminosity (HL)-LHC has the target of studying rare processes like double-Higgs production and achieving precision measurements of observables like the trilinear Higgs self-coupling, further pushing the LHC machine beyond its limits mainly through detector upgrades; the final goal would be to achieve a total integrated luminosity of 3000/4000~fb$^{-1}$ at a centre-of-mass energy of 14 TeV.
Prospects for the measurement of the trilinear Higgs self-coupling have been studied and reported in Reference~\cite{HL-LHC}.\newline
The HL-LHC is expected to be a Higgs-boson factory, producing more than $10^{5}$ double-Higgs pairs per experiment (ATLAS and CMS).\newline
Thus, a significant improvement of current limits on $\kappa_\lambda$ is expected together with an expected precision at the level of 50\% on $\kappa_\lambda$  at 68\% CL.
A combination between ATLAS and CMS projections has been made exploiting an Asimov dataset  generated under the SM hypothesis for a luminosity of 3000 fb$^{-1}$ at $\sqrt{s}=14$ TeV, and using the following channels:
\begin{itemize}
\item CMS: $b\bar{b}b\bar{b}$, $b\bar{b}W^+W^-$, $b\bar{b}\tau^+\tau^-$, $b\bar{b}\gamma \gamma$ and $b\bar{b}ZZ$;
\item ATLAS: $b\bar{b}b\bar{b}$, $b\bar{b}\tau^+\tau^-$ and $b\bar{b}\gamma \gamma$.
\end{itemize}
The channels are treated as uncorrelated, in particular because the systematic uncertainties, such as the theory uncertainties and the luminosity uncertainty, have little impact on the individual results.
The combined minimum negative-log-likelihoods are shown in Figure~\ref{HL_hh}.
The 68\% CL intervals for $\kappa_\lambda$ are 0.52 $\le \kappa_\lambda \le$ 1.5 and 0.57 $\le \kappa_\lambda \le$ 1.5 with and without systematic uncertainties respectively. The second minimum of the likelihood is excluded at 99.4\% CL~\cite{HL-LHC}.
 
\section{Higgs self-coupling through loop corrections of single-Higgs production and decay modes}
\label{theory_single}
\begin{figure}[htbp]
  \centering
\includegraphics[width =0.45\textwidth]{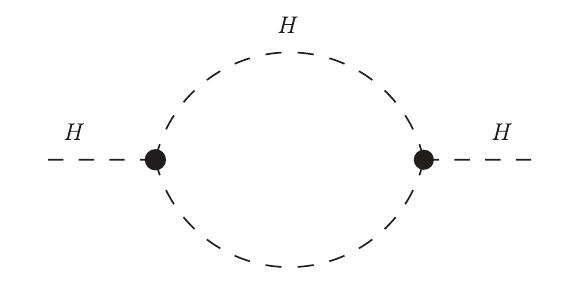}
 \caption{One-loop $\lambda_{HHH}$-dependent diagram in the Higgs self-energy~\cite{Degrassi}.}
  \label{fig:corrections_C2_kl}
\end{figure}

\begin{figure}[htbp]
  \centering
    \begin{subfigure}[b]{0.49\textwidth}
\includegraphics[width =\textwidth]{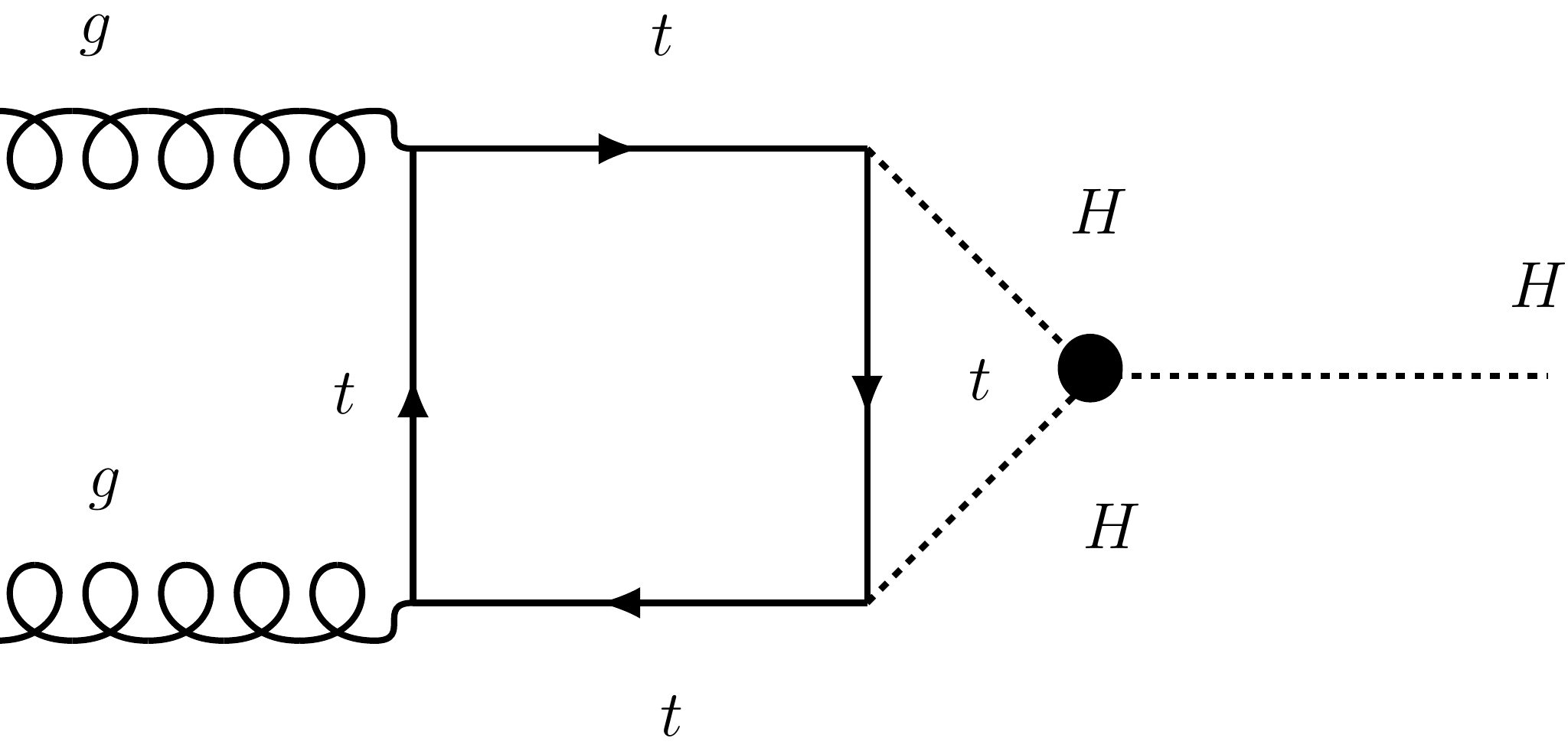}
 \caption{}
\end{subfigure}
  \begin{subfigure}[b]{0.49\textwidth}
\includegraphics[width =\textwidth]{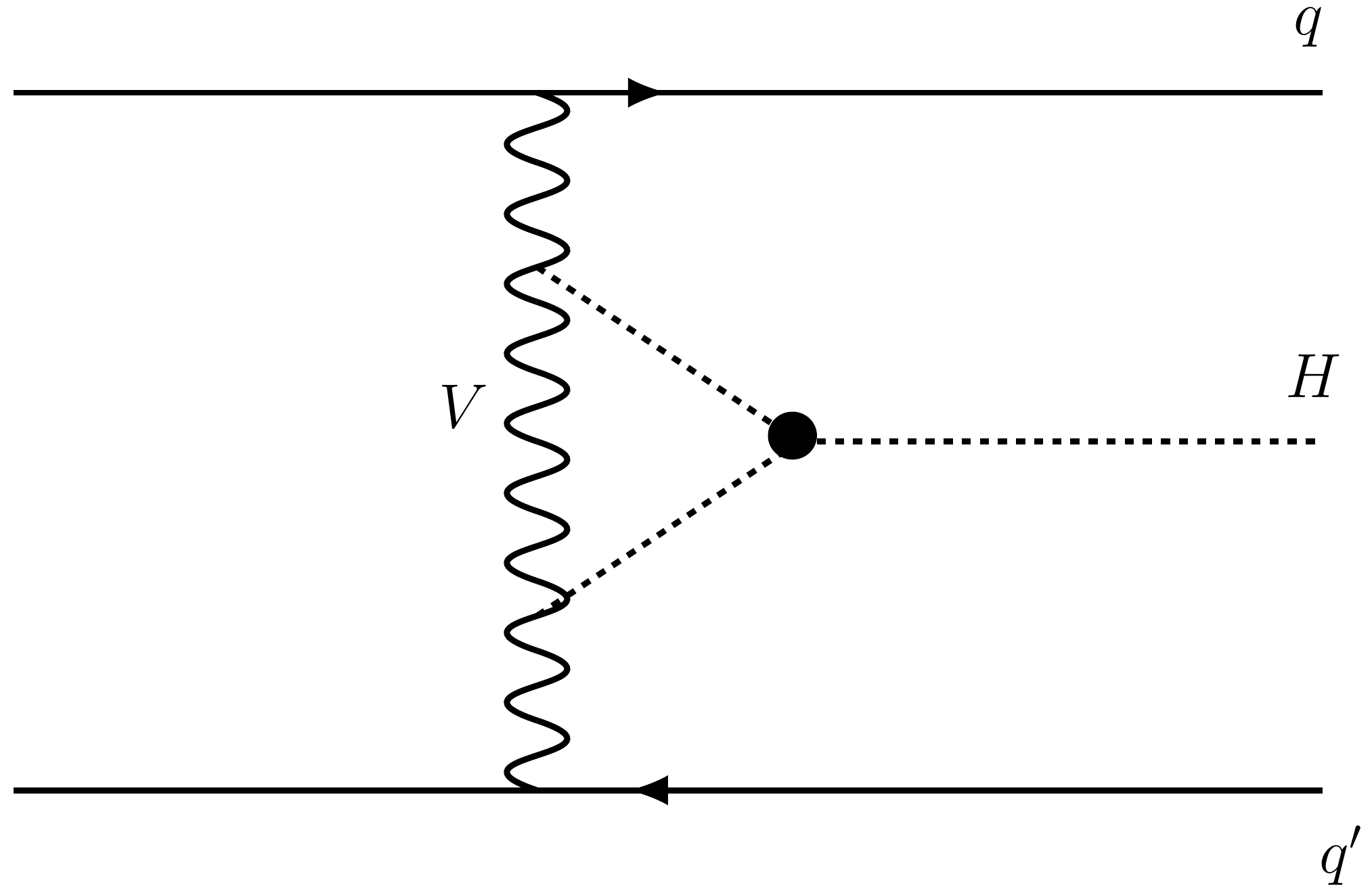}
 \caption{}
\end{subfigure}
  \begin{subfigure}[b]{0.45\textwidth}
\includegraphics[width =\textwidth]{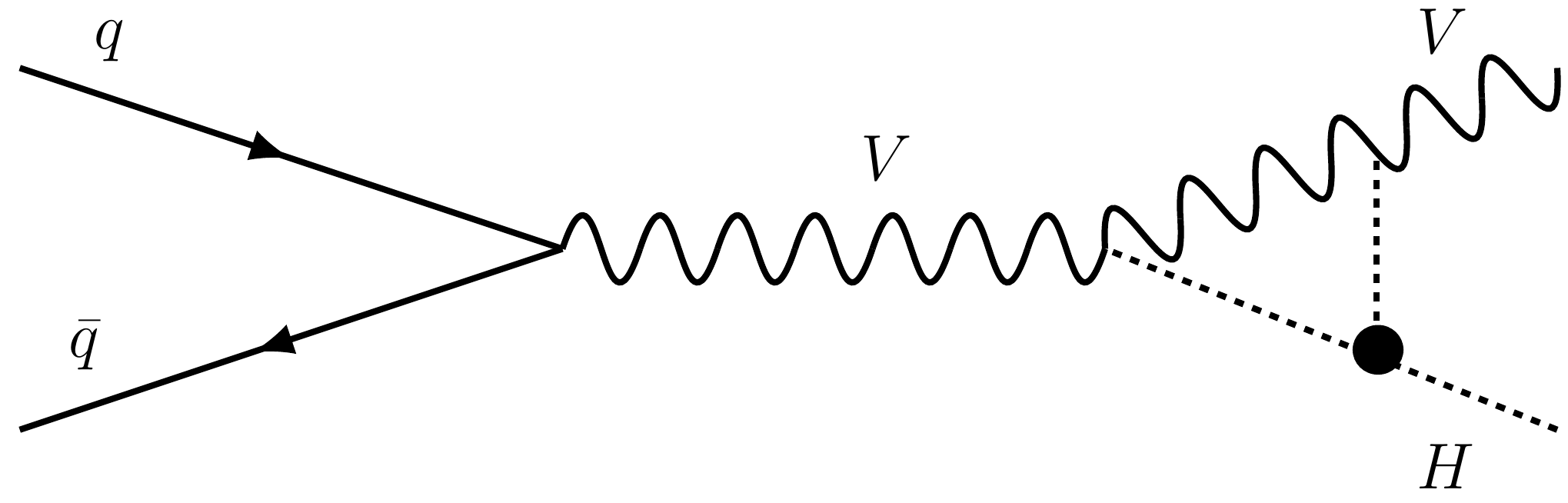}
 \caption{}
\end{subfigure}
\qquad
  \begin{subfigure}[b]{0.45\textwidth}
\includegraphics[width =\textwidth]{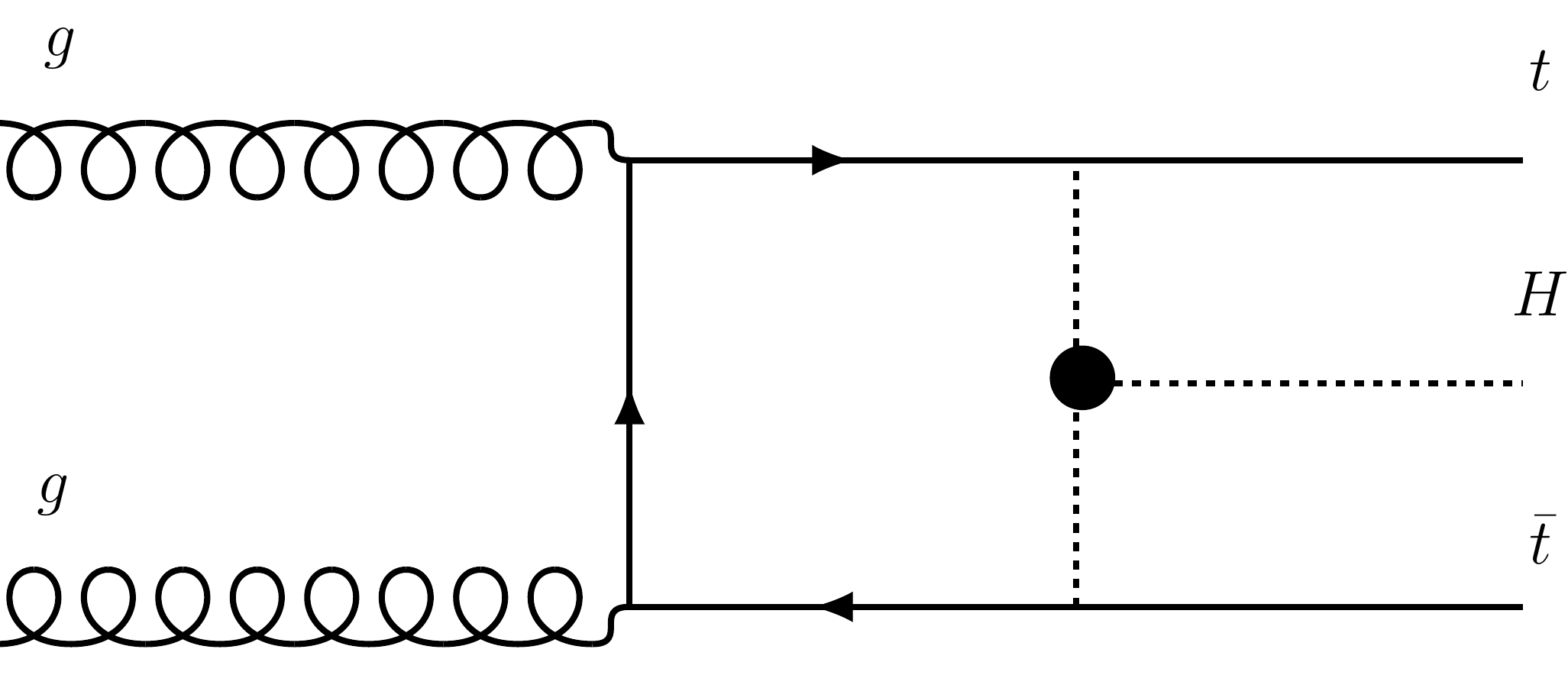}
 \caption{}
\end{subfigure}
    \caption{Diagrams contributing to the $C_1$ coefficients in the different Higgs-boson production modes, \ggF (a), \VBF (b), \VH (c) and $t\bar{t}H$ (d)~\cite{Degrassi}.}
  \label{fig:corrections_kl}
\end{figure}

\begin{figure}[htbp]
  \centering
    \begin{subfigure}[b]{0.49\textwidth}
\includegraphics[height=3.3 cm,width =0.8\textwidth]{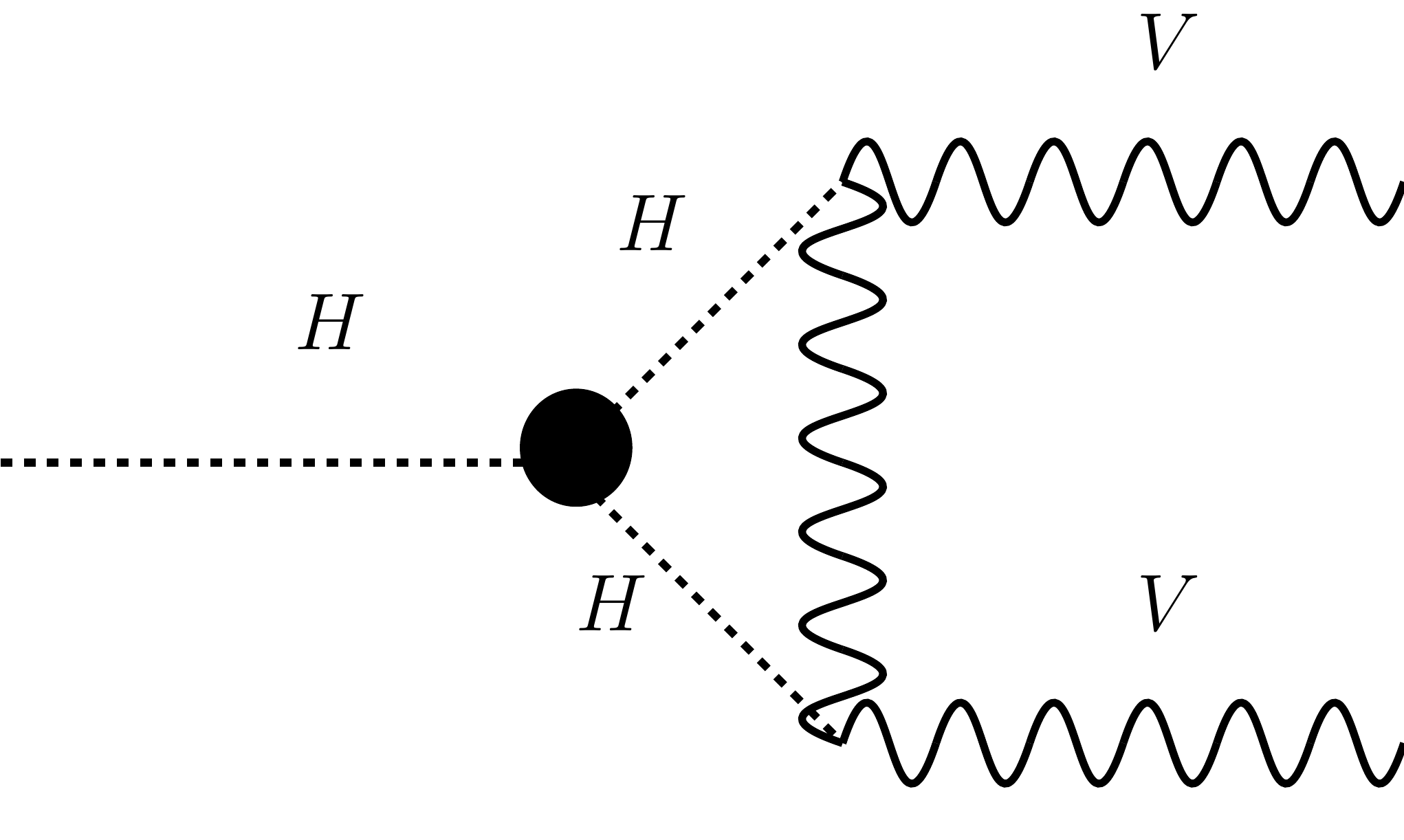}
 \caption{}
\end{subfigure}
  \begin{subfigure}[b]{0.49\textwidth}
\includegraphics[height=3.2 cm,width =1.04\textwidth]{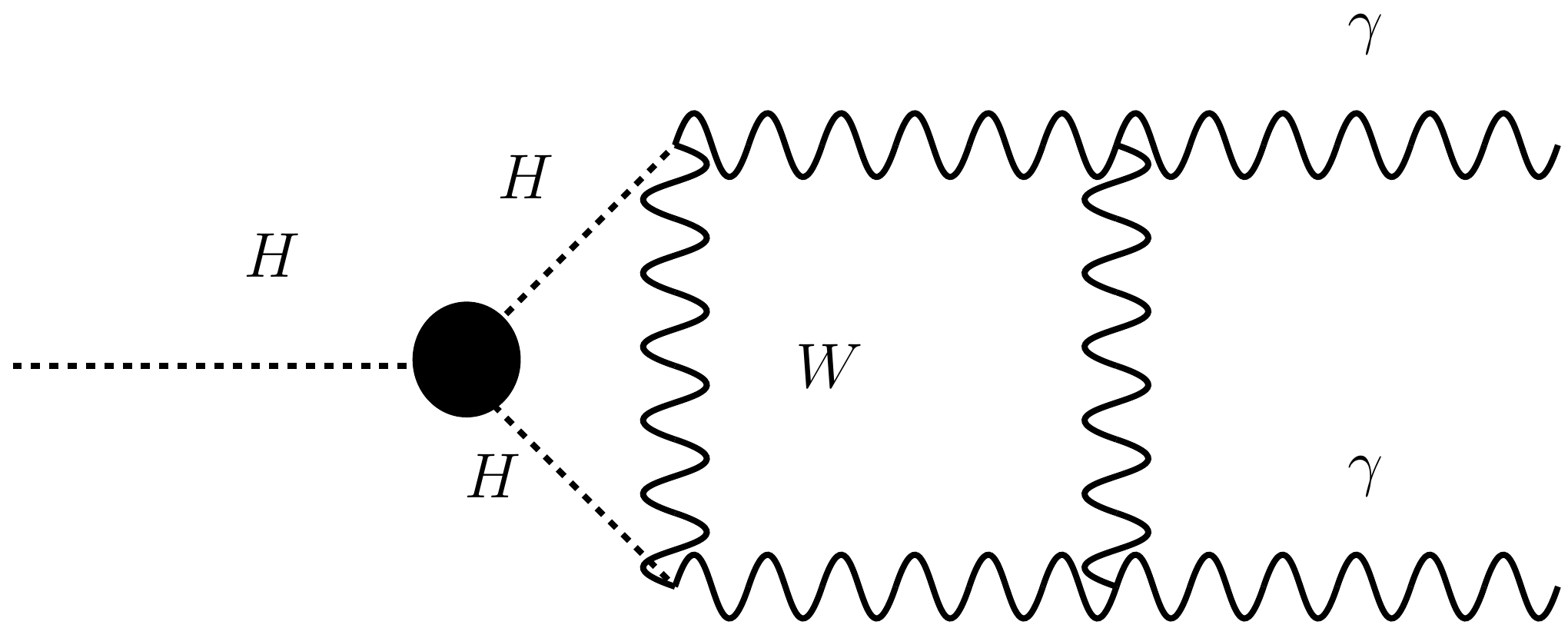}
 \caption{}
\end{subfigure}
    \caption{Examples of diagrams contributing to the $C_1$ coefficients in decay modes: $H\rightarrow VV$ (a), and $H\rightarrow \gamma \gamma$ (b)~\cite{Degrassi}.}
  \label{fig:corrections_kl_br}
\end{figure}

An alternative and complementary approach to constrain the Higgs-boson self-coupling has been proposed in References~\cite{Degrassi,Maltoni}, exploiting the fact that single-Higgs processes are sensitive to $\lambda_{HHH}$ at NLO in electroweak interactions via weak loops, while they depend on the couplings of the Higgs boson to the other particles of the SM at leading order; thus, it is possible to constrain $\kappa_\lambda$ through precise measurements of single-Higgs observables.
The effect of new physics at the weak scale are parameterised via a single parameter $\kappa_\lambda$, \ie the rescaling of the SM trilinear coupling, $\lambda_{HHH}^{SM}$. Thereby, the $H^3$ interaction in the potential, where $H$ is the physical Higgs field, is given by \cite{Degrassi}: 
\begin{equation}
V_{H^3}=\lambda_{HHH}\, \nu  H^3=\kappa_\lambda \, \lambda_{HHH}^{SM}\, \nu H^3 \quad,\quad \lambda_{HHH}^{SM}=\frac{G_\mu}{\sqrt{2}}\, m_H^2   \nonumber \, .
\end{equation}
All the calculations reported in this chapter are based on the assumption that new physics affects only the Higgs self-coupling or modifies in a negligible way the other Higgs couplings.
The NLO $\kappa_\lambda$-dependent corrections can be divided in two different contributions:
\begin{itemize}
\item a universal part, the $C_2$ coefficient, \ie\ common to all processes, quadratically dependent on $\lambda_{HHH}$, which originates from the diagram in the wave function renormalisation constant of the external Higgs field, whose Feynman diagram is shown in Figure~\ref{fig:corrections_C2_kl};
\item a process- and kinematic-dependent part, the $C_1$ coefficients, linearly proportional to $\lambda_{HHH}$, which is different for each process and kinematics; for each observable, the corresponding $C_1$ coefficient is identified as the contribution linearly proportional to $\lambda_{HHH}^{SM}$ in the NLO EW corrections, normalised to the LO result as evaluated in the SM. Figures~\ref{fig:corrections_kl} and~\ref{fig:corrections_kl_br} show examples of $\lambda_{HHH}$-dependent diagrams for the Higgs-boson production as well as for decay modes.
\end{itemize}
The range of validity of these perturbative calculations is taken as $|\kappa_\lambda|<$ 20, assumed in order to neglect $\mathcal{O}(\kappa_\lambda^3 \alpha^2)$ terms~\cite{Degrassi}. \newpage
In addition to the $\kappa_\lambda$-dependent corrections to single-Higgs processes, NLO EW corrections in the SM hypothesis (\ie\ $\kappa_\lambda=1$), have to be taken into account; they are introduced through the coefficients $K_{\mathrm{EW}}=$$\sigma_{\textrm{NLO}}^{\textrm{SM}}/\sigma_{\textrm{LO}}^{\textrm{SM}}$.
The $C_1$ and $K_{\mathrm{EW}}$ coefficients are decoupled because of the approach that has been followed in References~\cite{Degrassi,Maltoni}: the assumption that QCD corrections factorise $\lambda_{HHH}$ effects is reasonable while it is not true, in general, for NLO EW corrections.\newline
Assuming on-shell single Higgs production, the signal strength, defined in Chapter~\ref{sec:SM}, for the process $i\rightarrow H \rightarrow f$, where $i$ and $f$ are the initial and final states, respectively, can be expressed as:
\begin{equation}
\mu_i=1+\delta\sigma_{\lambda_{H^3}}(i) \, , \qquad \mu_f=1+\delta BR_{\lambda_{H^3}}(f) 
\end{equation}
where $\delta\sigma_{\lambda_H^3}(i)$ and $\delta BR_{\lambda_H^3}(f)$ are the deviations induced by an anomalous interaction, including the case of the trilinear Higgs self-coupling, to the production cross sections and branching fractions, respectively.\newline
Specifically, the signal strengths for the initial state $i$, \ie\ the production cross sections normalised to their SM expectations, can be written as:
\begin{equation}
\label{eq:mui}
\mu_i(\kappa_\lambda, \kappa_i) = \frac{\sigma^{\textrm{BSM}}
}{\sigma^{\textrm{SM}}} =
Z_{H}^{\textrm{BSM}}\left(\kappa_\lambda\right)\left[\kappa_i^2+\frac{(\kappa_\lambda-1)C_1^i}{K^i_{\mathrm{EW}}}\right] \, ,
\end{equation}
where $Z_{H}^{\textrm{BSM}}\left(\kappa_\lambda\right)$ is defined as:
\begin{equation}
Z_H^{\textrm{BSM}}\left(\kappa_\lambda\right) = \frac{1}{1-(\kappa_\lambda^2-1)\delta Z_H} \quad
\textrm{with} \quad \delta Z_H = -1.536 \times 10^{-3} \, ,
\label{eq:ZH}
\end{equation}
$K^i_{\mathrm{EW}}$ accounts for the complete NLO EW corrections of the production cross section for the process $i$ in the SM hypothesis, $C_1^{i}$ is the coefficient that provides the sensitivity of the measurement to $\kappa_\lambda$ for the $i$ process and $\kappa_i^2 =\sigma^{\textrm{BSM}}_{\textrm{LO,i}}/\sigma^{\textrm{SM}}_{\textrm{LO,i}}\,(\kappa_{\lambda}=1)$ represents multiplicative modifiers to other Higgs boson couplings, parameterised as in the LO $\kappa$-framework~\cite{k_framework}, taking into account additional BSM effects entering at LO; $\kappa_i$ can be taken equal to one when only variations of the Higgs self-coupling are considered. \newline
Indicating with $f$ a Higgs-boson final state, the decay branching fractions, normalised to their SM expectations, are modified as:
\begin{equation}
\label{eq:muf}
\mu_f(\kappa_\lambda, \kappa_f) = \frac{\textrm{BR}_f^{\textrm{BSM}}}{\textrm{BR}_f^{SM}} =
\frac{\kappa_f^2+(\kappa_\lambda-1)C_1^{f}}{\sum_{j}\textrm{BR}_j^{\textrm{SM}}\left[\kappa_j^2+(\kappa_\lambda-1)C^{j}_1   \right]} \, 
\end{equation}
where $\sum_{j}$ runs over all the Higgs-boson decay channels, $\textrm{BR}_{j}^{\textrm{SM}}$ is the Higgs-boson SM decay rate to
the $j$ final state, $C_1^{f}$ is the coefficient that provides the sensitivity of the measurement to $\kappa_\lambda$ for the $f$ final state, $\kappa_f$ is the branching fraction
modifier for the $f$ final state while $\kappa_j$ is the branching fraction modifier for all the Higgs-boson $j$ final states
$\left (\kappa_j^2 = \textrm{BR}^{\textrm{BSM}}_{\textrm{LO},j}/\textrm{BR}^{\textrm{SM}}_{\textrm{LO},j}
\right)$,  parameterised as in the LO $\kappa$-framework.
The model under discussion, as shown in Equation~\ref{eq:mui} and  Equation~\ref{eq:muf}, does not allow for any new physics beyond that encoded in the aforementioned $\kappa$ parameters.\newline
The process-independent factor $C_2$ can range from $C_2 =-1.536 \cdot 10^{-3}$  for $\kappa_\lambda=1$ up to $C_2=-9.514\cdot10^{-4}$ for $\kappa_\lambda=\pm 20$.
\clearpage 
The $C_1$ coefficients computed for different production and decay modes are reported in Tables~\ref{CKcoeff_theory} and~\ref{decays_C1}; it has been verified that, in the case of $H\rightarrow b\bar{b}$, $C_1^f=2.5 \times 10^{-5}$, so these coefficients have been set to zero for any $H\rightarrow f\bar{f}$ decay~\cite{Degrassi,Maltoni}.

\begin{table}[htbp]
\begin{center}
{\def\arraystretch{1.3}
\begin{tabular}{|c|ccccc|}
\hline 
Production mode & \ggF & \VBF & \ZH & \WH & $t\bar{t}H$ \\
\hline
$C_1^i\times 100$ & 0.66 & 0.63 & 1.19 & 1.03 & 3.52 \\
\hline
$K^i_{\textrm{EW}}$ & 1.049 & 0.932 & 0.947 & 0.93 & 1.014 \\
\hline
\end{tabular}
}
\end{center}
\caption{Values of $C^i_1$ and $K^i_\textrm{EW}$ coefficients for Higgs-boson production processes~\cite{Degrassi,Maltoni}.}
 \label{CKcoeff_theory}
\end{table} 
\begin{table}[htbp]
\begin{center}
{\def\arraystretch{1.3}
\begin{tabular}{|c|ccccc|}
\hline 
decay mode & $H\rightarrow \gamma \gamma$& $H\rightarrow WW^*$ & $H\rightarrow ZZ^*$ & $H\rightarrow f\bar{f}$ &$H\rightarrow gg$ \\
\hline
$C^f_1\times 100$ & 0.49 & 0.73 & 0.82 & 0 & 0.66 \\
\hline
\end{tabular}
}
\end{center}
\caption{Values of $C^f_1$ coefficients for different Higgs-boson decay modes~\cite{Degrassi,Maltoni}.} 
\label{decays_C1}
\end{table}

\begin{figure}[!htbp]
  \centering
  \begin{subfigure}[b]{0.49\textwidth}
\includegraphics[width=\textwidth]{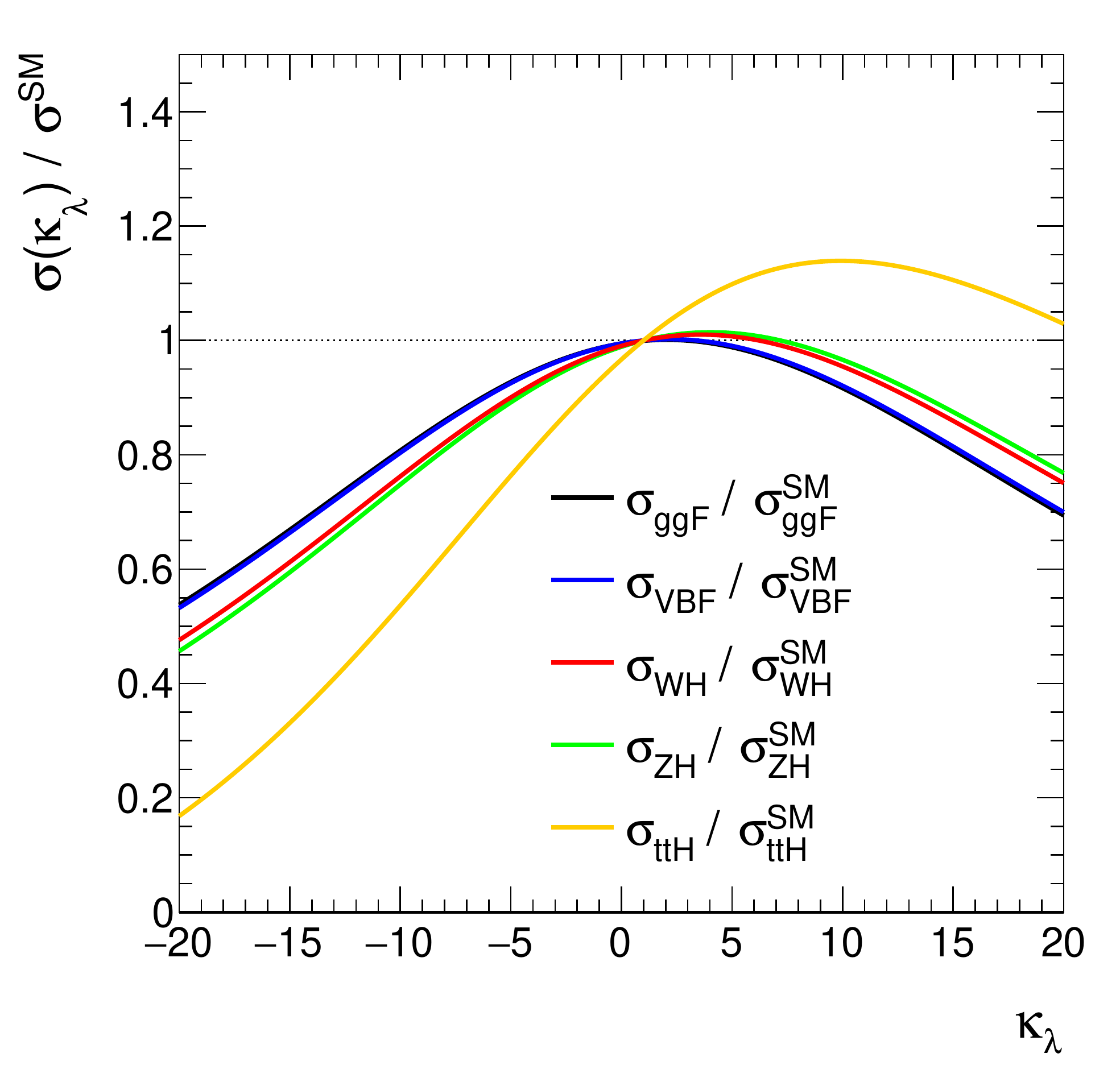}
 \caption{}
\end{subfigure}
  \begin{subfigure}[b]{0.49\textwidth}
\includegraphics[width=\textwidth]{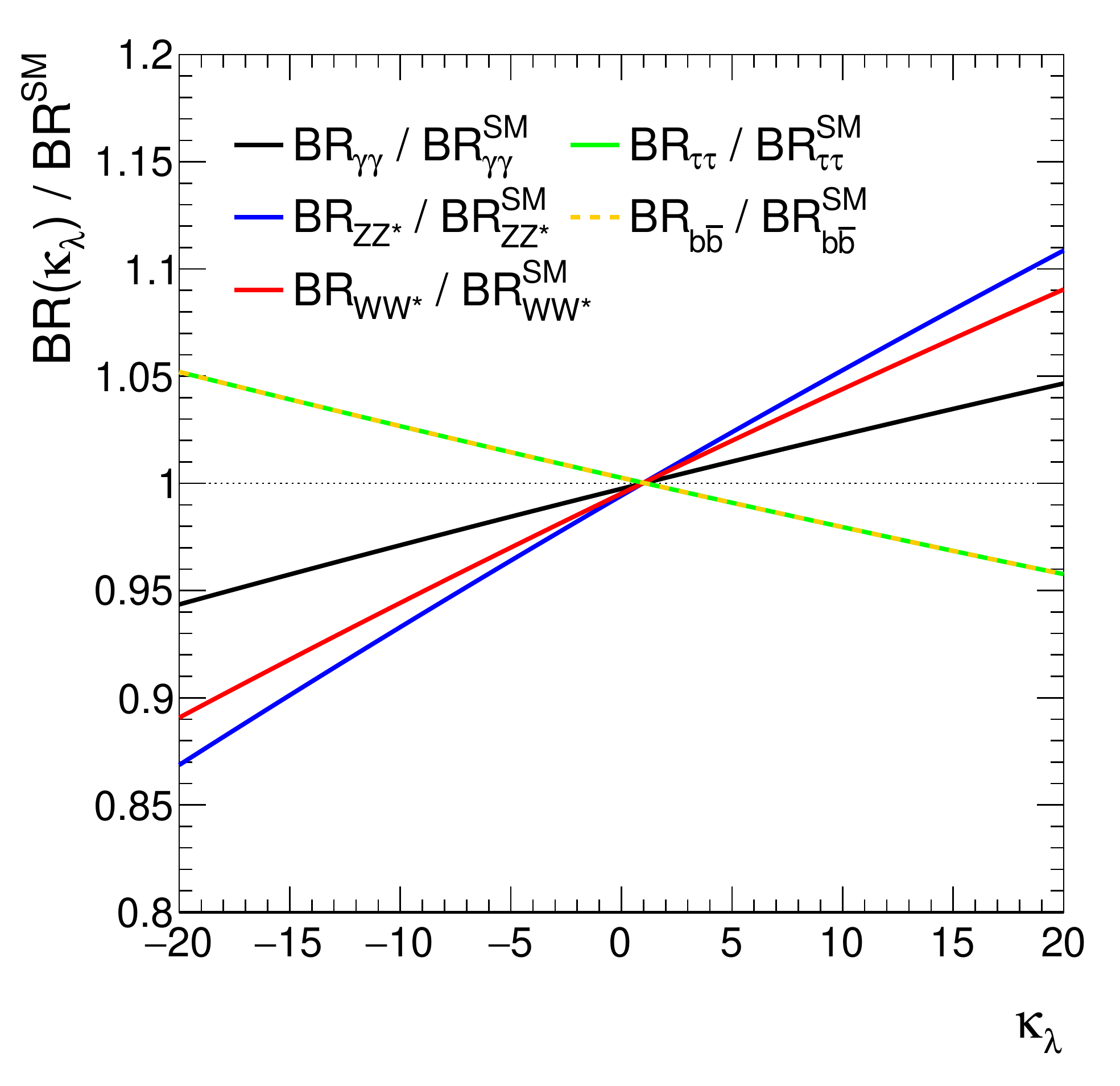}
 \caption{}
\end{subfigure}
    \caption{Variation of the cross sections (a) and branching fractions (b) as a function of the trilinear coupling modifier $\kappa_\lambda$~\cite{Degrassi,Maltoni}; given the fact that the $C_1^f$ coefficients are zero for all $H\rightarrow f\bar{f}$ decays, the $H\rightarrow \tau^+\tau^-$ (green solid line) and the $H\rightarrow b\bar{b}$ (yellow dashed line) lines are superimposed.}
  \label{fig:xs_br_kl}
\end{figure}

The $\kappa_\lambda$-dependent variations of the production cross sections and of the decay branching fractions are shown in Figure~\ref{fig:xs_br_kl}. 

\begin{figure}[htbp]
  \centering
  \begin{subfigure}[b]{0.49\textwidth}
\includegraphics[width=\textwidth]{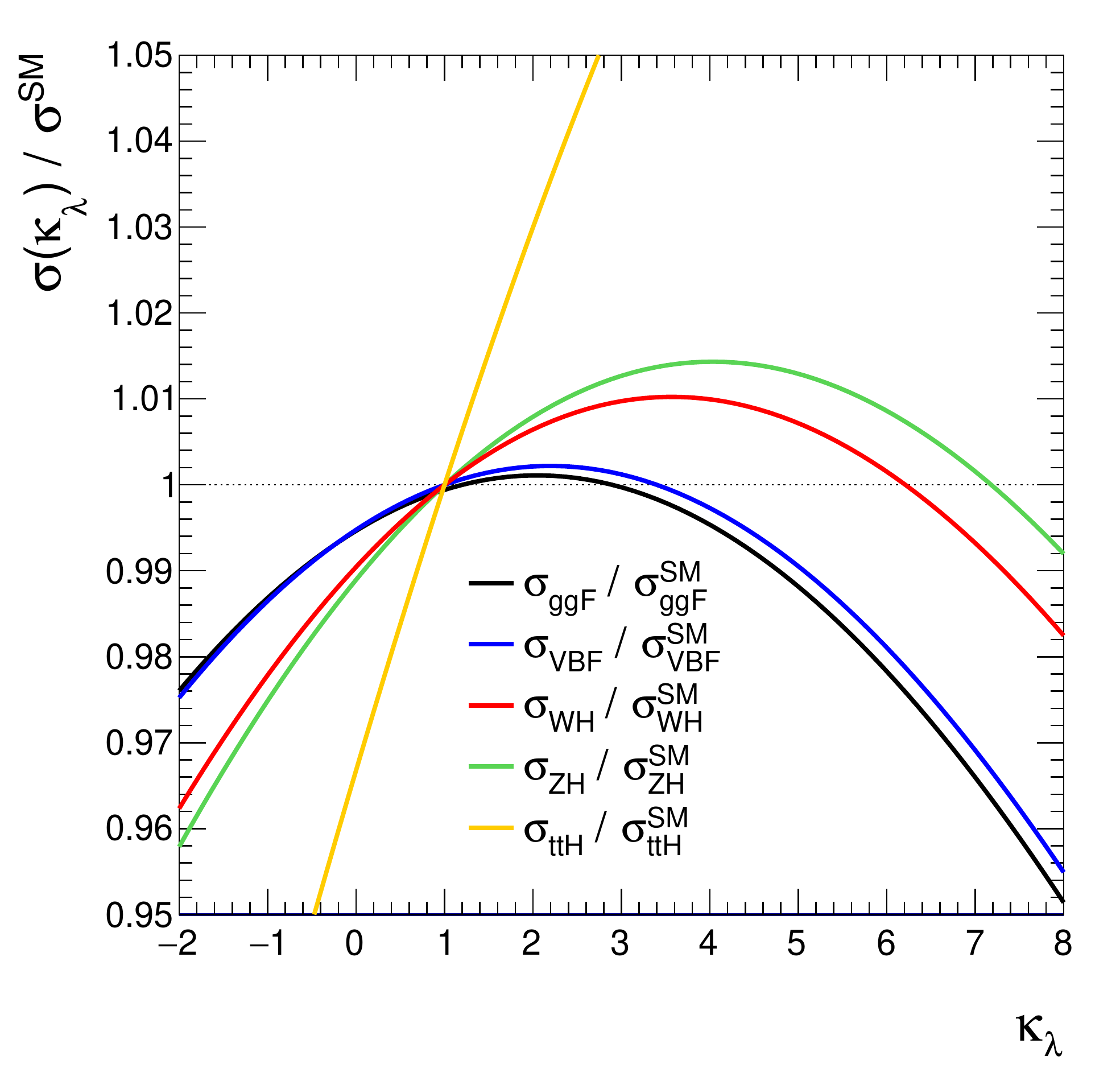}
 \caption{}
\end{subfigure}
  \begin{subfigure}[b]{0.49\textwidth}
\includegraphics[width=\textwidth]{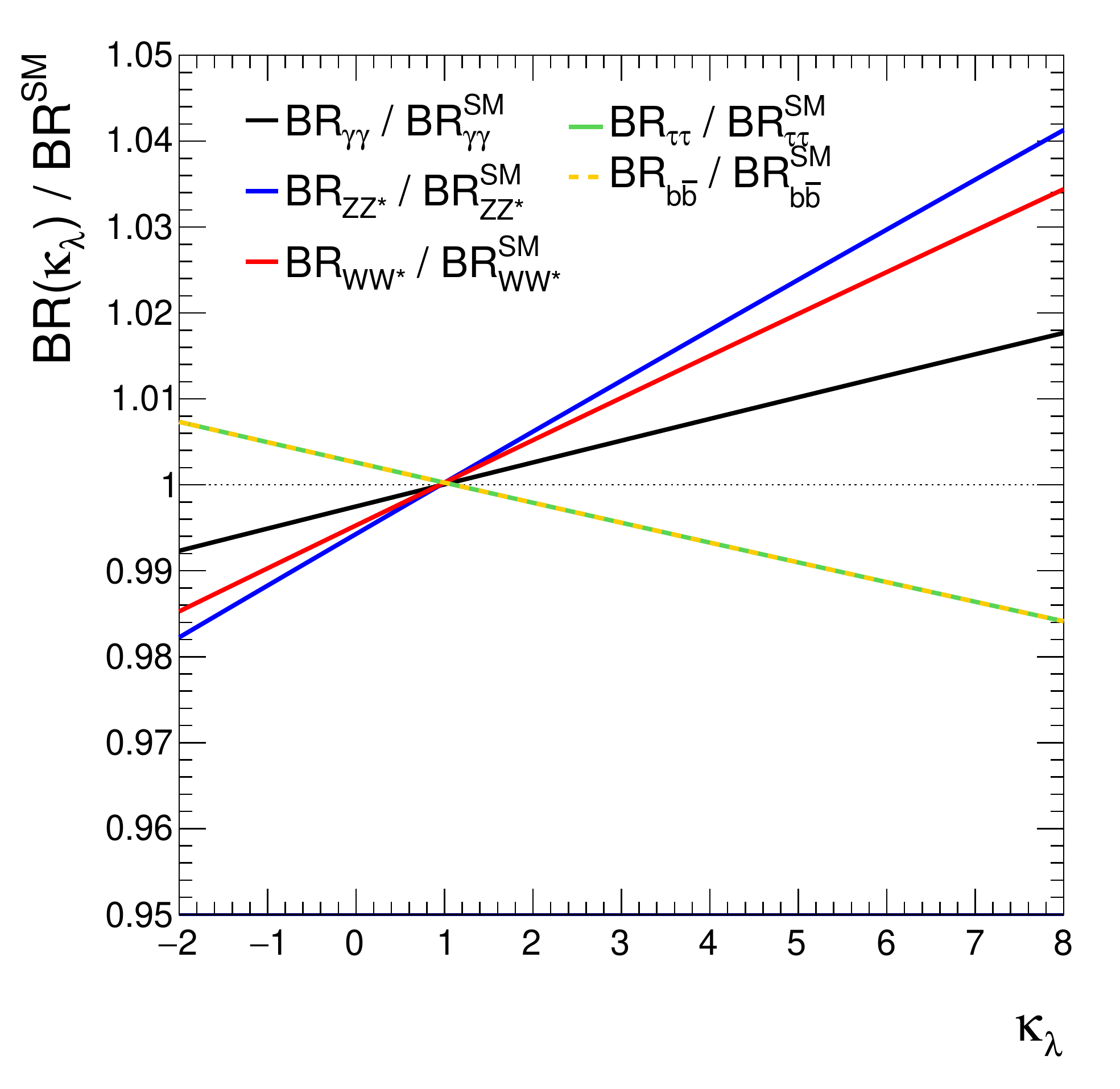}
 \caption{}
\end{subfigure}
    \caption{Variation of the cross sections (a) and branching fractions (b) as a function of the trilinear coupling modifier $\kappa_\lambda$ zoomed in the range $-2<\kappa_\lambda<8$~\cite{Degrassi,Maltoni}; given the fact that the $C_1$ coefficients are zero for all fermion decays, the $H\rightarrow \tau^+\tau^-$ (green solid line) and the $H\rightarrow b\bar{b}$ (yellow dashed line) lines are superimposed.}
  \label{zoom_br_sigma}
\end{figure}

The $t\bar{t}H$ production mode represents the process receiving much larger corrections ($\sim$10\% at $\kappa_\lambda=10$) with respect to the others, due to the fact that, being able to interact with another final-state particle, like \WH and \ZH production processes, it receives a Sommerfeld enhancement in the non-relativistic regime~\cite{Degrassi}. The corrections to the branching fractions, shown in Figure~\ref{fig:xs_br_kl} $(b)$, reach a maximum value of  $\sim$10\% in the $ZZ^*$ decay channel and seem to be much smaller than the corrections to production modes considering the whole $\kappa_\lambda$ validity interval; this effect comes from the linear dependence on $\kappa_\lambda$ entering in these corrections and from the fact that there is a partial cancellation in the ratio, given the same sign of the $C_1$ coefficients. However, in the range close to the SM predictions where corrections are within 5\% in absolute value for the production cross sections as it is shown in Figure~\ref{zoom_br_sigma}, the decay modes are more sensitive to $\kappa_\lambda$ than the production processes, apart from $t\bar{t}H$ production mode. 

\begin{figure}[!htbp]
  \centering
    \begin{subfigure}[b]{0.49\textwidth}
\includegraphics[height=7 cm,width =\textwidth]{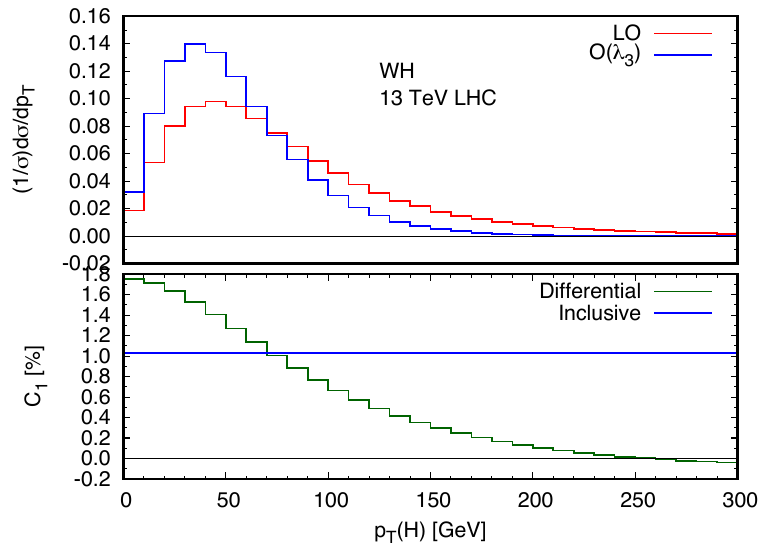}
 \caption{}
\end{subfigure}
  \begin{subfigure}[b]{0.49\textwidth}
\includegraphics[height=7.1 cm,width =\textwidth]{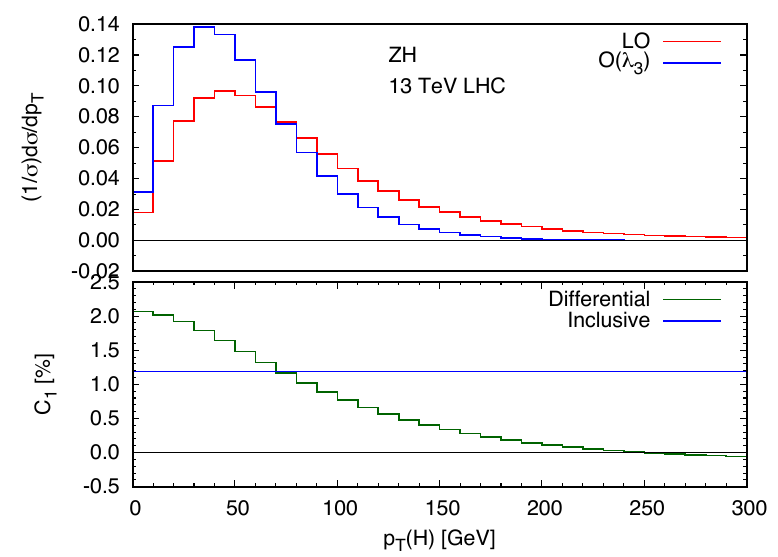}
 \caption{}
\end{subfigure}
      \caption{Effect of $\mathcal{O}(\lambda_{HHH})$ corrections to differential distributions ($p_T^H$) considering \WH (a) and \ZH (b) production modes at 13 TeV LHC. Upper panel: normalised distributions at LO (red) and at  $\mathcal{O}(\lambda_{HHH})$ (blue). Lower panel: $C_1$ at the differential (green) and inclusive (blue) level~\cite{Maltoni}.}
  \label{fig:differential_wh_zh}
\end{figure}

\begin{figure}[!htbp]
  \centering
  \begin{subfigure}[b]{0.49\textwidth}
\includegraphics[width=\textwidth]{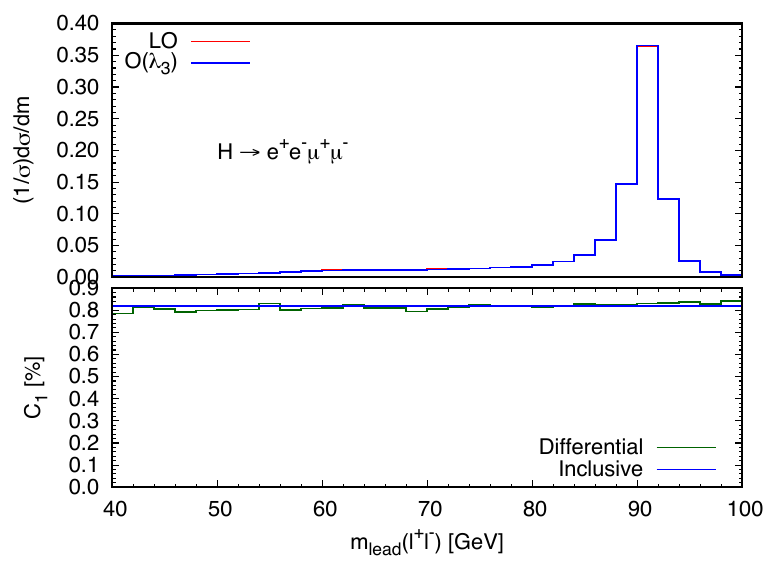}
 \caption{}
\end{subfigure}
  \begin{subfigure}[b]{0.49\textwidth}
\includegraphics[width=\textwidth]{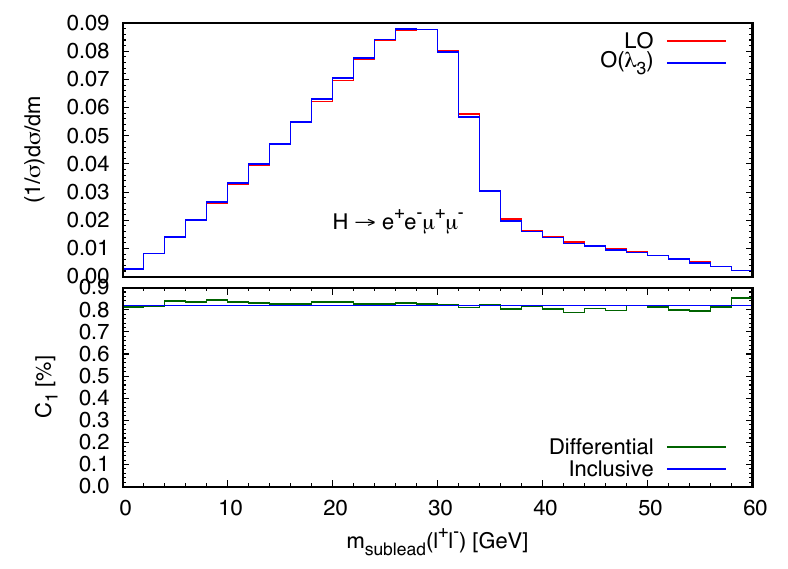}
 \caption{}
\end{subfigure}
    \caption{Leading (a) and subleading (b) opposite-sign same-flavour lepton pair invariant mass distributions in $H\rightarrow e^+e^-\mu^+\mu^-$. Upper panel: normalised LO (red) and $\mathcal{O}(\lambda_{HHH})$ (blue) distributions. Lower panel: $C_1$ for differential (green) and total decay width (blue)~\cite{Maltoni}.}
  \label{decay_diff}
\end{figure}

Variations of the Higgs self-coupling affect not only the inclusive production modes and decay channels, but, being the $C_1$ coefficients kinematic-dependent, they modify also the kinematics of the event. The largest modifications in kinematic distributions are expected in the \ZH, \WH and $t\bar{t}H$ production modes, due to the interaction of the final-state vectors or the top quark with the Higgs boson. \newline
Figure~\ref{fig:differential_wh_zh} shows the differential $C_1$ for \WH $(a)$ and \ZH $(b)$ production modes, considering the $p_T^H$ distribution, \ie\ the distribution of the transverse momentum of the Higgs boson.
The shapes of the LO distributions are compared to the shapes of the contributions induced by $C_1$~\cite{Maltoni}. $C_1$ coefficients at differential and inclusive level are also shown.
The $C_1$ coefficients are enhanced for high-$p_T^H$ regions where, however, the cross section is rather small.\newline
No significant modifications are expected for what concerns the Higgs-boson decay kinematics; in fact, the angular distribution of the decay products, coming from the two bodies decay of the Higgs boson, is fully determined by momentum conservation laws and by the rotational symmetry of the decay, having the Higgs boson a null spin, and cannot be affected by BSM effects. The only exception is represented by the decay to four fermions that is anyway characterised by an extremely small coupling of the Higgs boson to electrons and muons, thus leading to negligible differential $\kappa_\lambda$ contributions, as shown in Figure~\ref{decay_diff} for leading $(a)$ and subleading $(b)$ opposite-sign same-flavour lepton pair invariant mass distributions in the $H\rightarrow e^+e^-\mu^+\mu^-$ decay channel.
\clearpage
\section{HL-LHC projections for single-Higgs processes}
\label{sec:HL_LHC_theory}

\begin{figure}[htbp]
\begin{center}
\includegraphics[width=\textwidth]{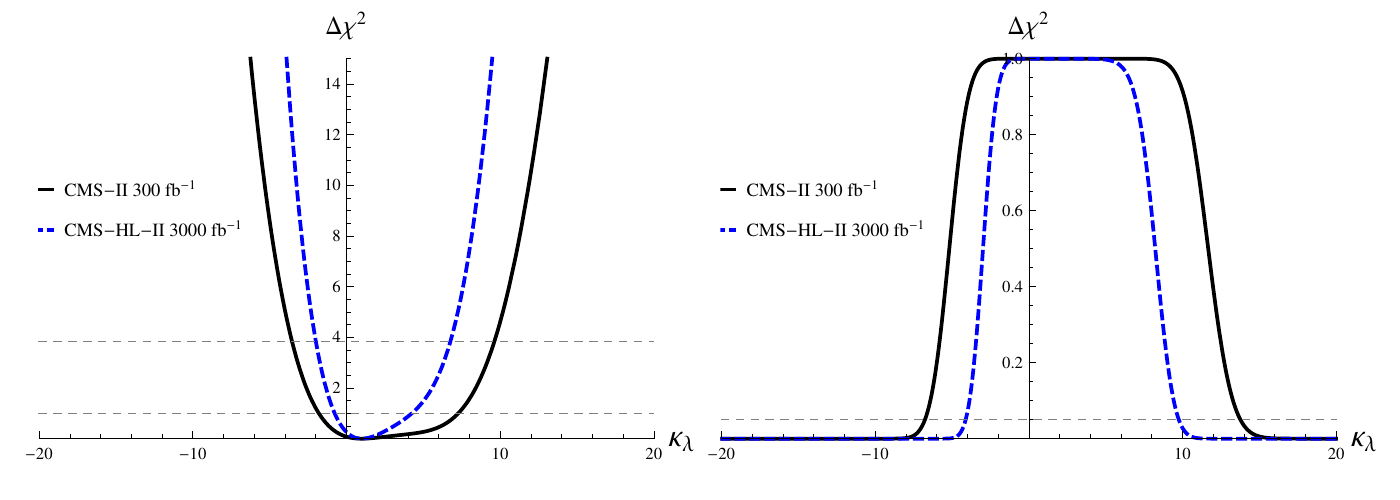}
\end{center}
\caption{Left: $\chi^2(\kappa_\lambda$) distribution as a function of $\kappa_\lambda$; right: $p-$value distribution as a function of $\kappa_\lambda$ for ``CMS-II$"$ (solid black line) and ``CMS-HL-II$"$ (blue dashed line)~\cite{Degrassi} where ``CMS-II$"$ (300 fb$^{-1}$) and ``CMS-HL-II$"$ (3000 fb$^{-1}$), are the scenarios presented in Table 1 of Reference~\cite{CMS_degrassi}.}
\label{HL_single_degrassi}
\end{figure}

Projections for the measurement of the trilinear Higgs self-coupling at HL-LHC have been made considering NLO-EW corrections depending on $\kappa_\lambda$ to single-Higgs processes; both the theoretical papers on the top of which the results of this thesis have been produced have performed these estimations, reported in detail in References~\cite{Degrassi,Maltoni}. \newline
An estimation of the improvement in constraining $\kappa_\lambda$ has been presented in Reference~\cite{Degrassi} exploiting projections of the CMS experiment at 300 fb$^{-1}$ and 3000 fb$^{-1}$, using the uncertainties reported in Table 1 of Reference~\cite{CMS_degrassi}; theoretical uncertainties are identical in the 3000~fb$^{-1}$ and in the 300~fb$^{-1}$ case, while experimental uncertainties are scaled with the square root of the ratio between the luminosities.
The $1\sigma$ and $2\sigma$ intervals are identified assuming a $\chi^2$ distribution.
Figure~\ref{HL_single_degrassi} reports the $\chi^2$ and $p-$value distributions as a function of $\kappa_\lambda$ assuming that the central value of the measurements in every channel coincides with the predictions of the SM for ``CMS-II$"$ (300 fb$^{-1}$) and ``CMS-HL-II$"$ (3000~fb$^{-1}$); experimental and theoretical uncertainties are included. The constraints that can be obtained using 3000 fb$^{-1}$ are~\cite{Degrassi}: 
\begin{equation}
\kappa_\lambda^{1\sigma} = [-0.7, 4.2] \qquad \text{and} \qquad  \kappa_\lambda^{2\sigma}= [-2.0, 6.8] \nonumber
\end{equation}
where $\kappa_\lambda^{1\sigma}$ and $\kappa_\lambda^{2\sigma}$ are the $1\sigma$ and $2\sigma$ intervals, respectively.

\begin{figure}[htbp]
\begin{center}
\includegraphics[height= 8 cm, width=\textwidth]{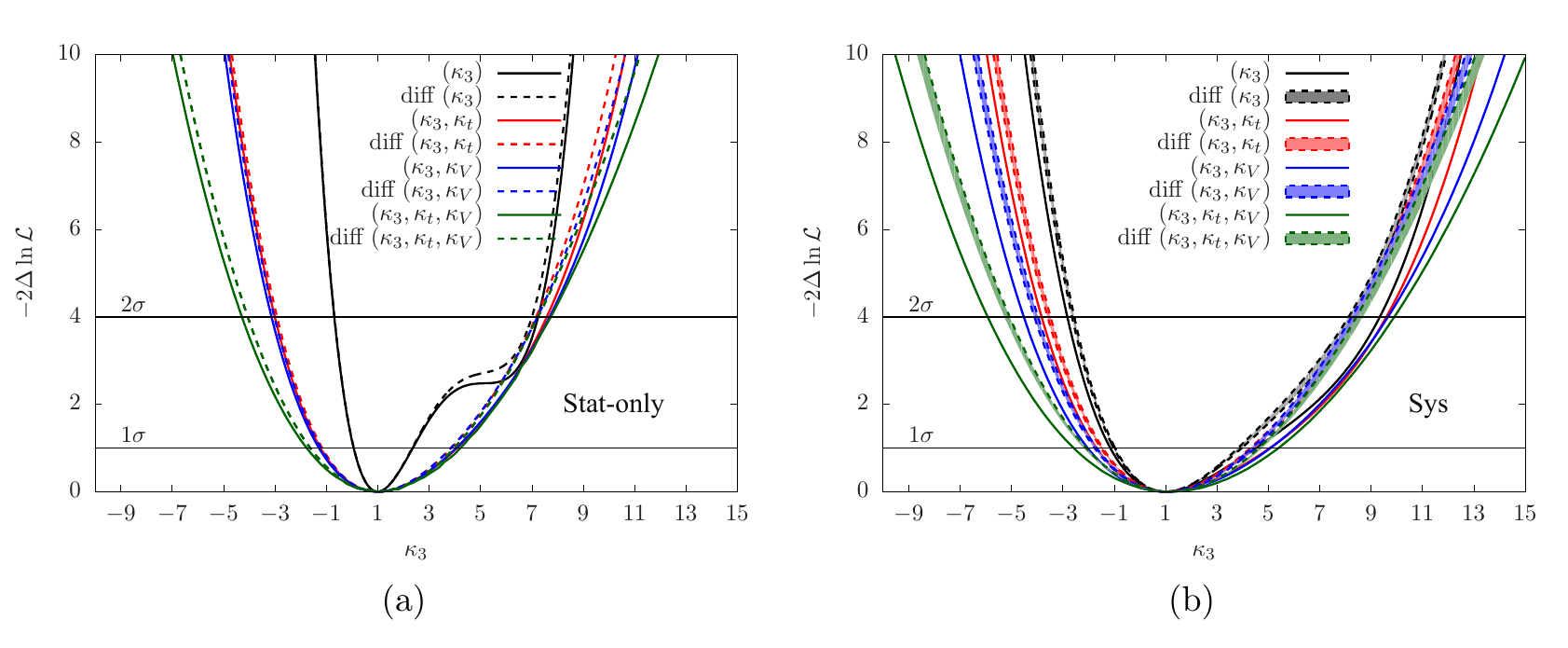}
\end{center}
\caption{Value of $-2 \ln{\Lambda(\kappa_\lambda)}$ as a function of $\kappa_\lambda$ for an Asimov dataset generated under the SM hypothesis considering a luminosity of 3000 fb$^{-1}$ at $\sqrt{s}=14$ TeV.  Two different scenarios are considered:  only statistical uncertainty are considered (Stat-only=S1) (a), experimental systematic uncertainty and theory uncertainty included (Run 2 sys=S2) (b). Different fit configurations have been tested: $\kappa_\lambda$-only model (black line), $\kappa_\lambda$-$\kappa_t$ model (red line), $\kappa_\lambda$-$\kappa_V$ model (blu line) and $\kappa_\lambda$-$\kappa_t$-$\kappa_V$, (green line). All the coupling modifiers that are not included in the fit are set to their SM predictions. The dotted horizontal lines show the $-2 \ln{\Lambda(\kappa_\lambda)}=1$ level that is used to define the $\pm 1\sigma$ uncertainty on $\kappa_\lambda$ as well as the $-2 \ln{\Lambda(\kappa_\lambda)}=4$ level used to define the $\pm 2\sigma$ uncertainty~\cite{Maltoni}.}
\label{HL_single_maltoni}
\end{figure}

A global fit to the likelihood function in order to extract $\kappa_\lambda$, based on the future projections of ATLAS-HL for single-Higgs production and decay at 14 TeV~\cite{Maltoni_hl, Maltoni_hl1}, has been performed in Reference~\cite{Maltoni}, assuming that the central value of the measurements in every channel coincides with the predictions of the SM. Two different scenarios concerning the uncertainties have been considered: in the first scenario (``Stat-only$"$), only the statistical uncertainty is included, describing an unrealistic scenario where theory and experimental systematic uncertainties are negligible; the second scenario (``Run 2 sys$"$), takes into account both theory and experimental systematic uncertainties. Differential information is included in the $VH$ and $t\bar{t}H$ production modes, for both the $C_1$ and the $K_{EW}$ coefficients. The same future scenario at 14 TeV (ATLAS-HL) considered in Reference~\cite{global} is exploited. Figure~\ref{HL_single_maltoni} shows the likelihood distribution after combining all the production channels for Scenarios ``Stat-only$"$ $(a)$ and ``Run 2 sys$"$ $(b)$, under different assumptions: i) only $\kappa_\lambda$ is anomalous, ii) $\kappa_\lambda$ and $\kappa_t$ or  $\kappa_\lambda$ and $\kappa_V$ are anomalous,  iii) all three parameters $\kappa_\lambda$, $\kappa_t$ and $\kappa_V$ are anomalous. Including additional degrees of freedom relaxes the limits in the region $\kappa_\lambda< 1$, even if they do not completely vanish, while the sensitivity to $\kappa_\lambda$ in the region $\kappa_\lambda >1$ is almost unaltered.
On the contrary, the role of differential information may be relevant, critically depending on the assumptions on the future experimental and theoretical uncertainties~\cite{Maltoni}.\newline

\begin{figure}[htbp]
\begin{center}
\includegraphics[width=0.7\textwidth]{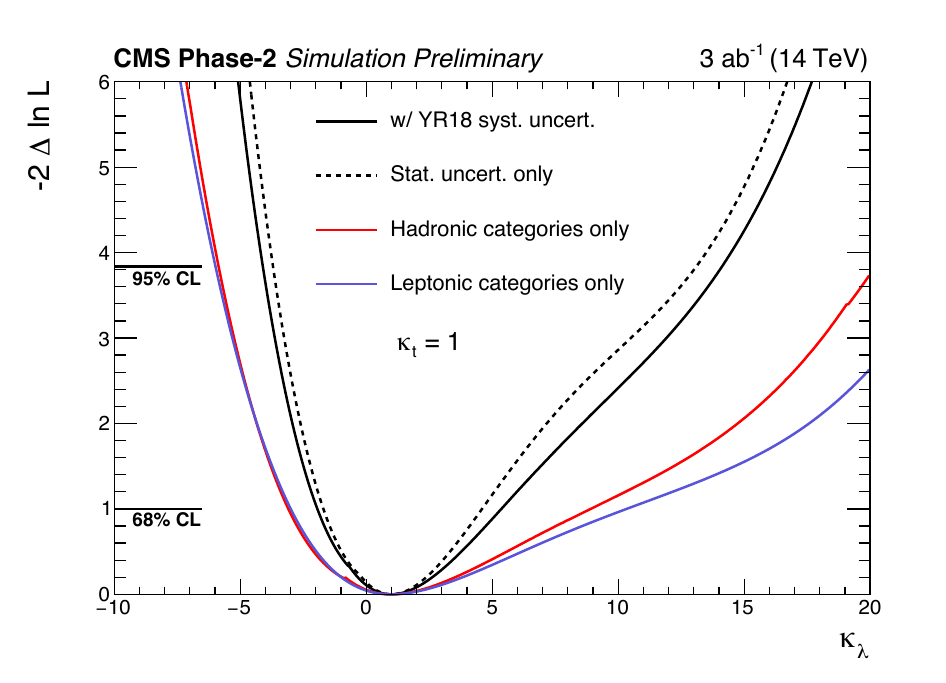}
\end{center}
\caption{Value of $-2 \ln{\Lambda(\kappa_\lambda)}$ as a function of $\kappa_\lambda$ for an Asimov dataset generated under the SM hypothesis considering a luminosity of 3000 fb$^{-1}$ at $\sqrt{s}=14$ TeV. The individual contributions of the statistical and systematic uncertainties are separated performing a likelihood scan with all systematics removed. Additionally, the contributions from the hadronic and leptonic channels have been separated, and are shown as red and purple solid lines, respectively \cite{HL-LHC}.}
\label{HL_CMS_ttH}
\end{figure}

Finally, in order to give an indication of the power in constraining $\kappa_\lambda$ coming from differential $t\bar{t}H$ measurements, a global likelihood fit considering $t\bar{t}H$ and $tH$ production modes together with $VH$, $H\rightarrow \gamma \gamma$, is reported, performed by the CMS collaboration~\cite{HL-LHC}. The $C_1$ coefficients have been computed for each bin of $p_T^H$ in the fiducial region. \newline
Figure~\ref{HL_CMS_ttH} shows the value of $-2 \ln{\Lambda(\kappa_\lambda)}$ as a function of $\kappa_\lambda$ for an Asimov dataset generated under the SM hypothesis considering a luminosity of 3000 fb$^{-1}$ at $\sqrt{s}=14$ TeV and  assuming all other couplings set to their SM values. For the purposes of constraining  $\kappa_\lambda$, theoretical uncertainties in the differential $t\bar{t}H$ + $tH$ cross section are included in the signal model. The results when only including the hadronic or leptonic categories are shown in addition to the results obtained from their combination.
From the differential cross-section measurement of a single Higgs-boson decay channel produced in association with top quarks, $\kappa_\lambda$ is constrained at 95\% CL in the interval  $-4.1< \kappa_\lambda < 14.1$~\cite{HL-LHC}.


\chapter{Constraints on the Higgs-boson self-coupling from double-Higgs production and decay measurements}
\label{sec:dihiggs}
This chapter presents the results of the extraction of constraints on $\kappa_\lambda$ from Higgs-boson pair production in the  $b\bar{b}b\bar{b}$, $b\bar{b}\tau^+\tau^-$ and $b\bar{b}\gamma\gamma$ channels. Data and input measurements as well as main features of the different channels are briefly described in Section~\ref{sec:data_hh}; the procedure exploited in order to simulate the signal samples used to extract $\kappa_\lambda$ results together with the implementation of the theoretical framework described in Chapter~\ref{sec:prob_self} are reported in Sections~\ref{sec:sim_hh} and~\ref{sec:theory_hh}, respectively. \newline
Section~\ref{sec:statistical_model_hh} reports details on the statistical model, on the construction of the likelihood function and on the different uncertainties that are included in the likelihood as nuisance parameters. Sections~\ref{sec:validation_hh} describes a validation of the inputs of this combination aiming at reproducing the latest results from the ATLAS experiment reported in Reference~\cite{Paper_hh}. Finally, Sections~\ref{sec:results_hh_channel} and~\ref{sec:results_hh_comb} present the constraints on $\kappa_\lambda$, starting from the double-Higgs single-channel constraints and then proceeding with the combination of the three double-Higgs decay channels.

\section{Data and input measurements}
\label{sec:data_hh}

The combination of searches for non-resonant Higgs-boson pair production exploits data collected by the ATLAS experiment in 2015 and 2016 from 13 TeV $pp$ collisions corresponding to an integrated luminosity of up to 36.1 fb$^{-1}$.
The double-Higgs analyses include the $b\bar{b}b\bar{b}$~\cite{4b}, the $b\bar{b}\tau^+\tau^-$~\cite{bbtautau} and the $b\bar{b}\gamma \gamma$~\cite{bbyy} decay channels.
The integrated luminosity of the datasets used in each double-Higgs analysis included in this combination is reported in Table~\ref{tab:lumi_hh}. 
\begin{table}[!htbp]
  \caption{Integrated luminosity of the datasets used for each input
    analysis to the double-Higgs combination. The last column provides references to
    publications describing each measurement included in detail.}
\begin{center}
{\def\arraystretch{1.4}
\begin{tabular}{|l|c|c|}
\hline 
Analysis & Integrated luminosity (fb$^{-1}$) & Reference \\
\hline
$HH\rightarrow b\bar{b}b\bar{b}$      & $27.5$         & \cite{4b} \\
$HH\rightarrow b\bar{b}\tau^+\tau^-$      & $36.1$         & \cite{bbtautau} \\
$HH\rightarrow b\bar{b} \gamma\gamma$      & $36.1$         & \cite{bbyy} \\
\hline 
\end{tabular}}
\end{center}
\label{tab:lumi_hh}
\end{table}
Each analysis separates the selected events into orthogonal kinematic and topological regions, called categories. The $b\bar{b}b\bar{b}$ categories are orthogonal to both the $b\bar{b}\tau^+\tau^-$ and $b\bar{b}\gamma \gamma$ categories by
definition, while a negligible overlap is present between the $b\bar{b}\tau^+\tau^-$ and $b\bar{b}\gamma \gamma$ analyses~\cite{Paper_hh}.\newline
The double-Higgs event selections are targeting double-Higgs production, but select also single-Higgs events that need to be included in the double-Higgs signal regions if their contribution is not negligible. Details on single-Higgs backgrounds included in the different channels are reported in Section~\ref{sec:sim_hh}.\newline
The double-Higgs analyses are categorised as in the following:
\begin{itemize}
\item the $b\bar{b}b\bar{b}$ analysis looks for final states with at least four small-$R$ $b$-tagged jets reconstructed using the anti-$k_t$ algorithm, as described in Chapter~\ref{sec:Reco}. The strategy exploited for the non-resonant search is to select two Higgs-boson candidates, each composed of two $b$-tagged jets, with invariant masses close to $m_H$. The invariant mass of the two-Higgs-boson-candidate system, $m_{4j}$, is used as the final discriminant between Higgs-boson pair production and the backgrounds, which are principally QCD multijets and $t\bar{t}$~\cite{4b}. The dataset is split according to the years 2015 and 2016, and then statistically combined taking into account the different trigger algorithms used in 2015 and 2016. In part of the 2016 data period, inefficiencies in the online vertex reconstruction affected $b$-jet triggers that were used in the analysis, reducing the total available integrated luminosity to 27.5 fb$^{-1}$. The shape of the $m_{HH}$ distribution has a strong dependence on $\kappa_\lambda$ as shown in Figure~\ref{shape_mhh} for various $\kappa_\lambda$ values;
\begin{figure}[htbp]
\centering
\begin{subfigure}[b]{0.49\textwidth}
\includegraphics[width =\textwidth]{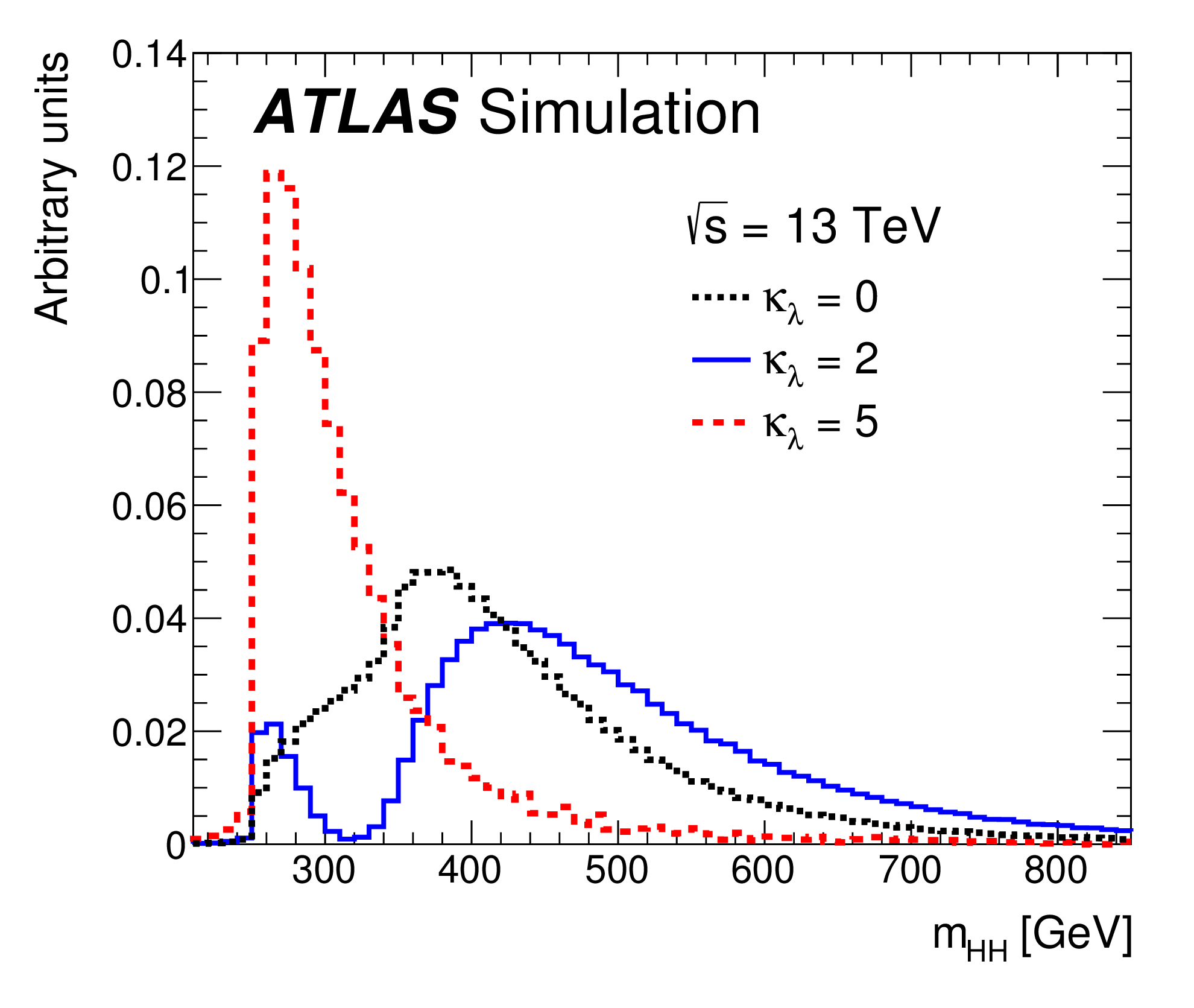}
 \caption{}
\end{subfigure}
\begin{subfigure}[b]{0.49\textwidth}
\includegraphics[width =\textwidth]{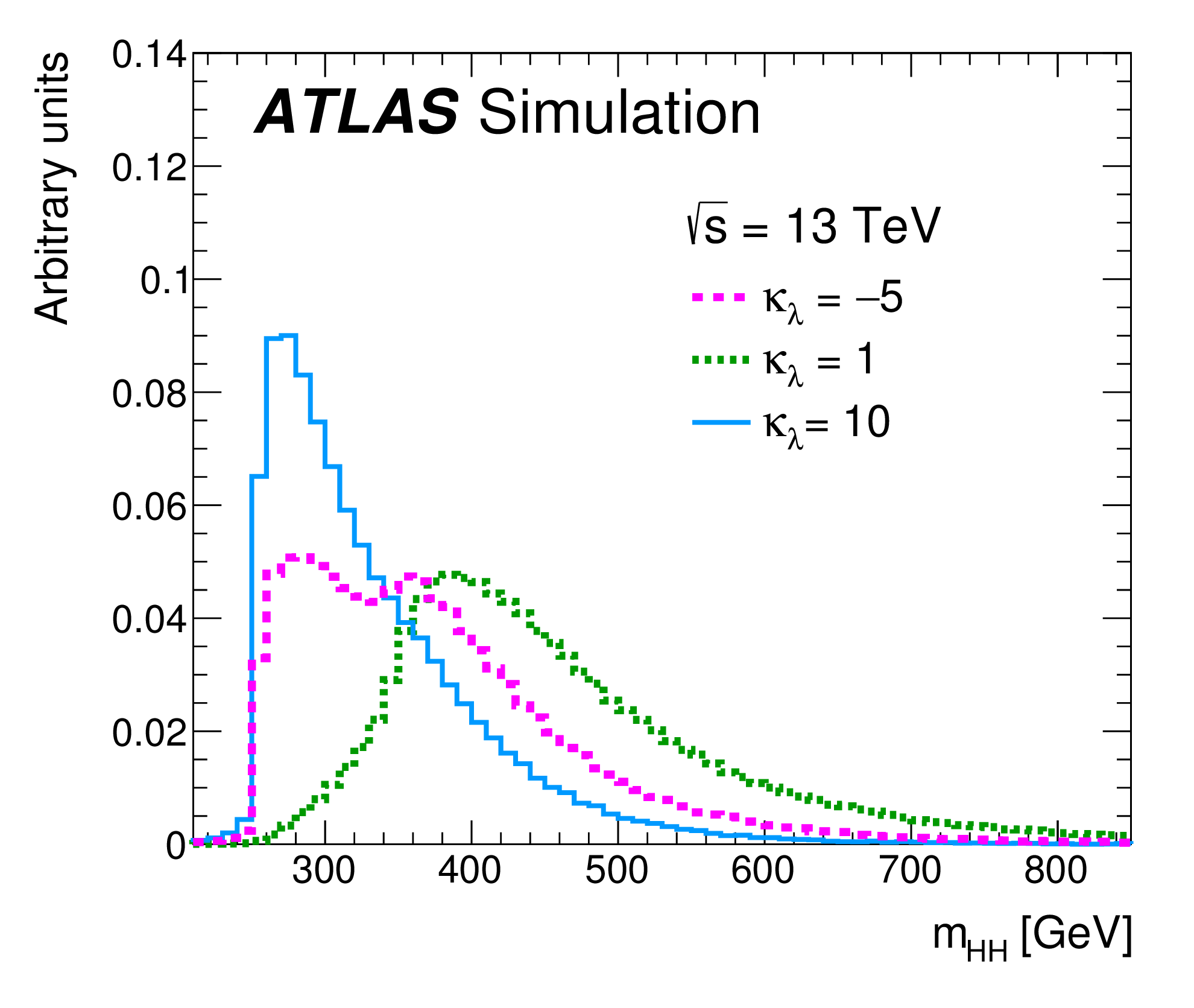}
 \caption{}
\end{subfigure}
\caption{Generator-level $m_{HH}$ distributions computed for various values of $\kappa_\lambda$ by linearly combining three LO samples produced with $\amc$. The $m_{HH}$ shape is affected by the interference pattern between the box diagrams and the triangle diagram~\cite{Paper_hh}.}     
\label{shape_mhh}
\end{figure}

so does the signal acceptance that varies by a factor 2.5 over the probed range of $\kappa_\lambda$-values as presented in Figure~\ref{acceptance}.
\begin{figure}[htbp]
\centering
\includegraphics[height=7 cm, width=8 cm]{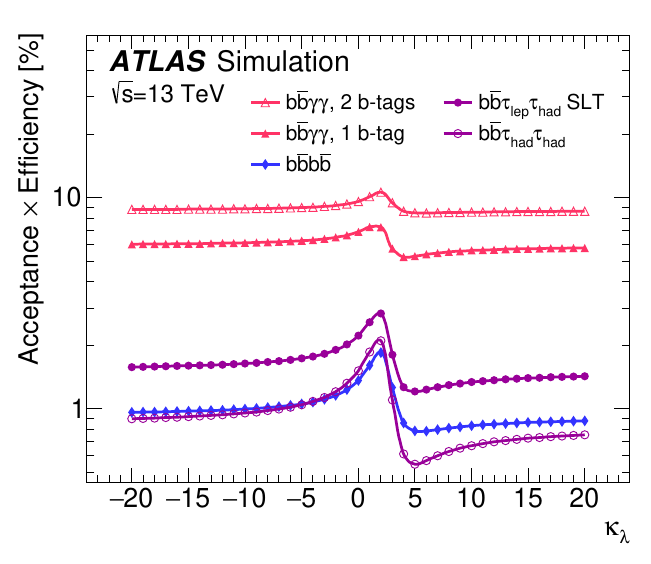}
 \caption{Signal acceptance times efficiency as a function of $\kappa_\lambda$ for the $b\bar{b}\tau^+\tau^-$,  $b\bar{b}b\bar{b}$ and $b\bar{b}\gamma\gamma$ channels~\cite{Paper_hh}.}     
\label{acceptance}
\end{figure}
\item The $b\bar{b}\tau^+\tau^-$ analysis looks for final states with two $R=0.4$ $b$-tagged jets reconstructed using the anti-$k_t$ algorithm and two $\tau$-leptons. Events are required to have at least one collision vertex reconstructed from at least two charged-particle tracks with transverse momentum $p_T>0.4$~GeV. The analysis is split into two categories: the $\tau_{\textrm{lep}} \tau_{\textrm{had}}$ channel, in which events are required to contain an electron or a muon from one of the two $\tau$-leptons decaying leptonically and a hadronically decaying $\tau$-lepton of opposite charge, and the $\tau_{\textrm{had}} \tau_{\textrm{had}}$ channel, in which events are required to contain two hadronically decaying $\tau$-leptons of opposite charge. BDTs, defined in Chapter~\ref{sec:Reco}, are used in the analysis to improve the separation of signal from background and, in order to compute the final results, the BDT score distributions, which have a dependence on $\kappa_\lambda$ through the shape variations of some variables, in particular $m_{HH}$ as shown in Figure~\ref{shape_mhh}, are used. In addition, the sensitivity of this analysis is affected by the variation of the signal acceptance by a factor 3 over the probed range of $\kappa_\lambda$, as shown in Figure~\ref{acceptance}.
\item The $b\bar{b}\gamma \gamma$ analysis looks for final states with two photons and one or two $R=0.4$ $b$-tagged jets. Particularly, two high-$p_T$ isolated photons, accompanied by two jets with dijet invariant mass, $m_{jj}$, compatible with the mass of the Higgs boson, \ie\ $80<m_{jj}<140$~\GeV, are required to have $E_T/m_{\gamma\gamma} > $ 0.35 and 0.25 respectively; the events are then analysed requiring a jet with $p_T > 40$ GeV and a second jet with $p_T > 25$ GeV. The signal consists of a narrow peak in the $m_{\gamma \gamma}$ distribution superimposed on a smoothly falling background. Events are subdivided into two categories according to the number of $b$-tagged jets. The $m_{\gamma\gamma}$ distribution dependence on $\kappa_\lambda$ has been examined by comparing the generated $m_{\gamma\gamma}$ spectrum in simulation using different $\kappa_\lambda$ values, and the one with $\kappa_\lambda = 1$ and finding an agreement within statistical uncertainties. Furthermore, being the shape of the diphoton mass distribution described by the double-sided CrystalBall function~\cite{dscb}, consisting of a Gaussian core with power-law tails on either side, the dependence on $\kappa_\lambda$ of the key shape parameters of this function has been tested, resulting in an almost flat behaviour against $\kappa_\lambda$ variations, however less influencing than the calibration uncertainties associated to these parameters. Thus the signal $m_{\gamma\gamma}$ distribution is implemented as a continuous function in the fit. The analysis acceptance, instead, varies by about 30\% over the probed range of $\kappa_{\lambda}$-values, as shown in Figure~\ref{acceptance}.
\end{itemize}

\section{Simulation of signal samples}
\label{sec:sim_hh}

Exploiting Equation~\ref{eq:pph}, it is possible to parameterise the signal distributions as a function of  $\kappa_\lambda/\kappa_{t}$. 
In the $b\bar{b}b\bar{b}$ and $b\bar{b}\tau^+\tau^-$ cases, three samples with different set of parameters $\kappa_\lambda/\kappa_t$ have been simulated and used to reproduce the signal distributions for any value of $\kappa_\lambda/\kappa_t$ through a linear combination method. Therefore signal samples have been generated choosing $\kappa_t=1$ for all samples and $\kappa_\lambda/\kappa_t = 0$, called sample S(1,0), $\kappa_\lambda/\kappa_t = 1$, called S(1,1), and $\kappa_{\lambda}/\kappa_t = 20$, called S(1,20). Thus the cross section, using Equation~\ref{eq:pph}, can be written in terms of these $\kappa_{\lambda}/\kappa_t$ values as:
\begin{eqnarray*}
\sigma(\kappa_t=1,\kappa_\lambda/\kappa_t = 0) & \sim  & |\mathcal{A}_1|^2 \\
\sigma(\kappa_t=1,\kappa_\lambda/\kappa_t = 1) & \sim & |\mathcal{A}_1|^2 + 2\Re(\mathcal{A}_1^*\mathcal{A}_2) + |\mathcal{A}_2|^2 \\
\sigma(\kappa_t=1, \kappa_\lambda/\kappa_t = 20) & \sim &  |\mathcal{A}_1|^2 + 2\cdot 20\Re(\mathcal{A}_1^*\mathcal{A}_2)   +
                                          20^2 |\mathcal{A}_2|^2 \, . 
\end{eqnarray*}
The solution of this system of equations provides the expression of $|\mathcal{A}_1|^2$, $\Re(\mathcal{A}_1^*\mathcal{A}_2$) and $|\mathcal{A}_2|^2$ as a function of the three arbitrary samples, leading to the following expression for the cross section, and the signal distributions:
\begin{equation}
\begin{split}
\sigma (\kappa_\lambda, \kappa_t) \sim {}&
 \kappa_t^2 \left ( \left( \kappa_t^2 + \frac{\kappa_\lambda^2}{20} - \frac{399}{380} \kappa_\lambda \kappa_t \right) |S(1, 0)|^2 + \left( \frac{40}{38} \kappa_\lambda \kappa_t - \frac{2}{38}\kappa_\lambda^2 \right ) |S(1, 1)|^2 \right ) + \\
& +  \left ( \left( \frac{\kappa_\lambda^2 - \kappa_\lambda \kappa_t}{380} \right ) |S(1, 20)|^2   \right ) \,. 
\end{split}
\label{eq:bases}
\end{equation}
The procedure followed in order to simulate the three signal samples with different $\kappa_\lambda/\kappa_t$ values that are included in the fit for the $b\bar{b} b\bar{b}$ and $b\bar{b}\tau^+\tau^-$ decay channels is described in Reference~\cite{white_paper}.

\section{Implementation of the theoretical model}
\label{sec:theory_hh}

The theoretical framework described in Chapter~\ref{sec:prob_self} is implemented in the double-Higgs channels taking into account that:
\begin{itemize}
\item double-Higgs kinematic distributions depend only on the ratio $\kappa_\lambda/\kappa_t$, and, consequently, the signal acceptance also depends only on $\kappa_\lambda/\kappa_t$, 
\item the $\kappa_t^4$ factor affects only the total cross section;
\item the self-coupling modifier $\kappa_\lambda$ can affect the Higgs-boson branching fractions and cross sections due to NLO-EW corrections; thus these corrections have to be included in the parameterisations of single-Higgs background production cross sections and $b\bar{b}$, $\gamma \gamma$ and $\tau^+\tau^-$ decay branching fractions.
\end{itemize}
Furthermore, the expression of $|\mathcal{A}_1|^2$, $\Re(\mathcal{A}_1^*\mathcal{A}_2$) and $|\mathcal{A}_2|^2$ reported in the previous section, are used in order to parameterise the three signal samples for the $b\bar{b}b\bar{b}$ and $b\bar{b} \tau^+\tau^-$ channels, included in the fit through the signal strengths associated to each of them $(\mu_0, \mu_1, \mu_{20})$, defined in Chapter~\ref{sec:SM}, that are parameterised as in the following:
\begin{eqnarray*}
\mu_0=(\mu_f(H\rightarrow b\bar{b}))^2 \times \left ( \kappa_t^4 + \frac{\kappa_t^2\kappa_\lambda^2}{20} - \frac{399}{380} \kappa_t^3\kappa_\lambda \right ) \nonumber , \\
\mu_1=(\mu_f(H\rightarrow b\bar{b}))^2 \times \left ( \frac{40}{38}\kappa_t^3\kappa_\lambda - \frac{2}{38} \kappa_t^2\kappa_\lambda^2  \right ) \nonumber , \\
\mu_{20}=(\mu_f(H\rightarrow b\bar{b}))^2 \times \left ( \frac{\kappa_\lambda^2\kappa_t^2-\kappa_t^3\kappa_\lambda}{380} \right ) \nonumber \\
\end{eqnarray*}
where $\mu_f$ describes the multiplicative corrections of each decay channel branching fraction ($\text{BR}_{\text{SM},f}$) as a function of the anomalous values of the trilinear Higgs self-coupling and of the couplings of the Higgs boson to the other particles of the SM. In the case of the $b\bar{b}\gamma \gamma$ channel, no signal sample has been used in the fitting procedure, but the signal $m_{\gamma \gamma}$ distribution is implemented as a continuous function in the fit because its shape shows a negligible dependence on $\kappa_{\lambda}$. The analysis acceptance, instead, depends on $\kappa_{\lambda}/\kappa_{t}$ and the dependence has been implemented as:
\begin{equation}
\text{Acceptance}=\frac{\text{Yield}\, (\kappa_\lambda/\kappa_t)}{\sigma(\kappa_\lambda/\kappa_t)\times BR (\kappa)\times \text{Luminosity}\times \text{Efficiency}}
\end{equation}
where $\kappa$ is a generic coupling modifier including both Higgs self-coupling and single-Higgs couplings, and $\sigma(\kappa_\lambda/\kappa_t) \times BR(\kappa)$ can be written as:
\begin{equation}
\sigma \times BR= \mu_{HH}=\mu_f(H\rightarrow b\bar{b}) \times \mu_f(H\rightarrow \gamma \gamma) \times \kappa_t^4 \times \sigma_{HH}(\kappa_\lambda/\kappa_t) \times 2 \times BR_{bb}^{SM} \times BR_{\gamma \gamma}^{SM} \nonumber 
\end{equation}
where $\sigma_{HH}$ is the SM double-Higgs cross section expressed as a function of $\kappa_\lambda/\kappa_t$ and $BR^{SM}$ are the branching fractions for a SM Higgs boson with $M_H$ = 125.09 $\GeV$, whose values are reported in Chapter~\ref{sec:SM}.\newline
The value of the SM double-Higgs production cross section used in this combination is: $\sigma^{\mathrm{SM}}_{\mathrm{ggF}} (pp \to HH) = 33.5^{+2.4}_{-2.8}$~fb at $\sqrt{s} = 13$~\TeV~\cite{Higgs_CS}, calculated at NLO in QCD with the measured value of the top-quark mass and corrected at NNLO in QCD matched to NNLL resummation using the heavy top-quark limit~\cite{Higgs_CS,Dawson:1998py,Borowka:2016ehy,Baglio:2018lrj, Bonciani:2018omm, deFlorian:2013jea,Shao:2013bz,deFlorian:2015moa}, consistently with the cross section used for 2015-2016 analyses~\cite{Paper_hh}. In addition to the signal samples, the dominant single-Higgs background processes have to be included in the double-Higgs channels if their contribution is not negligible. Thus they have been considered in the $b\bar{b}\gamma \gamma $ and $b\bar{b} \tau^+\tau^-$ channels and they are parameterised as:
\begin{equation}
b\bar{b}\tau^+\tau^-: VH=\mu_{ZH}^i \times \mu_f(H\rightarrow \tau \tau) \quad t\bar{t}H=\mu_{t\bar{t}H}^i \times \mu_f(H\rightarrow \tau \tau) ; \nonumber
\end{equation}
\begin{equation}
b\bar{b}\gamma \gamma : \mu_{XS\, ggF}=\mu_{t\bar{t}H}^i \quad \mu_{XS\,VBF}=\mu_{ZH}^i \nonumber
\end{equation}
where $\mu_i$ describes the multiplicative corrections of the expected SM Higgs production cross-sections ($\sigma_{\text{SM},i}$) as a function of the anomalous values of the trilinear Higgs self-coupling and of the couplings of the Higgs boson to the other SM particles, and the dominant production modes have been selected looking at the expected number of events in the single channels. \newline
$C_1$ inclusive coefficients, representing linear $\kappa_\lambda$-dependent corrections to single-Higgs production modes and decay channels, are shown in Table~\ref{CKcoeff_hh} for $ZH$ and $t\bar{t}H$ inclusive production modes together with the $\kappa$ modifiers at LO for the initial state $i$; the results of this thesis are presented exploiting the coupling modifiers $\kappa_t$, $\kappa_b$, $\kappa_{\ell}$, $\kappa_W$, $\kappa_Z$, describing the modifications of the SM Higgs-boson coupling to up-type quarks, to down-type quarks, to leptons and to $W$ and $Z$ vector bosons, respectively, in addition to the Higgs self-coupling. The values of $C^i_1$ and $K^i_{\textrm{EW}}$ are averaged over the full phase space of these processes.

\begin{table}[htbp]
\begin{center}
{\def\arraystretch{1.3}
\begin{tabular}{|c|c|c|}
\hline
Production mode & $ZH$ & $t\bar{t}H$ \\
\hline
$C_1^i\times 100$ & 1.19 & 3.52 \\
\hline
$K^i_{\textrm{EW}}$ & 0.947 & 1.014 \\
\hline
$\kappa_i^2$ & $\kappa_Z^2$ & $\kappa_t^2$ \\
\hline 
\end{tabular}
}
\end{center}
\caption{Values of the $C^i_1$ coefficients, representing linear $\kappa_\lambda$-dependent corrections to single-Higgs production modes (second row); values of the $K^i_\textrm{EW}$ coefficients~\cite{Degrassi, Maltoni}, taking into account NLO EW corrections in the SM hypothesis (third row); expressions of the initial state $\kappa$ modifiers at LO~\cite{Coupling_run2}, $\kappa_i^2$, for the Higgs boson production process included as background in double-Higgs analyses (fourth row).}
   \label{CKcoeff_hh}
\end{table} 
The coefficients for the decay channels, $C_{1}^f$,  and the expressions of the $\kappa$ modifiers at LO for the final state $f$ are reported in Table~\ref{tab:decays_hh} for all the analysed decay modes. 
\begin{table}[htbp]
\begin{center}
{\def\arraystretch{1.3}
\begin{tabular}{|c|ccc|}
\hline
Decay mode & $H\rightarrow\gamma\gamma$ & $H\rightarrow b\bar{b}$  & $H\rightarrow \tau\tau$ \\
\hline
$C^f_1\times 100$ & 0.49 & 0 & 0 \\
\hline
$\kappa_f^2$ & $1.59 \kappa_W^2+0.07\kappa_t^2 - 0.67 \kappa_W\kappa_t$ &$\kappa_b^2$& $\kappa_{\ell}^2$ \\
  \hline
\end{tabular}
}
\end{center}
\caption{Values of $C^f_1$~\cite{Degrassi, Maltoni} coefficients, representing linear $\kappa_\lambda$-dependent corrections to single-Higgs decay channels (second row); expressions of the final state $\kappa$ modifiers at LO~\cite{Coupling_run2}, $\kappa_f^2$, for each considered double-Higgs decay mode (third row).} 
\label{tab:decays_hh}
\end{table} 

\section{Statistical model}
\label{sec:statistical_model_hh}
The target of this chapter is to set constraints on the Higgs self-coupling and, possibly, on other single-Higgs couplings, looking first of all at the double-Higgs channels separately, and then proceeding with their combination in order to get a more stringent limit on the self-coupling. This target is pursued through the statistical tools described in Chapter~\ref{sec:stat} that are adapted to the double-Higgs analyses and categories described in the previous sections, exploiting the aforementioned parameterisations of the different observables to introduce the dependence on the parameters of interest, which are extracted using the profile-likelihood technique.\newline
The parameters of interest of the model, $\vec{\alpha}$, and the set of nuisance parameters, $\vec{\theta}$, including the systematic uncertainty contributions and background parameters that are constrained by side bands or control regions in data, are included in the global likelihood function, $L(\vec{\alpha},\vec{\theta})$, defined in Chapter~\ref{sec:stat}. For a combination of several channels and categories, the global likelihood function, $L(\vec{\alpha},\vec{\theta})$ is obtained as the product of the likelihoods of the input analyses, that are, in turn, products of likelihoods computed in the mutually orthogonal categories optimised in each analysis.
The number of signal events in each analysis category $j$ is defined as:
\begin{equation}
n^{\text{signal}}_j(\boldsymbol{\kappa}, \vec{\theta}) = \mathcal{L}_j(\vec{\theta}) \sum_i \sum_f \mu_{i}(\boldsymbol{\kappa}) \times \mu_{f}(\boldsymbol{\kappa}) (\sigma_{\text{SM},i}(\vec{\theta}) \times \text{BR}_{\text{SM},f}(\vec{\theta})) (\epsilon \times A)_{if,j}(\vec{\theta})
\label{eq:yields}
\end{equation} 
where the number of events is a function of the parameters of interest of the model, indicated by a generic $\boldsymbol{\kappa}$ standing for both the Higgs self-coupling and the single-Higgs couplings to other SM particles, and of the set of nuisance parameters $\vec{\theta}$, accounting for theoretical and experimental systematic uncertainties whose general features have been described in Chapter~\ref{sec:stat}, characterised in detail in the next section. The index $i$ runs over the double-Higgs production regions and the index $f$ includes all the considered decay channels, \ie\ $f= b\bar{b}b\bar{b}, b\bar{b}\tau^+\tau^-,  b\bar{b}\gamma\gamma$. $\mathcal{L}_j$ is the integrated luminosity of the dataset used in the $j$ category, and $(\epsilon \times A)_{if,j}$ represents the acceptance and efficiency estimation for the category $j$, the production process $i$ and the decay channel $f$. 
Finally, the term $\mu_i(\boldsymbol{\kappa})\times\mu_f(\boldsymbol{\kappa})$, where $\mu_i$ is defined in Equation~\ref{eq:mui} and $\mu_f$ in Equation \ref{eq:muf}, describes the dependence of the signal strengths for the initial and final state on the Higgs-boson self-coupling modifier $\kappa_\lambda$, and on the single-Higgs boson coupling modifiers; these modifiers represent potential deviations from the SM expectation of the self-coupling and of the other Higgs couplings, respectively. 
Confidence intervals for the parameters of interests are determined using, as test statistics, the profile-likelihood ratio, described in Chapter~\ref{sec:stat}:
\begin{equation}
q(\vec{\alpha})=-2 \ln \Lambda(\vec{\alpha})=-2 \ln \frac{L(\vec{\alpha},\hat{\hat{\vec{\theta}}}(\vec{\alpha}))}{L(\hat{\vec{\alpha}},\hat{\vec{\theta}})}
\end{equation}
where:
\begin{itemize}
\item in the numerator the nuisance parameters are set to their \textit{profiled} values $\hat{\hat{\vec{\theta}}}(\vec{\alpha})$, that maximise the likelihood for a given set of values of $\vec{\alpha}$;
\item in the denominator both the parameters of interest and the nuisance parameters are respectively set to the values $\hat{\vec{\alpha}}$ and $\hat{\vec{\theta}}$, that simultaneously maximise the likelihood $L(\vec{\alpha},\vec{\theta})$.
\end{itemize}
In the asymptotic limit, $-2\log{\Lambda(\vec{\alpha},\vec{\theta})}$ is approximately distributed as a $\chi^2$ statistic with $n$ degrees of freedom, where $n$ equals the number of parameters of interests in the model.\newline
The results presented in this thesis are based on the profile-likelihood evaluation, and 68\% as well as 95\% CL intervals are given in the asymptotic approximation~\cite{Cowan}.
\subsection{Systematic uncertainties}
\label{sec:statistical_model_hh_sys}

The systematic uncertainties included in the different analyses can be divided in two main categories: experimental uncertainties, related to object reconstruction and identifications algorithms or techniques, to data-taking conditions as well as detector response, to limited statistics in Montecarlo samples and data-driven background, and theoretical uncertainties, related to cross-section computations and to the modelling of signal and background processes.\newline
The ranking of the different uncertainties can be quantified looking at their impact on the final results. Considering as parameter of interest $\kappa_\lambda$ and setting all other single-Higgs couplings to their SM values, the uncertainties having the greatest impact can be identified.
The impact of each source of uncertainty is estimated by computing the maximum likelihood estimator of the parameter of interest, $\kappa_\lambda$, when the given uncertainty is fixed to its best-fit value $\pm 1\sigma$.
Thus the pre-fit impact represents the impact of the nuisance parameters on the parameter of interest as they enter in the global likelihood while the post-fit impact represents the impact of the nuisance parameters on the parameter of interest after they have been adjusted to better describe data.\newline

\subsubsection{Experimental uncertainties}
\textbf{Electron, photon, muon and tau uncertainties}\newline
Uncertainties related to electrons, muons and taus are considered in the $b\bar{b}\tau^+\tau^-$ channel, being this channel the only channel using $\tau$ objects; they include uncertainties on electron and muon trigger, identification and reconstruction efficiencies. Tau uncertainties are included to take into account the corrections in the Montecarlo samples to the energy scale, the tau-reconstruction and identification efficiency, as well as the corrections due to the tau-electron overlap removal, or trigger and isolation requirements. Furthermore, uncertainties related to electron and photon energy calibration and momentum scale are included. The uncertainties having the greatest impact on the results are reported in Table~\ref{sys1}.
\begin{table}[htbp]
\begin{center}
{\def\arraystretch{1.3}
\begin{tabular}{|c|c|}
\hline
NP name & Description \\
\hline
EG\_RESOLUTION\_ALL &  Electron and photon energy resolution uncertainty \\
\hline
TAU\_EFF\_ID\_TOTAL &  Tau identification efficiency uncertainty\\ \hline
EG\_SCALE\_ALLCORR &  Electron and photon energy scale uncertainty \\ \hline
TAU\_EFF\_RECO\_TOTAL & Tau reconstruction efficiency uncertainty \\ \hline
\end{tabular}
}
\end{center}
\caption{Electron, photon and tau uncertainties having the greatest impact on the results.}
   \label{sys1}
\end{table}

\textbf{Missing energy uncertainties}\newline
Uncertainties on the $E_T^{miss}$ are included only in the $b\bar{b}\tau^+\tau^-$ channel, using in the selections variables related to the $E_T^{miss}$. Uncertainties on the energy scale and resolution of the objects used to calculate the $E_T^{miss}$, such as electrons, muons, jets and taus are propagated to the calculation of the $E_T^{miss}$. Additional uncertainties on the scale, resolution, and reconstruction efficiency of tracks not associated to the reconstructed objects, are also included.\newline

\textbf{Jets and flavour tagging uncertainties}\newline
As it was explained in Chapter~\ref{sec:Reco}, jets, after being reconstructed, have to be calibrated to take into account several effects, like energy-scale corrections, energy-resolution differences between simulation and data, pile-up effects. These corrections are included as uncertainties in the three decay channels, being all characterised by the presence of jets. 
The jet uncertainties having the greatest impact on the results, reported in Table~\ref{sys2}, are those related to the energy resolution and energy scale.
Furthermore, flavour-tagging uncertainties are included in all channels, coming from correction factors that take into account flavour-tagging-efficiency differences between simulation and data; these factors are measured separately for $b$, $c$ and light-flavour jets and are decomposed into uncorrelated components, resulting in four uncertainties for $b$-jets, three uncertainties for $c$-jets and five uncertainties for light-flavour jets for all channels except for the $b\bar{b}\gamma \gamma$ channel, that has merged the flavour tagging NPs into one NP per flavour.\newline
The flavour-tagging uncertainties having the greatest impact on the results, reported in Table~\ref{sys2}, are those related to the flavour-tagging efficiency for $b$ and $c$-flavour jets coming from the $b\bar{b}\gamma \gamma$ channel.
\begin{table}[htbp]
\scalebox{0.92}{
{\def\arraystretch{1.5}
\begin{tabular}{|c|c|}
\hline
NP name & Description \\
\hline
JET\_GroupedNP\_3 & Jet energy scale uncertainty split in different components\\
\hline
JES\_EtaInter\_NonClosure & Non closure uncertainty of the $\eta$-intercalibration method\\ \hline
JES\_bbyy &  Merged jet energy scale uncertainty -  $b\bar{b}\gamma \gamma$ \\ \hline
JET\_Grouped NP\_2 & Jet energy scale uncertainty split in different components \\ \hline
JET\_Grouped NP\_1 & Jet energy scale uncertainty split in different components\\ \hline
FT\_EFF\_Eigen\_C\_WP70\_bbyy & Jet $c$-tagging uncertainty - 70\% working point - $b\bar{b}\gamma \gamma$\\ \hline
JER\_SINGLE\_NP & Jet energy resolution uncertainty\\ \hline
FT\_EFF\_Eigen\_B\_WP70\_bbyy & Jet $b$-tagging uncertainty - 70\% working point - $b\bar{b}\gamma \gamma$\\ \hline
\end{tabular}
}}
\caption{Jets and flavour uncertainties having the greatest impact on the results.}
   \label{sys2}
\end{table} 

\textbf{Luminosity and pile-up uncertainties}\newline
The uncertainty on the integrated luminosity that has been recorded by the ATLAS experiment in 2015--2016 is 2.1\%, derived using a methodology reported in Reference~\cite{lumi_param}, thus through a calibration of the luminosity scale using $x-y$ beam-separation scans.\newline
The nuisance parameters associated to the luminosity uncertainty are breakdown separately for the $b\bar{b}b\bar{b}$ channel considering 2015 and 2016 runs, given an inefficiency in the vertex reconstruction, and thereby $b$-tagging, at the trigger level during the 2016 data-taking, that led to an integrated luminosity not corresponding to the usual full dataset.\newline
Furthermore, an uncertainty related to the pile-up reweighting procedure, used in order to correctly reproduce the distribution of the number of $pp$ collisions per bunch crossing in data, is included.\newline

\textbf{Experimental uncertainties coming from data-driven backgrounds} \newline
Background estimation is carried out using data-driven methods in double-Higgs channels, with the dominant background represented by multijet events and $t\bar{t}$ events. Uncertainties arising from the fitting procedure, from the data samples used, from the correction factors applied, have to be included. These uncertainties represent the dominant contribution to the total experimental and theoretical uncertainties; the background uncertainties having the greatest impact on the results are reported in Table~\ref{sys3} and are those related to the $b\bar{b}b\bar{b}$ and $b\bar{b}\gamma \gamma$ channels.
\begin{table}[htbp]
\begin{center}
\scalebox{0.9}{
{\def\arraystretch{1.4}
\begin{tabular}{|c|l|}
\hline
NP name & Description \\
\hline
\multirow{2}{*}{bias\_2tag\_bbyy} & Uncertainty due to the background modelling (spurious signal)\\
&  for each $b\bar{b} \gamma \gamma$ category\\ 
\hline
\multirow{2}{*}{r16\_LowHtCR\_bbbb}  &  Background shape variation determined by the non-closure \\
& between Sideband and Control region derived models (2016) - $b\bar{b}b\bar{b}$\\ \hline
\multirow{2}{*}{r16\_HighHtCR\_bbbb} &   Background shape variation determined by the non-closure \\
& between Sideband and Control region derived models (2016) - $b\bar{b}b\bar{b}$\\ \hline
\multirow{2}{*}{r16\_norm\_NP2\_bbbb}  & Background fit uncertainty corresponding to the uncertainty on the\\
& non all-hadronic $t\bar{t}$ normalisations (2016) - $b\bar{b}b\bar{b}$\\ \hline
\multirow{2}{*}{bias\_1tag\_bbyy} &  Uncertainty due to the background modelling (spurious signal)\\
& for each $b\bar{b} \gamma \gamma$ category\\ \hline
\multirow{2}{*}{r15\_HighHtCR\_bbbb} &  Background shape variation determined by the non-closure \\
& between Sideband and Control region derived models (2015) - $b\bar{b}b\bar{b}$\\ \hline
\multirow{2}{*}{r15\_LowHtCR\_bbbb}  &  Background shape variation determined by the non-closure\\
& between Sideband and Control region derived models (2015) - $b\bar{b}b\bar{b}$\\ \hline
\multirow{2}{*}{r15\_norm\_NP2\_bbbb} & Background fit uncertainty corresponding to the uncertainty on the \\
& non all-hadronic $t\bar{t}$ normalisations (2015) - $b\bar{b}b\bar{b}$\\ \hline
\multirow{2}{*}{Sys1tag2tagTF\_bbtautau}  & \multirow{2}{*}{Multi-jet uncertainty from data-driven estimation - $b\bar{b}\tau\tau$} \\ 
& \\
\hline
\end{tabular}
}}
\end{center}
\caption{Experimental uncertainties coming from data-driven backgrounds having the greatest impact on the results.}
   \label{sys3}
\end{table} 
\subsubsection{Theoretical uncertainties}
The theoretical uncertainties come from the modelling of signal and background processes; they include uncertainties on the QCD scales, \ie\ renormalisation and factorisation scales, on the parton density function (PDF) used, on the modelling of the underlying events (UE) and parton shower (PS), on the running of the QCD coupling constant $\alpha_S$, on the single-Higgs processes and on the theoretical-cross-section prediction.
The theoretical uncertainties having the greatest impact on the results are reported in Table~\ref{sys4}, and are the uncertainties coming from the modelling of the dominant background both in the shape and the acceptance normalisation for the $b\bar{b}\tau^+\tau^-$ channel, the ones coming from the QCD scale for all channels, from parton shower for the $b\bar{b}b\bar{b}$ channel and from \ggF cross-section predictions.
\begin{table}[htbp]
\begin{center}
\scalebox{0.9}{
{\def\arraystretch{1.4}
\begin{tabular}{|l|l|}
\hline
NP name & Description \\
\hline
\multirow{2}{*}{BkgTheory\_SysZtautauMBB\_bbtautau} & Z+heavy flavour shape modelling \\
&  uncertainty - $b\bar{b}\tau\tau$ \\
\hline
\multirow{2}{*}{BkgTheory\_SysTTbarMBB\_bbtautau} & \multirow{2}{*}{$t\bar{t}$ shape modelling uncertainty - $b\bar{b}\tau\tau$} \\ 
& \\ \hline
\multirow{3}{*}{BkgTheory\_SysRatioHHSRZhfAcc2Tag\_bbtautau} &   Z+heavy flavour relative acceptance \\
& normalisation between control region  \\
& and signal region - $b\bar{b}\tau\tau$\\ \hline
\multirow{3}{*}{BkgTheory\_SysRatioHHSRTtbarAcc2Tag\_bbtautau} & $t\bar{t}$ relative acceptance \\
& normalisation between control region \\
& and signal region - $b\bar{b}\tau\tau$ \\ \hline
\multirow{3}{*}{BkgTheory\_SysRatioLHSRZhfAcc2Tag\_bbtautau}&  Z+heavy flavour relative acceptance \\
& normalisation between control region \\
& and signal region - $b\bar{b}\tau\tau$ \\ \hline
\multirow{2}{*}{TheorySig\_SIG\_PS\_bbbb}& Uncertainty due to modelling of the \\
& parton shower - $b\bar{b}b\bar{b}$ \\  \hline
\multirow{2}{*}{TheorySig\_QCDscale\_hh}  &  \multirow{2}{*}{QCD scale uncertainty - diHiggs}  \\ 
& \\ \hline
\multirow{2}{*}{TheorySig\_QCDscale\_ttH} & \multirow{2}{*}{QCD scale uncertainty - $t\bar{t}H$} \\
& \\ \hline
\multirow{2}{*}{TheorySig\_HF\_Higgs\_ggF} & Uncertainty associated to the heavy\\
& flavour content in \ggF computation\\  \hline
\end{tabular}
}}
\end{center}
\caption{Theoretical uncertainties having the greatest impact on the results.}
   \label{sys4}
\end{table} 
\subsubsection{Correlation between systematic uncertainties}
The correlation scheme adopted in the double-Higgs combination aiming at the extraction of $\kappa_\lambda$ follows these guidelines:
\begin{itemize}
\item detector systematics uncertainties, like those related to jet reconstruction, $b$-jet tagging, muon and photon reconstruction and identification, are correlated between the different decay channels;
\item uncertainties on the integrated luminosity are correlated among the different decay channels, even if, for the $b\bar{b}b\bar{b}$ decay channel, the nuisance parameters are not correlated with the luminosity NPs in the other channels, having this channel separate NPs for 2015 and 2016;
\item uncertainties on the signal acceptance are correlated among the different decay channels;
\item theoretical and modelling systematic uncertainties of the backgrounds are kept uncorrelated because of a negligible overlap between decay channels.
\end{itemize}
For the $HH\rightarrow b\bar{b}b\bar{b}$ channel, two different correlation schemes regarding flavour tagging (FT), jets (JET), parton shower (PS) and trigger uncertainties, have been considered and are described in details in Appendix~\ref{sec:appendix_correlation_hh}, Table~\ref{tab:correlation4b}:
\begin{itemize}
\item all NPs related to the signal samples S(1,0), S(1,1) and S(1,20) uncorrelated, \ie\ FT, JET, PS, trigger NPs kept split in the three signal samples (scheme 1);
\item all NPs related to the signal samples S(1,0), S(1,1) and S(1,20), \ie\ FT, JET, PS, trigger NPs, correlated to be consistent with the double-Higgs combination in Reference~\cite{Paper_hh} (scheme 2); the list of the correlated NPs is reported in Appendix~\ref{sec:appendix_correlation_hh}.
\end{itemize}

\begin{figure}[hbtp]
\centering
\begin{subfigure}[b]{0.49\textwidth}
\includegraphics[height=13 cm,width =\textwidth]{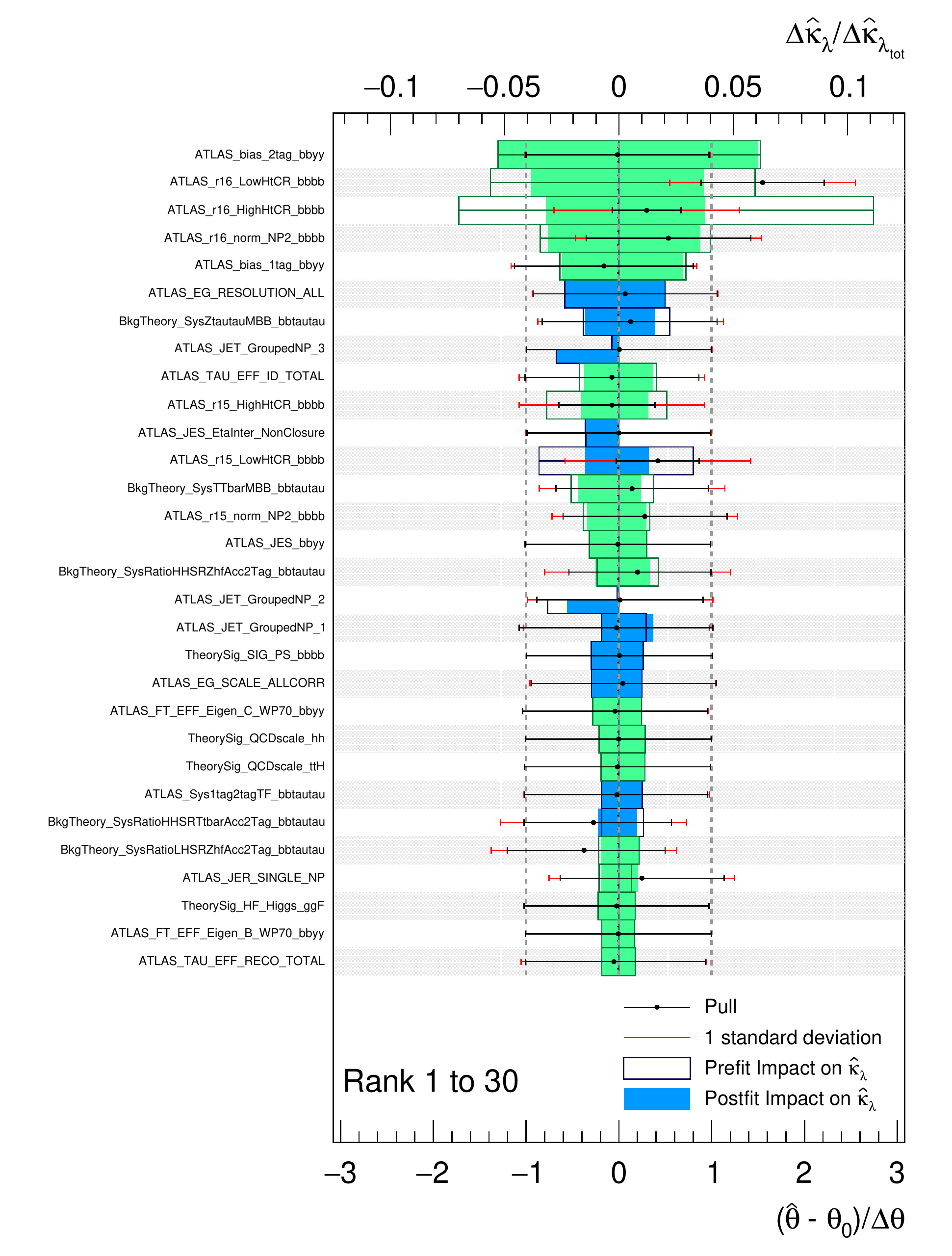}
 \caption{}
\end{subfigure}
\begin{subfigure}[b]{0.49\textwidth}
\includegraphics[height=13 cm,width =\textwidth]{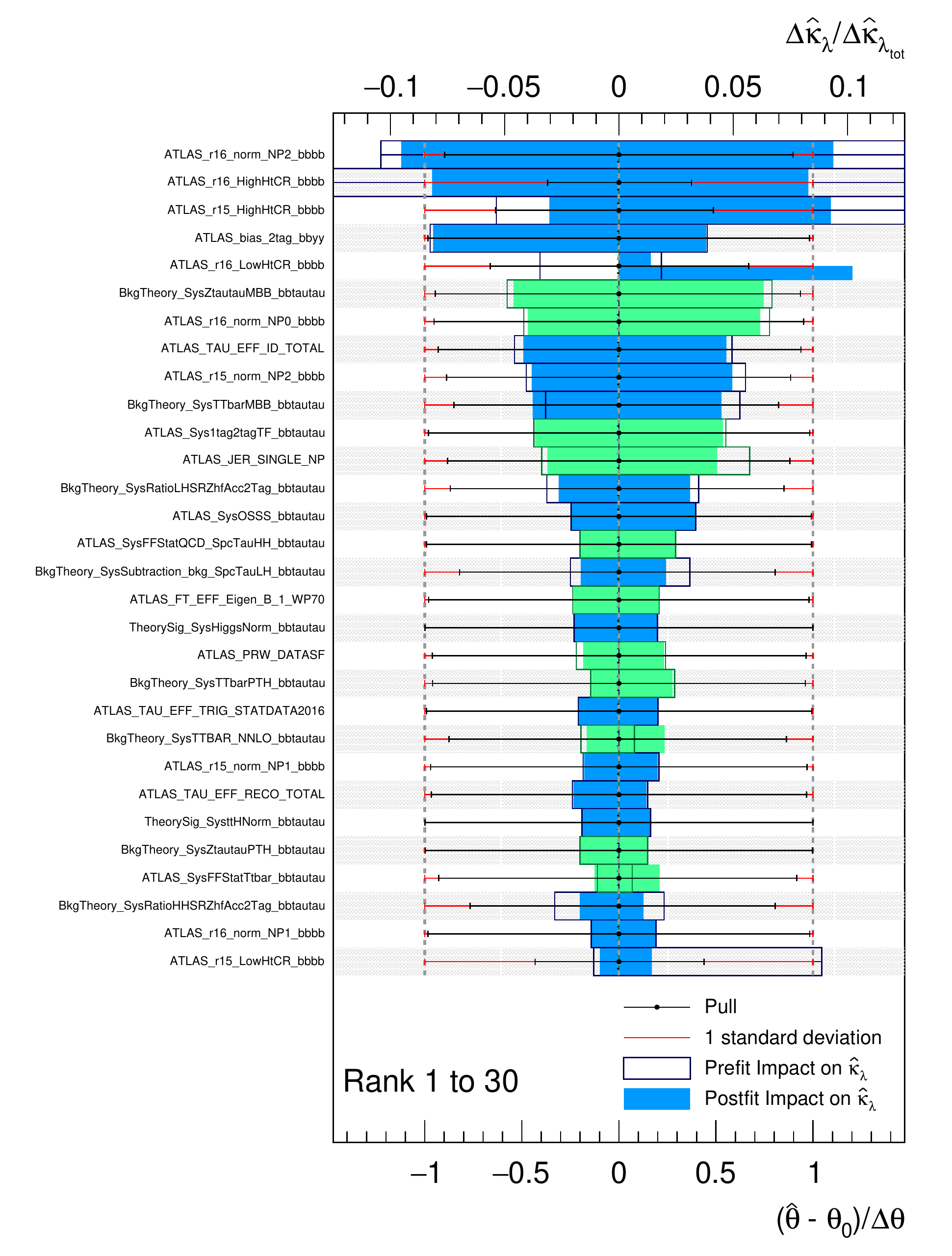}
 \caption{}
\end{subfigure}
\caption{Ranking of the top 30 systematic uncertainties in the double-Higgs combination for data (a) and for the Asimov dataset (b) generated under the SM hypothesis.}     
\label{ranking_hh}
\end{figure}

The list of the nuisance parameters ranked by their post-fit impact on the parameter of interest from the greatest (top) to the least (bottom) dominant systematic uncertainty is shown in the so called ``ranking plot$"$; the same plot shows also the nuisance parameter pulls, \ie\ the difference between the best-fit values of the nuisance parameters and the initial one ($\theta_0$), normalised to their pre-fit uncertainties; $\theta_0=0$ by construction of the likelihood function, so the compatibility of the pull with zero is a check of the robustness of the fit procedure. \newline
Figure~\ref{ranking_hh} shows the ranking plots for the double-Higgs combination considering the top 30 uncertainties for data $(a)$ and for the Asimov dataset $(b)$ generated in the SM hypothesis ($\kappa_\lambda$=1). The difference between the maximum likelihood estimator with or without varying the nuisance parameter is the $\Delta\hat{\kappa}_\lambda$ of the fit, that is normalised to the total error, $\Delta\hat{\kappa}_{\lambda_{tot}}$. Pre-fit and post-fit impacts of the different nuisance parameters on the central value $\kappa_\lambda$ are reported as white empty and cyan (green) filled bars corresponding to downward (upwards) systematic uncertainty variations, respectively. The points indicate how the parameter had to be pulled up or down to adjust data/MC agreement in the fit, and associated error bars show the best-fit values of the nuisance parameters and their post-fit uncertainties.  Most of the systematic uncertainties are within $1\sigma$ from the nominal (indicated by the dashed vertical lines) value, except for an experimental nuisance parameter, ``ATLAS\_r16\_LowHtCR\_bbbb$"$ related to the data-driven background modelling (mainly multijets) of  $HH\rightarrow b\bar{b}b\bar{b}$, that is also one of the nuisance parameter having the largest impact on $\kappa_\lambda$, together with other nuisance parameters all being related to the data-driven background modelling (mainly multijets) of  $HH\rightarrow b\bar{b}b\bar{b}$ and $HH\rightarrow b\bar{b}\gamma \gamma$ analyses.

\section{Validation of double-Higgs results}
\label{sec:validation_hh}
\begin{figure}[H]
\centering
\includegraphics[height=7 cm, width=11.5 cm]{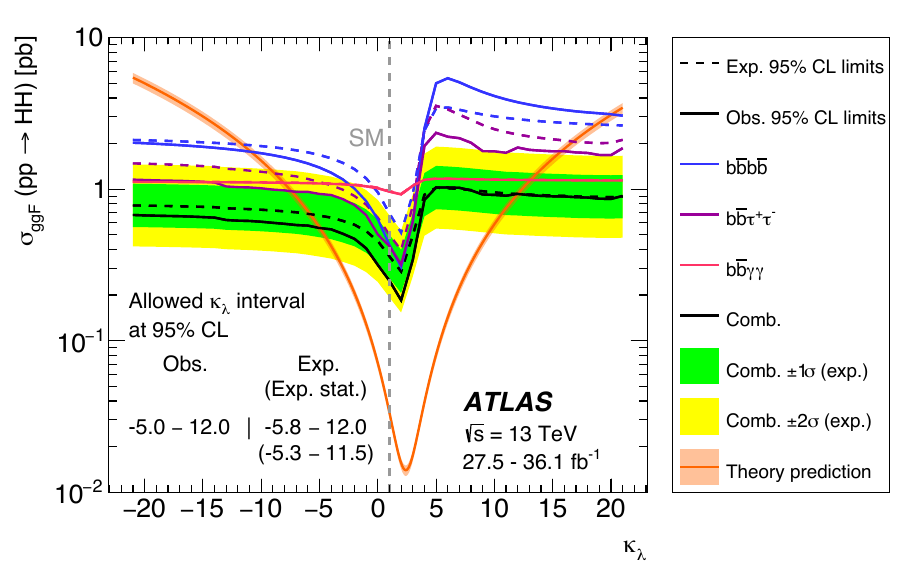}
\caption{Upper limits at 95\% CL on the cross section of the ggF non-resonant SM HH production as a function of $\kappa_\lambda$. The observed (expected) limits are shown as solid (dashed) lines. The $\pm 1\sigma$ and $\pm 2 \sigma$ bands are only shown for the combined expected limit. The orange solid line represents the theoretical prediction of the double-Higgs ggF cross section as a function of $\kappa_\lambda$~\cite{Paper_hh}.}     
\label{published_hh}
\end{figure}

\begin{figure}[htbp]
\begin{center}
\includegraphics[width=0.48\textwidth]{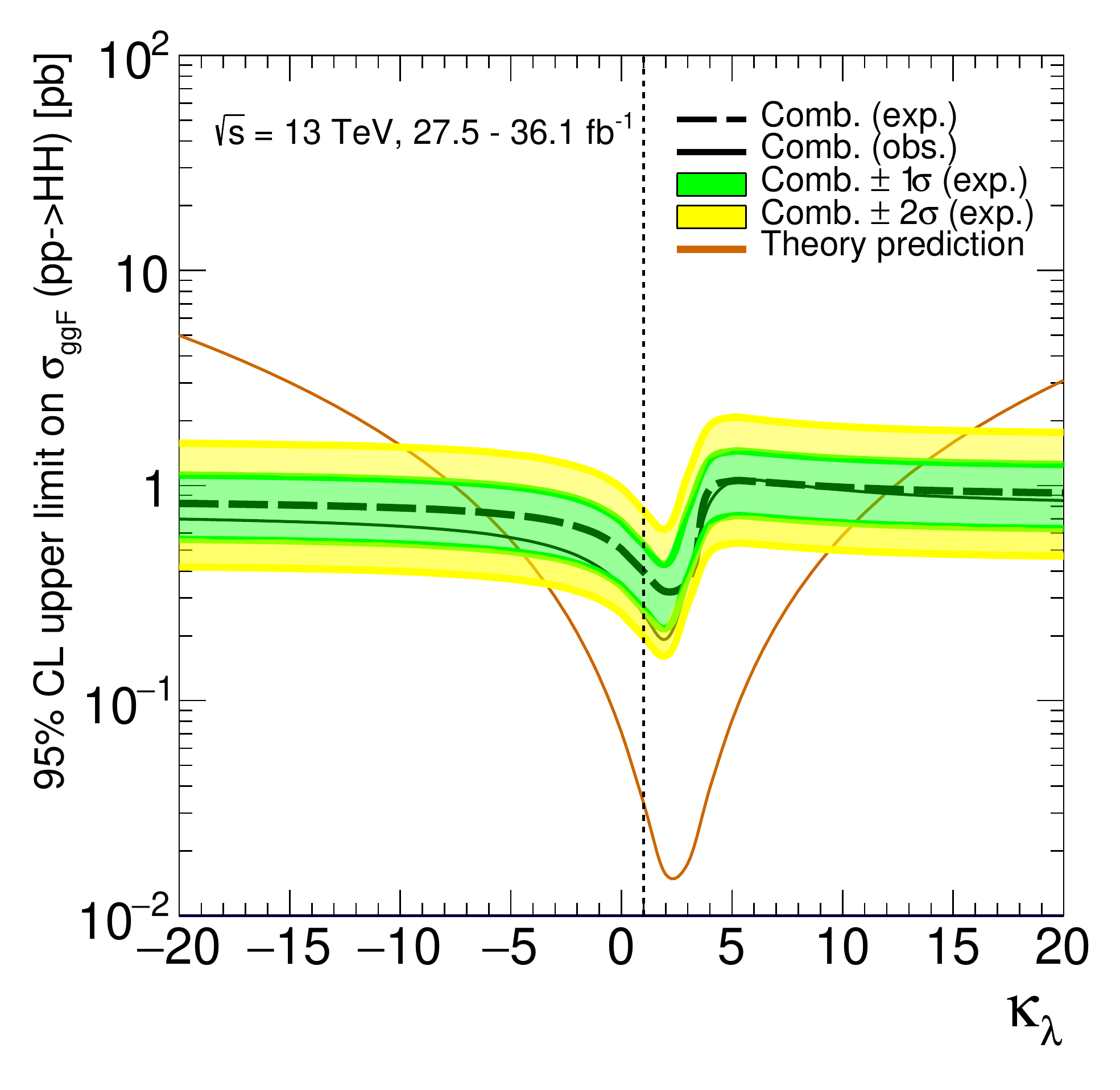}
\end{center} 
\caption{Upper limits at 95\% CL on the cross-section of the ggF non-resonant SM HH production as a function of $\kappa_\lambda$ obtained in order to validate the input of the double-Higgs analyses presented in this chapter. The observed (expected) limits are shown as solid (dashed) lines. The $\pm 1\sigma$ (green) and $\pm 2 \sigma$ (yellow) bands are only shown for the combined expected limit. The orange solid line represents the theoretical prediction of the HH ggF cross section as a function of $\kappa_\lambda$ used in order to extract $\kappa_\lambda$ limits.}  
\label{mu_limit1}
\end{figure}

The most recent constraints on the Higgs self-coupling from the ATLAS experiment, not considering the results of this thesis, come from the combination of the double-Higgs most sensitive final states that have been described in the previous sections, and are reported in Figure~\ref{published_hh}. In each analysis, the 95\% CL upper limits on the $\sigma_{ggF}(pp\rightarrow HH)$ cross-section were computed for different $\kappa_\lambda$ values; the intersection of the theoretical $\sigma_{ggF}(pp\rightarrow HH)$ as a function of $\kappa_\lambda$ with the measured cross section was used to indirectly extract the confidence intervals at 95\% for $\kappa_{\lambda}$.  Uncertainties on the theoretical $\sigma_{ggF}(pp\rightarrow$$ HH)$ cross section as well as the dependence of the Higgs-boson branching fractions and of the single-Higgs background on $\kappa_\lambda$, affecting both the double-Higgs signal and the single-Higgs-boson background, were neglected. This method will be referred to as ``$\kappa_\lambda$-scan method$"$.
The main differences between the method used in this chapter and the $\kappa_\lambda$-scan method through which double-Higgs direct limits have been produced, are the profile-likelihood technique used, where the limits and best-fit values are extracted after building a likelihood function as described in Section~\ref{sec:statistical_model_hh}, and the fact that the single-Higgs background and branching fractions are parameterised as a function of $\kappa_\lambda$. 
For the purpose of validating the approach followed in this combination, the published results have been reproduced exploiting the double-Higgs combined inputs used to produce the results of this chapter.
In order to be consistent with the aforementioned results, all couplings except for $\kappa_\lambda$, have been set to their SM values and both branching fractions and single-Higgs background have not been parameterised as a function of $\kappa_\lambda$. Furthermore, theory uncertainties have not been injected.
Thus the combined workspace has been used to measure the double-Higgs cross section as a function of $\kappa_{\lambda}$ and the 95\% CL of $\kappa_{\lambda}$ has been estimated to be $-5.1<\kappa_\lambda<11.9$ (observed) and $-5.9<\kappa_\lambda<12.0$ (expected). These results are comparable to the results reported in Reference~\cite{Paper_hh}, \text{i.e.}  $-5.0<\kappa_\lambda<12.0$ (observed) and  $-5.8<\kappa_\lambda<12.0$ (expected). Figure~\ref{mu_limit1} shows the upper limits on the double-Higgs cross section as a function of $\kappa_{\lambda}$: the solid black curve is the observed limit, while the dashed curve is the expected one. The green and yellow bands show the 1$\sigma$ and  2$\sigma$ intervals of the expected limit, respectively. The theoretical $\sigma_{ggF}(pp\rightarrow HH)$, used to find the limits on $\kappa_{\lambda}$, is represented by the orange curve.
Small differences with respect to the results reported in Reference~\cite{Paper_hh} mainly come from the $b\bar{b}\tau^+\tau^-$ channel, that uses a varying binning of the BDT distribution optimised for different $\kappa_\lambda$ values; the validation study, instead, uses a fixed binning (from $\kappa_\lambda=1$) differently from the original inputs.

\section{Results of fits to $\kappa_\lambda$ from individual channels}
\label{sec:results_hh_channel}
\begin{figure}[hbtp]
\centering
\begin{subfigure}[b]{0.48\textwidth}
\includegraphics[width =\textwidth]{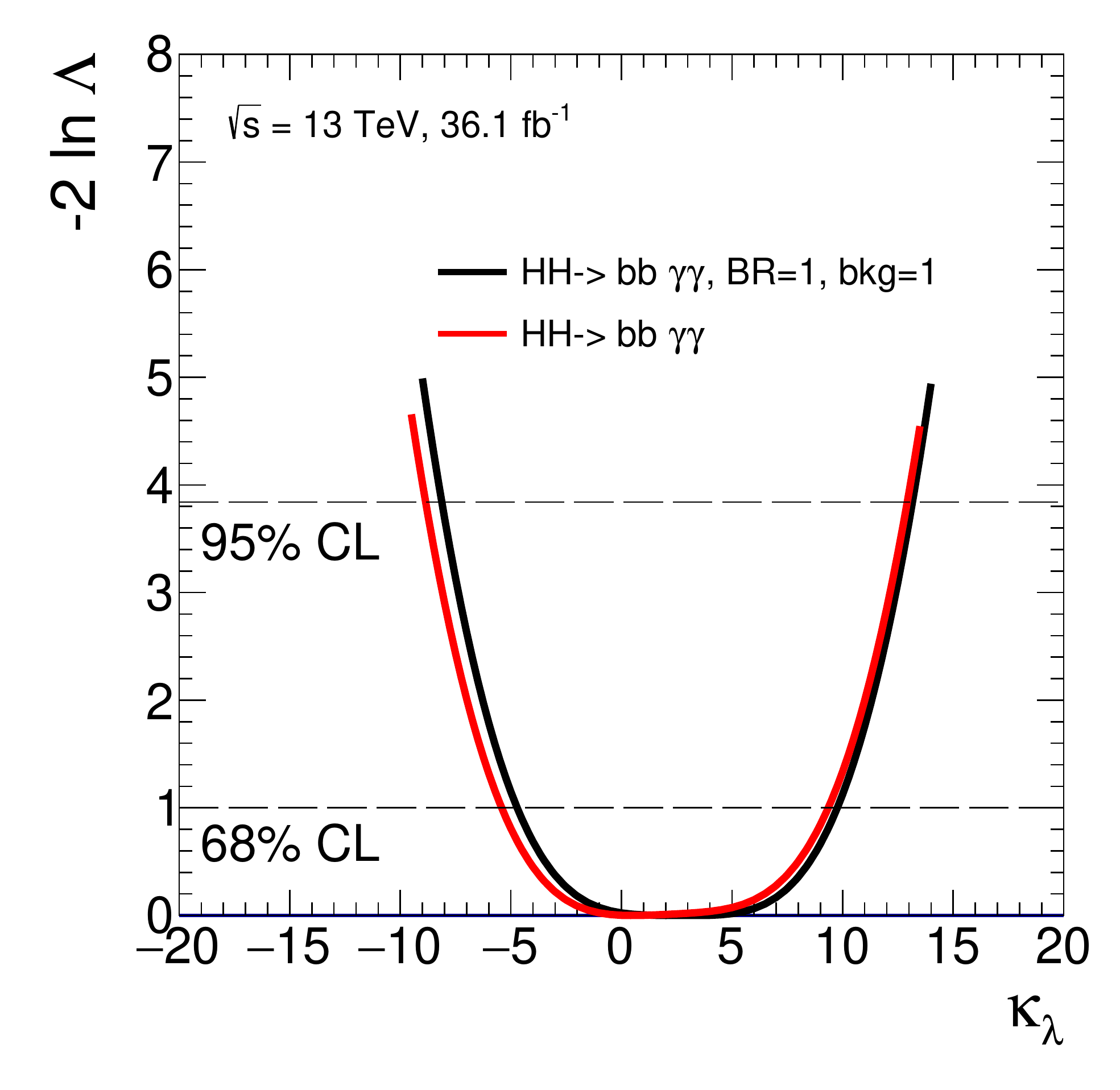}
 \caption{}
\end{subfigure}
\begin{subfigure}[b]{0.48\textwidth}
\includegraphics[width =\textwidth]{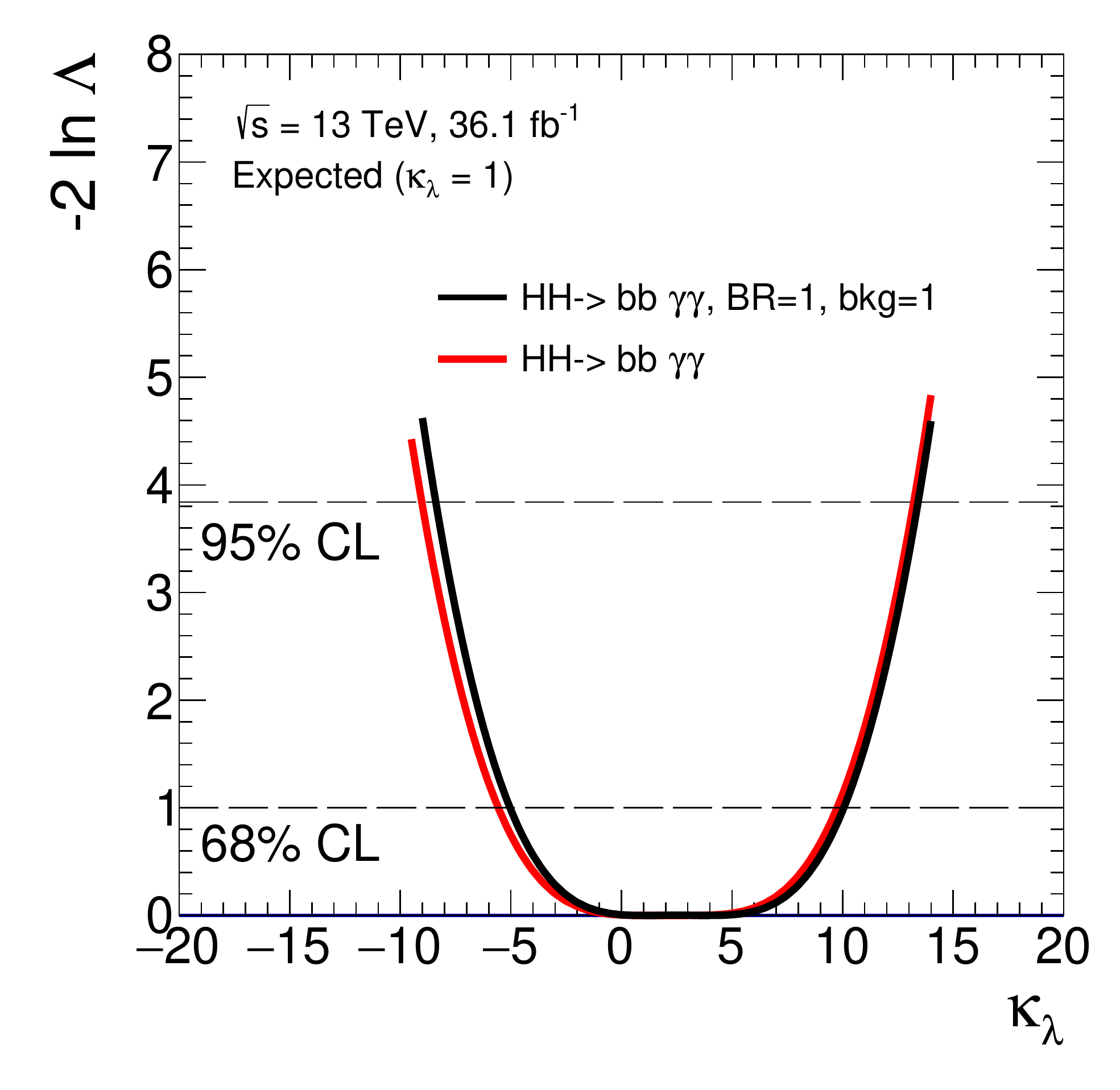}
 \caption{}
\end{subfigure}
\caption{Value of $-2 \ln{\Lambda(\kappa_\lambda)}$ as a function of $\kappa_\lambda$ for the $HH\rightarrow b\bar{b} \gamma \gamma$ channel considering two configurations. Black solid line: neither branching fractions nor single-Higgs background parameterised as a function of $\kappa_\lambda$; red solid line: branching fractions and single-Higgs background parameterised as a function of $\kappa_\lambda$. Likelihood distributions are reported for data (a) and for the Asimov dataset generated in the SM hypothesis (b). The dotted horizontal lines show the $-2 \ln{\Lambda(\kappa_\lambda)}=1$ level that is used to define the $\pm 1\sigma$ uncertainty on $\kappa_\lambda$ as well as the $-2 \ln{\Lambda(\kappa_\lambda)}=3.84$ level used to define the 95\% CL.}     
\label{scan_hh_bbyy}
\end{figure}

In this section, the main results of the double-Higgs analyses are presented, where a global likelihood function is built as described in Section~\ref{sec:statistical_model_hh} as the product of the likelihoods of each double-Higgs category, implementing, as the parameter of interest, the Higgs self-coupling, and as nuisance parameters the theoretical and experimental uncertainties described in Section~\ref{sec:statistical_model_hh_sys}. The profile-likelihood technique is used to constrain the value of the Higgs-boson self-coupling $\kappa_\lambda$ while leaving untouched other Higgs-boson couplings, taking as best-fit values of the unknown parameter of interest the values that maximise the likelihood function.\newline 
In order to check the impact of the Higgs-boson branching fractions and cross sections on $\kappa_\lambda$, results are presented either including or not including the single-Higgs-background and branching-fraction parameterisations as a function of $\kappa_\lambda$. The standard configuration includes these parameterisations.\newline 
Specifically, the $b\bar{b}b\bar{b}$ decay channel is only affected by branching-fraction variations, while the $b\bar{b}\tau^+\tau^-$ and the $b\bar{b}\gamma\gamma$ decay channels are affected by both branching-fraction and cross-section variations for dominant single-Higgs background ($ZH$ and $t\bar{t}H$).\newline
Results are always presented for data and for the Asimov dataset, a dataset in which all observed quantities are set equal to their expected values, that is practically generated from the likelihood distribution $\Lambda$ with nuisance parameters fixed to the best-fit values obtained on data and the parameter of interest fixed to the expected value, corresponding to the SM hypothesis (\ie\ $\kappa_\lambda = 1$).\newline
The $\kappa_\lambda$ self-coupling modifier is probed in the range $-20 < \kappa_\lambda < 20$, because outside this range the calculation in References~\cite{Degrassi,Maltoni} loses its validity.\newline
Constraints from individual double-Higgs channels are reported in the following. \newline
Starting from the $b\bar{b}\gamma \gamma$ channels, the value of $-2 \ln{\Lambda(\kappa_\lambda)}$ as a function of $\kappa_\lambda$ is shown in Figure~\ref{scan_hh_bbyy} for data and for the Asimov dataset, either including or not including the branching-fraction and single-Higgs parameterisations. 

\begin{figure}[hbtp]
\centering
\begin{subfigure}[b]{0.48\textwidth}
\includegraphics[width =\textwidth]{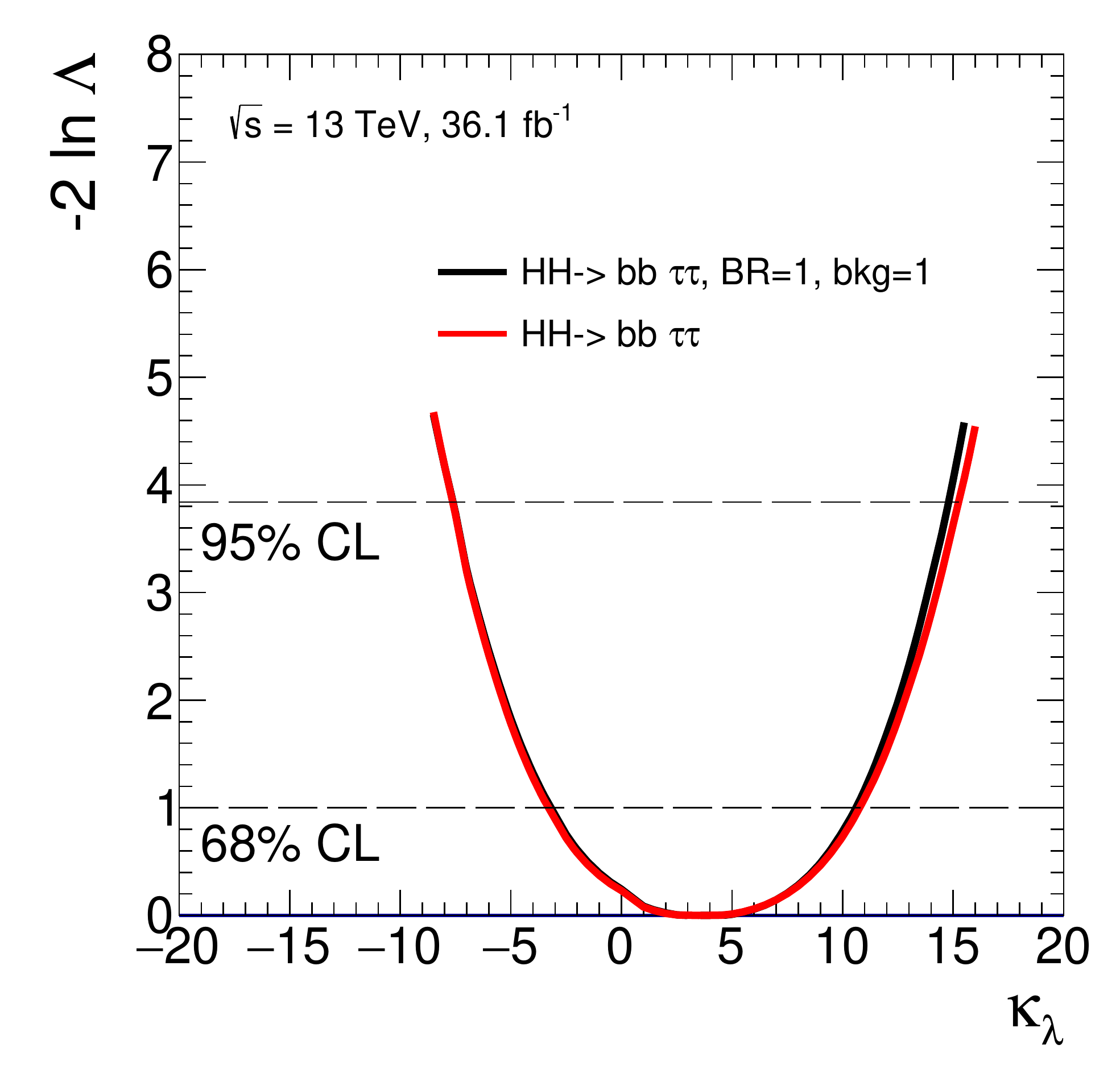}
 \caption{}
\end{subfigure}
\begin{subfigure}[b]{0.48\textwidth}
\includegraphics[width =\textwidth]{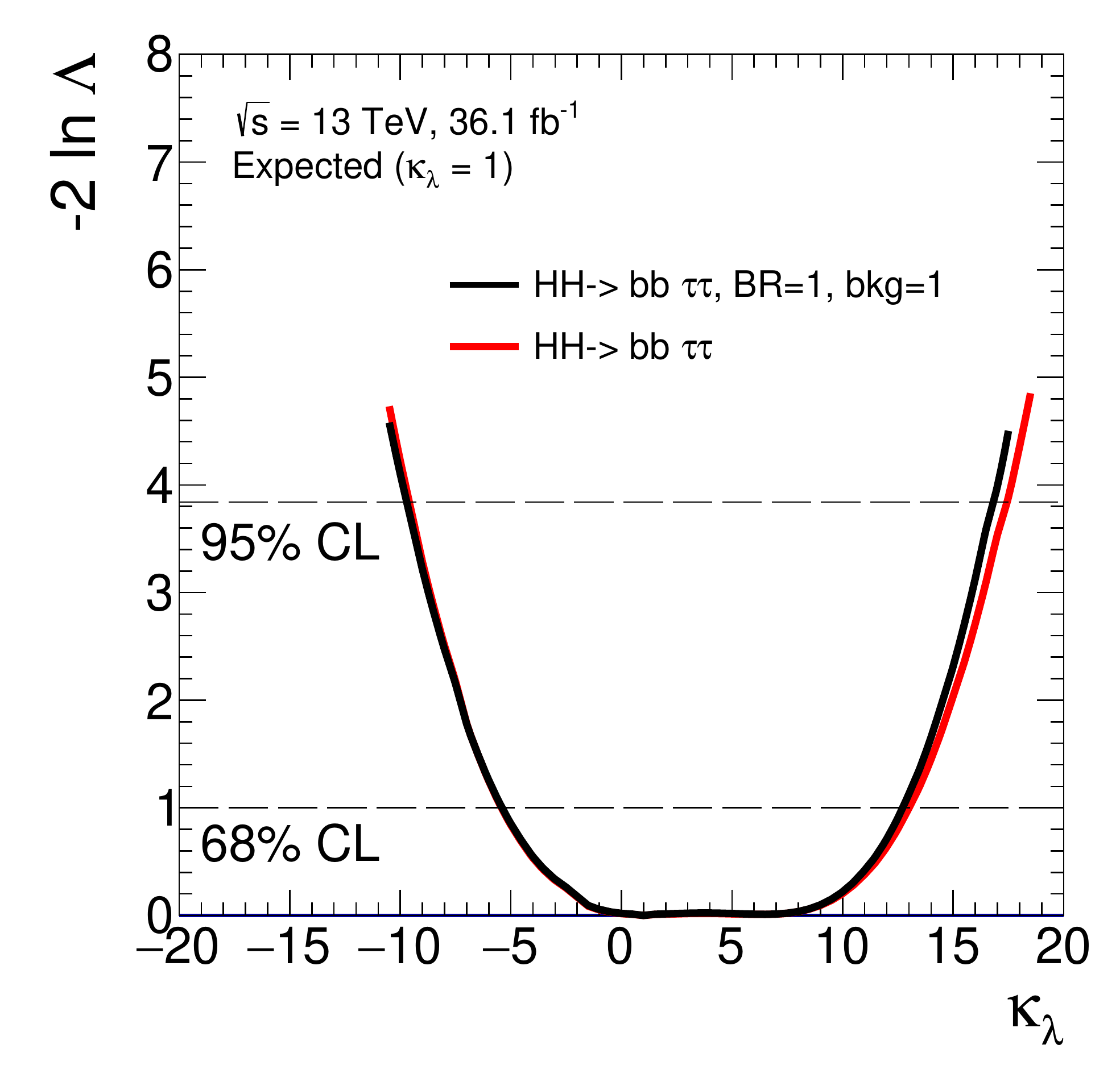}
 \caption{}
\end{subfigure}
\caption{Value of $-2 \ln{\Lambda(\kappa_\lambda)}$ as a function of $\kappa_\lambda$ for the $HH\rightarrow b\bar{b} \tau^+\tau^-$ channel considering two configurations. Black solid line: neither branching fractions nor single-Higgs background parameterised as a function of $\kappa_\lambda$; red solid line: branching fractions and single-Higgs background parameterised as a function of $\kappa_\lambda$. Likelihood distributions are reported for data (a) and for the Asimov dataset generated in the SM hypothesis (b). The dotted horizontal lines show the $-2 \ln{\Lambda(\kappa_\lambda)}=1$ level that is used to define the $\pm 1\sigma$ uncertainty on $\kappa_\lambda$ as well as the $-2 \ln{\Lambda(\kappa_\lambda)}=3.84$ level used to define the 95\% CL.}     
\label{scan_hh_bbtautau}
\end{figure}

The $\kappa_\lambda$ 95\% CL intervals from the $HH\rightarrow b\bar{b} \gamma \gamma$ channel are  $-8.9<\kappa_\lambda<12.9$ (observed) and  $-9.0<\kappa_\lambda<13.2$ (expected). 
The impact of the Higgs-boson branching fractions and cross sections on the allowed $\kappa_\lambda$ lower limit is $\sim$9\% while it is $\sim$2\% on the upper limit; as it is shown in the following lines, this channel is affected by the largest $\kappa_\lambda$-dependent corrections, given the fact that the single-Higgs contribution in the $b\bar{b}\gamma\gamma$ channel is larger than the contributions to other channels, and that the $C_1$ coefficients are different from 0 for the $\gamma \gamma$ channel.\newline
Even if a small deficit in data is present, as shown in Figure~\ref{ggF_hh} reporting the upper limits at 95\% CL on the double-Higgs ggF cross section, that would lead to more stringent observed limits with respect to the expected ones, the weaker dependence of the $b\bar{b}\gamma \gamma$ acceptance on $\kappa_\lambda$ with respect to the other channels, shown in Figure~\ref{acceptance}, leads to the fact that the observed and expected 95\% CL intervals are comparable.\newline
The best-final state for the non-resonant double-Higgs production is the $b\bar{b}\tau^+\tau^-$ channel. The value of $-2 \ln{\Lambda(\kappa_\lambda)}$ as a function of $\kappa_\lambda$ is shown in Figure~\ref{scan_hh_bbtautau} for data and for the Asimov dataset, generated in the SM hypothesis (\ie\ $\kappa_\lambda = 1$), either including or not including the branching-fraction and single-Higgs parameterisations.

\begin{figure}[hbtp]
\centering
\begin{subfigure}[b]{0.47\textwidth}
\includegraphics[height=7.5 cm,width =8.2 cm]{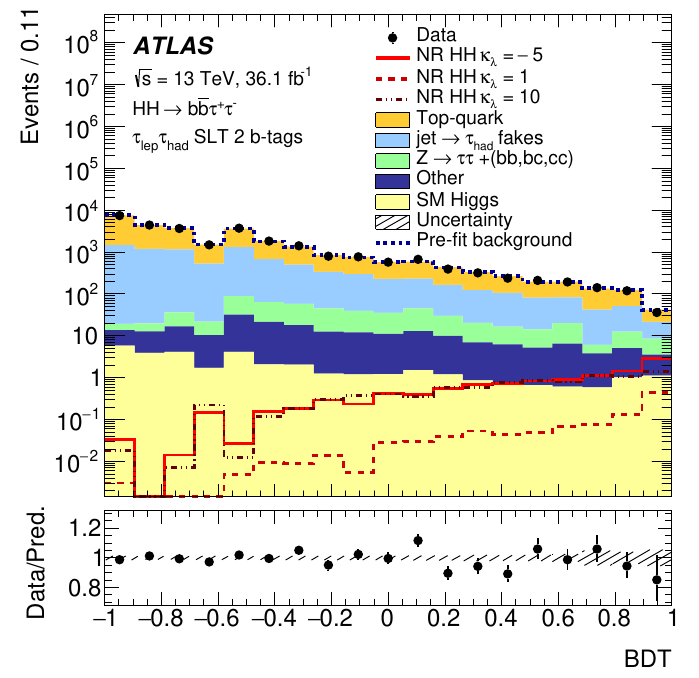}
 \caption{}
\end{subfigure}
\quad
\begin{subfigure}[b]{0.47\textwidth}
\includegraphics[height=7.5 cm,width =8.2 cm]{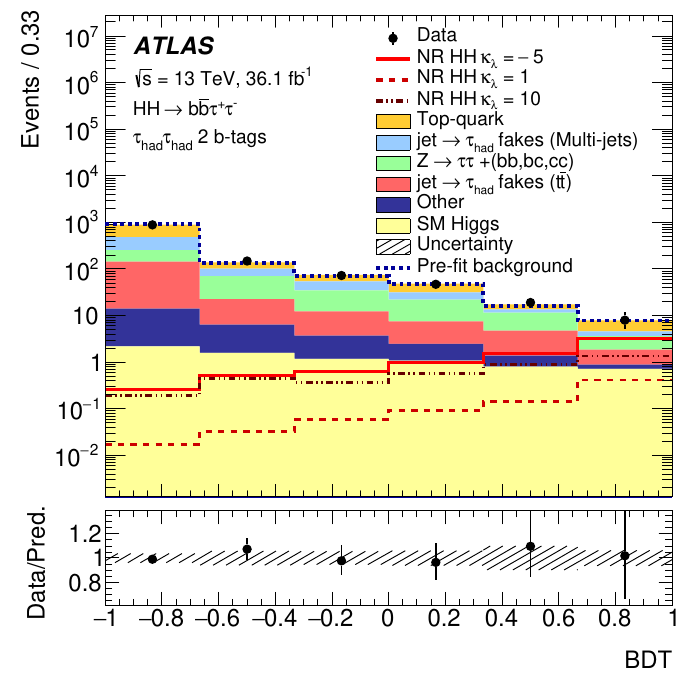}
 \caption{}
\end{subfigure}
\caption{BDT distributions used as final discriminants in the $\kappa_\lambda$-scan analysis for $b\bar{b}\tau^+\tau^-$ final state; (a): $\tau_{\textrm{lep}} \tau_{\textrm{had}}$ and (b) $\tau_{\textrm{had}} \tau_{\textrm{had}}$ channels~\cite{Paper_hh}.}     
\label{bdt_bbtautau}
\end{figure}

The $\kappa_\lambda$ 95\% CL intervals are  $-7.7<\kappa_\lambda<15.3$ (observed) and  $-9.7<\kappa_\lambda<17.5$ (expected). The impact of the Higgs-boson branching fractions and cross sections on the allowed $\kappa_\lambda$ lower limit is $<$1\% while it is $\sim$3\% on the upper limit; the dominant contribution affecting the upper limit of the interval comes from the parameterisations as a function on $\kappa_\lambda$ of the branching fractions. The observed limits are more stringent than the expected ones, as it is shown in Figure~\ref{ggF_hh}, over the whole range of $\kappa_\lambda$, due to a deficit of data relative to the background predictions at high values of the BDT score~\cite{Paper_hh}, as illustrated in Figure~\ref{bdt_bbtautau} reporting BDT distributions used as final discriminants both for the $\tau_{\textrm{lep}} \tau_{\textrm{had}}$ $(a)$ and the $\tau_{\textrm{had}} \tau_{\textrm{had}}$ $(b)$ channels.\newline

\begin{figure}[hbtp]
\centering
\begin{subfigure}[b]{0.48\textwidth}
\includegraphics[height=8 cm,width =8.2 cm]{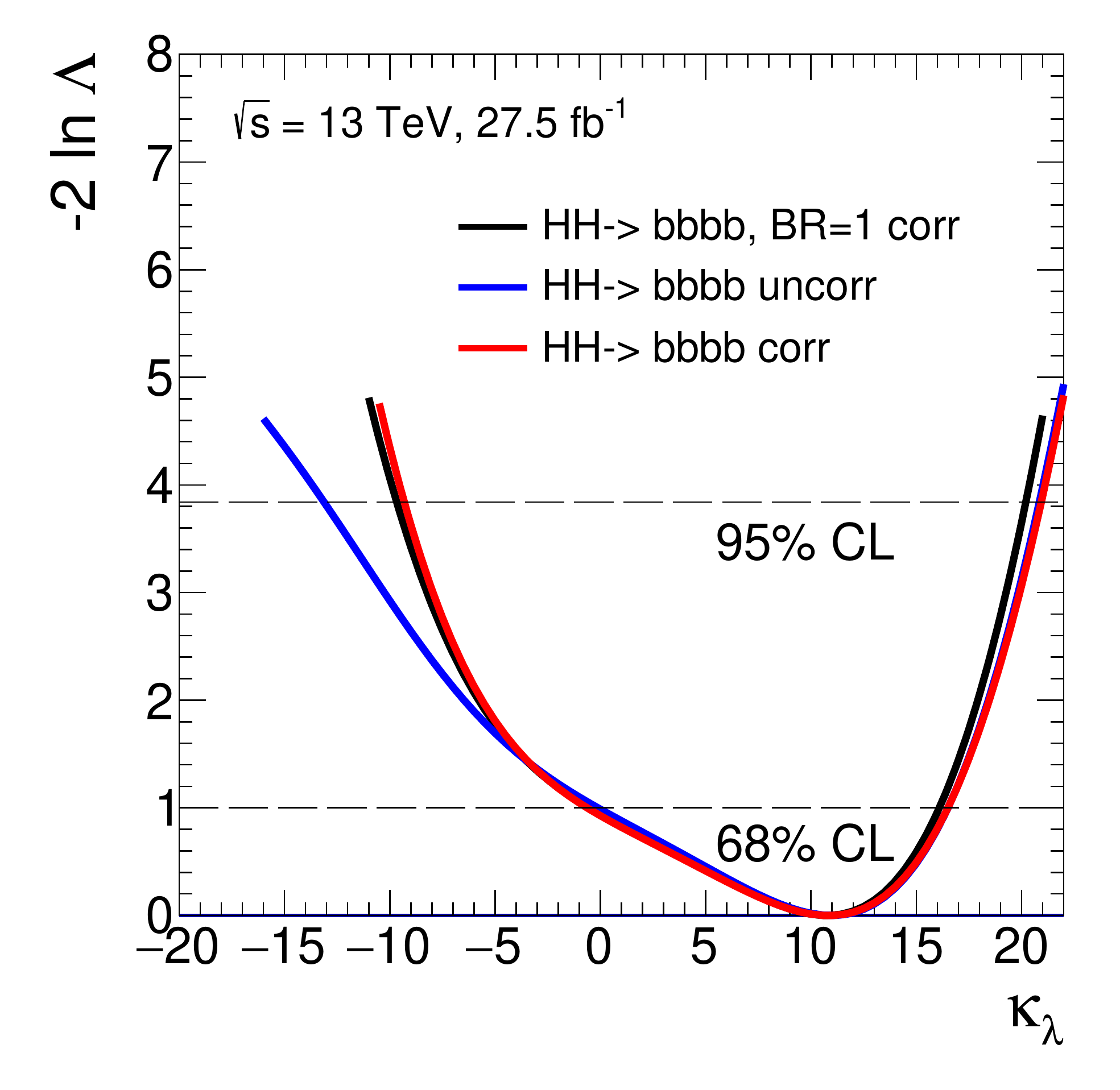}
 \caption{}
\end{subfigure}
\begin{subfigure}[b]{0.48\textwidth}
\includegraphics[height=8 cm,width =8.2 cm]{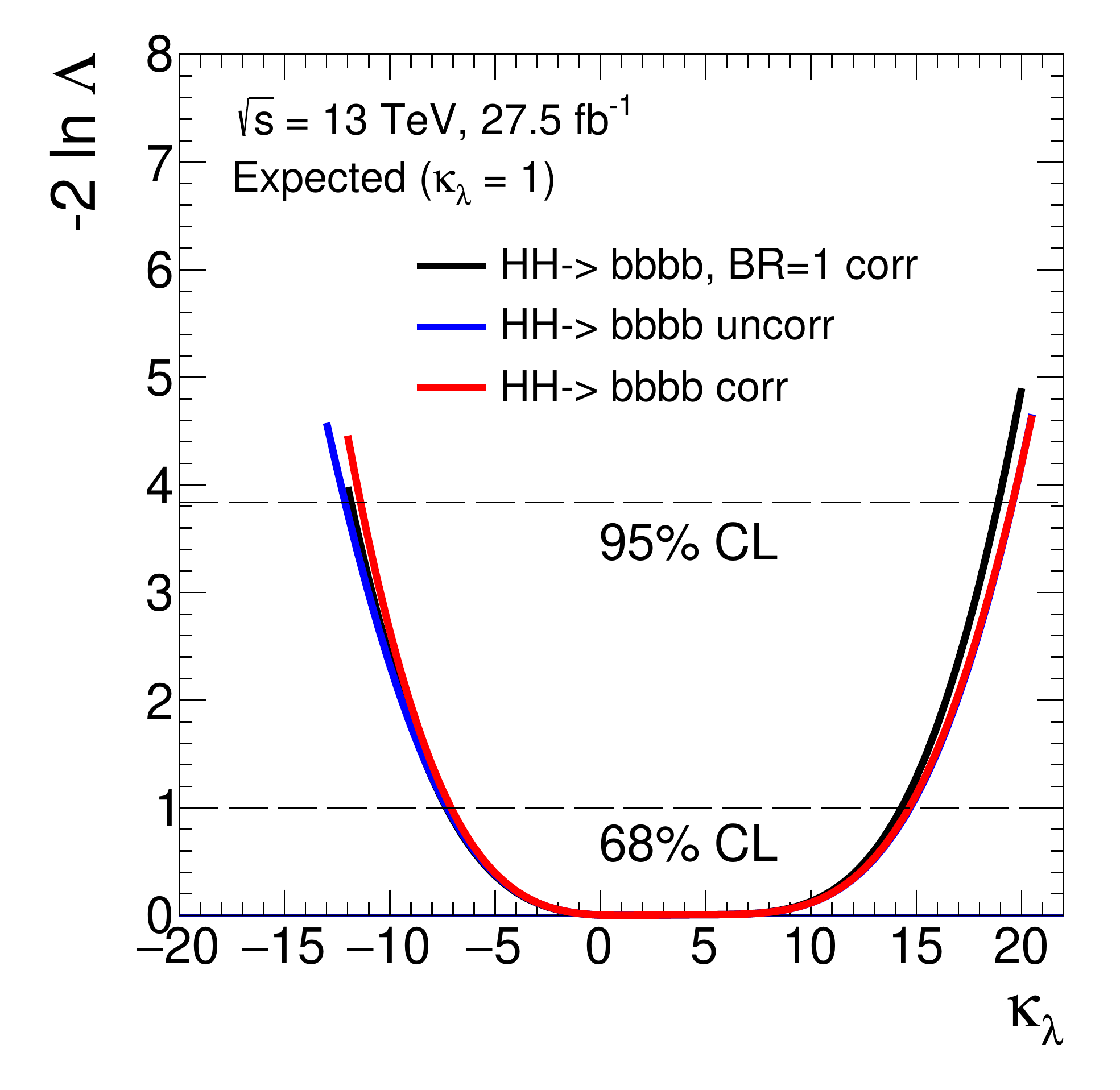}
 \caption{}
\end{subfigure}
\caption{Value of $-2 \ln{\Lambda(\kappa_\lambda)}$ as a function of $\kappa_\lambda$ for the $HH\rightarrow b\bar{b}b\bar{b}$ channel considering three configurations. Black solid line: branching fractions not parameterised as a function of $\kappa_\lambda$, NPs related to signal samples correlated; blue solid line: branching fractions and single-Higgs background parameterised as a function of $\kappa_\lambda$, NPs related to signal samples uncorrelated; red solid line: branching fractions and single-Higgs background parameterised as a function of $\kappa_\lambda$, NPs related to signal samples correlated. Likelihood distributions are reported for data (a) and for the Asimov dataset generated in the SM hypothesis (b). The dotted horizontal lines show the $-2 \ln{\Lambda(\kappa_\lambda)}=1$ level that is used to define the $\pm 1\sigma$ uncertainty on $\kappa_\lambda$ as well as the $-2 \ln{\Lambda(\kappa_\lambda)}=3.84$ level used to define the 95\% CL.}     
\label{scan_hh_4b}
\end{figure}

\begin{figure}[htbp]
\begin{center}
\includegraphics[width=0.47\textwidth]{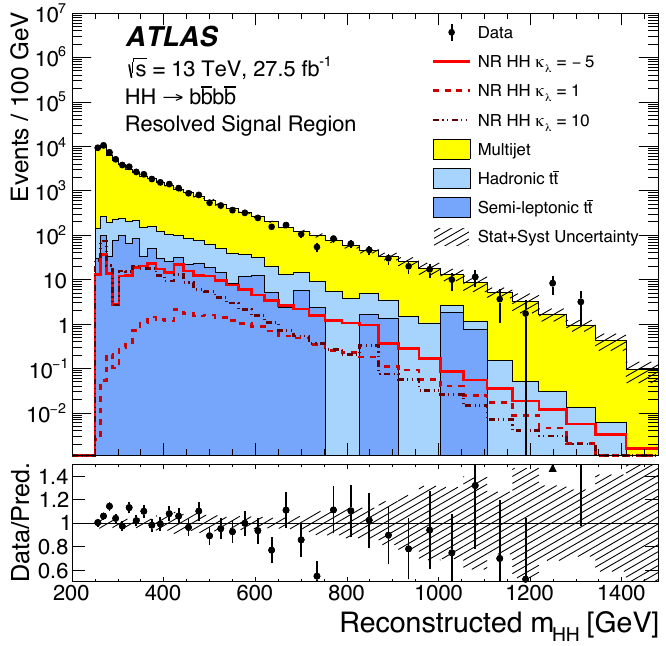}
\end{center}
\caption{The reconstructed $m_{HH}$ distribution used as final discriminant in the $\kappa_\lambda$-scan analysis for the $b\bar{b}b\bar{b}$ channel ~\cite{Paper_hh}.}     
\label{m_hh_4b}
\end{figure}

Finally, the fit has been performed exploiting the $b\bar{b}b\bar{b}$ channel and considering both correlation schemes introduced in Section~\ref{sec:statistical_model_hh}, \ie\ NPs related to signal samples correlated (scheme 1) or uncorrelated (scheme 2). The value of $-2 \ln{\Lambda(\kappa_\lambda)}$ as a function of $\kappa_\lambda$ is shown in Figure~\ref{scan_hh_4b} for data and for the Asimov dataset, generated in the SM hypothesis (\ie\ $\kappa_\lambda = 1$); the two schemes, labelled as ``corr$"$ and ``uncorr$"$, respectively, are considered. Furthermore, the likelihood distribution as a function of $\kappa_\lambda$ is shown for data and for the Asimov dataset not including the branching fraction parameterisations and considering NPs related to signal samples correlated among themselves. The single-Higgs background is not even included in the $b\bar{b}b\bar{b}$ analysis. The impact of these NLO-EW corrections on the allowed $\kappa_\lambda$ lower limit is $\sim$4\% while it is $\sim$3\% on the upper limit.\newline
The $\kappa_\lambda$ 95\% CL intervals for the $b\bar{b}b\bar{b}$ channel are  $-9.3<\kappa_\lambda<20.9$ (observed) and  $-11.4<\kappa_\lambda<19.6$ (expected).
The observed limits are more stringent than the expected ones at low values of $\kappa_\lambda$; in fact,  for these $\kappa_\lambda$ values, the signal $m_{HH}$ distributions have significant populations in the region above 400 GeV, where a deficit of data is observed, see Figure~\ref{m_hh_4b}.  An excess in the data below 300 GeV leads to the observed limits being less stringent than expected for high $\kappa_\lambda$ values. This non-significant excess tends to increase the $\kappa_\lambda=10$ contribution and causes the displacement of the minimum of the  $b\bar{b}b\bar{b}$ likelihood function towards higher values of $\kappa_\lambda$ as shown in Figure~\ref{scan_hh_bbyy}.
As expected, the constraint on $\kappa_\lambda$ is smaller when the NPs related to the signal samples are not correlated, reflecting a looser constraint on the NPs themselves.
This effect is enhanced in the fit to data, where NP post-fit values significantly differ from their nominal values.\newline
The scheme that has been chosen for the results reported in this thesis is ``scheme 1$"$, \ie\ the one where the nuisance parameters related to signal samples are correlated, in order to be consistent with the results reported in Reference~\cite{Paper_hh}.

\section{Results of fits to $\kappa_\lambda$ from the combination of double-Higgs channels}
\label{sec:results_hh_comb}
\begin{figure}[hbtp]
\centering
\begin{subfigure}[b]{0.49\textwidth}
\includegraphics[height=8 cm,width =8.2 cm]{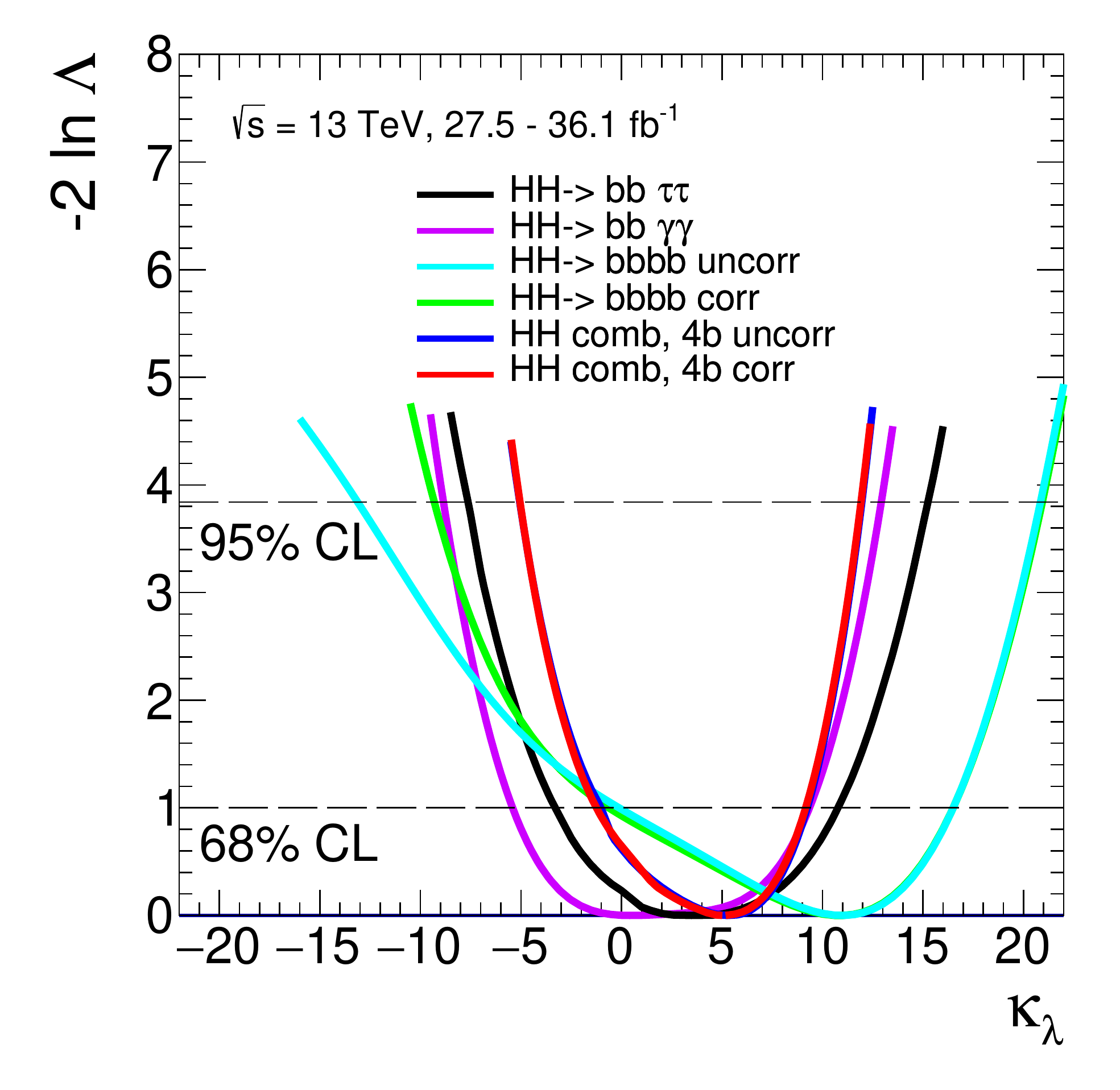}
 \caption{}
\end{subfigure}
\begin{subfigure}[b]{0.49\textwidth}
\includegraphics[height=8 cm,width =8.2  cm]{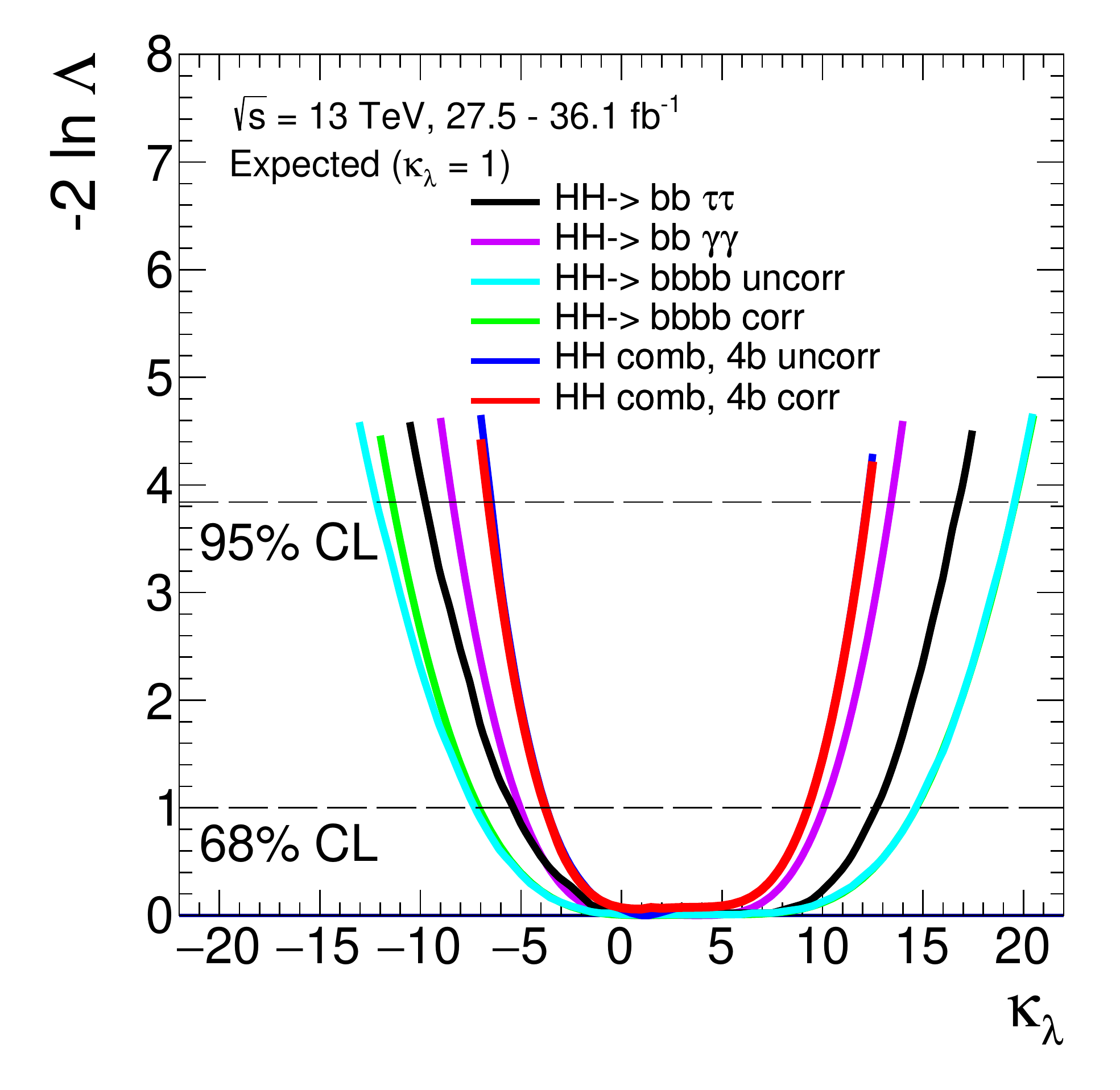}
 \caption{}
\end{subfigure}
\caption{Value of $-2 \ln{\Lambda(\kappa_\lambda)}$ as a function of $\kappa_\lambda$ for the three individual channels and for their combination; the $HH\rightarrow b\bar{b}b\bar{b}$ likelihood distribution has been reported both with NPs related to signal samples correlated and uncorrelated. Therefore, the HH combination too has been considered in the two cases. Negligible deviations are present in the combination with all other channels. Likelihood distributions are reported for data (a) and for the Asimov dataset generated in the SM hypothesis (b). The dotted horizontal lines show the $-2 \ln{\Lambda(\kappa_\lambda)}=1$ level that is used to define the $\pm 1\sigma$ uncertainty on $\kappa_\lambda$ as well as the $-2 \ln{\Lambda(\kappa_\lambda)}=3.84$ level used to define the 95\% CL.}     
\label{scan_comb_hh_allchannels}
\end{figure}

This section reports the results of the combined fit to $\kappa_\lambda$ exploiting the three most sensitive decay channels described in previous sections, whose individual results are presented in Section~\ref{sec:results_hh_channel}. \newline
The combined fit has been performed following the global correlation scheme described in Section~\ref{sec:statistical_model_hh} and adopting two different correlation schemes for the $b\bar{b}b\bar{b}$ channel, namely the scheme 1, labelled as ``corr$"$ on the plots, and the scheme 2, labelled as ``uncorr$"$. \newline
Figure~\ref{scan_comb_hh_allchannels} shows the value of $-2 \ln{\Lambda(\kappa_\lambda)}$ as a function of $\kappa_\lambda$ for the three individual channels and their combination for data and for the Asimov dataset, generated under the SM hypothesis (\ie\ $\kappa_\lambda = 1$). Negligible deviations are present adopting the different correlation schemes in the combination with all other channels.

\begin{figure}[H]
\begin{center}
\includegraphics[height=7 cm, width=8 cm]{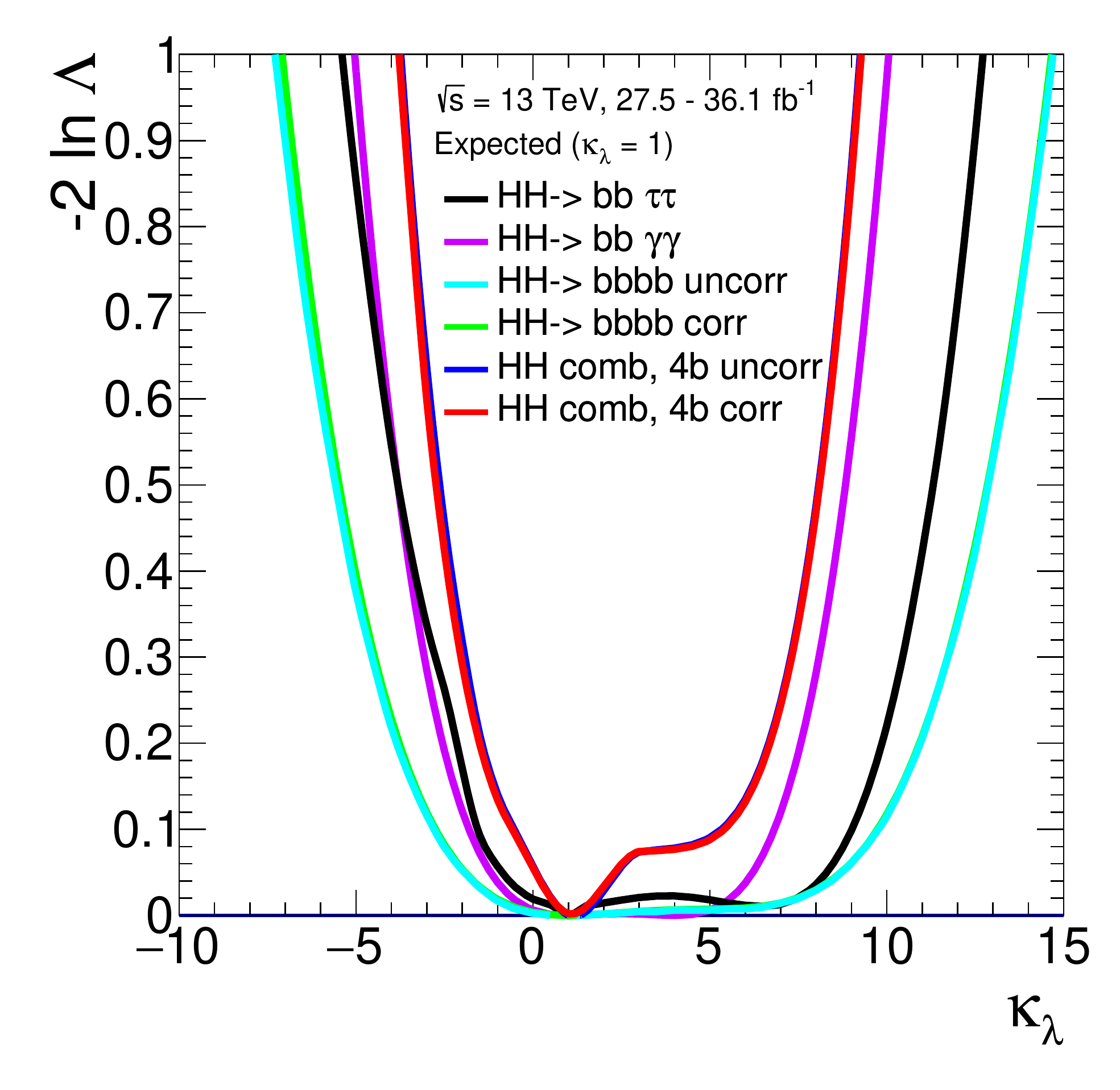}
\end{center}
\caption{Value of $-2 \ln{\Lambda(\kappa_\lambda)}$ as a function of $\kappa_\lambda$ for the three individual channels and their combination; the likelihood distribution is zoomed between $-2 \ln{\Lambda(\kappa_\lambda)}=0$ and $-2\ln{\Lambda(\kappa_\lambda)}=1$ in order to show the peculiar likelihood function characterised by two local minima related to the total cross section and kinematic properties dependence on $\kappa_\lambda$.}     
\label{zoom_asi_hh}
\end{figure}

As shown in Figure~\ref{zoom_asi_hh} the likelihood has a peculiar shape, characterised by two local minima, that is related to the dependence of the total cross section and double-Higgs kinematic properties on $\kappa_\lambda$, while the relative height of the two minima depends on the capability of the different analyses to access differential $m_{HH}$ information. 
The impacts of neglecting NLO-EW corrections depending on $\kappa_\lambda$, on the intervals extracted from the double-Higgs combination, have also been checked. \newline
The $\kappa_\lambda$ 95$\%$ CL intervals for the three different channels, $b\bar{b}\tau^+\tau^-$, $b\bar{b}\gamma\gamma$ and $b\bar{b}b\bar{b}$ and for their combination are summarised in Table~\ref{BR_tab1}, where different configurations have been considered:
\begin{itemize}
\item branching fractions (BRs) and single-Higgs (SH) background parameterised as a function of $\kappa_\lambda$;
\item neither branching fractions nor single-Higgs background parameterised as a function of $\kappa_\lambda$.
\end{itemize}

\begin{table}[htbp]
\centering
\scalebox{0.9}{
{\def\arraystretch{2}
\centering
\begin{tabular}{|c|c|c|c|}
 \hline
\small{Channel}  & \small{$\kappa_\lambda $ interval at 95$\%$ CL} &  \small{$\kappa_\lambda$ interval, BR=1, bkg=1} & NLO-EW corrections\\ 
\hline
$b\bar{b} \tau^+\tau^-$ (obs) & [-7.7 - 15.3]  &  [-7.7 - 14.8]& [$<1\%$, 3$\%$]\\
\hline
$b\bar{b} \tau^+\tau^-$ (exp) & [-9.7 - 17.5]  &  [-9.7 - 16.8]& [$<1\%$,  4$\%$]\\
\hline
$b\bar{b}b\bar{b}$ (obs) & [-9.3 - 20.9]   & [-9.7 - 20.2] &  [4$\%$, 3$\%$] \\
\hline
$b\bar{b}b\bar{b}$ (exp) & [-11.4 - 19.6]   & [-11.8 - 18.9] &  [3$\%$, 4$\%$] \\
\hline
$b\bar{b}\gamma \gamma$ (obs)& [-8.9 - 12.9]   & [-8.1 - 13.2] &  [9$\%$, $ 2\%$]\\
\hline
$b\bar{b}\gamma \gamma$ (exp)& [-9.0 - 13.2]  & [-8.4 - 13.4] &  [8$\%$, 1$\%$]\\
\hline
Comb (obs) & [-5.0 - 11.9]  & [-4.7 - 12.0] &  [$7\%$, $ < 1\%$]\\
\hline
Comb (exp)& [-6.6 - 12.2]  & [-6.4 - 12.3] &  [$3\%$, $ < 1\%$]\\
\hline
\end{tabular}}}
\caption{Allowed $\kappa_\lambda$ intervals at 95$\%$ CL with different configurations and impact of the Higgs-boson branching-fraction and cross-section corrections on $\kappa_\lambda$ limits for single channels and for the combination of the three channels.}
\label{BR_tab1}
\end{table}
The overall impact of the Higgs-boson branching fractions and cross sections on the allowed $\kappa_\lambda$ intervals has been estimated to be less than 10$\%$. These effects are taken into account in the present combination.\newline
The best-fit value and $\pm 1\sigma$ uncertainties are determined to be $\kappa_\lambda=5.2_{-6.4}^{+4.0}$ (observed) and $\kappa_\lambda=1.0^{+8.3}_{-4.7}$ (expected). As expected, the best-fit value is guided by the most sensitive channel, \ie\ the $b\bar{b}\tau^+\tau^-$ channel. \newline
The 95\% CL intervals for $\kappa_\lambda$ are  $-5.0<\kappa_\lambda<11.9$ (observed) and  $-6.6<\kappa_\lambda<12.2$ (expected) leading to a significant improvement of single channel performances as a result of the comparable sensitivity.\newline
Table~\ref{tab:klambda-summary_hh} reports a summary of $\kappa_\lambda$ fit values with $\pm 1 \sigma$ uncertainties and 95\% CL intervals for each decay channel entering the combination and for the combinations themselves.\newline
A further element that has been introduced, with respect to the results in Reference~\cite{Paper_hh}, is the injection of branching-fraction uncertainties and uncertainties on the double-Higgs cross section. As reported in Table~\ref{tab:klambda-summary_hh}, the overall impact of these modifications is really small.
\begin{table}[htbp]
\begin{center}

{\def\arraystretch{1.4}
\begin{tabular}{|c|c|c|}
\hline

Channels  & $\kappa_\lambda{}^{+1\sigma}_{-1\sigma}$ & $\kappa_\lambda$  [95\% CL] \\ 
\hline
\multirow{2}{*}{$HH\rightarrow b\bar{b}\tau^+\tau^-$}  & $3.6_{-7.0}^{+7.2}$ &  $[-7.7, 15.3]$ \\
                                         &        $1.0^{+12.1}_{-6.5}$ & $[-9.7, 17.5]$ \\ 
                                         \hline
\multirow{2}{*}{$HH\rightarrow b\bar{b}b\bar{b}$} & $10.9_{-11.5}^{+5.7}$ & $[-9.3, 20.9]$ \\
                                      &         $1.0_{-8.3}^{+13.7}$ &  $[-11.4, 19.6]$ \\ 
                                      \hline
                               
\multirow{2}{*}{$HH\rightarrow b\bar{b}\gamma \gamma$}  & $0.5_{-5.9}^{+8.9}$ & $[-8.9, 12.9]$ \\
                                      &           $1.0_{-6.6}^{+8.8}$ &  $[-9.0, 13.2]$ \\ 

 \hline
\multirow{2}{*}{Combination}   &  $5.2_{-6.4}^{+4.0}$ &  $[-5.0, 11.9]$ \\
                                      &            $1.0^{+8.3}_{-4.7}$ & $[-6.6, 12.2]$ \\ 
 \hline
\multirow{2}{*}{Combination injecting uncertainties}    &  $5.2_{-6.5}^{+4.0}$ &  $[-5.1, 12.0]$ \\
                                      &            $1.0^{+8.4}_{-4.9}$ & $[-6.8, 12.4]$ \\                                                                                                                
\hline
\end{tabular}
}
\caption{
Best-fit  $\kappa_\lambda$ values with $\pm 1 \sigma$ uncertainties. The first column shows each double-Higgs decay channel considered in the combination. The 95\% CL interval for $\kappa_\lambda$ are also reported.  For each fit result the upper row corresponds to the observed results, and the lower row to the expected results obtained using Asimov datasets generated under the SM hypothesis.}
\label{tab:klambda-summary_hh}
\end{center}
\end{table}

\begin{figure}[hbtp]
\centering
\begin{subfigure}[b]{0.49\textwidth}
\includegraphics[height=8 cm,width =8.2 cm]{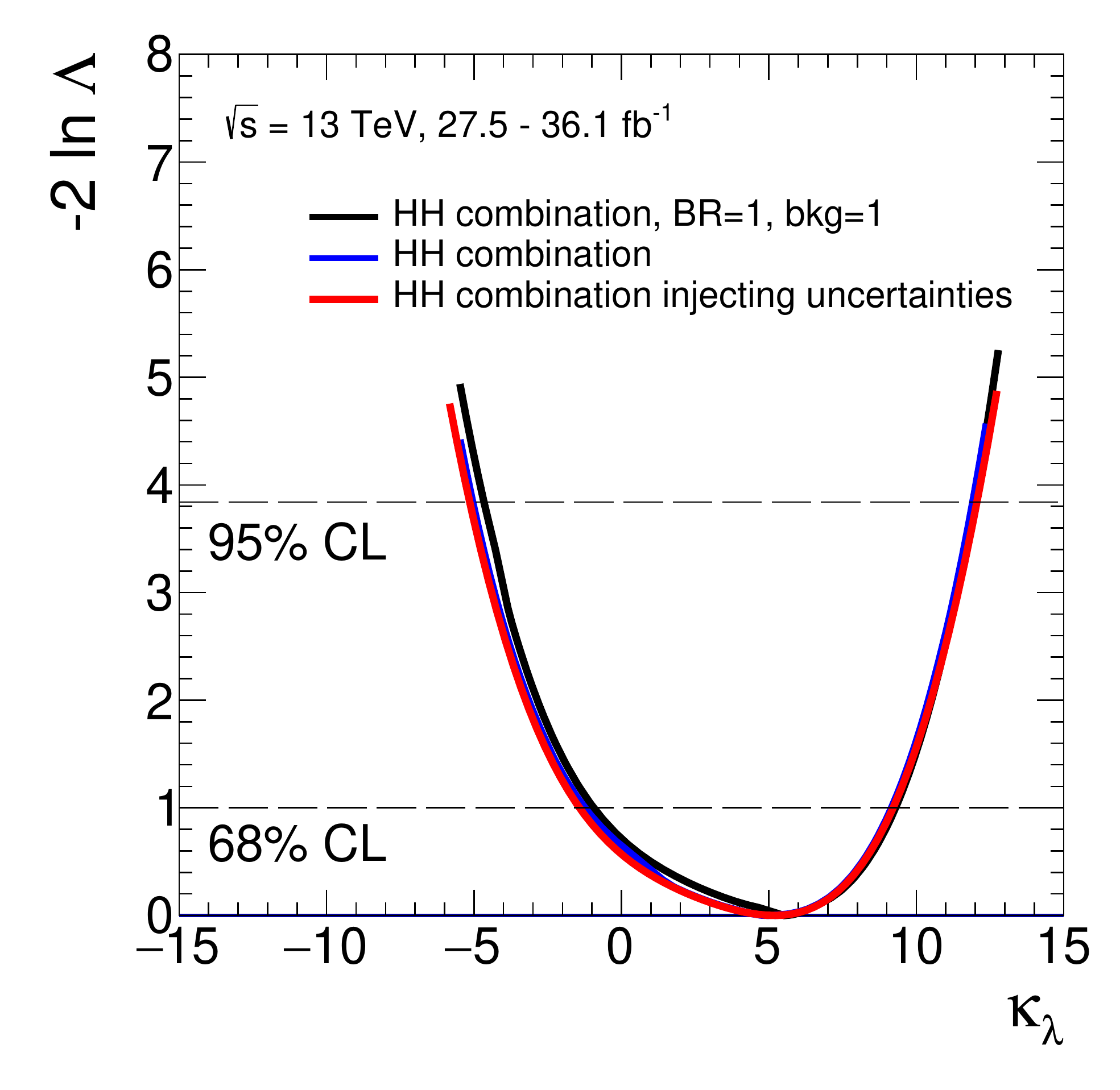}
 \caption{}
\end{subfigure}
\begin{subfigure}[b]{0.49\textwidth}
\includegraphics[height=8 cm,width =8.2 cm]{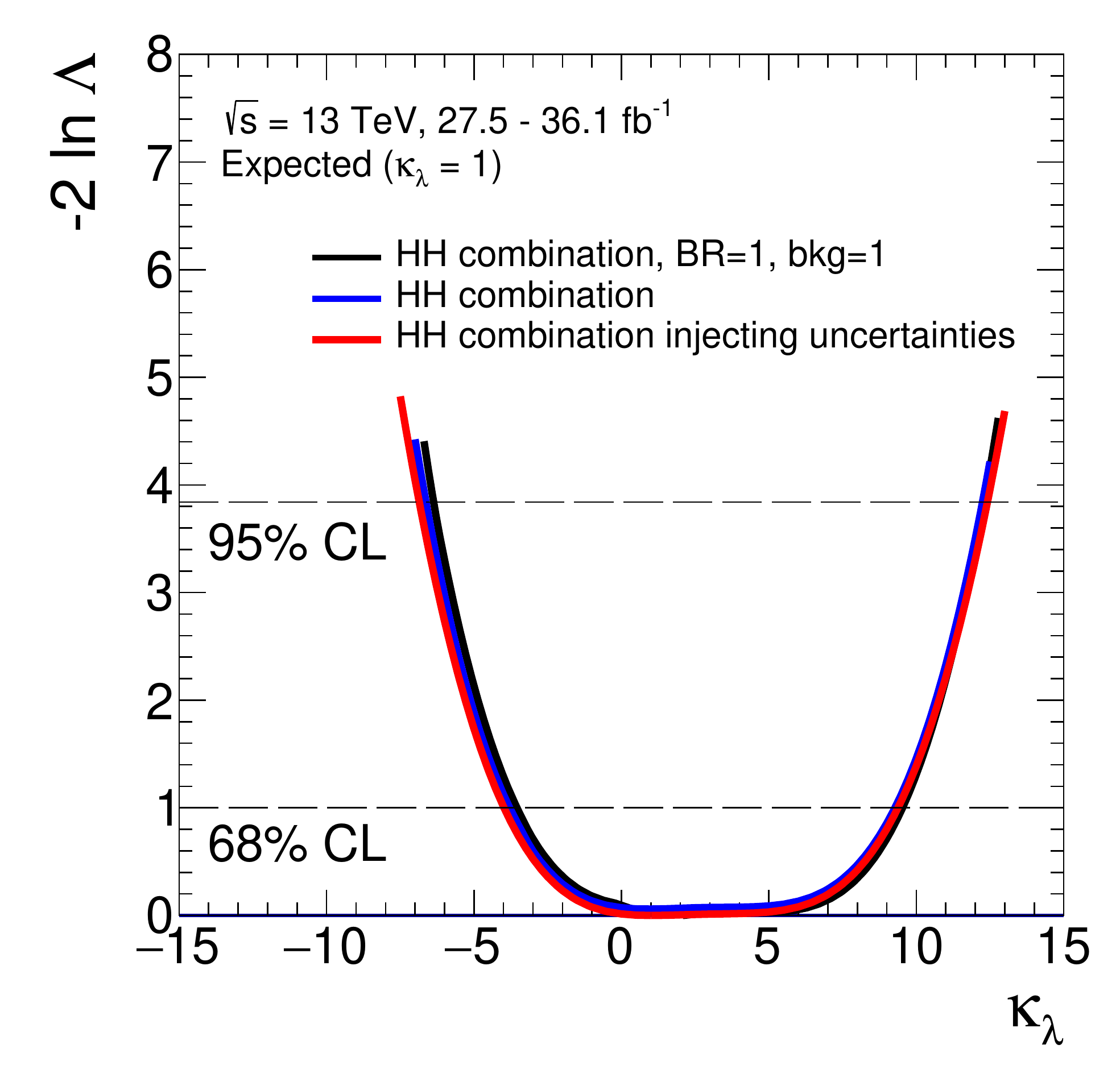}
 \caption{}
\end{subfigure}
\caption{Value of $-2 \ln{\Lambda(\kappa_\lambda)}$ as a function of $\kappa_\lambda$ considering different configurations. Black solid line: neither branching fractions nor single-Higgs background parameterised as a function of $\kappa_\lambda$; blue solid line: branching fractions and single-Higgs background parameterised as a function of $\kappa_\lambda$; red solid line: branching fractions and single-Higgs background parameterised as a function of $\kappa_\lambda$, theoretical uncertainties injected. Likelihood distributions are reported for data (a) and for the Asimov dataset generated in the SM hypothesis (b). The dotted horizontal lines show the $-2 \ln{\Lambda(\kappa_\lambda)}=1$ level that is used to define the $\pm 1\sigma$ uncertainty on $\kappa_\lambda$ as well as the $-2 \ln{\Lambda(\kappa_\lambda)}=3.84$ level used to define the 95\% CL.}     
\label{scan_comb_hh}
\end{figure}

A comparison of $\kappa_\lambda$ likelihood scans including or not branching fractions and single-Higgs background parameterisation as a function of  $\kappa_\lambda$, and including or not double-Higgs cross section theory uncertainties and branching-fraction uncertainties, is shown in Figure~\ref{scan_comb_hh}, in addition to the detailed results per channel reported in Table~\ref{BR_tab1}. The black solid line represents the likelihood distribution when neither branching fractions nor single-Higgs background are parameterised as a function of $\kappa_\lambda$; the blue solid line when the branching fractions and single-Higgs background are parameterised as a function of $\kappa_\lambda$, while the red solid line represents the case in which branching fractions and single-Higgs background are parameterised as a function of $\kappa_\lambda$ and theoretical uncertainties are injected. 

\begin{figure}[hbtp]
\centering
\begin{subfigure}[b]{0.49\textwidth}
\includegraphics[height=8 cm,width =\textwidth]{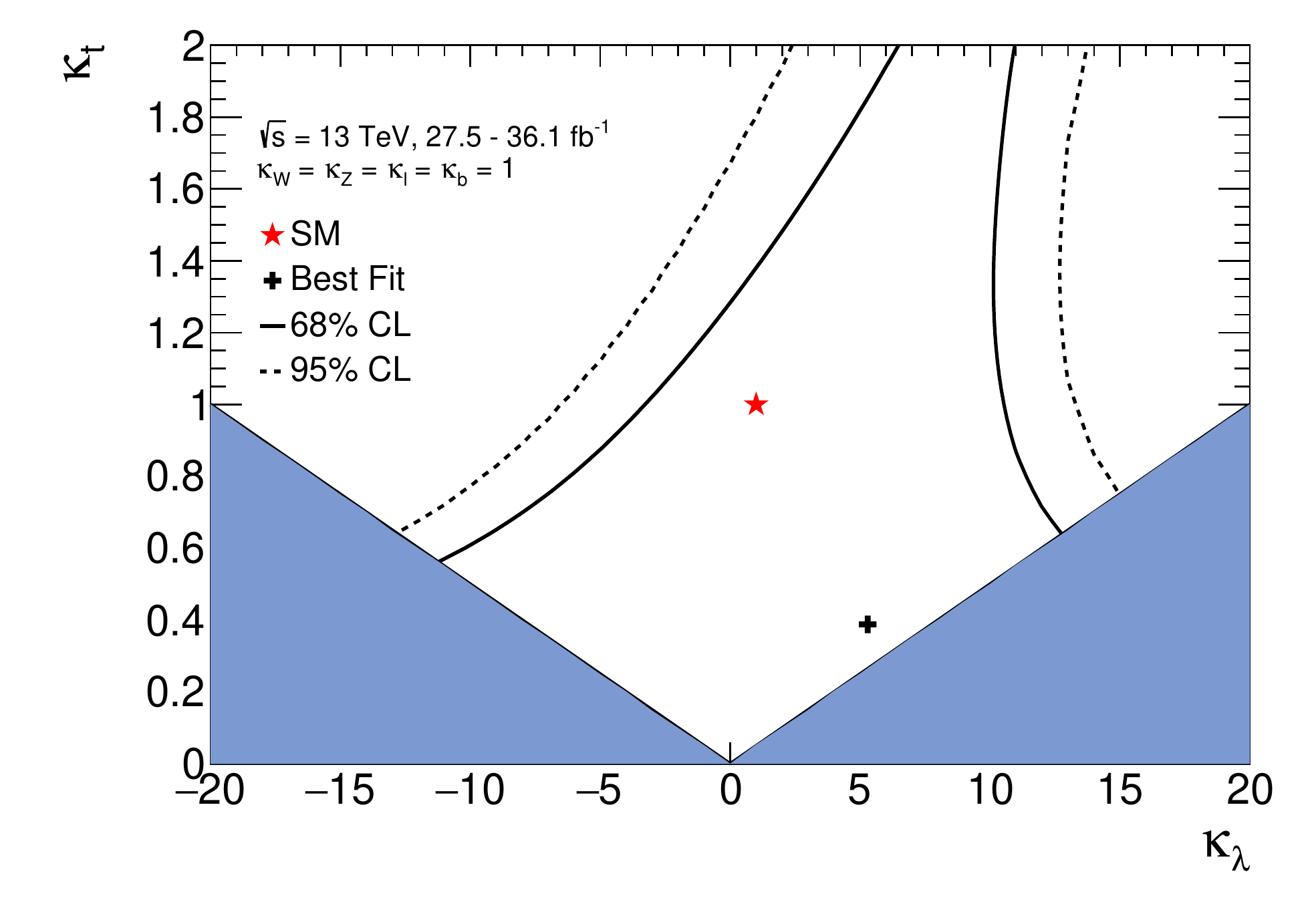}
 \caption{}
\end{subfigure}
\begin{subfigure}[b]{0.49\textwidth}
\includegraphics[height=8 cm,width =\textwidth]{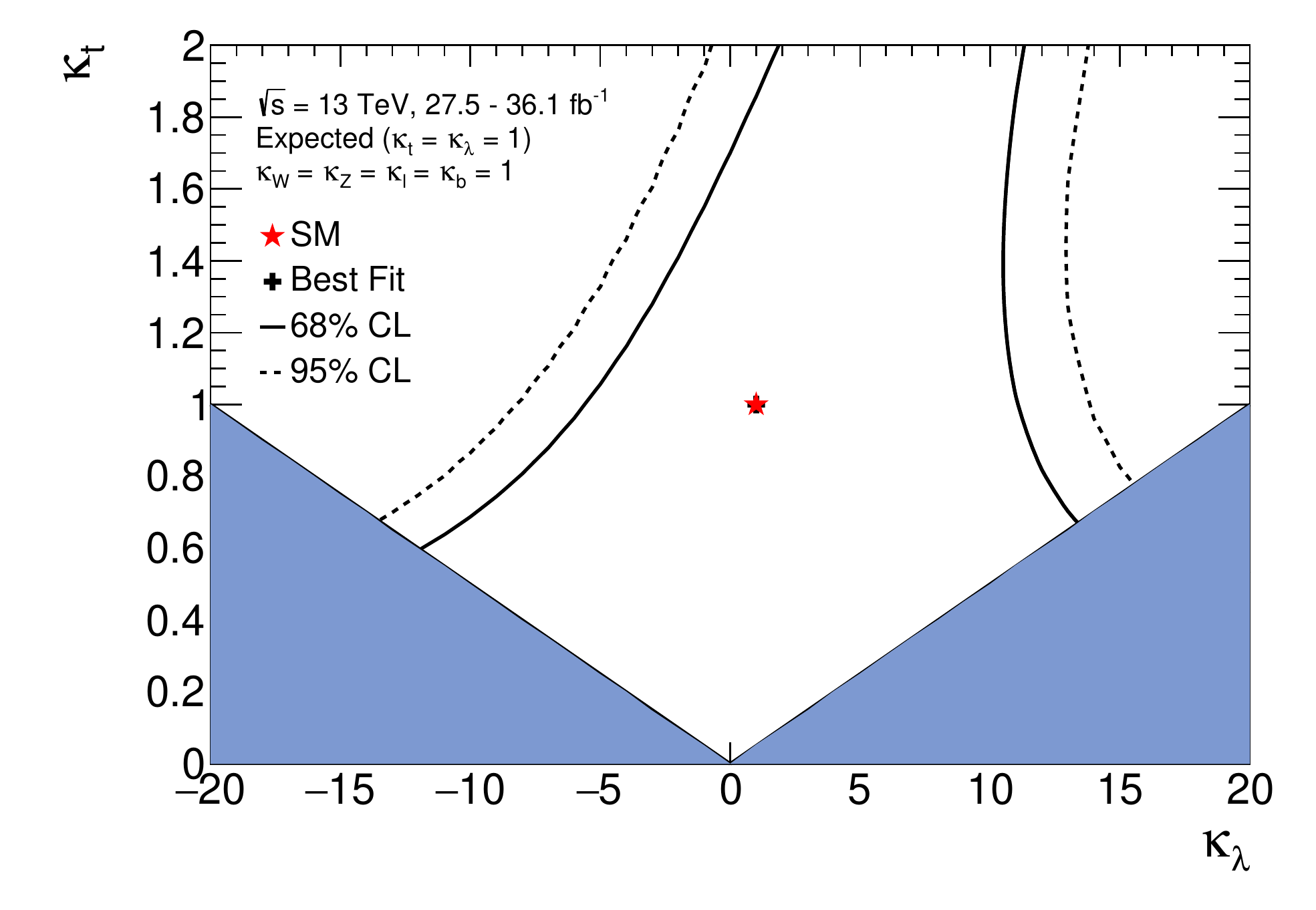}
 \caption{}
\end{subfigure}
\caption{Negative log-likelihood contours at 68\% and 95\% CL in the $(\kappa_\lambda,\kappa_t)$ plane on data (a) and on the Asimov dataset generated under the SM hypothesis (b). The best-fit value is indicated by a cross while the SM hypothesis is indicated by a star. The plot assumes that the approximations in References~\cite{Degrassi,Maltoni} are valid inside the shown contours. The degeneracy of measuring at the same time $\kappa_\lambda$ and $\kappa_t$ is shown; the coloured area corresponds to $| \kappa_\lambda /\kappa_t| < 20$, a constrain coming from the $HH\rightarrow b\bar{b} \gamma \gamma$ analysis.}     
\label{contour_kl_kt_hh_shaded}
\end{figure}

\begin{figure}[htbp]
\begin{center}
\includegraphics[height=7 cm,width =8 cm]{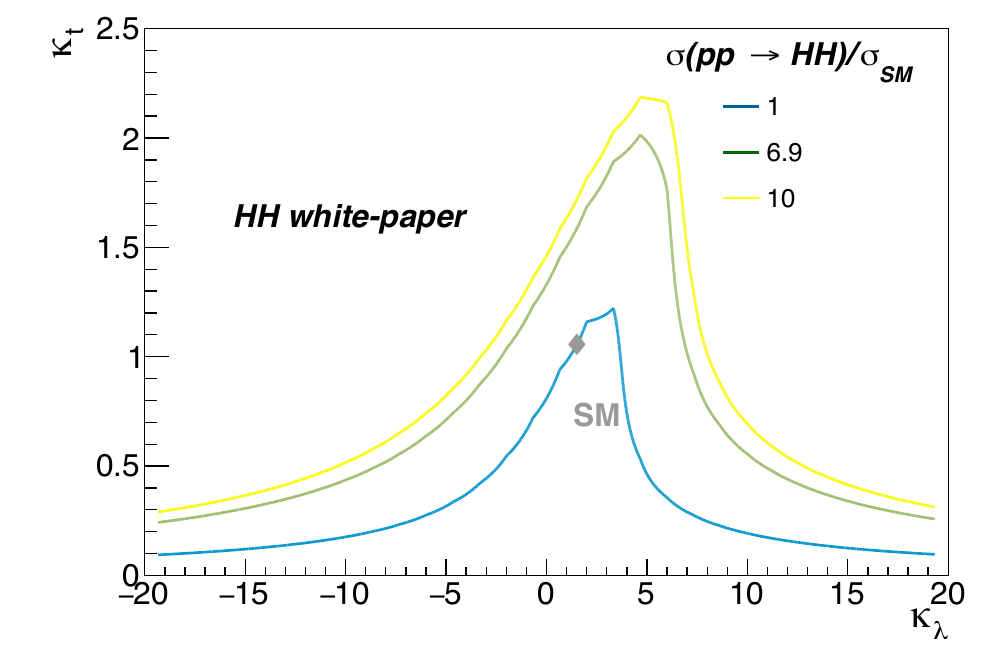}
\end{center}
\caption{Contour level of $\sigma(pp\rightarrow HH)/\sigma_{SM}$ as a function of $\kappa_t$ and $\kappa_\lambda$, under the assumption of no additional Higgs coupling vertices. The diamond indicates the SM predicted value. The reference values of 6.9 and 10 correspond to the best available observed and expected upper limits on the $\sigma_{ggF}(pp\rightarrow HH)$/$\sigma^{SM}_{ggF}(pp\rightarrow HH)$ as measured by the ATLAS experiment~\cite{white_paper}.}     
\label{contour_hh_theory}
\end{figure}

Furthermore, differences between the CLs method and a likelihood-based approach have been studied. Table~\ref{tab:hh_validate} summarises the different configurations that have been tested reporting the observed 95\% CL intervals in the aforementioned cases. The difference between the CLs method and the profile likelihood one is at the level of a few percents.
\begin{table}[htbp]
\begin{center}
{\def\arraystretch{1.3}
\begin{tabular}{|c|c|}
\hline
    Method &$\kappa_{\lambda}$ interval at 95\% CL (obs)\\ 
    \hline
    CLs&[-5.1, 11.9]\\ 
    \hline
    Likelihood scan &\multirow{2}{*}{[-4.7, 12.0]}\\
    (BR=1, SH=1, no theory uncertainties)&\\
    \hline
    Likelihood scan &\multirow{2}{*}{[-5.0, 11.9]}\\
    (BR and SH parameterised) & \\
    \hline
    Likelihood scan &\multirow{2}{*}{[-5.1, 12.0]}\\
    (BR and SH parameterised, theory uncertainties) & \\
\hline
\end{tabular}}
\caption{95\% CL comparisons for different approaches. The CLs approach is exploited in order to cross check double-Higgs publication results. In the likelihood approaches, the three different configurations listed in the first column have been tested.}
\label{tab:hh_validate}
\end{center}
\end{table}
\begin{figure}[hbtp]
\centering
\begin{subfigure}[b]{0.48\textwidth}
\includegraphics[width =\textwidth]{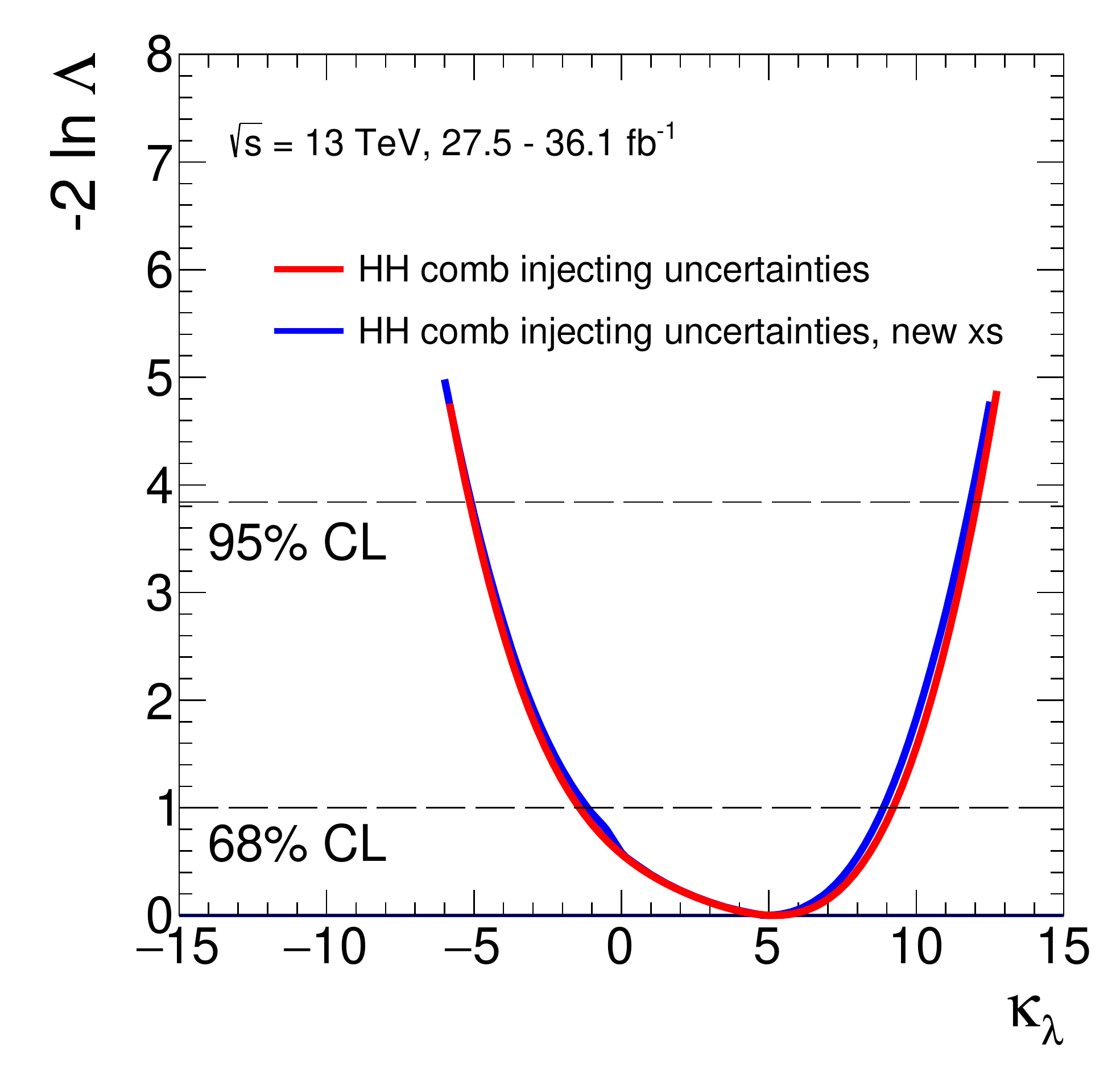}
 \caption{}
\end{subfigure}
\begin{subfigure}[b]{0.48\textwidth}
\includegraphics[width =\textwidth]{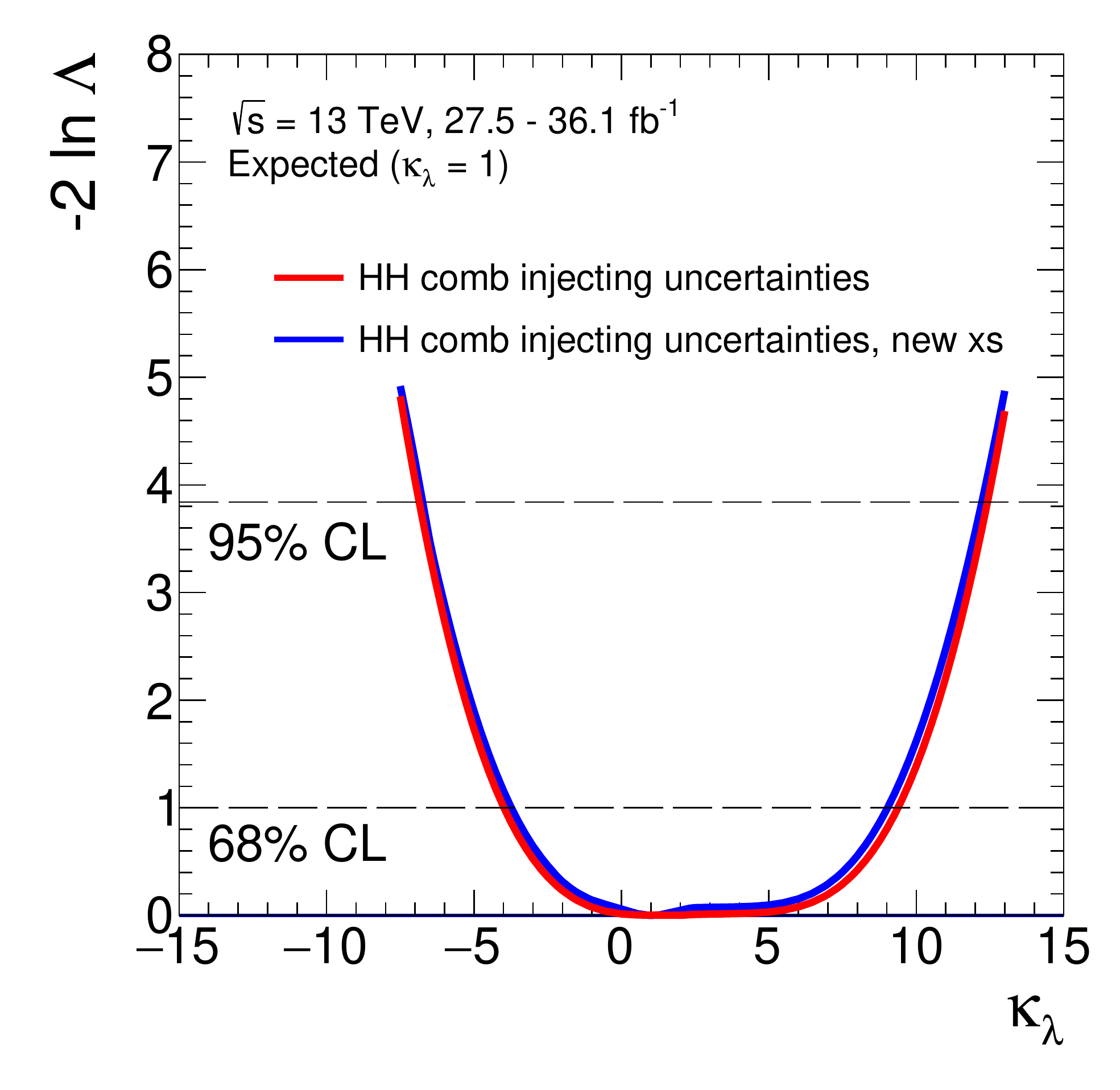}
 \caption{}
\end{subfigure}
\caption{Value of $-2 \ln{\Lambda(\kappa_\lambda)}$ as a function of $\kappa_\lambda$ for data (a) and for the Asimov dataset (b). Solid red line: HH combined likelihood distribution when BRs and SH background are parameterised as a function of $\kappa_\lambda$ and theory uncertainties are injected; blue solid line: likelihood distribution including recent computation of the HH SM cross section. The dotted horizontal lines show the $-2 \ln{\Lambda(\kappa_\lambda)}=1$ level that is used to define the $\pm 1\sigma$ uncertainty on $\kappa_\lambda$ as well as the $-2 \ln{\Lambda(\kappa_\lambda)}=3.84$ level used to define the 95\% CL.}     
\label{xs_corr_kl_hh}
\end{figure}
Recent computations of the double-Higgs production cross section have been performed, where QCD corrections are known up to next-to-leading order (NLO) and at next-to-next-to-leading order (NNLO) in the limit of heavy top quarks including partial finite top quark mass effects, leading to a small reduction of the SM cross section from 33.5 to 31.05 fb and a stronger dependence on $\kappa_\lambda$~\cite{xs_hh1,xs_hh2,white_paper}. Figure~\ref{xs_corr_kl_hh} shows the comparison between the likelihood distribution obtained exploiting the old recommendation and the likelihood distribution obtained including recent computation of the double-Higgs SM cross section for data and for the Asimov dataset.
In order to be consistent with the double-Higgs results reported in Reference~\cite{Paper_hh}, these recent calculations have not been used, but their impact on the Higgs-boson self-coupling 95\% CL interval has been evaluated to be less than 2\%.\newline
The strong dependence of the double-Higgs cross section $\sigma_{ggF}(pp\rightarrow HH)$ on $\kappa_t$ can be exploited through a likelihood fit performed to constrain at the same time $\kappa_\lambda$ and $\kappa_t$, setting all the other coupling modifiers to their SM values. Figure~\ref{contour_kl_kt_hh_shaded} shows negative log-likelihood contours on the $(\kappa_\lambda,\kappa_t)$ grid, where the coloured areas are not part of the allowed region because the acceptance of the $HH\rightarrow b\bar{b}\gamma \gamma$ analysis is not reliable for $|\kappa_\lambda/\kappa_t|\ge 20$. It is clear that there is no chance of measuring $\kappa_\lambda$ and $\kappa_t$  at the same time without assumptions on one of the two coupling modifiers, as it is also shown in the theoretical contours of Figure~\ref{contour_hh_theory} where the reference values of 6.9 and 10 correspond to the ATLAS observed and expected upper limits on the $\sigma_{ggF}(pp\rightarrow HH)$ cross section times the predicted SM cross section~\cite{Paper_hh}; furthermore $\kappa_t$ is superiorly limited and can be measured better than $\kappa_\lambda$, given the cross-section dependences on $\kappa_t^4$ and on $\kappa_\lambda /\kappa_t$.
\begin{figure}[H]
\begin{center}
\includegraphics[height=7 cm,width =8 cm]{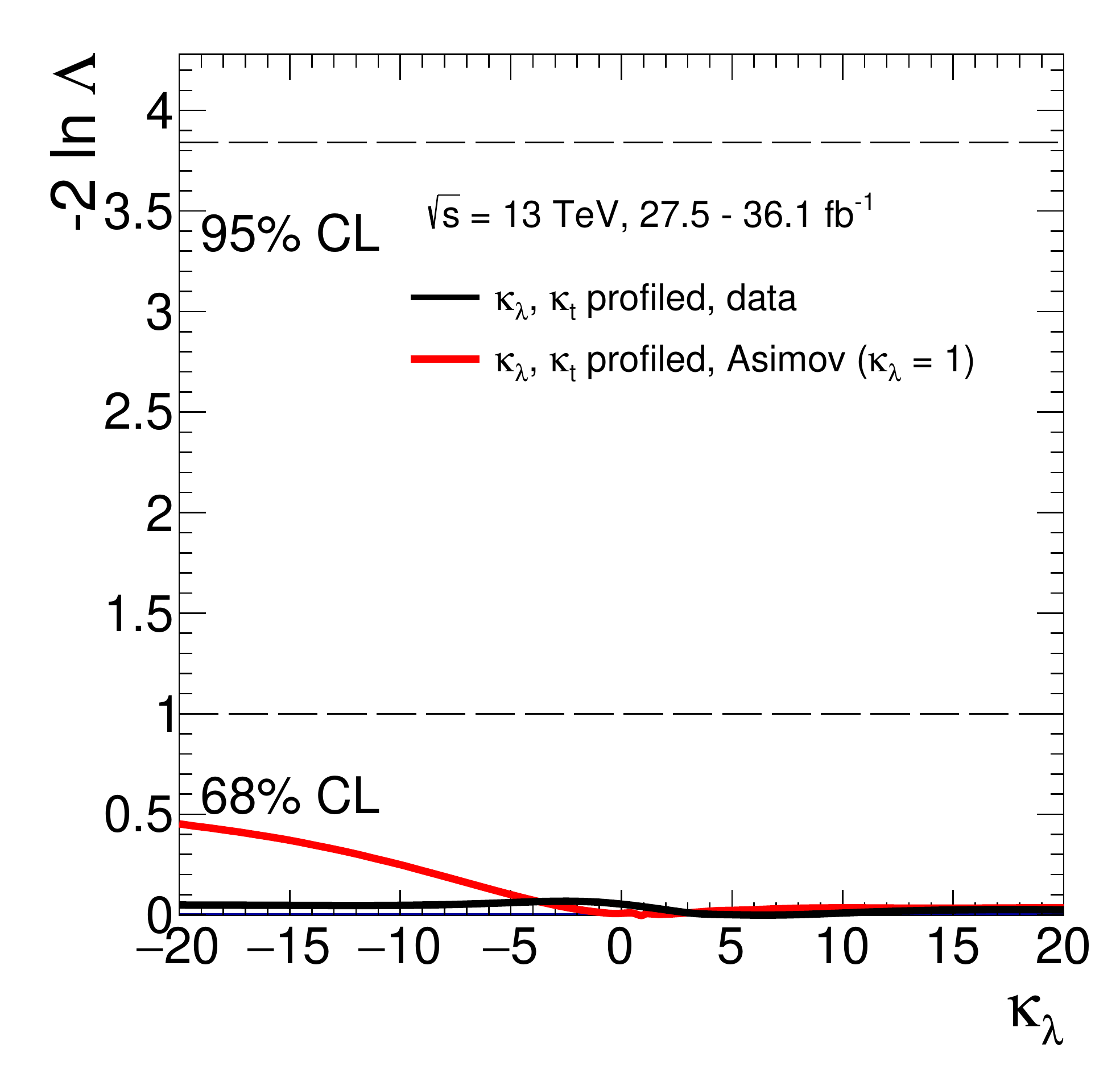}
\end{center}
\caption{Value of $-2 \ln{\Lambda(\kappa_\lambda)}$ as a function of $\kappa_\lambda$; the black solid line represents data while the red solid line represents the Asimov dataset generated in the SM hypothesis. The dotted horizontal lines show the $-2 \ln{\Lambda(\kappa_\lambda)}=1$ level that is used to define the $\pm 1\sigma$ uncertainty on $\kappa_\lambda$ as well as the $-2 \ln{\Lambda(\kappa_\lambda)}=3.84$ level used to define the 95\% CL.}     
\label{scan_kl_kt_asi_dati}
\end{figure}

\begin{figure}[H]
\begin{center}
\includegraphics[height=7.5 cm,width =8.2 cm]{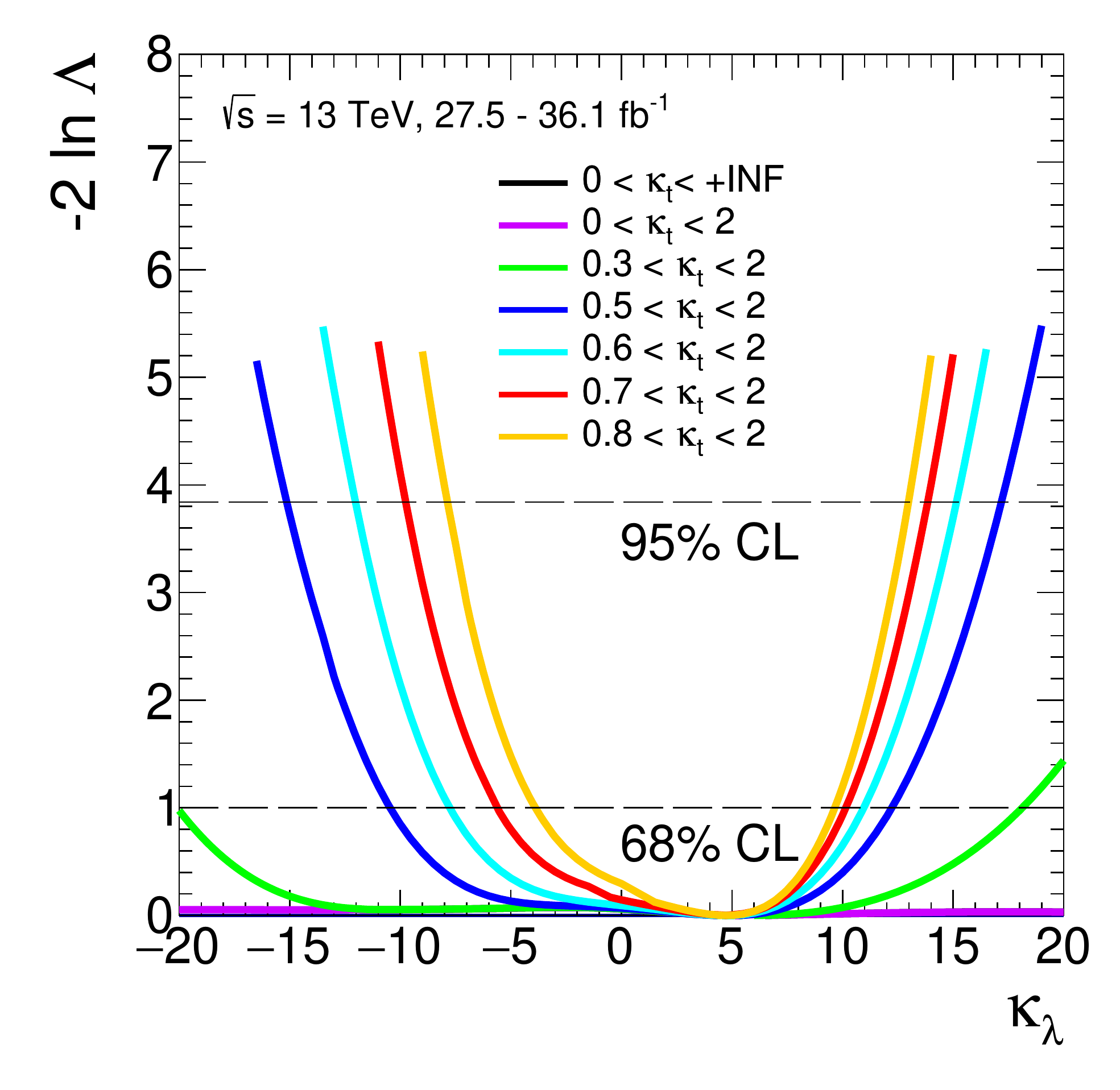}
\end{center}
\caption{Value of $-2 \ln{\Lambda(\kappa_\lambda)}$ as a function of $\kappa_\lambda$ varying $\kappa_t$ ranges for data. The dotted horizontal lines show the $-2 \ln{\Lambda(\kappa_\lambda)}=1$ level that is used to define the $\pm 1\sigma$ uncertainty on $\kappa_\lambda$ as well as the $-2 \ln{\Lambda(\kappa_\lambda)}=3.84$ level used to define the 95\% CL.}     
\label{scan_kl_ktrange}
\end{figure}

This effect is clear also looking at the 1D scan, shown in Figure~\ref{scan_kl_kt_asi_dati} both for data (black solid line) and for the Asimov dataset (red solid line). Due to the limited sensitivity of the double-Higgs analyses, even the $1\sigma$ interval cannot be reached in the $\kappa_\lambda$ likelihood scan profiling $\kappa_t$ and the curves are almost flat. The low constraining power, represented by a small raising of the likelihood distribution for negative $\kappa_\lambda$, comes from the parameterisation as a function of $\kappa_\lambda$ and $\kappa_t$ of the branching fractions and of the single-Higgs production cross sections in double-Higgs background, otherwise the curves become completely flat.\newline
From the 2D contours, it seems that, restricting $\kappa_t$ lower bound away from the best-fit value, greater values of the likelihood compared to the minimum one can be found and thus the $1\sigma$ interval or even 95\% CL can be reached. This effect is more evident looking at Figure~\ref{scan_kl_ktrange} where the value of $-2 \ln{\Lambda(\kappa_\lambda)}$ as a function of $\kappa_\lambda$ is shown varying $\kappa_t$ ranges for the fit to data.
It will be shown in Chapter~\ref{sec:combination} that the combination with the single-Higgs analyses allows to solve the degeneracy between $\kappa_\lambda$ and $\kappa_t$ and restores the ability of the double-Higgs analyses in constraining $\kappa_\lambda$.

\chapter{Constraints on the Higgs-boson self-coupling from single-Higgs production and decay measurements}
\label{sec:single}
This chapter presents the results of the extraction of $\kappa_\lambda$ limits from single-Higgs production and decay modes exploiting the complementary approach to measure the Higgs self-coupling described in Chapter~\ref{sec:prob_self}, applying next-to-leading order electroweak corrections depending on $\kappa_\lambda$ at these processes. Section~\ref{sec:data_single} introduces data and input measurements together with key elements of the analyses included in the combination. How the theoretical framework described in Chapter~\ref{sec:prob_self} is implemented in single-Higgs production and decay modes is briefly summarised in Section~\ref{sec:theory_single} while Section~\ref{sec:statistical_model_single} expands the statistical description reported in Chapter~\ref{sec:dihiggs} in order to introduce elements necessary to produce the results of this chapter. Constraints on $\kappa_\lambda$ in different fit configurations are reported in Sections~\ref{sec:results_single_kl} and~\ref{sec:results_single_otherfit} while Section~\ref{sec:HL_LHC_single} reports projections of single-Higgs results considering a luminosity of 3000 fb$^{-1}$.

\section{Data and input measurements}
\label{sec:data_single}
\begin{figure}[htbp]
\begin{center}
\includegraphics[height=6.5 cm,width=\textwidth]{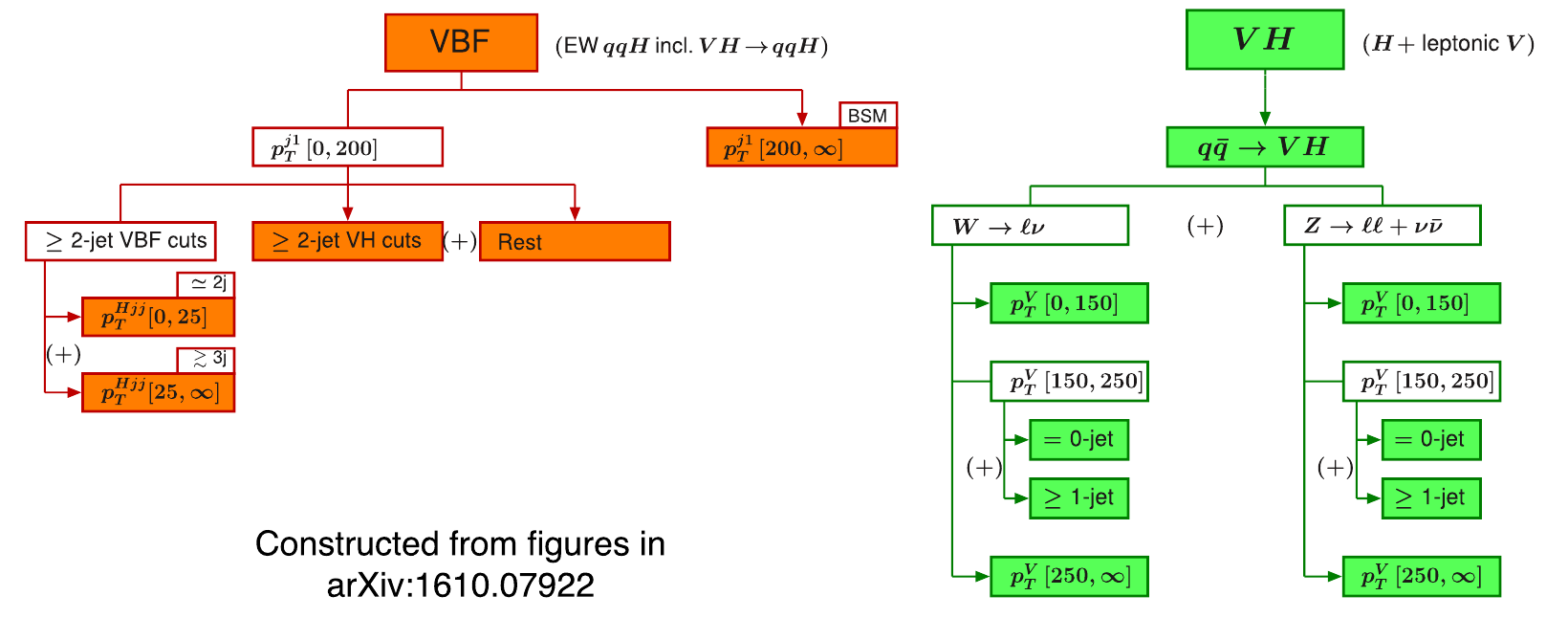}
\end{center}
\caption{Schematic diagram of the \VBF + V(had)H (left) and V(lep)H (right) STXS regions~\cite{PubNote}.}
\label{stxs_bin}
\end{figure}

The results presented in this chapter are obtained using data collected by the ATLAS experiment in 2015, 2016 and 2017 from 13 TeV $pp$ collision data corresponding to an integrated luminosity of up to 79.8 fb$^{-1}$. The integrated luminosity for each analysed decay channel is summarised in Table~\ref{lumi_single_ref}. The single-Higgs analyses exploited in order to make this combination include the $ggF$, \VBF, \WH, \ZH and $t\bar{t}H$ production modes and the $\gamma\gamma$, $WW^*$, $ZZ^*$, $b\bar{b}$ and $\tau^+\tau^-$ decay channels. 
\begin{table}[!htbp]
\scalebox{0.94}{
{\def\arraystretch{1.3}
\begin{tabular}{|l|c|c|}
\hline
Analysis & Integrated luminosity (fb$^{-1}$) & References \\
\hline
$H \rightarrow \gamma \gamma$\ (including $t\bar{t}H$, $H\rightarrow \gamma \gamma$)       & $79.8$         & \cite{yy,yy1,ttH_yy} \\
$H\rightarrow ZZ^*\rightarrow 4\ell$ (including $t\bar{t}H$, $H\rightarrow ZZ^*\rightarrow 4\ell$) & $79.8$         & \cite{ZZ,ZZ1} \\
$H\rightarrow WW^* \rightarrow e\nu \mu \nu$                        & $36.1$         & \cite{WW} \\
$H\rightarrow \tau\tau$                               & $36.1$         & \cite{tautau} \\
$VH$, $H\rightarrow b\bar{b}$                          & $79.8$         & \cite{VH_bb,VH_bb1} \\
$t\bar{t}H$, $H\rightarrow b\bar{b}$ and $t\bar{t}H$ multilepton  & $36.1$         & \cite{ttH,ttH_bb} \\
\hline
\end{tabular}}}
\caption{Integrated luminosity of the datasets used for each input analysis to the single-Higgs combination. The last column provides references to publications describing each measurement included in detail.}
\label{lumi_single_ref}
\end{table}
All single-Higgs regions are defined for a Higgs boson rapidity $y_H$ satisfying $|y_H | $< $ 2.5$. Jets are reconstructed from all stable particles with a lifetime greater than 10 ps, excluding the Higgs decay products, using the anti-$k_t$ algorithm with a jet radius parameter $R=0.4$, and must have a transverse momentum $p_{T,jet}>$30 GeV.\newline
The simplified template cross-section (STXS) framework is used, when available, for single-Higgs production modes in order to minimise the dependence on theoretical uncertainties that are directly folded into the measurements and to maximise experimental sensitivity to different processes. Several stages with an increasing number of bins are defined within this framework; in particular, the categories included in this combination are based on the stage-1 of the STXS framework within which, depending on the Higgs-boson production mode, the phase space is subdivided as follows~\cite{Higgs_CS,stxs_framework}:
\begin{itemize}
\item[-]the gluon-fusion production mode is subdivided in regions defined by
  jet multiplicity and transverse momentum of the Higgs boson,
  $p_\text{T}^H$. Additionally, two regions with \VBF-like kinematics,
  defined by the presence of two high-$\eta$ jets with large dijet mass, are
  considered. The $b\bar{b}H$ and $gg\rightarrow Z(\text{had})H$ production modes are considered as small additional contributions to the expected yields in each STXS bin;
  \item[-] the phase space of quark-initiated production processes $qq \rightarrow H qq$ is split using, as a variable, the transverse momentum of the highest-$p_T$ jet, called $p_T^{j1}$;  in fact, the lower $p_T^{j1}$ region is expected to be dominated by SM-like events, while the high-$p_T^{j1}$ region is sensitive to potential BSM contributions. Other regions are then defined, \ie\ two \VBF-topology regions by using the presence of two jets with $m_{jj} \ge$400~GeV and a pseudorapidity difference $|\Delta \eta_{jj}| \ge 2.8$, a region with two jets consistent with $V(\rightarrow qq)H$ production, and a region for the remaining events;
\item[-] $VH$ production mode is split according to the vector boson, \ie\ $W\rightarrow \ell \nu$ and $Z\rightarrow \ell \ell + \nu \bar{\nu}$ and a further split into bins of $p_T^V$ is made, aligned with the quantity used in the $H\rightarrow b\bar{b}$ analysis, representing one of the dominant contribution in the \VH bins. Bins depending on the number of jets, reflecting the different experimental sensitivity and with the target of avoiding the corresponding theory dependence, are also used;
\item[-] the $t\bar{t}H$ and $tH$ production modes are considered inclusively in one single region.
\end{itemize}
The stage-1 of the STXS framework exploited in this combination are reported in Figure~\ref{stxs_bin}.
The single-Higgs analysis categories are summarised in Tables~\ref{tab:category_single_ttH_VH} and~\ref{tab:category_single_VBF_ggF}~\cite{Coupling_run2}:
\begin{itemize}
\item[-] $t\bar{t}H$ production mode:
\begin{itemize}
\item seven categories are defined to select  $H\rightarrow \gamma \gamma$ decay channel, including both semileptonic and hadronic top-quark decay processes through various selections on the multiplicities and kinematics of leptons, jets, and jets tagged as containing $b$-hadrons;
\item two categories are defined to select $H\rightarrow ZZ^*\rightarrow 4\ell$ decay channel, with at least one $b$-tagged jet and three or more additional jets, or one additional lepton and at least two additional jets, with fully hadronic or semileptonic top-quark decays, respectively;
\item seven categories are defined to select $H\rightarrow WW^*, \,ZZ^* \,(\text{not}\, 4\ell),\, \tau^+\tau^-$ decay channels; they are categorised by the number and the flavour of reconstructed charged-lepton candidates: one lepton with two $\tau_{had}$ candidates, two same-charge leptons with zero or one $\tau_{had}$ candidates, two opposite-charge leptons with one $\tau_{had}$ candidate, three leptons with zero or one $\tau_{had}$ candidates, and four leptons. Events in all channels are required to have at least two jets, at least one of which must be $b$-tagged;
\item nineteen categories are defined to select $H\rightarrow b\bar{b}$ decay channel, events are classified into eleven (seven) orthogonal categories in the single-lepton (dilepton) channel, according to the jet multiplicity and to the values of the $b$-tagging discriminant for the jets. In the single-lepton channel, an additional category, referred to as ``boosted$"$, is designed to select events with large transverse momenta for the Higgs candidate ($p^H_T>$200~GeV) and one for the top-quark candidates ($p^t_T>$250~GeV). 
\end{itemize}
\item[-] \VH production mode:
\begin{itemize}
\item eight categories are defined to select  $\gamma \gamma$ decay channel: five categories are defined to select \WH and \ZH production modes with leptonic decays of the $W$ or of the $Z$ bosons, based on the presence of leptons and missing transverse momentum $E_T^{miss}$, one category requires the presence of two jets, with the leading jet transverse momentum $p^{j1}_T>$200~GeV and two categories select hadronic vector-boson decays by requiring two jets with an invariant mass compatible with the $W$ or $Z$ boson mass;
\item three categories are defined to select $H\rightarrow ZZ^*\rightarrow 4\ell$ decay channel, one category with leptonic vector-boson decays, and the other two categories with $0$-jet and $p_T^{4\ell}\ge$100~GeV, and at least two jets with a hadronically decaying vector boson for $m_{jj}<$120~GeV;
\item eight categories are defined to select $H\rightarrow b\bar{b}$ decay channel, where events are classified according to the charged-lepton multiplicity, the vector-boson transverse momentum, $p^V_T$, and the jet multiplicity. For final states with zero or one lepton, $p^V_T>$150~GeV is required. In two-lepton final states, two regions are considered, 75~GeV$< p^V_T<$150~GeV and $p^V_T>$150~GeV. Each of these regions is finally separated into a category with exactly two reconstructed jets and another with three or more. 
\end{itemize}
\item[-] \VBF production mode:
\begin{itemize}
\item four categories are defined for the $\gamma \gamma$ decay channel, in order to enrich VBF production by requiring forward jets in a VBF-like topology region;
\item two categories are defined to select $H\rightarrow ZZ^*\rightarrow 4\ell$ decay channel, in which the transverse momentum of the leading jet, $p^{j1}_T$, is required to be either above or below 200~$\GeV$;
\item one category is defined to select $H\rightarrow WW^*$ decay channel, containing events with the number of jets $\ge$ $2$, naturally sensitive to this production mode;
\item three categories are defined to select $H\rightarrow \tau^+\tau^-$ decay channel, one category is defined by requiring the transverse momentum of the $\tau^+\tau^-$ system, $p_T^{\tau\tau}$, to be above 140~GeV, for $\tau_{had}\tau_{had}$ events only, while the other two categories are defined for lower and higher values of $m_{jj}$.
\end{itemize}
\item[-] \ggF production mode:
\begin{itemize}
\item ten categories are defined for the $\gamma \gamma$ decay channel, separating events with 0, 1, and $\ge$ 2 jets and further classifying them according to the pseudorapidity of the two photons (for 0-jet events) or according to the transverse momentum of the diphoton system, $p_T^{\gamma \gamma}$, (for 1 and $\ge$ 2-jet events);
\item four categories are defined to select $H\rightarrow ZZ^*\rightarrow 4\ell$ decay channel, one category containing $0$-jet events and $p_T^{4\ell}<$100~GeV and the other three categories containing $1$-jet events with boundaries at $p_T^{4\ell}$=60~GeV and $p_T^{4\ell}$=120~GeV;
\item sixteen categories are defined to select $H\rightarrow WW^*$ decay channel, based on the flavour of the leading lepton ($e$ or $\mu$), in combination with two bins of the invariant mass of the dilepton system, $m_{\ell\ell}$, and with two bins of the transverse momentum of the subleading lepton, $p_T^{\ell_2}$;
\item two categories are defined to select $H\rightarrow \tau^+\tau^-$ decay channel, with selections on the angular separation between the $\tau$-leptons and requiring $p_T^{\tau\tau} >$ 140~GeV and $p_T^{\tau\tau} \le$ 140~GeV.
\end{itemize}
\end{itemize}
\begin{landscape}
\begin{table}[htbp]
\caption{Summary of the signal categories coming from the $t\bar{t}H$ and \VH production modes entering the combined measurements. The following conventions are adopted: $\ell$  refers to $e$ or $\mu$, $\ell_2$ to the lepton of lowest $p_\text{T}$, $E_\text{T}^{\textrm{miss}}$ to the missing transverse energy, $j$ to a light jet, $j1$ to the light jet of highest $p_\text{T}$; $p_\text{T}^{\ell+E_\text{T}^{\textrm{miss}}}$ is the $p_\text{T}$ of the $\ell,E_\text{T}^{\textrm{miss}}$ system, $p_\text{T}^{4\ell}$ the $p_\text{T}$ of the four lepton system and similarly for $p_\text{T}^{\gamma \gamma}$, $p_\text{T}^{\tau \tau}$ and $p_\text{T}^{\gamma \gamma jj}$. $p_\text{T}^{V}$ refers to the $p_\text{T}$ of the vector boson ($V$) in the $VH$ category, $m_{\ell \ell}$ is the invariant mass of the di-lepton system~\cite{PubNote}.}
\label{tab:category_single_ttH_VH}
\vspace*{+0.4cm}
\begin{center}
\resizebox{\columnwidth}{!}{
\footnotesize
{\def\arraystretch{1.4}
\begin{tabular}{|l|l|l|l|l|l|}
\hline
& $H\rightarrow \gamma \gamma$ & $H\rightarrow ZZ^*$ & $H\rightarrow WW^*$ & $H\rightarrow \tau \tau$ & $H\rightarrow b\bar{b}$ \\
\hline
\multirow{7}{*}{$t\bar{t}H$} 
 & $t\bar{t}H$ leptonic (3 categories) & \multicolumn{3}{l|}{$t\bar{t}H$ multilepton 1 $\ell$ + 2 $\tau_{had}$}                                   & $t\bar{t}H$ 1 $\ell$, boosted                  \\
 & $t\bar{t}H$ hadronic (4 categories) & \multicolumn{3}{l|}{$t\bar{t}H$ multilepton 2 opposite-sign $\ell$ + 1 $\tau_{had}$}                     & $t\bar{t}H$ 1 $\ell$, resolved (11 categories) \\
 &                               & \multicolumn{3}{l|}{$t\bar{t}H$ multilepton 2 same-sign $\ell$ (categories for $0$ or $1$ $\tau_{had}$)} & $t\bar{t}H$ 2 $\ell$ (7 categories)            \\
 &                               & \multicolumn{3}{l|}{$t\bar{t}H$ multilepton 3 $\ell$ (categories for $0$ or $1$ $\tau_{had}$)}           & \\
 &                               & \multicolumn{3}{l|}{$t\bar{t}H$ multilepton 4 $\ell$ (except $H\rightarrow ZZ^* \rightarrow 4 \ell$)}                            & \\
 &                               & \multicolumn{3}{l|}{$t\bar{t}H$ leptonic, $H\rightarrow ZZ^* \rightarrow 4 \ell$}                                                & \\
 &                               & \multicolumn{3}{l|}{$t\bar{t}H$ hadronic, $H\rightarrow ZZ^* \rightarrow 4 \ell$}                                                & \\
\hline
\multirow{8}{*}{$VH$} 
 & $VH$ 2 $\ell$                             & $VH$ leptonic                     & & & 2 $\ell$, $75 \leq p_T^V < 150$ \text{GeV}, $N_{jets}=2$      \\
 & $VH$ 1 $\ell$, $p_T^{\ell+E_T^{miss}}$ $\geq 150$ \text{GeV}    &                                   & & & 2 $\ell$, $75 \leq p_T^V <$ 150 \text{GeV}, $N_{jets} \geq 3$ \\
 & $VH$ 1 $\ell$, $p_T^{\ell+E_T^{miss}}$ $<$ 150 \text{GeV}           &                                   & & & 2 $\ell$, $p_T^V \geq 150$ \text{GeV}, $N_{jets}=2$           \\
 & $VH$ $E_T^{miss}$, $E_T^{miss}$ $\geq 150$ \text{GeV}           & 0-jet, $p_T^{4\ell} \geq 100$ \text{GeV}   & & & 2 $\ell$, $p_T^V \geq 150$ \text{GeV}, $N_{jets} \geq 3$      \\
 & $VH$ $E_T^{miss}$, $E_T^{miss}$ $<$ 150 \text{GeV}                  &                                   & & & 1 $\ell$  $p_T^V \geq 150$ \text{GeV}, $N_{jets}=2$           \\
 & $VH$+\VBF $p_T^{j1}$ $\geq 200$ \text{GeV}     &                                   & & & 1 $\ell$  $p_T^V \geq 150$ \text{GeV}, $N_{jets}=3$           \\
 & $VH$ hadronic (2 categories)              & 2-jet, $m_{jj} < 120$ \text{GeV}          & & & 0 $\ell$, $p_T^V \geq 150$ \text{GeV}, $N_{jets}=2$           \\
 &                                           &                                   & & & 0 $\ell$, $p_T^V \geq 150$ \text{GeV}, $N_{jets}=3$           \\
\hline
\end{tabular}}
}
\end{center}
\end{table}
\end{landscape}

\begin{landscape}
\begin{table}[htbp]
\caption{Summary of the signal categories coming from the \VBF and \ggF production modes entering the combined measurements. The following conventions are adopted: $\ell$  refers to $e$ or $\mu$, $\ell_2$ to the lepton of lowest $p_\text{T}$, $E_\text{T}^{\textrm{miss}}$ to the missing transverse energy, $j$ to a light jet, $j1$ to the light jet of highest $p_\text{T}$; $p_\text{T}^{\ell+E_\text{T}^{\textrm{miss}}}$ is the $p_\text{T}$ of the $\ell, E_\text{T}^{\textrm{miss}}$ system, $p_\text{T}^{4\ell}$ the $p_\text{T}$ of the four lepton system and similarly for $p_\text{T}^{\gamma \gamma}$, $p_\text{T}^{\tau \tau}$ and $p_\text{T}^{\gamma \gamma jj}$. $p_\text{T}^{V}$ refers to the $p_\text{T}$ of the vector boson ($V$) in the $VH$ category, $m_{\ell \ell}$ is the invariant mass of the di-lepton system~\cite{PubNote}.}
\label{tab:category_single_VBF_ggF}
\vspace*{+1.4cm}
\begin{center}
\resizebox{\columnwidth}{!}{
\footnotesize
{\def\arraystretch{1.4}
\begin{tabular}{|l|l|l|l|l|l|}
\hline
& $H\rightarrow \gamma \gamma$ & $H\rightarrow ZZ^*$ & $H\rightarrow WW^*$ & $H\rightarrow \tau \tau$ & $H\rightarrow b\bar{b}$ \\
\hline

\multirow{4}{*}{\VBF}
 & \VBF, $p_T^{\gamma\gamma j j }$ $\geq 25$ \text{GeV} (2 categories) & 2-jet \VBF, $p_T^{j1}\geq 200$ \text{GeV} & 2-jet \VBF & \VBF, $p_T^{\tau \tau}>140$ \text{GeV} &  \\
 & \VBF, $p_T^{\gamma\gamma}$$<$ 25 \text{GeV} (2 categories)        & 2-jet \VBF, $p_T^{j1}$ $< 200$ \text{GeV}       &            & \quad  ($\tau_{had}\tau_{had}$ only)       &  \\
 &                                              &                                        &            & \VBF high-$m_{jj}$          &           \\
 &                                              &                                        &            & \VBF low-$m_{jj}$           &                         \\
\hline
\multirow{9}{*}{$ggF$}
 & 2-jet, $p_T^{\gamma\gamma}$ $\geq 200$ \text{GeV}          & 1-jet, $p_T^{4\ell} \geq 120$ \text{GeV} & 1-jet, $m_{\ell\ell}< 30$ \text{GeV}, $p_T^{\ell_2}<20$ \text{GeV}      & Boosted, $p_T^{\tau \tau} > 140$ \text{GeV}  &  \\
 & 2-jet, 120 \text{GeV} $\leq$ $p_T^{\gamma\gamma}<$ 200 \text{GeV} & 1-jet, 60 \text{GeV}$\leq$ $p_T^{4\ell}$$<$ 120 \text{GeV}  & 1-jet, $m_{\ell\ell}< 30$ \text{GeV}, $p_T^{\ell_2}\geq 20$ \text{GeV}  & Boosted, $p_T^{\tau \tau} \leq 140$ \text{GeV}  & \\
 & 2-jet, 60 \text{GeV} $\leq$ $p_T^{\gamma\gamma}<$ 120 \text{GeV}  & 1-jet, $p_T^{4\ell} < 60$ \text{GeV}     & 1-jet, $m_{\ell\ell}\geq 30$ \text{GeV}, $p_T^{\ell_2}<$ 20 \text{GeV}      &                               &  \\
 & 2-jet, $p_T^{\gamma\gamma} <$ 60 \text{GeV}              & 0-jet, $p_T^{4\ell} < 100$ \text{GeV}    & 1-jet, $m_{\ell\ell}\geq 30$ \text{GeV}, $p_T^{\ell_2}\geq 20$ \text{GeV}  &                               &  \\
 & 1-jet, $p_T^{\gamma\gamma} \geq$ 200 \text{GeV}          &                                 & 0-jet, $m_{\ell\ell}<    30$ \text{GeV}, $p_T^{\ell_2}<20$ \text{GeV}      &                               &  \\
 & 1-jet, 120 \text{GeV} $\leq$ $p_T^{\gamma\gamma} <$ 200 \text{GeV} &                                 & 0-jet, $m_{\ell\ell}<    30$ \text{GeV}, $p_T^{\ell_2}\geq 20$ \text{GeV}  &                               &  \\
 & 1-jet, 60 \text{GeV} $\leq$ $p_T^{\gamma\gamma} <$ 120 \text{GeV}  &                                 & 0-jet, $m_{\ell\ell}\geq 30$ \text{GeV}, $p_T^{\ell_2}<$ 20 \text{GeV}      &                               &  \\
 & 1-jet, $p_T^{\gamma\gamma}  <$ 60 \text{GeV}              &                                 & 0-jet, $m_{\ell\ell}\geq 30$ \text{GeV}, $p_T^{\tau \tau}\geq 20$ \text{GeV}  &                               &  \\
 & 0-jet (2 categories)                  &                                 &                                                       &                               &  \\
\hline
\end{tabular}}
}
\end{center}
\end{table}
\end{landscape}

\section{Implementation of the theoretical model}
\label{sec:theory_single}

The theoretical framework described in Chapter~\ref{sec:prob_self} is implemented in the single-Higgs channels, \ie\  in the parameterisations of the signal strengths, defined in Chapter~\ref{sec:SM}; for the single-Higgs initial states, $i$:
\begin{equation}
\label{eq:mui_single}
\mu_i(\kappa_\lambda,\kappa_i) = \frac{\sigma^{\textrm{BSM}}
}{\sigma^{\textrm{SM}}} =
Z_{H}^{\textrm{BSM}}\left(\kappa_\lambda\right)\left[\kappa_i^2+\frac{(\kappa_\lambda-1)C_1^i}{K^i_{\mathrm{EW}}}\right] \, ,
\end{equation}
while the Higgs-boson decay final states, $f$, are modified as:
\begin{equation}
\label{eq:muf_single}
\mu_f(\kappa_\lambda,\kappa_f) = \frac{\textrm{BR}_f^{\textrm{BSM}}}{\textrm{BR}_f^{SM}} =
\frac{\kappa_f^2+(\kappa_\lambda-1)C_1^{f}}{\sum_{j}\textrm{BR}_j^{\textrm{SM}}\left[\kappa_j^2+(\kappa_\lambda-1)C^{j}_1  \right]} 
\end{equation}
being $\kappa_i$ and $\kappa_f$ the LO modifiers of the Higgs couplings to SM particles.\newline
Concerning the inclusive production modes, the values of the $C_1$ and $K_{EW}$ coefficients as well as the $\kappa$ modifiers at LO for the initial state $i$, are reported in Table~\ref{CKcoeff} for the $ggF$, \VBF, \ZH, \WH and $t\bar{t}H$ production modes; they are taken from References~\cite{Degrassi,Maltoni,Coupling_run2} and are averaged over the full phase space accessible through these processes. In this chapter, only the coupling modifiers $\kappa_F=\kappa_t=\kappa_b=\kappa_{\ell}$ and $\kappa_V=\kappa_W=\kappa_Z$ are considered, describing the modifications of the SM Higgs boson couplings to fermions and vector bosons, respectively. For small deviations of the coupling modifiers from one, the dependence of NLO-EW corrections on these coupling modifiers can be neglected.
\begin{table}[htbp]
\begin{center}
\scalebox{0.95}{
{\def\arraystretch{1.3}
\begin{tabular}{|c|c|c|c|c|c|}
\hline
Production mode & $ggF$ & \VBF & \ZH & \WH & $t\bar{t}H$ \\
\hline
$C_1^i\times 100$ & 0.66 & 0.63 & 1.19 & 1.03 & 3.52 \\
\hline
$K^i_{\textrm{EW}}$ & 1.049 & 0.932 & 0.947 & 0.93 & 1.014 \\
\hline
$\kappa_i^2$ &  $1.04\,\kappa_t^2+0.002\,\kappa_b^2-0.04\,\kappa_t\kappa_b$ & $0.73\,\kappa_W^2 + 0.27\,\kappa_Z^2$ & $\kappa_Z^2$ &$\kappa_W^2$& $\kappa_t^2$ \\
\hline  
\end{tabular} 
}}
\end{center}
\caption{Values of the $C^i_1$ coefficients, representing linear $\kappa_\lambda$-dependent corrections to single-Higgs production modes (second row); values of the $K^i_\textrm{EW}$ coefficients~\cite{Degrassi, Maltoni}, taking into account NLO EW corrections in the SM hypothesis (third row); expressions of the initial state $\kappa$ modifiers at LO, $\kappa_i^2$,~\cite{Coupling_run2} for Higgs-boson production processes (fourth row).}
   \label{CKcoeff}
\end{table} 

The $C_{1}^f$ coefficients and the expressions of the $\kappa$ modifiers at LO for the final state $f$, $\kappa_f^2$, are reported in Table~\ref{tab:decays} for all the analysed decay modes. 
\begin{table}[htbp]
\begin{center}
\scalebox{0.95}{
{\def\arraystretch{1.3}
\begin{tabular}{|c|c|c|c|c|c|}
\hline 
Decay mode & $H\rightarrow\gamma\gamma$ & $H\rightarrow WW^*$ & $H\rightarrow ZZ^*$ & $H\rightarrow b\bar{b}$  & $H\rightarrow \tau\tau$ \\
\hline
$C^f_1\times 100$ & 0.49 & 0.73 & 0.82 & 0 & 0 \\
\hline
$\kappa_f^2$ & $1.59 \kappa_W^2+0.07\kappa_t^2 - 0.67 \kappa_W\kappa_t$ & $\kappa_W^2$ & $\kappa_Z^2$ &$\kappa_b^2$& $\kappa_{\ell}^2$ \\
  \hline  
\end{tabular}
}}
\end{center}
\caption{Values of $C^f_1$~\cite{Degrassi, Maltoni} coefficients, representing linear $\kappa_\lambda$-dependent corrections to single-Higgs decay channels (second row); expressions of the final state $\kappa$ modifiers at LO, $\kappa_f^2$,~\cite{Coupling_run2} for Higgs-boson decay modes (third row).} 
\label{tab:decays}
\end{table} 
The interactions between the Higgs boson and the gluons and photons are resolved in terms of the coupling modifiers of the SM particles that enter in the loop-level diagrams, \ie\ in terms of the coupling modifiers $\kappa_b$=$\kappa_F$ and $\kappa_t$=$\kappa_F$, $\kappa_W$=$\kappa_V$ and $\kappa_t$=$\kappa_F$, respectively.\newline
The simulation of the single-Higgs signal samples is performed using Montecarlo simulations generated under the SM hypothesis. This is possible because at the lowest order in the electroweak expansion only one diagram
participates in the single-Higgs boson production, therefore the $\kappa$-modifiers ($\kappa_t$, $\kappa_b$, $\kappa_{lep}$, $\kappa_W$ and $\kappa_Z$) factorise completely the total cross section; this holds also for all decays.
The NLO-EW corrections depending on $\kappa_\lambda$, presenting quadratically and linearly dependent contributions, affect not only the inclusive rates of Higgs-boson production and decay processes, but also their kinematics, varying as a function of variables like $p_T^H$. In particular, the largest deviations in kinematic distributions with respect to the SM are expected in the \ZH, \WH, and $t\bar{t}H$ production modes, while in Higgs-boson decay kinematics, no significant modifications are expected, as shown in Chapter~\ref{sec:prob_self}.\newline
This dependence has been partially taken into account by exploiting cross-section measurements in the regions defined by the STXS stage-1 framework defined in the previous section. Not all the differential information available for different production modes is used; in fact, the gluon fusion production mode is subdivided in regions defined by jet multiplicity and transverse momentum of the Higgs boson, but differential corrections are not yet available, involving higher order calculations including two loop corrections; therefore the corresponding STXS bins share the same parameterisation and coefficients used for the inclusive $ggF$ production. The situation is reversed looking at $t\bar{t}H$ production mode that is considered inclusively in one single bin but because of the fact that, for this production mode, the STXS binned analyses are still under development. Furthermore, the $gg\rightarrow ZH$ cross section is not parameterised as a function of $\kappa_\lambda$ due to the missing theoretical computations and contributing mostly in high-$p_T^H$ regions where the sensitivity to $\kappa_\lambda$ is small, as shown in Figure~\ref{fig:differential_wh_zh} $(b)$.\newline
The parameterisation as a function of $\kappa_\lambda$ of the signal strengths for the production modes, shown in Equation~\ref{eq:mui_single}, can be adapted to describe the cross section in each STXS bin. This can be achieved re-deriving the value of the kinematic-dependent coefficients $C_1^i$ in each region defined in the measurement~\cite{PubNote}: the $C_1^i$ coefficients have thus been computed using samples of events generated at LO EW using \amc\ 2.5.5~\cite{C1_comp}, and reweighted on an event-by-event basis with the tool provided in Reference~\cite{C1_comp1}. The $C_1^{i}$ values for each STXS bin are reported in Table \ref{C1_stxs}.\newline
Modifications of the acceptances of the different production modes of the STXS framework have been tested, considering \ZH, \WH, and $t\bar{t}H$ production modes being characterised by a stronger kinematic-dependence on the Higgs self-coupling corrections with respect to the other modes. The largest variation is observed for $\kappa_\lambda <$-10 for the $t\bar{t}H \rightarrow \gamma \gamma$ channel, otherwise the acceptance is almost constant.\newline
Figure \ref{fig:fid_cros_kl} shows the ratio $\sigma_{BSM}/\sigma_{SM}$ for each \VBF, \WH and \ZH STXS bin normalised to the same ratio computed for the inclusive cross section of the corresponding production process. The STXS bins with the highest $p_T$ are the ones with the strongest dependence on $\kappa_\lambda$, otherwise there is a negligible dependence; the \WH and \ZH are the production modes showing the largest kinematic dependence (excluding the $t\bar{t}H$ that does not have STXS bins).
\begin{table}
\small
{\def\arraystretch{1.3}
\begin{center}
\begin{tabular}{|l|l|c|c|c|}
\hline
\multicolumn{2}{|c|}{\multirow{2}{*}{STXS region}}&\VBF &\WH & \ZH \\ \cline{3-5}
\multicolumn{2}{|c|}{}&\multicolumn{3}{c|}{$C_1^i \times 100$}\\
\hline
\multirow{5}{*}{\VBF+V$(\text{had})H$}&$VBF$-cuts $+\, p_\text{T}^{j1}<200$ GeV, ${}\le 2j$&0.63&0.91&1.07\\
&$VBF$-cuts $+\, p_\text{T}^{j1}<200$ GeV, ${}\ge 3j$ &0.61&0.85&1.04\\
&$VH$-cuts $+\, p_\text{T}^{j1}<200$ GeV&0.64&0.89&1.10\\
&no $VBF$/$VH$-cuts, $p_\text{T}^{j1}<200$ GeV &0.65&1.13&1.28\\
&$p_\text{T}^{j1} > 200$ GeV  &0.39&0.23&0.28\\
\hline
\multirow{4}{*}{$qq\to H \ell \nu$}&$p_\text{T}^V < 150$ GeV &&1.15&\\
&$150 < p_\text{T}^V < 250$ GeV, 0$j$ &&0.18&\\
&$150 < p_\text{T}^V <250$ GeV, ${}\ge 1j$ &&0.33&\\
&$p_\text{T}^V > 250$ GeV &&0&\\
\hline
\multirow{2}{*}{$qq\to H \ell \ell$}&$p_\text{T}^V < 150$ GeV &&&1.33\\
&$150 < p_\text{T}^V < 250$ GeV, 0$j$ &&&0.20\\
\multirow{2}{*}{$qq\to H \nu \nu$}&$150 < p_\text{T}^V <250$ GeV, ${}\ge 1j$ &&&0.39\\
&$p_\text{T}^V > 250$ GeV &&&0\\
\hline
\end{tabular}
\end{center}}
\caption{$C_1^i$ coefficients for each region of the STXS scheme for the \VBF, \WH and \ZH production modes. The same definition for STXS regions and production modes as in Tables~\ref{tab:category_single_ttH_VH} and~\ref{tab:category_single_VBF_ggF} is used. In the \VBF categories, ``\VBF-cuts''~\cite{Higgs_CS} indicates selections applied to target the \VBF dijet topology, with requirements on the dijet invariant mass ($m_{jj}$) and the difference in pseudorapidity between the two jets; the additional $\leq 2j$ and $\geq 3j$ region separation is performed indirectly by requesting $p_\text{T}^{Hjj}\lessgtr 25$~GeV. ``$VH$-cuts'' select the $W,Z \to jj$ decays, requiring an $m_{jj}$ value close to the vector boson mass~\cite{Higgs_CS}. The $C_1^i$ coefficients of the $p_\text{T}^V > 250$~GeV regions are negligible, $\mathcal{O}(10^{-6})$, and are set to $0$~\cite{PubNote}.}
\label{C1_stxs}
\end{table}

\begin{figure}[tb]
\begin{subfigure}[t]{\textwidth}
\includegraphics[width =1\textwidth]{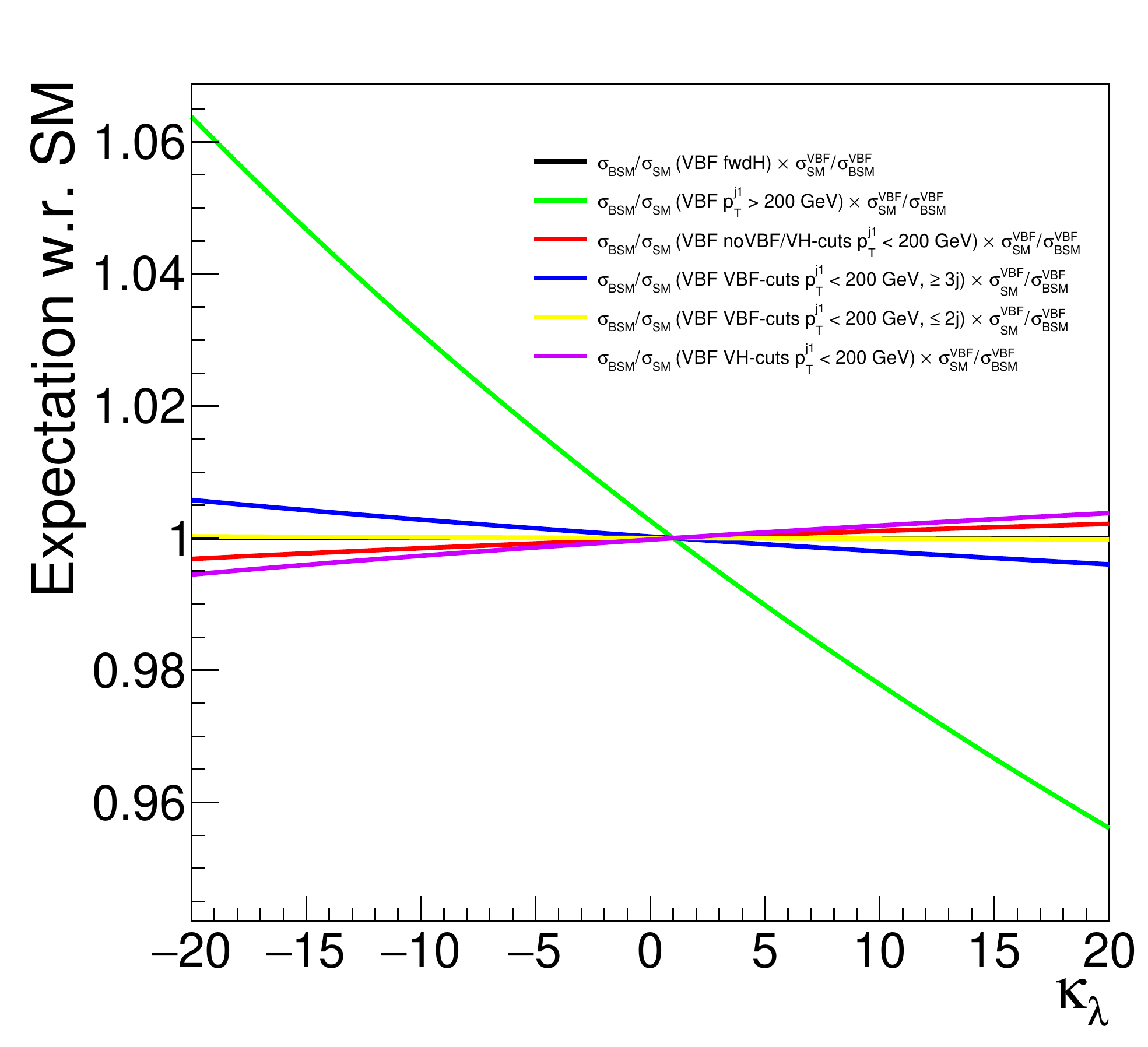}
 \caption{Variation of the fiducial cross section of the STXS categories normalised to the inclusive cross section of the corresponding production process: \VBF categories.}
\end{subfigure}
\end{figure}
\clearpage
    \begin{figure}[tb]\ContinuedFloat
  \begin{subfigure}[t]{1\textwidth}
  \includegraphics[width =\textwidth]{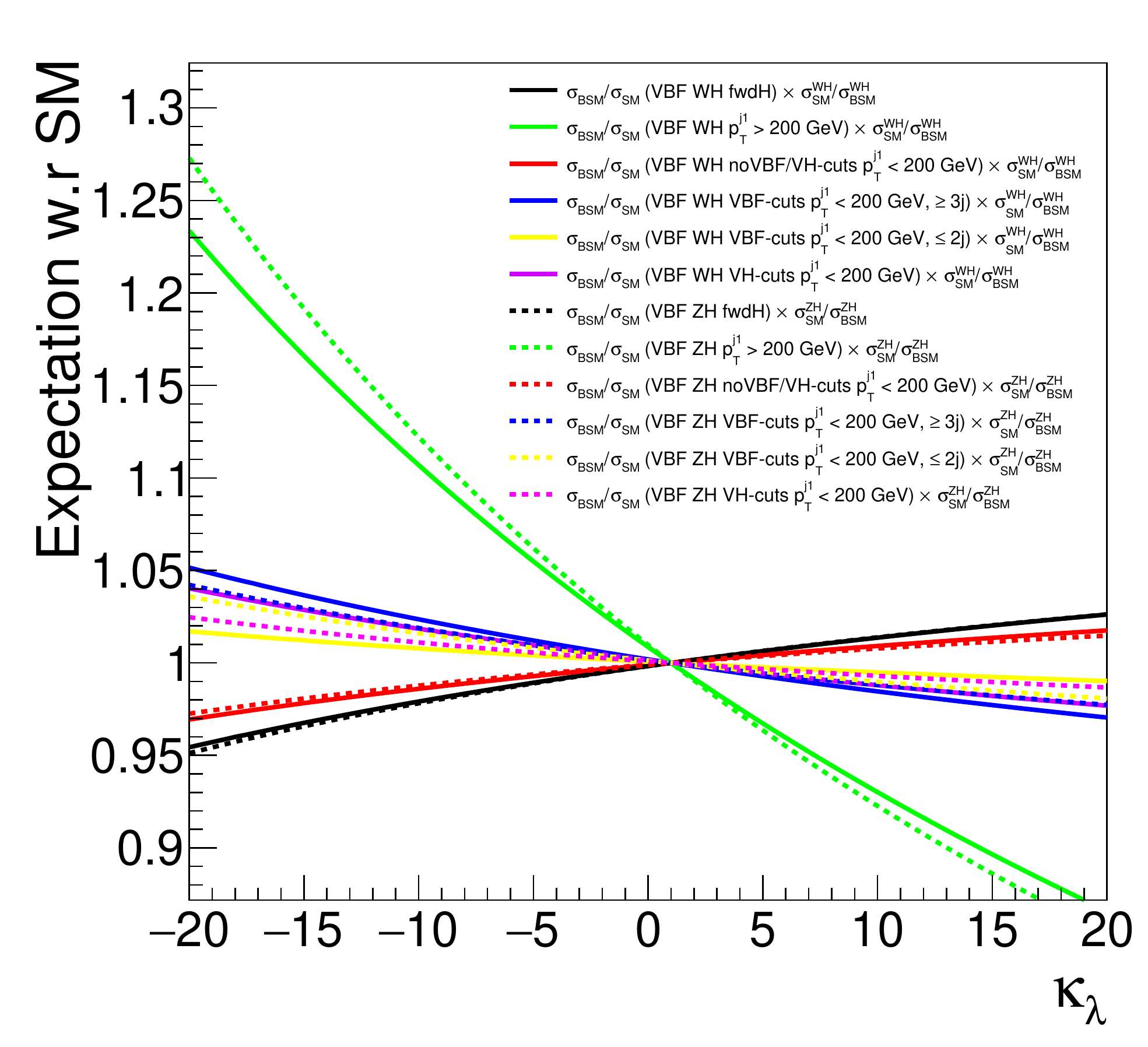}
   \caption{Variation of the fiducial cross section of the STXS categories normalised to the inclusive cross section of the corresponding production process: \WH/\ZH
  hadronic contributions in the \VBF categories.}
\end{subfigure}
\end{figure}
\clearpage
 \begin{figure}[tb]\ContinuedFloat
  \begin{subfigure}[t]{\textwidth}
  \includegraphics[width =1\textwidth]{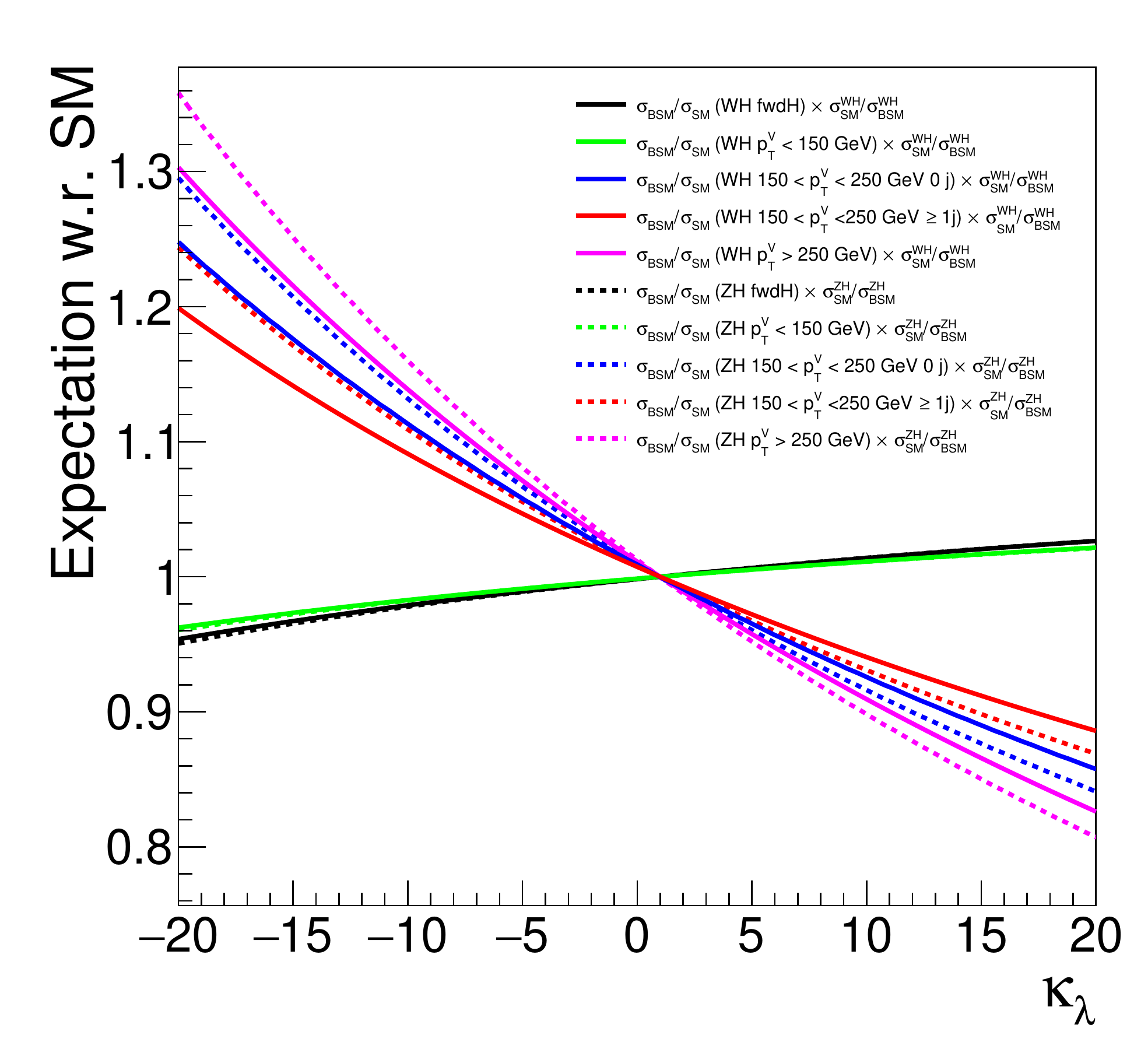}
   \caption{}
\end{subfigure}
\caption{Variation of the fiducial cross section of the STXS categories normalised to the inclusive cross section of the corresponding production process: (a) \VBF categories, (b) \WH/\ZH
  hadronic contributions in the \VBF categories, (c) \WH/\ZH categories.}
  \label{fig:fid_cros_kl}
\end{figure}
\clearpage
Even if electroweak corrections non-depending on $\kappa_\lambda$ can vary as a function of the kinematics of the process, inclusive $K_{EW}$ values, describing the ratio of NLO over LO cross-section in the SM case ($\kappa_\lambda=1$), have been used in the different phase-space regions.\newline
This approach has been followed after testing different $K_{EW}$ values for each STXS bin used in the analysis and ensuring that these corrections become more relevant for high value of the transverse momentum of the vector bosons, where the self-coupling correction is instead expected to be less significant; furthermore, the results from fits to the Asimov dataset and on data with the new $K_{EW}$ configuration differ by less than percent level with respect to the nominal results.

\section{Statistical model}
\label{sec:statistical_model_single}
The results presented in this chapter are obtained from a global likelihood function $L(\vec{\alpha},\vec{\theta})$, defined in Chapter~\ref{sec:stat}, where $\vec{\alpha}$ represents the vector of the parameters of interest of the model, \ie\ the coupling modifiers and the Higgs self-coupling, and $\vec{\theta}$ is the set of nuisance parameters. The global likelihood function is built as the product of the likelihood of the single-Higgs analyses that are themselves products of the likelihood of the different categories included in the analyses. Thus, the number of signal events in each analysis category $j$ is defined as:
\begin{equation}
\begin{split}
n^{\text{signal}}_j(\kappa_\lambda, \kappa_F, \kappa_V, \vec{\theta}) = {}&
\mathcal{L}_j(\vec{\theta}) \sum_i \sum_f \mu_{i}(\kappa_\lambda, \kappa_F, \kappa_V) \times \mu_{f}(\kappa_\lambda, \kappa_F, \kappa_V) \times \\
& \times (\sigma_{\text{SM},i}(\vec{\theta}) \times \text{BR}_{\text{SM},f}(\vec{\theta})) (\epsilon \times A)_{if,j}(\vec{\theta})
\end{split}
\label{eq:yields_single}
\end{equation} 
where, in this case, the index $i$ runs over all the production-process regions defined by the STXS stage-1 framework and the index $f$ over all the decay channels included in the combination, \ie\ $f= b\bar{b}, \gamma \gamma, \tau^+\tau^-, WW^*$ and $ZZ^*$.
The term $\mu_i(\kappa_\lambda, \kappa_F, \kappa_V)\times\mu_f( \kappa_\lambda,\kappa_F, \kappa_V)$, where $\mu_i$ is defined in Equation~\ref{eq:mui} and $\mu_f$ in Equation \ref{eq:muf}, describes the yield dependence on the Higgs-boson self-coupling modifier $\kappa_\lambda$, and on the single-Higgs boson coupling modifiers $\kappa_F$ and $\kappa_V$, representing potential deviations from the SM expectation of the self-coupling and of the couplings to vector bosons and to fermions, respectively.
A full description of other terms included in Equation~\ref{eq:yields_single} together with the test statistics used in order to determine confidence intervals for the parameters of interest are reported in Chapters~\ref{sec:stat} and~\ref{sec:dihiggs}. \newline
When presenting the results of the fit to $\kappa_\lambda$, its uncertainty is presented as decomposed in separate components that are:
\begin{itemize}
\item theoretical uncertainties affecting the background processes, \ie\ ``bkg. th.$"$; 
\item theoretical uncertainties affecting the Higgs-boson signal, \ie\ ``sig. th.$"$;
\item experimental uncertainties, \ie\ ``exp.$"$;
\item statistical uncertainties, \ie\ ``stat$"$.
\end{itemize}
The different components contributing to the total uncertainty are derived iteratively by fixing a given set of nuisance parameters to their best-fit values in the numerator and the denominator of the profile likelihood ratio repeating this procedure for each source of uncertainty following the order listed above. The value of each component is then evaluated as the quadratic difference between the resulting uncertainty at each step and the uncertainty obtained in the previous step, where for the initial step the total uncertainty is considered. The statistical uncertainty is evaluated as the last step, fixing to their best-fit values all the nuisance parameters except for the ones that are only constrained by data, such as the data-driven background normalisations.
\subsection{Systematic uncertainties}
\begin{figure}[hbtp]
\centering
\begin{subfigure}[b]{0.49\textwidth}
\includegraphics[height=12.9 cm,width =\textwidth]{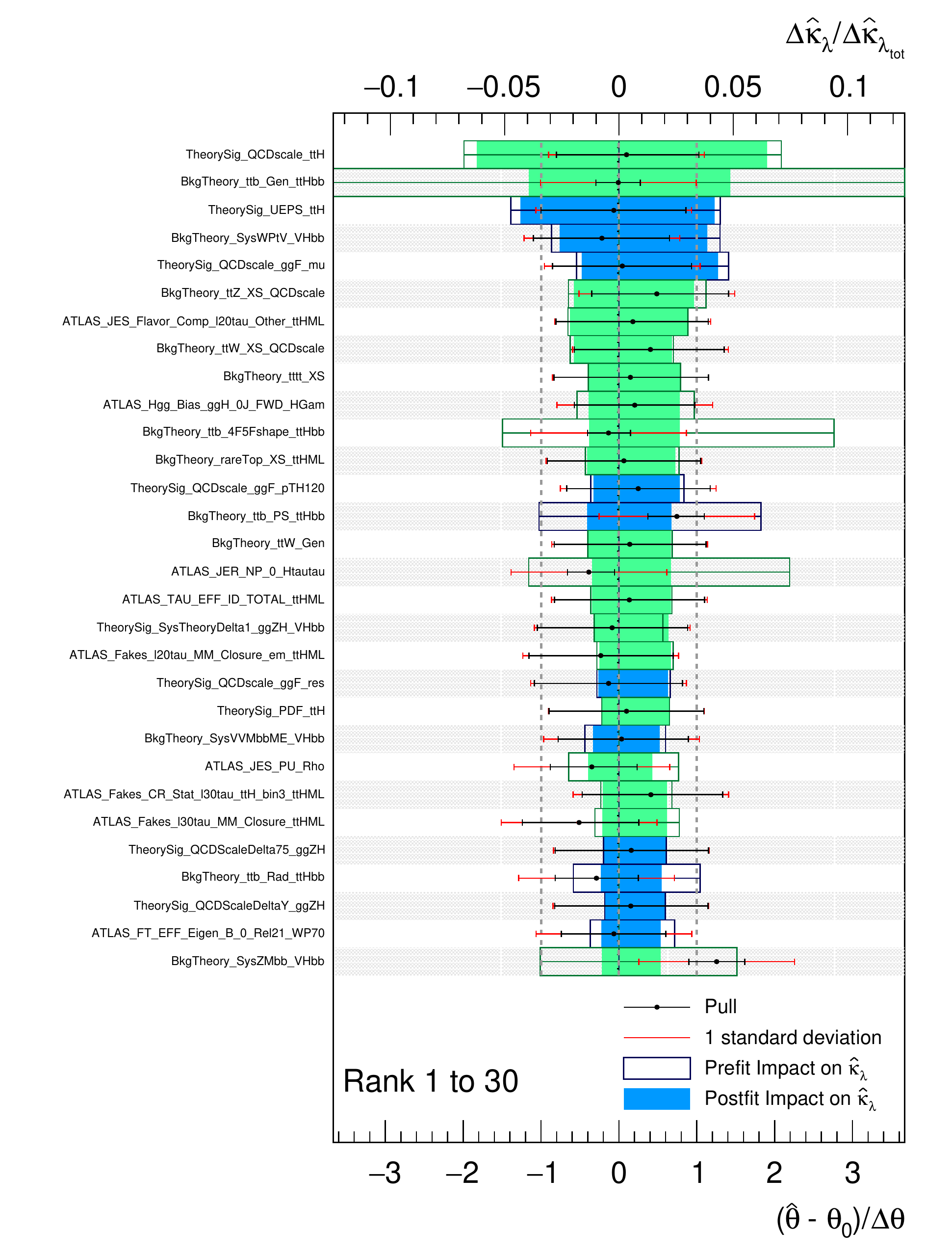}
 \caption{}
\end{subfigure}
\begin{subfigure}[b]{0.49\textwidth}
\includegraphics[height=12.9 cm,width =\textwidth]{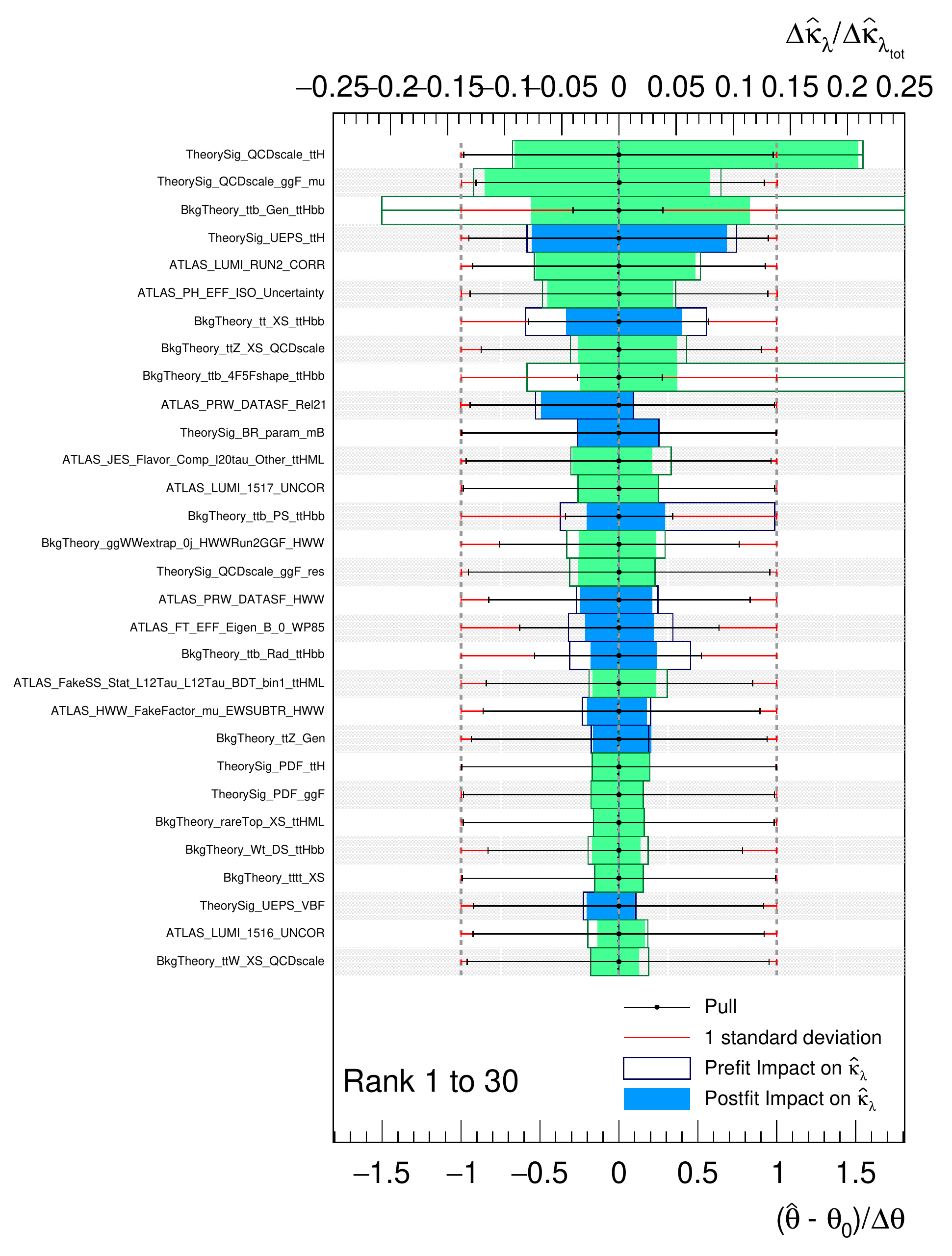}
 \caption{}
\end{subfigure}
\caption{Ranking of the top 30 systematic uncertainties in the single-Higgs combination for data (a) and for the Asimov dataset generated under the SM hypothesis.}     
\label{ranking_single}
\end{figure}

The systematic uncertainties included in the single-Higgs combination and the correlation scheme adopted are reported in the following~\cite{Coupling_run2}:
\begin{itemize}
\item the main theoretical uncertainties come from the limited precision reached by theoretical predictions for the signal and background processes (like QCD scale uncertainties) as well as the degree of knowledge of the parton distribution functions (PDF uncertainties) and the models used to simulate soft physics (like parton-shower uncertainties). Given the fact that different channels entering the combination have harmonised the evaluation of uncertainties, like the ones related to signal processes, they are modelled by a common set of nuisance parameter, namely they are correlated, in most of the channels.\newline
Looking at the uncertainties associated to the modelling of signal processes, the ones having the greatest impact on the results are the uncertainties related to the QCD scale for most of the production modes, to the PDFs and to underlying events and parton shower (UE/PS) on the signal acceptance, while for the background, they are the uncertainties coming from the modelling of the dominant background both in the normalisation and in the shape, as well as from the generators used.
They are reported in Tables~\ref{sys6} and~\ref{sys7}.
\begin{table}[htbp]
\begin{center}
\scalebox{0.92}{
{\def\arraystretch{1.3}
\begin{tabular}{|l|l|}
\hline
NP name & Description \\
\hline
\multirow{2}{*}{TheorySig\_QCDscale\_ttH} & \multirow{2}{*}{QCD scale uncertainty - $t\bar{t}H$}\\
&   \\ \hline
\multirow{2}{*}{TheorySig\_UEPS\_ttH} & Uncertainty on the choice of parton-shower \\
& and underlying event (PS / UE) model - $t\bar{t}H$ \\  \hline
\multirow{2}{*}{TheorySig\_QCDscale\_ggF\_mu} & QCD scale ggF uncertainty related to \\
& to the total cross section   \\
\hline
\multirow{2}{*}{TheorySig\_QCDscale\_ggF\_pTH120} &   QCD scale ggF variations related \\
& the Higgs $p_T$-shape uncertainties \\ \hline
\multirow{2}{*}{TheorySig\_SysTheoryDelta1\_ggZH\_VHbb}& QCD scale uncertainty on STXS bins\\
& for ggZH production - $VH\rightarrow b\bar{b}$ \\ \hline
\multirow{2}{*}{TheorySig\_QCDscale\_ggF\_res}& QCD scale ggF uncertainty related to\\
& the total cross section \\ \hline
\multirow{2}{*}{ TheorySig\_PDF\_ttH}  & Uncertainty on parton distribution function \\
& - $t\bar{t}H$  \\ \hline
\multirow{2}{*}{TheorySig\_QCDScaleDelta75\_ggZH} & QCD scale uncertainty on STXS bins for\\
& ggZH production \\ \hline
\multirow{2}{*}{TheorySig\_QCDScaleDeltaY\_ggZH} & QCD overall scale uncertainty \\
& for ggZH production \\ \hline
\end{tabular}
}}
\end{center}
\caption{Theoretical uncertainties, signal related, having the greatest impact on the results.}
   \label{sys6}
\end{table} 

\begin{table}[htbp]
\begin{center}
\scalebox{0.93}{
{\def\arraystretch{1.3}
\begin{tabular}{|l|l|}
\hline
NP name & Description \\
\hline
\multirow{2}{*}{BkgTheory\_ttb\_Gen\_ttHbb} & Uncertainty on the choice of MC generator \\
& - $t\bar{t}H \rightarrow b\bar{b}$\\ \hline
\multirow{2}{*}{BkgTheory\_SysWPtV\_VHbb} & W+jet modelling uncertainty   \\
& - $VH\rightarrow b\bar{b}$  \\  \hline
\multirow{2}{*}{BkgTheory\_ttZ\_XS\_QCDscale} & Uncertainty on the cross section and QCD scale   \\
&   - $t\bar{t}Z$ modelling\\
\hline
\multirow{2}{*}{BkgTheory\_ttW\_XS\_QCDscale} &  Uncertainty on the cross section and QCD scale  \\
&  - $t\bar{t}W$ modelling \\ \hline
\multirow{2}{*}{BkgTheory\_tttt\_XS}& Uncertainty on the cross section  \\
&  - $t\bar{t}t\bar{t}$ modelling\\ \hline
\multirow{2}{*}{BkgTheory\_ttb\_4F5Fshape\_ttHbb}& Uncertainty associated with the choice of NLO \\
& generator  -  $t\bar{t}H \rightarrow b\bar{b}$  \\ \hline
\multirow{2}{*}{BkgTheory\_rareTop\_XS\_ttHML}  &  Uncertainty on the cross section due to rare \\
& background contributions  - $t\bar{t}H$ multilepton \\ \hline
\multirow{2}{*}{BkgTheory\_ttb\_PS\_ttHbb} & Uncertainty due to the choice of parton shower \\
& model - $t\bar{t}H\rightarrow b\bar{b}$ \\ \hline
\multirow{2}{*}{BkgTheory\_ttW\_Gen} & Uncertainty on the matrix-element MC generator\\
&  - $t\bar{t}W$ modelling \\ \hline
\multirow{2}{*}{BkgTheory\_SysVVMbbME\_VHbb} & Diboson modelling uncertainty  \\
& - $VH\rightarrow b\bar{b}$ \\ \hline
\multirow{2}{*}{BkgTheory\_ttb\_Rad\_ttHbb} &  Uncertainty on the modelling of initial and final state \\
& radiation - $t\bar{t}H\rightarrow b\bar{b}$ \\ \hline
\multirow{2}{*}{BkgTheory\_SysZMbb\_VHbb} &  Z+jet modelling uncertainty\\
& -  $VH\rightarrow b\bar{b}$\\ \hline
\end{tabular}
}}
\end{center}
\caption{Theoretical uncertainties, background related, having the greatest impact on the results.}
   \label{sys7}
\end{table} 

\item The branching fraction uncertainty is correlated between the input channels.
\item Experimental uncertainties are usually grouped into subsets of uncertainties related to the procedure followed in order to identify and calibrate all the objects of the analyses defined in Chapter~\ref{sec:Reco}, like electron and photon calibration, jet energy scale and  resolution, flavour tagging, etc... The main issue concerning experimental uncertainties is related to the different releases of the ATLAS software, \ie\ Release 20.7 exploited by 2015-2016 analyses and Release 21 exploited by the analyses that include also the 2017 dataset. The experimental uncertainties that are correlated among different releases are the luminosity, the jet energy scale, the electron and photon resolution and energy scale, and the electron reconstruction and identification efficiency.\newline
The experimental uncertainties having the greatest impact on the results are the uncertainties coming from photon isolation efficiency, from jet energy scale and resolution, from tau identification efficiency and from data-driven background estimations; they are reported in Table~\ref{sys9}.
\begin{table}[htbp]
\begin{center}
\scalebox{0.9}{
{\def\arraystretch{1.4}
\begin{tabular}{|c|l|}
\hline
NP name & Description \\
\hline
\multirow{2}{*}{TAU\_EFF\_ID\_TOTAL\_ttHML} &  Tau identification efficiency \\
&  - $t\bar{t}H$ multilepton\\ 
\hline
\multirow{2}{*}{JES\_Flavor\_Comp\_l20tau\_Other\_ttHML}  &   Jet energy scale uncertainty related to  \\
& flavour composition - $t\bar{t}H$ multilepton \\ \hline
\multirow{2}{*}{JER\_NP\_0\_Htautau} &  Jet energy resolution  \\
& - $H\rightarrow \tau\tau$ \\ \hline
\multirow{2}{*}{JES\_PU\_Rho}  &  Jet energy scale uncertainty related to \\
& the pileup (density $\rho$)\\ \hline
\multirow{2}{*}{FT\_EFF\_Eigen\_B\_0\_Rel21\_WP70} &   Jet $b$-tagging uncertainty  \\
& - 70\% working point \\ \hline
\multirow{2}{*}{Hgg\_Bias\_ggH\_0J\_FWD\_HGam} &  Uncertainty due to the background modelling\\
&  (spurious signal) for each $t\bar{t}H \rightarrow \gamma \gamma$ category \\ \hline
\multirow{2}{*}{Fakes\_l20tau\_MM\_Closure\_em\_ttHML}  &  Data-driven non-prompt/fake leptons and charge \\
& misassignment uncertainty - $t\bar{t}H$ multilepton\\ \hline
\multirow{2}{*}{Fakes\_CR\_Stat\_l30tau\_ttH\_bin3\_ttHML} &  Data-driven non-prompt/fake leptons and charge\\
&  misassignment uncertainty - $t\bar{t}H$ multilepton \\ \hline
\multirow{2}{*}{Fakes\_l30tau\_MM\_Closure\_ttHML}  & Data-driven non-prompt/fake leptons and charge\\
&  misassignment uncertainty - $t\bar{t}H$ multilepton \\ \hline
\end{tabular}
}}
\end{center}
\caption{Experimental uncertainties having the greatest impact on the results.}
   \label{sys9}
\end{table} 
\end{itemize}
\clearpage
Figure~\ref{ranking_single} shows the ranking plots, defined in Chapter~\ref{sec:dihiggs}, of the top 30 systematic uncertainties for the single-Higgs combination considering data $(a)$ and the Asimov dataset $(b)$. The impact is estimated by varying each nuisance parameter and computing the maximum likelihood estimator of the parameter of interest, $\kappa_\lambda$. The difference between the maximum likelihood estimator with or without varying the nuisance parameter is the $\Delta\hat{\kappa}_\lambda$ of the fit, that is normalised to the total error, $\Delta\hat{\kappa}_{\lambda_{tot}}$. Pre-fit and post-fit impacts of the different nuisance parameter on the central value $\kappa_\lambda$ are reported as white empty and cyan (green) filled bars corresponding to downward (upwards) systematic uncertainty variations, respectively. The points indicate how the parameter had to be pulled up or down during the fit, and associated error bars show the best-fit values of the nuisance parameters and their post-fit uncertainties.  Most of the systematic uncertainties are within $1\sigma$ from the nominal (indicated by the dashed vertical lines) value, except for the ``BkgTheory\_SysZMbb\_VHbb$"$, a nuisance parameter related to the background modelling of the $VH\rightarrow b\bar{b}$ analysis that is pulled consistently with what is found in the combination analysis. The dominant uncertainties arise from the theory modelling of the signal and background processes in simulation. 


\clearpage
\section{Results of fit to $\kappa_\lambda$}
\label{sec:results_single_kl}
\begin{figure}[htbp]
\begin{center}
\includegraphics[width=\textwidth]{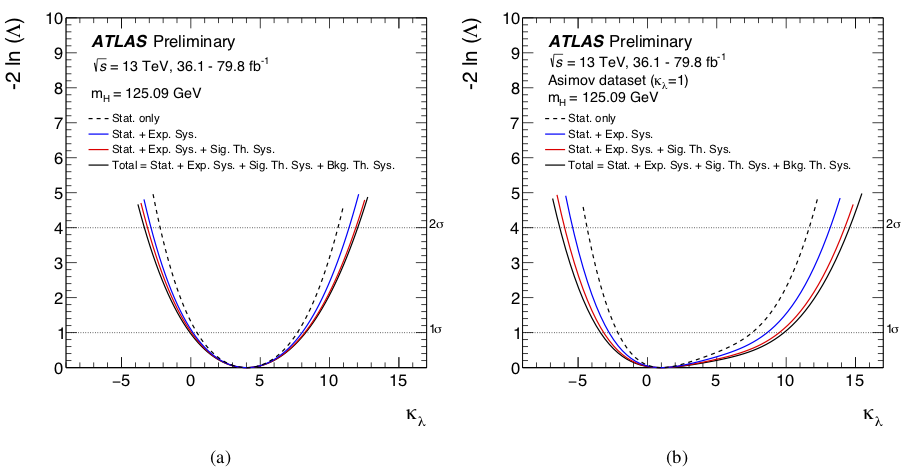}
\end{center}
\caption{Value of $-2 \ln{\Lambda(\kappa_\lambda)}$ as a function of $\kappa_\lambda$ for data (a) and for the Asimov dataset generated under the SM hypothesis (b). The solid black line shows the profile likelihood distributions obtained including all systematic uncertainties (``Total$"$). Results from a statistic only fit ``Stat. only$"$ (black dashed line), including the experimental systematics ``Stat. + Exp. Sys.$"$ (blue solid line), adding theory systematics related to the signal ``Stat.+ Exp. Sys.+ Sig. Th. Sys.$"$ (red solid line) are also shown. The dotted horizontal lines show the $-2 \ln{\Lambda(\kappa_\lambda)}=1$ level that is used to define the $\pm 1\sigma$ uncertainty on $\kappa_\lambda$ as well as the $-2 \ln{\Lambda(\kappa_\lambda)}=4$ level used to define the $\pm 2\sigma$ uncertainty~\cite{PubNote}.}
\label{scan_kl}
\end{figure}

This section presents the main results of the analysis exploiting the combination of the aforementioned single-Higgs channels as well as the differential information. A likelihood fit is performed in the theoretically allowed range $-20 < \kappa_\lambda <$ 20 to constrain the value of the Higgs-boson self-coupling $\kappa_\lambda$, setting all other Higgs-boson couplings to their SM values ($\kappa_F=\kappa_V=1$). This fit configuration targets scenarios and BSM models where new physics is expected to appear only as a modification of the Higgs-boson self-coupling, as for example the next-to-minimal supersymmetric extension of the Standard Model (NMSSM) in the alignment limit, where one of the neutral Higgs fields lies approximately in the same direction in field space as the doublet Higgs vacuum expectation value, and the observed Higgs boson is predicted to have Standard-Model-like properties~\cite{Carena}.\newline
Figure~\ref{scan_kl} shows the value of $-2 \ln{\Lambda(\kappa_\lambda)}$ as a function of $\kappa_\lambda$ for data and for the Asimov dataset, generated from the likelihood distribution $\Lambda$ with nuisance parameters fixed at the best-fit value obtained on data and the parameter of interest fixed to SM hypothesis (\ie\ $\kappa_\lambda = 1$). Likelihood distributions including either all or a selected part of the uncertainties are shown.

\begin{figure}[htbp]
\begin{center}
\includegraphics[width=0.53\textwidth]{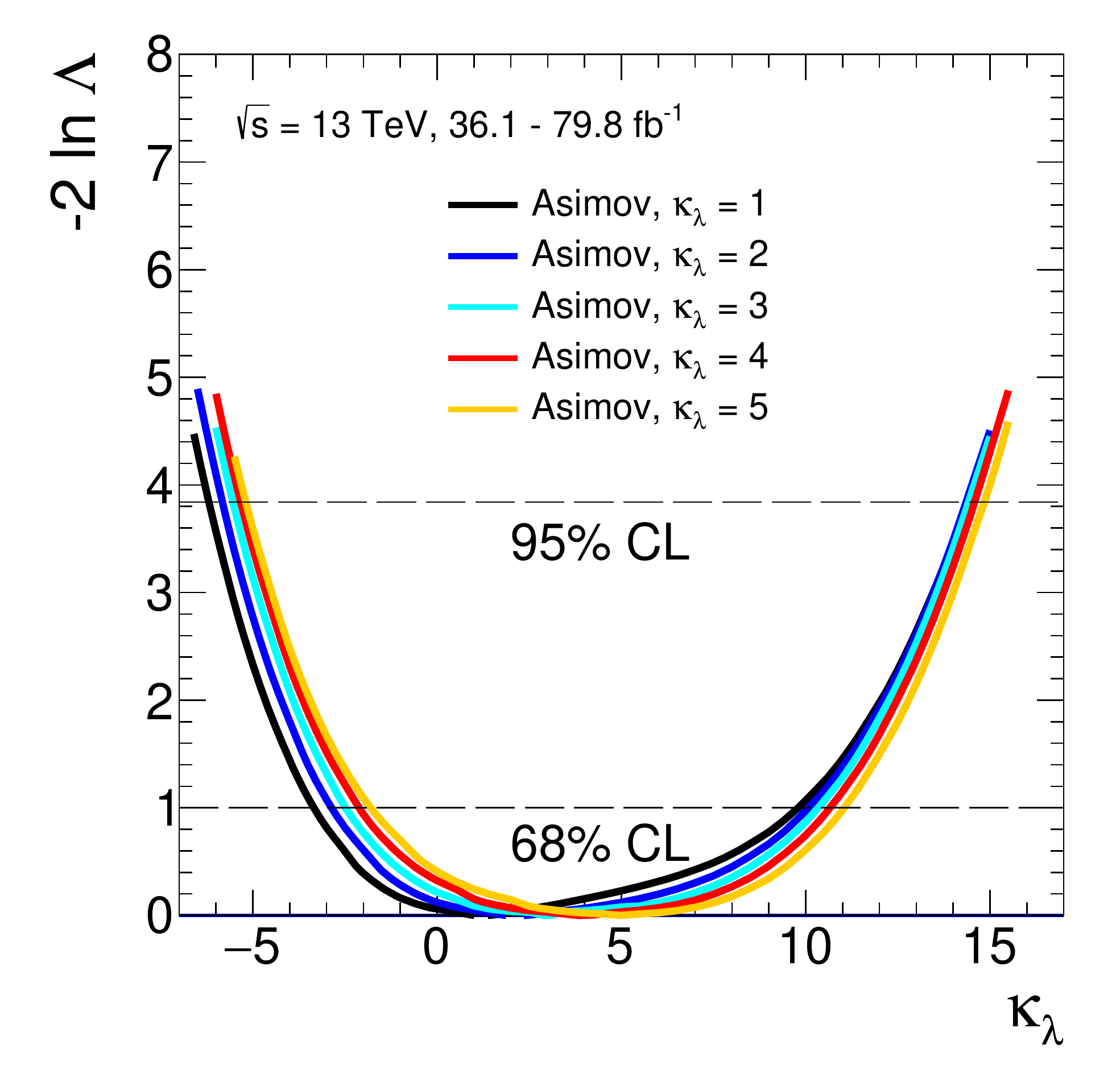}
\end{center}
\caption{Value of $-2 \ln{\Lambda(\kappa_\lambda)}$ as a function of $\kappa_\lambda$ for Asimov datasets generated with $\kappa_\lambda=1$ (black line), $\kappa_\lambda=2$ (blue line), $\kappa_\lambda=3$ (light blue line), $\kappa_\lambda=4$ (red line) and $\kappa_\lambda=5$ (orange line).  The dotted horizontal lines show the $-2 \ln{\Lambda(\kappa_\lambda)}=1$ level that is used to define the $\pm 1\sigma$ uncertainty on $\kappa_\lambda$ as well as the $-2 \ln{\Lambda(\kappa_\lambda)}=3.84$ level used to define the 95\% CL.}     
\label{asimov_bestfit_single}
\end{figure}

Differences in the shapes of the likelihood curves reported in Figures~\ref{scan_kl} between data and Asimov dataset are due to the non-linearity of the cross-section dependence on $\kappa_\lambda$, and the difference of the best-fit values of $\kappa_\lambda$ in the two cases. This effect has been thoroughly studied generating different Asimov datasets fixing $\kappa_\lambda$ to increasing values from 2 to 5. Figure~\ref{asimov_bestfit_single} presents the likelihood distributions for each Asimov dataset and for the Asimov dataset generated under the SM hypothesis showing that the shapes of the likelihood distribution, and thus the $1\sigma$ error and confidence interval around the best-fit value, are strictly dependent on the fitted value itself; the Asimov dataset generated at $\kappa_\lambda \simeq 4$ reproduces the symmetric shape as observed in data.

The central value and uncertainty of the $\kappa_\lambda$ modifier of the trilinear Higgs-boson self-coupling are determined to be:
\begin{equation*}
  \kappa_\lambda = 4.0^{+4.3}_{-4.1}=4.0 ^{+3.7}_{-3.6} \, (\text{stat.})\,^{+1.6}_{-1.5} \, (\text{exp.})\, {}^{+1.3}_{-0.9}\, (\text{sig. th.})\, ^{+0.8}_{-0.9} \, (\text{bkg. th.}) \text{(observed)}
\end{equation*}
\begin{equation*}
    \kappa_\lambda = 1.0^{+8.8}_{-4.4}=1.0 ^{+6.5}_{-3.1} \, (\text{stat.})\,^{+3.9}_{-2.0} \, (\text{exp.})\, {}^{+3.7}_{-1.7}\, (\text{sig. th.})\, ^{+2.4}_{-1.6} \, (\text{bkg. th.}) \text{(expected)}
\end{equation*}
where the total uncertainty is decomposed into components for statistical uncertainties, experimental systematic uncertainties, and theory uncertainties on signal and background modelling, following the procedure described in Section~\ref{sec:statistical_model_single}. The total uncertainty is dominated by the statistical component. \newline
Table~\ref{break_single} reports the detailed breakdown of the uncertainties affecting $\kappa_\lambda$ measurement; the procedure used to produce the numbers of the table is the following: in each case the corresponding nuisance parameters are fixed to their best-fit values, while other nuisance parameters are left free, and the resulting uncertainty is subtracted in quadrature from the total uncertainty.

\begin{table}[h]
\begin{center}
{\def\arraystretch{1.4}
\begin{tabular}{|l|c|}
\hline
    Uncertainty source & $\Delta \kappa_\lambda/\kappa_\lambda$ \%\\ 
    \hline
    Statistical uncertainty & 90\\ 
    \hline
     Systematic uncertainties  & 45 \\
       $\quad$Theory uncertainties & 30  \\
       $\qquad$Signal & 25\\
       $\qquad$Background & 21\\
       $\quad$Experimental uncertainties (excl. MC stat.) & 23\\
       $\quad$MC statistical uncertainty & 11\\
       \hline
       Total uncertainty & 106\\
       
\hline
\end{tabular}}
\caption{Summary of the relative uncertainties $\Delta \kappa_\lambda/\kappa_\lambda$ affecting the measurement of the combined $\kappa_\lambda$. In general, the sum in quadrature of systematic uncertainties from individual sources differs from the uncertainty evaluated for the corresponding group, due to the presence of small correlations among nuisance parameters describing the different sources.}
\label{break_single}
\end{center}
\end{table}
The observed 95\% CL interval of $\kappa_\lambda$ is  $-3.2< \kappa_\lambda<11.9$ while the expected interval is  $-6.2<\kappa_\lambda<14.4$, competitive with the intervals coming from double-Higgs measurements reported in Chapter~\ref{sec:dihiggs}. 
\begin{figure} [htbp]
\centering
\begin{subfigure}[b]{0.49\textwidth}
\includegraphics[width=\textwidth]{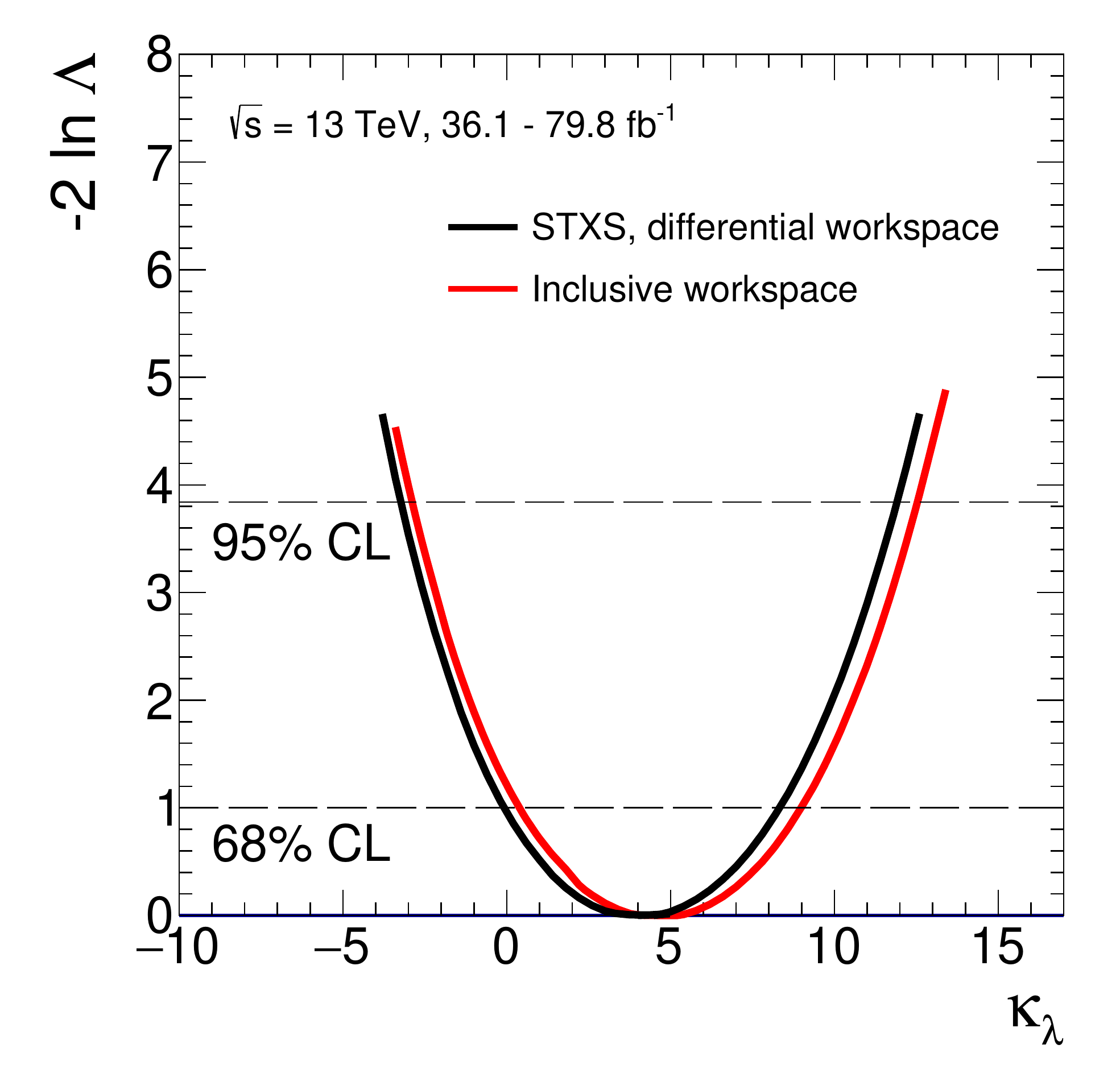}
 \caption{}
\end{subfigure}
\begin{subfigure}[b]{0.49\textwidth}
\includegraphics[width=\textwidth]{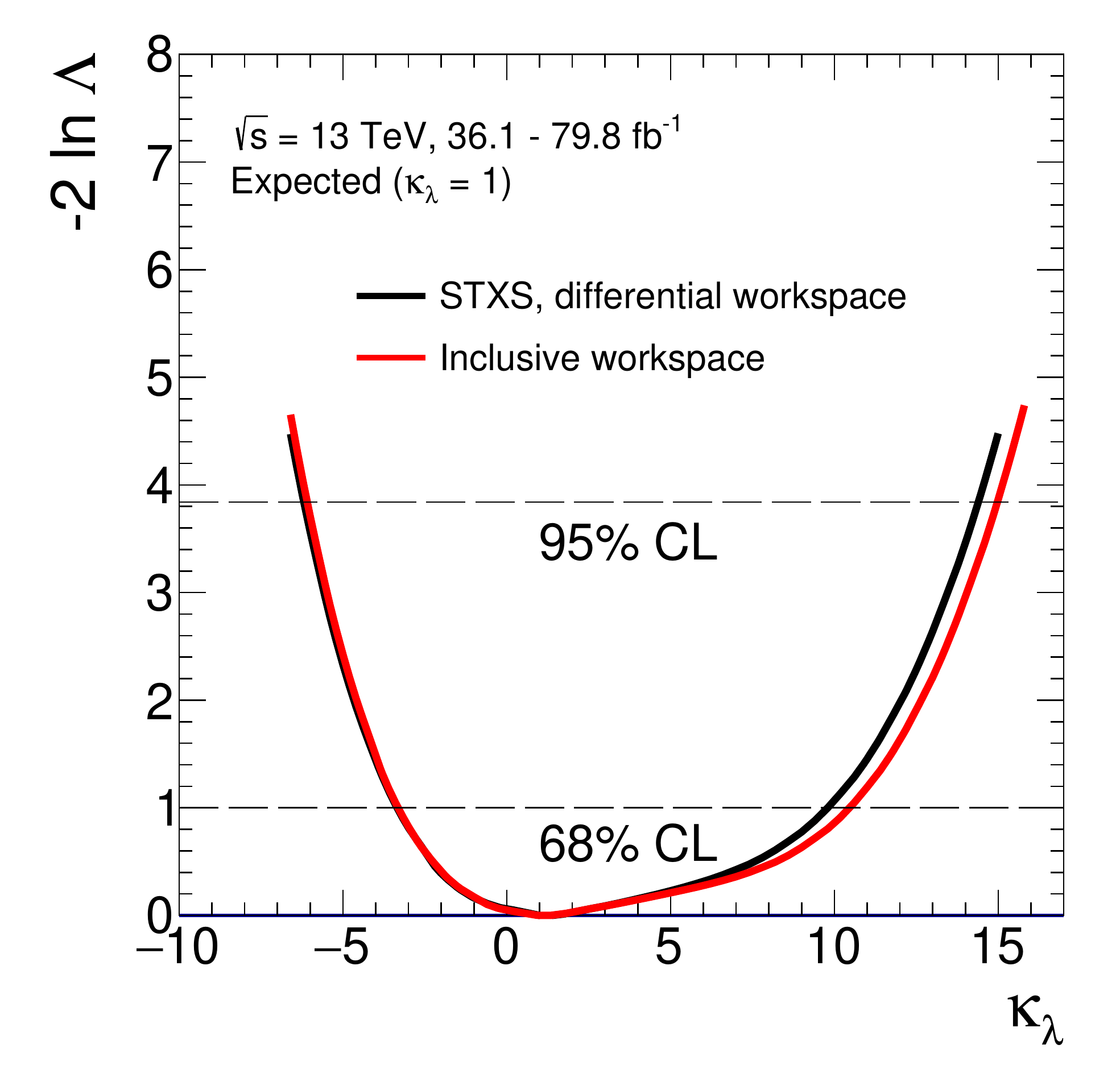}
 \caption{}
\end{subfigure}
\caption{Value of $-2 \ln{\Lambda(\kappa_\lambda)}$ as a function of $\kappa_\lambda$ for data (a) and for the Asimov dataset generated under the SM hypothesis (b). The solid black line shows the likelihood distribution using the $C_1$ coefficients computed for each STXS bin while the solid red line presents the likelihood distribution using inclusive $C_1$ coefficients for the \VBF, \ZH and \WH production modes. The dotted horizontal lines show the $-2 \ln{\Lambda(\kappa_\lambda)}=1$ level that is used to define the $\pm 1\sigma$ uncertainty on $\kappa_\lambda$ as well as the $-2 \ln{\Lambda(\kappa_\lambda)}=3.84$ level used to define the 95\% CL.}     
\label{scan_inclusive_diff}
\end{figure}

Inclusive corrections and cross-section measurements have been used in order to check the difference and the expected gain using the differential cross-section information contained in the STXS bins. Thus, the \VBF, \ZH and \WH production modes have been considered as inclusive regions. The differential information slightly improves the results as it is summarised in Table~\ref{tab:kl_inclu}, and shown in Figure~\ref{scan_inclusive_diff} reporting the value of $-2 \ln{\Lambda(\kappa_\lambda)}$ as a function of $\kappa_\lambda$ for the two different granularities.
\begin{table}[htbp]
\begin{center}
{\def\arraystretch{1.2}
\begin{tabular}{|c|c|c|c|c|c|}
\hline

POIs & Granularity & $\kappa_\lambda{}^{+1\sigma}_{-1\sigma}$ & $\kappa_\lambda$  [95\% CL] \\ 
\hline
\multirow{2}{*}{$\kappa_\lambda$} &\multirow{2}{*}{STXS}  & $4.0_{-4.1}^{+4.3}$ &  $[-3.2, 11.9]$ \\
                                      &     &       $1.0^{+8.8}_{-4.4}$ & $[-6.2, 14.4]$ \\ 
\hline
\multirow{2}{*}{$\kappa_\lambda$} &\multirow{2}{*}{inclusive}  & $4.6_{-4.2}^{+4.3}$ & $[-2.9, 12.5]$ \\
                                      &            & $1.0_{-4.3}^{+9.5}$ &  $[-6.1, 15.0]$ \\ 
                                      
\hline
\end{tabular}
}
\caption{Best-fit values for $\kappa_\lambda$ with $\pm 1 \sigma$ uncertainties and 95\% CL interval are reported for the inclusive and differential configurations. For each fit result the upper row corresponds to the observed results, and the lower row to the expected results obtained using Asimov datasets generated under the SM hypothesis.}
\label{tab:kl_inclu}
\end{center}
\end{table}

The global likelihood shape depends on how the contributions from different production and decay modes are combined: while for $\kappa_\lambda <$ 1 both the $\kappa_\lambda$ and $\kappa_\lambda^2$ terms induce negative contributions in the production signal strengths, for $\kappa_\lambda >$ 1, the interplay between positive $\kappa_\lambda$ and $\kappa_\lambda^2$ terms leads to cancellations that suppress the effect of $\kappa_\lambda$. 
\begin{figure}[htbp]
\centering
\begin{subfigure}[b]{0.49\textwidth}
\includegraphics[width=\textwidth]{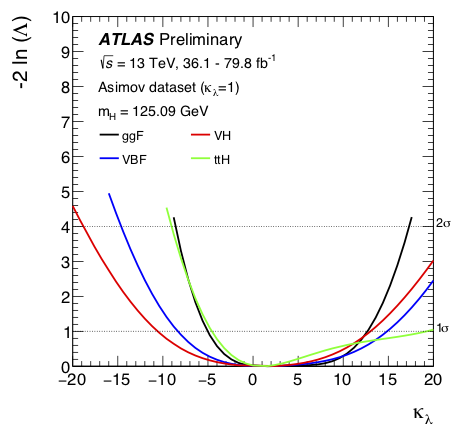}
 \caption{}
\end{subfigure}
\begin{subfigure}[b]{0.49\textwidth}
\includegraphics[width=\textwidth]{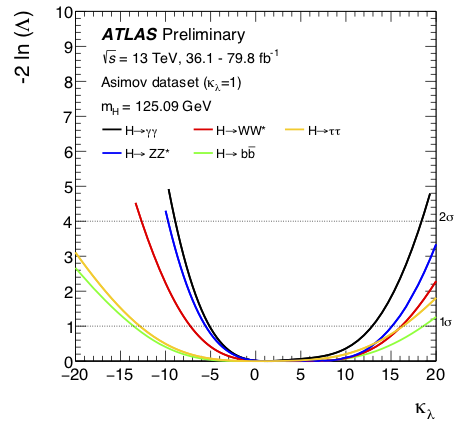}
 \caption{}
\end{subfigure}
\caption{Expected value of $-2 \ln{\Lambda(\kappa_\lambda)}$ as a function of $\kappa_\lambda$ for Asimov datasets generated under the SM hypothesis for Higgs-boson production modes (a) and decay channels (b). The $t\bar{t}H$ multi-lepton categories are excluded from the $H\rightarrow ZZ^*$, $H\rightarrow WW^*$, and $H\rightarrow \tau \tau$ fits~\cite{PubNote}.}     
\label{scan_production_decay}
\end{figure}
Likelihood distributions for each production and decay modes are presented in Figure~\ref{scan_production_decay}; concerning production modes, each curve is obtained fitting $\kappa_\lambda$ for one production mode, keeping the other production modes fixed to the SM; from the decay side, instead, the dataset is directly split in the different decay channels which are fitted one at a time independently.\newline
The dominant contributions to the $\kappa_\lambda$ sensitivity derive from the di-boson decay channels $\gamma \gamma$, $ZZ^*$, $WW^*$ and from the $ggF$ and $t\bar{t}H$  production modes. 
Differential information would improve even more the constraints on $\kappa_\lambda$ particularly including the most sensitive production modes, \ie\ $t\bar{t}H$ and $ggF$.
Given the fact that the dominant contribution to constrain $\kappa_\lambda$ comes from the $ggF$ production, a check has been made to verify that no significant bias on the final results is introduced by considering the \ggF production mode as inclusive regions of the phase space, \ie\ using a kinematic independent parameterisation as a function of $\kappa_\lambda$; a fit has been performed excluding the STXS bins where a great impact is expected, \ie\ the bins with Higgs-boson transverse momentum above 120 GeV. This study has been realised by introducing signal strength parameters for these STXS bins and profiling them independently in the fit. The result is a minimal change of the central value ($\sim$5\%) and uncertainty on $\kappa_\lambda$. \newline
In order to investigate the constraining power of each single-Higgs channel included in this combination, a test has been performed removing different categories, corresponding to the different single-Higgs decay channels, from the combined fit and checking the $\kappa_\lambda$ 95\%~CL interval obtained using the remaining channels. In order to avoid statistics fluctuations, Asimov datasets are used. Results are reported in Table~\ref{tab:channelRank_single} showing the ranking of the different channels in constraining $\kappa_{\lambda}$: each row shows the 1$\sigma$ interval and the 95\%~CL $\kappa_{\lambda}$ interval obtained by removing the specific channel listed in the first column. 
\begin{table}[htbp]
\begin{center}
{\def\arraystretch{1.3}
\begin{tabular}{|c|c|c|}
\hline
Channels &$\kappa_\lambda{}^{+1\sigma}_{-1\sigma}$ & $\kappa_\lambda$  [95\% CL]\\
\hline
$H\rightarrow \gamma\gamma$&$1.00_{-5.0}^{+10.0}$&[-7.2, 15.9]\\ 
\hline
$t\bar{t}H$ multilepton&$1.00_{-7.3}^{+9.3}$&[-6.6, 14.5]\\ 
\hline
$H\rightarrow ZZ^*$&$1.00_{-6.8}^{+8.9}$&[-6.8, 14.8]\\
\hline
$VH\rightarrow b\bar{b}$&$1.00_{-6.3}^{+9.6}$&[-6.2, 15.2]\\
\hline
$H\rightarrow WW^*$&$1.00_{-6.4}^{+8.9}$&[-6.5, 14.7]\\ 
\hline
$H\rightarrow \tau^+\tau^-$&$1.00_{-6.1}^{+9.1}$&[-6.2, 14.8]\\ 
\hline
$t\bar{t}H\rightarrow b\bar{b}$&$1.00_{-6.4}^{+8.8}$&[-6.3, 14.3]\\ 
\hline
Nominal expected result&$1.00_{-4.4}^{+8.8}$&[-6.2, 14.4]\\
\hline
\end{tabular}
}
\caption{Ranking of single-Higgs channels in constraining $\kappa_{\lambda}$ using the Asimov dataset. Each row shows the 1$\sigma$ interval and 95\% CL of $\kappa_{\lambda}$ obtained by removing one specific channel.}
\label{tab:channelRank_single}
\end{center}
\end{table}

The more important a channel is in constraining $\kappa_{\lambda}$, the larger would be the expected CL interval obtained removing this channel with respect to the nominal results reported in the last row of Table~\ref{tab:channelRank_single}. Keeping in mind that the analyses use different integrated luminosities, similar results are obtained.

\section{Results of fit to $\kappa_\lambda$ and either $\kappa_F$ or $\kappa_V$}
\label{sec:results_single_otherfit}
\begin{figure}[hbtp]
\centering
\begin{subfigure}[b]{0.49\textwidth}
\includegraphics[width=\textwidth]{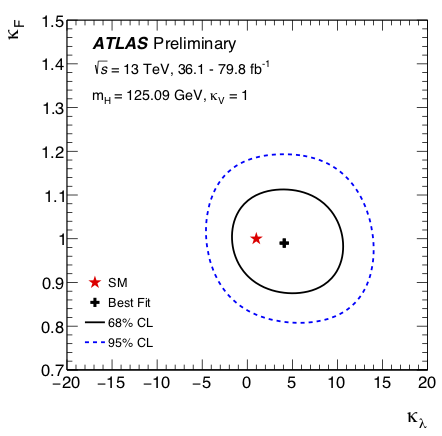}
 \caption{}
\end{subfigure}
\begin{subfigure}[b]{0.49\textwidth}
\includegraphics[width=\textwidth]{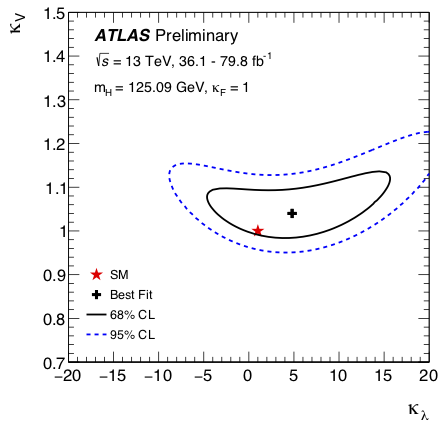}
 \caption{}
\end{subfigure}
\caption{Negative log-likelihood contours at 68\% and 95\% CL in the $(\kappa_\lambda,\kappa_F)$ plane under the assumption of $\kappa_V$ =1 (a) and in the $(\kappa_\lambda,\kappa_V)$ plane under the assumption of $\kappa_F$ =1 (b). The best-fit value is indicated by a cross while the SM hypothesis is indicated by a star. The plot assumes that the approximations in References~\cite{Degrassi,Maltoni} are valid inside the shown contours.}     
\label{contour_kl_kv_kf}
\end{figure}

The constraints on $\kappa_\lambda$, comparable to the double-Higgs limits in the case of an exclusive $\kappa_\lambda$ fit, become weaker including additional degrees of freedom to the fit, \ie\ when BSM modifications of the single-Higgs couplings are taken into account. Additional fit configurations are tested, \ie\ a simultaneous fit is performed to ($\kappa_\lambda$, $\kappa_F$) and ($\kappa_\lambda$, $\kappa_V$). These fits target BSM physics scenarios where new physics could affect only the Yukawa type terms ($\kappa_V$= 1) of the SM or only the couplings to vector bosons ($\kappa_F$= 1), in addition to the Higgs-boson self-coupling ($\kappa_\lambda$)~\cite{coupling_bsm}. Figures~\ref{contour_kl_kv_kf} (a) and (b) show observed negative log-likelihood contours on the ($\kappa_\lambda$, $\kappa_F$) and ($\kappa_\lambda$, $\kappa_V$) grids obtained from fits performed for the $\kappa_V$=1 or $\kappa_F$=1 hypothesis, respectively. 
The sensitivity is not much degraded when simultaneously fitting $\kappa_\lambda$ and $\kappa_F$ while it is degraded by 50\% in the $\kappa_V$ case (expected lower limit of 95\% CL interval).

\begin{figure}[htbp]
\centering
\begin{subfigure}[b]{0.49\textwidth}
\includegraphics[width=\textwidth]{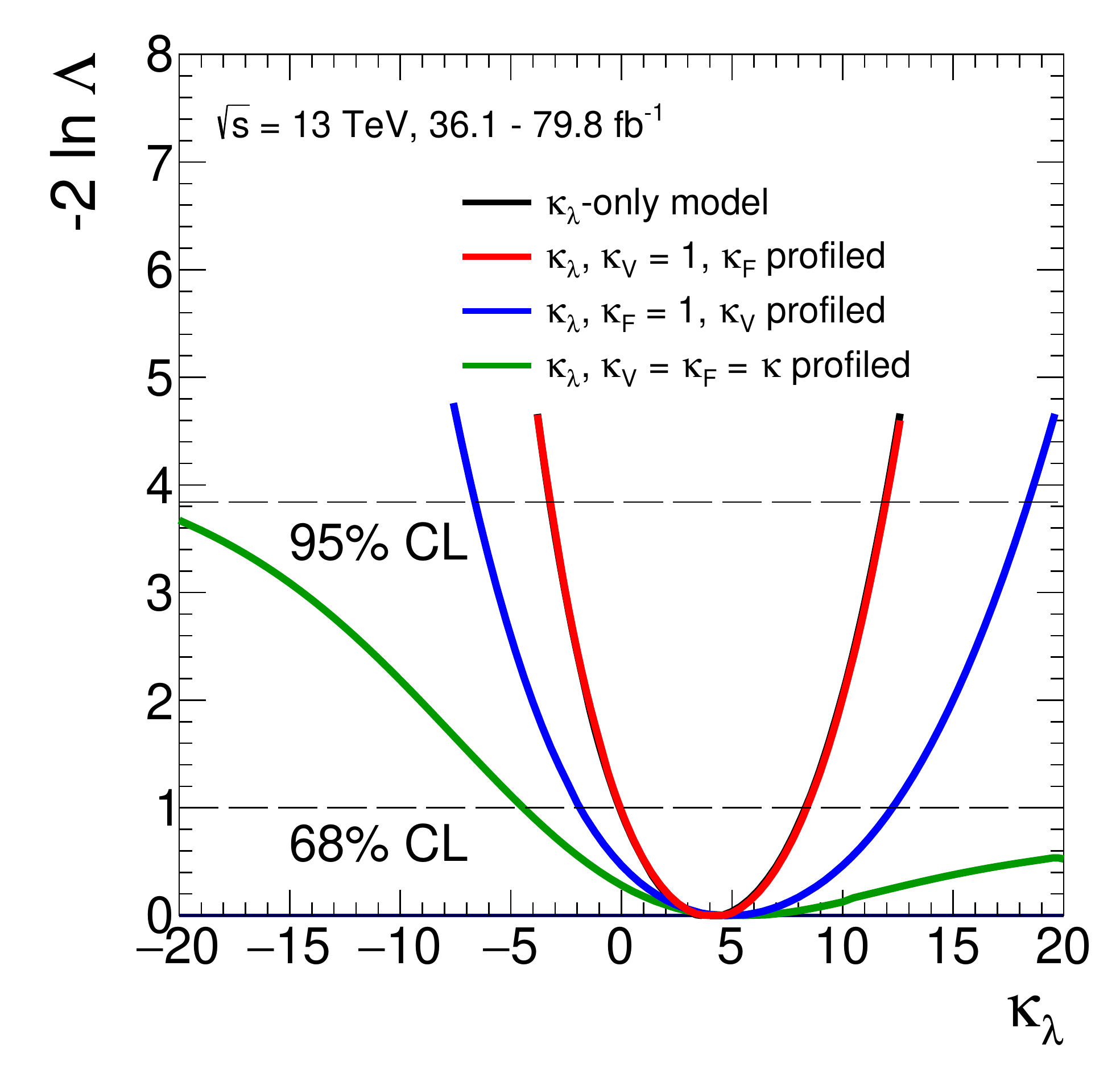}
 \caption{}
\end{subfigure}
\begin{subfigure}[b]{0.49\textwidth}
\includegraphics[width=\textwidth]{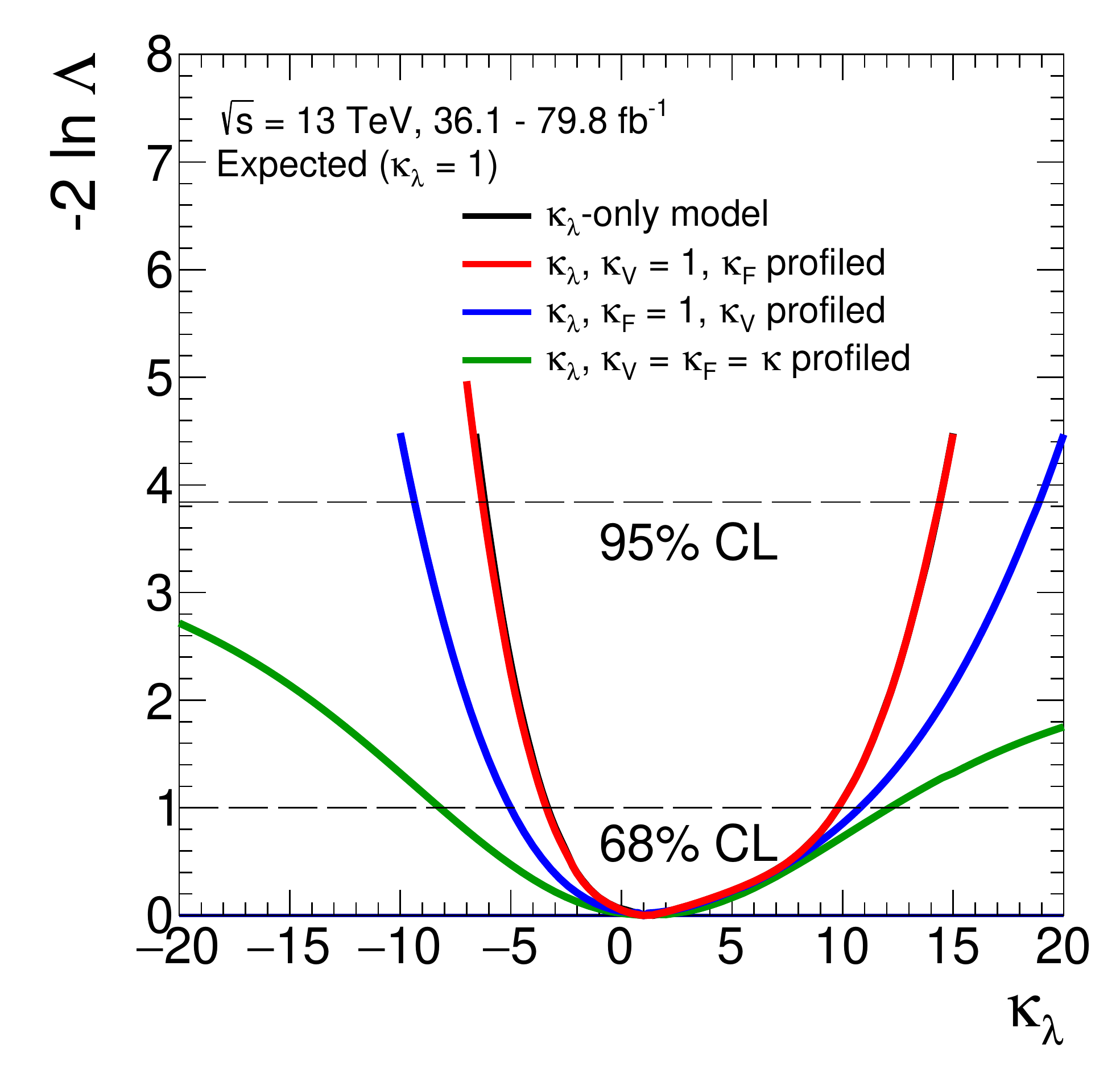}
 \caption{}
\end{subfigure}
\caption{Value of $-2 \ln{\Lambda(\kappa_\lambda)}$ as a function of $\kappa_\lambda$ for data (a) and for the Asimov dataset generated under the SM hypothesis (b). Different fit configurations have been tested: $\kappa_\lambda$-only model (black line), $\kappa_\lambda$-$\kappa_F$ model (red line), $\kappa_\lambda$-$\kappa_V$ model (blu line) and $\kappa_\lambda$-$\kappa$, where $\kappa$ stands for a common single Higgs-boson coupling modifier, $\kappa= \kappa_V =  \kappa_F$, (green line) All the coupling modifiers that are not included in the fit are set to their SM predictions. The dotted horizontal lines show the $-2 \ln{\Lambda(\kappa_\lambda)}=1$ level that is used to define the $\pm 1\sigma$ uncertainty on $\kappa_\lambda$ as well as the $-2 \ln{\Lambda(\kappa_\lambda)}=3.84$ level used to define the 95\% CL.}     
\label{scan_single_otherfit} 
\end{figure}

\begin{figure}[htbp]
  \centering
  \begin{subfigure}[b]{0.49\textwidth}
\includegraphics[height= 7 cm,width =\textwidth]{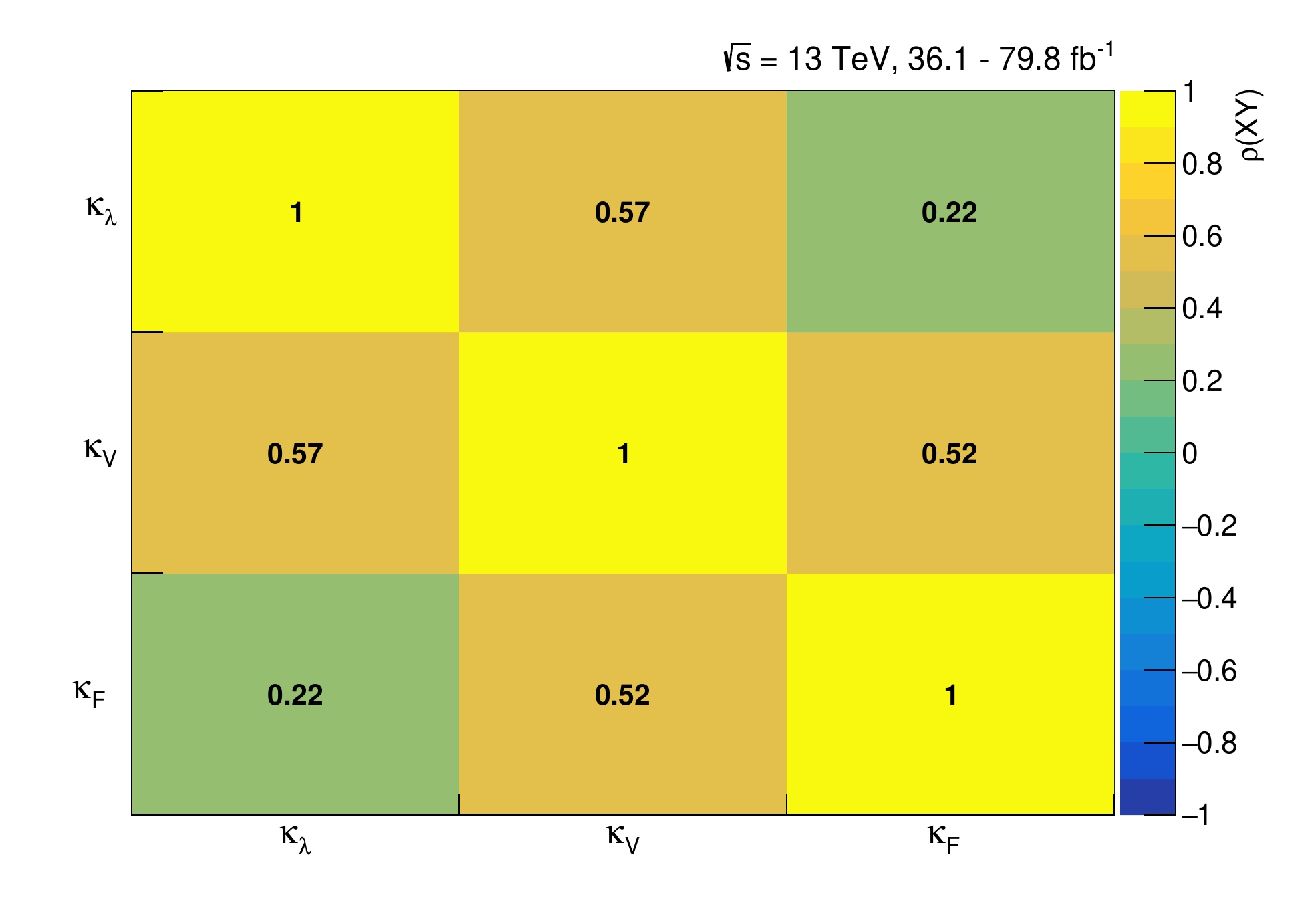}
 \caption{}
\end{subfigure}
  \begin{subfigure}[b]{0.49\textwidth}
\includegraphics[height=7 cm,width =\textwidth]{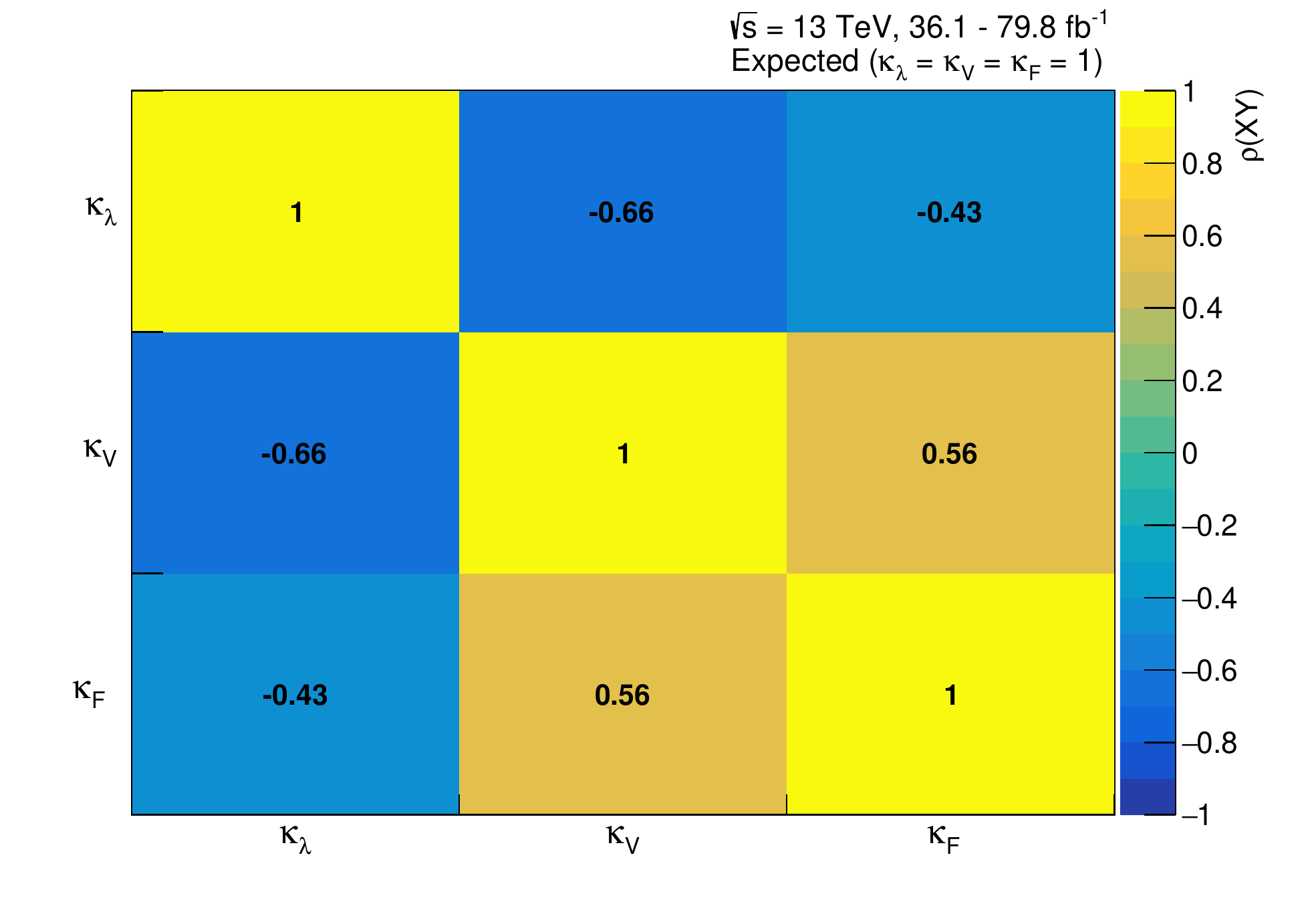}
 \caption{}
\end{subfigure}
    \caption{Correlations between the parameters of interest, \ie\ $\kappa_\lambda$, $\kappa_F$ and $\kappa_V$, for data (a) and for thee Asimov dataset generated under the SM hypothesis (b).}
  \label{correlation_kl_kF_kV_single}
\end{figure}

An even less constrained fit, performed by simultaneously fitting $\kappa_\lambda$ and $\kappa$, where $\kappa$ stands for a common Higgs-boson coupling modifier ($\kappa=\kappa_F=\kappa_V$), results in nearly no sensitivity to $\kappa_\lambda$ within the theoretically allowed range of $| \kappa_\lambda | <$ 20 as it is shown in Table~\ref{tab:klambda-kF_kV} and in Figure~\ref{scan_single_otherfit}, summarising all the fit configurations tested.
\begin{table}[htbp]
\begin{center}
{\def\arraystretch{1.2}
\begin{tabular}{|c|c|c|c|c|c|}
\hline

POIs & Granularity& $\kappa_F{}^{+1\sigma}_{-1\sigma}$& $\kappa_V{}^{+1\sigma}_{-1\sigma}$ & $\kappa_\lambda{}^{+1\sigma}_{-1\sigma}$ & $\kappa_\lambda$  [95\% CL] \\ 
\hline

\multirow{2}{*}{$\kappa_\lambda$, $\kappa_V$ } &\multirow{2}{*}{STXS}  &\multirow{2}{*}{1}  & $1.04^{+0.05}_{-0.04}$  & $4.8^{+7.4}_{-6.7}$ & $[-6.7,18.4]$ \\
                                     &   &   &   $1.00^{+0.05}_{-0.04}$ &  $1.0_{-6.1}^{+9.9}$ & $[-9.4, 18.9]$ \\   
\hline
\multirow{2}{*}{$\kappa_\lambda$, $\kappa_F$ } &\multirow{2}{*}{STXS} &$0.99^{+0.08}_{-0.08}$  & \multirow{2}{*}{1} & $4.1^{+4.3}_{-4.1}$  & $[-3.2, 11.9]$ \\
                                    &    &       $1.00^{+0.08}_{-0.08}$ &  & $1.0_{-4.4}^{+8.8}$ & $[-6.3, 14.4]$ \\
\hline 
\multirow{2}{*}{$\kappa_\lambda$-$\kappa=\kappa_F=\kappa_V$} &\multirow{2}{*}{STXS}   &$1.05^{+0.58}_{-0.05}$ &$1.05^{+0.58}_{-0.05}$  & $5.3_{-9.7}^{>14.7}$ &  $[<-20, >20]$ \\
                                      &  & $1.00^{+0.10}_{-0.04}$ &  $1.00^{+0.10}_{-0.04}$    &  $1.0^{+10.7}_{-9.3}$ & $[<-20, >20]$ \\
 \hline     
\end{tabular}
}
\caption{
Best-fit values for $\kappa$ modifiers with $\pm 1 \sigma$ uncertainties. The first column shows the parameter(s) of interest in each fit configuration, where the other coupling modifiers are kept fixed to the SM prediction. The 95\% CL interval for $\kappa_\lambda$ is also reported. For each fit result the upper row corresponds to the observed results, and the lower row to the expected results obtained using Asimov datasets generated under the SM hypothesis.}
\label{tab:klambda-kF_kV}
\end{center}
\end{table}
\clearpage
The correlations among the parameters of interest, \ie\ $\kappa_\lambda$, $\kappa_F$ and $\kappa_V$, are shown in Figure~\ref{correlation_kl_kF_kV_single} for data $(a)$ and for the Asimov dataset $(b)$. A strong correlation is present between $\kappa_F$ and $\kappa_V$.

\subsection{Cross-check on the validity of the theoretical approximations}
\begin{figure}[htbp]
  \centering
  \begin{subfigure}[b]{0.35\textwidth}
\includegraphics[height=4 cm,width=1.1\textwidth]{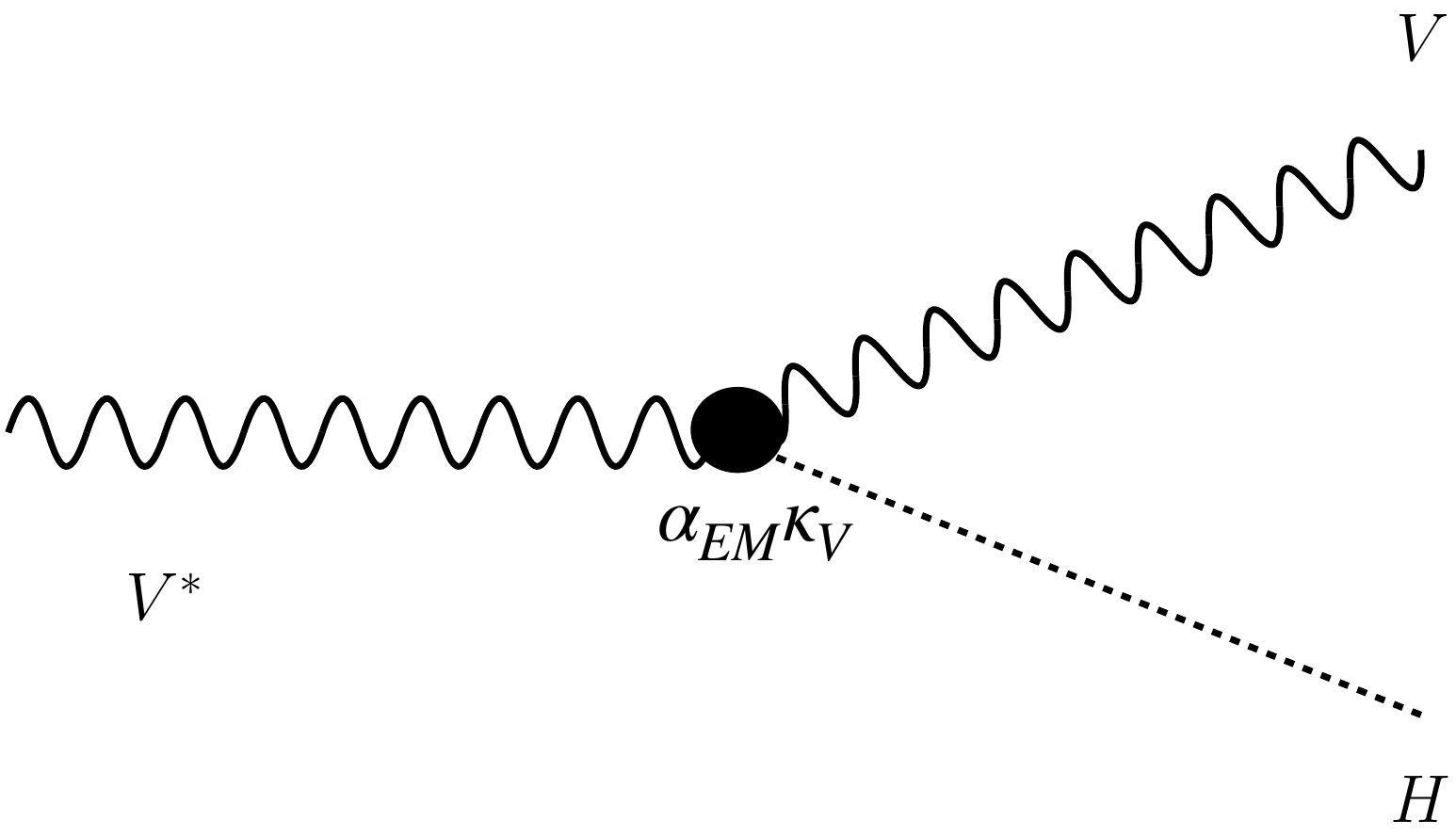}
 \caption{}
\end{subfigure}
\qquad
  \begin{subfigure}[b]{0.4\textwidth}
  \includegraphics[height=4 cm,width=1.1\textwidth]{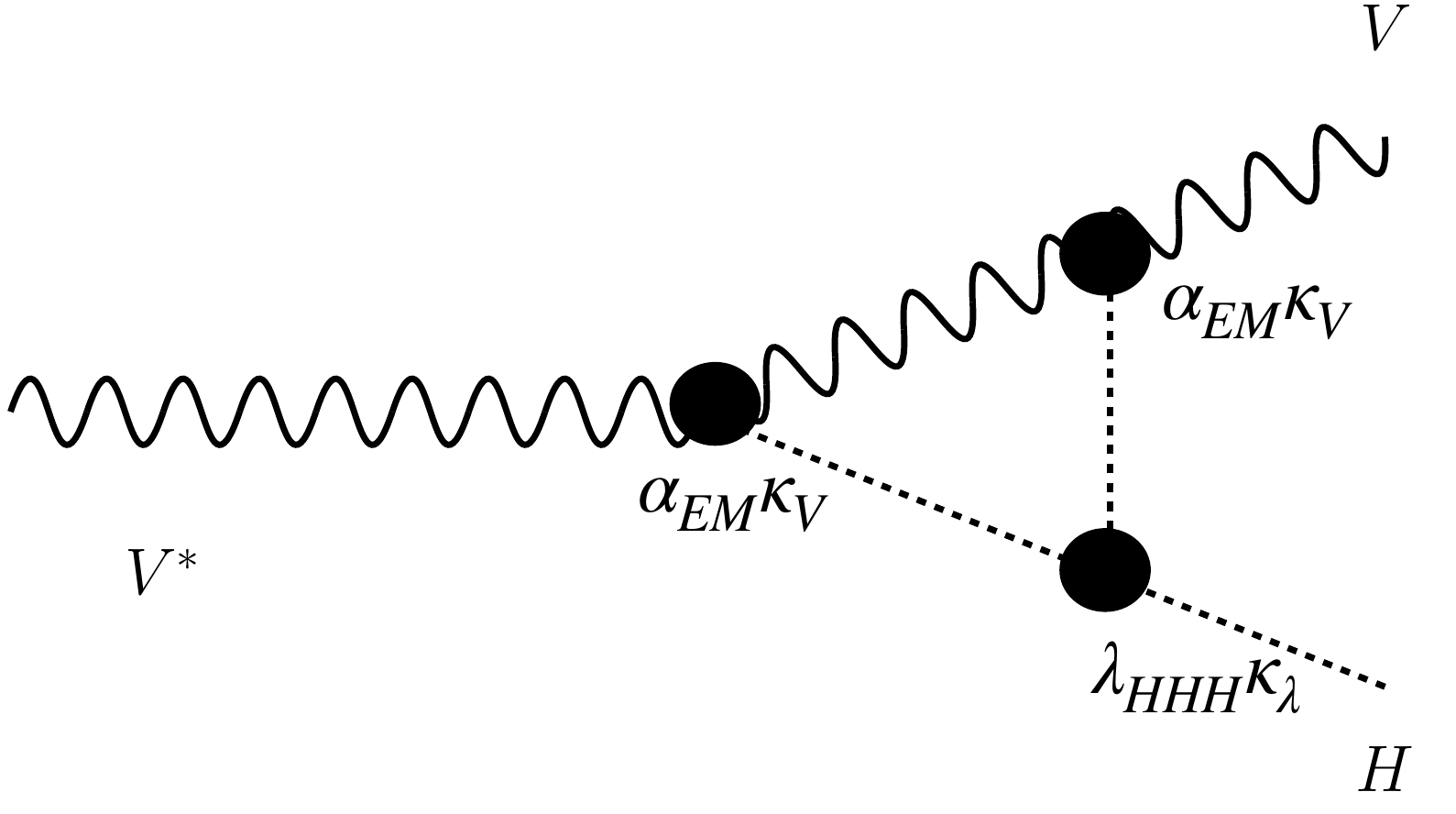}
   \caption{}
\end{subfigure}
    \caption{VH tree-level and one-loop diagrams.}
  \label{fig:diagram-interference}
\end{figure}

\begin{figure}[htbp]
\centering
\begin{subfigure}[b]{0.49\textwidth}
\includegraphics[width=\textwidth]{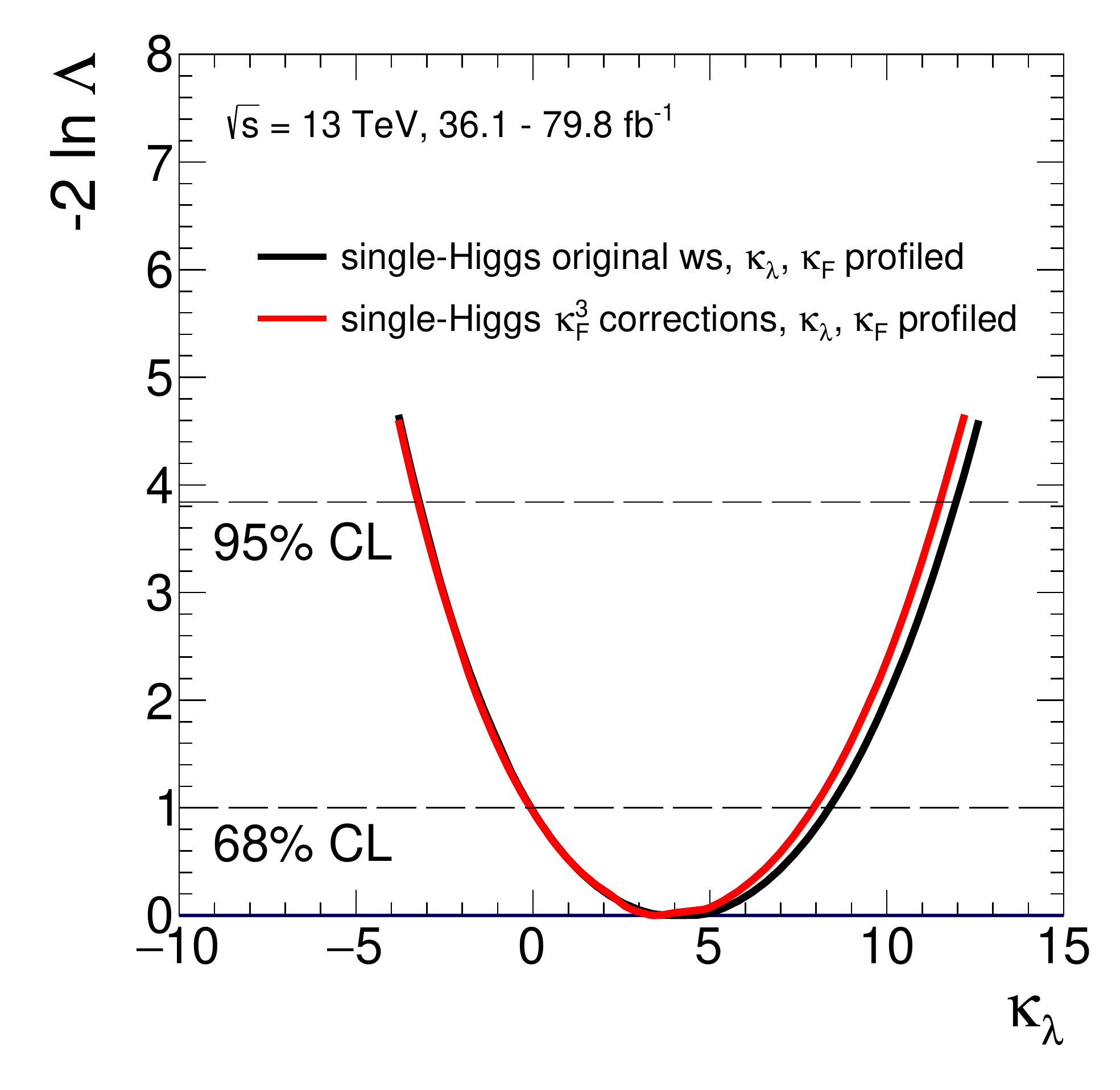}
 \caption{}
\end{subfigure}
\begin{subfigure}[b]{0.49\textwidth}
\includegraphics[width=\textwidth]{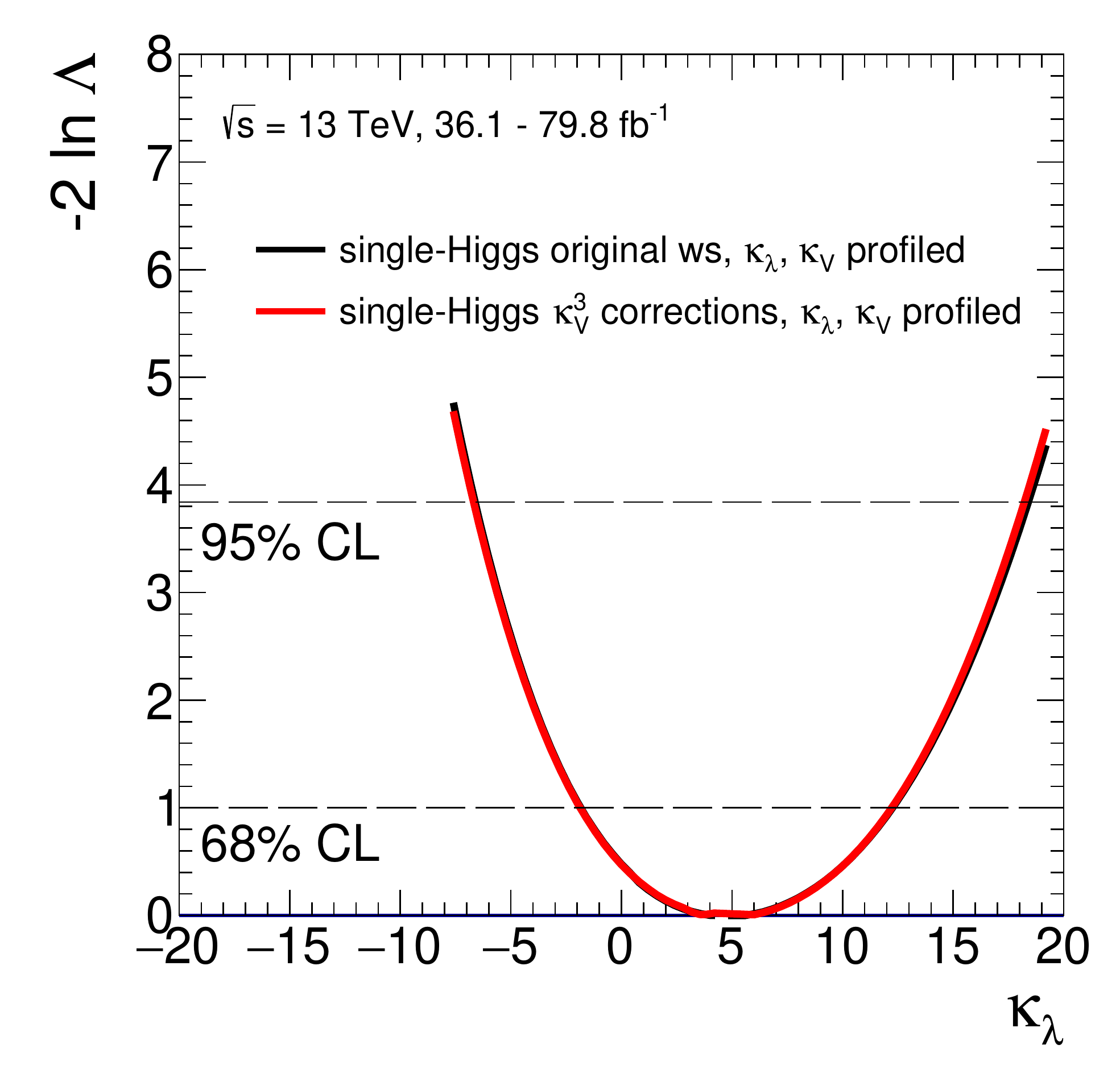}
 \caption{}
\end{subfigure}
\caption{Value of the observed $-2 \ln{\Lambda(\kappa_\lambda)}$ as a function of $\kappa_\lambda$ with $\kappa_{F}$ profiled under the assumption of $\kappa_V=1$ (a) and with $\kappa_V$ profiled under the assumption of $\kappa_F=1$ (b). 
The solid black lines show the nominal approximation in the two fit configurations while the solid red lines show the likelihood distribution including higher order corrections.
The dotted horizontal lines show the $-2 \ln{\Lambda(\kappa_\lambda)}=1$ level that is used to define the $\pm 1\sigma$ uncertainty on $\kappa_\lambda$ as well as the $-2 \ln{\Lambda(\kappa_\lambda)}=3.84$ level used to define the 95\% CL.}     
\label{corrections_scan}
\end{figure}

\begin{figure}[htbp]
\centering
\begin{subfigure}[b]{0.49\textwidth}
\includegraphics[height=7 cm, width=\textwidth]{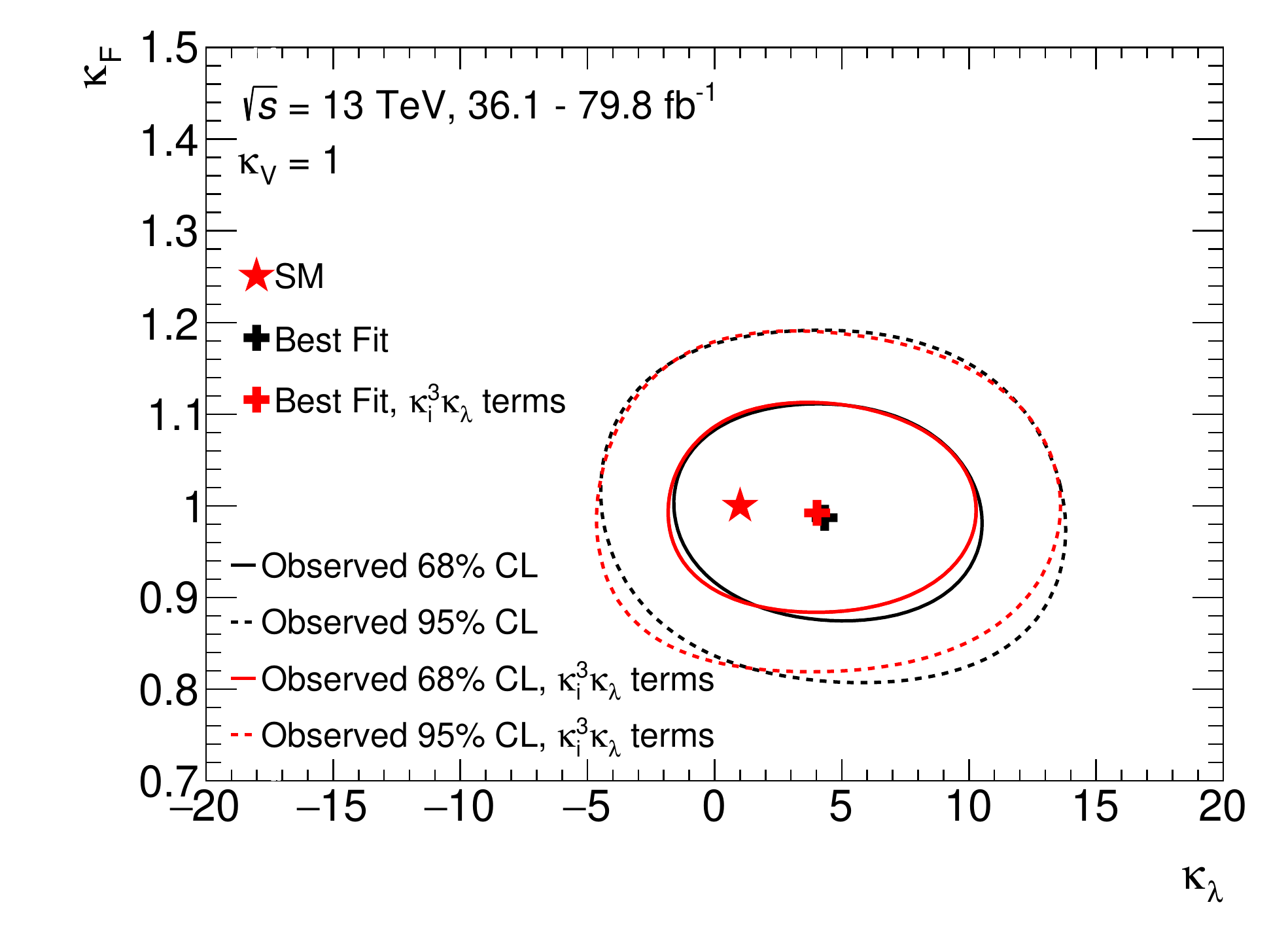}
 \caption{}
\end{subfigure}
\begin{subfigure}[b]{0.49\textwidth}
\includegraphics[height=7 cm, width=\textwidth]{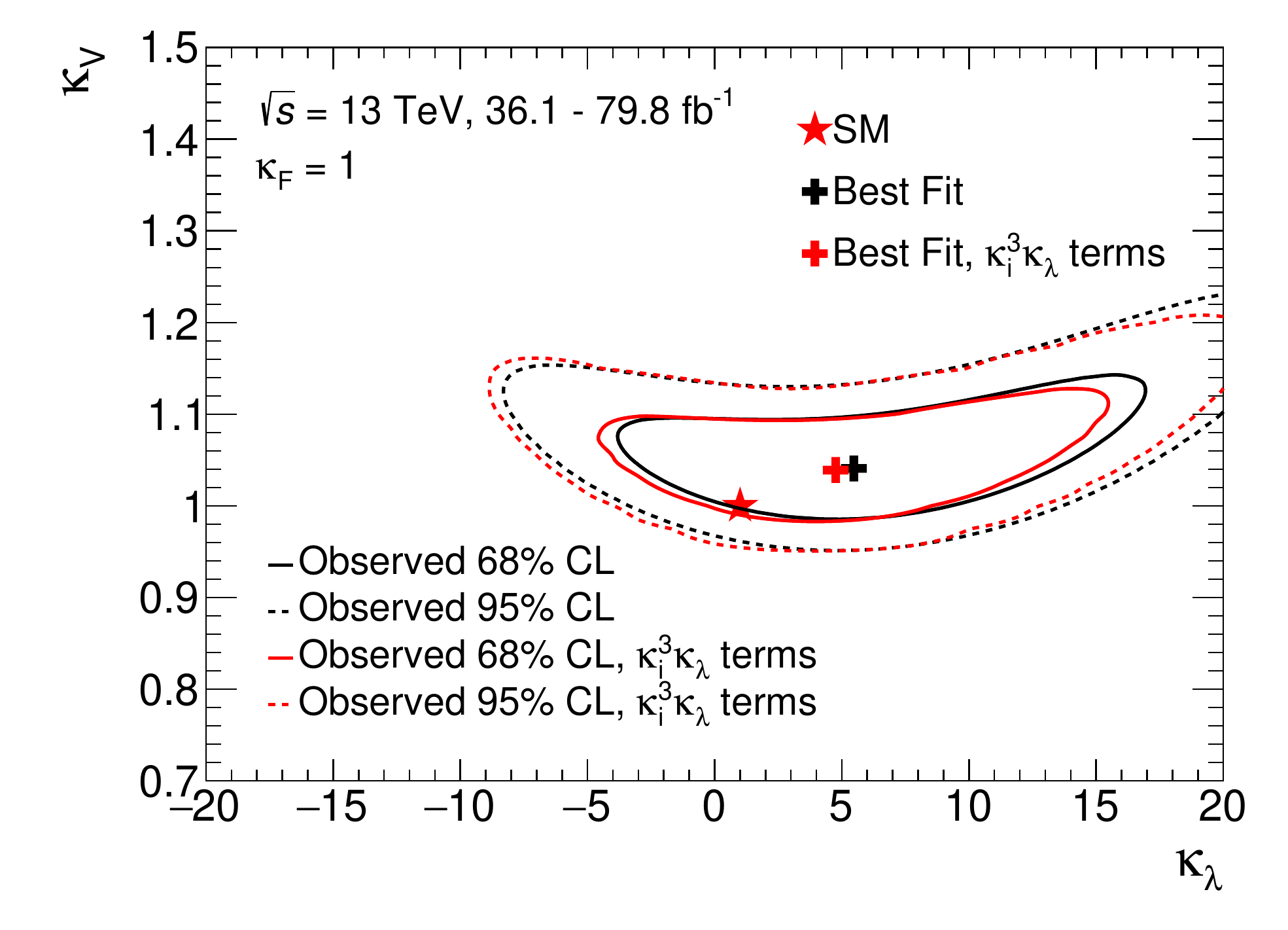}
 \caption{}
\end{subfigure}
\caption{Negative log-likelihood contours at 68\% and 95\% CL in the $(\kappa_\lambda,\kappa_F)$ plane under the assumption of $\kappa_V=1$ (a), and in the $(\kappa_\lambda,\kappa_V)$ plane under the assumption of $\kappa_F=1$ (b). The best-fit value is indicated by a cross while the SM hypothesis is indicated by a star. The red contour is produced using the nominal input corrected with $\kappa_i^3$ terms, where $\kappa_i=\kappa_F$ in (a) and $\kappa_i=\kappa_V$ in (b), while the black contour represents the nominal results. Negligible discrepancies are present adding these correction terms.}     
\label{corrections_contours}
\end{figure}

In order to ascertain the robustness of the nominal approach and of the approximations of the theoretical framework described in Chapter~\ref{sec:prob_self}, against higher-order terms regarding modifications of the LO coupling modifiers $\kappa_i$ or $\kappa_f$ included in Equations~\ref{eq:mui_single} and~\ref{eq:muf_single}, a test has been made exploiting a different approach with respect to Reference~\cite{Maltoni}. \newline
Indeed, the aforementioned paper used an ``additive approach$"$ to include $\kappa_i$ or $\kappa_f$ coupling modifications with respect to $\kappa_\lambda$ (except for the multiplicative term $Z_H^{BSM}\kappa_i^2$); a pure ``multiplicative approach$"$ was not possible without guesses on higher orders, that can be only treated with a full Effective-Field-Theory (EFT) approach. \newline
Thus, as a check, the authors of Reference~\cite{Maltoni} suggested to add additional higher-order $\kappa_i$ and $\kappa_f$ terms, like $\kappa_i^3$ or $\kappa_f^3$ terms, coming from the interference between tree-level and one-loop diagrams like the ones shown in Figure~\ref{fig:diagram-interference}.
The study including additional $\kappa_i^3$ terms to Equations~\ref{eq:mui} is here reported, where $\kappa_i$ modifies also the loops together with $\kappa_\lambda$, like it is shown in Equation~\ref{eq:mui-test}:
\begin{equation}
\label{eq:mui-test}
\mu_i(\kappa_\lambda, \kappa_i) = \frac{\sigma^{\textrm{BSM}}
}{\sigma^{\textrm{SM}}} =
Z_{H}^{\textrm{BSM}}\left(\kappa_\lambda\right)\left[\kappa_i^2+\frac{(\kappa_\lambda-1)\kappa_i^3C_1^i}{K^i_{\mathrm{EW}}}\right] \, .
\end{equation}
Results of the fit to $\kappa_\lambda$ and either $\kappa_F$ or $\kappa_V$ with the remaining Higgs-boson modifier set to its SM value, are summarised in Table~\ref{tab:klambda_test_ki3}, reporting best-fit observed values and 95\% CL $\kappa_\lambda$ intervals considering both the nominal configuration and the configuration testing higher order corrections. Small discrepancies with respect to the nominal results have been found; this effect is clear looking at Figure~\ref{corrections_scan}, reporting a comparison of the observed $-2 \ln{\Lambda(\kappa_\lambda)}$ as a function of $\kappa_\lambda$ profiling $\kappa_F$ ($\kappa_V=1$) $(a)$ and  $\kappa_V$ ($\kappa_F=1$) $(b)$, and Figure~\ref{corrections_contours} showing negative log-likelihood contours at 68\% and 95\% CL on the $(\kappa_\lambda,\kappa_F)$ $(a)$  and $(\kappa_\lambda,\kappa_V)$ $(b)$ planes; the black solid lines both for the scans and for the contours represent the nominal configuration, while the red solid lines show the modified configuration. 
\begin{table}[htbp]
\begin{center}
{\def\arraystretch{1.4}
\begin{tabular}{|l|c|c|c|c|}
\hline
POIs &  $\kappa_F{}^{+1\sigma}_{-1\sigma}$& $\kappa_V{}^{+1\sigma}_{-1\sigma}$ & $\kappa_\lambda{}^{+1\sigma}_{-1\sigma}$ &  $\kappa_\lambda$  [95\% CL] \\ 
\hline
\multicolumn{5}{|c|}{Nominal}\\
\hline
 $\kappa_\lambda$, $\kappa_V$    &1 & $1.04^{+0.05}_{-0.04}$  & $4.8^{+7.4}_{-6.7}$ & $[-6.7,18.4]$ \\
 \hline
 $\kappa_\lambda$, $\kappa_F$  &$0.99^{+0.08}_{-0.08}$  & 1 & $4.1^{+4.3}_{-4.1}$  & $[-3.2, 11.9]$ \\
\hline 

\multicolumn{5}{|c|}{Modified with $\kappa_i^3$ terms}\\
\hline
 $\kappa_\lambda$, $\kappa_V$  &1  & $1.04^{+0.05}_{-0.04}$  & $4.8^{+7.4}_{-6.7}$ & $[-6.7,18.3]$ \\
\hline
 $\kappa_\lambda$, $\kappa_F$   &$0.99^{+0.08}_{-0.07}$  & 1 & $4.0^{+4.3}_{-4.1}$  & $[-3.3, 11.9]$ \\
\hline
\end{tabular}
}
\caption{
Best-fit observed values for $\kappa$ modifiers with $\pm 1 \sigma$ uncertainties. The 95\% CL interval for $\kappa_\lambda$ is also reported. The first column shows the parameters of interest included in each fit configuration, where the other coupling modifiers are kept fixed to the SM prediction. The set of rows on the top of the table shows fit results obtained under the nominal assumption while the set on the bottom shows the results obtained introducing the $\kappa_i^3$ corrections.}
\label{tab:klambda_test_ki3}
\end{center}
\end{table}

\clearpage
\section{HL-LHC projections}
\label{sec:HL_LHC_single}

Projections for the measurement of the trilinear Higgs self-coupling at HL-LHC have been made considering NLO-EW corrections depending on $\kappa_\lambda$ to single-Higgs processes.\newline
The parameterisations described at the beginning of this chapter and in Chapter~\ref{sec:prob_self} have been exploited in order to give an estimation of ATLAS HL-LHC projections in constraining $\kappa_\lambda$; the differential information has not been used because of the fact that only inclusive inputs are available for these HL-LHC studies; these inputs have been obtained starting from the inclusive inputs tested in Section~\ref{sec:results_single_kl} and studying different systematic scenarios. An Asimov dataset generated under the SM hypothesis considering a luminosity of 3000 fb$^{-1}$ at $\sqrt{s}=14$ TeV has been used. Concerning systematic uncertainties, two different scenarios have been considered: a ``Run 2 sys$"$ scenario, where the uncertainties are assumed to be equal to the Run 2 ones and a ``Reduced sys$"$ scenario, considering reduced systematics uncertainties obtained thanks to the much larger amount of data collected: theory uncertainties are halved with respect to the Run 2 uncertainties and other systematic uncertainties are scaled as the statistical errors. Table~\ref{kl_kf_kv_HL_LHC} summarises all the configurations tested in the two systematic scenarios.
\begin{table}[h]
\begin{center}
{\def\arraystretch{1.1}
\begin{tabular}{|c|c|c|c|c|c|}
\hline

POIs & Systematic scenarios & $\kappa_F{}^{+1\sigma}_{-1\sigma}$& $\kappa_V{}^{+1\sigma}_{-1\sigma}$ & $\kappa_\lambda{}^{+1\sigma}_{-1\sigma}$ & $\kappa_\lambda$  [95\% CL] \\ 
\hline
\multirow{2}{*}{$\kappa_\lambda$ } & Run 2 sys   &\multirow{2}{*}{1}  & $\multirow{2}{*}{1} $  & $1.0^{+3.6}_{-2.3}$ & $[-3.0, 9.0]$ \\
                                     &  Reduced sys &   &    &  $1.0_{-1.7}^{+2.3}$ & $[-2.0, 6.5]$ \\   
\hline

\multirow{2}{*}{$\kappa_\lambda$, $\kappa_V$ } & Run 2 sys   &\multirow{2}{*}{1}  & $1.00^{+0.02}_{-0.02}$  & $1.0^{+4.2}_{-2.8}$ & $[-3.9, 8.9]$ \\
                                     &  Reduced sys &   &   $1.00^{+0.01}_{-0.01}$ &  $1.0_{-2.1}^{+2.6}$ & $[-2.7, 6.5]$ \\   
\hline
\multirow{2}{*}{$\kappa_\lambda$, $\kappa_F$ } &Run 2 sys &$1.00^{+0.02}_{-0.02}$  & \multirow{2}{*}{1} & $1.0^{+3.8}_{-2.3}$  & $[-3.0, 9.3]$ \\
                                    & Reduced sys    &      $1.00^{+0.02}_{-0.02}$ &  & $1.0_{-1.7}^{+2.3}$ & $[-2.0, 6.6]$ \\
\hline 
\multirow{2}{*}{$\kappa_\lambda-\kappa_F-\kappa_V$} & Run 2 sys & $1.00^{+0.03}_{-0.03}$ &$1.00^{+0.03}_{-0.02}$  & $1.0_{-3.5}^{+4.1}$ &  $[-6.0, 10.4]$ \\
                                      & Reduced sys  & $1.00^{+0.02}_{-0.02}$ &  $1.00^{+0.02}_{-0.02}$    &  $1.0^{+2.6}_{-2.4}$ & $[-3.8, 6.6]$ \\
 \hline     
\end{tabular}
}
\caption{
Best-fit values for $\kappa$ modifiers with $\pm 1 \sigma$ uncertainties using an Asimov dataset generated under the SM hypothesis considering a luminosity of 3000 fb$^{-1}$ at $\sqrt{s}=14$ TeV. The first column shows the parameter(s) of interest in each fit configuration, where the other coupling modifiers are kept fixed to the SM prediction. The 95\% CL interval for $\kappa_\lambda$ is also reported.}
\label{kl_kf_kv_HL_LHC}
\end{center}
\end{table}

\begin{figure}[!htbp]
\centering
\begin{subfigure}[b]{0.49\textwidth}
\includegraphics[height=7 cm, width=\textwidth]{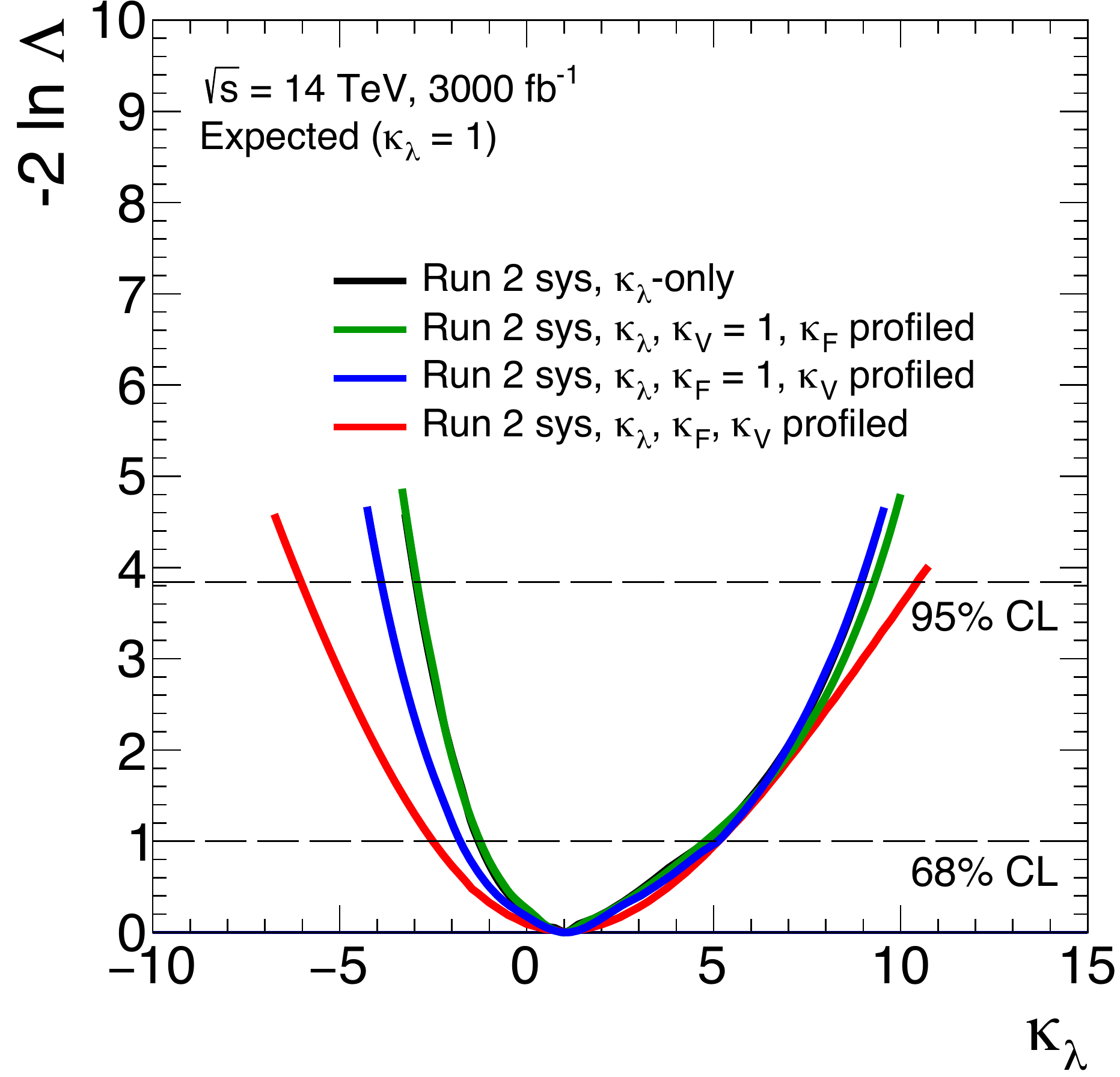}
 \caption{}
\end{subfigure}
\begin{subfigure}[b]{0.49\textwidth}
\includegraphics[height=7 cm, width=\textwidth]{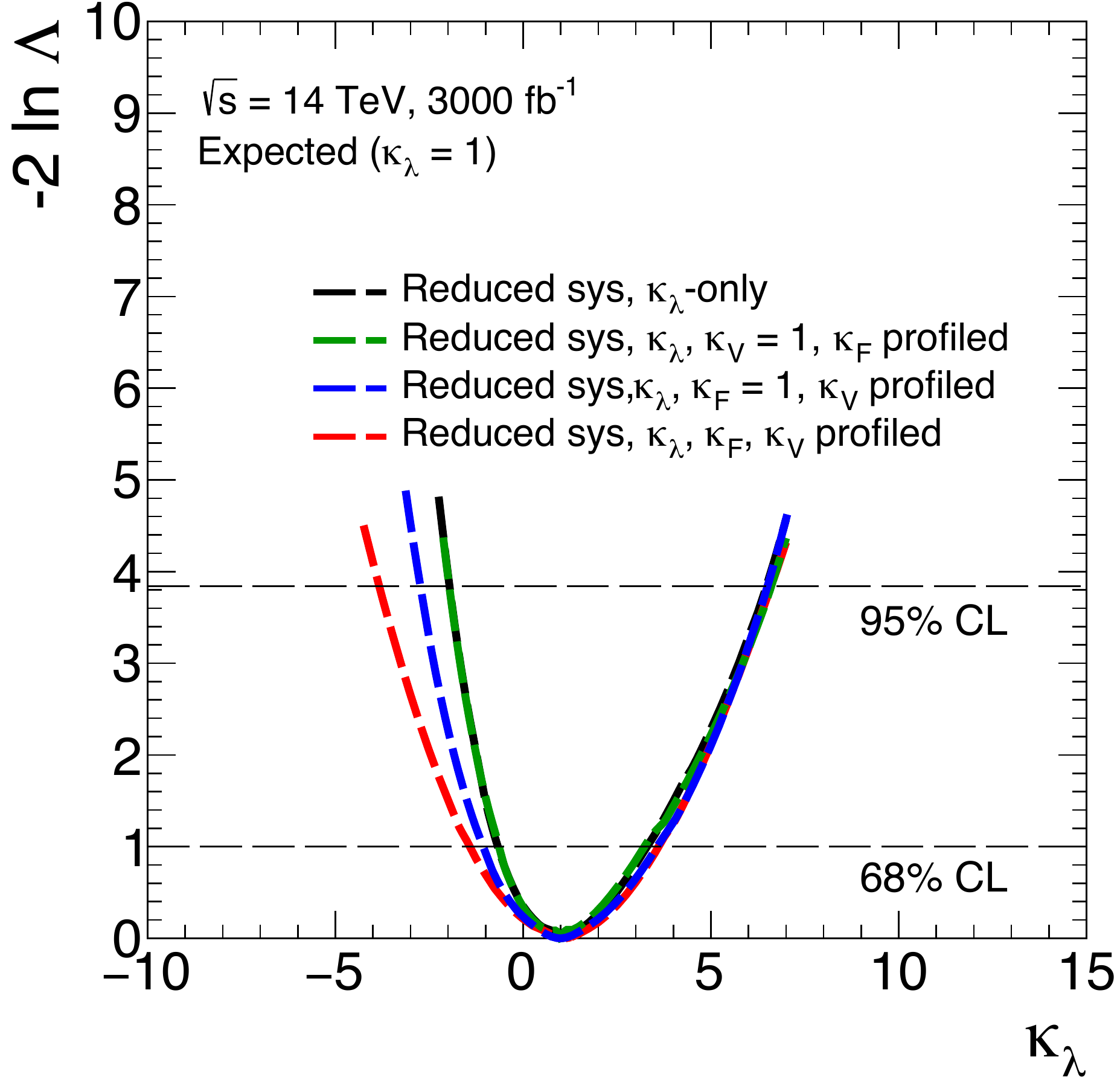}
 \caption{}
\end{subfigure}
\caption{Value of $-2 \ln{\Lambda(\kappa_\lambda)}$ as a function of $\kappa_\lambda$ for an Asimov dataset generated under the SM hypothesis considering a luminosity of 3000 fb$^{-1}$ at $\sqrt{s}=14$ TeV.  Two different scenarios are considered: ``Run 2 sys$"$ (a) and ``Reduced sys$"$ (b). Different fit configurations have been tested: $\kappa_\lambda$-only model (black line), $\kappa_\lambda$-$\kappa_F$ model (green line), $\kappa_\lambda$-$\kappa_V$ model (blu line) and $\kappa_\lambda$-$\kappa_F$-$\kappa_V$ model, (red line). All the coupling modifiers that are not included in the fit are set to their SM predictions. The dotted horizontal lines show the $-2 \ln{\Lambda(\kappa_\lambda)}=1$ level that is used to define the $\pm 1\sigma$ uncertainty on $\kappa_\lambda$ as well as the $-2 \ln{\Lambda(\kappa_\lambda)}=3.84$ level used to define the 95\% CL.}
\label{HL_single_1}
\end{figure}

\begin{figure}[!htbp]
\begin{center}
\includegraphics[width=0.5\textwidth]{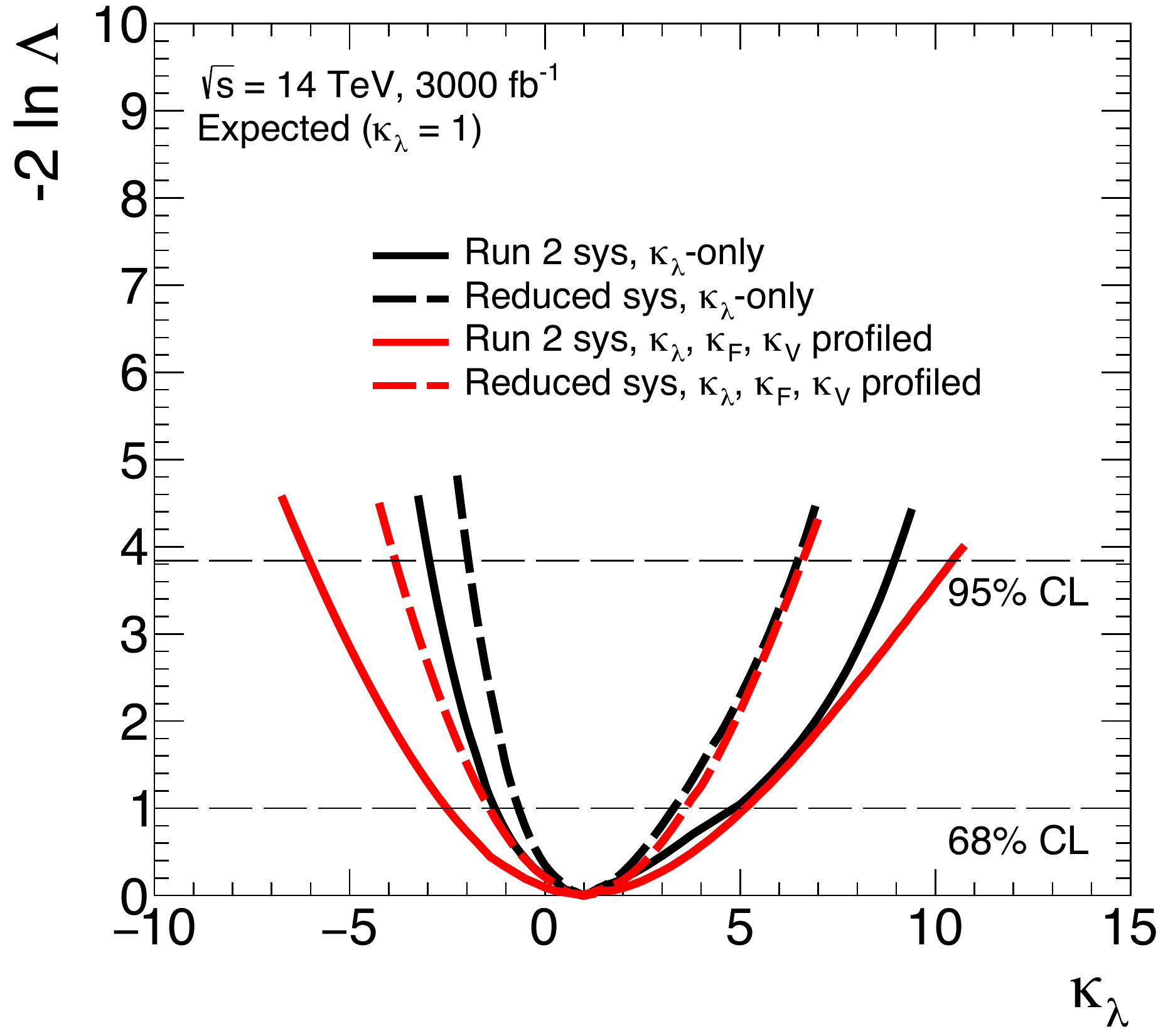}
\end{center}
\caption{Value of $-2 \ln{\Lambda(\kappa_\lambda)}$ as a function of $\kappa_\lambda$ for an Asimov dataset generated under the SM hypothesis considering a luminosity of 3000 fb$^{-1}$ at $\sqrt{s}=14$ TeV.  Two different scenarios are considered: ``Run 2 sys$"$ (solid lines) and ``Reduced sys$"$ (dashed lines). Two different fit configurations are reported: $\kappa_\lambda$-only model (black line) and $\kappa_\lambda$-$\kappa_F$-$\kappa_V$ model, (red line). All the coupling modifiers that are not included in the fit are set to their SM predictions. The dotted horizontal lines show the $-2 \ln{\Lambda(\kappa_\lambda)}=1$ level that is used to define the $\pm 1\sigma$ uncertainty on $\kappa_\lambda$ as well as the $-2 \ln{\Lambda(\kappa_\lambda)}=3.84$ level used to define the 95\% CL.}
\label{HL_single_2}
\end{figure}

The value of $-2 \ln{\Lambda(\kappa_\lambda)}$ as a function of $\kappa_\lambda$ for an Asimov dataset generated under the SM hypothesis is shown in Figures~\ref{HL_single_1} and~\ref{HL_single_2} for the different fit configurations listed in Table~\ref{kl_kf_kv_HL_LHC} and for the different systematic scenarios. Solid and dotted lines represent the ``Run 2 sys$"$ and the ``Reduced sys$"$ scenarios in the different fit configurations, respectively. The expected 95\% CL intervals of $\kappa_\lambda$, considering just modifications of the self-coupling, are $-2.0<\kappa_\lambda < 6.5$ and $-3.0< \kappa_\lambda < 9.0$ for the Reduced Scenario and for the Run 2 systematic scenario, respectively.
The ``Run 2 sys$"$ interval, considering just modifications of the Higgs self-coupling ($\kappa_\lambda$-only), is comparable to the interval obtained in Reference~\cite{Maltoni}, ``Run 2 sys$"$ scenario, shown in Figure~\ref{HL_single_maltoni}, and the likelihood shapes of the additional fit configurations are also compatible. The ``Reduced sys$"$ $\kappa_\lambda$-only interval is, instead, comparable to the one obtained in Reference~\cite{Degrassi}, shown in Figure~\ref{HL_single_degrassi}, considering the same uncertainty scenario.

\chapter{Constraining the Higgs-boson self-coupling combining single- and double-Higgs production and decay measurements}
\label{sec:combination}
This chapter presents the final results of this dissertation, providing the most stringent constraint on $\kappa_\lambda$ from experimental measurements to date, through the combination of the double- and single-Higgs analyses whose details have been described in Chapters~\ref{sec:dihiggs} and~\ref{sec:single}. The chapter is organised as follows: Section~\ref{sec:comb_data} presents a brief summary of the analyses exploited in this combination; Sections~\ref{sec:comb_theory} and~\ref{sec:comb_stat} revise the theoretical framework that has been implemented and the statistical model including nuisance parameter correlations between single- and double-Higgs analyses, respectively. Overlap studies aiming to understand and take into account overlaps between categories of the different analyses are shown in Section~\ref{sec:comb_overlap}, while a validation of the inputs of the combination performed separately for single- and double-Higgs analyses is presented in Section~\ref{sec:comb_validation}.
Finally, Sections~\ref{sec:comb_results_kl},~\ref{sec:comb_results_kl_kt} and~\ref{sec:comb_results_kl_other} summarise the results of this combination exploiting different fit configurations ranging from a $\kappa_\lambda$-only model to a generic model.

\section{Data and input measurements}
\label{sec:comb_data}

The results presented in this chapter are obtained using data collected by the ATLAS experiment in 2015, 2016 and 2017 from 13 TeV $pp$ collision data corresponding to a luminosity of up to 79.8 fb$^{-1}$. The integrated luminosity for each analysed decay channel is summarised in Table~\ref{tab:lumi_comb}. The combination takes as inputs the double-Higgs analyses described in Chapter~\ref{sec:dihiggs}, \ie\ $b\bar{b}\tau^+\tau^-$, $b\bar{b}\gamma\gamma$ and $b\bar{b}b\bar{b}$ final states, as well as the single-Higgs analyses described in Chapter~\ref{sec:single}, including the $ggF$, \VBF, \WH, \ZH and $t\bar{t}H$ production modes and the $\gamma\gamma$, $WW^*$, $ZZ^*$, $b\bar{b}$ and $\tau^+\tau^-$ decay channels. Details on the individual channels are already reported in the aforementioned chapters.\newline
The single-Higgs and double-Higgs categories are not all orthogonal to each other. The overlap between these categories has been studied and is described in Section~\ref{sec:comb_overlap}; only categories with negligible overlap have been included in this combination: thus the $ t\bar{t}H \rightarrow \gamma \gamma$ categories included in the results of Chapter~\ref{sec:single} have been removed from this combination as they show large overlap with the $HH \rightarrow b \bar{b} \gamma \gamma$ categories and have an impact which is significantly smaller than the impact coming from $HH\rightarrow b\bar{b} \gamma\gamma$ categories.

\begin{table}[!htbp]
  \caption{Integrated luminosity of the datasets used for each input
    analysis to the $H+HH$ combination. The last column provides references to
    publications describing each measurement included in detail.}
\begin{center}
\scalebox{0.94}{
{\def\arraystretch{1.4}
\begin{tabular}{|l|c|c|}
\hline 
Analysis & Integrated luminosity (fb$^{-1}$) & Reference \\
\hline
$HH\rightarrow b\bar{b}b\bar{b}$      & $27.5$         & \cite{4b} \\
$HH\rightarrow b\bar{b}\tau^+\tau^-$      & $36.1$         & \cite{bbtautau} \\
$HH\rightarrow b\bar{b} \gamma\gamma$      & $36.1$         & \cite{bbyy} \\
\hline 
$H \rightarrow \gamma \gamma$\ (excluding $t\bar{t}H$, $H\rightarrow \gamma \gamma$)       & $79.8$         & \cite{yy,yy1,ttH_yy} \\
$H\rightarrow ZZ^*\rightarrow 4\ell$ (including $t\bar{t}H$, $H\rightarrow ZZ^*\rightarrow 4\ell$) & $79.8$         & \cite{ZZ,ZZ1} \\
$H\rightarrow WW^* \rightarrow e\nu \mu \nu$                        & $36.1$         & \cite{WW} \\
$H\rightarrow \tau\tau$                               & $36.1$         & \cite{tautau} \\
$VH$, $H\rightarrow b\bar{b}$                          & $79.8$         & \cite{VH_bb,VH_bb1} \\
$t\bar{t}H$, $H\rightarrow b\bar{b}$ and $t\bar{t}H$ multilepton  & $36.1$         & \cite{ttH,ttH_bb} \\
\hline
\end{tabular}}}
\end{center}
\label{tab:lumi_comb}
\end{table}

\section{Theoretical model}
\label{sec:comb_theory}
The theoretical models described in Chapter~\ref{sec:prob_self} are exploited in order to implement the dependence on $\kappa_\lambda$ and on the other coupling modifiers in double- and single-Higgs analyses. Details on how this dependence has been implemented are reported in the chapters describing the corresponding analyses, Chapters~\ref{sec:dihiggs} and~\ref{sec:single}. The values of the $C_{1}$ coefficients for both the initial, $i$, and the final, $f$, states and the $K_{EW}$ coefficients are reported in Chapter~\ref{sec:prob_self}, while the $\kappa$ modifiers at LO for the initial and final states are reported in Chapters~\ref{sec:dihiggs} and~\ref{sec:single}.
For single-Higgs analyses, the differential information is exploited in the regions defined by the STXS stage-1 framework, particularly in the \VBF, \ZH and \WH production modes.\newline
The results of this chapter are presented exploiting the coupling modifiers $\kappa_t$, $\kappa_b$, $\kappa_{\ell}$, $\kappa_W$, $\kappa_Z$. They describe the modifications of the SM Higgs boson couplings to up-type quarks, to down-type quarks, to
leptons and to $W$ and $Z$ vector bosons, respectively; alternatively, the coupling modifiers $\kappa_F$=$\kappa_t$=$\kappa_b$=$\kappa_{\ell}$ and $\kappa_V$=$\kappa_W$=$\kappa_Z$ are used, describing modifications of SM Higgs boson couplings to fermions and vector bosons, respectively. 

\section{Statistical model}
\label{sec:comb_stat}

The main principles used in order to extract the final results have been reported in both Chapters~\ref{sec:dihiggs} and~\ref{sec:single}. In this section elements that have been added in order to produce the results of the combination are reported, together with the basic principles of the correlation scheme used to combine single- and double-Higgs analyses.
The results of the $H+HH$ combination are obtained from a likelihood function $L(\vec{\alpha},\vec{\theta})$, where $\vec{\alpha}$ represents the vector of POIs of the model and $\vec{\theta}$ is the set of nuisance parameters, including the systematic uncertainty contributions and background parameters that are constrained by side bands or control regions in data. The number of signal events in each analysis category $j$ is defined as:
\begin{equation}
n^{\text{signal}}_j(\kappa_\lambda, \kappa_t, \kappa_b, \kappa_\ell, \kappa_W, \kappa_Z, \vec{\theta}) = \mathcal{L}_j(\vec{\theta}) \sum_i \sum_f \mu_{i}(\kappa) \times \mu_{f}(\kappa) (\sigma_{\text{SM},i}(\vec{\theta}) \times \text{BR}_{\text{SM},f}(\vec{\theta})) (\epsilon \times A)_{if,j}(\vec{\theta})
\label{eq:yields}
\end{equation}

where the index $i$ runs over all the production regions defined by the STXS stage-1 framework and all the double-Higgs regions and the index $f$ includes all the considered decay channels, \ie $f= \gamma\gamma, ZZ^*, WW^*, \tau^+\tau^-, b\bar{b}$ for the single-Higgs part while $f= b\bar{b}b\bar{b}, b\bar{b}\tau^+\tau^-,  b\bar{b}\gamma\gamma$ for the double-Higgs part. $\mathcal{L}_j$ is the integrated luminosity of the dataset used in the $j$ category, and $(\epsilon \times A)_{if,j}$ represents the acceptance and efficiency estimation for the category $j$, the production process $i$ and the decay channel $f$. All these terms depend also on a set of nuisance parameters $\vec{\theta}$, that account for theoretical and systematic uncertainties that can affect the luminosity, the cross-section and branching fraction prediction, the efficiency estimation, and the background estimation. Finally, $\mu_i(\kappa)\times \mu_f(\kappa)=\mu_i(\kappa_\lambda, \kappa_t, \kappa_b, \kappa_\ell, \kappa_W, \kappa_Z)\times \mu_f( \kappa_\lambda,\kappa_t, \kappa_b, \kappa_\ell, \kappa_W, \kappa_Z)$, describes the yield dependence on the Higgs-boson self-coupling modifier $\kappa_\lambda$, and on the single Higgs boson coupling modifiers $\kappa_t$, $\kappa_b$, $\kappa_\ell$, $\kappa_W$ and $\kappa_Z$, representing potential deviations from the SM expectation.\newline 
Confidence intervals for the POIs are determined using the profile likelihood ratio, described in previous chapters, as the test statistic, using the 68\% as well as the 95\% CL intervals in the asymptotic limit~\cite{Cowan}.\newline
The correlations between the systematic uncertainties within the single-Higgs and double-Higgs individual combinations are described in the chapters corresponding to the two analyses, \ie\ Chapters~\ref{sec:dihiggs} and~\ref{sec:single}. The correlation of the systematic uncertainties between single- and double-Higgs analyses has also been investigated and is taken into account in this combination, as described in the following section.
\subsection{Correlation scheme between single- and double-Higgs analyses}

The correlation scheme adopted in order to make the $H+HH$ combination is driven by the following guidelines~\cite{Confnote_comb}:
\begin{itemize}
\item experimental uncertainties have been correlated whenever relevant, like it was made in the case of the uncertainty on the integrated luminosity;
\item experimental uncertainties that are related to the same physics object but determined with different methodologies or implemented with different parameterisations have been kept uncorrelated, like in the case of flavour tagging uncertainties;
\item theory uncertainties related to signal processes have been kept uncorrelated;
\item theory uncertainties on the decay branching fractions have been correlated;
\item data-driven background uncertainties in double-Higgs analyses are not correlated with single-Higgs analyses.
\end{itemize}

\begin{figure}[hbtp]
\centering
\begin{subfigure}[b]{0.49\textwidth}
\includegraphics[height=13 cm,width =\textwidth]{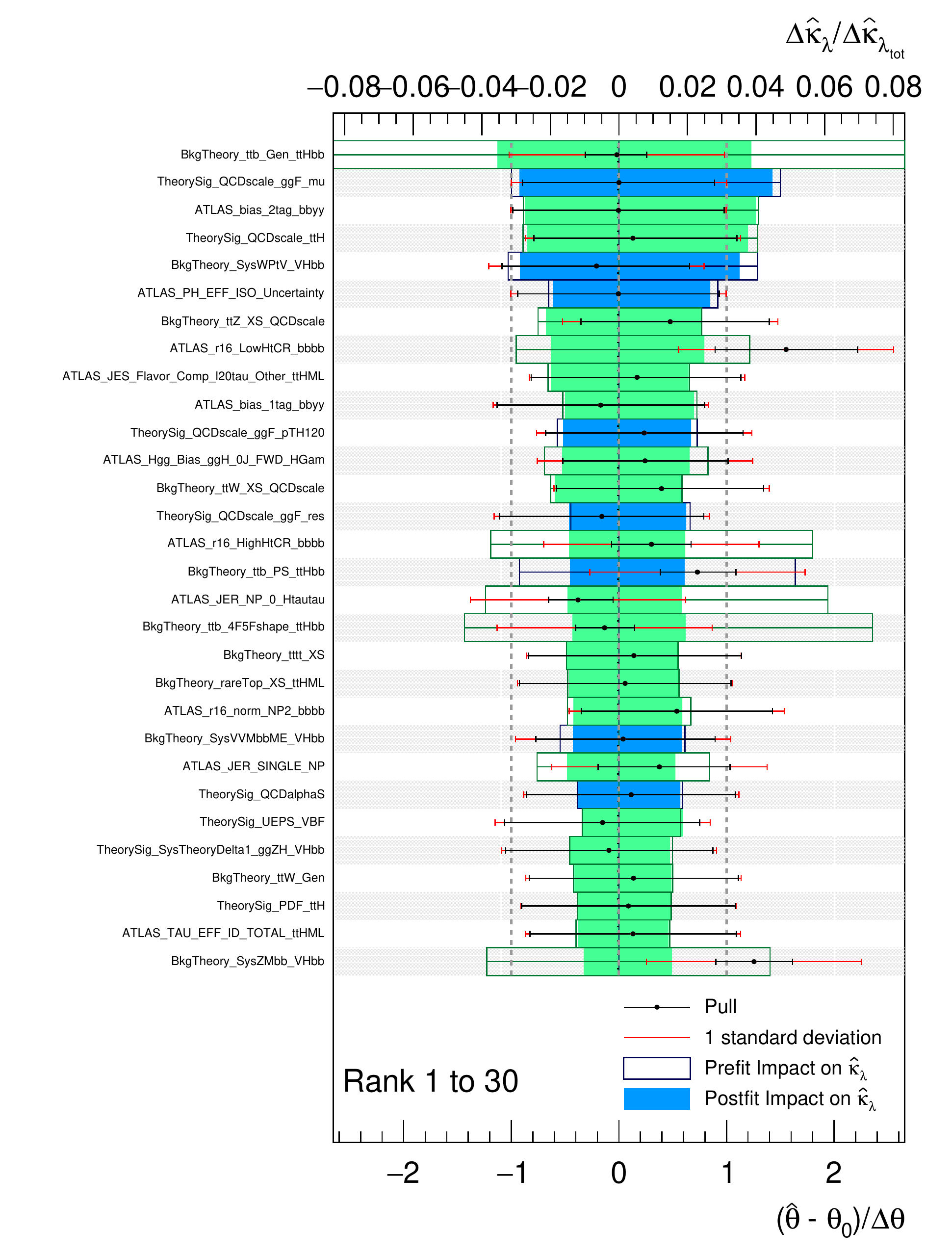}
 \caption{}
\end{subfigure}
\begin{subfigure}[b]{0.49\textwidth}
\includegraphics[height=13 cm,width =\textwidth]{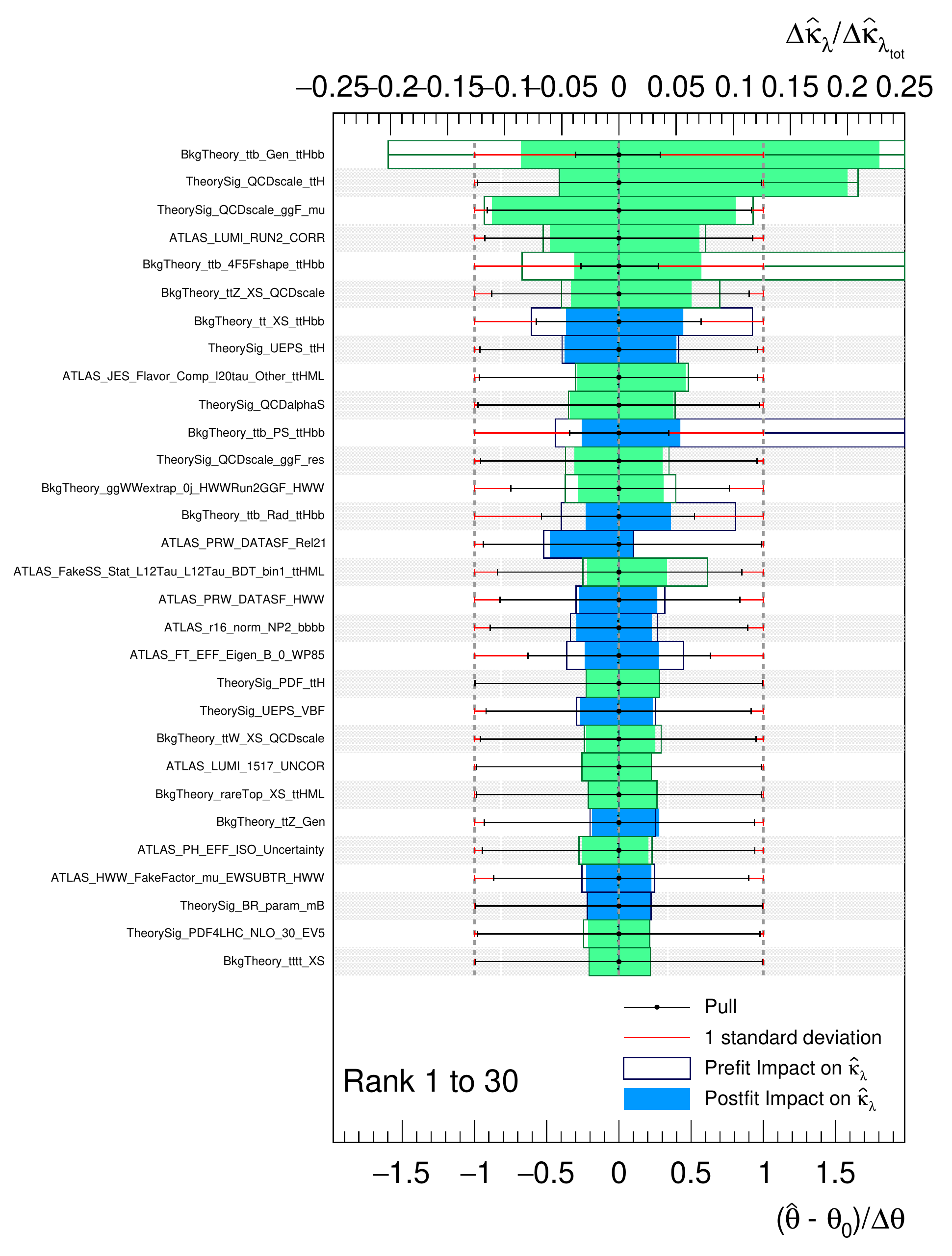}
 \caption{}
\end{subfigure}
\caption{Ranking of the top 30 systematic uncertainties in the $H+HH$ combination for data (a) and for the Asimov dataset (b) generated under the SM hypothesis.}     
\label{ranking_comb}
\end{figure}

Several studies have been made in order to assess the impact of correlating different uncertainties, that are kept uncorrelated in the nominal configuration; the strategy that has been adopted has been not to correlate uncertainties whose correlation has a negligible or a null impact on the results:
\begin{itemize}
\item in the configuration scheme adopted for the combination:
\begin{itemize}
\item the theoretical $\alpha_S$ uncertainties and pile-up reweighting (PRW) uncertainties are uncorrelated among single-Higgs and double-Higgs analyses and among $HH\rightarrow b\bar{b}\gamma \gamma$ and other channels, respectively; the impact on $\kappa_\lambda$ results of correlating these uncertainties is negligible as reported in Table~\ref{tab:correlateAlphaSPRW} where a comparison with the nominal correlation scheme adopted in this combination is made;
\begin{table}[h]
\begin{center}
{\def\arraystretch{1.3}
\begin{tabular}{|c|c|c|}
\hline
Correlation scheme&1 $\sigma$&95\% CL\\
\hline
\multirow{2}{*}{Nominal}&$4.6_{-3.8}^{+3.2}$&[-2.3, 10.3]\\
&$1.0_{-3.8}^{+7.3}$&[-5.1, 11.2]\\
\hline
\multirow{2}{*}{Correlating $\alpha_S$, PRW}&$4.6_{-3.8}^{+3.2}$&[-2.3, 10.3]\\
&$1.0_{-3.8}^{+7.3}$&[-5.1, 11.2]\\
\hline
\end{tabular}}
\caption{Comparison of the $\kappa_{\lambda}$ measurement nominal results with the results obtained correlating $\alpha_S$ and PRW uncertainties. Both $\kappa_\lambda$ $\pm$1$\sigma$ uncertainty and 95\% CL interval are reported. The first row shows the observed results while the second row shows the expected results.}
\label{tab:correlateAlphaSPRW}
\end{center}
\end{table}
\item the experimental uncertainties related to the identification and energy scale of $\tau$ leptons, \ie\ ``TAU\_EFF\_ID\_HIGHPT$"$, ``TAU\_EFF\_ID\_TOTAL TAU\_TES\\\_DET$"$ and ``TAU\_TES\_INSITU, TAU\_TES\_MODEL$"$, are uncorrelated among $t\bar{t}H\rightarrow$ multilepton and the double-Higgs channels; the impact on $\kappa_\lambda$ of correlating these uncertainties is negligible as reported in Table~\ref{tab:correlateTAU} where a comparison with the nominal correlation scheme adopted in this combination is made;
\begin{table}[h]
\begin{center}
{\def\arraystretch{1.3}
\begin{tabular}{|c|c|c|}
\hline
Correlation scheme &1 $\sigma$&95\% CL\\
\hline
\multirow{2}{*}{Nominal}&$4.6_{-3.8}^{+3.2}$&[-2.3, 10.3]\\
&$1.0_{-3.8}^{+7.3}$&[-5.1, 11.2]\\
\hline
\multirow{2}{*}{Correlating TAU}&$4.6_{-3.8}^{+3.2}$&[-2.3, 10.3]\\
&$1.0_{-3.8}^{+7.3}$&[-5.1, 11.2]\\
\hline
\end{tabular}}
\caption{Comparison of the $\kappa_{\lambda}$ measurement nominal results with the results obtained correlating the TAU-related experimental uncertainties. Both $\kappa_\lambda$ $\pm$1$\sigma$ uncertainty and 95\% CL interval are reported. The first row shows the observed results while the second row shows the expected results.}
\label{tab:correlateTAU}
\end{center}
\end{table}
\item flavour tagging uncertainties, \ie\ ``FT\_EFF\_Eigen\_X\_N (X = B, C, Light, \\ N = 0, 1, 2, 3, 4)$"$, ``FT\_EFF\_extrapolation$"$, ``FT\_EFF\_extrapolation\_from\\\_charm$"$, are uncorrelated since single-Higgs analyses use the 85\% efficiency $b$-tagging working point, for 2015-2016 analyses using Release 20.7, while double-Higgs analyses use the 70\% efficiency working point for the same release. No significant changes with respect to the nominal results are found correlating these uncertainties, as shown in Table~\ref{tab:correlateFT};
\begin{table}[h]
\begin{center}
{\def\arraystretch{1.3}
\begin{tabular}{|c|c|c|}
\hline
Correlation scheme &1 $\sigma$&95\% CL\\
\hline
\multirow{2}{*}{Nominal}&$4.6_{-3.8}^{+3.2}$&[-2.3, 10.3]\\
&$1.0_{-3.8}^{+7.3}$&[-5.1, 11.2]\\
\hline
\multirow{2}{*}{Correlating FT}&$4.6_{-3.8}^{+3.2}$&[-2.3, 10.3]\\
&$1.0_{-3.8}^{+7.3}$&[-5.1, 11.2]\\
\hline
\end{tabular}}
\caption{$\kappa_{\lambda}$ measurement results by correlating FT uncertainties with different WPs. Both $\kappa_\lambda$ $\pm$1$\sigma$ uncertainty and 95\% CL interval are reported. The first row shows the observed results while the second row shows the expected results.}
\label{tab:correlateFT}
\end{center}
\end{table}
\end{itemize}
\item due to the fact that, in single-Higgs and double-Higgs analyses, ggF QCD scale and PDF uncertainties use different schemes and number of nuisance parameters, they are not correlated.
\end{itemize}
The list of the NPs included in single-Higgs analyses, in double-Higgs analyses and in the combination of the two of them is reported in Appendix~\ref{sec:appendix_correlation_comb}.\newline
Figure~\ref{ranking_comb} shows the ranking plots, defined in Chapter~\ref{sec:dihiggs}, of the top 30 systematic uncertainties for the $H+HH$ combination considering data $(a)$ and considering the Asimov dataset $(b)$. The nuisance parameters having the largest impact on $\kappa_\lambda$ are the ones related to the theory modelling of signal and background processes in simulation coming from the single-Higgs analyses, to the data-driven background modelling (mainly multi-jets) of $HH\rightarrow b\bar{b}b\bar{b}$ and $HH\rightarrow b\bar{b}\gamma \gamma$ analyses and the experimental nuisance parameters related to photons and jets, consistently with what has been found in double- and single-Higgs combinations. The different nuisance parameters ranked in Figure~\ref{ranking_comb} have been described in Chapters~\ref{sec:dihiggs} and~\ref{sec:single}. Being the measurement statically dominated, this impact is small compared to the statistical uncertainties as it is shown in the result section.

\clearpage
\section{Overlap studies}
\label{sec:comb_overlap}
The event selection of the double-Higgs categories is not orthogonal by construction to the event selection of all the single-Higgs categories included in the combined fit. Thus some events might pass both selections and might be double counted. Overlap checks have been performed in order to quantify the expected fraction of shared events among overlapping categories normalised to the total number of events that pass the single-Higgs or the double-Higgs selections. In cases where a non-negligible overlap was found between the double-Higgs and single-Higgs signal regions, a fit has been performed to constrain $\kappa_\lambda$, with all other single-Higgs couplings set to their SM values, and to constrain $\kappa_\lambda$ and $\kappa_t$, setting all other couplings to their SM values except for $\kappa_t$, in order to exploit the dependence on $\kappa_\lambda$ and $\kappa_t$ of the double-Higgs analyses. The impact of the overlap on the combined $\kappa_\lambda$-only and $\kappa_\lambda-\kappa_t$ results has thus been checked removing the overlapping single-Higgs categories from the combined fit. 
\subsection{$HH\rightarrow b\bar{b} \gamma\gamma$ and $H\rightarrow\gamma\gamma$ overlap}
The single-Higgs $H\rightarrow \gamma \gamma$ analysis applies the following event selection:
\begin{itemize}
\item two isolated (FixedCutLoose isolation criterion, defined in Chapter~\ref{sec:Reco}) and identified (Tight identification criterion, defined in Chapter~\ref{sec:Reco}) photons; 
\item $p_T^\gamma/m_{\gamma\gamma}$> 0.35 and 0.25 for leading and subleading photons, respectively;
\item $105 \ \text{GeV}<m_{\gamma\gamma}<160 \ \text{GeV}$.
\end{itemize}
The double-Higgs $HH\rightarrow b\bar{b} \gamma\gamma$ event selection adds to the listed requirements, the following:
\begin{itemize}
\item 1-tag:
\begin{itemize}
\item single-Higgs selection;
\item two jets $|\eta|<2.5$;
\item one $b$-tagged jet (at the 60\% $b$-tagging efficiency working point);
\item the highest-$p_T$ jet is required to have $p_T^{j,1}>40$ GeV and the next-highest-$p_T$ jet must satisfy $p_T^{j,2}>25$ GeV;
\item $80<m_{jj}<140$ GeV.
\end{itemize}
\item 2-tag:
\begin{itemize}
\item single-Higgs selection;
\item two jets $|\eta|<2.5$;
\item exactly two $b$-tagged jets (at the 70\% $b$-tagging efficiency working point, as defined in Chapter~\ref{sec:Reco});
\item $p_T^{j,1}>40$ GeV and $p_T^{j,2}>25$ GeV;
\item $80<m_{jj}<140$ GeV.
\end{itemize}
\end{itemize}
The overlap between $b\bar{b} \gamma\gamma$ and $H\rightarrow \gamma \gamma$ analyses, defined as the number of $b\bar{b} \gamma\gamma$ signal events in $H\rightarrow \gamma \gamma$ categories normalised to the number of $b\bar{b} \gamma\gamma$ signal events, is 100\%, being the $HH\rightarrow b\bar{b} \gamma\gamma$ analysis a subsample of the $H\rightarrow \gamma \gamma$ analysis.
A comparison between the data events passing the $HH \rightarrow b\bar{b} \gamma \gamma$ event selection and the data events passing the event selection of each $H \rightarrow \gamma \gamma$ category has been made in order to identify the most overlapping regions and quantify the size of this overlap; thus the run number, assigned uniquely to each data taking run that starts after the declaration of stable beams, and the event number that, combined with the run number, uniquely identifies an event, of these double-Higgs and single-Higgs events have been compared. The fraction of overlapping events is negligible for all the processes, at the level of a few percent, except for the $t\bar{t}H \rightarrow \gamma \gamma$ production mode, considering both hadronic and leptonic categories, given the absence of any lepton veto at the selection level of this double-Higgs analysis: indeed, in $t\bar{t}H$ production mode, there are categories where up to 50\% of the selected $t\bar{t}H \rightarrow \gamma \gamma$ events are also selected by the $HH\rightarrow b\bar{b} \gamma\gamma$ analysis.\newline
The impact of the overlap on the results has been checked performing the combined fit removing the single-Higgs most overlapping categories. Tables~\ref{tab:ttH_yy_impact_exp} and~\ref{tab:ttH_yy_impact_obs} report the expected and observed fit results either excluding or including all $t\bar{t}H\rightarrow \gamma \gamma$ categories or all $HH\rightarrow b\bar{b} \gamma \gamma$ categories as well as excluding overlapped events in the $H+HH$ combination; considering the $\kappa_\lambda$-only fit, where all the other Higgs couplings are set to their SM values, the impact on the results removing different categories/events is summarised as follows:
\begin{itemize}
\item removing all $t\bar{t}H$ categories: 2.5\% (2\%) impact on the 95\% CL interval considering data (Asimov);
\item removing overlapped events: 1\% (2\%) impact on the 95\% CL interval considering data (Asimov);
\item removing all $b\bar{b}\gamma \gamma$ categories: 10\% (14\%) impact on the 95\% CL interval considering data (Asimov).
\end{itemize}
All the $t\bar{t}H\rightarrow \gamma \gamma$ categories have been removed from this combination as they show large overlap with the $HH \rightarrow b \bar{b} \gamma \gamma$ categories; such a decision has been made after also checking that the impact on the expected and observed combined limits of removing these categories is smaller with respect to removing $HH\rightarrow b\bar{b} \gamma \gamma$ categories; in fact these categories, as it will be shown in Section~\ref{sec:comb_results_kl}, represent one of the dominant contributions in order to constrain $\kappa_\lambda$.

\begin{landscape}
\begin{table}
\begin{center}
{\def\arraystretch{1.4}
\begin{tabular}{|c|c|c|c|}
 \hline
Analysis  & Fit configuration & $\kappa_\lambda$ interval at 95$\%$ CL (exp) & $\kappa_t$ interval at 95$\%$ CL (exp)   \\ 
\hline
 \multirow{2}{*} {single-Higgs v9 (no $ t\bar{t}H \rightarrow \gamma \gamma$) +   HH} & $\kappa_\lambda$ only & [-5.107 - 11.232] & -- \\
 & $\kappa_\lambda - \kappa_t$ & [-5.534 - 11.280] & [0.872 - 1.139]\\ 
 \hline
  \multirow{2}{*} {single-Higgs v9 (no overlapped events) +   HH} & $\kappa_\lambda$-only & [-4.830 - 10.933] & -- \\
 & $\kappa_\lambda - \kappa_t$ & [-5.297 - 10.966] & [0.874 - 1.138]\\ 
 \hline
   \multirow{2}{*} {single-Higgs v9  +   HH (no $b\bar{b}\gamma\gamma$ categories)} & $\kappa_\lambda$ only & [-5.349 - 12.910] & -- \\
 & $\kappa_\lambda - \kappa_t$ & [-6.241 - 13.139] &  [0.872 - 1.147]\ \\ 
 \hline
\multirow{2}{*} {single-Higgs v9 +   HH } &  $\kappa_\lambda$-only & [-4.893 - 11.122] & -- \\
 & $\kappa_\lambda - \kappa_t$ &[-5.393 - 11.159] & [0.871 - 1.139]\\ 
\hline
\end{tabular}}
\caption{95\% CL expected intervals for $\kappa_\lambda$ only and $\kappa_\lambda$, $\kappa_t$ fit configurations exploiting $H+HH$ combination. The fit has been performed \\ either excluding or including the categories listed in the first column.}
\label{tab:ttH_yy_impact_exp}
\end{center}
\end{table}
\end{landscape}

\begin{landscape}
\begin{table}
\begin{center}
{\def\arraystretch{1.4}
\begin{tabular}{|c|c|c|c|}

 \hline
Analysis  & Fit configuration & $\kappa_\lambda$ interval at 95$\%$ CL (obs) & $\kappa_t$ interval at 95$\%$ CL (obs)   \\ 
\hline
 \multirow{2}{*} {single-Higgs  v9 (no $ t\bar{t}H \rightarrow \gamma \gamma$) +   HH} & $\kappa_\lambda$-only & [-2.302 - 10.287] & -- \\
 & $\kappa_\lambda - \kappa_t$ & [-2.882 - 10.642] & [0.912 - 1.169]\\ 
 \hline
  \multirow{2}{*} {single-Higgs  v9 (no overlapped events) +   HH} & $\kappa_\lambda$-only & [-2.677 - 9.744] & -- \\
 & $\kappa_\lambda - \kappa_t$ & [-3.264 - 10.042] & [0.910 - 1.161]\\ 
 \hline
  \multirow{2}{*} {single-Higgs v9  +   HH (no $b\bar{b}\gamma\gamma$ categories)} & $\kappa_\lambda$-only & [-2.236 - 11.261] & -- \\
 & $\kappa_\lambda - \kappa_t$ & [-3.073 - 11.992] & [0.914 - 1.174] \\ 
 \hline
\multirow{2}{*} {single-Higgs  v9 +   HH} &  $\kappa_\lambda$-only & [-2.186 - 10.099] & -- \\
 & $\kappa_\lambda - \kappa_t$ &[-2.833 - 10.433] & [0.913 - 1.166]\\ 
\hline
\end{tabular}
}
\caption{95\% CL observed intervals for $\kappa_\lambda$ only and $\kappa_\lambda$, $\kappa_t$ fit configurations exploiting $H+HH$ combination. The fit has been performed\\ either excluding or including the categories listed in the first column.}
\label{tab:ttH_yy_impact_obs}
\end{center}
\end{table}
\end{landscape}

\subsection{$HH\rightarrow  b\bar{b} \tau^+\tau^-$ overlap with $H\rightarrow\tau^+\tau^-$, $t\bar{t}H\rightarrow b\bar{b}$, $t\bar{t}H$ multilepton, $VH\rightarrow b\bar{b}$}
The $HH\rightarrow  b\bar{b} \tau^+\tau^-$ analysis selects events using two categories, the $\tau_{\textrm{had}} \tau_{\textrm{had}}$ category and the $\tau_{\textrm{lep}} \tau_{\textrm{had}}$ category, described in detail in Chapter~\ref{sec:dihiggs}.\newline
Thus, looking at single-Higgs similar final states and event selections, possible overlapping analyses have been identified as the $H\rightarrow \tau^+ \tau^-$, $t\bar{t}H\rightarrow b\bar{b}$, $t\bar{t}H\rightarrow$ multilepton and $VH\rightarrow b\bar{b}$ single-Higgs analyses.
In order to quantify the overlap between signal regions, the run and event number of the data events passing the $HH \rightarrow b\bar{b} \tau^+ \tau^-$ event selection have been compared to the ones of the data events passing the event selection of the single-Higgs signal regions listed above. As the $HH\rightarrow  b\bar{b} \tau^+\tau^-$ analysis uses a BDT distribution as the final discriminant of the analysis, being the last BDT bin the most sensitive, the check comparing run and event number of double- and single-Higgs analyses has been performed also looking exclusively at this bin. The fraction of data events that passes both the $HH\rightarrow  b\bar{b} \tau^+\tau^-$ selections and the selections of a certain single-Higgs category, normalised to the number of double-Higgs events, gives an estimation of the overlap:
\begin{itemize}
\item $H\rightarrow \tau^+ \tau^-$: the overlap is present only between $HH\rightarrow  b\bar{b} \tau^+\tau^-$ ($\tau_{\textrm{had}} \tau_{\textrm{had}}$) category and $H\rightarrow \tau^+ \tau^-$ HadHad boosted category; the fraction of overlapping events is quantified as $3.4 \times 10^{-3}$ of full $HH\rightarrow  b\bar{b} \tau^+\tau^-$ ($\tau_{\textrm{had}} \tau_{\textrm{had}}$) signal region and $1.4 \times 10^{-1}$ in last BDT bin;
\item $t\bar{t}H\rightarrow b\bar{b}$: the overlap is present between $HH\rightarrow  b\bar{b} \tau^+\tau^-$ ($\tau_{\textrm{lep}} \tau_{\textrm{had}}$) category and $t\bar{t}H\rightarrow b\bar{b}$ single lepton (6j SR2, 6j SR3, 5j SR, boosted) categories; the fraction of overlapping events is quantified as $10^{-4}$ of full $HH\rightarrow  b\bar{b} \tau^+\tau^-$ ($\tau_{\textrm{lep}} \tau_{\textrm{had}}$) signal region and 0 in last BDT bin;
\item $t\bar{t}H\rightarrow$ multilepton; the categories are orthogonal, so no overlap is present;
\item $VH\rightarrow b\bar{b}$: the overlap is present between $HH\rightarrow  b\bar{b} \tau^+\tau^-$ and $VH\rightarrow b\bar{b}$ categories (1lep 2btags $\ge$ 3jets) and (0lep 2btags $\ge$ 3jets), while other categories are orthogonal: the fraction of overlapping events is quantified as:
\begin{itemize}
\item 1lep 2btags $\ge$ 3 jets: overlap of $1.7 \times 10^{-2}$ of full $HH\rightarrow  b\bar{b}\tau^+\tau^-$ ($\tau_{\textrm{lep}} \tau_{\textrm{had}}$) signal region and $5.3 \times 10^{-2}$ in last BDT bin;
\item 0lep 2btags $\ge$ 3 jets: overlap of $10^{-3}$ of full $HH\rightarrow  b\bar{b}\tau^+\tau^-$ ($\tau_{\textrm{had}} \tau_{\textrm{had}}$) signal region and 0 in last BDT bin.
\end{itemize}
\end{itemize}
The impact of the overlap on the results has been checked performing the combined fit removing the single-Higgs categories showing an overlap with double-Higgs categories above 1\%. Table \ref{tab:H_tautau_impact} reports the expected fit results either excluding or including the $H \rightarrow \tau \tau$ (Had Had boosted category) in the $H+HH$ combination and either excluding or including the $VH \rightarrow bb$ (2 btags 1 lep 3 jets) category in the $H+HH$ combination.\newline
The choice of keeping these categories in the combination arises from the fact that the overlap is relatively small and their impact on the $\kappa_\lambda$ extraction is smaller than 1\% so the overlapping events are not biasing it; furthermore, the approach of not removing categories that, although not having a significant impact on $\kappa_\lambda$, can have an impact on the other Higgs couplings, has been followed.
\begin{landscape}
\begin{table}
\scalebox{0.88}{
{\def\arraystretch{1.4}
\begin{tabular}{|c|c|c|c|}
 \hline
Analysis  & Fit configuration & $\kappa_\lambda$ interval at 95$\%$ CL (exp) & $\kappa_t$ interval at 95$\%$ CL (exp)   \\ 
\hline
\multirow{2}{*} {single-Higgs v8 (excluding H $\rightarrow \tau\tau$ categories) + HH} & $\kappa_\lambda$ only & [-4.78 - 11.00] & -- \\
 & $\kappa_\lambda - \kappa_t$ & [-5.15 - 11.02] & [0.90 - 1.11]\\ 
 \hline
 \multirow{2}{*} {single-Higgs v8 (excluding $VH\rightarrow bb$ category) + HH} & $\kappa_\lambda$ only & [-4.77 - 11.03] & -- \\
 & $\kappa_\lambda - \kappa_t$ & [-5.14 - 11.03] & [0.88 - 1.12]\\ 
 \hline
\multirow{2}{*} {single-Higgs v8 +  HH} &  $\kappa_\lambda$ only & [-4.75 - 11.00] & -- \\
 & $\kappa_\lambda - \kappa_t$ &[-5.17 - 11.00] & [0.88 - 1.12]\\ 
\hline
\end{tabular}}
}
\caption{95\% CL expected intervals for $\kappa_\lambda$ only and $\kappa_\lambda$, $\kappa_t$ fit configurations exploiting $H+HH$ combination. The fit has been performed \newline either excluding or including $H \rightarrow \tau\tau$ (had had boosted) category or $VH \rightarrow bb$ (2 btags 1 lep 3 jets) category.}
\label{tab:H_tautau_impact}
\end{table}
\end{landscape}
\subsection{$HH\rightarrow  b\bar{b} b\bar{b}$ and $t\bar{t}H \rightarrow  b\bar{b}$ overlap}
The $HH\rightarrow  b\bar{b} b \bar{b}$ analysis selects events with two Higgs boson candidates, each composed of two $b$-tagged anti-$k_t$ R=0.4 jets, with invariant masses near $m_H$, as described in Chapter~\ref{sec:dihiggs}. Possible overlaps can be found with $t\bar{t}H\rightarrow b\bar{b}$ signal regions. To check this overlap, the run and event number of the data events passing the $HH \rightarrow b\bar{b} b\bar{b}$ event selection have been compared to the ones of the data events passing the $t\bar{t}H\rightarrow b\bar{b}$ event selection. It is found that there is overlap with some of the $t\bar{t}H\rightarrow b\bar{b}$ categories with fractions of overlapping events between $1.2 \times 10^{-4}$ and $1.5 \times 10^{-3}$. These categories are kept in this combination as the overlap is small and cannot affect the results.

\subsection{Signal contamination in double-Higgs and single-Higgs channels}

The double-Higgs event selections, in addition to targeting double-Higgs processes, also select a number of single-Higgs signal events that have to be taken into account in double-Higgs signal regions if their contribution is not negligible.
Thus this contribution has been taken into account both for $HH\rightarrow  b\bar{b} \tau^+\tau^-$ and $HH\rightarrow  b\bar{b} \tau^+\tau^-$ channels, while it is negligible for the $HH\rightarrow  b\bar{b} b\bar{b}$ channel.
Events from $ZH$ and $t\bar{t}H$ single-Higgs production modes passing the $HH\rightarrow  b\bar{b} \gamma \gamma$ event selection are not negligible and are therefore included in the $HH\rightarrow b\bar{b} \gamma \gamma$ signal regions.
Double-Higgs contribution into $H\rightarrow \gamma \gamma$ categories has been checked: for high negative $\kappa_\lambda$ values, double-Higgs contribution becomes comparable to single-Higgs expectation, particularly looking at $t\bar{t}H\rightarrow \gamma \gamma$ hadronic categories where the double-Higgs contribution becomes even three times bigger than the one expected from single Higgs. \newline
Events from $ZH$ and $t\bar{t}H$ single-Higgs production modes passing the $HH\rightarrow  b\bar{b} \tau^+\tau^-$ event selection are not negligible and are therefore included in the $HH\rightarrow  b\bar{b} \tau^+\tau^-$ signal regions. 
The highest contamination, given in terms of acceptance times efficiency, is $10^{-2}$ coming from the contamination between double-Higgs $\tau_{\textrm{lep}} \tau_{\textrm{had}}$ and the $Z(\rightarrow \tau\tau) H (\rightarrow b\bar{b})$ signal regions.

\section{Validation of single- and double-Higgs inputs}
\label{sec:comb_validation}

Both $\kappa_{\lambda}$ measurement results reported in Chapter~\ref{sec:single} and in Chapter~\ref{sec:dihiggs} have been cross-checked before combining the different analyses.
\subsection{Single-Higgs input validation}

Some updates have been made with respect to the input used in the single-Higgs results reported in Chapter~\ref{sec:single}, that will be named as ``Nominal input$"$, while the single-Higgs input entering this combination will be named as ``Updated input$"$. The Updated input has the following modifications:
\begin{itemize}
\item the sign of the $\alpha_{S}$ in the $H\rightarrow ZZ$ channel has been flipped to be consistent with other channels.
\item theoretical Underlying Events and Parton Shower (UE/PS) uncertainties on the signal acceptance, ``UEPS\_WH/ZH/ggZH$"$ and ``UEPS\_VH$"$, have been merged in the $H\rightarrow \gamma \gamma$ channel.
\end{itemize}
These updates have been validated to have negligible impacts on the Higgs-coupling estimations of Reference~\cite{Coupling_run2}. Besides those, an additional modification having a negligible impact has been made:
\begin{itemize}
\item the branching fractions of $H\rightarrow s\bar{s}$ (0.04\%) is included in the parameterisation of the Higgs total width in Equation~\ref{eq:muf}.
\end{itemize}
The negligible differences including these updates are clear looking at Table~\ref{tab:validateHK3} reporting $\kappa_{\lambda}$ best-fit values, together with 1$\sigma$ interval and 95\% CL intervals for both data and the Asimov dataset generated under the SM hypothesis.

\begin{table}[h]
 \renewcommand{\arraystretch}{1.4}
\begin{center}
\begin{tabular}{|c|c|c|}
\hline
    Input workspaces &$\kappa_\lambda{}^{+1\sigma}_{-1\sigma}$ & $\kappa_{\lambda}$ interval at 95\% CL\\ 
    \hline
    \multirow{2}{*}{Nominal input}&$4.0_{-4.1}^{+4.3}$&[-3.2, 11.9]\\
    &$1.0_{-4.4}^{+8.8}$&[-6.2, 14.4]\\
    \hline
    \multirow{2}{*}{Updated input}&$4.0_{-4.1}^{+4.3}$&[-3.3, 11.9]\\
    &$1.0_{-4.4}^{+8.9}$&[-6.3, 14.5]\\
\hline
\end{tabular}
\caption{1$\sigma$ interval and 95\% CL results of $\kappa_{\lambda}$ considering different single-Higgs inputs. Observed results are reported in the first row, while expected results are reported in the second row.}
\label{tab:validateHK3}
\end{center}
\end{table}

After the studies reported in Section~\ref{sec:comb_overlap}, it was decided to exclude $t\bar{t}H\rightarrow \gamma \gamma$ categories from this combination; the impacts on the $\kappa_\lambda$ single-Higgs best-fit value and 95\% CL interval on data are determined to be  8\% and 4\%, respectively; a 2\% difference is found in the expected 95\% CL interval with respect to the nominal configurations. 
The impacts on the global signal strength, $\mu$, and on the signal strength for the $t\bar{t}H$ production mode, $\mu_{t\bar{t}H}$, fitting simultaneously the signal strengths of the other production modes, have been checked and the values of the signal strengths considering the different configurations are reported in Table~\ref{mu_ttH}. The changes on the global and ``local$"$ signal strengths are ~1\% and 8\%, respectively. 
\begin{table}[htbp]
\begin{center}
{\def\arraystretch{1.3}
\begin{tabular}{|c|c|c|}
 \hline
Signal strength & Nominal input & Removing $t\bar{t}H\rightarrow \gamma \gamma$  \\ 
\hline
$\mu$ & $1.11^{+0.09}_{-0.08}$ & $1.10^{+0.09}_{-0.08}$   \\
\hline
$\mu_{t\bar{t}H}$ & $1.09^{+0.29}_{-0.25}$ & $1.18^{+0.34}_{-0.30}$\\
\hline
\end{tabular}}
\end{center}
\caption{Global signal strengths coming from single-Higgs combined input either including or excluding $t\bar{t}H\rightarrow\gamma \gamma$ (second row); signal strengths associated to the $t\bar{t}H$ production mode either including or excluding $t\bar{t}H\rightarrow\gamma \gamma$ (third row).}
\label{mu_ttH}
\end{table}

\begin{figure}[hbtp]
\centering
\begin{subfigure}[b]{0.49\textwidth}
\includegraphics[height=8 cm,width =8 cm]{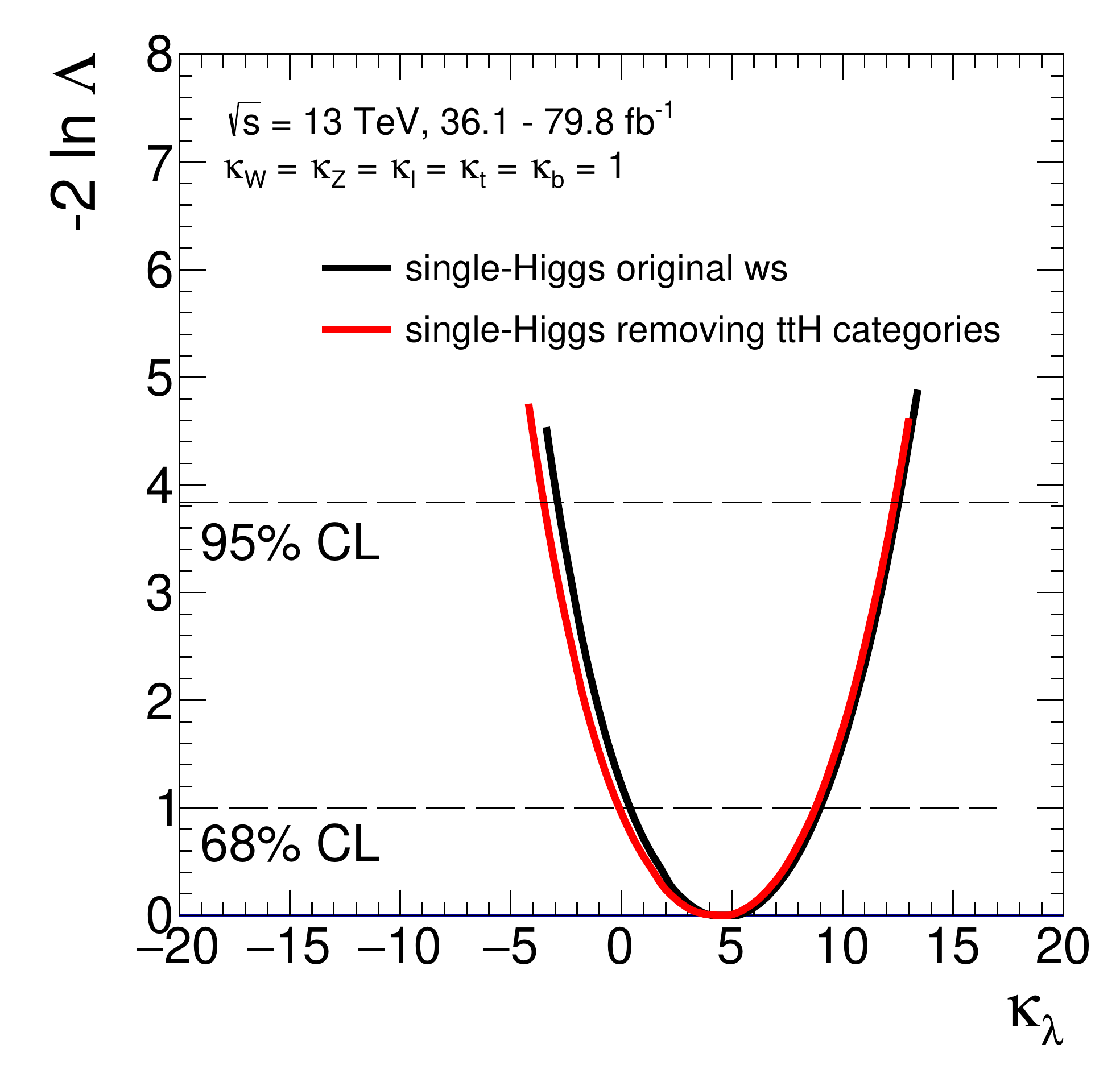}
 \caption{}
\end{subfigure}
\begin{subfigure}[b]{0.49\textwidth}
\includegraphics[height=8 cm,width =8 cm]{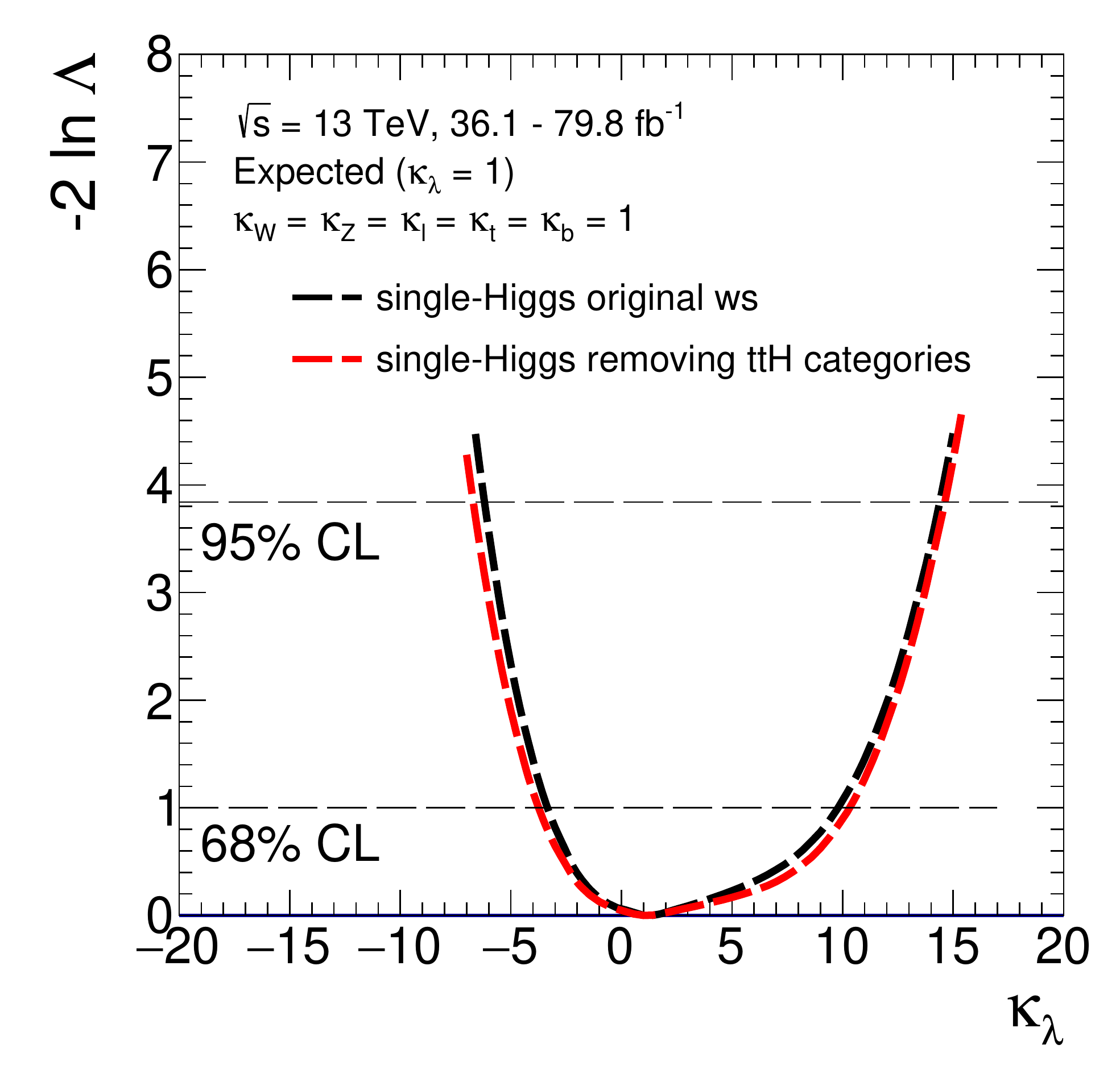}
 \caption{}
\end{subfigure}
\caption{Value of $-2 \ln{\Lambda(\kappa_\lambda)}$ as a function of $\kappa_\lambda$ for single-Higgs analysis comparing the inputs used to extract $\kappa_\lambda$ results reported in Chapter~\ref{sec:single} and removing $t\bar{t}H \rightarrow \gamma \gamma$ categories; the likelihood distribution for data is reported in (a) while (b) is for the Asimov dataset generated under the SM hypothesis. The dotted horizontal lines show the $-2 \ln{\Lambda(\kappa_\lambda)}=1$ level that is used to define the $\pm 1\sigma$ uncertainty on $\kappa_\lambda$ as well as the $-2 \ln{\Lambda(\kappa_\lambda)}=3.84$ level used to define the 95\% CL.}     
\label{scan_kl_nottH}
\end{figure}

Figure~\ref{scan_kl_nottH} reports the value of $-2 \ln{\Lambda(\kappa_\lambda)}$ as a function of $\kappa_\lambda$ for single-Higgs analyses comparing the inputs either including or excluding $t\bar{t}H\rightarrow \gamma \gamma$ categories both for data $(a)$ and for the Asimov dataset $(b)$ generated under the SM hypothesis.
\subsection{Double-Higgs input validation}
\begin{figure}[htbp]
\begin{center}
\includegraphics[width=0.5\textwidth]{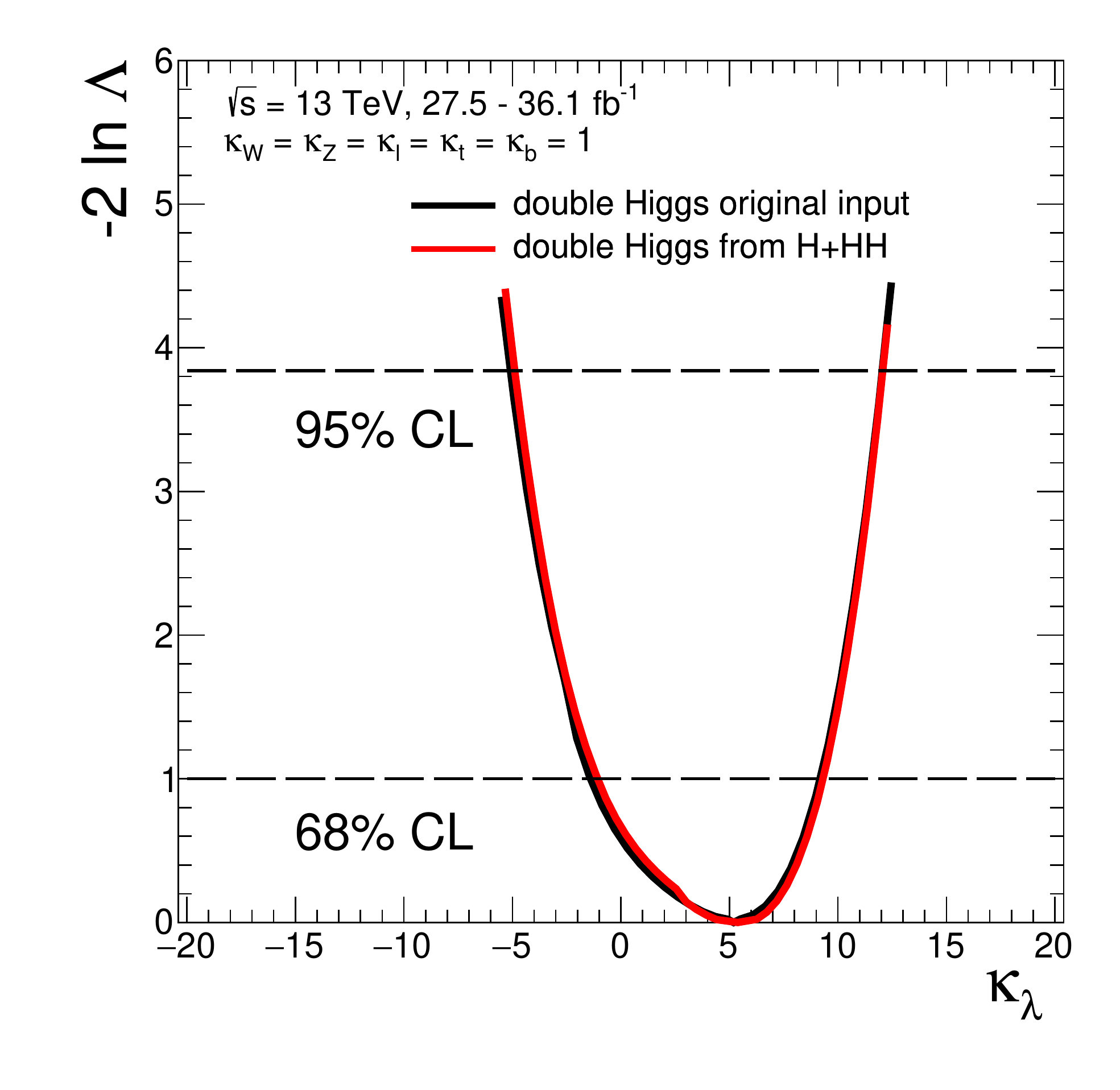}
\end{center}
\caption{Value of $-2 \ln{\Lambda(\kappa_\lambda)}$ as a function of $\kappa_\lambda$ for data considering double-Higgs analyses; the black solid line shows the original double-Higgs workspace used in order to produce the results of Chapter~\ref{sec:dihiggs} while the red solid line represents the double-Higgs input extracted from the combined $H+HH$ analysis.}   
\label{comparison_dihiggs}
\end{figure}

In order to check that the $H+HH$ combination is behaving correctly, the results coming from the double-Higgs workspace extracted from the combined $H+HH$ workspace and the ones coming from the combination of double-Higgs channels, reported in Chapter~\ref{sec:dihiggs}, have been compared, exploiting a global normalisation floating for single-Higgs analysis which does not depend on $\kappa_{\lambda}$.
Figure~\ref{comparison_dihiggs} shows the comparison between the two double-Higgs $\kappa_{\lambda}$ measurements; they are not completely identical due to the correlations of systematic uncertainties.

\clearpage
\section{Results of fit to $\kappa_\lambda$}
\label{sec:comb_results_kl}

In this section the main result of the analysis is presented, where a likelihood fit is performed to constrain the value of the Higgs-boson self-coupling $\kappa_\lambda$, while setting all other Higgs boson couplings to their SM values ($\kappa_t=\kappa_b=\kappa_\ell=\kappa_W=\kappa_Z=1$).\newline 
The constraints on $\kappa_\lambda$, derived exploiting the NLO EW $\kappa_\lambda$-dependent corrections to single-Higgs processes, can be directly compared to the constraints set by double-Higgs production analyses and the sensitivity gain from their combination can be evaluated.\newline
The $\kappa_\lambda$ self-coupling modifier is probed in the range $-20 < \kappa_\lambda < 20$.
The central value and uncertainty of the $\kappa_\lambda$ modifier of the trilinear Higgs-boson self-coupling for the combination of single- and double-Higgs analyses are determined to be:
\begin{equation*}
   \kappa_\lambda = 4.6^{+3.2}_{-3.8}=4.6 ^{+2.9}_{-3.5} \, (\text{stat.})\,^{+1.2}_{-1.2} \, (\text{exp.})\, {}^{+0.7}_{-0.5}\, (\text{sig. th.})\, ^{+0.6}_{-1.0} \, (\text{bkg. th.}) \text{(observed)}
\end{equation*}
\begin{equation*}
    \kappa_\lambda = 1.0^{+7.3}_{-3.8}=1.0 ^{+6.2}_{-3.0} \, (\text{stat.})\,^{+3.0}_{-1.7} \, (\text{exp.})\, {}^{+1.8}_{-1.2}\, (\text{sig. th.})\, ^{+1.7}_{-1.1} \, (\text{bkg. th.}) \text{(expected)}
\end{equation*}
where the total uncertainty is decomposed into components for statistical uncertainties, experimental systematic uncertainties, and theory uncertainties on signal and background modelling. The total uncertainty is dominated by the statistical component.
\begin{figure}[hbtp]
\centering
\begin{subfigure}[b]{0.49\textwidth}
\includegraphics[height=8 cm,width =8 cm]{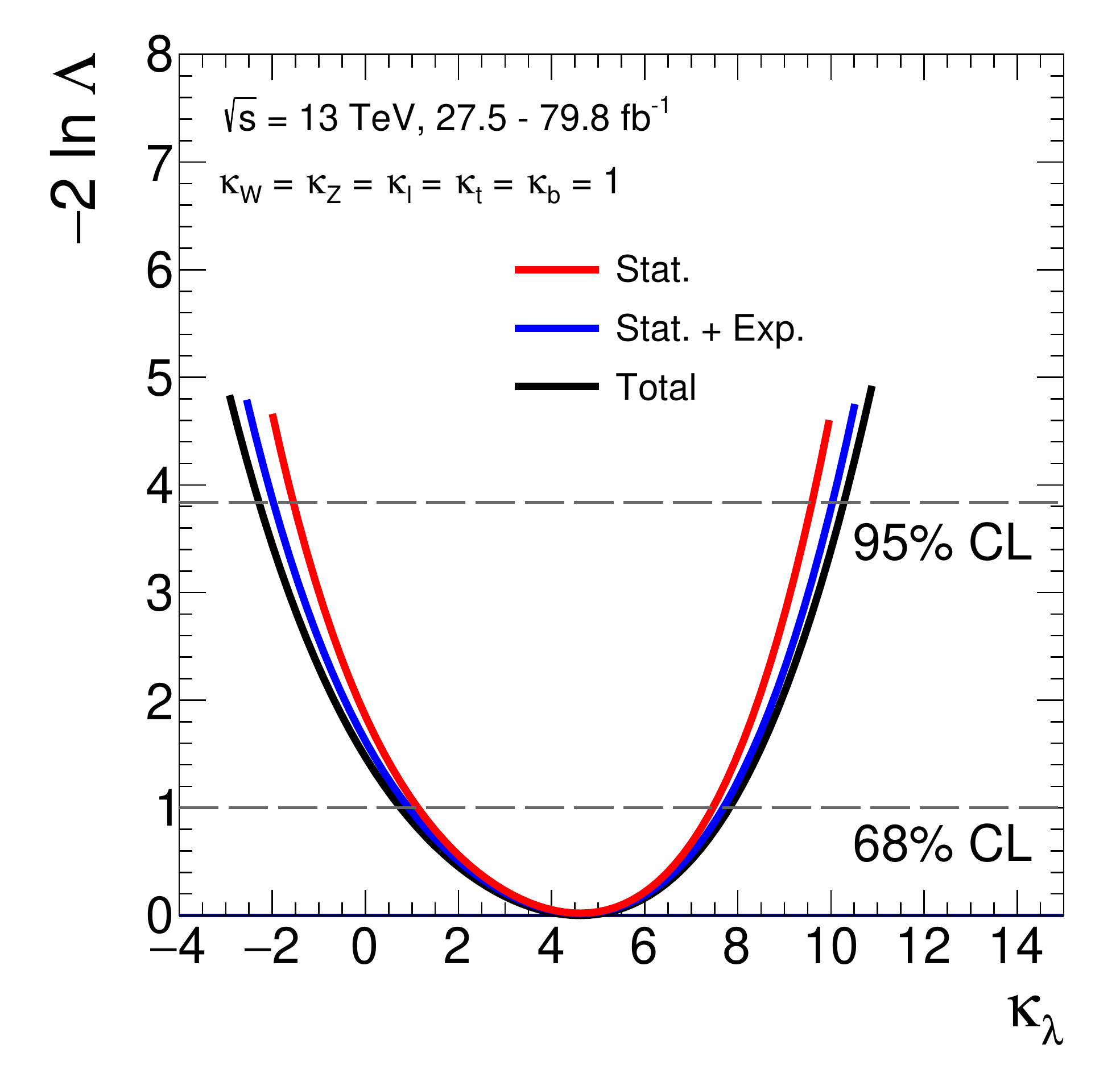}
 \caption{}
\end{subfigure}
\begin{subfigure}[b]{0.49\textwidth}
\includegraphics[height=8 cm,width =8 cm]{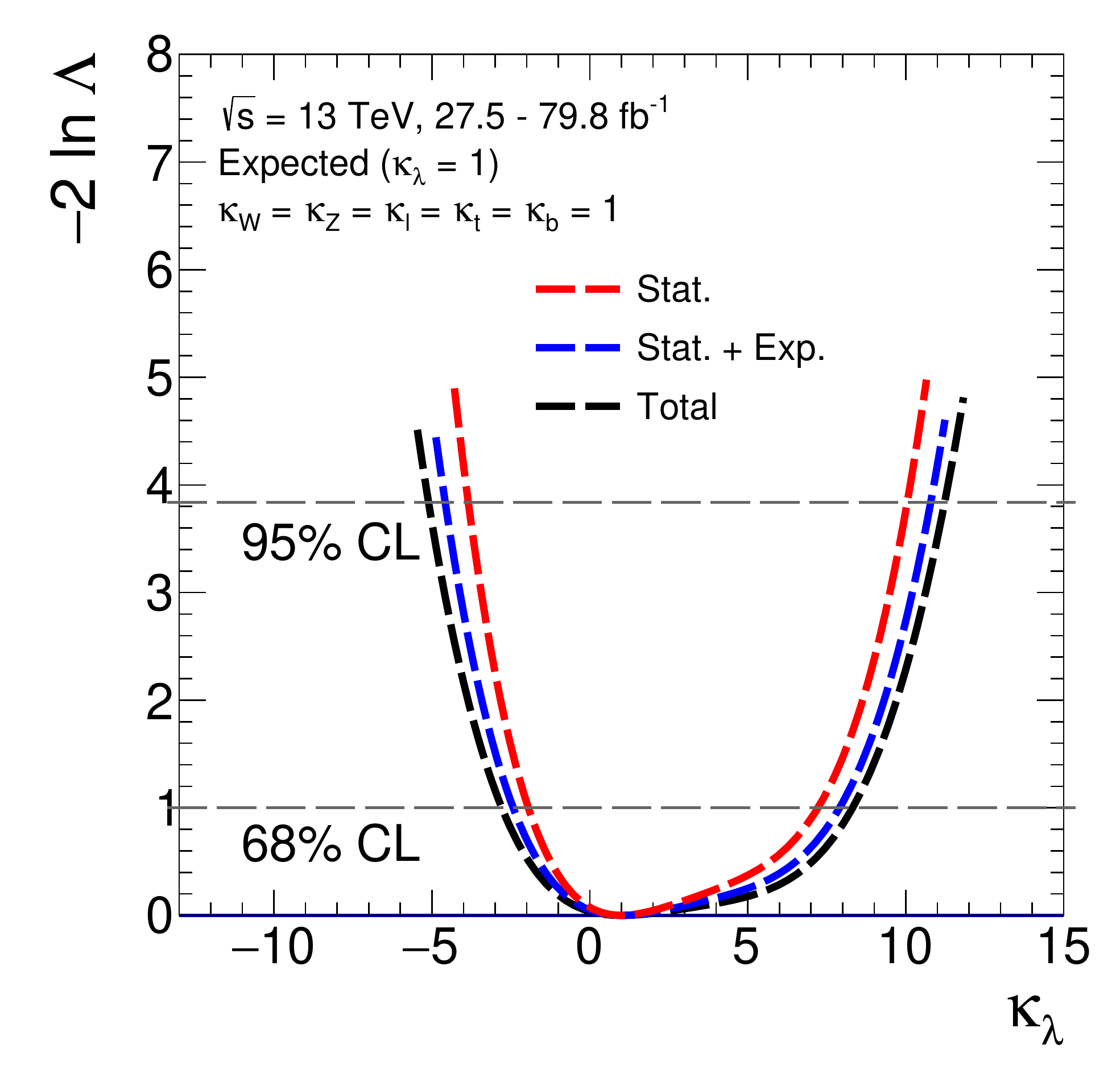}
 \caption{}
\end{subfigure}
\caption{Value of $-2 \ln{\Lambda(\kappa_\lambda)}$ as a function of $\kappa_\lambda$ for data (a), solid lines, and for the Asimov dataset generated in the SM hypothesis (b), dashed lines. The dotted horizontal lines show the $-2 \ln{\Lambda(\kappa_\lambda)}=1$ level that is used to define the $\pm 1\sigma$ uncertainty on $\kappa_\lambda$ as well as the $-2 \ln{\Lambda(\kappa_\lambda)}=3.84$ level used to define the 95\% CL. The black lines show the profile likelihood distributions obtained including all systematic uncertainties, ``Total$"$. Results from a statistics only fit, ``Stat.$"$, (red lines) and including the experimental systematics, ``Stat. + Exp.$"$, (blue lines) are also shown.}     
\label{decomposition}
\end{figure}

The value of $-2 \ln{\Lambda(\kappa_\lambda)}$ as a function of $\kappa_\lambda$ is shown in Figure~\ref{decomposition} for data and for the Asimov dataset, generated from the likelihood distribution $\Lambda$ with nuisance parameters fixed at the best-fit value obtained on data and the parameter of interest fixed to SM hypothesis (\ie\ $\kappa_\lambda = 1$). The profile likelihood distribution is obtained including statistical uncertainties, statistical and experimental systematic uncertainties, and including all statistical and systematic uncertainties.\newline
The 95\% CL intervals for $\kappa_\lambda$ are  $-2.3<\kappa_\lambda<10.3$ (observed) and  $-5.1<\kappa_\lambda<11.2$ (expected).\newline
Table~\ref{break_comb} reports the detailed breakdown of the uncertainties affecting the measurement of the combined $\kappa_\lambda$; the procedure used to produce the numbers of the table is the following: in each case the corresponding nuisance parameters are fixed to their best-fit values, while other nuisance parameters are left free, and the resulting uncertainty is subtracted in quadrature from the total uncertainty.
\begin{table}[h]
\begin{center}
{\def\arraystretch{1.4}
\begin{tabular}{|l|c|}
\hline
    Uncertainty source & $\Delta \kappa_\lambda/\kappa_\lambda$ \%\\ 
    \hline
    Statistical uncertainty & 70\\ 
    \hline
     Systematic uncertainties  & 33 \\
       $\quad$Theory uncertainties &  20\\
       $\qquad$Signal & 12\\
       $\qquad$Background & 15\\
       $\quad$Experimental uncertainties (excl. MC stat.) & 18\\
       $\quad$MC statistical uncertainty & 8\\
       \hline
       Total uncertainty & 77\\
       
\hline
\end{tabular}}
\caption{Summary of the relative uncertainties $\Delta \kappa_\lambda/\kappa_\lambda$ affecting the measurement of the combined $\kappa_\lambda$. The sum in quadrature of systematic uncertainties from individual sources differs from the uncertainty evaluated for the corresponding group in general, due to the presence of small correlations between nuisance parameters describing the different sources.}
\label{break_comb}
\end{center}
\end{table}

Table~\ref{compare_kl} presents the comparison of $\kappa_\lambda$ intervals at 95$\%$ CL for single-Higgs analyses, double-Higgs analyses and for the $H+HH$ combination; a sensitivity gain of more than 20\% is achieved in the combination with respect to single- and double-Higgs analyses alone.
\begin{table}[h]
\begin{center}
\begin{tabular}{|c|c|c|}
 \hline
Analysis  & $\kappa_\lambda$  interval at 95$\%$ CL(obs) & $\kappa_\lambda$ interval at 95$\%$ CL(exp)  \\ 
\hline
Single-Higgs & [-3.5 - 12.3]  & [-6.7 - 14.6]\\
Double-Higgs  & [-5.1 - 12.1]  & [-6.9 - 12.4]\\
$H+HH$ & [-2.3 - 10.3]  & [-5.1 - 11.2]\\
\hline
\end{tabular}
\end{center}
\caption{Comparison of $\kappa_\lambda$ interval at 95$\%$ CL for single-Higgs analyses, double-Higgs analyses and for the combination of the two of them.}
\label{compare_kl}
\end{table}

\begin{figure}[hbtp]
\centering
\begin{subfigure}[b]{0.49\textwidth}
\includegraphics[height=8 cm,width =8 cm]{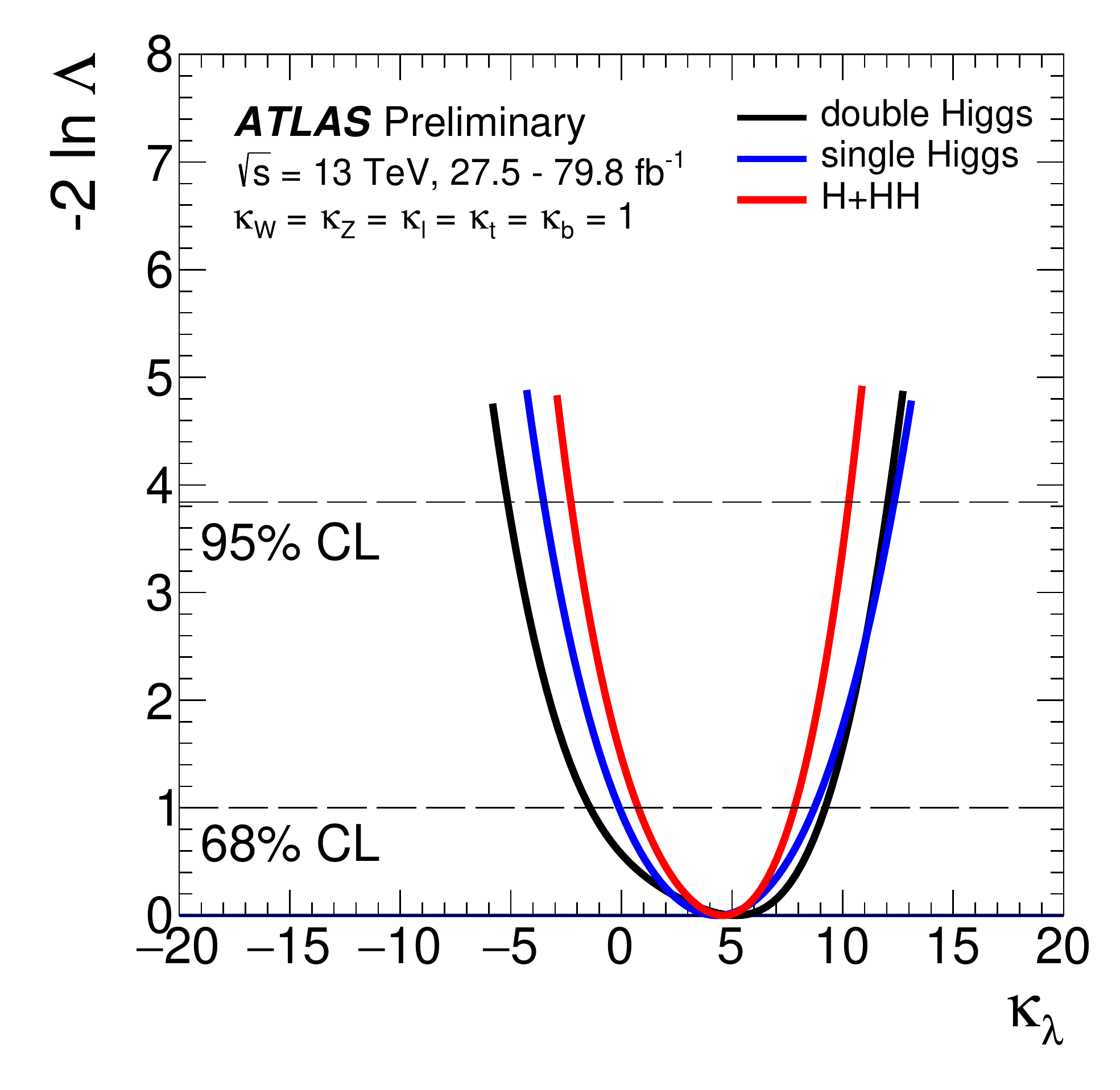}
 \caption{}
\end{subfigure}
\begin{subfigure}[b]{0.49\textwidth}
\includegraphics[height=8 cm,width =8 cm]{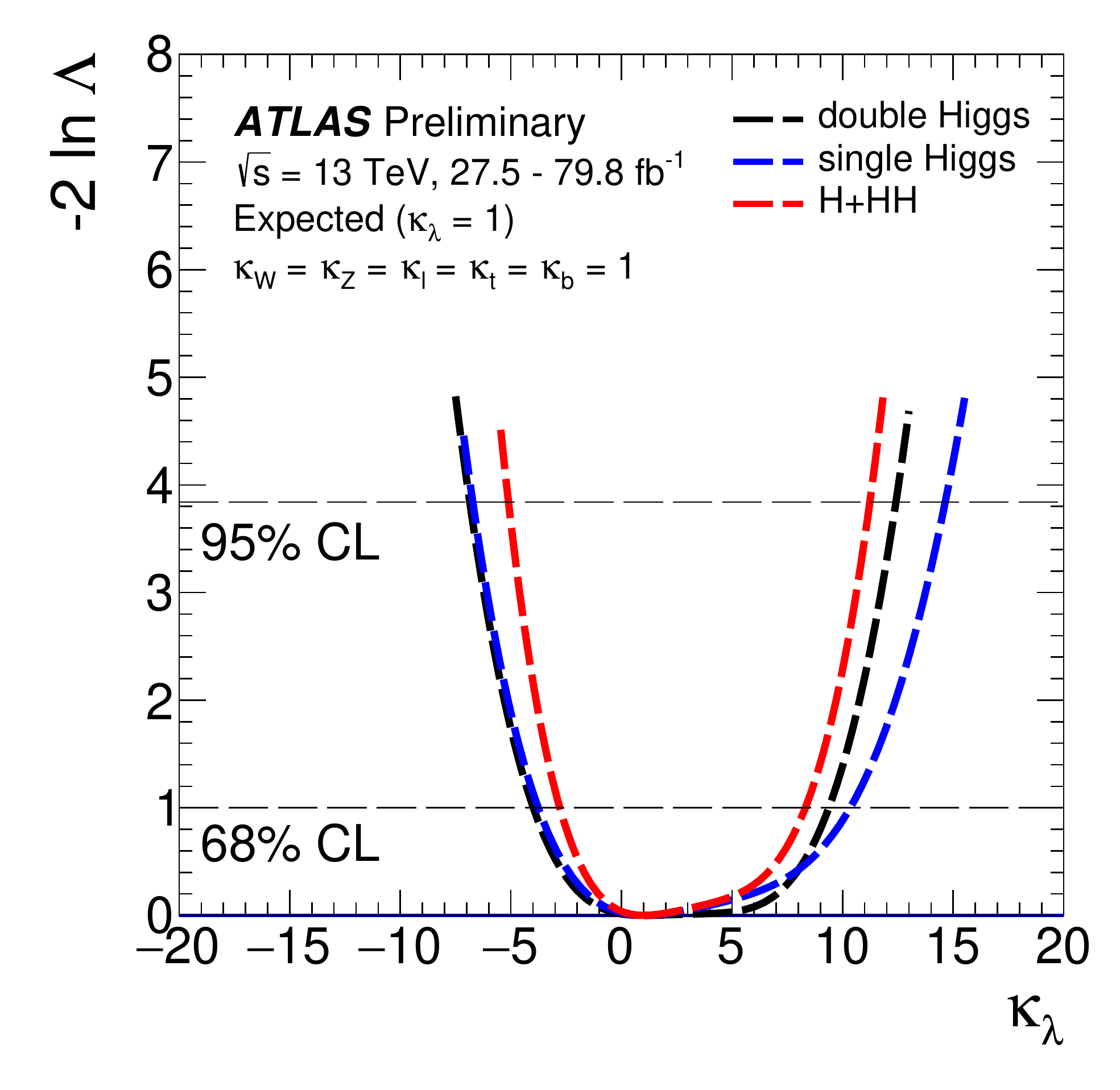}
 \caption{}
\end{subfigure}
\caption{Value of $-2 \ln{\Lambda(\kappa_\lambda)}$ as a function of $\kappa_\lambda$ for single and double-Higgs analyses separately and for the combination of the two analyses: (a) is for data and (b) is for the Asimov dataset. The dotted horizontal lines show the $-2 \ln{\Lambda(\kappa_\lambda)}=1$ level that is used to define the $\pm 1\sigma$ uncertainty on $\kappa_\lambda$ as well as the $-2 \ln{\Lambda(\kappa_\lambda)}=3.84$ level used to define the 95\% CL.}     
\label{scan_kl_all_analyses}
\end{figure}

Figure~\ref{scan_kl_all_analyses} shows the value of $-2 \ln{\Lambda(\kappa_\lambda)}$ as a function of $\kappa_\lambda$ for single- and double-Higgs analyses separately and for the combination of the two analyses. The double-Higgs analyses are more sensitive than the single-Higgs measurement for $\kappa_\lambda \gg$ 1 and show similar sensitivity for negative $\kappa_\lambda$. 

\begin{figure}[htbp]
\begin{center}
\includegraphics[height=8.5 cm, width=9.0 cm]{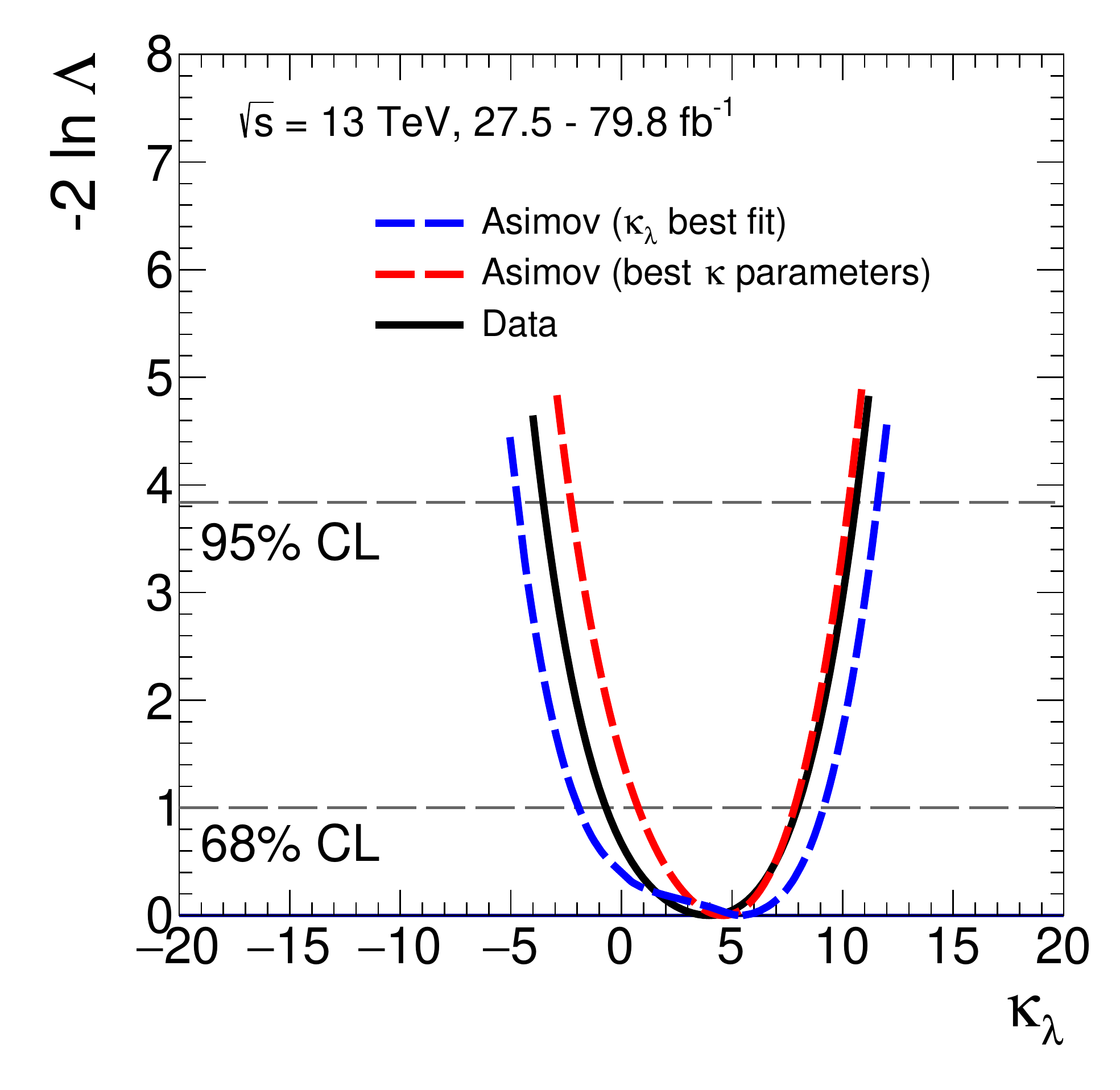}
\end{center}
\caption{Value of $-2 \ln{\Lambda(\kappa_\lambda)}$ as a function of $\kappa_\lambda$ comparing data (black solid line), and the two Asimov dataset generated with $\kappa_\lambda$ fixed to the generic model best-fit value (blue solid line) and all $\kappa$ parameters fixed to the best-fit values (red solid line). The dotted horizontal lines show the $-2 \ln{\Lambda(\kappa_\lambda)}=1$ level that is used to define the $\pm 1\sigma$ uncertainty on $\kappa_\lambda$ as well as the $-2 \ln{\Lambda(\kappa_\lambda)}=3.84$ level used to define the 95\% CL.}     
\label{asimov_bestfit}
\end{figure}

Differences in the shapes of the likelihood curves, reported in Figures~\ref{decomposition} and~\ref{scan_kl_all_analyses}, in the best-fit values of $\kappa_\lambda$ between data and Asimov dataset are due to the non-linearity of the cross-section dependence from $\kappa_\lambda$, \ie\ due to the interplay between $\kappa_\lambda$ and $\kappa_\lambda^2$ terms in Equations~\ref{eq:mui} and~\ref{eq:muf}, and due to the fact that the measured yields from single-Higgs and double-Higgs processes are slightly different than the expectation.  
This effect has been investigated generating two different Asimov datasets, the first one fixing $\kappa_\lambda$ to the generic model best-fit value, $\kappa_\lambda$=5.5, and all other couplings to their SM values, while the second one is generated fixing all $\kappa$ to their observed values. While in the first Asimov dataset, data and Asimov likelihood shapes still have large differences due to a tension between data and the model assumption, after fixing the other couplings to the observed values, the likelihoods have similar shapes. Figure~\ref{asimov_bestfit} shows the comparison between the likelihood curves for data (black solid line) and for the two Asimov datasets (blue and red shaded lines). Furthermore, the Asimov dataset generated fixing all the $\kappa$ to their observed values, has been used in order to check the asymmetric uncertainty decomposition of the Asimov dataset generated under the SM hypothesis ($\kappa_\lambda=1$), thus leading to a $\kappa_\lambda$ central value and uncertainty:
\begin{equation*}
    \kappa_\lambda= 4.0^{+4.0}_{-4.7}=4.0 ^{+3.5}_{-4.0} \, (\text{stat.})\,^{+1.6}_{-1.8} \, (\text{exp.})\, {}^{+0.9}_{-1.0}\, (\text{sig. th.})\, ^{+0.6}_{-1.2} \, (\text{bkg. th.}) \text{(expected)}
\end{equation*}
where the total uncertainty is decomposed into components for statistical uncertainties, experimental systematic uncertainties, and theory uncertainties on signal and background modelling. It is clear that the shapes of the likelihood distribution, the $1\sigma$ errors and 95\%~CL intervals around the best-fit value, are strictly dependent on the fitted value itself.

\begin{figure}[htbp]
\begin{center}
\includegraphics[ width=0.6\textwidth]{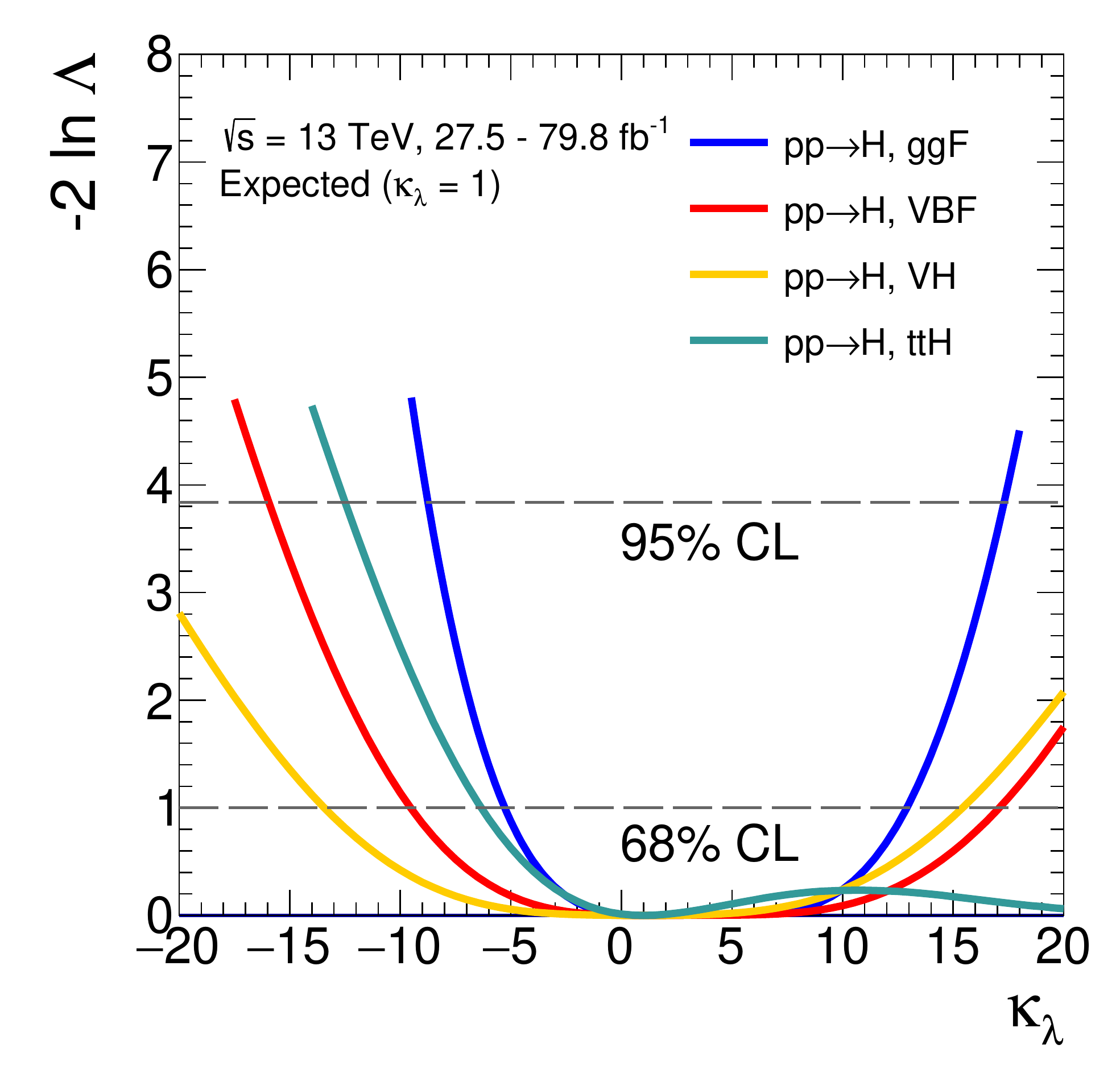}
\end{center}
\caption{Expected value of $-2 \ln{\Lambda(\kappa_\lambda)}$ as a function of $\kappa_\lambda$ in the $\kappa_\lambda$-only model with all other couplings set to their SM values obtained in the $\kappa_\lambda$=1 hypothesis for each single-Higgs production mode.}     
\label{production_mode}
\end{figure}

\begin{figure}[htbp]
\begin{center}
\includegraphics[width=0.6\textwidth]{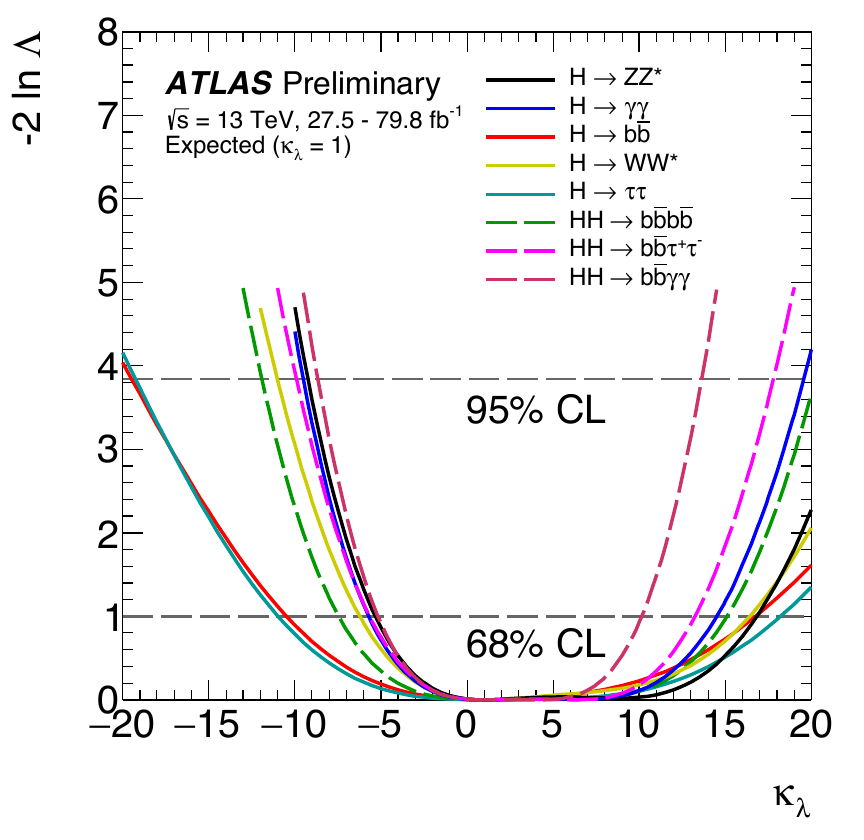}
\end{center}
\caption{Expected value of $-2 \ln{\Lambda(\kappa_\lambda)}$ as a function of $\kappa_\lambda$ with all other couplings set to their SM values obtained in the $\kappa_\lambda$=1 hypothesis for the single-Higgs and double-Higgs decay modes~\cite{Confnote_comb}.}     
\label{decay_channels}
\end{figure}

\begin{figure}[hbtp]
\centering
\begin{subfigure}[b]{0.49\textwidth}
\includegraphics[height=8 cm,width =8 cm]{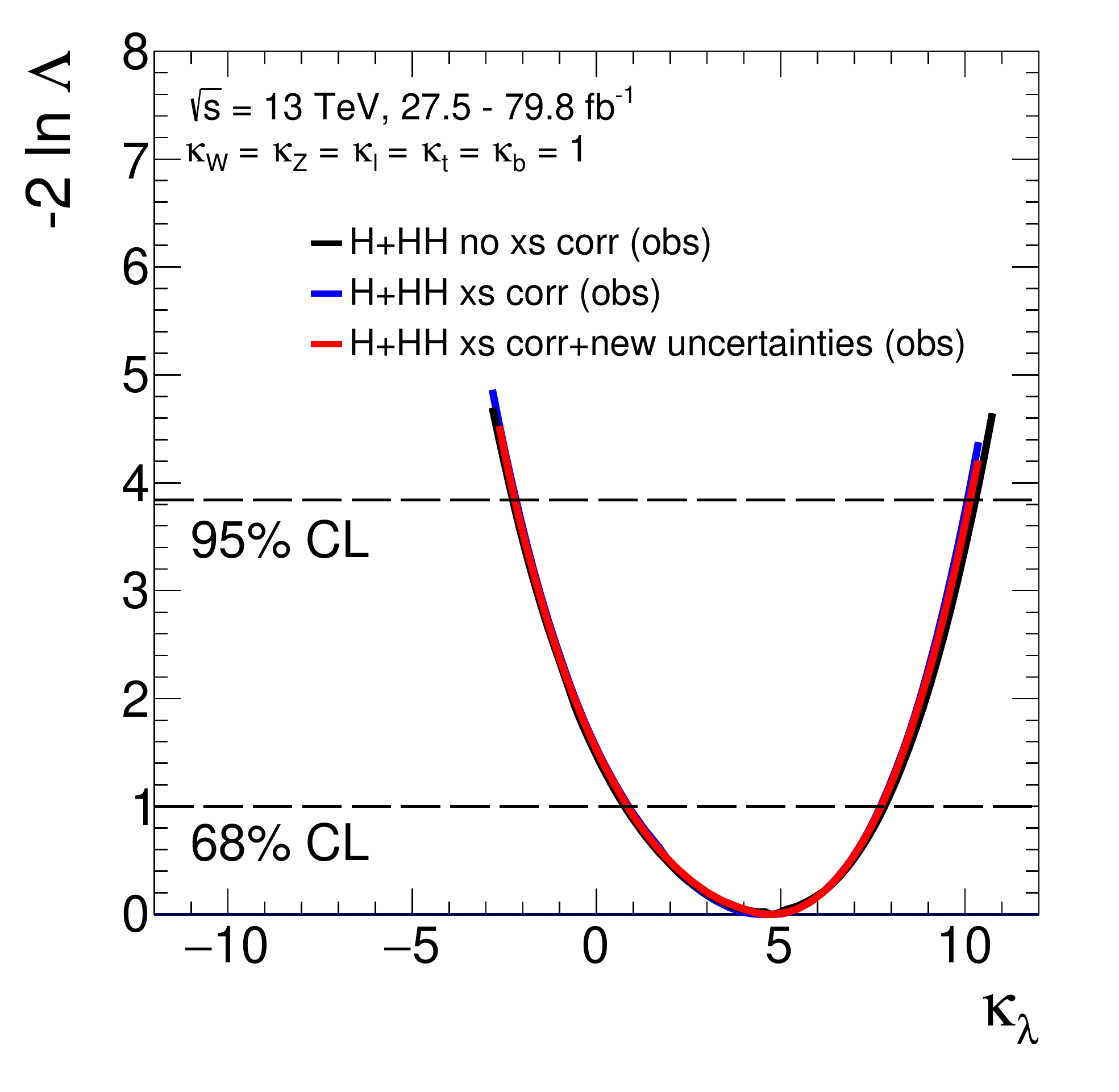}
 \caption{}
\end{subfigure}
\begin{subfigure}[b]{0.49\textwidth}
\includegraphics[height=8 cm,width =8 cm]{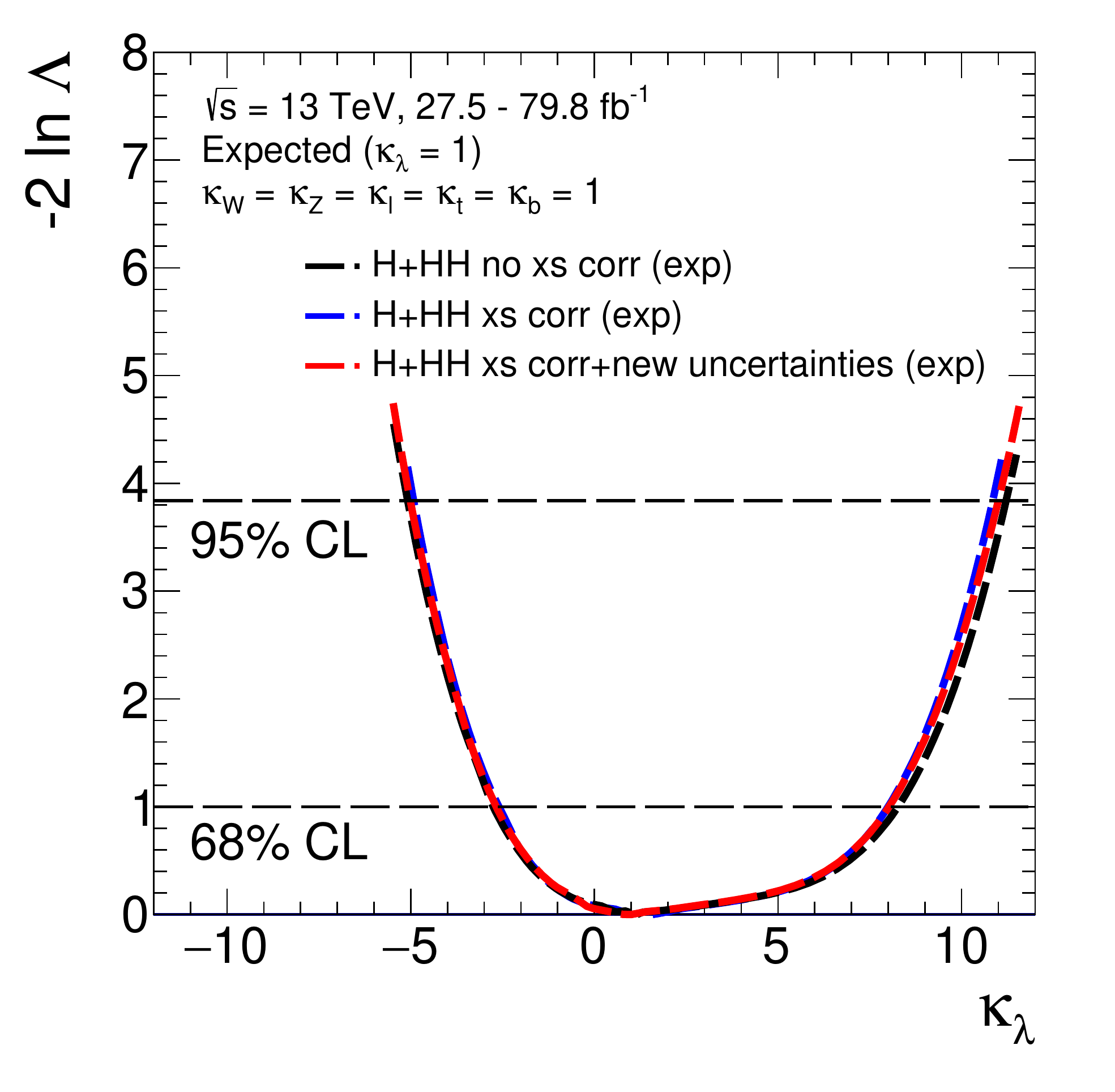}
 \caption{}
\end{subfigure}
\caption{Value of $-2 \ln{\Lambda(\kappa_\lambda)}$ as a function of $\kappa_\lambda$ for data (a) and for the Asimov dataset (b). The solid black line shows the combined likelihood distribution used to extract $\kappa_\lambda$ public results; the blue solid line represents the same likelihood distribution including recent computations of the double-Higgs SM cross section that are not exploited for the results presented in this thesis, while the red solid line takes into account also the uncertainties on this recent computation. The dotted horizontal lines show the $-2 \ln{\Lambda(\kappa_\lambda)}=1$ level that is used to define the $\pm 1\sigma$ uncertainty on $\kappa_\lambda$ as well as the $-2 \ln{\Lambda(\kappa_\lambda)}=3.84$ level used to define the 95\% CL.}     
\label{xs_corr_kl}
\end{figure}

In order to investigate the constraining power of the several channels included in this combination, the test explained in Chapter~\ref{sec:single} has been performed removing different categories, corresponding for example to the different double-Higgs decay channels, from the combined fit and checking the $\kappa_\lambda$ 95\% CL intervals obtained using the remaining channels. In order to avoid statistics fluctuations,  Asimov datasets are used.\newline
Table~\ref{tab:channelRank} shows the ranking of the different channels in constraining $\kappa_{\lambda}$: each row shows the 1$\sigma$ interval and 95\% CL of $\kappa_{\lambda}$ obtained by removing one specific channel. The $b\bar{b}\gamma \gamma$ channel is the most sensitive one.\newline
Figures~\ref{production_mode} and~\ref{decay_channels} show the expected dependence of $-2 \ln{\Lambda(\kappa_\lambda)}$ on $\kappa_\lambda$ for the $\kappa_\lambda$-only model (obtained in the $\kappa_\lambda$ = 1 hypothesis) considering different production modes and decay channels; when showing a specific production or decay channels, all the $\kappa_\lambda$ involved in the parameterisation of the corresponding signal yields (including those entering in the parametrisation of the branching ratios) are correlated in the $-2 \ln{\Lambda(\kappa_\lambda)}$ scan, while all the others are profiled. Looking at the single-Higgs, the di-boson decay channels $\gamma \gamma$, $ZZ^*$, $WW^*$ and \ggF and $t\bar{t}H$ production modes represent the dominant contributions.
The $t\bar{t}H$ production mode is not sensitive for $\kappa_\lambda >$ 0 because of the degeneracy in the cross-section.
\begin{table}[htbp]
\begin{center}
{\def\arraystretch{1.6}
\begin{tabular}{|c|c|c|}
\hline
Channels &$\kappa_\lambda{}^{+1\sigma}_{-1\sigma}$ & $\kappa_\lambda$  [95\% CL]\\
\hline
$HH\rightarrow b\bar{b}\gamma \gamma$ &$1.00_{-4.05}^{+8.68}$&[-5.60, 13.05]\\ 
\hline
$HH\rightarrow b\bar{b}\tau^+\tau^-$ &$1.00_{-4.10}^{+7.44}$&[-5.55, 11.53]\\ 
\hline
$H\rightarrow \gamma\gamma$&$1.00_{-3.88}^{+7.47}$&[-5.28, 11.48]\\ 
\hline
$t\bar{t}H$ multilepton&$1.00_{-3.98}^{+7.63}$&[-5.33, 11.36]\\ 
\hline
$H\rightarrow ZZ^*$&$1.00_{-3.97}^{+7.38}$&[-5.38, 11.29]\\
\hline
$HH\rightarrow b\bar{b}b\bar{b}$ &$1.00_{-3.91}^{+7.41}$&[-5.30, 11.39]\\
\hline
$VH\rightarrow b\bar{b}$&$1.00_{-3.79}^{+7.57}$&[-5.10, 11.49]\\
\hline
$H\rightarrow WW^*$&$1.00_{-3.88}^{+7.37}$&[-5.25, 11.32]\\ 
\hline
$H\rightarrow \tau^+\tau^-$&$1.00_{-3.76}^{+7.46}$&[-5.07, 11.34]\\ 
\hline
$t\bar{t}H\rightarrow b\bar{b}$&$1.00_{-3.84}^{+7.39}$&[-5.17, 11.25]\\ 
\hline
Nominal expected result&$1.00_{-3.79}^{+7.31}$&[-5.11, 11.23]\\
\hline
\end{tabular}
}
\caption{Ranking of channels in constraining $\kappa_{\lambda}$. Each row shows the 1$\sigma$ interval and 95\% CL of $\kappa_{\lambda}$ obtained by removing one specific channel. The $HH\rightarrow b\bar{b}\gamma \gamma $ channel is the most sensitive one.}
\label{tab:channelRank}
\end{center}
\end{table}
\clearpage
Recent computations of the double-Higgs production cross section have been performed leading to a small reduction of the SM cross section from 33.5 to 31.05 fb and a stronger dependence on $\kappa_\lambda$, as it was described in Chapter~\ref{sec:dihiggs}. In order to be consistent with the double-Higgs results in Reference~\cite{Paper_hh}, these recent calculations have not been used, but their impact on the self-coupling combined interval at 95\% CL has been evaluated to be less than 2\%. Figure~\ref{xs_corr_kl} shows the comparison between the likelihood distribution presented in Figure~\ref{decomposition} and the likelihood distribution obtained including recent computation of the HH SM cross section that are not included in the results presented in this thesis and taking into account the uncertainties on this recent computation both for data and for the Asimov dataset.

\begin{figure}[htbp]
\begin{center}
\includegraphics[height=8.5 cm, width=9 cm]{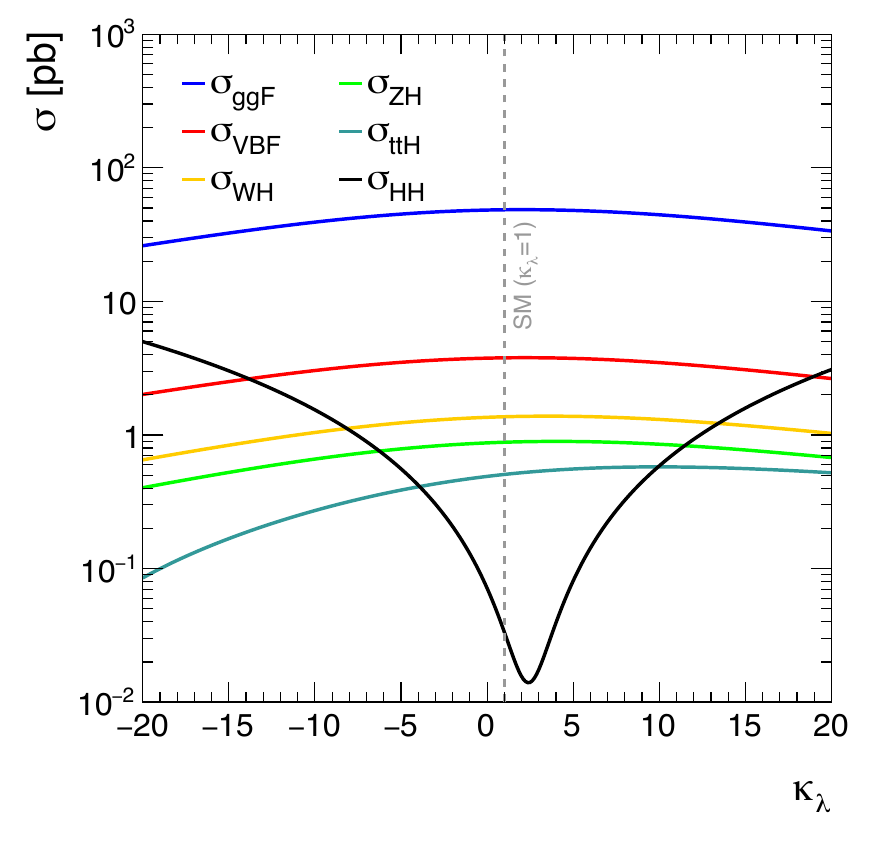}
\end{center}
\caption{Single-Higgs and double-Higgs production cross section as a function of $\kappa_\lambda$~\cite{Degrassi,Maltoni,Higgs_CS}. The dashed line intercepts the values corresponding to the SM hypothesis ($\kappa_\lambda=1$)~\cite{Confnote_comb}.}     
\label{production_xs}
\end{figure}

The single-Higgs and double-Higgs production cross sections are shown in Figure~\ref{production_xs} for all the production modes included in the combination and for the gluon-gluon fusion $pp\rightarrow HH$ production mode as a function of $\kappa_\lambda$~\cite{Degrassi,Maltoni,Higgs_CS}.

\clearpage
\section{Results of fit to $\kappa_\lambda$ and $\kappa_t$}
\label{sec:comb_results_kl_kt}

\begin{figure}[hbtp]
\centering
\begin{subfigure}[b]{0.49\textwidth}
\includegraphics[height=8 cm,width =8 cm]{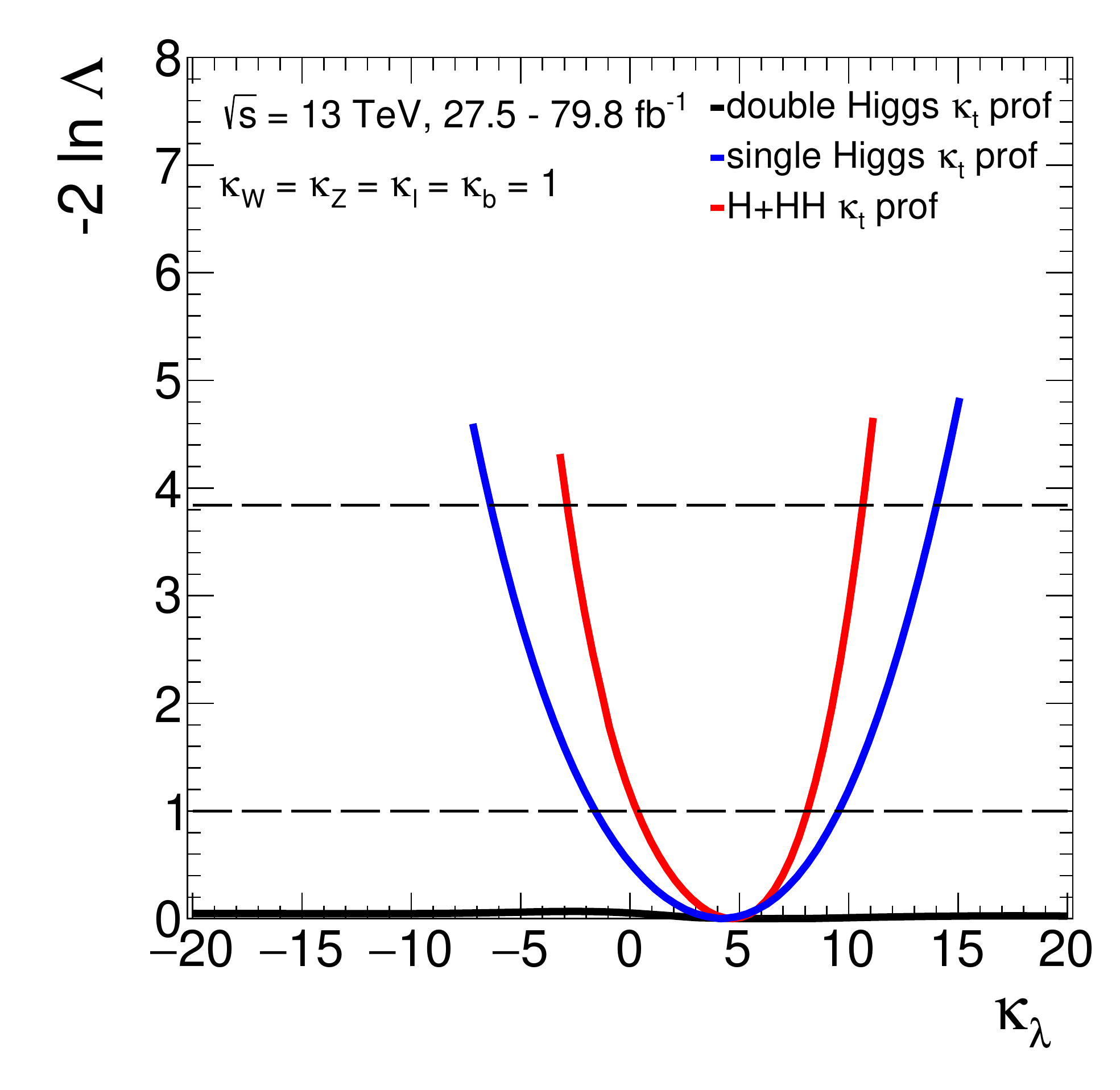}
 \caption{}
\end{subfigure}
\begin{subfigure}[b]{0.49\textwidth}
\includegraphics[height=8 cm,width =8 cm]{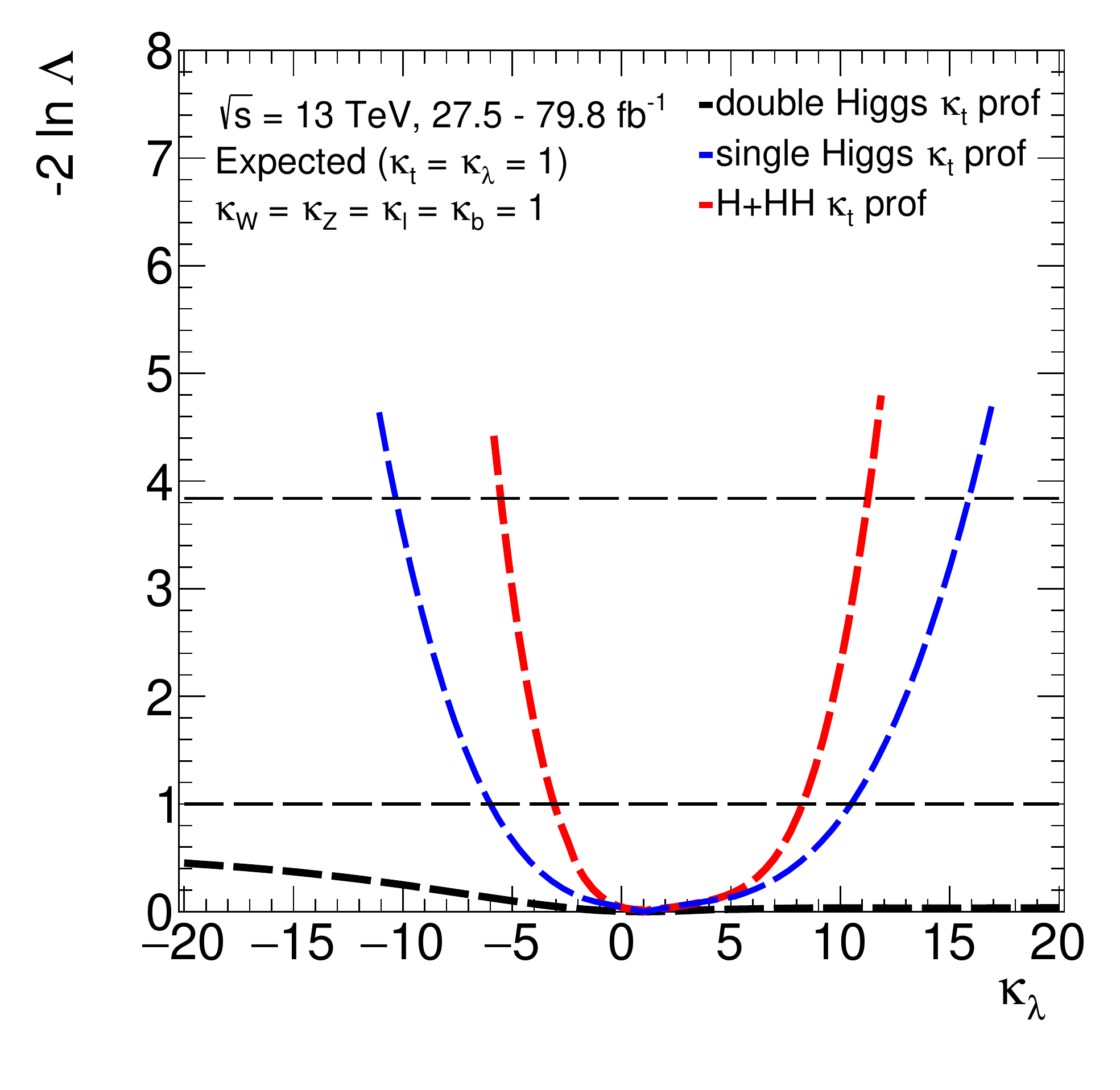}
 \caption{}
\end{subfigure}
\caption{Value of $-2 \ln{\Lambda(\kappa_\lambda)}$ as a function of $\kappa_\lambda$ (with $\kappa_{t}$ profiled) for single and double-Higgs analyses separately and for the combination of the two analyses; for data (a) and for the Asimov dataset generated in the SM hypothesis (b). The dotted horizontal lines show the $-2 \ln{\Lambda(\kappa_\lambda)}=1$ level that is used to define the $\pm 1\sigma$ uncertainty on $\kappa_\lambda$ as well as the $-2 \ln{\Lambda(\kappa_\lambda)}=3.84$ level used to define the 95\% CL. The double-Higgs analyses have almost no sensitivity to $\kappa_\lambda$ profiling $\kappa_t$.}     
\label{scan_kl_kt_all_analyses}
\end{figure}

In order to exploit the sensitivity of the double-Higgs production mechanism and the strong dependence of the double-Higgs cross section $\sigma(pp\rightarrow HH)$ on $\kappa_t$, a likelihood fit is performed to constrain at the same time $\kappa_\lambda$ and $\kappa_t$, setting all the other coupling modifiers to their SM values.\newline
The value of $-2 \ln{\Lambda(\kappa_\lambda)}$ as a function of $\kappa_\lambda$ is shown in Figure~\ref{scan_kl_kt_all_analyses} for data and for the Asimov dataset, generated under the SM hypothesis (\textit{i.e.} $\kappa_\lambda = \kappa_t=1$); single and double-Higgs analyses are shown both separately and combined in order to extract $\kappa_\lambda$ best-fit values and 95\% CL intervals. While in single-Higgs analyses a significant loss in sensitivity is present when fitting both $\kappa_\lambda$ and $\kappa_t$, the constraining power of the combined $H+HH$ measurement is only slightly worse than in the $\kappa_\lambda$-only model ($\kappa_t=1$). The small constraining power, represented by a small raising of the likelihood distribution for negative $\kappa_\lambda$ in the double-Higgs analyses, comes from the parameterisation as a function of $\kappa_\lambda$ and $\kappa_t$ of the Higgs decay channels and single-Higgs productions in double-Higgs background, otherwise the curve becomes flat as it has been shown in Chapter~\ref{sec:dihiggs}. Due to the limited sensitivity of the double-Higgs analyses, the $1\sigma$ interval cannot be reached in the $\kappa_\lambda$ scan profiling $\kappa_t$.

\begin{figure}[hbtp]
\centering
\begin{subfigure}[b]{0.49\textwidth}
\includegraphics[width=\textwidth]{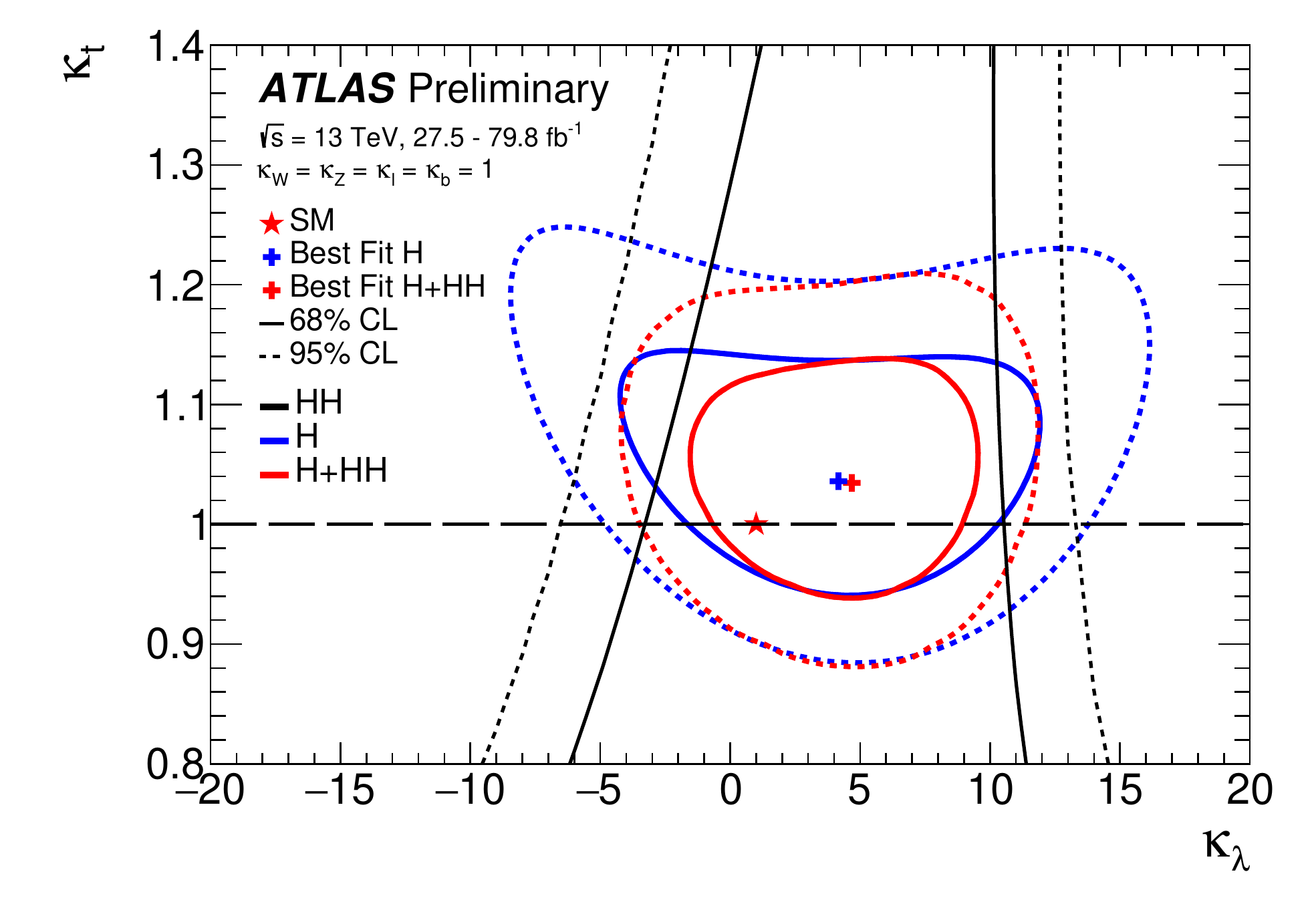}
 \caption{}
\end{subfigure}
\begin{subfigure}[b]{0.49\textwidth}
\includegraphics[width=\textwidth]{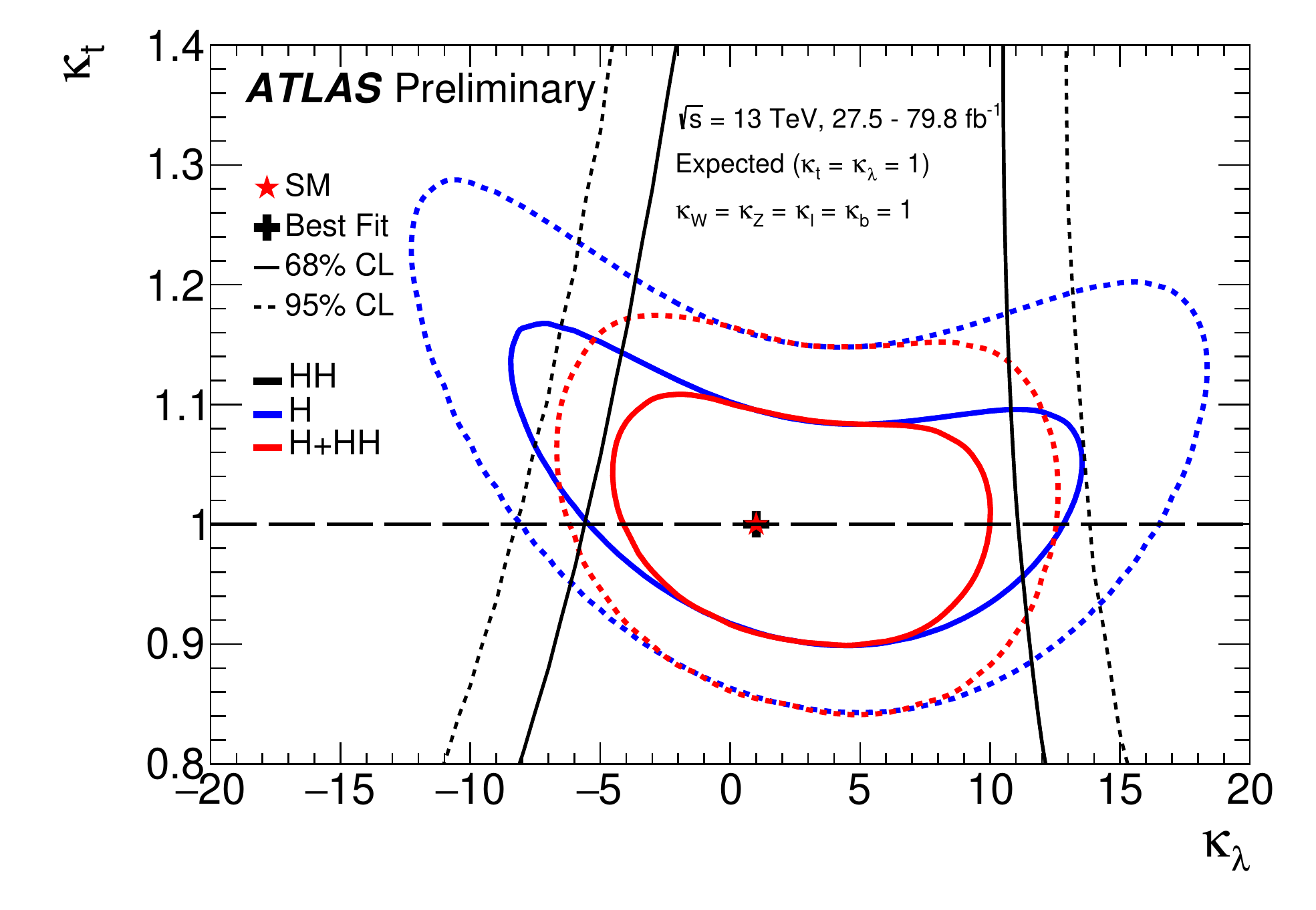}
 \caption{}
\end{subfigure}
\caption{Negative log-likelihood contours at 68\% and 95\% CL in the $(\kappa_\lambda,\kappa_t)$ plane on data (a) and on the Asimov dataset generated under the SM hypothesis (b). The best-fit value  is indicated by a cross while the SM hypothesis is indicated by a star. The dotted horizontal line at $\kappa_t$=1 shows the $\kappa_\lambda$-only fit result. The plot assumes that the approximations in References~\cite{Degrassi,Maltoni} are valid inside the shown contours.}     
\label{contour_kl_kt_all_analyses}
\end{figure}

Table~\ref{tab:k3kt} reports a summary of fit results in the $\kappa_\lambda$-only and the $\kappa_\lambda -\kappa_t$ fit configurations. The $\kappa_t$ best-fit value is compatible with the SM prediction.
\begin{table}[htbp]
\begin{center}
{\def\arraystretch{1.4}
\begin{tabular}{|c|c|c|c|c|c|c|c|}
\hline
 POIs & $\kappa_{W}{}^{+1\sigma}_{-1\sigma}$&$\kappa_{Z}{}^{+1\sigma}_{-1\sigma}$& $\kappa_{t}{}^{+1\sigma}_{-1\sigma}$ & $\kappa_b{}^{+1\sigma}_{-1\sigma}$&$\kappa_{\ell}{}^{+1\sigma}_{-1\sigma}$& $\kappa_\lambda{}^{+1\sigma}_{-1\sigma}$ & $\kappa_\lambda$  [95\% CL] \\ 
\hline
  \multirow{2}{*}{$\kappa_\lambda$}&  \multirow{2}{*}{1} &  \multirow{2}{*}{1} & \multirow{2}{*}{1} &\multirow{2}{*}{1} &\multirow{2}{*}{1} &$4.6_{-3.8}^{+3.2}$ &  $[-2.3, 10.3]$ \\
    & &   &      & & &  $1.0^{+7.3}_{-3.8}$ & $[-5.1, 11.2]$ \\ 
     \hline
  \multirow{2}{*}{$\kappa_\lambda$-$\kappa_t$}&  \multirow{2}{*}{1}  &  \multirow{2}{*}{1} & $1.03_{-0.06}^{+0.07}$ &\multirow{2}{*}{1} &\multirow{2}{*}{1} &$4.7_{-4.3}^{+3.4}$ &  $[-2.9, 10.6]$ \\
    &&   &  $1.00_{-0.07}^{+0.07}$     & & &  $1.0^{+7.3}_{-4.1}$ & $[-5.5, 11.3]$ \\ 
\hline
\end{tabular}}
\caption{Best-fit values for $\kappa$ modifiers with $\pm 1 \sigma$ uncertainties. The 95\% CL interval for $\kappa_\lambda$ is also reported. For the fit result the upper row corresponds to the observed results, and the lower row to the expected results obtained using Asimov datasets generated under the SM hypothesis.}
\label{tab:k3kt}
\end{center}
\end{table}
\newpage
Figure~\ref{contour_kl_kt_all_analyses} shows negative log-likelihood contours on the $(\kappa_\lambda,\kappa_t)$ grid obtained from fits performed in the $\kappa_b=\kappa_{\ell}=\kappa_W=\kappa_Z = 1$ hypothesis for single and double-Higgs analyses separately and for the combination of the two analyses. 
Considering the log-likelihood contour of the double-Higgs analyses, it is clear that $\kappa_\lambda$ and $\kappa_t$  cannot be constrained at the same time, as it has already been shown in Chapter~\ref{sec:dihiggs}, Figure~\ref{contour_hh_theory}. The combination with the single-Higgs measurements allows, even for $\kappa_\lambda$ values deviating from the SM prediction, the determination of $\kappa_t$ to a sufficient precision to restore most of the ability of the double-Higgs analyses to constrain $\kappa_\lambda$~\cite{Confnote_comb}. Furthermore, the combination allows to retain sensitivity to $\kappa_\lambda$ and solve the degeneracy even when introducing additional degrees of freedom.
In both single-Higgs and combined measurements, $\kappa_t$ is roughly constrained away from its best-fit value coming from double-Higgs analyses, in a range where greater values of the double-Higgs likelihood compared to the minimum one can be found and thus the $1\sigma$ interval or even 95\% CL can be reached: in this range the combination can get benefits from double-Higgs measurements.

\begin{figure}[hbtp]
\centering
\begin{subfigure}[b]{0.49\textwidth}
\includegraphics[height=8.5 cm,width =8 cm]{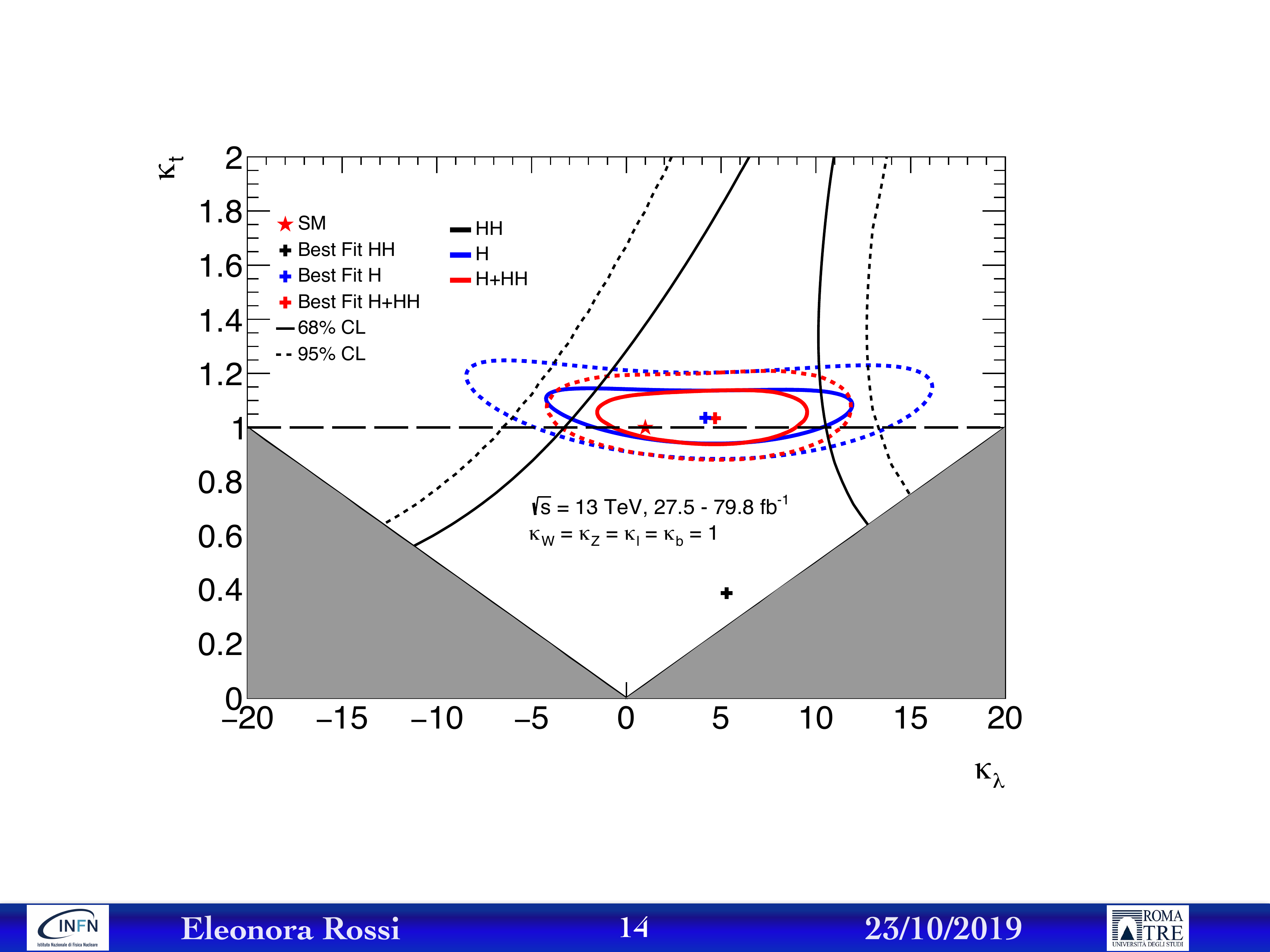}
 \caption{}
\end{subfigure}
\begin{subfigure}[b]{0.49\textwidth}
\includegraphics[height=8.5 cm,width =8 cm]{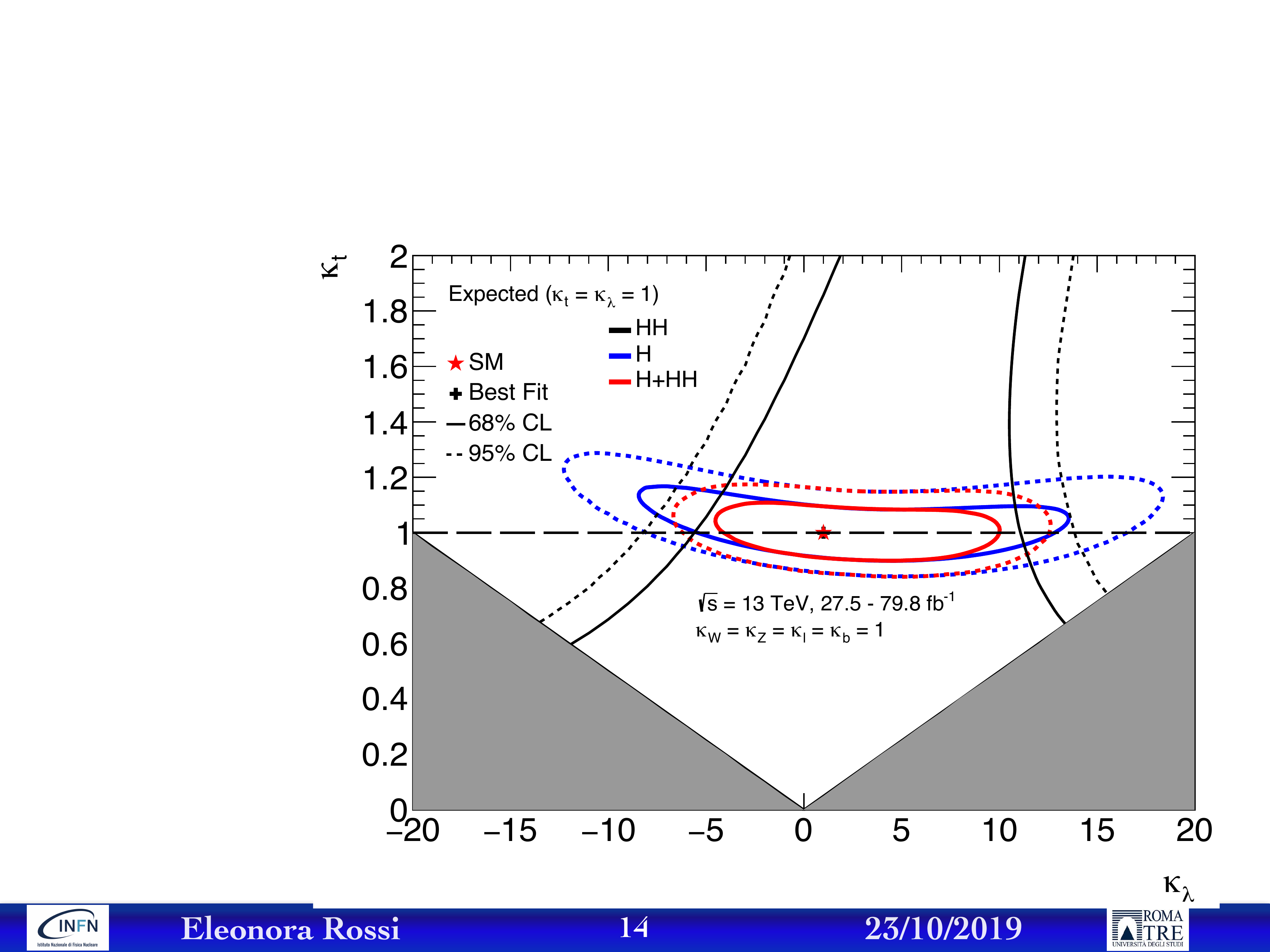}
 \caption{}
\end{subfigure}
\caption{Negative log-likelihood contours at 68\% and 95\% CL in the $(\kappa_\lambda,\kappa_t)$ plane on data (a) and on the Asimov dataset generated under the SM hypothesis (b).  The best-fit value is indicated by a cross while the SM hypothesis is indicated by a star. The dotted horizontal line at $\kappa_t$=1 shows the $\kappa_\lambda$-only fit result. The plot assumes that the approximations in References~\cite{Degrassi,Maltoni} are valid inside the shown contours. The extended range of the contour makes the double-Higgs $\kappa_\lambda$ best-fit value visible, showing also the degeneracy of measuring at the same time $\kappa_\lambda$ and $\kappa_t$; the shaded area corresponds to $| \kappa_\lambda /\kappa_t| < 20$, a constrain coming from the $HH\rightarrow b\bar{b} \gamma \gamma$ analysis.}     
\label{contour_kl_kt_all_analyses_shaded}
\end{figure}

The same negative log-likelihood contours on the $(\kappa_\lambda,\kappa_t)$ grid are shown in Figure~\ref{contour_kl_kt_all_analyses_shaded} in an extended range, where the degeneracy of the double-Higgs analyses when fitting simultaneously $\kappa_\lambda$ and $\kappa_t$ is more evident. The coloured areas are not part of the allowed region because the acceptance of the $HH\rightarrow b\bar{b}\gamma \gamma$ analysis is not reliable for $|\kappa_\lambda/\kappa_t|\ge 20$.
 
\section{Results of fit to more generic models}  
\label{sec:comb_results_kl_other}
\begin{figure}[htbp]
  \centering
  \begin{subfigure}[b]{0.49\textwidth}
\includegraphics[height=7 cm,width =8 cm]{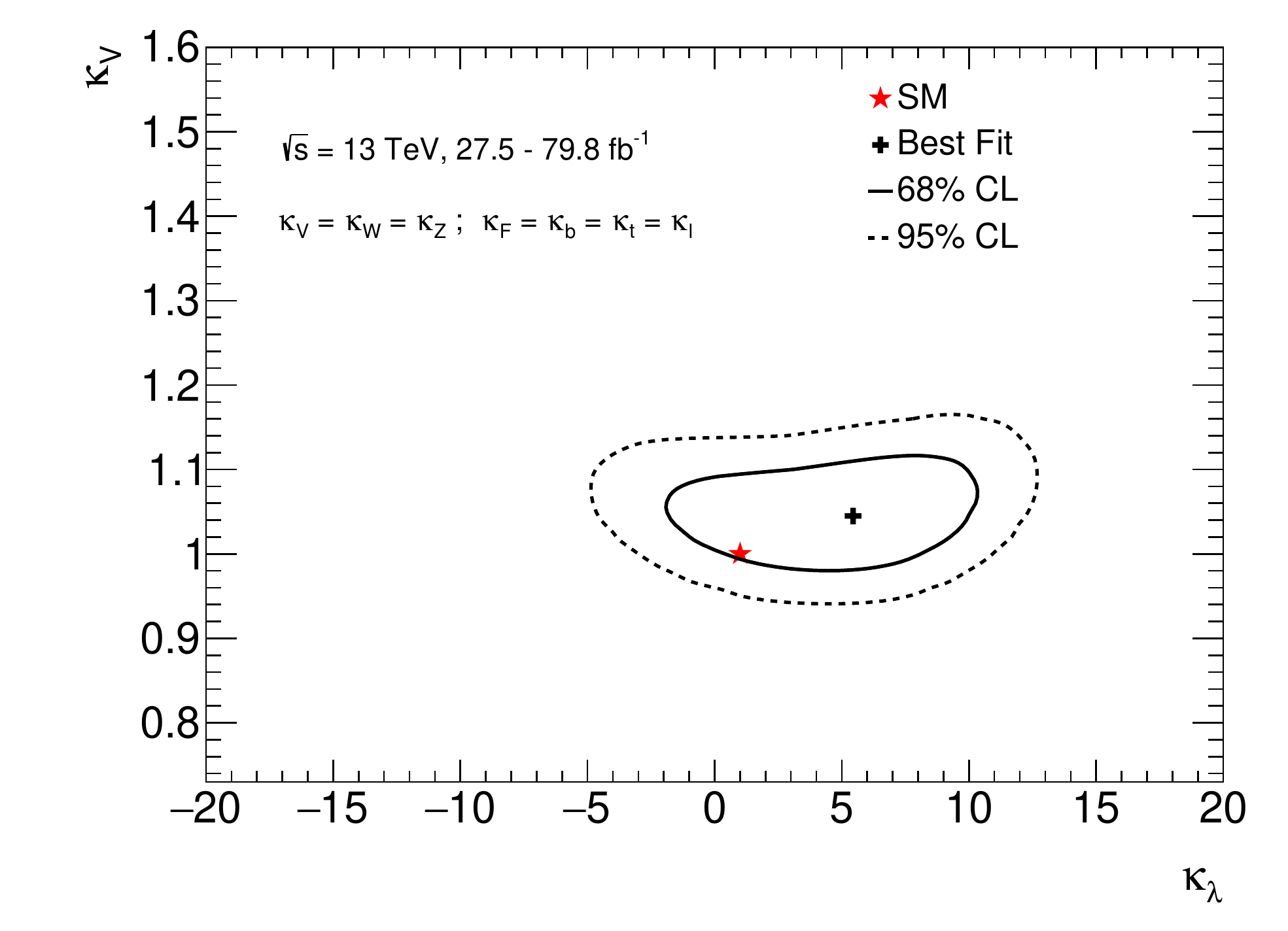}
 \caption{}
\end{subfigure}
  \begin{subfigure}[b]{0.49\textwidth}
\includegraphics[height=7 cm,width =8 cm]{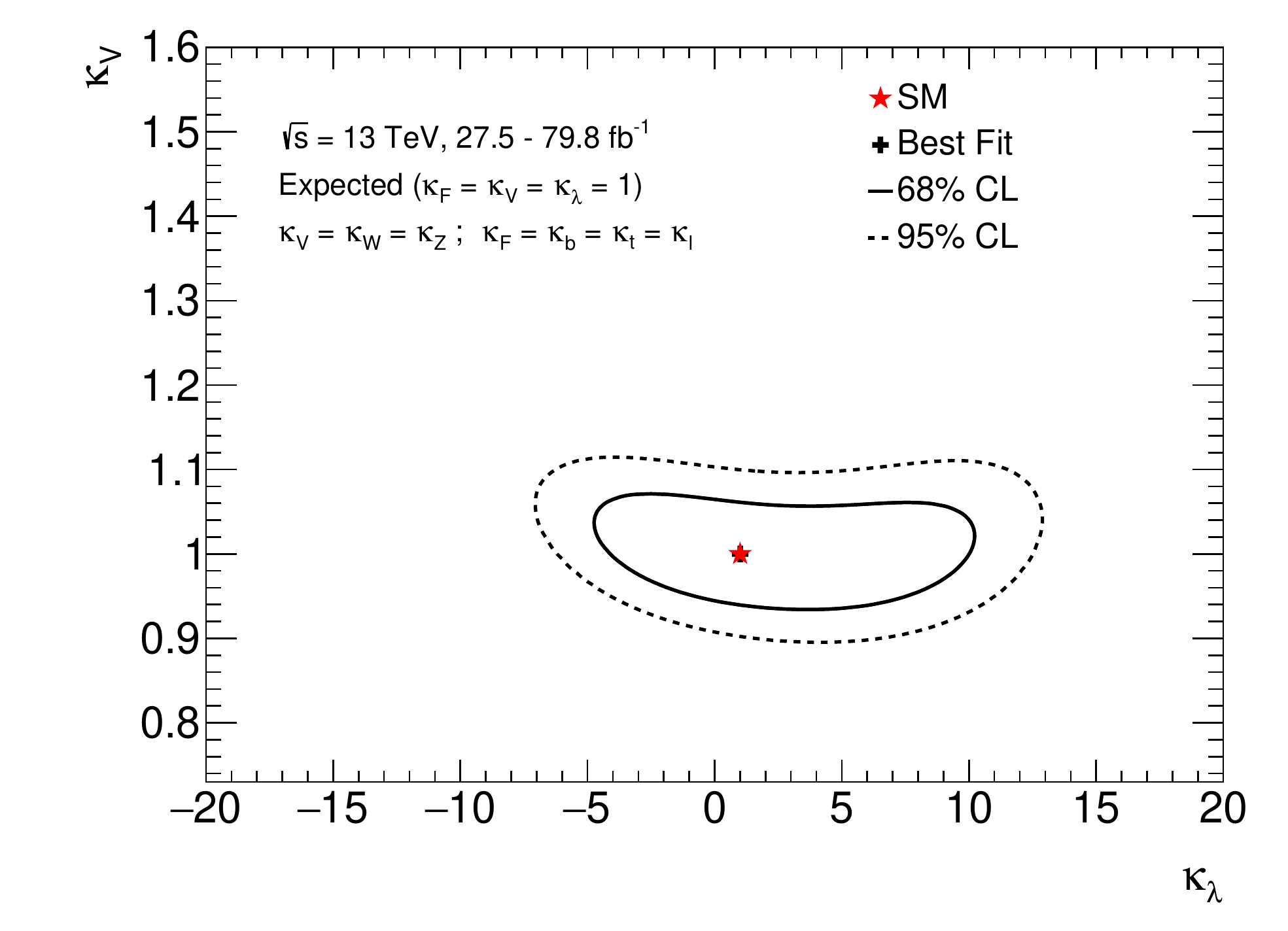}
 \caption{}
\end{subfigure}
    \caption{Negative log-likelihood contours at 68\% and 95\% CL in the $(\kappa_\lambda,\kappa_V)$ plane ($\kappa_{F}$ profiled) on data (a) and on the Asimov dataset generated under the SM hypothesis (b). The best-fit value is indicated by a cross while the SM hypothesis is indicated by a star. The plot assumes that the approximations in References~\cite{Degrassi,Maltoni} are valid inside the shown contours.}
  \label{fig:k3kVContour}
\end{figure}

\begin{figure}[htbp]
  \centering
  \begin{subfigure}[b]{0.49\textwidth}
\includegraphics[height=7 cm,width =8 cm]{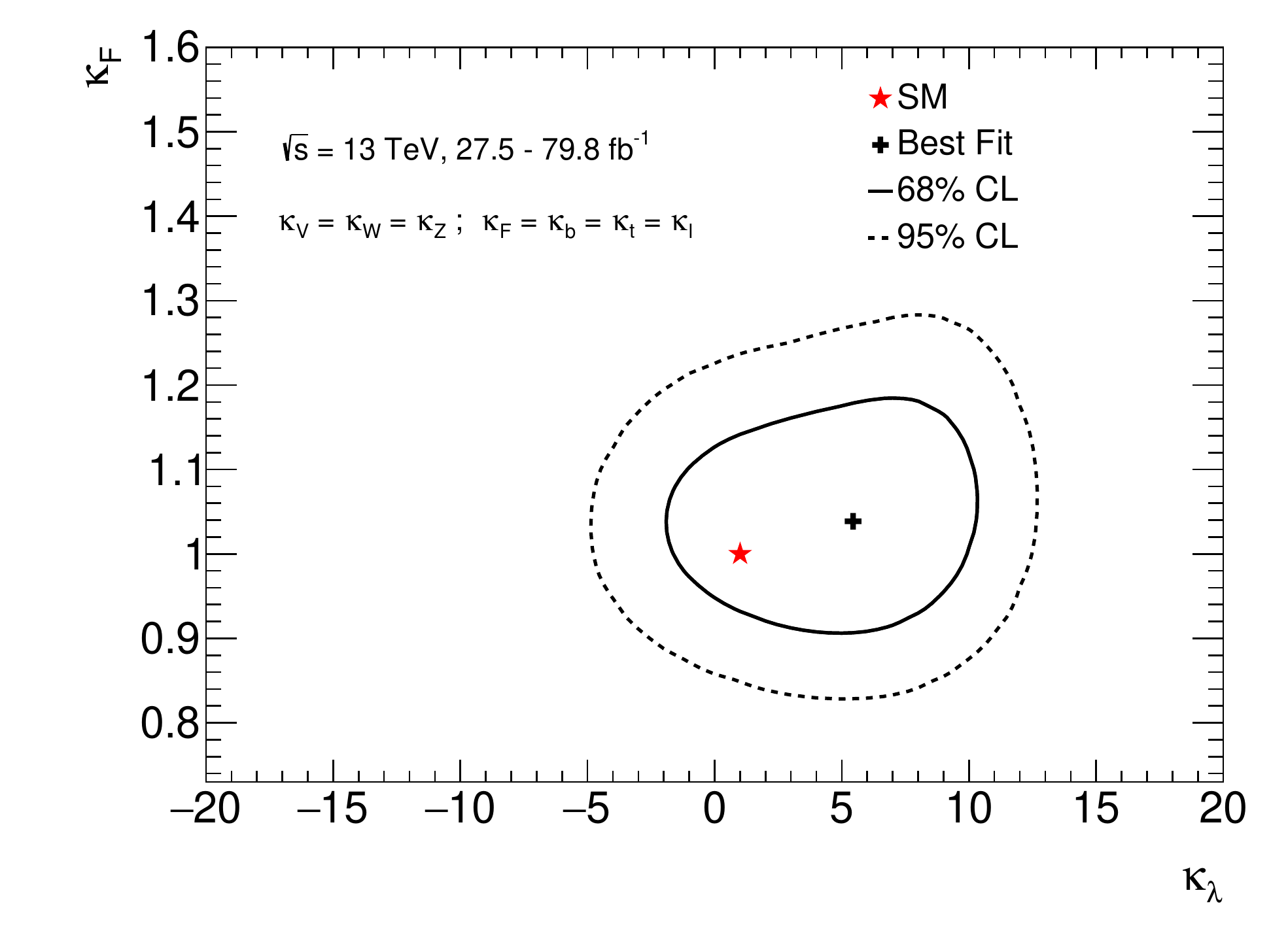}
 \caption{}
\end{subfigure}
  \begin{subfigure}[b]{0.49\textwidth}
\includegraphics[height=7 cm,width =8 cm]{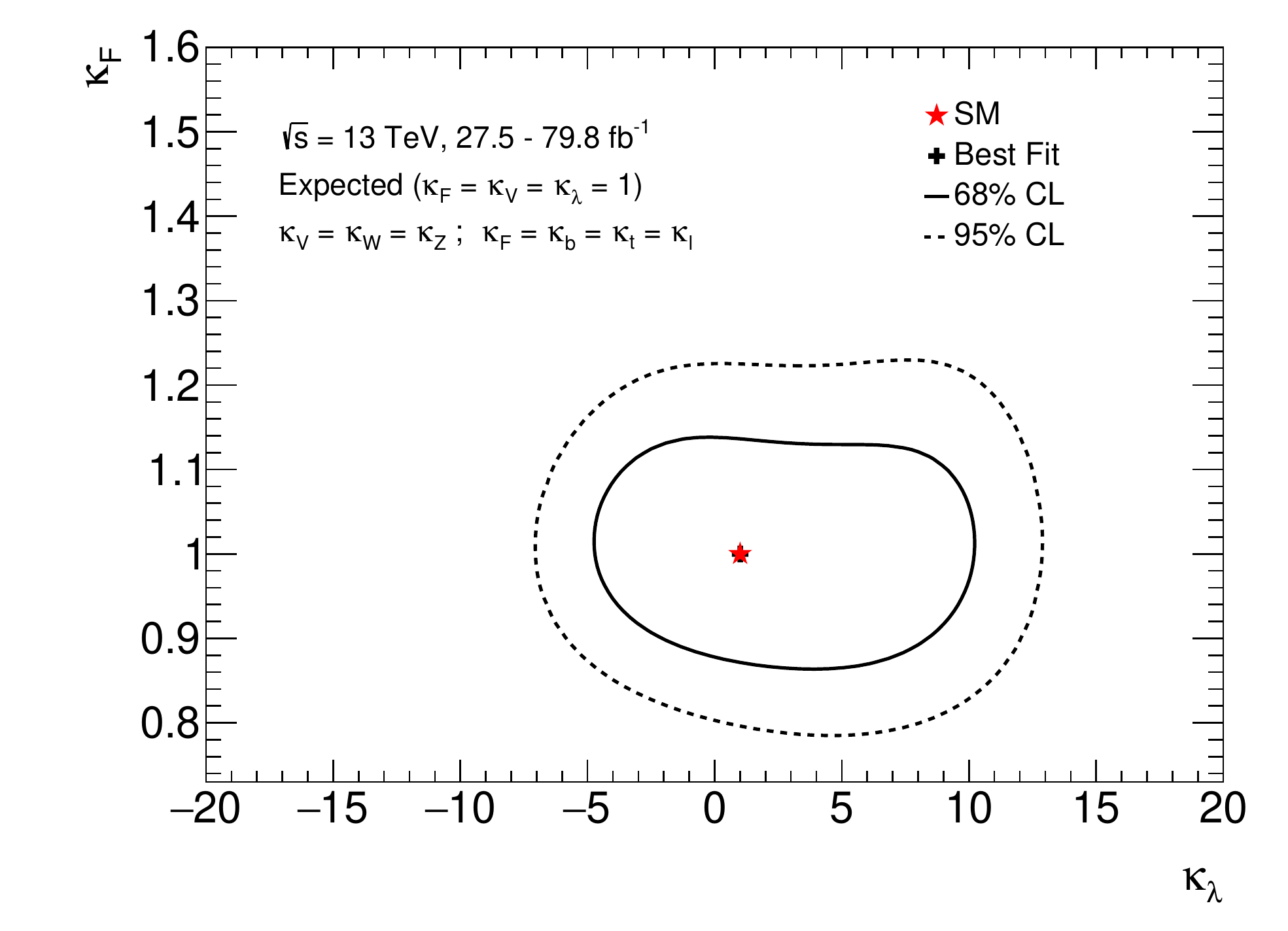}
 \caption{}
\end{subfigure}
    \caption{Negative log-likelihood contours at 68\% and 95\% CL in the $(\kappa_\lambda,\kappa_F)$ plane ($\kappa_{V}$ profiled) on data (a) and on the Asimov dataset generated under the SM hypothesis (b). The best-fit value  is indicated by a cross while the SM hypothesis is indicated by a star. The plot assumes that the approximations in References.~\cite{Degrassi,Maltoni} are valid inside the shown contours.}
  \label{fig:k3kFContour}
\end{figure}

\begin{figure}[htbp]
  \centering
  \begin{subfigure}[b]{0.49\textwidth}
\includegraphics[height=7 cm,width =8 cm]{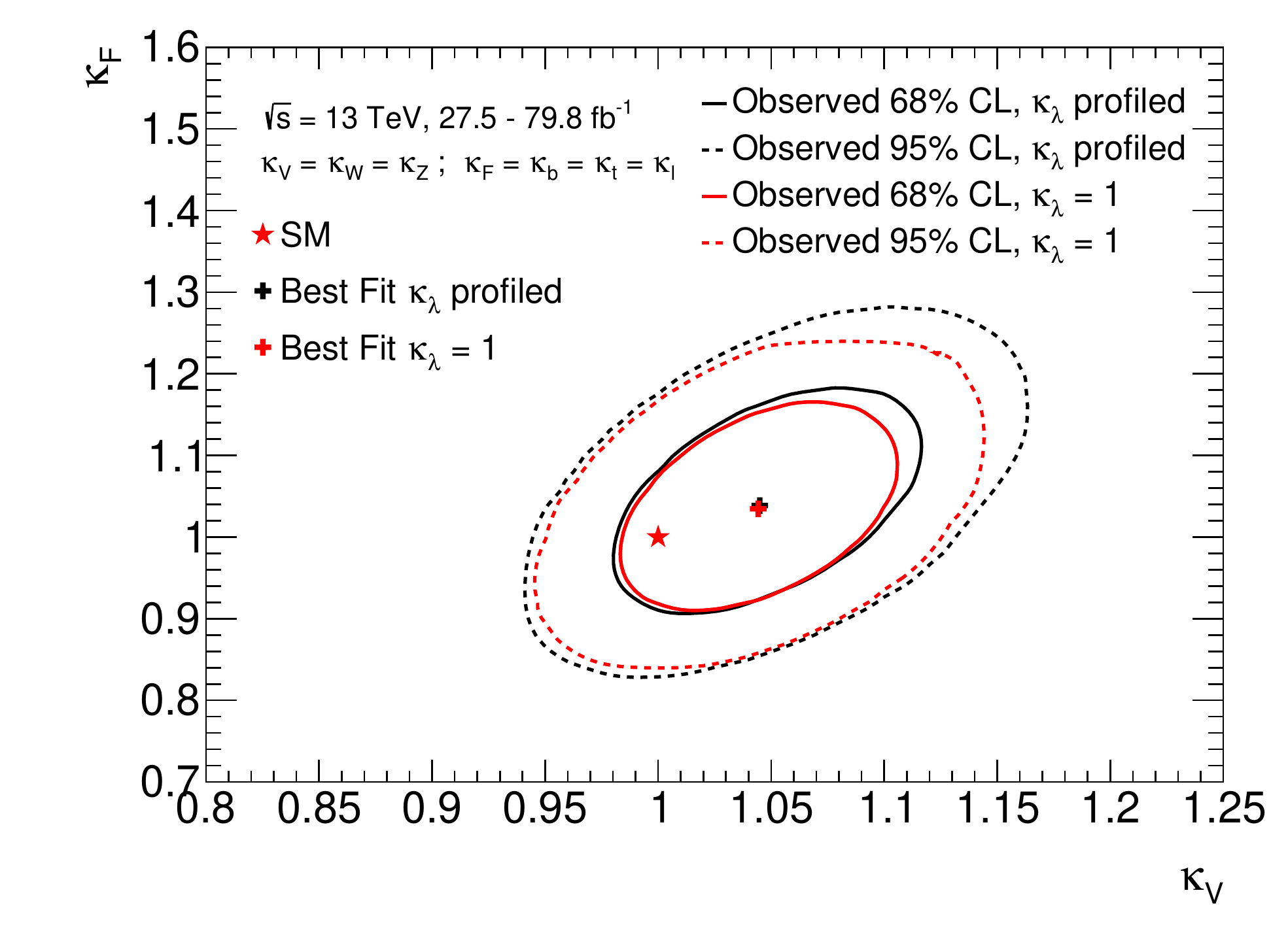}
 \caption{}
\end{subfigure}
  \begin{subfigure}[b]{0.49\textwidth}
\includegraphics[height=7 cm,width =8 cm]{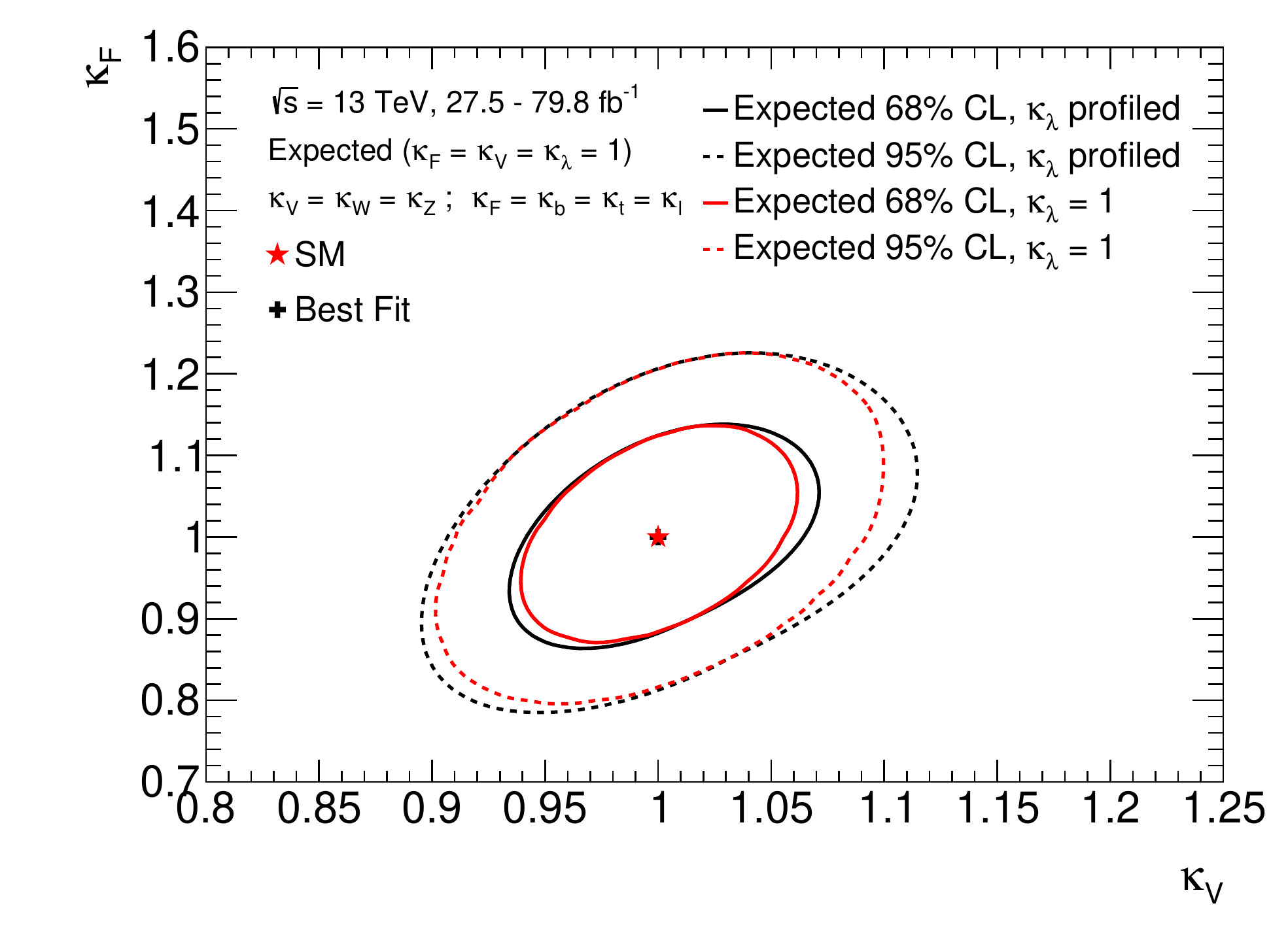}
 \caption{}
\end{subfigure}
    \caption{Negative log-likelihood contours at 68\% and 95\% CL in the $(\kappa_V,\kappa_F)$ plane with $\kappa_\lambda$ profiled (black line) and $\kappa_\lambda$=1 (red line) on data (a) and on the Asimov dataset generated under the SM hypothesis (b). The best-fit value is indicated by a cross while the SM hypothesis is indicated by a star. The plot assumes that the approximations in References~\cite{Degrassi,Maltoni} are valid inside the shown contours.}
  \label{fig:kVkFContour}
\end{figure}

The combination with double-Higgs analyses allows to include further degrees of freedom without loosing too much power in constraining $\kappa_\lambda$. Thus, additional fit configurations have been tested, in which a simultaneous fit is performed to constrain $\kappa_\lambda$, $\kappa_F=\kappa_t=\kappa_b=\kappa_{\ell}$ for fermions and $\kappa_V=\kappa_W=\kappa_Z$ for vector bosons ($\kappa_\lambda-\kappa_F-\kappa_V$ model) and $\kappa_\lambda$, $\kappa_{W}$ for $W$ boson, $\kappa_{Z}$ for $Z$ boson, $\kappa_{t}$ for up type quarks, $\kappa_{b}$ for down type quarks and $\kappa_{\ell}$ for charged leptons.
\begin{figure}[hbtp]
\centering
\begin{subfigure}[b]{0.49\textwidth}
\includegraphics[height=8 cm,width =8 cm]{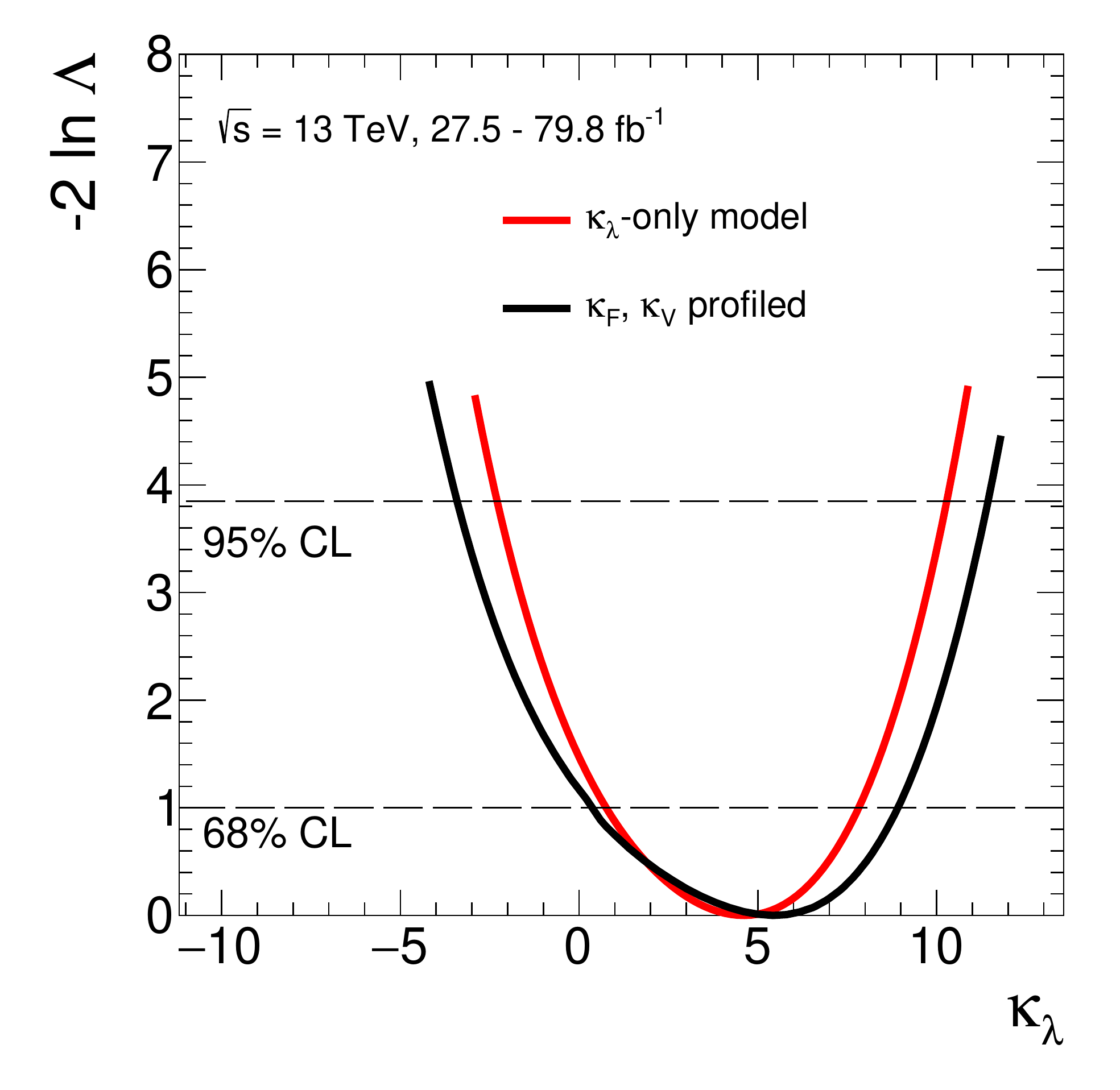}
 \caption{}
\end{subfigure}
\begin{subfigure}[b]{0.49\textwidth}
\includegraphics[height=8 cm,width =8 cm]{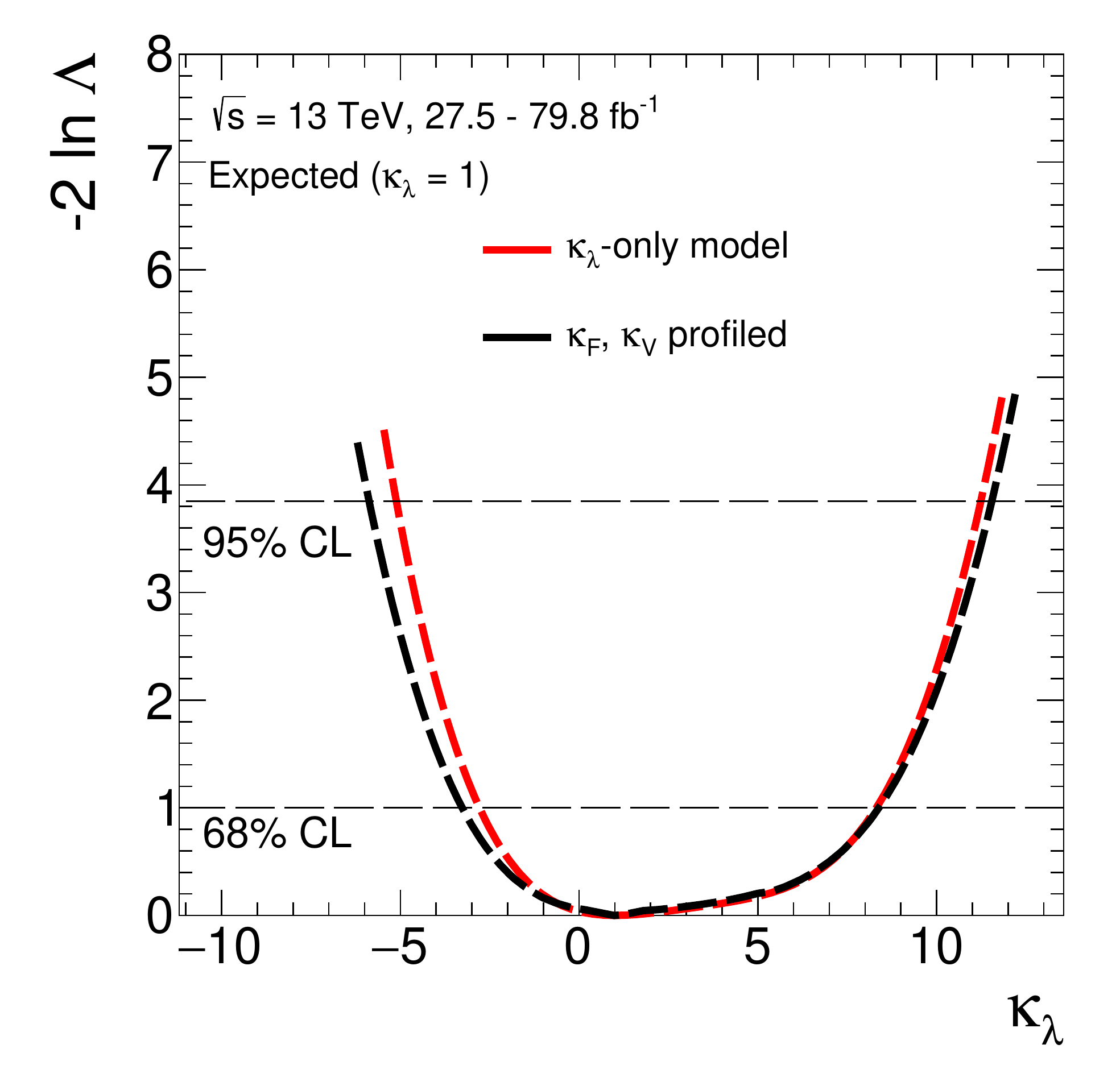}
 \caption{}
\end{subfigure}
\caption{Value of $-2 \ln{\Lambda(\kappa_\lambda)}$ as a function of $\kappa_\lambda$ (with $\kappa_{F}$ and $\kappa_V$ profiled); for data (a) and for the Asimov dataset generated in the SM hypothesis (b). The dotted horizontal lines show the $-2 \ln{\Lambda(\kappa_\lambda)}=1$ level that is used to define the $\pm 1\sigma$ uncertainty on $\kappa_\lambda$ as well as the $-2 \ln{\Lambda(\kappa_\lambda)}=3.84$ level used to define the 95\% CL. .}     
\label{scan_kl_kv_kF_comb}
\end{figure}
The value of $-2 \ln{\Lambda(\kappa_\lambda)}$ as a function of $\kappa_\lambda$ for the $\kappa_\lambda-\kappa_F-\kappa_V$ model is shown in Figure~\ref{scan_kl_kv_kF_comb} for data and for the Asimov dataset, and it is compared to the curves obtained in the $\kappa_\lambda$-only model. The sensitivity is degraded at most by $\sim$20\% in the $\kappa_\lambda-\kappa_F-\kappa_V$ model. Figures~\ref{fig:k3kVContour} - \ref{fig:kVkFContour} show negative log-likelihood contours in the $(\kappa_\lambda,\kappa_{V})$, $(\kappa_\lambda,\kappa_{F})$ and $(\kappa_{V},\kappa_{F})$ planes obtained profiling the remaining coupling, $\kappa_F$, $\kappa_V$ and $\kappa_\lambda$, respectively.

\begin{figure}[htbp]
  \centering
  \begin{subfigure}[b]{0.49\textwidth}
\includegraphics[height=7 cm,width =8 cm]{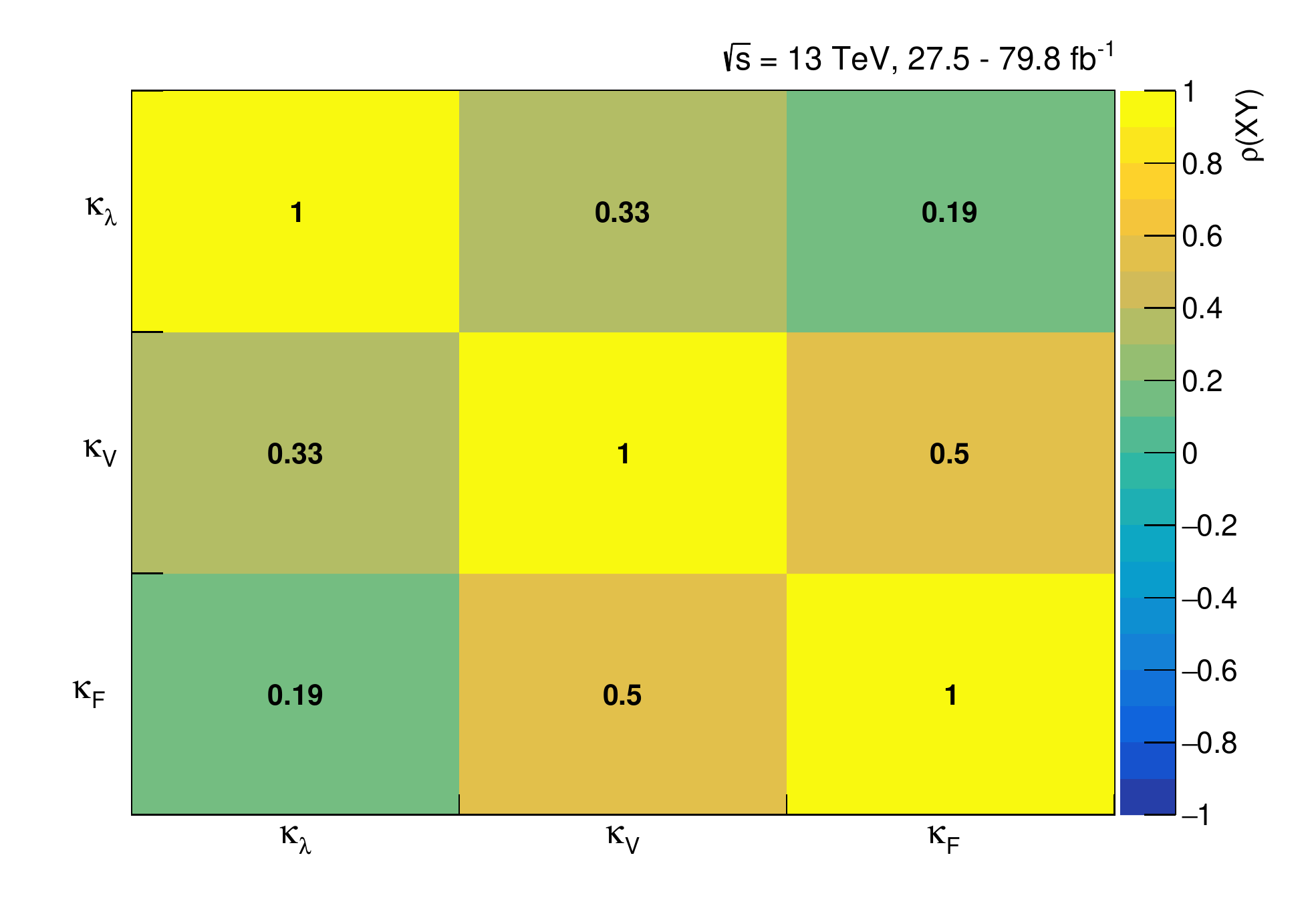}
 \caption{}
\end{subfigure}
  \begin{subfigure}[b]{0.49\textwidth}
\includegraphics[height=7 cm,width =8 cm]{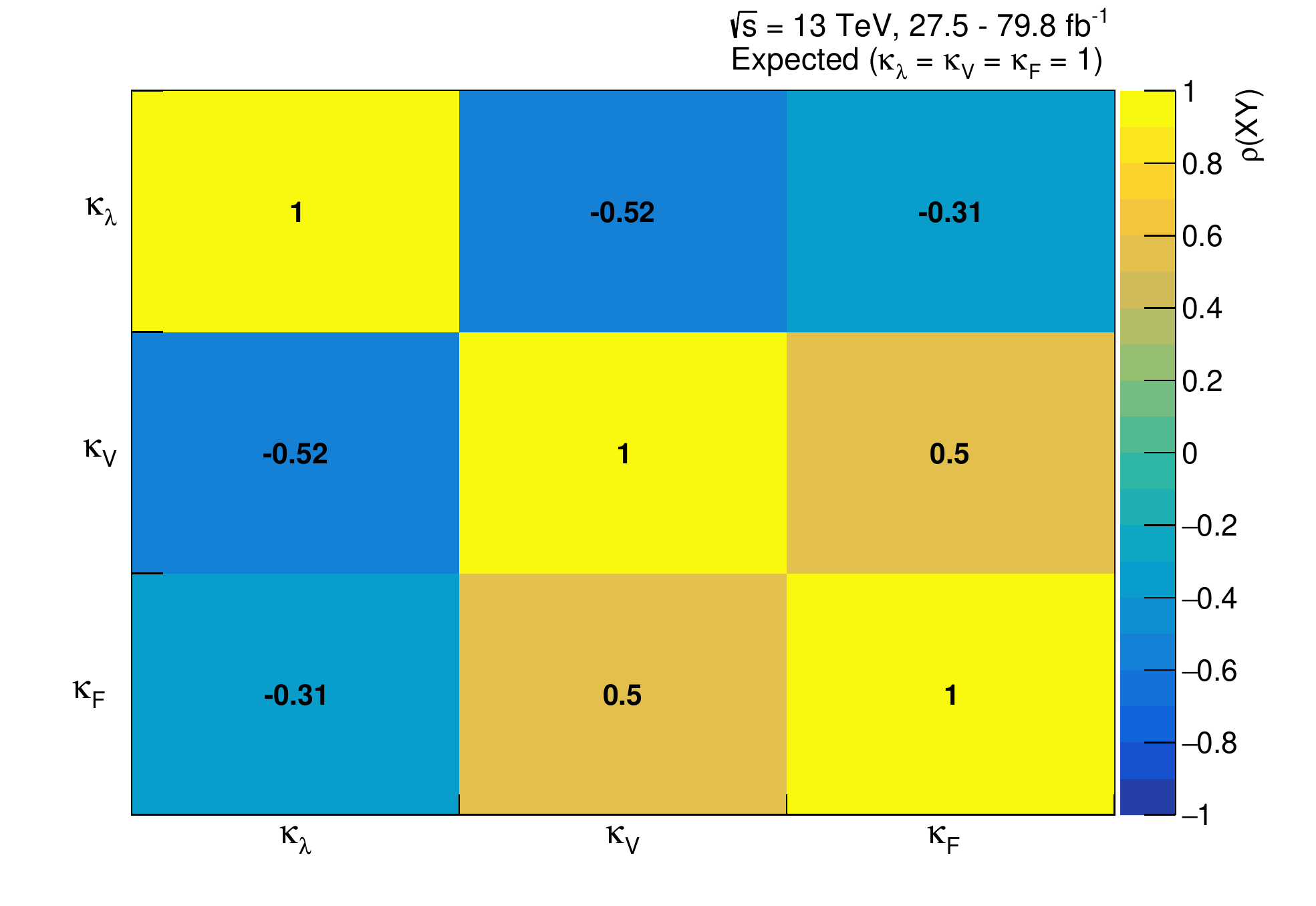}
 \caption{}
\end{subfigure}
    \caption{Correlations between the parameters of interest, \ie\ $\kappa_\lambda$, $\kappa_F$ and $\kappa_V$, for (a) data and for (b) Asimov dataset generated under the SM hypothesis.}
  \label{correlation_kl_kF_kV}
\end{figure}

The correlations among the parameters of interest, \ie\ $\kappa_\lambda$, $\kappa_F$ and $\kappa_V$ are shown in Figure~\ref{correlation_kl_kF_kV} for data $(a)$ and for the Asimov dataset. A weaker correlation between $\kappa_F$ and $\kappa_V$ with respect to the single-Higgs combination is present given the introduction of further terms from the double-Higgs analyses.

A more generic model has been considered, performing a likelihood fit to constrain simultaneously $\kappa_\lambda$, $\kappa_{W}$, $\kappa_{Z}$, $\kappa_{t}$, $\kappa_{b}$ and $\kappa_{\ell}$; this generic model represents an intermediate parameterisations with respect to a very general EFT parameterisation, like the one presented in Reference~\cite{global} thus targeting BSM models than can modify at the same time the Higgs-boson self-coupling and other SM couplings.\newline
Table~\ref{tab:k3kVkF} reports a summary of fit results in the different configurations. The best-fit values for all the couplings are compatible with the SM prediction.
\begin{table}[htbp]
\begin{center}
{\def\arraystretch{1.3}
\begin{tabular}{|c|c|c|c|c|c|c|c|}
\hline

    POIs & $\kappa_{W}{}^{+1\sigma}_{-1\sigma}$&$\kappa_{Z}{}^{+1\sigma}_{-1\sigma}$& $\kappa_{t}{}^{+1\sigma}_{-1\sigma}$ & $\kappa_b{}^{+1\sigma}_{-1\sigma}$&$\kappa_{\ell}{}^{+1\sigma}_{-1\sigma}$& $\kappa_\lambda{}^{+1\sigma}_{-1\sigma}$ & $\kappa_\lambda$  [95\% CL] \\ 
\hline
  \multirow{2}{*}{$\kappa_\lambda$}&  \multirow{2}{*}{1} &  \multirow{2}{*}{1} & \multirow{2}{*}{1} &\multirow{2}{*}{1} &\multirow{2}{*}{1} &$4.6_{-3.8}^{+3.2}$ &  $[-2.3, 10.3]$ \\
    & &   &      & & &  $1.0^{+7.3}_{-3.8}$ & $[-5.1, 11.2]$ \\ 
     \hline
  \multirow{2}{*}{$\kappa_\lambda$-$\kappa_F$-$\kappa_V$}& $1.04_{-0.04}^{+0.05}$  & $1.04_{-0.04}^{+0.05}$ & $1.04_{-0.09}^{+0.09}$ &$1.04_{-0.09}^{+0.09}$ &$1.04_{-0.09}^{+0.09}$ &$5.4_{-5.1}^{+3.5}$ &  $[-3.4, 11.4]$ \\
    &$1.00_{-0.04}^{+0.05}$& $1.00_{-0.04}^{+0.05}$  &  $1.00_{-0.09}^{+0.09}$     & $1.00_{-0.09}^{+0.09}$ & $1.00_{-0.09}^{+0.09}$&  $1.0^{+7.4}_{-4.3}$ & $[-5.9, 11.6]$ \\ 
\hline 
    \multirow{2}{*}{$\kappa_\lambda$ generic} &$1.03_{-0.08}^{+0.08}$ &  $1.10_{-0.09}^{+0.09}$& $1.00_{-0.11}^{+0.12}$&$1.03_{-0.18}^{+0.20}$&$1.06_{-0.16}^{+0.16}$&$5.5_{-5.2}^{+3.5}$ &  $[-3.7, 11.5]$ \\
    &$1.00_{-0.08}^{+0.08}$&  $1.00_{-0.08}^{+0.08}$& $1.00_{-0.12}^{+0.12}$&$1.00_{-0.19}^{+0.21}$&$1.00_{-0.15}^{+0.16}$&$1.0_{-4.5}^{+7.6}$ & $[-6.2, 11.6]$ \\ 
\hline
\end{tabular}
}
\caption{
Best-fit values for $\kappa$ modifiers with $\pm 1 \sigma$ uncertainties for the different fit configurations listed in the first column. The 95\% CL interval for $\kappa_\lambda$ is also reported. For the fit result the upper row corresponds to the observed results, and the lower row to the expected results obtained using Asimov datasets generated under the SM hypothesis.}
\label{tab:k3kVkF}
\end{center}
\end{table}

\begin{figure}[hbtp]
\centering
\begin{subfigure}[b]{0.49\textwidth}
\includegraphics[height=8 cm,width =8 cm]{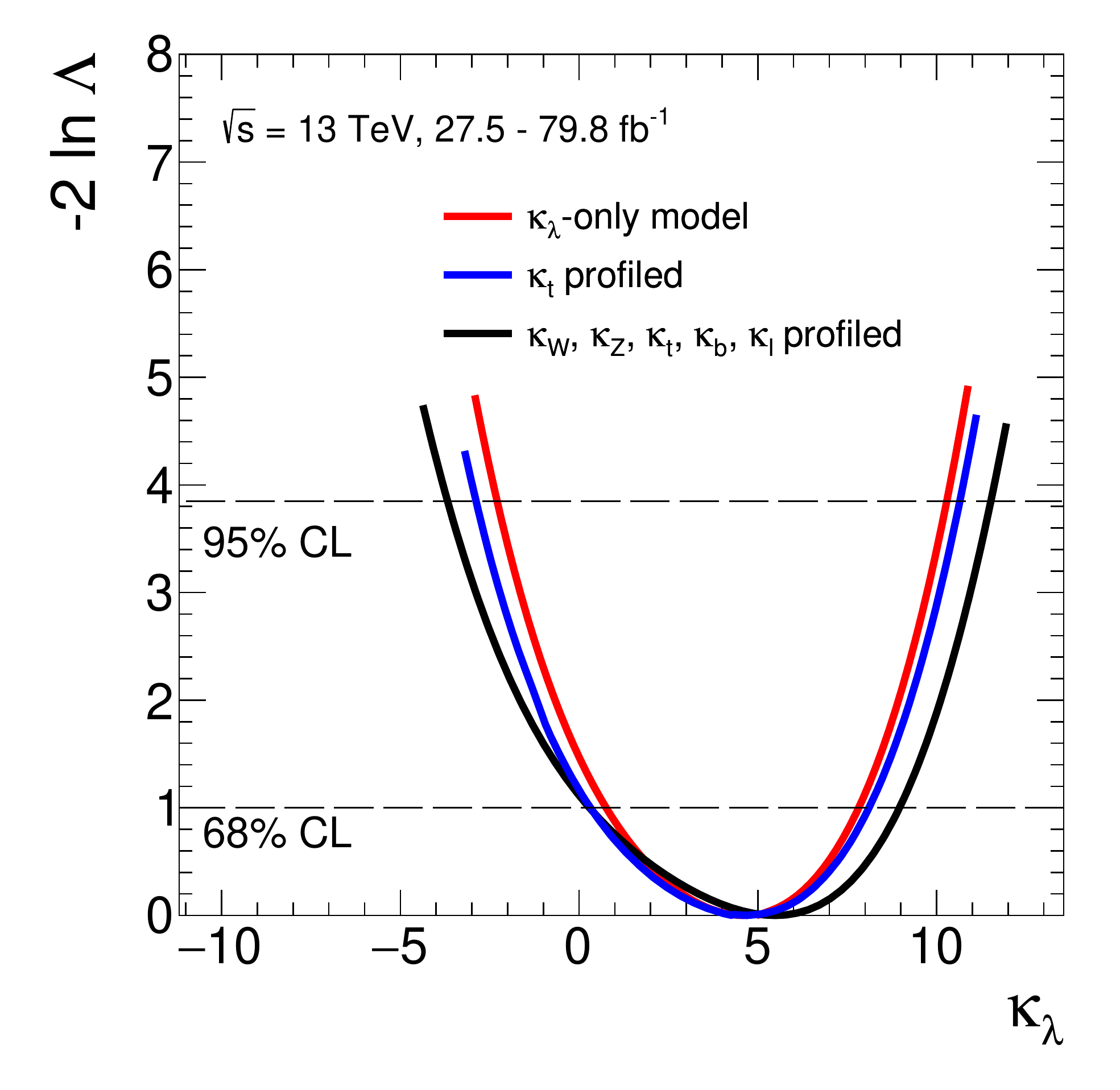}
 \caption{}
\end{subfigure}
\begin{subfigure}[b]{0.49\textwidth}
\includegraphics[height=8 cm,width =8 cm]{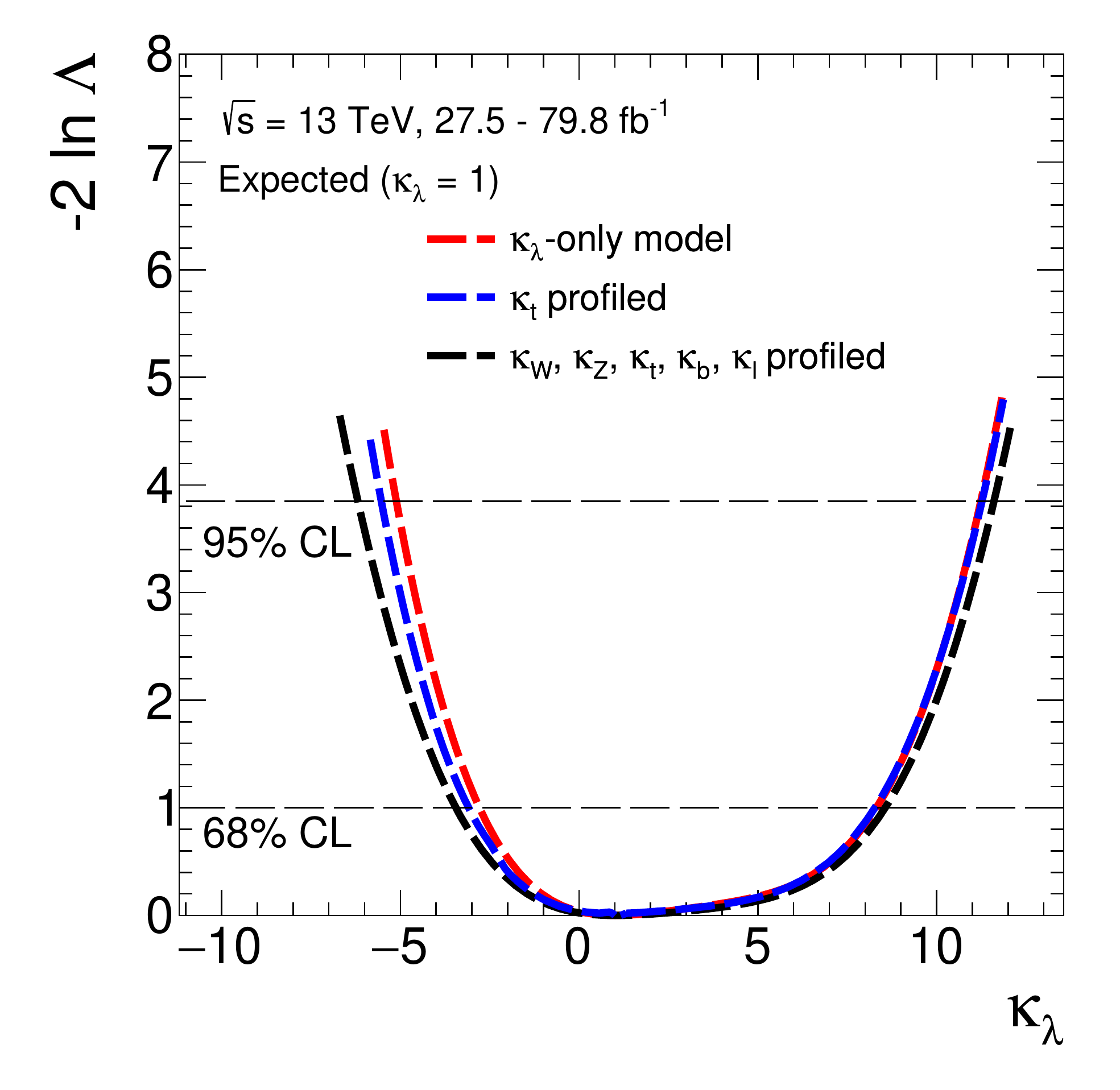}
 \caption{}
\end{subfigure}
\caption{Value of $-2 \ln{\Lambda(\kappa_\lambda)}$ as a function of $\kappa_\lambda$ with $\kappa_{W}$, $\kappa_{Z}$, $\kappa_{t}$, $\kappa_{b}$, $\kappa_{lep}$ profiled for data (a) and for the Asimov dataset (b), generated under the SM hypothesis. The generic model curves both for data and for the Asimov dataset are compared to the corresponding curves for the $\kappa_\lambda$-only model where all the couplings, except for $\kappa_\lambda$, are set to their SM values and $\kappa_\lambda$-$\kappa_t$ model where all the other couplings are set to their SM values. The dotted horizontal lines show the $-2 \ln{\Lambda(\kappa_\lambda)}=1$ level that is used to define the $\pm 1\sigma$ uncertainty on $\kappa_\lambda$ as well as the $-2 \ln{\Lambda(\kappa_\lambda)}=3.84$ level used to define the 95\% CL.}     
\label{scan_generic}
\end{figure}

The value of $-2 \ln{\Lambda(\kappa_\lambda)}$ as a function of $\kappa_\lambda$ profiling all the other couplings is shown in Figure~\ref{scan_generic} for data and for the Asimov dataset, and it is compared to the curves obtained in the $\kappa_\lambda$-only model and in the $\kappa_\lambda$-$\kappa_t$ model. The sensitivity is degraded at most by $\sim$20\% going from the 95\% CL interval for the $\kappa_\lambda$-only model to the 95\% CL interval for the generic model. Concerning other couplings, their best-fit values are compatible with the SM prediction. The combination allows to put sizeable constraints also in this generic model, despite the number of degrees of freedom introduced.

\begin{figure}[htbp]
  \centering
  \begin{subfigure}[b]{0.49\textwidth}
\includegraphics[height=8 cm,width =8.3 cm]{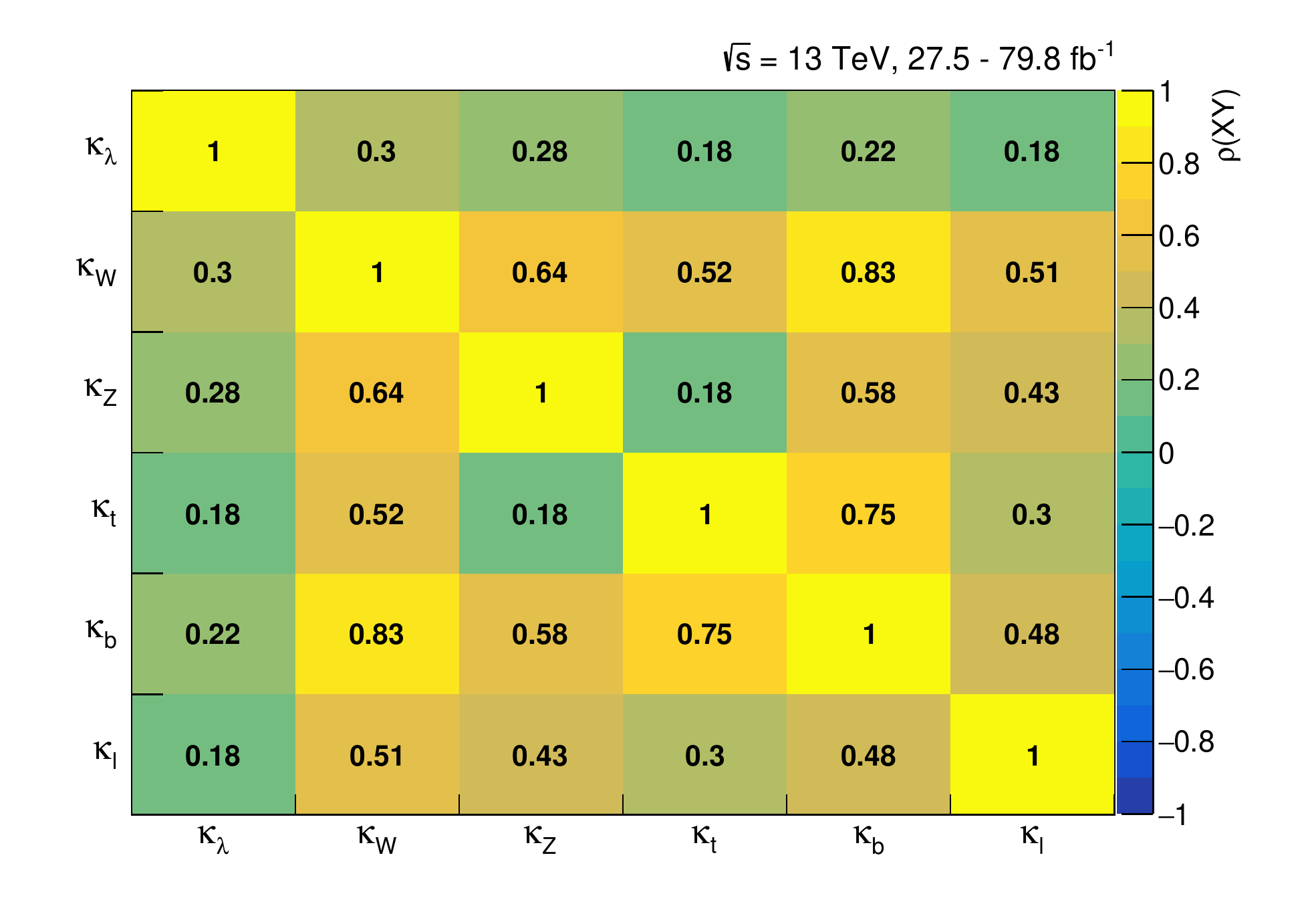}
 \caption{}
\end{subfigure}
  \begin{subfigure}[b]{0.49\textwidth}
\includegraphics[height=8 cm,width =8.3 cm]{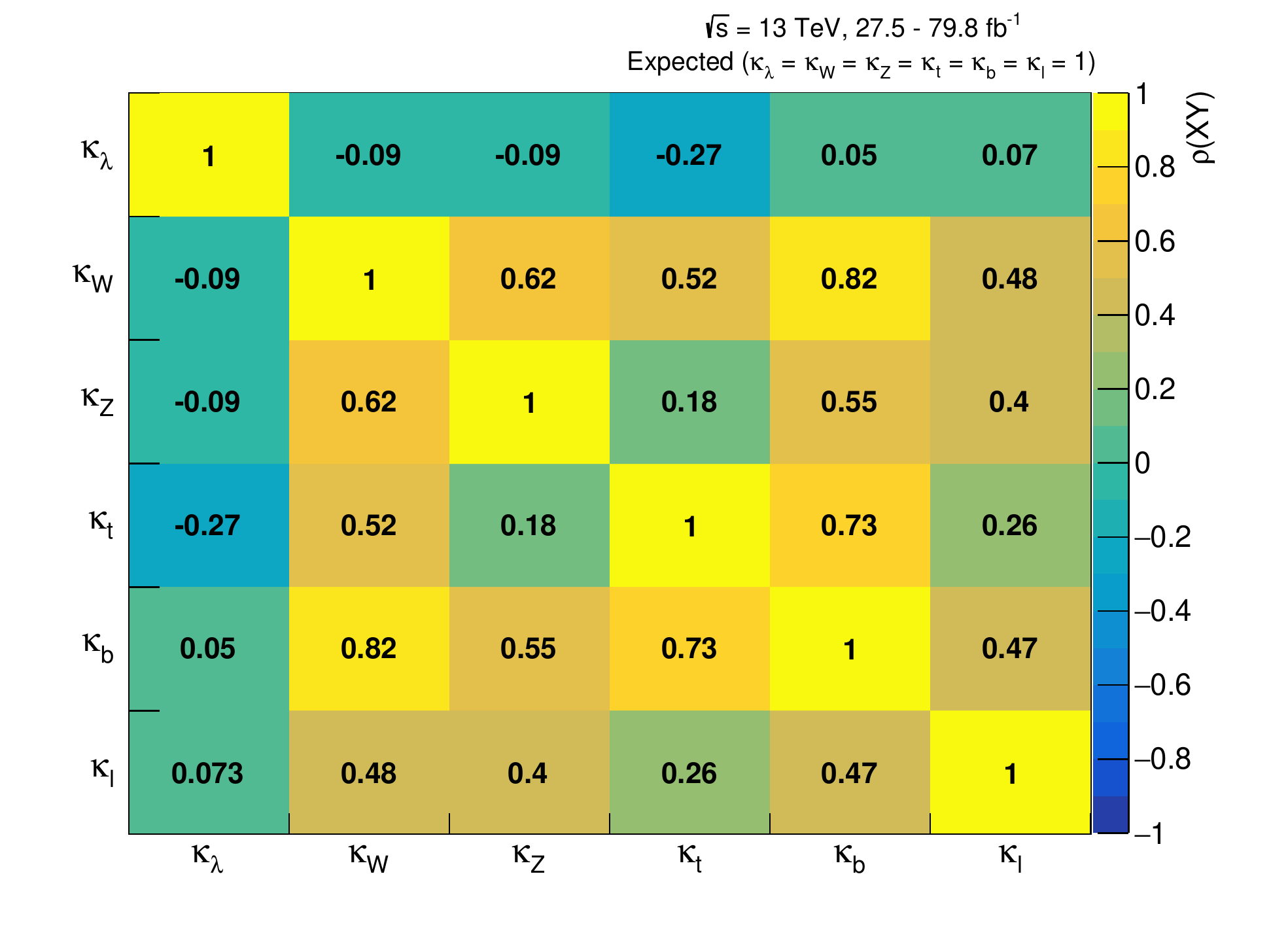}
 \caption{}
\end{subfigure}
    \caption{Correlations between the parameters of interest, \ie\ $\kappa_\lambda$, $\kappa_W$, $\kappa_Z$, $\kappa_t$, $\kappa_b$ and $\kappa_\ell$, for (a) data and for (b) Asimov dataset generated under the SM hypothesis.}
  \label{correlation_kl_generic}
\end{figure}

The correlations between the parameters of interest, \ie\ $\kappa_\lambda$, $\kappa_W$, $\kappa_Z$, $\kappa_t$, $\kappa_b$ and $\kappa_\ell$, are shown in Figure~\ref{correlation_kl_generic} for data $(a)$ and for the Asimov dataset. A strong correlation is found between $\kappa_W$, $\kappa_Z$ and $\kappa_b$ being mostly constrained by the $VH\rightarrow b\bar{b}$ channel as well as $\kappa_t$ and $\kappa_b$ constrained by the $t\bar{t}H \rightarrow b\bar{b}$ channel.

\clearpage

\addcontentsline{toc}{chapter}{Conclusion}
\markboth{Conclusion}{Conclusion} 
\chapter*{Conclusion}
\label{sec:conclusion}
The Higgs-boson self-coupling has been constrained through single-Higgs processes, exploiting the next-to-leading order dependence of these processes on $\kappa_\lambda$ via weak loops; this approach has been applied to the combination of analyses targeting  single-Higgs production and decay modes, \ie\ \ggF, \VBF, \ZH, \WH and $t\bar{t}H$ production modes together with $WW^*$, $ZZ^*$, $\tau^+\tau^-$, $\gamma \gamma$ and $b\bar{b}$ decay channels, on data collected with the ATLAS experiment using up to 79.8 fb$^{-1}$ of LHC proton-proton collisions.\newline
In the simplified assumption that all deviations from the SM expectation have to be interpreted as modifications of the trilinear coupling of the
Higgs boson, the Higgs-boson self-coupling modifier $\kappa_\lambda=\lambda_{HHH}/\lambda_{HHH}^{SM}$, extracted with a global fit procedure, is determined to be $\kappa_\lambda=4.0^{+4.3}_{-4.1}$, excluding at the 95\% CL values outside the interval
$-3.2 < \kappa_\lambda < 11.9$. Additional results, including the simultaneous determination of the Higgs self-coupling and single-Higgs couplings to either fermions or bosons, have been derived.  The constraints on $\kappa_\lambda$ become weaker when further degrees of freedom are introduced, to the point where no sensitivity to $\kappa_\lambda$ is found when considering modifications to the self-coupling and to a common single-Higgs coupling modifier.\newline
Being the limits on $\kappa_\lambda$ competitive with the ones coming from double-Higgs measurements, a combination between the most sensitive double-Higgs channels, $b\bar{b}\tau^+\tau^-$, $b\bar{b}\gamma\gamma$ and $b\bar{b}b\bar{b}$ exploiting data collected with the ATLAS experiment using up to 36.1 fb$^{-1}$ of LHC proton-proton collisions, and the aforementioned single-Higgs analyses has been performed. 
The dependence of the double-Higgs cross section on both the coupling of the Higgs boson to the top quark and the Higgs self-coupling has been taken into account.
Under the assumption that new physics affects only the Higgs-boson self-coupling, the combined best-fit value of the coupling modifier is $\kappa_\lambda = 4.6^{+3.2}_{-3.8}$, excluding values outside the interval $-2.3<\kappa_\lambda<10.3$ at 95\% CL. This result represents a significant improvement in constraining $\kappa_\lambda$ with respect to single-Higgs and double-Higgs analyses alone.
Moreover, the single- and double-Higgs combination allows to decouple the self-coupling and top-Yukawa coupling as well as other couplings. Thus, sizeable constraints on $\kappa_\lambda$ can be set also when less model dependent parameterisation are considered, including coupling modifiers for the Higgs-boson self-coupling, for the up- and down-type quarks, for leptons and for $W$ and $Z$ bosons. In this more generic configuration, the self-coupling modifier has been constrained at the 95\% CL value to the interval $-3.7<\kappa_\lambda<11.5$. All other coupling modifiers are compatible with the SM predictions. 
Being the measurements statistically dominated, both from single-Higgs and double-Higgs side, a significant improvement in constraining $\kappa_\lambda$ is expected at the High-Luminosity LHC.


\printbibliography

\clearpage
\appendix
\addcontentsline{toc}{part}{Appendices}
  \chapter{Correlations between double-Higgs analyses}
\label{sec:appendix_correlation_hh}
For the $HH\rightarrow b\bar{b}b\bar{b}$ channel, two different correlation schemes regarding flavour tagging, jet, parton shower and trigger uncertainties, have been considered and are described in Table~\ref{tab:correlation4b}:
\begin{itemize}
\item all NPs related to the signal samples uncorrelated, \ie\ keeping  FT, JET, PS, trigger NPs split in the three signal samples (scheme 1);
\item all NPs related to the signal samples correlated to be consistent with the published double-Higgs combination~\cite{Paper_hh} (scheme 2).
\end{itemize}


\begin{longtable}{|c|c|}
\hline
 Scheme 1 & Scheme 2\\ \hline
\endfirsthead
\hline
 Scheme 1 & Scheme 2\\ \hline
\endhead 
\endlastfoot 
\hline

alpha\_FT\_EFF\_Eigen\_B\_0\_lhh00 & \multirow{3}{*}{alpha\_FT\_EFF\_Eigen\_B\_0} \\
alpha\_FT\_EFF\_Eigen\_B\_0\_lhh01 & \\
alpha\_FT\_EFF\_Eigen\_B\_0\_lhh20 & \\
\hline
alpha\_FT\_EFF\_Eigen\_B\_1\_lhh00 & \multirow{3}{*}{alpha\_FT\_EFF\_Eigen\_B\_1} \\
alpha\_FT\_EFF\_Eigen\_B\_1\_lhh01& \\
alpha\_FT\_EFF\_Eigen\_B\_1\_lhh20 & \\
\hline
alpha\_FT\_EFF\_Eigen\_B\_2\_lhh00 & \multirow{3}{*}{alpha\_FT\_EFF\_Eigen\_B\_2} \\
alpha\_FT\_EFF\_Eigen\_B\_2\_lhh01 & \\
alpha\_FT\_EFF\_Eigen\_B\_2\_lhh20 & \\
\hline
alpha\_FT\_EFF\_Eigen\_B\_3\_lhh00 & \multirow{3}{*}{alpha\_FT\_EFF\_Eigen\_B\_3} \\
alpha\_FT\_EFF\_Eigen\_B\_3\_lhh01 & \\
alpha\_FT\_EFF\_Eigen\_B\_3\_lhh20 & \\
\hline
alpha\_FT\_EFF\_Eigen\_B\_4\_lhh00 & \multirow{3}{*}{alpha\_FT\_EFF\_Eigen\_B\_4} \\
alpha\_FT\_EFF\_Eigen\_B\_4\_lhh01 & \\
alpha\_FT\_EFF\_Eigen\_B\_4\_lhh20 & \\
\hline
alpha\_FT\_EFF\_extrapolation\_lhh00 & \multirow{3}{*}{alpha\_FT\_EFF\_extrapolation} \\
alpha\_FT\_EFF\_extrapolation\_lhh01& \\
alpha\_FT\_EFF\_extrapolation\_lhh20 & \\
\hline
alpha\_JET\_EtaIntercalib\_Nonclos\_lhh00 & \multirow{3}{*}{alpha\_JET\_EtaIntercalib\_Nonclos} \\
alpha\_JET\_EtaIntercalib\_Nonclos\_lhh01 & \\
alpha\_JET\_EtaIntercalib\_Nonclos\_lhh20 & \\
\hline
alpha\_JET\_GroupedNP\_1\_lhh00 & \multirow{3}{*}{alpha\_JET\_GroupedNP\_1} \\
alpha\_JET\_GroupedNP\_1\_lhh01 & \\
alpha\_JET\_GroupedNP\_1\_lhh20& \\
\hline
alpha\_JET\_GroupedNP\_2\_lhh00 & \multirow{3}{*}{alpha\_JET\_GroupedNP\_2} \\
alpha\_JET\_GroupedNP\_2\_lhh01& \\
alpha\_JET\_GroupedNP\_2\_lhh20 & \\
\hline
alpha\_JET\_GroupedNP\_3\_lhh00 & \multirow{3}{*}{alpha\_JET\_GroupedNP\_3} \\
alpha\_JET\_GroupedNP\_3\_lhh01 & \\
alpha\_JET\_GroupedNP\_3\_lhh20 & \\
\hline
alpha\_JET\_JER\_lhh00 & \multirow{3}{*}{alpha\_JET\_JER} \\
alpha\_JET\_JER\_lhh01 & \\ 
alpha\_JET\_JER\_lhh20 & \\
\hline
alpha\_Theoretical\_lhh00 & \multirow{3}{*}{alpha\_Theoretical} \\
alpha\_Theoretical\_lhh01 & \\
alpha\_Theoretical\_lhh20 & \\
\hline
alpha\_trig\_r15\_lhh00 & \multirow{3}{*}{alpha\_trig\_r15} \\
alpha\_trig\_r15\_lhh01 & \\
alpha\_trig\_r15\_lhh20 & \\
\hline
alpha\_trig\_r16\_lhh00 & \multirow{3}{*}{alpha\_trig\_r16} \\
alpha\_trig\_r16\_lhh01 & \\
alpha\_trig\_r16\_lhh20 & 
  \label{tab:correlation4b}
                                \\ \hline

\end{longtable}

\clearpage
 \chapter{Correlations between single- and double-Higgs analyses}
\label{sec:appendix_correlation_comb}
The nuisance parameters correlation scheme adopted in the H+HH combination, in addition to the correlations between single-Higgs and double-Higgs individual combination described in Chapters~\ref{sec:dihiggs} and~\ref{sec:single}, is reported in the following tables. 
Experimental uncertainties have been correlated whenever relevant, like in the case of the integrated luminosity and detector related uncertainties. Experimental uncertainties that are related to the same physics object but determined with different methodologies or implemented with different parameterisations, like the flavour tagging uncertainties, have been kept uncorrelated. Signal theory uncertainties have been kept uncorrelated while the systematic uncertainties on the decay branching ratios have been correlated. Theoretical uncertainties on the $pp\rightarrow HH$ ggF cross section have been included in the double-Higgs analyses.
The correlation between the systematic uncertainties is implemented in the final fit procedure by associating different uncertainties to the same nuisance parameter in the combined likelihood function.
\begin{table}
\scalebox{0.85}{
{\def\arraystretch{1.1}
\begin{tabular}{|l|c|c|}
\hline
NP name & single Higgs & double Higgs\\ 
\hline
\multicolumn{3}{|c|}{Experimental uncertainties}\\ \hline
\multicolumn{3}{|c|}{Luminosity uncertainties}\\ \hline
                                     ATLAS\_LUMI\_1516\_UNCOR & $\circ$ & $\circ$ \\ \hline
                                     ATLAS\_LUMI\_1517\_UNCOR & $\circ$ &         \\ \hline
                                     ATLAS\_LUMI\_RUN2\_CORR & $\circ$ & $\circ$ \\ \hline
                                     ATLAS\_r15\_Luminosity\_bbbb &         & $\circ$ \\ \hline
                                     ATLAS\_r16\_Luminosity\_bbbb &         & $\circ$ \\
\hline

\end{tabular}
}
}
\label{tab:correlation1}
\end{table}

\begin{table}
\scalebox{0.9}{
\begin{tabular}{|l|c|c|}
\hline
NP name & single Higgs & double Higgs\\ 
\hline
\multicolumn{3}{|c|}{Experimental uncertainties}\\ \hline
\multicolumn{3}{|c|}{Electron and photon calibration uncertainties}\\ \hline
ATLAS\_EG\_RESOLUTION\_ALL & $\circ$ & $\circ$ \\ \hline
ATLAS\_EG\_RESOLUTION\_MATERIALCALO & $\circ$ &         \\ \hline
ATLAS\_EG\_RESOLUTION\_MATERIALCRYO & $\circ$ &         \\ \hline
ATLAS\_EG\_RESOLUTION\_MATERIALGAP & $\circ$ &         \\ \hline
ATLAS\_EG\_RESOLUTION\_MATERIALIBL & $\circ$ &         \\
\hline
\end{tabular}
}
\label{tab:correlation4}
\end{table} 

\begin{table}
\begin{center}
\scalebox{0.75}{
{\def\arraystretch{1.1}
\begin{tabular}{|l|c|c|}
\hline
NP name & single Higgs & double Higgs\\ 
\hline
\multicolumn{3}{|c|}{Experimental uncertainties}\\ \hline
\multicolumn{3}{|c|}{Muon uncertainties}\\ \hline
                          ATLAS\_MUON\_EFF\_ISO\_STAT\_Rel21 & $\circ$ &         \\ \hline
                          ATLAS\_MUON\_EFF\_ISO\_SYS\_Rel21 & $\circ$ &         \\ \hline
                          ATLAS\_MUON\_EFF\_RECO\_STAT\_LOWPT\_Rel21 & $\circ$ &         \\ \hline
                            ATLAS\_MUON\_EFF\_RECO\_STAT\_Rel21& $\circ$ &         \\ \hline
                           ATLAS\_MUON\_EFF\_RECO\_SYS\_LOWPT\_Rel21 & $\circ$ &         \\
\hline
\end{tabular}}
}
\end{center}
\label{tab:correlation6}
\end{table} 
\begin{small}
\begin{longtable}{|l|c|c|}
\hline
NP name & single Higgs & double Higgs\\ \hline
\multicolumn{3}{|c|}{Experimental uncertainties}\\ \hline
\multicolumn{3}{|c|}{Jet uncertainties }\\ 
\endfirsthead
\hline
NP name & single Higgs & double Higgs\\ \hline
\multicolumn{3}{|c|}{Experimental uncertainties}\\ \hline
\multicolumn{3}{|c|}{Jet uncertainties }\\ 
\endhead 
\endlastfoot 
\hline

                                                   ATLAS\_BJES & $\circ$ &         \\ \hline
                                       ATLAS\_JER\_CR2\_3j\_dl\_ttHbb & $\circ$ &         \\ \hline
                                ATLAS\_JER\_CROSSCALIBFWD\_Htautau & $\circ$ &         \\ \hline
                                ATLAS\_JER\_NOISEFWD\_Htautau & $\circ$ &         \\ \hline
                                ATLAS\_JER\_NP\_0\_Htautau & $\circ$ &         \\ \hline
                                ATLAS\_JER\_NP\_1\_Htautau & $\circ$ &         \\ \hline
                                ATLAS\_JER\_NP\_2\_Htautau & $\circ$ &         \\ \hline
                                ATLAS\_JER\_NP\_3\_Htautau & $\circ$ &         \\
\hline
                                
\end{longtable} 
\end{small}

\begin{small}
\begin{longtable}{|l|c|c|}
\hline
NP name & single Higgs & double Higgs\\ \hline
\multicolumn{3}{|c|}{Experimental uncertainties}\\ \hline
\multicolumn{3}{|c|}{Flavour tagging uncertainties }\\
\endfirsthead
\hline
NP name & single Higgs & double Higgs\\ \hline
\multicolumn{3}{|c|}{Experimental uncertainties}\\ \hline
\multicolumn{3}{|c|}{Flavour tagging uncertainties }\\ 
\endhead 
\endlastfoot 
\hline

                                                 ATLAS\_FTAG\_B0 & $\circ$ &         \\ \hline
                                                 ATLAS\_FTAG\_B1 & $\circ$ &         \\ \hline
                                                ATLAS\_FTAG\_B10 & $\circ$ &         \\ \hline
                                                ATLAS\_FTAG\_B11 & $\circ$ &         \\ \hline
                                                ATLAS\_FTAG\_B12 & $\circ$ &         \\ \hline
                                                ATLAS\_FTAG\_B13 & $\circ$ &         \\ \hline
                                                ATLAS\_FTAG\_B14 & $\circ$ &         \\ \hline
                                                ATLAS\_FTAG\_B15 & $\circ$ &         \\
\hline

\end{longtable} 
\end{small}

\begin{longtable}{|l|c|c|}
\hline
NP name & single Higgs & double Higgs\\ \hline
\multicolumn{3}{|c|}{Theoretical uncertainties}\\ \hline
\multicolumn{3}{|c|}{Signal uncertainties}\\
\endfirsthead
\hline
NP name & single Higgs & double Higgs\\ \hline
\multicolumn{3}{|c|}{Theoretical uncertainties}\\ \hline
\multicolumn{3}{|c|}{Signal uncertainties}\\ 
\endhead 
\endlastfoot 
\hline
                        TheorySig\_BR\_HiggsDecayWidthTHU\_hVV & $\circ$ & $\circ$ \\ \hline
                        TheorySig\_BR\_HiggsDecayWidthTHU\_hbb & $\circ$ & $\circ$ \\ \hline
                        TheorySig\_BR\_HiggsDecayWidthTHU\_hgg & $\circ$ & $\circ$ \\ \hline
                        TheorySig\_BR\_HiggsDecayWidthTHU\_htautau & $\circ$ & $\circ$ \\ \hline
                        TheorySig\_BR\_HiggsDecayWidthTHU\_hyy & $\circ$ & $\circ$ \\ \hline
                        TheorySig\_BR\_Others\_ttHML & $\circ$ &         \\ \hline
                        TheorySig\_BR\_Others\_ttHbb & $\circ$ &         \\ \hline
                        TheorySig\_BR\_param\_mB & $\circ$ & $\circ$ \\
\hline
\end{longtable}

\end{document}